\documentclass[12pt,a4paper,fleqn,notitlepage]{report}

\RequirePackage[l2tabu, orthodox]{nag}

\usepackage[top=20mm,bottom=20mm,left=25mm,right=25mm]{geometry}

\usepackage{amsmath, amssymb, amsfonts, amscd, latexsym, empheq}

\usepackage{esint}

\usepackage[fleqn]{nccmath}

\usepackage{makeidx}
\makeindex
\usepackage[nottoc]{tocbibind}

\usepackage{amsthm}
\usepackage{thmtools}

\theoremstyle{definition}
\newtheorem*{theorem}{Theorem}

\newtheoremstyle{break}{9pt}{9pt}{\itshape}{}{\bfseries}{}{\newline}{}
\theoremstyle{break}    
\newtheorem{exo}{Exercise}[chapter]

\makeatletter
\def\ll@exo{%
  \protect\numberline{\csname the\thmt@envname\endcsname}%
  \ifx\@empty\thmt@shortoptarg
    \thmt@thmname
  \else
    \thmt@shortoptarg
  \fi}
\def\l@thmt@exo{} 
\makeatother

\usepackage{tikz}
\usetikzlibrary{knots}
\usetikzlibrary{calc}
\usetikzlibrary{cd}
\usetikzlibrary{decorations.pathreplacing} 
\usetikzlibrary{decorations.markings}
\usetikzlibrary{arrows}
\usetikzlibrary{arrows.meta}
\tikzset{->-/.style={decoration={
  markings,
  mark=at position .6 with {\arrow{latex}}},postaction={decorate}}}
 
\usepackage[vcentermath]{youngtab}

\usepackage{ifthen}

\numberwithin{equation}{section}

\usepackage[colorlinks=true, linktoc=all, linkcolor=black, citecolor=red, urlcolor=green!50!black, backref=page]{hyperref}

\usepackage{etoolbox}
\makeatletter
\patchcmd{\BR@backref}{\newblock}{\newblock(}{}{}
\patchcmd{\BR@backref}{\par}{)\par}{}{}
\makeatother

\DeclareMathOperator*{\res}{Res}
\DeclareMathOperator{\Tr}{Tr}

\title{\bfseries Random matrices}

\author{by Bertrand Eynard, Taro Kimura and Sylvain Ribault,
\vspace{2mm}
\\ based on lectures by Bertrand Eynard at IPhT, Saclay
\vspace{2mm}
\\ \small \texttt{bertrand.eynard@ipht.fr, taro.kimura@ube.fr, sylvain.ribault@ipht.fr}
}

\begin{document}

\maketitle

\begin{abstract}
We provide a self-contained introduction to random matrices. 
While some applications are mentioned, our main emphasis is on three different approaches to random matrix models: the Coulomb gas method and its interpretation in terms of algebraic geometry, loop equations and their solution using topological recursion, orthogonal polynomials and their relation with integrable systems.
Each approach provides its own definition of the spectral curve, a geometric object which encodes all the properties of a model.
We also introduce the two peripheral subjects of counting polygonal surfaces, and computing angular integrals.
\end{abstract}

\begin{center}
 \textbf{Keywords}
\end{center}
\noindent Random matrix, eigenvalue distribution, ribbon graph, large size limit, random surface, spectral curve, loop equations, topological recursion, orthogonal polynomials, integrable systems
\vspace{1cm}

\tableofcontents

\hypersetup{linkcolor=blue}

\addtocounter{chapter}{-1}

\chapter{Preliminaries}

\section{Scope}

% NB: Say that our algebraic / geometric approach differs from the probabilistic approach of the rest of the literature? We use analyticity of integrands to full extent. Normal matrices are fundamental.

This text is based on a series of six lectures given at IPhT Saclay in early 2015, which amounted to an introduction to random matrices. 
Notes taken during the lectures were then completed and expanded, with the aim of introducing some of the most important approaches to random matrices. 
Still, this text is far from being a survey of the whole topic of random matrices, as entire areas of activity are left out. 
Some of the missing areas are covered in the following books or articles:
\begin{itemize}

 \item  \cite{Mehta:2004RMT} M. L. Mehta, {\em Random Matrices}

The bible of random matrices, one of the first books on the topic, it covers orthogonal polynomials very well.

 \item \cite{Bleher:2001random} P. Bleher and A. Its, eds.,
       {\em Random Matrix Models and their Applications}

This book mostly deals with the large $N$ asymptotic analysis, using the Riemann--Hilbert method.       
      
 \item \cite{Anderson:2009RMT} G. W. Anderson, A. Guionnet, and O. Zeitouni, {\em An Introduction to Random Matrices}
       
This book deals with probabilistic methods, such as large deviations, concentration, and diffusion equations.              

 \item \cite{Forrester:2010log} P. J. Forrester,
       {\em Log-Gases and Random Matrices}

This book deals with the Coulomb gas approach, which leads to many interesting results, in particular formulas involving orthogonal or Jack polynomials.             

 \item \cite{Akemann:2011RMT} G. Akemann, J. Baik, and P. Di Francesco, eds.,
       {\em The Oxford Handbook of Random Matrix Theory}

This is a topical review book, collecting contributions from many authors in random matrix theory and their applications in mathematics and physics.              

 \item \cite{Harnad:2011RM} J. Harnad, ed.,
       {\em Random Matrices, Random Processes and Integrable Systems}

This book focuses on the relationships of random matrices with integrable systems, fermion gases, and Grassmannians. 

\item \cite{Kazakov:1996} V. Kazakov, M. Staudacher and T. Wynter,
       {\em Character expansion methods for matrix models of dually weighted
	graphs}

This article introduces the character expansion method.

\item \cite{Efetov:1996SUSY} K. Efetov, {\em Supersymmetry in Disorder and Chaos}

This book on condensed-matter physics introduces the supersymmetric approach to random matrices. 

\item \cite{Tao:2010CMP} T. Tao and V. Vu, {\em Random matrices: Universality of local eigenvalue statistics up to the edge}

This article %book
presents a mathematical treatment of the universality of eigenvalue statistics.

\item \cite{Brezin:2016eax}  E. Br\'ezin and S. Hikami, {\em Random Matrix Theory with an External Source}

This book focuses on the Gaussian random matrix model with a deterministic external matrix source, and its applications to moduli spaces of curves.
\end{itemize}
% NB: In the BIB file there is also a book by Deift.
% NB: a few more math books
% \cite{Bai:2010,Pastur:2011,Tao:2012,Erdoos:2017}
The present text is intended for theoretical physicists and mathematicians. 
Graduate-level knowledge of linear algebra and complex analysis on the plane is assumed. 
Other mathematical subjects that appear, and on which some background would be welcome although not strictly necessary, are Lie groups, Riemann surfaces, and basic combinatorics.   

\section{Plan of the text}

The introductory \autoref{chap:intro} explains why it is useful and interesting to study random matrices, introduces the fundamental definitions of the random matrix ensembles, and describes  
important universal features of random matrix models. 
These features are given without proofs, not only for the sake of conciseness, but also because they have several different proofs, depending on which approach is used. 

The remaining chapters are rather independent, and can be read separately.

Chapters \ref{chap:SaddlePoint}-\ref{chap:ortho} introduce three different approaches to random matrices: the Coulomb gas method with its reliance on algebraic geometry, the loop equations and the powerful topological recursion technique, and the orthogonal polynomials with their relation to integrable systems.
Each approach will provide its own definition of the spectral curve -- the fundamental geometrical object which characterizes a matrix model. 

The two outlier chapters are devoted to specific topics.
In \autoref{chap:FeynmanGraph}, matrix integrals are considered as formal integrals, and used for studying the combinatorics of graphs. 
In \autoref{chap:AngularIntegral}, altogether different types of matrix models are studied, where there is no invariance under conjugation. 

The Bibliography is relatively short, as the cited works are selected for their potential usefulness to today's readers, leaving out many works of historical importance. 
The Index refers to the definitions or first occurrences of the listed terms. 
A few terms have several different definitions. 
We have often refrained from defining standard terms, in particular when adequate definitions are found in Wikipedia.

\section{Highlighted topics}

%NB Split and merge rules?

This text includes a number of ideas and results that, while not new, are often underappreciated or not as well-known as they would deserve, in particular:
\begin{itemize}
 \item Normal matrix ensembles include both Gaussian and circular ensembles as particular cases. (See \autoref{sectionnormaloncontour}.)
 \item There is a fundamental distinction between convergent matrix integrals and formal matrix integrals. 
 For the combinatorics of polygonal surfaces, we need formal integrals, not convergent integrals. (See \autoref{sec:fmi}.) 
 \item For a given potential, the space of convergent normal matrix integrals and the space of solutions of the loop equations have the same dimension, which is given by a simple formula \eqref{eq:dimh1}.
 \item The duality $\beta \leftrightarrow \frac{4}{\beta}$, which also acts nontrivially on the matrix size $N$, leaves the loop equations invariant. (See \autoref{sec:loop}.)
 \item Loop equations can be solved perturbatively using topological recursion. (See \autoref{sec:toprec}.) 
 Non-perturbative contributions to matrix integrals can also be computed systematically, and are also universal. (See \autoref{sec:non-perturbative}.)
 \item Not only orthogonal polynomials, but also their Hilbert transforms, and the corresponding self-reproducing kernel, have expressions as minors of infinite matrices. 
 These quantities moreover have graphical representations in terms of Motzkin paths.
 (See \autoref{sec:ortho}.)
 \item Random matrix models lead to integrable systems with polynomial spectral curves, which are therefore both simple and capable of approximating arbitrary curves. As prototypes of integrable systems, matrix models are therefore superior to spin chains. (See \autoref{sec:ais}.)
 \item Itzykson--Zuber integrals obey Calogero--Moser, duality and recursion equations, and can be expressed in terms of Jack polynomials. (See \autoref{sec:izi}.)
\end{itemize}

\section{Acknowledgements}

We are grateful to IPhT, Saclay for hosting the lectures, and to the members of the audience for their interest and questions. 
We wish to thank Ga\"etan Borot for comments on the draft of this text.

The work of TK was supported in part by the French ``Investissements d'Avenir'' program, Project ISITE-BFC (No.~ANR-15-IDEX-0003), and EIPHI Graduate School (No.~ANR-17-EURE-0002).

\chapter{Introduction} \label{chap:intro}

In this chapter, we define and motivate the various random matrix ensembles that we will study in detail in the other chapters.
In particular, we define the integration measures, partition functions and expectation values that are associated to these ensembles.
We then describe some essential properties of these objects, in particular their behaviour in the limit of large matrix size, and their universal properties.
Proofs and precise statements of these properties are postponed to later chapters.

\section{Motivations, history, and applications}

Matrices are everywhere in physics and mathematics -- they appear as soon as there are linear relations between variables. 
Randomness is also everywhere. 
So it is only natural that random matrices should arise in many places. 
Random matrices have however originally been introduced in order to deal with rather specific problems.

\subsection{First appearances of random matrices in mathematics}

Random matrices first appeared in statistics, when Wishart~\cite{Wishart:1928Bio}
generalized the chi-squared law to multivariate random data. 
Let us consider $N\geq p$ samples of a random  $p$-component vector, in the form of a rectangular matrix $X$ made of $N$ columns of size $p$,
\begin{align}
 X = \left( X_1 \ X_2 \ \cdots \ X_N \right)
 \, , \qquad
 X_i = \left( x_{i1} \ x_{i2} \ \cdots \ x_{ip} \right)^\mathrm{ T}
 \, , \qquad x_{ij}\in\mathbb{R}\ .
\end{align}
The $X_i$s are independent, identically distributed (i.i.d.) random vectors with a
multivariate normal distribution $N_p(0,V)$ with zero mean and a covariance matrix $V$ of size $p$:
\begin{equation}
N_p(0,V)(X_i) = \frac{e^{- \frac12 \operatorname{Tr} X_i^\mathrm{ T} V^{-1} X_i}}{(2\pi)^{p/2} \sqrt{\det V}}  \prod_{j=1}^p dx_{ij} .
\end{equation}
We are interested in the properties of the correlation matrix  
\begin{equation}
 M =  X X^\mathrm{ T} \, ,
\end{equation}
which is a symmetric square matrix of size $p$.
%We assume that $X$ is a random matrix, such that each row obeys a 
%$p$-dimensional
%multivariate normal distribution $N_p(0,V)$ with zero mean and a covariance matrix $V$ of size $p$.
Then the probability law for the correlation matrix $M$ is given by the
\textbf{Wishart distribution}\index{Wishart distribution}, 
\begin{align}
 P(M)dM  = \frac{1}{2^{\frac{Np}{2}}(\det V)^{\frac{N}{2}}\Gamma_p(\frac{N}{2})}
 \left( \det M \right)^{\frac{N-p-1}{2}}
 e^{-\frac{1}{2} \operatorname{Tr}V^{-1}M}
 \prod_{i,j=1}^p dM_{ij} \, .
\end{align}
After this work by Wishart on real, symmetric matrices, Ginibre studied other matrix ensembles in the
1960s, not necessarily assuming reality or even Hermiticity~\cite{Ginibre:1965zz}.
(See Sections~\ref{sec:other_examples}, \ref{sec:non-Hermitian_oriented}, \ref{sec:real_quaternionic_ribbons} and \ref{sec:cmplx_harmonic}.)

Later, in 1967, Marchenko
and Pastur~\cite{Marchenko:1967MS} studied the large $N$ behaviour of the size $p$ matrix $M=\frac{1}{N}X X^\mathrm{ T} $, where the entries of $X$ are i.i.d. variables with a normal distribution $N(0,\sigma^2)$ (i.e. the case $V=\sigma^2\text{Id}$ of Wishart). %independent, identically distributed 
In the limit $N \to \infty$ with $u= \frac{p}{N}\leq 1$ fixed, the eigenvalue density of $M$ converges to the \textbf{Marchenko--Pastur law}\index{Marchenko--Pastur law},
\begin{align}
 \bar{\rho}(x) \propto \frac{1}{2\pi u\sigma^2}
 \frac{\sqrt{(a_+-x)(x-a_-)}}{x}  \  \, ,
 \qquad
 a_\pm = \sigma^2 (1\pm \sqrt{u})^2 \ ,
 \label{MP-law}
\end{align}
whose graph is
 
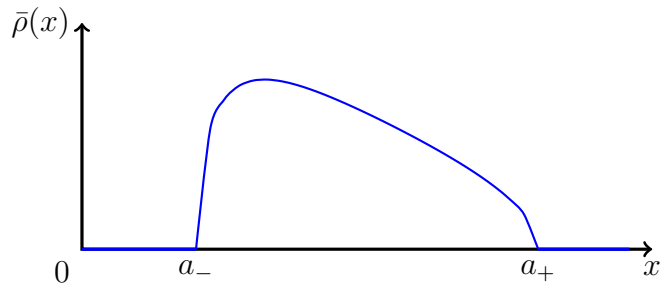
\begin{align}
  \begin{tikzpicture}[scale = 1.5,baseline=(current  bounding  box.center)]
\draw[very thick, latex-latex] (0, 2) node[left]{$\bar{\rho}(x)$} -- (0, 0) node[below left]{$0$} -- (5, 0) node[below]{$x$};
\draw[domain=1:4,smooth,variable=\x,blue, thick] 
   plot({\x}, {2*sqrt( (1 - \x) * (\x - 4))/\x} );
   \draw (1,0) node [below] {$a_-$};
   \draw (4,0) node [below] {$a_+$};
   \draw[very thick, blue] (0,0) -- (1,0);
   \draw[very thick, blue] (4,0) -- (4.8,0);
 \end{tikzpicture}
\end{align}
The eigenvalues are concentrated in a compact interval whose bounds $a_-, a_+$ are called the spectral edges.

\subsection{Heavy nuclei and Wigner's idea}

In the 1950s, Wigner was interested in the energy spectrums of heavy nuclei such as thorium and uranium.
While theoretically much more complicated than the hydrogen nucleus, these nuclei are accessible to neutron scattering experiments. 
The nuclei have large numbers of energy levels, which appear in experimental data as peaks of the diffusion rate of neutrons as a function of the energy:
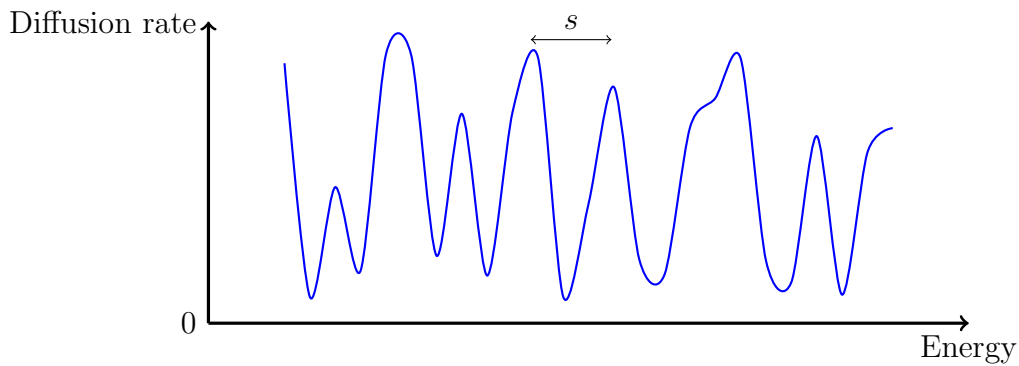
\begin{align}
\begin{tikzpicture}[
    baseline=(current  bounding  box.center),
    scale=2,
    axis/.style={very thick, ->},
    ]
\draw[axis] (0,0) node[left]{0} -- (5,0) node[below]{Energy}; 
\draw[axis] (0,0) -- (0,2) node[left]{Diffusion rate};
\draw[latex-latex] (2.12,1.88) -- node[above]{$s$} (2.65,1.88) ;
\draw[domain=.5:4.5,smooth,variable=\x,blue, thick] plot ({\x},{(1+.8*sin((9*\x+5*sin(5*\x r)) r)+.07*sin(7*\x r))});
\end{tikzpicture}
\end{align}
The interesting observable which Wigner studied is the statistical
distribution of the distance $s$ between neighbouring energy levels. If
the energy levels were uncorrelated random numbers, the variable $s$
would be governed by the \textbf{Poisson distribution}\index{Poisson
distribution}, with probability density
\begin{equation}
 P(s) = e^{-s} \, .
\end{equation}
But the experimentally observed probability density looks quite different, and is 
very well approximated (within 1\%) by the \textbf{Wigner surmise}\index{Wigner surmise}
\begin{equation}
 P(s) = C_\beta s^\beta e^{-a_\beta s^2} \, ,
  \label{W_distWignerbetasurmise}
\end{equation}
where the parameter $\beta \in \{1, 2, 4\}$ is determined by the nucleus' symmetries under time reversal and spin rotation, and the values of $C_\beta$ and $a_\beta$ are such that $\int_0^\infty ds \, P(s) = \int_0^\infty ds \, s \, P(s) = 1$.
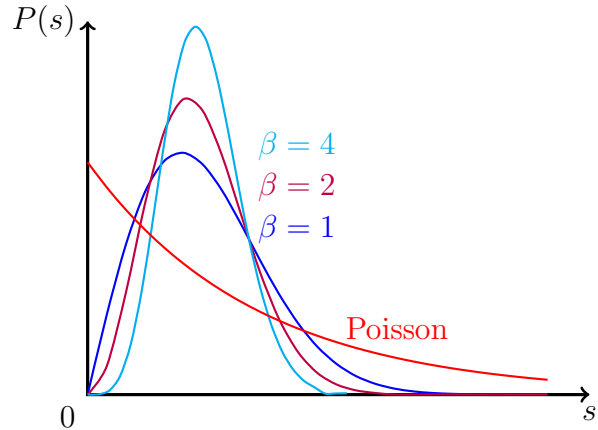
\begin{align}
 \begin{tikzpicture}[scale = 1.1,baseline=(current  bounding  box.center)]
\draw[very thick, latex-latex] (0, 4.5) node[left]{$P(s)$} -- (0, 0) node[below left]{$0$} -- (6, 0) node[below]{$s$};
\draw[domain=0:5.5,smooth,variable=\x,blue, thick] plot({\x},{4.3*\x *exp(-.4*\x *\x)}) ;
\draw (2.5, 2) node[blue]{$\beta=1$};
\draw[domain=0:5.5,smooth,variable=\x,purple, thick] plot({\x},{6.8*\x^2 *exp(-.7*\x ^2)}) ;
\draw (2.5, 2.5) node[purple]{$\beta=2$};
\draw[domain=0:3.1,smooth,variable=\x,cyan, thick] plot({\x},{11.8*\x^4 *exp(-1.2*\x ^2)}) ;
\draw (2.5, 3) node[cyan]{$\beta=4$};
\draw[domain=0:5.5,smooth,variable=\x,red, thick] plot({\x},{2.8*exp(-.5*\x)});
\draw (3.7, .8) node[red]{Poisson};
 \end{tikzpicture}
\end{align}
The Wigner surmise is also a very good approximation of the large size limit of the probability density for the distance between consecutive eigenvalues of random matrices.
Actually, this density is universal, to the extent that it does not depend on the precise measure of the random matrices, but only on the choice of a matrix ensemble, with real, complex and quaternionic matrices leading to $\beta=1,2,4$ respectively. (See \autoref{sec:laws}.)
This shows that complicated Hamiltonians can be accurately modelled by random matrices.

On the other hand, the large size limit of the eigenvalue density (as opposed to their distances) is not universal.
In the case of Gaussian random matrices, it is given by the \textbf{Wigner semi-circle law}\index{Wigner semi-circle law},
\begin{equation}
 \bar{\rho}(x) = 
  \begin{cases}
   \frac{1}{2\pi} \sqrt{4 - x^2} & (|x| \le 2) \\
   0 & (|x| > 2)
  \end{cases}
  \, .
  \label{sc_law}
\end{equation}
whose name comes from the shape of its graph,
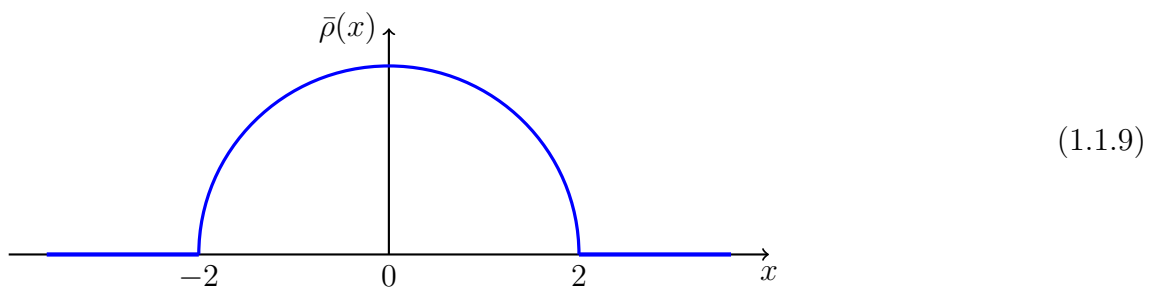
\begin{equation}
 \begin{tikzpicture}[baseline=(current  bounding  box.center)]
  \draw[thick, -latex] (-5,0) -- (5,0) node[below]{$x$};
  \draw[thick, -latex] (0,0) node[below]{$0$} -- (0,3) node[left]{$\bar{\rho}(x)$};
  \draw[very thick, blue] (2.5,0) arc[radius = 2.5, start angle = 0, end angle = 180];
  \draw[ultra thick, blue] (-4.5, 0) -- (-2.5, 0);
  \draw[ultra thick, blue] (2.5, 0) -- (4.5, 0);
  \draw (2.5, 0) node[below]{$2$};
  \draw (-2.5, 0) node[below]{$-2$};
 \end{tikzpicture}
\end{equation}
For other non--Gaussian random matrix measures, the limit eigenvalue density $\bar{\rho}(x)$ is often an algebraic function and is supported on finite intervals, as we shall see in \autoref{chap:SaddlePoint}.

\subsection{Quantum chaos in billiards}

Besides heavy nuclei, other examples of systems whose Hamiltonians have very complicated spectrums are provided by quantum mechanics of a particle in 2d chaotic billiards. 
The particle is free, its only interaction is being reflected by the boundary.
That a billiard is chaotic can already be seen at the level of classical trajectories, which are then ergodic. 
Billiards with rather simple shapes can be chaotic: this is the case with  Sinai's billiard, and stadium-shaped (rectangle + 2 half discs) billiards:
\begin{align}
 \begin{tikzpicture}[baseline=(current bounding box.center)]
\draw[thick] (-2,-2) rectangle (2,2);
\draw[thick] (0,0) circle [radius = .8];
\draw[blue, -latex] (1.7, .5) -- (1.1, 1.98) -- (.55, .6) -- (2, 1.4) -- (.9, 1.98) -- (-1.1, .8);
  \draw (0,-2.5) node [below] {Sinai};
  \draw[thick] (5,1.5) -- + (3,0);
  \draw[thick] (5,-1.5) -- + (3,0);
  \draw[thick] (8,1.5) arc (90:-90:1.5);
  \draw[thick] (5,1.5) arc (90:270:1.5);
  \draw (6.5,-2.5) node [below] {Stadium};
  \draw[blue,-latex] (7,1) -- ++(-10:2.4) -- ++(260:1.7) -- ++(190:1.6);
 \end{tikzpicture}
\end{align}
If a billiard is classically integrable (for example a rectangle or an ellipse), then the spacings between eigenvalues of the corresponding quantum Hamiltonian follow a Poisson law, and there are many level crossings. 
If it is chaotic, then the eigenvalues have the same statistical properties as eigenvalues of random matrices, and they don't cross, ``repelling'' each other instead.
This can be seen in plots of energy levels as functions of a parameter such as a magnetic field $B$:
\begin{align}
\begin{tikzpicture}[scale = .7,
declare function = {f1(\x) = 1+ .4*cos(\x + .1*\x^2 + 2*sin(1.2*\x r) r) + .4* cos(.62*\x + 1/(1+\x) r) ;},
declare function = {f2(\x) = .6*(.4+ sin((.6+1.2*\x) r)^2) ;},
declare function = {f3(\x) = .7*(.3+ cos((.9*\x + 1/(2+\x)) r)^2);},
baseline=(current bounding box.center)
]
\draw[thick, latex-latex] (0,6) node[left]{$E$} -- (0,0) -- (6,0) node[below]{$B$};
\draw[thick, latex-latex] (9,6) node[left]{$E$} -- (9,0) -- (15,0) node[below]{$B$};
\node at (3,-1.5){Integrable billiard};
\node at (12,-1.5){Chaotic billiard};
\foreach \y in {1,2,3}
\foreach \z in {1,2,3,4}
\draw[domain=0:6,smooth,variable=\x,blue] plot (\x,{\y + .1*\z*\x*exp(.06*\x)});
\draw[domain=9:15,smooth,variable=\x,blue] plot (\x,{f1(9+\x)});
\draw[domain=9:15,smooth,variable=\x,blue] plot (\x,{f1(9+\x)+f2(9+\x)});
\draw[domain=9:15,smooth,variable=\x,blue] plot (\x,{.6+f1(9+\x)+f3(8+\x)});
\draw[domain=9:15,smooth,variable=\x,blue] plot (\x,{.7+ f1(5+\x)+f3(8+\x)+f2(2+\x)});
\draw[domain=9:15,smooth,variable=\x,blue] plot (\x,{1.55+ f1(2+\x)+f3(4+\x)+f2(2+\x)});
\draw[domain=9:15,smooth,variable=\x,blue] plot (\x,{1.55+ f1(2+\x)+f3(4+\x)+f2(2+\x) + .2+ (f1(\x)-1)^2});
\end{tikzpicture}
\end{align}
The appearance of random matrix statistics was first observed in experiments of Berry and Tabor~\cite{Berry:1977PRS}, and then raised to a major quantum chaos conjecture by Bohigas, Giannoni, and Schmit in 1984~\cite{Bohigas:1984PRL}. 
It has been so well established experimentally, that random matrix statistics are considered as experimental signatures of quantum chaos.
However, this relation between quantum chaos and random matrices is still not really understood theoretically.

\subsection{Quantum chromodynamics and random surfaces} \label{ssec:qcd}

Quantum chromodynamics (QCD) is the theory of strong nuclear interactions. It is a gauge quantum field theory, whose gauge field, which describes gluons, has the structure of a unitary matrix of size $N_c$. 
The integer $N_c$ is called the number of colors and is set to $N_c=3$ in QCD. 
However, 't Hooft proposed to study a generalization of QCD in the limit
$N_c \rightarrow \infty$~\cite{'tHooft:1973jz}, where important
simplifications occur, and then recover $N_c=3$ by perturbation
theory. 
Moreover, the $N_c\times N_c$ matrix is random, because of
quantum mechanical dynamics.
So, the $N_c \rightarrow \infty$ generalization of QCD is a theory of large random matrices.

In Feynman graphs of gluons, the matrix structure can be represented by having double lines in propagators, where each line carries one of the two indices of the matrix:
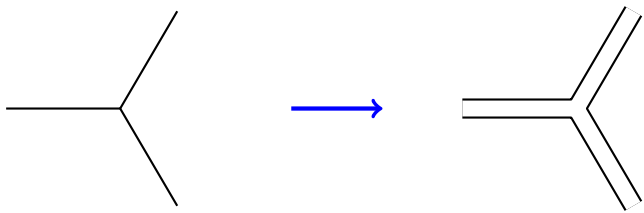
\begin{align}
 \begin{tikzpicture}[scale = 1.5,baseline=(current bounding box.center)]
  \draw[thick] (-1,0) -- (0,0) -- (.5, .86);
  \draw[thick] (0,0) -- (.5, -.86);
  \draw[ultra thick,blue,-latex] (1.5,0) -- (2.3,0);
  \draw [thick,double,double distance=6pt]
  plot coordinates{(3,0) (4,0) (4.5, .86)}
  plot coordinates{(4,0) (4.5, -.86)};
 \end{tikzpicture}
\end{align}
Such a graph with double lines is called a \textbf{ribbon graph}\index{ribbon graph}.
A trivalent
ribbon graph is dual to a triangulation of a surface:
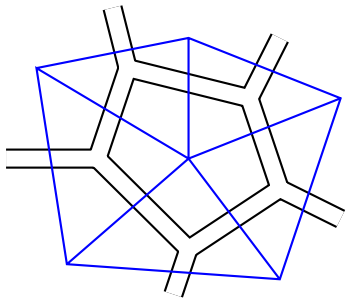
\begin{align}
 \begin{tikzpicture}[scale = .4,baseline=(current bounding box.center)]
  \draw [thick,double,double distance=6pt]
  plot coordinates{(0,0)(1,3)(5,2)(6,-1)(3,-3)(0,0)}
  plot coordinates{(0,0)(-3,0)}
  plot coordinates{(1,3)(.5,5)}
  plot coordinates{(5,2)(6,4)}
  plot coordinates{(6,-1)(8,-2)}
  plot coordinates{(3,-3)(2.5,-4.5)}
  ;
  \draw [thick,blue]
  plot coordinates{(8,2)(3,0)(-1,-3.5)}
  plot coordinates{(3,0)(6,-4)}
  plot coordinates{(-2,3)(3,4)(8,2)(6,-4)(-1,-3.5)(-2,3)(3,0)(3,4)};
 \end{tikzpicture}
\end{align}
QCD probability amplitudes can be reformulated as sums over ribbon graphs, that is dynamics of surfaces.
It was initially believed that this surface dynamics would be the expected missing link between QCD and string theory, but it turned out to be more complicated.
However, 't Hooft's approach was the inspiration for studying random surfaces using random matrices.
This was initiated by Br\'ezin, Itzykson, Parisi and Zuber~\cite{Brezin:1977sv}, and led to the random matrix theory of 2d quantum gravity in the 1980s and 1990s. 

Another link between QCD and random matrices was found by considering the Dirac operator (instead of the gauge field) as a  large random matrix. This led to a random matrix model with additional chiral symmetry, called the chiral random matrix, which provides a good approximation of Dirac operator spectrums at low energy~\cite{Verbaarschot:2000dy}.

\subsection{Other applications}

\begin{enumerate}
 \item \textbf{Transport in disordered systems:} 
 A mesoscopic conductor is at the same time macroscopic in that it has a large number of states $N$, and microscopic (at low temperatures) in that its size is smaller than the quantum decoherence length.
 It follows that quantum effects are relevant, so that the conductance doesn't obey Ohm's law.
 The conductor scatters incoming electronic states $i$ (from the left) or $o'$ (from the right) into outgoing states $i'$ and $o$:
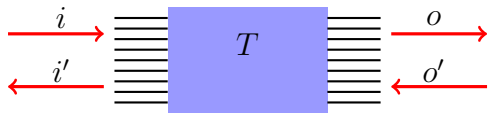
 \begin{align}
  \begin{tikzpicture}[baseline=(current  bounding  box.center),thick,scale=.7]
  \filldraw [fill=blue,opacity=0.4,draw=none] (-1.5,-1) rectangle (1.5,1);  
  \draw (-2.5,-.8) -> (-1.5,-.8);
  \draw (-2.5,-.6) -> (-1.5,-.6);
  \draw (-2.5,-.4) -> (-1.5,-.4);
  \draw (-2.5,-.2) -> (-1.5,-.2);
  \draw (-2.5,0) -> (-1.5,0);
  \draw (-2.5,.8) -> (-1.5,.8);
  \draw (-2.5,.6) -> (-1.5,.6);
  \draw (-2.5,.4) -> (-1.5,.4);
  \draw (-2.5,.2) -> (-1.5,.2);
  \draw (2.5,-.8) -> (1.5,-.8);
  \draw (2.5,-.6) -> (1.5,-.6);
  \draw (2.5,-.4) -> (1.5,-.4);
  \draw (2.5,-.2) -> (1.5,-.2);
  \draw (2.5,0) -> (1.5,0);
  \draw (2.5,.8) -> (1.5,.8);
  \draw (2.5,.6) -> (1.5,.6);
  \draw (2.5,.4) -> (1.5,.4);
  \draw (2.5,.2) -> (1.5,.2);
  \node at (0,.3) {$T$};
  \draw[->, very thick, red] (-4.5,.5) -> (-2.7,.5);
  \draw[latex-, very thick, red] (-4.5,-.5) -> (-2.7,-.5);
  \node at (-3.5,.8) {$i$};
  \node at (-3.5,-.2) {$i'$};
  \draw[latex-, very thick, red] (4.5,.5) -> (2.7,.5);
  \draw[->, very thick, red] (4.5,-.5) -> (2.7,-.5);
  \node at (3.5,.8) {$o$};
  \node at (3.5,-.2) {$o'$};
\end{tikzpicture}
 \end{align}
This scattering is described by a $2N\times 2N$ transfer matrix $T$ such that
\begin{align}
 \begin{pmatrix} o \cr o' \end{pmatrix}
= T  \begin{pmatrix} i \cr i' \end{pmatrix}\ .
\end{align}
The four blocks of $T$ can be written in terms of size $N$ reflection and transmission matrices $r,r',t,t'$ such that $o= t i + r' o'$ and $i'=r i + t' o'$. Current conservation $|o|^2-|o'|^2 = |i|^2-|i'|^2$ means $T\in SU({N,N})$. 
The conductance is $G
 =\frac{2 e^2}{h} \operatorname{Tr} t^\dagger t
=\frac{4 e^2}{h} \operatorname{Tr} \frac{T^\dagger T}{(1+T^\dagger T)^2}  $.

Describing $T$ as a random matrix in a subgroup of $SU({N,N})$ correctly explains the statistics of the conductance, which is related to the eigenvalues statistics of $T^\dagger T$. These eigenvalue statistics are described by a diffusion equation called the Dorokov--Mellow--Pereira--Kumar (DMPK) equation. 
Assembling conductors in series amounts to multiplying their transfer matrices, and an important problem is to determine the properties of spectrums of products of such random matrices.

Moreover, Efetov \cite{Efetov:1996SUSY} has developed a supersymmetric method for dealing with Gaussian random matrices in arbitrary ensembles, including non-compact ensembles. These ensembles then have applications to solid-state physics, and even a hope of describing high-temperature superconductors.

 \item \textbf{String theory:} 
 There are several ways to relate string theory to random matrices. All of them are rather indirect.
 Historically, the first relation came from the identification of the string worldsheet 
 with a random surface, through the BIPZ \cite{Brezin:1977sv} approach. 
 Unfortunately this approach was limited to a target space dimension $d\leq 1$, 
 thus unable to describe string theory in $4$ or $10$ or $26$ dimensions.
Then, in 2002, Dijkgraaf and Vafa~\cite{Dijkgraaf:2002fc} observed that the tree-level prepotential in some string theories is analogous to  a random matrix model partition function, which allowed them to perform some non-perturbative computations in string theory.

Another relation came from Kontsevich's work~\cite{Kontsevich:1992ti}, who used a matrix integral, now known as Kontsevich's integral, to prove Witten's conjecture on 2d quantum gravity: Kontsevich's method consisted in cutting the moduli space of Riemann surfaces into cells parametrized by ribbon graphs, identifying those ribbon graphs with the Feynman graph expansion of a random matrix as in BIPZ \cite{Brezin:1977sv}, and using the fact that a random matrix partition function is the Tau function of an integrable system as we will see in \autoref{chap:ortho}.

Recently, other relations were found between topological string theories
and matrix models. 
In particular, in 2008, the
Bouchard--Klemm--Mari\~{n}o--Pasquetti (BKMP)
conjecture~\cite{Bouchard:2007ys} (since then
proved~\cite{Eynard:2012nj}) claimed that topological string
amplitudes, which coincide with the Gromov--Witten invariants of the string's target space, obey the same recursion relations 
as matrix models' correlation functions.
These recursion relations will appear in \autoref{chap:LoopEq} under the name of topological recursion relations.

 \item \textbf{Knot theory:} 
 Knots are characterized by topological invariants such as Jones polynomials, Alexander polynomials, and HOMFLY polynomials. 
 Like the topological string amplitudes of the BKMP conjecture, these invariants are found to obey the same recursion relations as matrix integrals. 
 In the special case of torus knots, such as the trefoil knot, the HOMFLY polynomials are actually expressed as matrix integrals. (See \autoref{sec:gvi}.)
The Rosso--Jones formula and its generalizations can be recast in writing the Jones polynomial of a torus knot as a matrix integral~\cite{Brini:2011}. (See \autoref{exo:jones}.) 
% NB: SR: Removed definition of torus knots, as this is common knowledge, see Wikipedia. On the other
% hand, the Rosso-Jones formula is not so well-known, it would be nice to say what it is. Is it 
% an explicit formula for the Jones polynomial of a torus knot? 
 
 \item \textbf{Conformal field theory in two dimensions:}
 Penner-type random
matrix integrals~\cite{Penner:1988JDG} are formally identical to the Dotsenko--Fateev integrals of two-dimensional conformal field theory~\cite{Dotsenko:1984nm,Dotsenko:1984ad}.
This technical agreement has no deep explanation so far, but it implies that random matrix techniques can compute certain CFT correlation functions. 

 \item \textbf{Integrable systems:}
In a random matrix model, the eigenvalue distribution is very often related to the
Tau function of some integrable system.
The Tau function in question plays the role of the partition function of the system, and obeys important relations such as Hirota equations or Sato relations.

For example, the level spacing distribution of the Gaussian Unitary Ensemble is the Tau function of the Painlev\'e~V integrable system, also known as the Fredholm determinant of the sine kernel. The Tracy--Widom law is the Tau function of the Painlev\'e~II integrable system, also known as the Fredholm determinant of the Airy kernel.

Random matrix  models are good prototypes of integrable systems, in the sense that
many relations which can be derived in random matrix theory, actually extend to all integrable systems. This general principle has been extremely fruitful so far, and is one of the focuses of this text.
This is a strong motivation for studying universal structures in random
matrix models.
See \autoref{chap:ortho}.

 \item \textbf{Crystal growth:}
 A simple model of a crystal is a 3d partition, i.e. a pile of small cubic boxes. It has been experimentally and numerically observed that the statistics of cubes on a growing crystal obey random matrix laws.
 In fact, there is a matrix integral which coincides with the partition function of uniformly distributed 3d partitions in a box \cite{Eynard:2009nd}. Similar results hold for 2d partitions, and lead to the Logan--Shepp--Vershik--Kerov profile~\cite{Logan:1977AM,Vershik:1977SMD}. See \autoref{exo:plancherel}.
  
 \item \textbf{Hele--Shaw problem:}\index{Hele--Shaw problem}
 This is a model for the growth of a pocket of a non-viscous liquid (such as water) in a highly viscous liquid (such as oil) in two dimensions. 
 The problem is to predict the evolution of the shape of the pocket, knowing that the high viscosity keeps all its moments except the area fixed. Unexpectedly,  the Hele--Shaw problem was solved using random matrices~\cite{MineevWeinstein:2000ie,Kostov:2000ed}, and it was found that the pocket has the shape of the large size spectrum (thus a domain of $\mathbb C$) of a complex random matrix. (See \autoref{sec:cmmh}.)
 
 \item \textbf{Number theory:}
In 1973, in a discussion between  Montgomery and Dyson~\cite{Montgomery:1973ANT}, it was observed numerically that the distribution of the non-trivial zeros of the Riemann zeta function on the axis $\Re (s)=\frac{1}{2}$ agrees extraordinarily well with the eigenvalue distribution of a random matrix in the Circular Unitary Ensemble~\cite{Odlyzko:1987MC}.
A proof of this observation would not only imply the Riemann hypothesis, but also have deep implications for the statistical distribution of prime numbers.
This is considered a very promising approach in number theory.
 
 \item \textbf{Telecommunication:} Telecommunication signals can be modelled  as  random vectors, and a group of signals arriving at an antenna forms a large random matrix. 
 Real-time treatment of signals may mean real time inverting or diagonalizing such huge matrices. Random matrix theory is helpful for addressing this challenge. 
Moreover, one way to optimize transmission is to ensure that signal packages repel one another, in the same way as random matrix eigenvalues repel one another.
 
 \item \textbf{Biology, etc:}
Ribbon graphs are a simple model of RNA's secondary  structure.
Random matrix models, which generate ribbon graphs, can therefore help predict the shape of RNA. Some of the best RNA shape codes were developed using random matrix techniques~\cite{Orland:2001ba}.

Moreover, random matrix statistics were observed to describe the distributions of synapses in neurons, trees in a rainforest, waiting times between buses, prices in a stock exchange, correlations in large random networks, etc.

\end{enumerate}

\section{Random matrix models} \label{sec:laws}

% \subsection{Definition}

A random matrix is a matrix whose elements are randomly distributed.
A random matrix model is characterized by
\begin{itemize}
 \item a matrix ensemble $E$,
 \item an event set, which is the set of measurable subsets of $E$,
 \item a \textbf{measure}\index{measure (matrix model)} $d\mu(M)$ for $M\in E$, also called the random matrix law.
\end{itemize}
In most cases of interest, and in all what follows, $E$ is a closed subset of $M_N(\mathbb C)$. The event set is the Borel event set, i.e. the sigma algebra generated by open subsets of $E$ by taking unions, intersections, and complements.

The \textbf{partition function}\index{partition function} or matrix integral is then defined as 
\begin{align}
 \boxed{\mathcal{Z} = \int_E d\mu(M)}\ .
\end{align}
In general the measure is not normalized i.e. $\mathcal{Z}\neq 1$, and not positive. 
Nevertheless, it allows us to define the \textbf{expectation values}\index{expectation value} of measurable functions $f:E \to \mathbb C$, also called the \textbf{moments}\index{moment} of the measure $d\mu(M)$,
\begin{equation}
 \frac{1}{\mathcal Z} \int_{E} f(M) d\mu(M)
\overset{\mathrm{ physics}}{=} \langle f(M) \rangle 
\overset{\mathrm{ maths}}{=} \mathbb E\left(f(M)\right)\  .
\end{equation}
(We will adopt the physics notation.) 
The measure can be characterized by its moments, in other words by the linear form $f\mapsto \int_E  f(M)  d\mu(M)$ on an algebra of functions on $E$, for example the algebra of polynomials. With the Borel event set, any continuous function is measurable.

\subsection{Gaussian and circular ensembles}\label{sec:gce}

\subsubsection{Matrix ensembles}

The three matrix ensembles whose eigenvalue spacing distributions are approximated by the Wigner surmise are called the Gaussian ensembles or Wigner ensembles:
\begin{itemize}
 \item the \textbf{G}aussian \textbf{O}rthogonal \textbf{E}nsemble 
       \textbf{GOE}\index{Gaussian ensemble!GOE}
       of real symmetric matrices,
\item the \textbf{G}aussian \textbf{U}nitary \textbf{E}nsemble 
      \textbf{GUE}\index{Gaussian ensemble!GUE}
      of complex Hermitian matrices,
\item and the \textbf{G}aussian \textbf{S}ymplectic \textbf{E}nsemble 
      \textbf{GSE}\index{Gaussian ensemble!GSE}
      of quaternionic Hermitian matrices.
\end{itemize}
Eigenvalue spacing distributions happen to be universal, that is (almost) independent of the measure. 
So they were originally studied using Gaussian measures, and the ensembles were themselves called Gaussian. This is however a potentially confusing abuse of language, as these ensembles are also often considered with non-Gaussian measures.

A matrix $M$ in one of Wigner's three ensembles has real eigenvalues, and can be diagonalized as 
\begin{align}
 \boxed{ M = U\Lambda U^{-1} \quad \text{with} \quad \Lambda = \operatorname{diag}(\lambda_1, \cdots \lambda_N) \in \mathbb{R}^N }\ .
\label{eq:diagonalize}
 \end{align}
This is called the \textbf{angular-radial decomposition}\index{angular-radial decomposition}, where $\Lambda$ is the radial and $U$ the angular part. 
The matrix $U$ belongs to a compact Lie group, which we call the corresponding circular ensemble:
\begin{itemize}
 \item the \textbf{C}ircular \textbf{O}rthogonal \textbf{E}nsemble \textbf{COE}\index{Circular ensemble!COE} of real orthogonal matrices,
 \item the \textbf{C}ircular \textbf{U}nitary \textbf{E}nsemble \textbf{CUE}\index{Circular ensemble!CUE} of complex unitary matrices,
 \item the \textbf{C}ircular \textbf{S}ymplectic \textbf{E}nsemble \textbf{CSE}\index{Circular ensemble!CSE} of complex symplectic matrices.
\end{itemize}
We will denote the Gaussian ensembles as $E_N^\beta$, and the corresponding circular ensembles as $U_N^\beta$, with $\beta\in\{1, 2, 4\}$:
\begin{equation}
\renewcommand{\arraystretch}{1.3}
\text{
\begin{tabular}{|cc|cc|cc|} \hline \hline
 $\beta$ & Ensemble type & Gaussian ensemble & $E_N^\beta$ & Circular ensemble & $U_N^\beta$
 \\
 \hline
 $1$ & orthogonal & GOE & $S_N$ & COE & $O_N$
 \\
 $2$ & unitary & GUE & $H_N$ & CUE & $U_N$
 \\
 $4$ & symplectic & GSE & $Q_N$ & CSE & $Sp_{2N}$
\\ \hline \hline
 \end{tabular}
 }
\end{equation}
Each one of these ensembles can be realized as a space of square matrices of size $N$, whose coefficients are real if $\beta=1$, complex if $\beta=2$, and \textbf{quaternionic}\index{quaternion} if $\beta=4$. 
In all cases, the coefficients can be written as elements of a $\beta$-dimensional \textbf{Clifford algebra}\index{Clifford algebra} over $\mathbb{R}$ with generators $(\mathbf{e}_\alpha) = (\mathbf{e}_0=\mathbf{1},\mathbf{e}_1,\dots,\mathbf{e}_{\beta -1})$ and relations ($i,j,k\neq 0$)
\begin{align}
 \mathbf{e}_i^2  = -\mathbf{1}\ , \quad 
 \mathbf{e}_i\mathbf{e}_j =\epsilon_{ijk} \mathbf{e}_k\ \qquad (\epsilon_{i,j,k}=\text{completely antisymmetric tensor}) .
\end{align}
The Clifford algebra also comes with the conjugation 
\begin{align}
 q= \sum_{\alpha = 0}^{\beta-1} q^{(\alpha)} \mathbf{e}_\alpha \quad \implies \quad \bar q = q^{(0)} {\mathbf 1} - \sum_{i=1}^{\beta-1} q^{(i)} \mathbf{e}_i\quad \text{where} \quad q^{(\alpha)} \in \mathbb{R}\ ,
\end{align}
and norm
\begin{align}
|q|^2 = q \bar q = 
  \sum_{\alpha = 0}^{\beta-1} (q^{(\alpha)})^2\ .
\end{align}
Defining the conjugate of a matrix by $M^\dagger = \bar{M}^\mathrm{T}$,  
Gaussian ensembles are then defined by the constraint $M^\dagger = M$, and circular ensembles by the orthonormality constraint $MM^\dagger = \operatorname{Id}$.
The two quaternionic ensembles have alternative realizations in terms of complex matrices of size $2N$, which are obtained by replacing the generators $\mathbf{e}_i$ with size two Pauli matrices. 

The diagonalization \eqref{eq:diagonalize} of $M\in E_N^\beta$ is not unique. 
In the case $\beta=2$, the set $\{\lambda_1, \cdots \lambda_N\}$ of eigenvalues is left unchanged if we multiply $U$ on the right by elements of
\begin{itemize}
 \item the set $(U_1)^N\subset U_N$ of diagonal unitary matrices, which is a maximal Abelian subgroup of $U_N$,
 \item and the set $\mathfrak{S}_N\subset U_N$ of permutation matrices.
\end{itemize}
If the eigenvalues are not all distinct, there is a larger subgroup of $U(N)$ that leaves the set of eigenvalues unchanged. We can ignore this case because the subset of matrices whose eigenvalues are not all distinct is of measure $0$. Locally, in the neighbourhood of any matrix with distinct eigenvalues, $H_N$ looks like the quotient
$H_N \simeq \frac{{\frac{U_N}{(U_1)^N} \times \mathbb{R}^N}}{\mathfrak{S}_N}$.

%\begin{equation}
%H_N \simeq \frac{{\frac{U_N}{(U_1)^N} \times \mathbb{R}^N}}{\mathfrak{S}_N}\ .
%\label{huur}
%\end{equation}

\subsubsection{Measures}

On each one of the three Gaussian ensembles $E_N^\beta$, there is a
\textbf{Lebesgue measure}\index{Lebesgue measure}, which is the product of the Lebesgue measures of all real components of the matrix $M$.
Writing the real components of $M_{i,j}$ as $M^{(\alpha)}_{i,j}$, the Lebesgue measure is
\begin{align}
 dM = \prod_i dM_{i,i}\prod_{i<j} \prod_{\alpha=0}^{\beta -1} dM_{i,j}^{(\alpha)}\ .
 \label{eq:betaLebesguemeasure}
\end{align}
The Lebesgue measure coincides with the volume form of the canonical metric
\begin{equation}
||dM||^2 = \operatorname{Tr} dM \ dM^\dagger,
\label{dm2}
\end{equation}
which is the restriction to $E_N^\beta$ of the canonical metric of $\mathbb C^{\beta N^2}$.
The Lebesgue measure is invariant under changes of bases, and thus invariant under conjugation by elements of the corresponding circular ensemble.

Interesting measures are usually built from $dM$ by multiplication with a function of $M$. 
For example, given a function $V$ called the \textbf{potential}\index{potential}, we can consider the measure
       \begin{align}
	\boxed{ d\mu(M)  =  e^{-\operatorname{Tr} V(M)} dM }
	\ .
	\label{eq:potentialmeasure}
       \end{align}
The potential $V$ is often chosen to be a polynomial.
Gaussian statistics are obtained when $V$ is a polynomial of degree two.
Generalizing polynomial potentials, we can consider \textbf{rational potentials}\index{potential!rational---}, which we define as potentials $V$ such that 
$V'$ is a rational function. (In the literature, rational potentials are sometimes called semi-classical potentials.)
Far more exotic potentials can also be considered, and product of traces can be added in the exponential.
See, for example, \autoref{exo:plancherel}.

% NB BE : attention Haar on the wrong group !

The canonical measure on a circular ensemble is the \textbf{Haar measure}\index{Haar measure}, which is characterized (up to a normalization) as being invariant under the left action of the corresponding Lie group on itself.
The circular ensembles are compact and therefore unimodular groups, which implies that their left invariant Haar measures are also right-invariant. The Haar measure is the restriction to $U_N^{\beta}$ of the volume form of the canonical metric \eqref{dm2}, since that metric is invariant under left multiplication $M\to UM$ with $U^\dagger U=\text{Id}$.
Using
\begin{equation}
dM^\dagger = d(M^{-1}) = - M^{-1} \ dM \ M^{-1} = - M^\dagger \ dM \ M^\dagger,
\end{equation}
we rewrite the canonical metric on $U_N^{\beta}$ as
\begin{equation}
||dM||^2 = - \operatorname{Tr} \Omega(M)^2,
\end{equation}
where we introduced the \textbf{Maurer-Cartan form}\index{Maurer-Cartan form}:
\begin{equation}
\Omega(M) = M^\dagger dM.
\end{equation}
The Haar measure on $U_N^\beta$ is thus
\begin{equation}
dM_{\text{Haar}(U_N^\beta)} = \prod_i \Omega(M)_{i,i} \prod_{i<j}\prod_{\alpha=0}^{\beta-1} \Omega(M)_{i,j}^{(\alpha)},
\end{equation}
and the Haar measure on $U_N^\beta/U_1^N$ is
\begin{equation}
dM_{\text{Haar}(U_N^\beta/U_1^N)} = \prod_{i<j}\prod_{\alpha=0}^{\beta-1} \Omega(M)_{i,j}^{(\alpha)}.
\end{equation}

\subsubsection{Reduction to eigenvalue integrals}

Let us use the invariance of the measure $d\mu(M)$ \eqref{eq:potentialmeasure} under conjugations, and reduce the corresponding matrix integrals to eigenvalue integrals. (See \autoref{chap:AngularIntegral} for examples of measures that are not invariant under conjugations.)
Under diagonalization \eqref{eq:diagonalize}, the Lebesgue measure $dM$ on $E_N^\beta$ can be rewritten in terms of measures on $\Lambda$ and $U$.
As we will shortly show, we have 
\begin{align}
 \boxed{ dM  = \left|\Delta(\Lambda)\right|^\beta d\Lambda\, dU_\text{Haar} } \ ,
 \label{eq:dM}
\end{align}
where $dU_\text{Haar}$ is the Haar measure on $U^\beta_N/(U_1^\beta)^N$, and  $d\Lambda = \prod_{i=1}^N
d\lambda_i$ is the Lebesgue measure on $\mathbb{R}^N$, and the Jacobian is written in terms of the 
\textbf{Vandermonde determinant}\index{Vandermonde determinant}, 
\begin{align}
 \boxed{ \Delta(\Lambda) =  \prod_{1\leq i<j\leq N} (\lambda_i -\lambda_j) =\det_{1\leq i,j\leq N} \lambda_i^{j-1}} \ .
\label{eq:vandermonde}
\end{align}
The partition function associated to the measure $d\mu(M)$ \eqref{eq:potentialmeasure} becomes 
\begin{align}
 \boxed{ \mathcal{Z} = 
 %\operatorname{Vol}\left(\frac{E_N^\beta}{\mathbb{R}^N}\right)
 \mathcal V_N^\beta 
  \int_{\mathbb{R}^N} d\Lambda\, \left|\Delta(\Lambda)\right|^\beta e^{-\operatorname{Tr} V(\Lambda)} } \ ,
\label{eq:zeigenvalues}
\end{align}
where the prefactor is
\begin{align}
\mathcal V_N^\beta  =
 \operatorname{Vol}\left(\frac{E_N^\beta}{\mathbb{R}^N}\right) 
 = 
 \frac{1}{N!}
 \operatorname{Vol}\left(\frac{U_N^\beta}{(U_1^\beta)^N}\right) 
% \ .
%% \frac{1}{\left|\text{Weyl}(U_N^\beta)\right|}
%% \operatorname{Vol}\left(\frac{U_N^\beta}{T_N^\beta}\right) \ .
%\end{align}
%Explicitly,
%\begin{align}
% \operatorname{Vol}\left(\frac{E_N^\beta}{\mathbb{R}^N}\right) 
% &\ 
 = \ \pi^{\frac{N(N-1)\beta}{4}} \prod_{j=1}^{N} \frac{\Gamma(1+\frac\beta{2})}{ \Gamma(1+j\frac{\beta}{2})}\ .
\label{eq:VolENbeta}
\end{align}
(See \autoref{exo:gaussian} and \autoref{exo:gaussian2} for computing $\mathcal{Z}$ in the case $N=2$ with respectively $\beta=2$ and $\beta$ arbitrary.)
One way to compute this prefactor is to consider the Gaussian potential $V(x)=\frac12 x^2$. In this case, the original matrix integral is simply $\mathcal{Z} = (2\pi)^{\frac{N}{2}} \pi^{\frac12 \beta \frac{N(N-1)}{2}}$, whereas the integral over eigenvalues is a Selberg--Mehta integral whose expression is also known \cite{Mehta:2004RMT}.
Another way to compute the prefactor is to use 
\begin{equation}\label{eq:VolUNbeta}
 \operatorname{Vol}\left(U_N^\beta\right) 
 = 2 \left(\frac\beta 2\right)^N  \pi^{\frac{N(N+1)\beta}{4}}  \frac{N!}{ \prod_{j=1}^{N}\Gamma(1+j\frac{\beta}{2})}\ .
\end{equation}
(See \autoref{exo:Haarvolu2} for the volume of $U_2$.)
% NB: Derivation of U_\beta to be added here or exercise, as suggested by the referee?

\subsubsection{Proof of Eq.~\eqref{eq:dM}}

The Lebesgue measure $dM$ is invariant under conjugation of $M$ by a circular matrix, and the Haar measure $dU_\text{Haar}$ is invariant under right and left multiplication, therefore the Jacobian  $\frac{dM}{d\Lambda dU_\text{Haar}}$ in Eq.~\eqref{eq:dM} is invariant under $U_N^\beta$ conjugations, and can without loss of generality be computed at $U=\operatorname{Id}$.
Denoting $\delta M$ the differential of $M$, differentiating Eq.~\eqref{eq:diagonalize} yields
\begin{align}
U^{-1}\delta M U  =  \delta \Lambda + [\Omega(U), \Lambda],
\end{align}
where $\Omega(U)$ is the Maurer--Cartan form. At $U=\operatorname{Id}$, this implies
\begin{align}
 (\delta M)_{ij} = \delta_{ij}\delta\lambda_i  + (\lambda_j-\lambda_i)\Omega(U)_{ij} \ .
\end{align}
Therefore, the Jacobian is diagonal, in the sense that $\frac{(\delta M)_{ij}}{(\delta U)_{i'j'}} \neq 0 \implies (i,j)=(i',j')$. The determinant is simply the product of the diagonal elements,
\begin{align}
 \frac{dM}{d\Lambda dU_\text{Haar}}= \prod_{i} \left(\frac{\delta M_{ii}}{\delta \lambda_{i}}\right) \prod_{i< j}\prod_{\alpha}\left( \frac{\delta M^{(\alpha)}_{ij}}{\delta U^{(\alpha)}_{ij}}\right)
 =
 \prod_{i} 1 \prod_{i< j}\prod_{\alpha}\left( \lambda_i-\lambda_j\right)
 =
 \Delta(\Lambda)^\beta\ .
\end{align}

\subsection{\texorpdfstring{$\beta$}{beta}-matrix models}

The three Gaussian matrix ensembles $E_N^\beta$ are distinguished by a parameter $\beta$ that takes the three values $\beta=1,2,4$. However, when written in terms of eigenvalues, the Lebesgue measure \eqref{eq:dM} makes sense for any complex value of $\beta$. We thus define a \textbf{\boldmath $\beta$-matrix model}\index{beta-matrix model@$\beta$-matrix model} by allowing an arbitrary complex value of $\beta$. 
This definition is only useful for quantities that can be written in terms of eigenvalues.
Nevertheless, $\beta$-matrix models can be defined beyond eigenvalue integrals, using for example 
the split and merge rules of \autoref{chap:LoopEq}, or equivalently the Calogero--Moser equation of \autoref{chap:AngularIntegral}.
% NB: These two definitions are not independent, since the derivation of the Calogero--Moser equation uses the split rule.

\subsubsection{Tridiagonal realization}

In 2002, I.~Dumitriu and A.~Edelman~\cite{DUMITRIU2002,DUMITRIU2007587}
discovered a matrix model whose measure depends on $\beta\in\mathbb{R}_+$, and whose eigenvalue distribution agrees with that of a Gaussian $\beta$-matrix model. 
This ensemble is made of symmetric real, tridiagonal matrices, whose diagonal and upper triangular elements are independent and follow respectively a normal Gaussian distribution $N(0,1)$ and a Chi-distribution $\chi_r$. These distributions are defined by the probability measures
\begin{equation}
N(0,1):\ \frac{1}{\sqrt{2\pi}} e^{-\frac{x^2}{2}}\,dx \qquad , \qquad \chi_r:\ \frac{2}{\Gamma(\frac{r}{2})} x^{r-1} e^{-x^2}\,dx\ .
\end{equation}
Then, the measures on the elements of our size $N$ symmetric tridiagonal matrices are given  by
\begin{equation}
% OK, verified for N=1,2
\begin{pmatrix}
% OK verified for N=1,2
N(0,1) & \chi_{(N-1)\beta} & & & \cr
\chi_{(N-1)\beta} & N(0,1) & \chi_{(N-2)\beta}  & & \cr
&\ddots & \ddots & \ddots & \cr
& & \chi_{2\beta} & N(0,1) & \chi_{\beta}  \cr
& & & \chi_{\beta} & N(0,1)  \cr
\end{pmatrix} \ 
\end{equation}
where the line below the diagonal, is the symmetric of the line above (it is not sampled independantly).
The resulting eigenvalue distribution is 
\begin{align}
 d\mu(\Lambda) = (2\pi)^{-\frac{N}{2}} \left(\prod_{j=1}^N \frac{\Gamma(1+\frac\beta 2)}{\Gamma(1+j \frac\beta 2)}\right)
|\Delta(\Lambda)|^\beta e^{-\frac12 \operatorname{Tr} \Lambda^2} d\Lambda\ .
\label{dml}
\end{align}
(For the proof in the case $N=2$, see \autoref{exo:DumiEdel}.)
Numerically, this realization is very efficient at sampling $\beta$-matrix models.

\subsubsection{Nekrasov variables and the duality $\beta \leftrightarrow \frac{4}{\beta}$}

Instead of the variables $N$ and $\beta$, let us introduce the Nekrasov variables~\cite{Nekrasov:2002qd}
\begin{equation}
\epsilon_1 = \frac{1}{N}
\qquad , \qquad
\epsilon_2 = -\frac{2}{N\beta}\ .
\label{Nekrasov_notation}
\end{equation}
These variables originate from supersymmetric gauge theory, and their identification with 
our matrix model variables is suggested by the AGT relation between gauge theory and two-dimensional conformal field theory~\cite{Alday:2009aq}, together with the relation between certain matrix integrals and the Dotsenko--Fateev integrals of conformal field theory. 
Now, in gauge theory, the variables $\epsilon_1,\epsilon_2$ can take arbitrary complex values. This suggests that $N$ and $\beta$ can be continued beyond their integer values.

Moreover, the gauge theory is invariant under the duality $\epsilon_1\leftrightarrow \epsilon_2$. 
In conformal field theory, this corresponds to the invariance of the central charge
\begin{align}
 c 
 = 1 - 6 \left( \sqrt{\frac{\beta}{2}} - \sqrt{\frac{2}{\beta}} \right)^2
 \, ,
\end{align}
under $\beta \leftrightarrow \frac{4}{\beta}$.
However, in matrix models, the meaning of this duality is less clear, as it corresponds to
\begin{align}
\boxed{ (\beta,N) \ \leftrightarrow  \
\left( \frac{4}{\beta}, -\frac{N\beta}{2}\right) }
\, .
\label{beta_duality}
\end{align}
Nevertheless, there is a natural matrix model quantity which is invariant under this duality: the number of independent real variables in the Gaussian ensembles $E_N^\beta$, 
\begin{equation}
N+\frac{N(N-1)}{2}\beta
= \frac{\epsilon_1+\epsilon_2-1}{\epsilon_1 \epsilon_2}
\, ,
\end{equation}
where $N$ variables come from the diagonal, and  $\frac{N(N-1)}{2}\beta$ variables from off-diagonal matrix elements.

Many matrix integrals are conjectured to be invariant under this duality, see for example \autoref{exo:duality}. These conjectures are supported by the invariance of loop equations, see \autoref{sec:loop}. In the case $\beta=2$, the duality transformation reduces to $N\to -N$. Hermitian matrix integrals are therefore even functions of $N$, and their large $N$ asymptotics involve expansions in $\frac{1}{N^2}$ rather than the generic $\frac{1}{N}$, which gives them stronger convergence properties.

\subsection{Normal matrices} \label{sectionnormaloncontour}

Normal matrices are matrices that can be diagonalized by a unitary transformation
\begin{equation}
M = U \Lambda U^\dagger \ , \qquad \text{with} \qquad  U\in U_N, \ \Lambda\in \mathbb C^N\ .
\end{equation}
Equivalently, a normal matrix commutes with its adjoint:
\begin{equation}
[M,M^\dagger]=0.
\end{equation}
Given a subset $\gamma\subset \mathbb{C}$, we define the  ensemble of \textbf{normal matrices}\index{normal matrix} with eigenvalues in $\gamma$ as the set of matrices that can be diagonalized by a unitary matrix, and have eigenvalues in $\gamma$: 
\begin{align}
 \boxed{ H_N(\gamma) \overset{\text{def}}{=} \left\{M=U\Lambda U^{-1} \middle| U\in U_N,\ \Lambda=\mathrm{diag}(\lambda_1,\dots,\lambda_N),\ \lambda_i\in\gamma\right\} }\ .
\end{align}
The set $\gamma$ should be a measurable subset of $\mathbb C$, for example a domain or a Jordan arc. A good technical assumption is that $\gamma$ is a Suslin space, i.e. the image of a Polish space by a continuous map. (Jordan arcs are Suslin spaces.)

Normal matrices are in particular useful in the Coulomb gas method of \autoref{chap:SaddlePoint}. This method involves deforming an initial integration contour $\gamma_0$ to a steepest descent contour $\gamma$ passing through a saddle point. 
Even in cases where $\gamma_0 = \mathbb{R}$, the saddle point is often complex, and $\gamma$ may not be the real line, but another open Jordan arc, typically going from $\infty$ to $\infty$.

Normal matrices generalize both Hermitian and unitary matrices:
\begin{align}
H_N(\mathbb R) = H_N \qquad , \qquad
H_N(S^1) = U_N\ .
\end{align}
%Generalizing the Gaussian ensembles' angular-radial decomposition \eqref{huur}, we have
%\begin{align}
%H_N(\gamma) \simeq  \frac{\frac{U_N}{(U_1)^N}\times \gamma^N}{\mathfrak{S}_N}\ .
%\end{align}

\subsubsection{Measure}

Assuming that $\gamma = f([0,1])$ is a Jordan arc of class $C^1$ with a parametrization $f:[0,1]\mapsto \mathbb{C}$, we have the curvilinear measure on $\gamma$,
\begin{equation}
d\lambda = f'(t)dt  \qquad \text{for}\qquad  \lambda=f(t)\ .
\end{equation}
Using this curvilinear measure, we define the canonical measure on $H_N(\gamma)$ as 
\begin{equation}
dM \overset{\text{def}}= \Delta(\Lambda)^2 d\Lambda\, dU_\text{Haar}\qquad \text{with} \qquad d\Lambda =\prod_{i=1}^N d\lambda_i\ .
\label{eq:dMnormal}
\end{equation}
The measure $dM$ is in general neither positive, nor even real.

In the case $\gamma=S^1$, the resulting measure on $U_N=H_N(S^1)$ differs from the Haar measure. To compare these two measures, we diagonalize $M\in U_N$ as $M = U\Lambda U^\dag$, with eigenvalues $\lambda_i = e^{i\theta_i}\in S^1$ with $\theta_i\in \mathbb R/2\pi\mathbb Z$.
The Haar measure for $M$ is given in terms of the Haar measure for $U\in \frac{U_N}{(U_1)^N}$ by Eq.~\eqref{eq:dM} with $\beta=2$, with $d\lambda_i$ replaced by the canonical measure $d\theta_i$ on $S^1$:
\begin{align}
 dM_\text{Haar} = dU_\text{Haar}\prod_{i<j} |\lambda_i-\lambda_j|^2 \prod_{i=1}^N d\theta_i\ .
 \label{eq:dmhaar}
\end{align}
In contrast, the measure $dM$ \eqref{eq:dMnormal} involves $(\lambda_i-\lambda_j)^2 =-\lambda_i\lambda_j |\lambda_i-\lambda_j|^2$, and $d\lambda_i = i \lambda_id\theta_i$. As a result, the relation between the normal and Haar measures is
\begin{equation}
 dM_\text{Haar}
 = (-1)^{\frac{N(N-1)}{2}}(i\det M)^{-N} dM
 = (-1)^{\frac{N(N-1)}{2}}i^{-N} e^{-N\operatorname{Tr}\log M} dM\ .
  \label{eq:circular_Haar}
\end{equation}
This relates the positive but non-analytic Haar measure, with the analytic but not positive measure $dM$.
This is useful, because an integral with $dM$ can be computed by deforming integration contours.
It often happens that the only poles come from the prefactor $(\det M)^{-N}$, in which case the integral reduces to a sum of residues at $\lambda_i=0$. See \autoref{exo:volu2} for a simple example, and \autoref{exo:volun} for applications.

\subsubsection{Partition function}

The partition function for normal matrices is a generalization of the partition function \eqref{eq:zeigenvalues} for Hermitian matrices:
\begin{align}
 \boxed{\mathcal{Z}(\gamma^N) = \mathcal V_N^2  \ \int_{\gamma^N} d\Lambda\, \Delta(\Lambda)^2 e^{-\operatorname{Tr} V(\Lambda)} } \ .
\label{eq:znormal}
\end{align}
In contrast to the Hermitian case, the factor $\Delta(\Lambda)^2$ does not come with an absolute value or modulus. If the potential $V$ is analytic, the integrand is analytic, and the partition function is invariant under homotopic deformations of the integration domain $\gamma^N$.

More generally, for any complex value of $\beta$, we define the partition function
\begin{align}
 \boxed{ \mathcal{Z}(\Gamma)\propto \int_{\Gamma} d\Lambda\, \Delta(\Lambda)^\beta e^{-\operatorname{Tr} V(\Lambda)} } \ ,
\label{eq:znormalbeta}
\end{align}
for a half--dimensional submanifold $\Gamma\subset \mathbb C^N$, such that the integrand is well-defined, and the integral is absolutely convergent. 
In particular, for the factor $\Delta(\Lambda)^\beta$ to be well-defined, there must be cuts ending at the hyperplanes $\{\lambda_i=\lambda_j\}$, and $\Gamma$ should not intersect these cuts. 
Due to the analyticity of the integrand, the partition function is unchanged under homotopic deformations of $\Gamma$, and therefore only depends on $\Gamma$ through its homology class. 
For rational values of $\beta$, and rational potentials, the homology space of possible integration domains is finite-dimensional, as we will now see in more detail for $\beta=2$.

\subsubsection{Integration contours}

We have just seen the interest of integrating matrix eigenvalues on fairly arbitrary contours in the complex plane.
Let us now determine which contours give rise to absolutely convergent integrals. 

For $\beta=2$ and for a rational potential such that $V'\in\mathbb C(x)$ and $\lim_{|x|\to\infty}(|V(x)|-2\log{|x|})=+\infty$, the convergence of the partition function $\mathcal{Z}(\gamma^N)$ is equivalent to the convergence of the one-dimensional integral
\begin{align}
 z(\gamma) = \int_\gamma e^{-V(\lambda)}d\lambda\ .
\end{align}
Let us study this integral near a singularity $\alpha$ of $V'$: either $\infty$, or a pole of $V'$. 
We define the degrees of $V$ and $V'$ at their singularities by 
\begin{align}
 \deg_\alpha \frac{1}{(\lambda-\alpha)^n} = n \quad , \quad \deg_\alpha \log(\lambda-\alpha) = 0 \quad , \quad \deg_\infty \lambda^n = n\ .
\end{align}
The total \textbf{degree}\index{degree (of a rational function)} of $V'$ is 
\begin{align}
\boxed{ d = \deg V' = \sum_\alpha \deg_\alpha V' }\ ,
\end{align}
and a rational function $V'$ of degree $d$ has $d+1$ independent parameters.
An integration contour $\gamma$ can end at a singularity $\alpha$ that is not a simple pole, provided $\lim_{\lambda\in\gamma, \lambda\to\alpha}\Re V(\lambda)= +\infty$.
This defines $\deg_\alpha V$ allowed sectors around $\alpha$ where $\Re V(\lambda)\to +\infty$, and $\deg_\alpha V$ forbidden sectors where $\Re V(\lambda)\to -\infty$.
For example, in the case $\deg_\alpha V=4$ with $\alpha = \infty$ (left) or $\alpha\neq \infty$ (right), we draw the four forbidden sectors in green, 
and an example of an allowed integration contour in red:
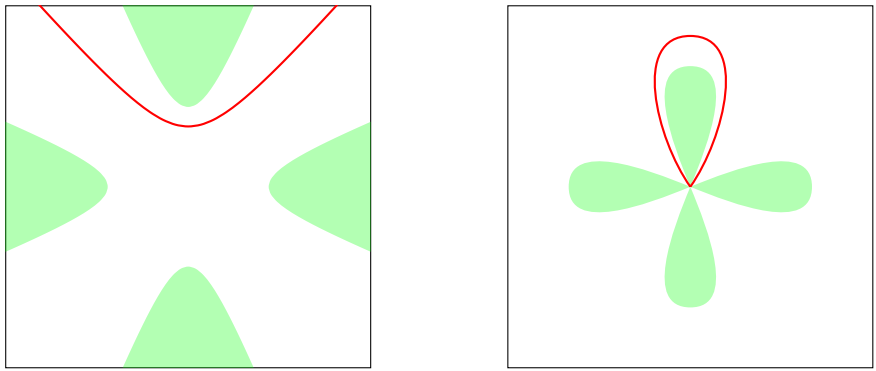
\begin{align}
\begin{tikzpicture}[scale = .4, baseline=(current  bounding  box.center)]
 \draw [clip] (-6,-6) -- (-6,6) -- (6,6) -- (6,-6) -- cycle;
 \foreach \y in {0,1,2,3} {
 \begin{scope}[rotate = \y*90]
  \fill [green, opacity=.3] plot [domain = -3:3, smooth] ({\x},{sqrt((2.5*\x)^2+7)});
  \end{scope}
  }
  \draw [red, thick] plot [domain = -5:5, smooth] ({\x},{sqrt(1.35*(\x)^2 + 4)});
\end{tikzpicture}
\qquad \qquad
\begin{tikzpicture}[scale = .4, baseline=(current  bounding  box.center)]
\draw [clip] (-6,-6) -- (-6,6) -- (6,6) -- (6,-6) -- cycle;
 \foreach \y in {0,1,2,3} {
 \begin{scope}[rotate = \y*90]
 \fill [green, opacity=.3] (0, 0) to [out = 22.5, in = 90] (4,0) to [out = -90, in = -22.5] (0, 0);
 \end{scope}
 }
 \draw [red, thick] (0, 0) to [out = 55, in = 0] (0,5) to [out = 180, in = 125] (0,0);
\end{tikzpicture}
\end{align}
%When it comes to finding integration contours, these two examples are actually equivalent, and it does not matter whether $\alpha = \infty$ or not.
In a case with three singularities such that $\deg_\alpha V = 2,3,4$, let us draw a set of homologically independent contours:
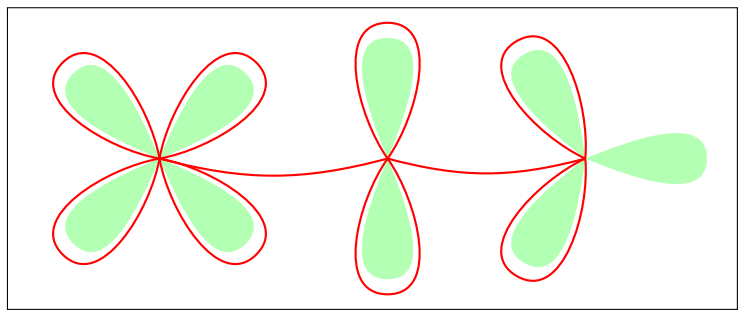
\begin{align}
 \begin{tikzpicture}[scale = .4, baseline=(current  bounding  box.center)]
\draw [clip] (-5,-5) -- (-5,5) -- (19,5) -- (19,-5) -- cycle;
\newcommand{\aleaf}[2]{
\begin{scope}[shift = {(#1, 0)}, rotate = #2]
 \fill [green, opacity=.3] (0, 0) to [out = 67.5, in = 0] (0,4) to [out = 180, in = 112.5] (0, 0);
 \draw [red, thick] (0, 0) to [out = 55, in = 0] (0,4.5) to [out = 180, in = 125] (0,0);
 \end{scope}
}
 \foreach \y in {0,1,2,3} {
 \aleaf{0}{\y*90+45}
 }
 \foreach \y in {0,1} {
 \aleaf{7.5}{\y*180}
 }
 \foreach \y in {1,2} {
 \aleaf{14}{\y*120-90}
 }
 \begin{scope}[shift = {(14,0)}, rotate = -90]
  \fill [green, opacity=.3] (0, 0) to [out = 67.5, in = 0] (0,4) to [out = 180, in = 112.5] (0, 0);
 \end{scope}
 \draw [red, thick] (0,0) to [out = -15, in = 195] (7.5,0) to [out = -15, in = 195] (14, 0);
\end{tikzpicture}
\end{align}
In general, each extra singularity $\alpha$ gives rise to $\deg_\alpha V + 1$ extra independent contours.
However, the first singularity gives rise to only $\deg_\alpha V - 1$ independent contours. 
If by convention we consider that first singularity to be $\alpha = \infty$, then we have $\deg_\alpha V'$ independent contours per singularity, and the total number of independent contours is the total degree $d$ of $V'$. 
In other words, if we define $\mathbb{C}_\text{allowed}$ as the complex plane minus the forbidden sectors,
the rank of the first fundamental group of $\mathbb{C}_\text{allowed}$ is
\begin{equation}\label{eq:rankpi1allowed}
\operatorname{rank} \pi_1(\mathbb C_{\text{allowed}} ) = d = \deg V'\ .
\end{equation}
Arbitrary complex linear combinations of these contours form the $d$-dimensional \textbf{homology space}\index{homology space!of a measure} for the measure $e^{-V(\lambda)}d\lambda$
\begin{equation}
H_1(e^{-V(\lambda)}d\lambda) = \left\{\gamma = \sum_{i=1}^d c_i \gamma_i,\ c_i\in \mathbb C \right\}\ 
\quad , \quad
\dim H_1(e^{-V(\lambda)}d\lambda) = d=\deg V' \ .
\end{equation}
These results also hold if $V'$ has simple poles. (See \autoref{exo:dotsenko} for the example of Dotsenko--Fateev integrals.)
For $\alpha$ a simple pole of $V'$, the integrand behaves as $e^{-V(\lambda)} \underset{\lambda \to \alpha}{\propto} (\lambda - \alpha)^{-t_\alpha}$ with $t_\alpha = \operatorname{Res}_\alpha V'$. A simple pole always gives rise to $\deg_\alpha V'=1$ extra contour: either around $\alpha$ if $\Re t_\alpha >0$, or ending at $\alpha$ if $\Re t_\alpha <0$.
(The condition for integrability would actually be $\Re t_\alpha <1$, but we need $e^{-V(\alpha)} = 0$ for integrating by parts.)
Possible branch cuts do not modify the counting of contours. 
There is a branch cut ending at $\alpha$ if $t_\alpha \notin \mathbb{Z}$. 
A branch cut, if present, must end at another singularity, possibly at infinity: this is always possible, because $\sum_\alpha t_\alpha =0$.

\subsubsection{Filling fractions}

The normal matrix ensemble $H_N(\gamma)$, and the corresponding convergent matrix integral $\mathcal{Z}(\gamma^N)$, can be defined not only for individual contours, but also for their linear combinations $\gamma\in H_1(e^{-V(\lambda)}d\lambda)$. 
We now relax the requirement that the $N$-dimensional integration domain be symmetric under permutations, and consider the contours
\begin{equation}\label{intnormalduple}
 \vec\gamma ^{\vec n} = \prod_{i=1}^d \gamma_i^{n_i} 
 \quad \text{with} \quad 
\vec n=(n_1,n_2,\dots,n_d)
 \quad \text{and} \quad 
 \sum_{i=1}^d n_i = N\ ,
\end{equation}
where $n_i$ eigenvalues are integrated over the contour $\gamma_i$. 
The quantities
\begin{equation}
\epsilon_i = \frac{n_i}{N}
\end{equation}
are then called the \textbf{filling fractions}\index{filling fraction} associated with our basis $\gamma_i$ of contours. 
%(%NB BE: wrong !! A change of basis $\gamma_i \to \sum_{j} C_{ij} \gamma_j$ acts linearly on filling fractions, 
%$\epsilon_i \to \sum_{j} C_{ij} \epsilon_j$.)
The corresponding integral is called $\mathcal{Z}(\vec\gamma ^{\vec n})$. 
Linearly combining $N$-dimensional contours, we define 
\begin{equation}
\mathcal{Z}\left(\sum_{\vec n}c_{\vec n}\vec\gamma ^{\vec n}\right) = \sum_{\vec n}c_{\vec n}\mathcal{Z}(\vec\gamma ^{\vec n})\ , \quad \text{with}\ c_{\vec n}\in \mathbb C\ .
\end{equation}
The dimension of the space of convergent integrals, equivalently of the homology space of the one-form $\prod_{i=1}^N e^{-V(\lambda_i)} d\lambda_i$, is therefore the number of $d$-uples $\vec{n}$ whose sum is $N$,
\begin{equation}
\boxed{ d_N = \dim H_1\left(\prod_{i=1}^N e^{-V(\lambda_i)} d\lambda_i\right) = \binom{N+d-1}{N} }\ .
\label{eq:dimh1}
 \end{equation}
For any contour $\gamma=\sum_{i=1}^d c_i\gamma_i\in H_1(e^{-V(\lambda)}d\lambda)$, the normal matrix integral $\mathcal{Z}(\gamma^N)$ can be decomposed as 
\begin{equation}
\frac{1}{N!}\mathcal{Z}(\gamma^N) = \sum_{\vec n} \left(\prod_{i=1}^d \frac{c_i^{n_i}}{n_i!}\right)\mathcal{Z}(\vec\gamma^{\vec n})\ .
\label{eq:zgn}
\end{equation}
In the case of a Gaussian potential, we have $\deg V'=d=1$ and therefore $d_N=1$, so that all allowed integration domains are homologically equivalent.

Let us now comment on integration contours and on the space of convergent integrals if $\beta \neq 2$. 
If $\beta\notin \mathbb Z $, then the integrand is singular at coinciding points, and the integral depends on a choice of cuts that end at coinciding points.
If however $\beta \in \mathbb Q$ is rational, then the homology space for the measure $\prod_{i<j} (\lambda_i-\lambda_j)^\beta \ \prod_i e^{- V(\lambda_i)} d\lambda_i $ nevertheless remains finite-dimensional, because repeatedly crossing a cut only produces finitely many independent contours. 
If $\beta$ is not rational the dimension of the homology space is usually infinite.

\subsection{Other examples}\label{sec:other_examples}

Let us give a number of interesting matrix ensembles and/or measures. More examples are obtained by combining or generalizing them.

\subsubsection{Complex matrix model}\index{matrix model!complex---}

Let us consider the ensemble of complex matrices $M_N(\mathbb C)$, with the Lebesgue measure $dM=\prod_{i,j=1}^N d\Re M_{i,j} \, d\Im M_{i,j} $. 
Let us consider a matrix integral of the form 
\begin{equation}
 \boxed{ \mathcal{Z} = \int_{M_N(\mathbb{C})} dM\, e^{-\operatorname{Tr} V(M, M^\dag)}} \ .
\end{equation}
with $V(M,M^\dagger ) =V(UMU^\dagger, U M^\dagger U^\dagger)$ chosen to be invariant under conjugations of $M$ by unitary matrices.
Using such conjugations, we cannot diagonalize $M$, but only bring it to a triangular form
\begin{equation}
\renewcommand{\arraystretch}{1.3}
M = U(\Lambda+T) U^{-1}\ , \quad 
\left\{\begin{array}{l} 
U\in \frac{U_N}{(U_1)^N}\ , \\ \Lambda\in \mathbb{C}^N\ , \\ T\in B_N(\mathbb{C})\ ,
\end{array} \right.
\end{equation}
where the Borel subalgebra $B_N(\mathbb{C})$ of $U_N$ is the vector space of strictly upper  triangular complex matrices.
This amounts to writing our complex matrix ensemble as 
\begin{equation}
M_N(\mathbb C) 
 \simeq 
 \frac{\frac{U_N}{(U_1)^N} \times \mathbb{C}^N \times B_N(\mathbb{C})}
      {\mathfrak{S}_N}
 \ .
\end{equation}
Under this decomposition, the 
Lebesgue measure on $M_N(\mathbb{C})$ becomes 
\begin{equation}
\boxed{ dM = |\Delta(\Lambda)|^2|d\Lambda|^2 |dT|^2 dU_{\text{Haar}} }\ ,
\end{equation}
where $|d\Lambda|^2$ and $|dT|^2$ are the Lebesgue measures on $\mathbb{C}^N$ and $B_N(\mathbb{C})$.  

Let us now specialize to the Gaussian complex matrix model, 
\begin{align}
 \operatorname{Tr} V(M,M^\dagger) = \operatorname{Tr} MM^\dagger = \operatorname{Tr}\left(\Lambda\bar{\Lambda} + TT^\dagger\right)\ .
\end{align}
Let us show how the calculation of moments can be reduced to integrals over eigenvalues. 
This is straightforward for moments that only involve eigenvalues, such as 
\begin{align}
\left< \operatorname{Tr} M^k \right>
&= \frac{1}{ \mathcal{Z}} \int_{M_N(\mathbb C)} dM\ e^{-\operatorname{Tr} MM^\dagger} \operatorname{Tr}M^k  \nonumber
\\
&= \frac{\int |d\Lambda|^2  \prod_{i<j} |\lambda_i-\lambda_j|^2 \prod_i e^{-|\lambda_i|^2} \left( \sum_i \lambda_i^k\right) }{\int |d\Lambda|^2 \prod_{i<j} |\lambda_i-\lambda_j|^2 \prod_i e^{-|\lambda_i|^2} }\ .
\end{align}
This is more complicated for moments that also involve the triangular matrix $T$, such as
\begin{align}
\left<\operatorname{Tr} M^2 M^{\dagger^2}\right>
= \left<\operatorname{Tr}\left( \Lambda^2 \bar\Lambda^2
 + \Lambda \bar\Lambda T T^\dagger
 + \Lambda \bar\Lambda T^\dagger T
 + \bar\Lambda T \Lambda T^\dagger
 + \Lambda  T \bar\Lambda T^\dagger
 + T^2 T^{\dagger^2} \right)\right> \ .
\end{align}
The Gaussian integral over $T$ can be performed using Wick's theorem:
\begin{multline}
 \int |dT|^2 e^{- \operatorname{Tr} T T^\dagger}
 \operatorname{Tr} M^2 M^{\dagger^2}
 \\
 = \left(\int |dT|^2 e^{- \operatorname{Tr} T T^\dagger} \right)
 \left(
\sum_i |\lambda_i|^4
 + (N-1) \sum_{i} |\lambda_i|^2 
+ \sum_{i\neq j} \lambda_i \bar\lambda_j  
 + \binom{N}{3}
 \right)\ .
\end{multline}
This shows that $\left< \operatorname{Tr} M^2 M^{\dagger^2} \right>$ can be written in terms of expectation values of functions of the eigenvalues $\lambda_i$.

\subsubsection{Multi-matrix models}\index{matrix model!multi---} 
 
The matrix models considered so far could be called \textbf{one-matrix models}\index{matrix model!one---}, as the corresponding integrals involved only one matrix. 
A natural generalization is to consider integrals over multiple matrices, and the corresponding multi-matrix models. 
For example, 
a \textbf{two-matrix model}\index{matrix model!two---} can be defined from the ensemble $E=H_N\times H_N$ and the measure
      \begin{equation}
	\boxed{ d\mu(M_1,M_2) =
	 e^{-\operatorname{Tr}\left( V_1(M_1) + V_2(M_2) - M_1 M_2 \right)}
	 dM_1 dM_2 }\ ,
       \end{equation}
where the functions $V_1$ and $V_2$ are the potentials.
Further generalizations include normal two-matrix ensembles $E= H_N(\gamma_1)\times H_N(\tilde \gamma_2)$ where $\gamma_1,\tilde \gamma_2$ are two contours, and linear combinations (homology classes) of such ensembles.
Another generalization is the \textbf{matrix chain}\index{matrix chain} with an ensemble $E=(H_N)^k$ or $E=\prod_{i=1}^k H_N(\gamma_i)$, and a measure
      \begin{equation}
	\boxed{ d\mu(M_1,\dots,M_k) =
	 e^{-\operatorname{Tr} \sum_{i=1}^k V_i(M_i) }
	 e^{  \operatorname{Tr} \sum_{i=1}^k M_i M_{i+1} }
	 \prod_{i=1}^k dM_i }\ .
       \end{equation}
The fixed matrix $M_{k+1}$ is called an \textbf{external field}\index{external field}, and breaks the invariance under conjugation.
See \autoref{chap:AngularIntegral} for calculations in the presence of an external field in the case $k=1$. 

\subsubsection{More general measures}

Interesting measures include measures of the type
\begin{equation}\label{eq:defRmeasures}
d\mu(M) = \prod_{i<j} R(\lambda_i,\lambda_j) \prod_i e^{-V(\lambda_i)} d\Lambda dU_\text{Haar}\ ,
\end{equation}
where $R(\lambda,\tilde \lambda)= R(\tilde\lambda,\lambda)$ is a symmetric function that vanishes at coinciding points with an exponent $\beta$:
\begin{equation}
R(\lambda,\tilde{\lambda}) \sim (\lambda-\tilde{\lambda})^{\beta} \left(1+O(\lambda-\tilde{\lambda})\right)\ .
\end{equation}
The case $R(\lambda,\tilde{\lambda}) = \frac{(\lambda-\tilde\lambda)^2}{(\lambda+\tilde\lambda)^n}$ yields the $O(n)$ matrix model of \autoref{sec:multi-matrix}, with the measure
\begin{multline}
 d\mu(M) 
 = e^{-\operatorname{Tr}V(M)}\det(\operatorname{Id}\otimes M + M\otimes \operatorname{Id})^{-\frac{n}{2}}dM\ ,
 \\
 = dU \,\frac{\prod_{i<j}(\lambda_i-\lambda_j)^2}{\prod_{i,j}(\lambda_i+\lambda_j)^{\frac n 2}} \prod_i e^{-V(\lambda_i)} d\lambda_i\ .
\label{eq:mon}
\end{multline}
Models for which $\frac{\partial}{\partial \lambda} \log R(\lambda,\tilde{\lambda})$ is a rational function of  $\lambda,\tilde \lambda$ usually have simple algebraic properties.
Another feature of $R(\lambda,\tilde{\lambda})$ that helps with model solving is when it is convex, i.e. when the following quadratic form on the space of positive measures $d\nu(\lambda)$ on $\gamma$ is positive definite:
\begin{equation}
 -\int_\gamma\int_\gamma  \log{\left| R(\lambda,\tilde \lambda) \right|} d\nu(\lambda)d\nu(\tilde \lambda)\ .
\end{equation}
This is because convexity implies the existence of a unique minimum.

%\begin{equation}
%  d\mu(M) = e^{-\operatorname{Tr}V(M)}\det(\Gamma(M))dM\ ,
%\end{equation}
%which appears in crystal growth and string theory. Here $\Gamma(M)$ is Euler's Gamma function, or some rational combination built from it.

\subsubsection{Matrix models for knot theory}

A \textbf{torus knot}\index{torus knot} $\mathfrak K_{P,Q}$, is a knot that can be drawn without self-intersection on a torus. 
It is characterized by two integers $(P,Q)$, representing the winding numbers around the longitude and the meridian of the torus.
These integers must be relatively prime, otherwise they do not characterize a knot, but a link with $\mathrm{ gcd}(P,Q)$ connected components.
$\mathfrak K_{1,1}$ is called the unknot, and $\mathfrak K_{3,2}$ is called the trefoil knot.
Let $q\in \mathbb C^*$ with $|q|<1$.

Let us consider the matrix model with the ensemble $H_N$ and the measure
\begin{align}\label{HOMFLY}
  d\mu_{P,Q}(M) 
  = e^{\frac{\operatorname{Tr} M^2}{2PQ\log q}}
  \det \left(
  f_P(\operatorname{Id}\otimes M,M\otimes \operatorname{Id})
  \right)
  \det \left(
  f_Q(\operatorname{Id}\otimes M,M\otimes \operatorname{Id})
  \right) dM\ ,
\end{align}
where
\begin{align}
f_P(x,x') = \frac{\sinh{\frac{x-x'}{2P}}}{x-x'}\ .
\end{align}
This measure is of the type \eqref{eq:defRmeasures}, with $R(\lambda,\tilde{\lambda}) = \sinh\frac{\lambda-\tilde\lambda}{2P} \sinh\frac{\lambda-\tilde\lambda}{2Q}$.
% The corresponding kernel $R(\lambda,\tilde\lambda) = (\lambda-\tilde\lambda)^2 f_P(\lambda,\tilde\lambda) f_Q(\lambda,\tilde\lambda) $ is always convex.
Then the expectation values of Schur polynomials $s_\mu$ of the exponentiated eigenvalues of our random matrix turn out to be the colored HOMFLY polynomials of the torus knot $\mathfrak K_{P,Q}$ \cite{Brini:2011},
\begin{align}\label{eq:HOMFLY}
 \frac{\left< s_\mu(e^{\lambda_1},\dots,e^{\lambda_N})\right>_{P,Q}}{\left< s_\mu(e^{\lambda_1},\dots,e^{\lambda_N})\right>_{1,1}}
 = \mathrm{HOMFLY}_\mu(\mathfrak K_{P,Q},SU_N,q)
\ ,
\end{align}
where the partition $\mu$ labels a representation of $SU_N$,
and the average $\left< \, \cdot \, \right>_{P,Q}$ is taken with respect to the measure $d\mu_{P,Q}(M)$.
For $N=2$, colored HOMFLY polynomials reduce to colored Jones polynomials. 
(See \autoref{exo:jones}.)

The invariants of more general knots can also be written as Vassiliev--Kontsevich matrix integrals, which become substantially more complicated when the knots are not torus knots.

\subsubsection{Algebraic submanifolds of $M_N(\mathbb{C})$ as matrix ensembles}

Many normal matrix ensembles (including the unitary ensemble) are algebraic submanifolds of $M_N(\mathbb{C})$, because conditions on eigenvalues can be written in terms of the characteristic polynomial. 
More general algebraic submanifolds can be used as matrix ensembles.

\subsubsection{Supermatrix models}\index{matrix model!super---} 

These models involve
matrices with fermionic entries~\cite{Alvarez-Gaume:1991ozb,Yost:1991ht} (see also~\cite{Efetov:1996SUSY,Kimura:2023iup}).
The relevant ensembles are a supermanifolds, for example the ensemble of super-Hermitian matrices.
The corresponding circular ensembles are supergroups.

\subsubsection{Classification by involutions: symmetric spaces}

There were many attempts of classifying matrix ensembles using $U_N$ symmetry together with a finite group of involutions such as 
\begin{itemize}
 \item symmetry: $M=M^\mathrm{T}$,
 \item conjugation: $M=\bar{M}$,
 \item chirality and spin symmetries, that we will not detail.
\end{itemize}
For example, the proposed classification~\cite{Caselle:2003qa,Zirnbauer:2011RMT,Bernard:2002SFT}
corresponds 
to the classification of symmetric spaces, and involves Dynkin diagrams.
The three Wigner ensembles belong to this classification, with the GUE corresponding to the Dynkin diagram $A_N$.
The classification of \cite{Bernard:2002SFT} contains 43 ensembles.

\section{Expectation values and correlation functions}
\label{sec:observables}

Let us consider random variables on matrix ensembles, which are called \textbf{observables}\index{observable} by physicists.
Some of the most useful observables in random matrix theory are expectation values, including in particular correlation functions.
It is useful to have observables that take values in $N$-independent spaces, so that we can study their large $N$ limits.

Observables can be built from eigenvectors and/or eigenvalues of random matrices.
Eigenvectors can be expressed algebraically in terms of eigenvalues, indeed if $\lambda= \text{eigenvalue of }M$, an eigenvector $v=(v_1,\dots,v_N)$ is given by Cramer's rule on a submatrix:
\begin{equation}
v_i = (-1)^i\operatorname{Minor}_{n,i}(\lambda \text{Id}-M).
\end{equation}
For this reason, research has mostly focused on eigenvalues for a long time. 
Nowadays, applications to big-data science and neural networks have shifted the focus towards eigenvectors.
In this text, we mostly concentrate on observables built from eigenvalues.

\subsection{Polynomial observables}
\label{sec:powersums}

\subsubsection{Power sum and Schur polynomials}

An important class of observables are the invariant polynomials of the matrix $M$. 
Functions of $M$ that are invariant under conjugations are symmetric functions of the eigenvalues of $M$.
Invariant polynomials of $M$ are symmetric polynomials of its eigenvalues.

For a \textbf{partition}\index{partition} $\mu=(\mu_1,\dots,\mu_\ell)$, i.e. a sequence of integers such that $\mu_1\geq \mu_2\geq \dots \geq \mu_\ell \ge 0$, we define
\begin{equation}
p_\mu(M)  = \operatorname{Tr} M^{\mu_1}  \operatorname{Tr} M^{\mu_2} \dots \operatorname{Tr} M^{\mu_\ell}\ .
\end{equation}
The corresponding function of the eigenvalues $\lambda_1,\dots \lambda_N$ is the  \textbf{power sum
polynomial}\index{power sum polynomial}
\begin{align}
 p_\mu(\lambda_1,\cdots \lambda_N)
  \overset{\text{def}}{=} \prod_{j=1}^\ell \sum_{i=1}^N \lambda_i^{\mu_j}\ .
\end{align}
The \textbf{length}\index{length (partition)} of $\mu$ is $\ell(\mu) = \#\{ i \mid \mu_i>0 \} = \mu^\mathrm{T}_1$ where $\mu^\mathrm{T}$ is the transposed partition, and its \textbf{weight}\index{weight (partition)} is $|\mu|=\sum_{i=1}^\ell \mu_i$.
Partitions are  graphically represented using \textbf{Young
diagrams}\index{Young diagram}:
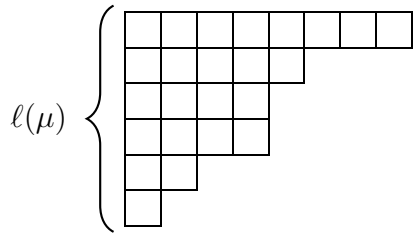
\begin{align}
 \begin{tikzpicture}[baseline=(current bounding box.center)]
  \node at (2,0) [right] {$\yng(8,5,4,4,2,1)$};
  \node at (.5,0) [right] {$\ell(\mu)$};
  \draw [decorate,decoration={mirror,brace,amplitude=10pt},thick]
  (2,1.5) -- (2,-1.5);
 \end{tikzpicture}
\end{align}
In this example we represented the partition
 $\mu=(8,5,4,4,2,1)$, with $\ell(\mu)=6$ (number of rows), and $|\mu|=24$ (number of boxes).
Let $\mathcal P$ be the set of all partitions, and $\mathcal P_N=\{\mu\in\mathcal P \mid \ell(\mu)\leq N\}$ the set of partitions with at most $N$ rows.
 
Power sum polynomials  $(p_\mu)_{\mu \in\mathcal{P}_N}$ provide a basis of the $\mathbb C$-vector space $\mathcal V_N$ of symmetric polynomials of $N$ variables. Another basis of this infinite-dimensional vector space is given by the
 \textbf{Schur polynomials}\index{Schur polynomial}
\begin{equation}
s_{\mu}(\lambda_1,\dots,\lambda_N)
 = \frac{\det_{1\leq i,j\leq N} \lambda_i^{\mu_j-j+N}}{\det_{1\leq i,j\leq N} \lambda_i^{N-j}}\ ,
\end{equation}
where we completed the partition $\mu$ by $\mu_j=0$ if $\ell<j\leq N$, and $\ell(\mu)>N \implies s_\mu = 0$.
Both the numerator and the denominator are antisymmetric polynomials, in particular the denominator is the Vandermonde determinant.
Expectation values of Schur polynomials appear in knot invariants \eqref{eq:HOMFLY},
see \autoref{exo:jones}.
 
\subsubsection{Cumulants}

Let us define a two-point cumulant to be the covariance
\begin{equation}
\left< \operatorname{Tr} M^{\mu_1} \operatorname{Tr} M^{\mu_2}\right\rangle_\text{c}
\stackrel{\text{def}}{=} \left< \operatorname{Tr} M^{\mu_1} \operatorname{Tr} M^{\mu_2}\right\rangle - \left< \operatorname{Tr} M^{\mu_1}\right\rangle \left< \operatorname{Tr} M^{\mu_2}\right\rangle\ .
\end{equation}
The covariance can be generalized, and we define $n$-point \textbf{cumulants}\index{cumulant} $\left<p_\mu\right>_\text{c}$ by the triangular system of equations
\begin{equation}
\left< p_{(\mu_1,\dots,\mu_\ell)} \right\rangle 
\stackrel{\text{def}}{=} \sum_{k=1}^\ell\ \sum_{\sqcup_{j=1}^k I_j=\{\mu_1,\dots,\mu_\ell\}}\
\prod_{j=1}^k \left< p_{I_j} \right>_\text{c} \ .
\end{equation}
In words, the expectation values of a product of factors $p_\mu = \prod_{j=1}^\ell \operatorname{Tr} M^{\mu_j}$, is the sum over partitions of the product into sub-products, of products of cumulants:
\begin{subequations}
\begin{align}
 \langle p_{(\mu_1)} \rangle &= \langle p_{(\mu_1)} \rangle_\text{c}\ ,
 \\
 \langle p_{(\mu_1,\mu_2)} \rangle &= \langle p_{(\mu_1,\mu_2)} \rangle_\text{c} + \langle p_{(\mu_1)} \rangle \langle p_{(\mu_2)} \rangle\ ,
\end{align}
\begin{multline}
 \langle p_{(\mu_1,\mu_2,\mu_3)}\rangle =
 \langle p_{(\mu_1,\mu_2,\mu_3)}\rangle_\text{c} 
 + \langle p_{(\mu_1)}\rangle \langle p_{(\mu_2,\mu_3)}\rangle_\text{c} + \langle p_{(\mu_2)}\rangle \langle p_{(\mu_1,\mu_3)}\rangle_\text{c} + \langle p_{(\mu_3)}\rangle \langle p_{(\mu_1,\mu_2)}\rangle_\text{c}
 \\
+ \langle p_{(\mu_1)} \rangle \langle p_{(\mu_2)} \rangle \langle p_{(\mu_3)} \rangle\ .
\end{multline}
\end{subequations}
This system of equations is triangular and can be inverted, for example for $n=3$, the inverse relation reads
\begin{multline}
\langle p_{(\mu_1,\mu_2,\mu_3)}\rangle_\text{c}
= \langle p_{(\mu_1,\mu_2,\mu_3)} \rangle
- \langle p_{(\mu_1)}\rangle \langle p_{(\mu_2,\mu_3)}\rangle - \langle p_{(\mu_2)}\rangle \langle p_{(\mu_1,\mu_3)}\rangle - \langle p_{(\mu_3)}\rangle \langle p_{(\mu_1,\mu_2)}\rangle
\\
+2 \langle p_{(\mu_1)} \rangle \langle p_{(\mu_2)} \rangle \langle p_{(\mu_3)} \rangle\ .
\end{multline}

\subsection{Correlation functions and eigenvalue density}\label{sec:cfed}

\subsubsection{Correlation functions}

While polynomial observables depend on integers $\mu_j$, the same information can be encoded in analytic functions of auxiliary variables $x_j$ called \textbf{spectral parameters}\index{spectral parameter}. (Other possible names are catalytic variables or coupling parameters.)
For example, the traces $\operatorname{Tr}M^\mu$ are encoded in a generating function, called the \textbf{resolvent}\index{resolvent},
\begin{align}
W_1(x) 
=  \left< \sum_{i=1}^N \frac{1}{x - \lambda_i} \right>  
= \left< \operatorname{Tr}\frac{1}{x-M} \right> = \sum_{\mu=0}^\infty x^{-\mu-1} \left< \operatorname{Tr}M^\mu \right> \ ,
 \label{eq:trsum}
\end{align}
where $\lambda_1, \cdots \lambda_N$ denote the eigenvalues of $M$, and the last equality holds if $|x|>\max|\lambda_i|$ almost surely.
If $M\in H_N(\gamma)$,
the resolvent is analytic in the domain $x\in\mathbb C\setminus\gamma$.

More generally, we define the (disconnected) \textbf{correlation functions}\index{correlation function}, and their cumulants called the \textbf{connected correlation functions}\index{correlation function!connected---},
\begin{subequations}
\begin{empheq}[marginbox = \fbox]{align}
  W_n(x_1,\cdots x_n) & =
 \left(\frac{\beta}{2}\right)^\frac{n}{2} \left< 
 \operatorname{Tr} \frac{1}{x_1-M} \cdots  
 \operatorname{Tr} \frac{1}{x_n-M}
 \right>_\text{c} 
\label{wn_conn}
 \, , \\
\widehat{W}_n(x_1,\cdots x_n) & =
 \left(\frac{\beta}{2}\right)^\frac{n}{2} \left< 
 \operatorname{Tr} \frac{1}{x_1-M} \cdots  
 \operatorname{Tr} \frac{1}{x_n-M}
 \right> 
\label{wn_disconn}
 \, .
\end{empheq}
\end{subequations}
For later use we included a $\beta$-dependent prefactor that is nontrivial if $\beta\neq 2$, although for the moment we assume $\beta =2$.
In particular we have $\widehat{W}_1(x) = W_1(x)=\left<\operatorname{Tr}\frac{1}{x-M}\right> $, and 
\begin{align}
 \widehat{W}_2(x_1,x_2) = W_1(x_1)W_1(x_2) + W_2(x_1,x_2)\ .
\end{align}
The definition of correlation functions determines their leading behaviour for $x_i\to \infty$,
\begin{equation}
\boxed{ W_{n}(x_1,\dots,x_n) \underset{x_1\to\infty}{=} \frac{N\delta_{n,1}}{x_1} + O\left(\frac{1}{x_1^2}\right)  } \ .
\label{eq:wninf}
\end{equation}
Polynomial observables can then be recovered from large $x$ expansions, for example
\begin{subequations}
\begin{align}
 \left\langle \operatorname{Tr} M^\mu \right\rangle
 & = -\underset{x \to \infty}{\operatorname{Res}} \, W_1(x) x^\mu dx\ ,
\\
 \left\langle p_\mu(M) \right\rangle
 & = (-1)^{\ell(\mu)}
 \underset{x \to \infty}{\operatorname{Res}} \cdots \underset{x \to \infty}{\operatorname{Res}} \,
 \widehat{W}_{\ell(\mu)}(x_1, \dots ,x_{\ell(\mu)}) x_1^{\mu_1} \dots x_{\ell(\mu)}^{\mu_{\ell(\mu)}} dx_1 \dots dx_{\ell(\mu)}\ , 
\\
 \left\langle p_\mu(M) \right\rangle_\text{c}
 & = (-1)^{\ell(\mu)}
 \underset{x \to \infty}{\operatorname{Res}} \cdots \underset{x \to \infty}{\operatorname{Res}} \,
 W_{\ell(\mu)}(x_1, \dots ,x_{\ell(\mu)}) x_1^{\mu_1} \dots x_{\ell(\mu)}^{\mu_{\ell(\mu)}} dx_1 \dots dx_{\ell(\mu)}\ .
\end{align}
\end{subequations}

\subsubsection{The eigenvalue density and its Stieltjes transform}

We define the \textbf{eigenvalue density}\index{eigenvalue density} as the expectation value of the empirical density
\begin{align}
 \boxed{ \rho (x)  = 
 \left<
  \frac{1}{N} \sum_{i=1}^N \delta(x - \lambda_i)
 \right> }\ ,
 \label{den_dist01}
\end{align}
so that $\rho(x)dx$ is the expected probability of having eigenvalues in an interval $dx$ centered at $x$.
Up to a prefactor $N$, the resolvent coincides with the 
 \textbf{Stieltjes transform}\index{Stieltjes transform} $W(x)$ of the eigenvalue density,
\begin{equation}
W_1(x) = N W(x) \quad \text{with} \quad 
 W(x) = \int_{\operatorname{supp} \rho} \frac{\rho(x') dx'}{x-x'}  \, .
\end{equation}
Since the eigenvalue density obeys $\int_{\operatorname{supp} \rho} \rho(x') dx'=1$, its Stieltjes transform has the asymptotic behaviour
\begin{equation}
 W(x) \underset{x\to \infty}{=} \frac{1}{x} + O\left(\frac{1}{x^2}\right)\, .
 \label{eq:omegainf}
\end{equation}
The resolvent $W_1(x)$, and actually all resolvent correlation functions, are analytic on $\mathbb{C}\backslash\operatorname{supp} \rho$, and singular on $\operatorname{supp} \rho$. 
The eigenvalue density can then be recovered from the behaviour of its Stieltjes transform at its singularities. 
In particular, a simple pole $\lambda_i$ of $W_1(x)$ corresponds to a contribution $\delta(x - \lambda_i)$ to the eigenvalue density.
For any closed contour $\mathcal C$ we have 
\begin{align}
\frac{1}{2\pi i}\oint_{\mathcal C} W_1(x) dx 
 = \Big< \# \text{eigenvalues enclosed by }\ \mathcal C \Big>  \, .
\end{align}
And if the resolvent is discontinuous on a contour $\gamma$, then $\gamma\subset \operatorname{supp} \rho$, and the eigenvalue density along $\gamma$ is proportional to the discontinuity of its Stieltjes transform
\begin{equation}
\boxed{ \rho(x) \underset{x\in\gamma}{=} -\frac{1}{2\pi i N} \left( W_1(x+i0) - W_1(x-i0) \right)}\ .
\label{eq:rewo}
\end{equation}
This relation amounts to writing $\delta(x-\lambda_i) = \mp \frac{1}{\pi} \Im \frac{1}{x-\lambda_i \pm i 0}$ for each eigenvalue $\lambda_i$.
The Stieltjes transform is much more regular than the eigenvalue density: the former is an analytic function, the latter can a priori be a distribution. Accordingly, in the large $N$ limit, the eigenvalue density converges only in law, whereas its Stieltjes transform converges pointwise.

If the eigenvalues are viewed as electric charges, then the two-dimensional vector $\frac{1}{N}(\Im W_1(x),-\Re W_1(x))$ is the electrostatic field produced by the charge distribution at point $x$. 
For example, let us draw the electrostatic field in the case where the potential $V$ is quadratic, for which the eigenvalue distribution is the semi-circle law \eqref{sc_law}. In the large $N$ limit, this field is given by Eq.~\eqref{eq:bwsc}, and the eigenvalue density is supported on the red segment $[-2,2]$:
% electrostatic field
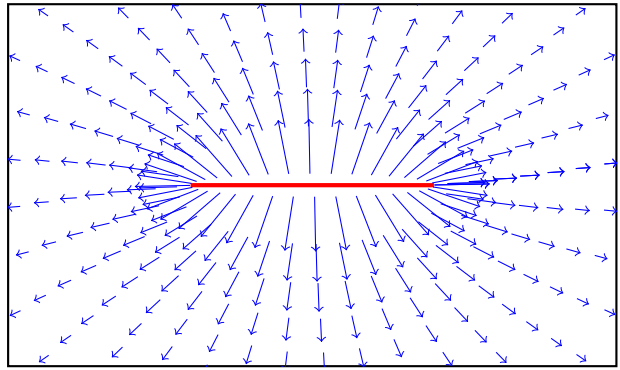
\begin{align}
 \begin{tikzpicture}[scale = .8,baseline=(current bounding box.center)]
   \draw [thick] (-5,-3) rectangle(5,3);
\clip (-5,-3) rectangle(5,3);
\draw[ultra thick, red] (-2,0) -- (2,0);
\foreach \r in {1.1,1.6,...,5.5}{
\foreach \a in {0,10,...,360}{
\draw[->,blue] ({(\r+1/\r)*cos(\a+\r)},{(\r-1/\r)*sin(\a+\r)})  -> ({(\r+1/\r)*cos(\a+\r)+cos(\a+\r)/\r},{(\r-1/\r)*sin(\a+\r)+sin(\a+\r)/\r});
}
}
 \end{tikzpicture}
\end{align}

\subsection{Wave functions}
\label{sec:charpol}

Let the \textbf{wave function}\index{wave function} $\psi(x)$ be the expectation value of the characteristic polynomial,
\begin{equation}
\psi(x) = \langle \det (x-M) \rangle \ .
\end{equation}
Using $\det(x-M) = x^N \exp {\operatorname{Tr} \log(1-\frac{M}{x})} = x^N \exp {\int_{\infty}^x dx'\operatorname{Tr}\left(\frac{1}{x'-M}-\frac{1}{x'}\right) }$, 
we can rewrite $\det(x-M)$ in terms of resolvents,
\begin{equation}
\psi(x)
 = x^N \left\{ 1 + \sum_{n=1}^\infty \frac{1}{n!}\left(\prod_{i=1}^n\int_{\infty}^x dx'_i\right) \left\langle \prod_{i=1}^n\operatorname{Tr} \left( \frac{1}{x'_i-M}-\frac{1}{x'_i}\right)  \right\rangle \right\}\ .
\end{equation}
This linear combination of disconnected correlation functions $\widehat{W}_n$ can be rewritten as the exponential of a linear combination of connected correlation functions $W_n$, after regularizing $W_1$ by subtracting a term $\frac{N}{x'_1}$,
\begin{equation}\label{eq:psiexpWn1}
\boxed{
\psi(x)
 = x^N \exp  \sum_{n=1}^\infty \frac{1}{n!}\left(\prod_{i=1}^n\int_{\infty}^x dx'_i\right)  \left(W_n(x'_1,\dots,x'_n)-\delta_{n,1}\frac{N}{x'_1} \right)\, .
}\end{equation}
This can be generalized to products or ratios of characteristic polynomials. In particular, in combinations with the same numbers of polynomials in the numerator as in the denominator, the infinite integration bound and the associated regularizing terms disappear. The simplest combination of this type is 
\begin{equation}
\psi(x;y) = 
\left\langle \frac{\det (x-M)}{\det(y-M)} \right\rangle
 = \exp\left\{  \sum_{n=1}^\infty \frac{1}{n!}\left(\prod_{i=1}^n\int_{y}^x dx'_i\right) W_n(x'_1,\dots,x'_n) \right\}\, .
\end{equation}
The more general combination
\begin{align}
\psi(x_1,\dots,x_k;y_1,\dots,y_k)
=\left\langle \prod_{j=1}^k \frac{\det (x_j-M)}{\det(y_j-M)} \right\rangle \ ,
\end{align}
is obtained from our expression for $\psi(x;y)$ by the replacement $\int_y^x \to \sum_{j=1}^k \int_{y_j}^{x_j}$.
Even more generally, consider a \textbf{divisor}\index{divisor} $D=\sum_{i} \alpha_i\cdot  [x_i] $, i.e. a formal linear combination of finitely many points $x_i\in\mathbb{C}$, whose degree is 
$\deg D=\sum_i \alpha_i$. Assuming $\deg D=0$, the integral
\begin{equation}
\int_D f(x')dx' \stackrel{\mathrm{ def}}{=} \sum_i \alpha_i \int_{x'=o}^{x_i} f(x')dx'\ ,
\end{equation}
does not depend on the point $o$. We then define the wave function 
\begin{align}
\psi(D)
 \overset{\text{def}}{=}\left\langle \prod_{j} \det (x_j-M)^{\alpha_j} \right\rangle  
= \exp\left\{  \sum_{n=1}^\infty \frac{1}{n!}\left(\prod_{i=1}^n\int_D dx'_i\right)  W_n(x'_1,\dots,x'_n)\right\}\ .
\label{eq:psid}
\end{align}
Values of $\psi(D)$ for $\deg D\neq 0$ are obtained in a limit $x_i\to \infty$, where the integral of $W_1$ should be regularized as in the case of $\psi(x)$ \eqref{eq:psiexpWn1}.

Conversely, one can also recover correlation functions from wave functions. To do this, we should first recover the resolvent from the characteristic polynomial, using $\operatorname{Tr} \frac{1}{x-M} = \underset{\epsilon =0}{\operatorname{Res}} \frac{1}{\epsilon^2} \frac{\det(x+\epsilon-M)}{\det(x-M)}$. This leads to 
\begin{equation}
W_1(x) = \underset{\epsilon= 0}{\operatorname{Res}}\, \frac{1}{\epsilon^2} \psi([x+\epsilon]-[x])\ ,
\end{equation}
and in general to
\begin{equation}\label{eq:Wnhatpsi}
\boxed{\widehat{W}_n(x_1,\dots, x_n) = \underset{\epsilon_1= 0}{\operatorname{Res}}\, \frac{1}{\epsilon_1^2}\dots 
\underset{\epsilon_n= 0}{\operatorname{Res}}\, \frac{1}{\epsilon_n^2}
\psi\left(\textstyle{\sum_{i=1}^n} ([x_i+\epsilon_i] - [x_i]) \right)
}\ .
\end{equation}
The relations \eqref{eq:Wnhatpsi} and \eqref{eq:psiexpWn1} between correlation functions and wave functions 
are inverses of one another. 
These relations are equivalent to the Hubbard--Stratonovich boson-fermion correspondence, considering resolvents as bosonic and characteristic polynomials as fermionic.
  
\subsection{Other observables}

We mention a few more useful observables, that we will however not discuss in detail:

\begin{itemize}

\item The Fourier transforms $\left\langle \operatorname{Tr} e^{i p_1 M} \dots \operatorname{Tr} e^{i p_n M}\right\rangle$ are trigonometric analogs of moments, and  
can be very useful for circular ensembles, whose eigenvalues lie on a circle.

\item Expectation values of determinants of functions of the matrix,
see \autoref{exo:volun} for the case of a Theta function.

\item Multi-matrix observables
include expectation values of mixed traces $\operatorname{Tr} (M_1^k M_2^l)$ and of the corresponding resolvents $\operatorname{Tr} \frac{1}{x_1-M_1}\frac{1}{x_2-M_2}$, more generally
$\operatorname{Tr} \prod_j \frac{1}{x_j-M_{i_j}}$. These expressions cannot be written in terms of the eigenvalues of the matrices $M_i$, they involve angular integrals introduced in \autoref{chap:AngularIntegral}.

\end{itemize}

\section{Universal properties}\label{sec:defuni}

A property is called \textbf{universal}\index{universality} if it only depends on the matrix ensemble, and not (or almost not) on the measure. 

Let us focus on the case of Hermitian matrices, with a measure built from a potential $V$ as in 
\eqref{eq:potentialmeasure}.  
If a property is universal, it is independent of the potential $V$, thus it can be studied in the Gaussian matrix integral, whose potential is quadratic.

More generally, we say that a property is universal, if it depends only on a class of potentials and is independent of $V$ within the class. For example, all $V\in\mathbb R[x] $ convex on $\mathbb R$ will be in the same 1-cut class.

Universality will arise in limits where the matrix size $N$ is large, under specific assumptions on the large $N$ behaviour of the potential $V$. 
From now on, we rescale  $V\to N V $, and rewrite our matrix integral as 
\begin{align}
 \boxed{ \mathcal{Z} = \int_{H_N} dM\ e^{-N\operatorname{Tr}V(M)} } \ .
\end{align}
This rescaling of the potential is such that
both factors $dM$ and 
$e^{-N\operatorname{Tr} V(M)}$ behave as $O(e^{N^2})$: the Lebesgue measure $dM$ \eqref{eq:betaLebesguemeasure} is the product of Lebesgue measures for the $N^2$ real components of $M$, the trace
is a sum of $N$ terms, and $N\operatorname{Tr} V(M) = O(N^2)$. 
This allows us to expect an interesting and nontrivial large $N$ behaviour, as the entropy and energy are of the same order of magnitude.
If we chose a larger scaling for the potential, the matrix $M$ would freeze at the bottom of the potential wells. If we chose a smaller scaling, the matrix $M$ would flow away until reaching a region where the potential is of order $N$.
There are however exotic potentials that require prefactors such as $N^\alpha (\log N)^\beta$ with $\alpha \neq 1$ and/or $\beta\neq 0$. In supermatrix models, $dM$ effectively has fewer components due to supersymmetric cancellations, and it can happen that no prefactor is needed.

Most known universal properties of matrix models are related to
integrable systems. 
For example, universal eigenvalue spacing distributions are typically related to Tau functions of integrable systems. 
This is remarkable, because systems described by matrix models, such as chaotic billiards, are typically not integrable. 
A tentative explanation is that the statistical averaging, thanks to ergodicity, has the effect of integrating out all non-integrable degrees of freedom, and leaving only integrable degrees of freedom.
This would explain why integrable systems, rather than being exceptional, are in fact commonly found in nature.

Let us now give an overview of the most important universal large $N$ regimes.

\subsection{Macroscopic limits} \label{sec:macro}

\subsubsection{Equilibrium density and spectral curve}

The eigenvalue density often has a finite limit
\begin{equation}
 \boxed{ \bar{\rho}(x)
  = \lim_{N \to \infty} \rho(x)} \, ,
  \label{den_dist02}
\end{equation}
called the \textbf{equilibrium density}\index{equilibrium density}.
If the potential $V$ is rational, the equilibrium density exists under some conditions on the behaviour of $V$ near
its singularities.
In particular, it exists if $V$ is real and more confining than $\log|x|^2$ at infinity, i.e. $ \lim_{x\to\infty} \left(V(x)-\log |x|^2\right)  = +\infty$.
The convergence of the equilibrium density is in law rather than pointwise, in other words the equilibrium density is only a weak limit: 
$ \lim_{N \to \infty} \int f(x)\rho(x) dx = \int f(x) \bar{\rho}(x)dx
$ for any bounded, continuous, $N$-independent test function $f(x)$.
On the other hand, the resolvent is an integral transform of the eigenvalue density, and does have a pointwise large $N$ limit. 

The equilibrium density is not universal. 
It depends on the potential $V$, and on the eigenvalue integration paths.
In the case of a Gaussian potential $V(M) =
\frac{1}{2} M^2$, it is given by Wigner's semi-circle law.
The semi-circle law is the limit of an eigenvalue density that has oscillations with frequency $O(N)$ and amplitude $O(\frac{1}{N})$ on the semi-circle's support, and decreases exponentially outside the support:
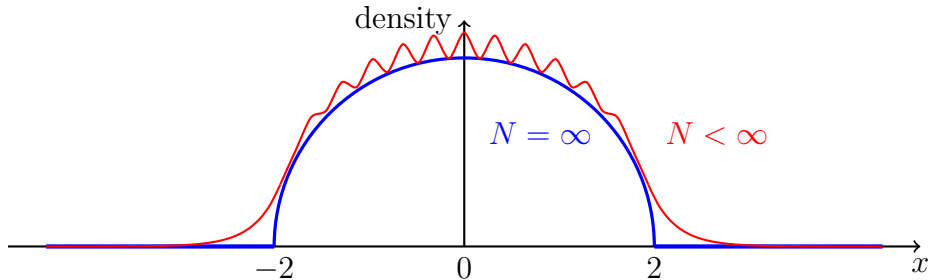
\begin{align}
 \begin{tikzpicture}[baseline=(current  bounding  box.center)]
  \draw[thick, -latex] (-6,0) -- (6,0) node[below]{$x$};
  \draw[thick, -latex] (0,0) node[below]{$0$} -- (0,3) node[left]{density};
  \draw[very thick, blue] (2.5,0) arc[radius = 2.5, start angle = 0, end angle = 180];
  \draw[ultra thick, blue] (-5.5, 0) -- (-2.5, 0);
  \draw[ultra thick, blue] (2.5, 0) -- (5.5, 0);
  \draw (2.5, 0) node[below]{$2$};
  \draw (-2.5, 0) node[below]{$-2$};
 \draw[domain=-2.42:2.42,smooth,variable=\x,red, thick] plot ({\x},{sqrt(6.25-\x*\x) + (1+cos((15*\x) r))/6});
 \draw[domain=-5.5:-2.42,smooth,variable=\x,red, thick] plot ({\x},{.75*exp(3*(\x+2.45))});
 \draw[domain=2.42:5.5,smooth,variable=\x,red, thick] plot ({\x},{.75*exp(3*(-\x+2.45))});
  \draw (1,1.5) node [blue] {$N = \infty$};
  \draw (2.5,1.5) node [red, right] {$N < \infty$};
 \end{tikzpicture}
\end{align}

More generally, for rational potentials, the equilibrium density is usually the square root of a polynomial $\bar\rho(x) = \sqrt{U(x)}$. It has a compact support where $U(x)>0$, and boundaries called the \textbf{spectral edges}\index{spectral edge} where $U(x)=0$.
Then $y=2\pi i\bar{\rho}(x)$ 
obeys an algebraic equation of the type
\begin{align}
P(x,y) = 0 \quad \text{with}  \quad P\in \mathbb{R}[x,y] \quad \text{and} \quad     \deg_yP=2\ .
 \label{eq:spectralcurve}
\end{align}
(Multi-matrix models can actually give rise to polynomials $P$ with $\deg_y P>2$.)
Now, instead of $(x,y)\in\mathbb{R}^2$, consider the zero locus in $\mathbb{C}^2$ of the same algebraic equation 
\begin{equation}
\{(x,y) \, |\, P(x,y)=0 \} \subset \mathbb C^2 \ .
\end{equation}
This defines a Riemann surface immersed in $\mathbb C^2$,  called the
\textbf{spectral curve}\index{spectral curve} of the model.
The spectral curve can be described using algebraic geometry, and plays a key role in random matrix theory, and more generally in integrable systems.

\subsubsection{Correlation functions}

Let us then consider the connected two-point
function in the large $N$ limit,
\begin{subequations}
\begin{align}
 \rho_2(x_1, x_2)
 & =
 \left<
  \sum_{i,j}^N \delta(x_1 - \lambda_i) \delta(x_2 - \lambda_j)
 \right>
 -
 \left< \sum_{i=1}^N \delta(x_1 - \lambda_i) \right>
 \left< \sum_{i=1}^N \delta(x_2 - \lambda_i) \right> \, , \\
  \bar{\rho}_2 (x_1,x_2)
 & = \lim_{N \to \infty} \rho_2(x_1, x_2) \, .
 \label{2pt_func01}
\end{align}
\end{subequations}
In many cases, this limit exists as a weak limit, and is an algebraic function.
More specifically,
as we will see in Chapters \ref{chap:SaddlePoint} and \ref{chap:LoopEq}, in the algebraic geometry language, $\bar{\rho}_2 (x_1,x_2)$ is the fundamental second kind differential on the spectral curve.
This only depends on the complex
structure of the spectral curve, and not on its ($V$-dependent) immersion into $\mathbb C\times \mathbb C$.
In this sense, the two-point function is a universal quantity.

Let us consider the large $N$ connected $n$-point function
\begin{align}
 \bar{\rho}_n (x_1, \ldots, x_n)
  = \lim_{N \to \infty}
 N^{n-2}
 \left<
  \prod_{k=1}^n
  \left(
   \sum_{i=1}^N \delta(x_k - \lambda_i)
  \right)
 \right>_\text{c}
 \, .
\end{align}
The scaling factor $N^{n-2}$ ensures the existence of a large $N$ weak limit in many cases.
This factor
is determined by order counting in Feynman graphs. (See \autoref{chap:FeynmanGraph}.)
The large $N$ connected $n$-point function, when it exists, has a universal expression in terms of algebro-geometric properties of the spectral curve. This expression can be computed using the topological recursion equation. (See \autoref{chap:LoopEq}.)

\subsubsection{Asymptotic expansions}

In many cases, the density function (\ref{den_dist01}) has not only a large ${N}$ limit, but also an asymptotic expansion of the type
\begin{align}
 \rho(x) 
 & =
 \sum_{g=0}^\infty N^{-2g} \bar{\rho}^{(g)}(x)
 + \sum_k N^{-a_k} e^{-N A_k(x)}
 \, .
\end{align}
The terms with powers of $\frac{1}{N}$ are called \textbf{perturbative}\index{perturbative term}, while the terms with exponentials are called non-perturbative.
That only even powers of $\frac{1}{N}$ appear is a feature of the Hermitian ensemble: odd powers would appear in the real symmetric or  quaternionic cases.
This can be understood in terms of the duality \eqref{beta_duality} of $\beta$-matrix models, which amounts to $H_N \leftrightarrow H_{-N}$ in the case $\beta=2$, and to $S_{-2N} \leftrightarrow Q_N$ in the cases $\beta=1, 4$.

In the expansion, all perturbative and non-perturbative terms except $\bar{\rho}(x) = \bar{\rho}^{(0)}(x)$ are universal, and have universal expressions in terms of the spectral curve. 
For example, as we will see in \autoref{chap:LoopEq} the functions $A_k(x)$ are periods of the spectral curve.
These functions must obey $\Re A_k\geq 0 $. If $A_k \in i \mathbb R$, then the corresponding term is called \textbf{oscillatory}\index{oscillatory term}.

Similarly, the expansions of higher correlation functions typically take the form
\begin{equation}
\rho_n(x_1,\dots,x_n) = \sum_{k=0}^\infty N^{-k} \bar{\rho}_n^{(k)}(x_1,\dots,x_n) + \sum_k N^{-a_{k}} e^{-N A_{n,k}(x_1,\dots,x_n)}\ ,
\end{equation}
where $\bar{\rho}_n^{(k)}$ is an algebraic function of $n$ variables, which again has a  universal expression in terms of the large $N$ one- and two-point functions, as we will see in \autoref{chap:LoopEq}.

Let us now discuss the large $N$ behaviour of the partition function.
In many cases, it is not the partition function itself, but the \textbf{free energy}\index{free energy} $F=-\log\mathcal{Z}$, that has a nice expansion of the form 
\begin{equation}
F = N^2 F_0 + N^0 F_1 + N^{-2} F_2 + \sum_{g=3}^\infty N^{2-2g} F_g + \sum_k C_k N^{-a_k} e^{-N f_k}\ .
\label{eq:fene}
\end{equation}
The leading term $F_0$ is not universal.
However, all the higher order terms are universally expressed in terms of the geometry of the spectral curve.
For example $F_1$ is proportional to the logarithm of the Laplacian determinant on the spectral curve.

\subsection{Microscopic limits}\label{sec:micro}

\subsubsection{Bulk microscopic limit}\index{microscopic limit!bulk---}

For a point $x\in \operatorname{supp} \bar\rho $ such that $\bar{\rho}(x)\neq 0$, we define the bulk microscopic limit of the two-point function as 
\begin{equation}
 \bar{\rho}_2^\text{ micro, bulk}(x, s) = \frac{1}{\bar\rho(x)^2} \lim_{N\to \infty} \rho_2\left(x, x+\frac{s}{N}\right)\ .
\end{equation}
This has a universal expression in terms of the equilibrium density, 
\begin{align}
 \bar{\rho}_2^\text{ micro, bulk}(x,s)
 =
 -\left(\frac{\sin \pi s \bar{\rho}(x)}{\pi s \bar{\rho}(x)}\right)^2
  \, .
\end{align}
This expression is related to the \textbf{sine kernel}\index{kernel!sine---},
\begin{align}
 K(x_1, x_2) = \frac{\sin(x_1-x_2)}{x_1-x_2}\ ,
\end{align}
from which the eigenvalue spacing distribution $P(s)$ can be computed via the \textbf{Fredholm determinant}\index{Fredholm determinant} of a functional operator:
\begin{align}
 P(s) \sim - \frac{\partial^2}{\partial s^2}\det(\operatorname{Id}-\hat{K}_s)\ , \quad 
 \hat{K}_s(f)(x) = \int_{-s}^sK(x,x')f(x')dx'\ .
\end{align}
This provides the exact expression for $P(s)$, of which the Wigner surmise gives an approximation.
Now the sine kernel is an \textbf{integrable kernel}\index{kernel!integrable---}, in the sense that it can be written as 
\begin{equation}
K(x_1,x_2) = \frac{\sum_{i,j} A_{i,j} f_i(x_1)\tilde f_j(x_2)}{x_1-x_2}\ ,
\end{equation}
for some matrix $A$ of finite size (here $2$) and some functions $f_i$. 
And Fredholm determinants of integrable kernels are Tau functions of integrable systems.
In particular, the Fredholm determinant of the sine kernel is the Tau function of the Painlev\'e V integrable system.
(See \autoref{chap:ortho}.)

\subsubsection{Edge microscopic limits}\index{microscopic limit!edge---}

Let us consider limits of correlation functions whose arguments $x_i$ are close to a spectral edge $a$. 
The definition of an edge microscopic limit will depend 
on the behaviour of the equilibrium density $\bar{\rho}(x)$, which is characterized by the algebraic equation \eqref{eq:spectralcurve}.
For a generic potential $V$, we have \textbf{regular edges}\index{regular edge}, such that $\bar\rho(x)\underset{x\to a}{\propto} \sqrt{x-a}$. 
For special choices of the potential, we can have $\bar\rho(x) \underset{x\to a}{\propto} (x-a)^\frac{p}{2}$ for some positive integer $p$.
The half integer exponent $\frac{p}{2}$ comes from the quadratic nature of matrix model's spectral curve. In more general multi-matrix models, the equilibrium density $\bar{\rho}(x)$ may satisfy a higher degree equation \eqref{eq:spectralcurve},
and we may obtain a \textbf{critical edge}\index{critical edge} with an arbitrary rational exponent
\begin{align}\label{eq:critedgebehavior}
 \bar{\rho}(x)\underset{x\to a}{\propto} (x-a)^\frac{p}{q}\ ,
\end{align}
where $p$ and $q$ are positive integers:
 
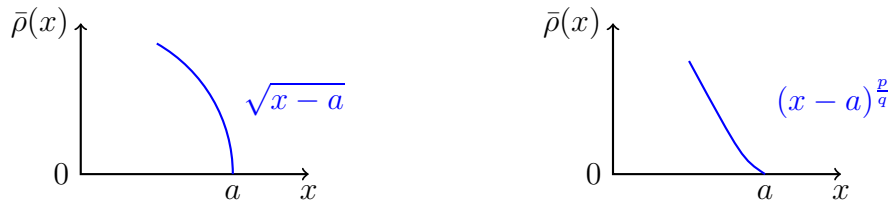
\begin{align}
  \begin{tikzpicture}[baseline=(current bounding box.center)]   
   % x-axis
   \draw [thick,latex-latex] (-5, 2) node[left] {$\bar{\rho}(x)$} -- (-5,0) node[left] {$0$} -- (-2,0) node [below] {$x$};
   \draw [thick,latex-latex] (2, 2) node[left] {$\bar{\rho}(x)$} -- (2,0) node[left] {$0$} -- (5,0) node [below] {$x$};
   % edges
   \draw (-3,0) node [below] {$a$} ;
   \draw (4,0) node [below] {$a$};
   % semi-circle
   \draw [thick, blue] (-3,0) arc (0:60:2);
   % (p,q) curve
   \draw [thick, blue] (4,0) .. controls (3.7,.2) .. (3,1.5);
   % behaviours
   \node at (-3, 1) [right, blue] {$\sqrt{x-a}$};
   \node at (4, 1) [right, blue ] {$(x-a)^\frac{p}{q}$};
  \end{tikzpicture}
 \end{align}
There are also matrix models for which the critical exponent $\bar{\rho}(x)\underset{x\to a}{\propto} (x-a)^\nu \ $ is non-rational.
For example, the edge critical exponent $\nu$ of the $O(n)$ matrix model is such that $n=-2\cos \pi \nu$.
 
Near a critical edge, correlation functions usually have universal limits in terms of the rescaled coordinate
\begin{equation}
\xi_i = ( x_i - a) N^{\frac{1}{1+\nu}}\ .
\end{equation}
In the case of a regular edge $\nu=\frac{p}{q}=\frac{1}{2}$, we have $x_i - a \underset{N\to \infty}{\propto} N^{-\frac23} \xi_i$, and the edge microscopic limit of the connected correlation function is 
\begin{align}
 \bar{\rho}_2^\text{ micro, edge}(\xi_1,\xi_2)
 \sim - K_\text{Airy}(\xi_1,\xi_2)^2\ , 
 \end{align}
 where the \textbf{Airy kernel}\index{kernel!Airy---} is written in terms of the Airy function $\operatorname{Ai}(x)$ as
 \begin{align}
 K_\text{Airy}(\xi_1,\xi_2) = 
 \frac{\operatorname{Ai}'(\xi_1) \operatorname{Ai}(\xi_2) 
       - \operatorname{Ai}(\xi_1) \operatorname{Ai}'(\xi_2)}
      {\xi_1 - \xi_2} \, .
\end{align}
The Airy kernel, which is an integrable kernel, plays an important role
in deriving the Tracy--Widom law for
the largest eigenvalue distribution, via the related Fredholm determinant. (See \autoref{sec:spacing}.)
As a result, the Tracy--Widom law can be expressed in terms of the Tau function of the Painlev\'e II integrable system.

For a more general critical edge of the type \eqref{eq:critedgebehavior}, the Airy function is replaced with a Baker--Akhiezer function $\psi_{p,q}(x)$, solution of a differential equation of order $q$ whose coefficients are polynomials of degrees $p$ or less, which generalizes the Airy differential equation 
$
\operatorname{Ai}''(x) = x \operatorname{Ai}(x)
$.
The Baker--Akhiezer functions $\psi_{p,q}(x)$ are associated with the KP ($q\geq 3$) and KdV ($q=2$) integrable hierarchies.
These integrable hierarchies are related to the $(p, q)$ minimal models of conformal field theory, whose central charges are 
\begin{equation}
c= 1-6\frac{(p-q)^2}{pq}\ .
\label{eq:charge}
\end{equation}
In particular, the Airy case $(p, q) = (1,2)$ corresponds to $c=-2$, whereas the pure gravity or Painlev\'e I case $(p, q) = (3,2)$ corresponds to $c=0$, and the Ising case $(p,q)=(4,3)$ to $c=\frac{1}{2}$.

In non-unitary Gaussian ensembles, 
the bulk and edge microscopic behaviours are still universal.
However, in contrast to the sine or Airy kernels, the corresponding kernels are no longer of the integrable type, at least not in the classical sense of integrability.
Rather, these kernels are related to quantum integrable systems, which are far less known at the present time.

\subsubsection{Merging and birth of edges}

Universal microscopic behaviour is also observed near a point $a$ where two edges merge:
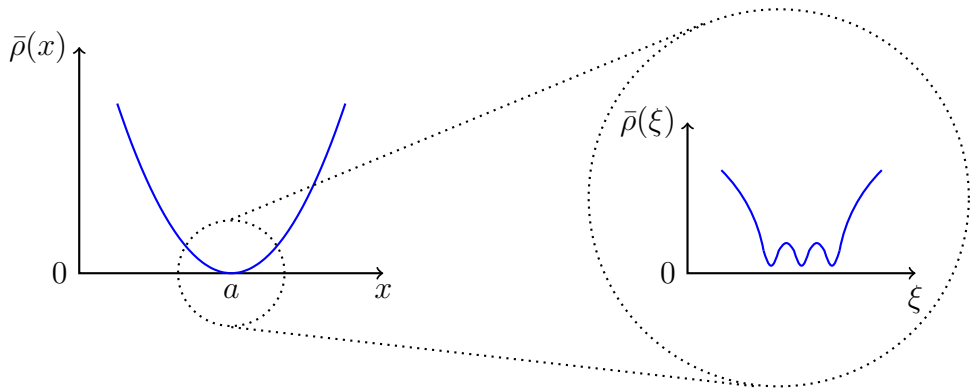
\begin{align}
 \begin{tikzpicture}[baseline=(current bounding box.center)]
   % x-axis
   \draw [thick,latex-latex] (-6, 3) node[left] {$\bar{\rho}(x)$} -- (-6,0) node[left] {$0$} -- (-2,0) node [below] {$x$};
   \draw [thick,latex-latex] (2, 2) node[left] {$\bar{\rho}(\xi)$} -- (2,0) node[left] {$0$}-- (5,0) node [below] {$\xi$};
  \draw [domain = -5.5:-2.5, smooth, variable = \x, thick,blue] plot ({\x}, {(\x+4)^2});
\node at (-4,0) [below] {$a$};
%
  % focusing
   \node at (3.2, 1) [draw,circle,dotted, thick] (big) [minimum size=5cm] {};
    \node at (-4, 0) [draw,circle,dotted, thick] (small) [minimum size=1.4cm] {};
    \draw [dotted, thick] (small.south) -- (tangent cs:node=big,point={(small.south)});
    \draw [dotted, thick] (small.north) -- (tangent cs:node=big,point={(small.north)},solution=2);
  \draw [thick,blue] (3,.3) arc (10:45:2);
  \draw [thick,blue] (4,.3) arc (170:135:2);
  \draw [thick,blue] (3,.3) sin (3.1,.1) cos (3.2,.3) sin (3.3,.4) cos
  (3.4,.3) sin (3.5,.1) cos (3.6,.3) sin (3.7,.4) cos (3.8,.3) sin
  (3.9,.1) cos (4,.3);
 \end{tikzpicture}
\end{align}
For generic choices of the potential $V$, the equilibrium density vanishes quadratically at this point, $\bar \rho(x) \propto (x-a)^2$. 
So the microscopic limit of the two-point function is the $(p, q) = (2,1)$ kernel, which is the Fourier transform of the $(p,q)=(1,2)$ Painlev\'e II kernel.
This is a manifestation of the $(p, q)\leftrightarrow (q,p)$ duality, which is apparent in the formula \eqref{eq:charge} for the central charge of the corresponding conformal field theory. For special choices of the potential $V$, we can have 
$\bar \rho(x) \propto (x-a)^{2m}$, which leads to the $(p, q) =(2m,1)$ kernel.

Similarly, when we vary $V$, we can observe the birth of a new edge far from the already existing edges:
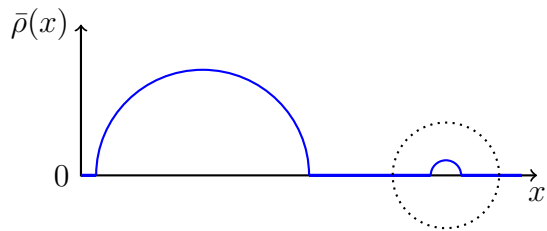
\begin{align}
 \begin{tikzpicture}[baseline=(current bounding box.center)]
  \draw [thick, latex-latex] (0, 2) node[left]{$\bar{\rho}(x)$} -- (0,0) node[left]{$0$} -- (6,0) node[below]{$x$};
  \draw [thick, blue] (3,0) arc (0:180:1.4);
  \draw [thick, blue] (5,0) arc (0:180:.2);
  \draw [very thick, blue] (0,0) -- (.2,0);
\draw [very thick, blue] (3,0) -- (4.6, 0);
\draw [very thick, blue] (5,0) -- (5.8, 0);
 \draw [dotted,thick] (4.8,0) circle [radius = .7];
 \end{tikzpicture}
\end{align}
This gives rise to another universal microscopic regime, called the birth of a cut. 
Generically, the equilibrium density behaves as a square root near the newborn edge, 
and this regime is governed by the Hermite kernel. 
Other possible behaviours of the equilibrium density give rise to other integrable kernels \cite{Eynard:2006:JSM,Bertola:2009CA}.

\section{Exercises}

\begin{exo}[$2\times 2$ Gaussian Hermitian matrices] 
~\label{exo:gaussian}
Consider  $2\times 2$ Hermitian matrices with the potential $V(x)=\frac{x^2}{2}$, or 
equivalently the space  $\mathbb R^2$ of two real eigenvalues $\lambda_1,\lambda_2$ with the measure
\begin{equation}
d\mu(\lambda_1,\lambda_2) = |\lambda_1-\lambda_2|^{2} e^{-\frac12 (\lambda_1^2+\lambda_2^2)} \,d\lambda_1 d\lambda_2 \ .
\end{equation}
Compute
\begin{enumerate}
\item the partition function $\mathcal Z = \int_{\mathbb R\times \mathbb R} d\mu(\lambda_1,\lambda_2)$\, ,
\item the eigenvalue density $\rho(x)$,
\item the probability density function $p(s)=\left<\delta(s-|\lambda_1-\lambda_2|)\right>$ that $s\in\mathbb R_+$ is the distance between the two eigenvalues. Compare with the Wigner surmise.
\end{enumerate}
\end{exo}

\begin{exo}[$2\times 2$ Gaussian matrices with arbitrary $\beta$]
~\label{exo:gaussian2}
For an arbitrary $\beta\in\mathbb{C}$, consider the space  $\mathbb R^2$ of two real eigenvalues $\lambda_1,\lambda_2$ with the measure
\begin{equation}
d\mu(\lambda_1,\lambda_2) = |\lambda_1-\lambda_2|^{\beta} e^{-\frac12 (\lambda_1^2+\lambda_2^2)} \,d\lambda_1 d\lambda_2\ .
\end{equation}
Compute
\begin{enumerate}
\item the partition function $\mathcal Z = \int_{\mathbb R\times \mathbb R} d\mu(\lambda_1,\lambda_2)$\, ,
\item the probability density function $p(s)=\left<\delta(s-|\lambda_1-\lambda_2|)\right>$ that $s\in\mathbb R_+$ is the distance between the two eigenvalues. Compare with the Wigner surmise.
\end{enumerate}
\end{exo}

\begin{exo}[Haar volume of $U_2$]
~\label{exo:Haarvolu2}
Let us compute the volume of $U_2$ with the Haar measure.
\begin{enumerate}
 \item Compute the integral over Hermitian matrices $\mathcal{Z}=\int_{H_2} dM e^{-\frac12\Tr M^2}$ by performing the Gaussian integrals over the elements of $M$.
 \item Compute the same integral by diagonalizing $M$, and deduce $\mathrm{Vol}(U_2)=4\pi^3$.
\end{enumerate}
\end{exo}

\begin{exo}[Haar measure on $U_2$]
~\label{exo:volu2}
Consider the Haar measure $dM_\text{Haar}$ on $U_2$, given by Eq.~\eqref{eq:dmhaar}.
\begin{enumerate}
 \item Write a manifestly analytic expression for $dM_\text{Haar}$, and check that you recover the case $N=2$ of the relation \eqref{eq:circular_Haar}.
 \item Use the analytic expression for computing the ratio $\frac{\int_{U_2} dM_\text{Haar}}{\int_{U_2/U_1^2} dU_\text{Haar}}$. What is the geometrical interpretation of this ratio?
\end{enumerate}
\end{exo}

\begin{exo}[Dumitriu--Edelman tridiagonal matrices, case $N=2$]
~\label{exo:DumiEdel}
Let us compute the eigenvalue distribution of a random symmetric matrix
\begin{align}
M = \begin{pmatrix}
a & u \cr 
u & b
\end{pmatrix}
\quad \text{with the law} \quad  \begin{pmatrix}
N(0,1) & \chi_{\beta} \cr 
\chi_\beta & N(0,1)
\end{pmatrix}
\ .
\end{align}
\begin{enumerate}
\item Show that the measure is
\begin{align}
d\mu(M)= \frac{1}{\pi \Gamma(\frac{\beta}{2})} e^{-\frac12 (a^2+b^2) - u^2} u^{\beta-1} \ da \wedge db \wedge du\ .
\end{align}
\item Let $\lambda_1,\lambda_2$ be the eigenvalues of $M$. Check that
\begin{align}
a^2+b^2+2u^2 = \lambda_1^2+\lambda_2^2
\quad , \quad
(a-b)^2 = (\lambda_1-\lambda_2)^2 - 4 u^2\ ,
\end{align}
and deduce
\begin{align}
da \wedge db \wedge du 
= \frac{|\lambda_1-\lambda_2|}{|a-b|} d\lambda_1 \wedge d\lambda_2 \wedge du\ .
\end{align}

\item Renaming $u = \frac12 |\lambda_1-\lambda_2| y$, show that
\begin{align}
 d\mu(M)
=  \frac{1}{2^\beta \pi \Gamma(\frac{\beta}{2})} |\lambda_1-\lambda_2|^\beta e^{-\frac12 (\lambda_1^2+\lambda_2^2) }   \frac{ y^{\beta-1}}{\sqrt{1-y^2}} \ d\lambda_1 \wedge d\lambda_2 \wedge dy\ .
\end{align}

\item
Conclude that the eigenvalue distribution is
\begin{align}
  \int_{y\in(0,1)}d\mu(M) =
\frac{1}{2^\beta\sqrt\pi \Gamma(\frac{\beta+1}{2})}   |\lambda_1-\lambda_2|^\beta e^{-\frac12 (\lambda_1^2+\lambda_2^2) }
 \ d\lambda_1 \wedge d\lambda_2 \ ,
\end{align}
and show that this agrees with the $N=2$ case of Eq.~\eqref{dml}.

\end{enumerate}

\end{exo}

\begin{exo}[Dotsenko--Fateev integrals in conformal field theory.]
~\label{exo:dotsenko}
Given $n$ distinct points $z_1,\dots,z_n\in\mathbb{C}$, and $\alpha_1,\dots,\alpha_n\in\mathbb{C}$, consider the integral
\begin{equation}
\mathcal Z = \int_{\mathbb C^N} \prod_{i=1}^N dx_id\bar x_i \,\prod_{i<j} |x_i-x_j|^{2\beta} \prod_{i=1}^N\prod_{j=1}^n |x_i-z_j|^{-4b\alpha_j} 
\, , \qquad
\beta=-2 b^2\ .
\end{equation}

\begin{enumerate}

\item
Show that $\mathcal{Z}$ is covariant under a Möbius transformation $z\mapsto f(z)=\frac{Az+B}{Cz+D}$ (with $AD-BC=1$), applied to all $z_j$s, provided
\begin{equation}
 \Xi = 0 \quad \text{where} \quad \Xi = \sum_{j=1}^n \alpha_j - \frac{1}{b} + (N-1)b\ .
 \label{eq:saqnb}
\end{equation}
Under this condition, $\mathcal{Z}$ is a correlation function $\left\langle  V_{\alpha_1}(z_1) \dots  V_{\alpha_n}(z_n) \right\rangle$ of the Liouville conformal field theory on the plane, with the central charge $c= 1+6(b+\frac{1}{b})^2$ and the momentums $\alpha_i$. Then $N$ is called the number of screening charges.

\item Let us assume $\beta=2$, so that the condition \eqref{eq:saqnb} becomes $\sum_j 2b\alpha_j =2 N$. Let us further assume $2b\alpha_j\in\mathbb{N}$.
Consider the following integral, which is obtained from $\mathcal{Z}$ by taking holomorphic factors and integrating over $x_1$ only:
\begin{equation}
\mathcal{Z}_\gamma= \int_{\gamma} dx_1  \,\prod_{j>1} (x_1-x_j)^{\beta} \prod_{j=1}^n (x_1-z_j)^{-2b\alpha_j}\ .
\end{equation}
Show that the vector space of homology paths $\gamma$ on which this integral converges is generated by small circles around $z_1,\dots,z_{n-1}$,
so that the integral is a combination of residues at $x_1=z_j$.
Deduce that $\mathcal{Z}$ is a combination of residues at $x_i=z_j$.

\item Let us further specialize to the case $n=2$.
Compute $\mathcal{Z}$, and check that
\begin{equation}
\renewcommand{\arraystretch}{1.5}
\mathcal Z = \left\{\begin{array}{ll} (N!)^2 (2\pi)^{2N}\,|z_1-z_2|^{-2N^2} & \mathrm{if\ } \alpha_1=\alpha_2=\frac{N}{2b}\ ,
                     \\
                     0 & \mathrm{otherwise\ }.
                    \end{array} \right.
\end{equation}
\end{enumerate}
\end{exo}

\begin{exo}[Duality $\beta \leftrightarrow \frac{4}{\beta}$]
~\label{exo:duality}
Compute the Gaussian matrix integral
\begin{equation}
\mathcal Z(N,\beta) = 2^{-\frac{N}{2}}\int_{E_N^\beta} dM\, e^{-N\sqrt{\frac \beta 2} \operatorname{Tr} M^2} \ ,
\end{equation}
where $dM$ \eqref{eq:betaLebesguemeasure} is the Lebesgue measure on the Gaussian ensemble $E_N^\beta$.
Show that in terms of the Nekrasov variables \eqref{Nekrasov_notation}, the result is
\begin{equation}
\mathcal Z(N,\beta)
= \left( \frac{\pi}{2} \sqrt{-\epsilon_1 \epsilon_2 }  \right)^{\frac{\epsilon_1+\epsilon_2-1}{2\epsilon_1 \epsilon_2}} \ .
\end{equation}
Conclude that $\mathcal Z(N,\beta)$ is invariant under the duality \eqref{beta_duality}.
\end{exo}

\begin{exo}[Jones and HOMFLY polynomials of torus knots]
~\label{exo:jones}
To a torus knot $\mathfrak K_{P,Q}$ and a finite-dimensional representation of $SU_n$ labelled by a partition $\lambda=(\lambda_1,\lambda_2,\dots,\lambda_n)$ with $n$ rows, we associate a couloured HOMFLY polynomial: a polynomial function of some fractional power of a parameter $q=e^{-g_s}$. The Rosso--Jones--Mari\~no formula for the coloured HOMFLY polynomial is
\begin{equation}
\mathrm{HOMFLY}_{\mathfrak K_{P,Q},\lambda}(q) = \frac{\mathcal Z_{\mathfrak K_{P,Q},\lambda}(q)}{\mathcal Z_{\mathfrak K_{P,Q},\emptyset}(q)}\ ,
\end{equation}
where $\emptyset = (0,0,\dots, 0)$, and
\begin{multline}
\mathcal Z_{\mathfrak K_{P,Q},\lambda}(q)
= \int_{\mathbb R^n} du_1 \dots du_n\ e^{-\frac{PQ}{2g_s} \sum_i u_i^2} \,
\frac{\det_{1\leq i,j\leq n} e^{(\lambda_i-i+n) P Q u_j}}{\det_{1\leq i,j\leq n} e^{(n-i)P Q u_j}}
\\
\times \prod_{i<j} 4\sinh\tfrac{P}{2}(u_i-u_j)\sinh\tfrac{Q}{2}(u_i-u_j)\ .
\end{multline}
For $N\in\mathbb{N}^*$, the coloured Jones polynomial is $J_{\mathfrak K_{P,Q},N}(q)=\mathrm{HOMFLY}_{\mathfrak K_{P,Q},(N-1,0)}(q)$. The Jones polynomial is $J_{\mathfrak K_{P,Q}}(q) = \frac{q^{PQ}J_{\mathfrak K_{P,Q},2}(q)}{qJ_{\mathfrak K_{1,1},2}(q)}$, corresponding to the fundamental representation of $SU_2$. The knot $\mathfrak K_{1,1}$ is called the unknot.

In this exercise we compute these polynomials explicitly in a number of special cases. For convenience, we introduce the notation
\begin{align}
 G(a) := \int_{\mathbb{R}} du \, e^{-\frac{PQ}{2 g_s} u^2} \, e^{a u}= \sqrt{\frac{2\pi g_s}{PQ}} q^{-\frac{a^2}{2PQ}} \ .
\end{align}

\begin{enumerate}
\item Compute the normalization factor $\mathcal Z_{\mathfrak K_{P,Q},\emptyset}(q)$ in the case $n=2$.

\item Compute the Jones polynomial, and check that you find
\begin{equation}
J_{\mathfrak K_{P,Q}}(q) = q^{\frac12 (P-1)(Q-1)}\ \frac{q^{P+Q}+1-q^{P+1}-q^{Q+1}}{1-q^2} \ .
\end{equation}

\item In the case of the unknot $P=Q=1$, simplify the integral expression for $\mathcal Z_{\mathfrak K_{1,1},\lambda}(q)$ by using hyperbolic factors for cancelling the Vandermonde determinant in the denominator.

\item Compute the coloured Jones polynomial of the unknot, and check that it is a polynomial of a fractional power of $q$.

\item Compute the coloured HOMFLY polynomial of the unknot in the case $n=3$ with $\lambda = (N-1,0,0)$.
\end{enumerate}

\end{exo}

\chapter{Ribbon graphs}\label{chap:FeynmanGraph}

According to Wick's theorem, every Gaussian integral can be rewritten as a combinatorial sum of Feynman graphs.
In the case of Gaussian matrix integrals, the resulting Feynman graphs are ribbon graphs -- graphs that can be drawn on surfaces. 
Interpreting these graphs as geometrical objects gives rise to applications of matrix models to quantum gravity, string theory, statistical physics on random lattices, and combinatorics of maps.

\section{General properties}

\subsection{Wick's theorem}\label{sec:Wick_thm}

Consider the Gaussian $N$-dimensional real random vector $X= (x_1,\dots,x_N)$ with the probability measure
\begin{equation}
 d\mu(X)=\sqrt{\det A} (2\pi)^{-\frac{N}{2}} e^{-\frac{1}{2}\sum_{i,j} x_i A_{i,j} x_j}\prod_i dx_i
  \, ,
\end{equation}
which is normalized as $\int d\mu(X) = 1$.
Here  $A$ is a positive definite symmetric matrix, whose inverse is called the \textbf{propagator}\index{propagator}:
\begin{equation}
B_{i,j} = (A^{-1})_{i,j}\ .
\end{equation}

\begin{theorem}[Wick]
The expectation value of a product of Gaussian random variables is:
\begin{equation}
\left<x_{i_1}\dots x_{i_n}\right> = 
\left\{\begin{array}{ll}
0 & \qquad \text{if $n$ is odd,}\\
B_{i_1,i_2}& \qquad \text{if $n=2$,} \\
\sum_{\text{pairings of }(i_1,\dots,i_n) } \prod_{\text{pairs }(i,j)} B_{i,j} & \qquad \text{if $n\geq 2$ is even.}
\end{array}\right.
\end{equation}
\end{theorem}
We emphasize that $i_1,\dots,i_n$ need not be distinct, and in fact Wick's theorem is particularly useful when some indices are repeated.
In the case of the expectation value $\left<x_{i_1}^{p_1}\dots x_{i_n}^{p_n}\right>$, Wick's theorem's pairings can be interpreted as graphs with $n$ vertices of valencies $p_1,\dots,p_n$, whose weights are determined by the propagators that correspond to their edges:
\begin{equation}
\left<x_{i_1}^{p_1}\dots x_{i_n}^{p_n}\right>
= \sum_{\text{graphs $G$ with $n$ vertices of valencies $p_k$}}\ \prod_{(i,j) \text{ edge of } G} B_{i,j}\ . 
\end{equation}
However, in this sum, many graphs may actually have the same weight, because of the symmetries between edges attached to the same vertex, and possibly between vertices with the same valency and index.
The group $R$ of relabelings of half-edges that conserve vertices, is a subgroup of permutations of all half-edges, and it has cardinal $\#R = \prod_{k=1}^n p_k  . \prod_{j} n_j!$ (where $n_j=\#$vertices of degree $j$). A graph $G$ is a gluing of half-edges, thus an involution $\sigma$ of half-edges with no fixed point ($\sigma^2=1$ and $\nexists i , \ \sigma(i)=i$).
The number of different gluings corresponding to the same topological graph $G$, is the orbit of $\sigma$ under the action of $R$ by conjugation.
The Automorphisms group $\operatorname{Aut}(G)$ is the subgroup of $R$ that leaves $\sigma$ invariant by conjugation, i.e. the stabilizer of $\sigma$.
The famous orbit/stabilizer theorem in group theory implies
\begin{align}
\#\text{gluings} = \frac{\# R}{\# \operatorname{Aut}(G)}\ .
\end{align}
We can thus rewrite Wick's theorem as a sum over orbits of $G$:
\begin{equation}
\frac{1}{\# R} \left<x_{i_1}^{p_1}\dots x_{i_n}^{p_n}\right>
= \sum_{\text{equivalence classes of graphs $G$}} \frac{1}{\# \mathrm{ Aut}(G)}\prod_{(i,j) \text{ edge of } G} B_{i,j}\ .
\end{equation}
We are now ready to see how Wick's theorem gives rise to ribbon graphs in the case of random matrices.

\subsection{Ribbon graphs}

Let us consider the Hermitian Gaussian matrix model whose partition function is
\begin{equation}
{\mathcal{Z}_0} = \int_{H_N} dM  
    e^{-\frac{N}{2} \operatorname{Tr} M^2}  = 2^{\frac{N}{2}}\left(\frac{\pi}{N}\right)^{\frac{N^2}{2}}\,
    .
\end{equation}

\subsubsection{An example of a correlation function}

Let us compute the correlation function
\begin{align}
 \Big< \operatorname{Tr} M^4 \Big>
 = \frac{1}{\mathcal{Z}_0} \int_{H_N} dM  
    e^{-\frac{N}{2} \operatorname{Tr} M^2}  \operatorname{Tr}M^4 \, ,
\end{align}
using Wick's
theorem. The propagator
\begin{align}
 \big< M_{ij} M_{kl} \big> = \frac{1}{N} \delta_{il} \delta_{jk}
 \ ,
\end{align}
is the inverse of the quadratic form $N\operatorname{Tr}M^2$.
Thus we obtain
\begin{subequations}
\begin{align}
 \Big< N \operatorname{Tr} M^4 \Big>
 & = \sum_{i,j,k,l}
 N \Big< M_{ij} M_{jk} M_{kl} M_{li} \Big>\ ,
   \\
 & = \frac{1}{N} \sum_{i,j,k,l}
 \left(
   \delta_{ik} \delta_{jj} \delta_{ki} \delta_{ll}
 + \delta_{ii} \delta_{jl} \delta_{jl} \delta_{kk}
 + \delta_{il} \delta_{jk} \delta_{ji} \delta_{kl}
 \right)\ ,
   \\
 & = N^2 + N^2 + 1 \ .
\end{align}
\end{subequations}
This computation can be understood in terms of Feynman graphs.
Our $N\operatorname{Tr} M^4$ corresponds to a four-leg vertex,
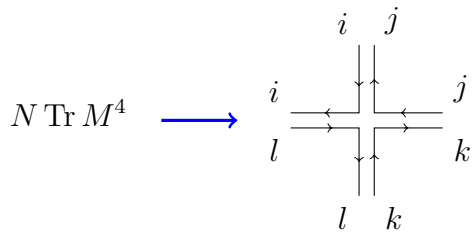
\begin{align}
 \begin{tikzpicture}[baseline=(current bounding box.center)]
 \node at (0,0.1) [left] {$\displaystyle N\operatorname{Tr} M^4$};
 \draw[very thick,blue,-latex] (0.3,0) -- (1.3,0);
%
  % vertex
 \draw (2,.1) 
  node [above left] {$i$} -- (2.9,.1) -- (2.9,1) node [above left] {$i$};
 \draw (2,-.1) 
  node [below left] {$l$} -- (2.9,-.1) -- (2.9,-1) node [below left] {$l$};
 \draw (3.1,1) 
  node [above right] {$j$} -- (3.1,.1) -- (4,.1) node [above right] {$j$};
 \draw (3.1,-1) 
  node [below right] {$k$} -- (3.1,-.1) -- (4,-.1) node [below right] {$k$};
%
  % arrows
  \draw [-latex] (3.1,-.5) -- +(0,.05);
  \draw [-latex] (2.9,-.5) -- +(0,-.05);
  \draw [-latex] (3.1,.5) -- +(0,.05);
  \draw [-latex] (2.9,.5) -- +(0,-.05);
  \draw [-latex] (3.5,.1) -- +(-.05,0);
  \draw [-latex] (3.5,-.1) -- +(.05,0);
  \draw [-latex] (2.5,.1) -- +(-.05,0);
  \draw [-latex] (2.5,-.1) -- +(.05,0);
 \end{tikzpicture}
\end{align}
Using Wick's theorem amounts to joining the four legs with one another in all possible ways. 
Since the propagator is a product of Kronecker deltas, joining legs forms flat ribbons:
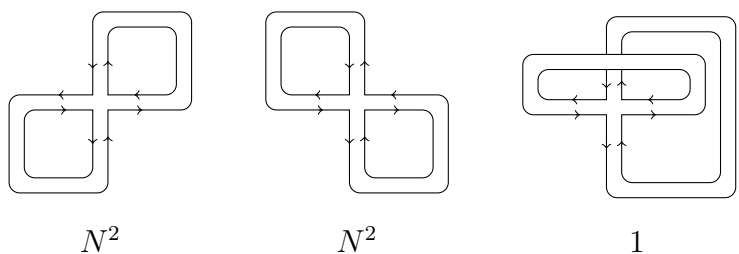
\begin{align}
 \begin{tikzpicture}[baseline=(current bounding box.center)]
  % contraction 1
  \draw (1.1,.1) [rounded corners] --(1.1,1) -- (2,1) -- (2,.1) 
  [sharp corners] -- cycle;
  \draw (.9,-.1) [rounded corners] -- (.9,-1) -- (0,-1) -- (0,-.1)
  [sharp corners] -- cycle;
  \draw (.9,.1) [rounded corners] -- (.9,1.2) -- (2.2,1.2) -- (2.2,-.1)
  [sharp corners] -- (1.1,-.1) [rounded corners] -- (1.1,-1.2) --
  (-.2,-1.2) -- (-.2,.1) [sharp corners] -- cycle;
%
  % arrows
  \draw [-latex] (1.1,-.5) -- +(0,.05);
  \draw [-latex] (.9,-.5) -- +(0,-.05);
  \draw [-latex] (1.1,.5) -- +(0,.05);
  \draw [-latex] (.9,.5) -- +(0,-.05);
  \draw [-latex] (1.5,.1) -- +(-.05,0);
  \draw [-latex] (1.5,-.1) -- +(.05,0);
  \draw [-latex] (.5,.1) -- +(-.05,0);
  \draw [-latex] (.5,-.1) -- +(.05,0);
%
  % order
  \node at (1,-1.5) [below] {$N^2$};
 \end{tikzpicture}
 \qquad
 \begin{tikzpicture}[baseline=(current bounding box.center)]
  % contraction 2
%  
  \draw (-1.1,.1) [rounded corners] -- (-1.1,1) -- (-2,1) -- (-2,.1) 
  [sharp corners] -- cycle;
  \draw (-.9,-.1) [rounded corners] -- (-.9,-1) -- (0,-1) -- (0,-.1)
  [sharp corners] -- cycle;
  \draw (-.9,.1) [rounded corners] -- (-.9,1.2) -- (-2.2,1.2) -- (-2.2,-.1)
  [sharp corners] -- (-1.1,-.1) [rounded corners] -- (-1.1,-1.2) --
  (.2,-1.2) -- (.2,.1) [sharp corners] -- cycle;
%
  % arrows
  \draw [-latex] (-1.1,-.5) -- +(0,-.05);
  \draw [-latex] (-.9,-.5) -- +(0,.05);
  \draw [-latex] (-1.1,.5) -- +(0,-.05);
  \draw [-latex] (-.9,.5) -- +(0,.05);
  \draw [-latex] (-1.5,.1) -- +(-.05,0);
  \draw [-latex] (-1.5,-.1) -- +(.05,0);
  \draw [-latex] (-.5,.1) -- +(-.05,0);
  \draw [-latex] (-.5,-.1) -- +(.05,0);
%
  % order
  \node at (-1,-1.5) [below] {$N^2$};
 \end{tikzpicture}
 \qquad
 \begin{tikzpicture}[baseline=(current bounding box.center)]
  % contraction 3
  \draw (.9,.5) -- (.9,.1) [rounded corners] -- (0,.1) -- (0,.5) --
  (2,.5) -- (2,.1) [sharp corners] -- (1.1,.1) -- (1.1,.5);
  \draw (.9,.7) [rounded corners] -- (.9,1.2) -- (2.6,1.2) -- (2.6,-1.2)
  -- (.9,-1.2) [sharp corners] -- (.9,-.1) [rounded corners] --
  (-.2,-.1) -- (-.2,.7) -- (2.2,.7) -- (2.2,-.1) [sharp corners] --
  (1.1,-.1) [rounded corners] -- (1.1,-1) -- (2.4,-1) -- (2.4,1) -- (1.1,1)
  -- (1.1,.7);
%
  % arrows
  \draw [-latex] (1.1,-.5) -- +(0,.05);
  \draw [-latex] (.9,-.5) -- +(0,-.05);
  \draw [-latex] (1.1,.3) -- +(0,.05);
  \draw [-latex] (.9,.3) -- +(0,-.05);
  \draw [-latex] (1.5,.1) -- +(-.05,0);
  \draw [-latex] (1.5,-.1) -- +(.05,0);
  \draw [-latex] (.5,.1) -- +(-.05,0);
  \draw [-latex] (.5,-.1) -- +(.05,0);
%  
  % order
  \node at (1.3,-1.5) [below] {$1$};
 \end{tikzpicture}
 \label{880}
\end{align}

\subsubsection{Sums over ribbon graphs}

Let us count the powers of $N$ for a given ribbon graph $G$. 
Summing over the indices produces a factor $N$ for each single line, and the number of single lines is also the number of faces of the graph, i.e. the number of connected components of the graph's complement when drawn as a cellular graph on a surface. (A cellular graph is a graph whose faces are topological discs.)
Each edge corresponds to a propagator that is proportional to $N^{-1}$, and each vertex $\displaystyle N\operatorname{Tr} M^4$ carries a factor $N$.
So the power of $N$ is 
\begin{equation}
\chi(G) =\#\mathrm{ vertices} - \#\mathrm{ edges}+\#\mathrm{ faces}\ .
\end{equation} 
This coincides with the \textbf{Euler characteristic}\index{Euler characteristic}, a topological invariant of graphs on surfaces.
The Euler characteristic of a disconnected graph is the sum of Euler characteristics of its connected components.
The Euler characteristic of a connected graph $G$ is related to the genus $g$ of the lowest genus surface where $G$ can be drawn as a cellular graph by 
\begin{equation}
 \chi(G) = 2-2g \qquad \text{with} \qquad g \in \mathbb{N} %\mathbb{N}
  \ .
\label{chig}
\end{equation}
So all connected graphs obey $\chi(G)\leq 2$, and $\chi(G)=2$ if and only if $G$ is a \textbf{planar
graph}\index{planar graph}, a graph that can be drawn on the plane, such as the first two graphs in Eq.~\eqref{880}. The third graph can be drawn on a torus of genus $g=1$, and has $\chi(G)=0$.
This can be generalized to arbitrary correlation functions of traces of powers of $M$:
\begin{theorem}['t Hooft~\cite{'tHooft:1973jz}, Br\'ezin--Itzykson--Parisi--Zuber~\cite{Brezin:1977sv}]
 Expectation values in the Gaussian Hermitian matrix model, are sums over weighted ribbon graphs:
  \begin{align}
    \boxed{   \frac{1}{ \prod_{j} n_j!} \left< 
    \prod_{k=1}^n N \frac{\operatorname{Tr} M^{p_k}}{p_k}
   \right>	     
   =
   \sum_{\substack{\text{graphs $G$} \\ 
   \text{with $n$ vertices of valency $p_k$}}}
   \frac{N^{\chi(G)}}{\# \operatorname{Aut} (G)} }\, ,
   \label{top_exp01}
  \end{align}
 where $\operatorname{Aut} (G)$ is the  stabilizer of the graph $G$ in the relabelling symmetry group $R$ of cardinal $\# R = \prod_{k=1}^n p_k \prod_j n_j!$ (see \cite[Eq.~(1.1.1)]{EynBook}.)
\end{theorem}
Since the contribution of a disconnected graph is the product of the contributions of its connected components, 
classical theorems of combinatorics ensure that summing over connected graphs yields 
cumulants:
  \begin{align}
\frac{1}{ \prod_{j} n_j!}   \left< 
    \prod_{k=1}^n N \frac{\operatorname{Tr} M^{p_k}}{p_k}
   \right>_\text{c}	     
   =
   \sum_{\substack{\text{connected graphs $G$} \\ 
   \text{with $n$ vertices of valency $p_k$}}}
   \frac{N^{\chi(G)}}{\# \operatorname{Aut} (G)} \, .
  \end{align}

\subsubsection{Sums over polygonal surfaces}

Since the dual of a vertex with $p_k$ legs is a polygon with $p_k$
sides, the dual graph of a ribbon graph is a discretized surface made of polygons -- an object called a \textbf{map}\index{map (combinatorics)} in combinatorics:
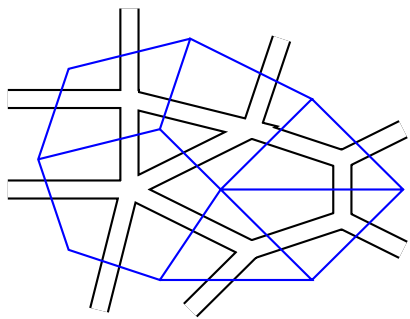
\begin{align}
 \begin{tikzpicture}[scale = .4,baseline=(current bounding box.center)]
  \draw [thick,double,double distance=6pt]
  plot coordinates{(-1,4)(3,4)(2,0)}
  plot coordinates{(-1,7)(3,7)(3,4)(7,2)(5,0)}
  plot coordinates{(3,10)(3,7)(7,6)(3,4)}
  plot coordinates{(8,9)(7,6)(10,5)(12,6)}
  plot coordinates{(10,5)(10,3)(12,2)}
  plot coordinates{(10,3)(7,2)}
  ;
  \draw [thick,blue]
  plot coordinates{(5,9)(1,8)(0,5)(1,2)(4,1)(6,4)(4,6)(0,5)}
  plot coordinates{(4,6)(5,9)(9,7)(6,4)(9,1)(4,1)}
  plot coordinates{(9,1)(12,4)(9,7)}
  plot coordinates{(12,4)(6,4)};
 \end{tikzpicture}
\end{align}
The theorem can be reformulated in terms of polygonal surfaces,
  \begin{align}
  \boxed{ \frac{1}{ \prod_{j} n_j!}\left< 
    \prod_{k=1}^n N \frac{\operatorname{Tr} M^{p_k}}{p_k}
   \right>	     
    =
   \sum_{\substack{\text{surfaces $\Sigma$} \\ 
   \text{with $n$ polygons of size $p_k$}}}
   \frac{N^{\chi(\Sigma)}}{\# \mathrm{Aut} (\Sigma)} }\ ,
  \end{align}
  where the Euler characteristic of a polygonal surface coincides with the characteristic of the dual graph, $\chi(\Sigma)=\chi(G)$. 
Cumulants are sums over connected polygonal surfaces,
  \begin{align}
 \frac{1}{ \prod_{j} n_j!}\left< 
    \prod_{k=1}^n N \frac{\operatorname{Tr} M^{p_k}}{p_k}
   \right>_\text{c}	     
    =
   \sum_{\substack{\text{connected surfaces $\Sigma$} \\ 
   \text{with $n$ polygons of size $p_k$}}}
   \frac{N^{\chi(\Sigma)}}{\# \mathrm{Aut} (\Sigma)} \ .
  \end{align}

\subsection{Formal matrix integrals}\label{sec:fmi}

Given the potential
\begin{align}
 V(M) = \frac12  M^2 - \sum_{k=3}^\infty \frac{t_k}{k}  M^k\ ,
 \label{eq:vom}
\end{align}
let us define a \textbf{formal matrix integral} \index{formal matrix integral} \cite{EynBook}  by treating the non-Gaussian factors of $e^{-N\operatorname{Tr} V(M)}$ as formal Taylor series, 
\begin{align}
 \frac{1}{\mathcal{Z}_0} \int_{\text{formal}} dM \, e^{-N\operatorname{Tr} V(M)}
 \stackrel{\text{def}}{=}
 \sum_{n=0}^\infty \frac{1}{n!}
 \sum_{k_1, \ldots, k_n\geq 3}
   \left< 
    \prod_{m=1}^n N \frac{\operatorname{Tr} M^{k_m}}{k_m} t_{k_m}
   \right>   \in \mathbb{Q}[[t_3,t_4,\dots]] .
\end{align}
We are therefore exchanging the matrix integral with Taylor series expansions.
The coefficients of the resulting formal series are 
Gaussian correlation functions that can be computed using the formula (\ref{top_exp01}), and therefore: 
\begin{theorem}['t Hooft~\cite{'tHooft:1973jz}, Br\'ezin--Itzykson--Parisi--Zuber~\cite{Brezin:1977sv}]
 \begin{align}
 \boxed{ \frac{1}{\mathcal{Z}_0} \int_{\text{formal}} dM \, e^{-N\operatorname{Tr} V(M)}
 =
 \sum_G \frac{N^{\chi(G)}}{\#\mathrm{Aut}(G)} 
 \prod_k t_k^{\#k\text{-vertices}} }
 \, . \label{eq:sumg}
\end{align}
The sum over connected graphs yields the 
logarithm of the formal matrix integral.
\begin{align}
 \boxed{ 
 \log{\left(\frac{1}{\mathcal{Z}_0} \int_{\text{formal}} dM \, e^{-N\operatorname{Tr} V(M)} \right)}
 =
 \sum_{\mathrm{ connected}\,G} \frac{N^{\chi(G)}}{\#\mathrm{Aut}(G)} 
 \prod_k t_k^{\#k\text{-vertices}} }
 \, . \label{eq:sumgc}
\end{align} 
\end{theorem}
Formal matrix integrals provide efficient methods for computing generating series of graphs, and algebraically encoding the combinatorial relationships among graphs.
For instance, Tutte's recursion relations on the number of edges in the graphs correspond to loop equations in the formal matrix models \cite{EynBook}.
Deriving loop equations by integrating by parts in the formal integral is easier than finding bijections among sets of graphs. (See \autoref{chap:LoopEq}.)

\subsubsection{Expectation values}

Let us provide a diagrammatic interpretation of the expectation values that come with our non-Gaussian formal matrix integral. 
We will use the expression of our matrix integral \eqref{eq:sumg} as a sum over polygonal surfaces $\Sigma$,
\begin{align}
 \frac{1}{\mathcal{Z}_0} \int_{\text{formal}} dM \, e^{-N\operatorname{Tr} V(M)}
 =
 \sum_{\Sigma} \frac{N^{\chi(\Sigma)}}{\#\mathrm{Aut}(\Sigma)} 
 \prod_k t_k^{\#k\text{-gons}}
 \, .
\label{eq:iformal}
\end{align}
We now denote $\left<f(M)\right>$ the insertion of $f(M)$ in this formal integral, rather than in a Gaussian integral.
In particular, inserting the operator $\operatorname{Tr} M^{\mu_i}$ now amounts to inserting a marked $\mu_i$-gon, and we have 
\begin{align}
 \left<
  \frac{N}{\mu_1}\operatorname{Tr}M^{\mu_1} \cdots   \frac{N}{\mu_n}\operatorname{Tr}M^{\mu_n}
 \right>
 & =
 \sum_{\Sigma \in \tilde{\Sigma}_{\mu_1, \cdots \mu_n}}
 \frac{N^{\chi(\Sigma)}}{\# \operatorname{Aut}(\Sigma)}
 \prod_k t_k^{\#k\text{-gons}}
 \, ,
\label{eq:gmarked}
\end{align}
where $\tilde{\Sigma}_{\mu_1,\cdots \mu_n}$ is the set of all discrete surfaces, connected or not, with $n$ labelled marked faces, and an arbitrary number of unmarked faces. (Only unmarked faces are counted in $\#k\text{-gons}$, and all connected components must carry at least one marked face.)

Let us now multiply Eq.~\eqref{eq:gmarked} with $N^{-n}\prod_i \mu_i$.
The factor $N^{-n}$ amounts to replacing $\chi(\Sigma)$ with the \textbf{Euler charactaristic}\index{Euler characteristic!---with boundaries} $\chi(\Sigma)-n$ of the graph that would be obtained by replacing the $n$ marked faces with $n$ holes, thereby producing a surface with $n$ boundaries.
The factor $\mu_i$ amounts to counting the $i$-th marked face $\mu_i$ times, which is equivalent to choosing a marked edge on its boundary. 
This greatly simplifies the combinatorics of graphs, as 
the presence of a face with a marked edge ensures that a connected graph has no nontrivial automorphism, thus $\# \operatorname{Aut}(\Sigma)=1$.
We nevertheless keep the factor $\frac{1}{\# \operatorname{Aut}(\Sigma)}$, because it can be nontrivial in the case $n=0$. Therefore we have
\begin{align}
 \Big<
  \operatorname{Tr}M^{\mu_1} \cdots \operatorname{Tr}M^{\mu_n}
 \Big>
 & =
 \sum_{\Sigma \in \hat{\Sigma}_{\mu_1, \cdots \mu_n}}
 \frac{N^{\chi(\Sigma)}}{\# \operatorname{Aut}(\Sigma)}
 \prod_k t_k^{\#k\text{-gons}}
 \, ,
 \label{eq:nsd}
\end{align}
where $\hat{\Sigma}_{\mu_1, \cdots \mu_n}$ is the set of discrete surfaces with $n$ marked faces with marked edges, and where marked faces are counted as holes for computing the Euler characteristic.
Computing cumulants amounts to
restricting the sum to the set $\Sigma_{\mu_1, \cdots \mu_n}$ of connected discrete surfaces
\begin{align}
\boxed{ \Big<
  \operatorname{Tr}M^{\mu_1} \cdots \operatorname{Tr}M^{\mu_n}
 \Big>_\text{c}
  =
 \sum_{\Sigma \in \Sigma_{\mu_1, \cdots \mu_n}}
 \frac{N^{\chi(\Sigma)}}{\# \operatorname{Aut}(\Sigma)}
 \prod_k t_k^{\#k\text{-gons}}
 } \, .
\label{eq:nsg}
\end{align}
In terms of connected correlation functions, this becomes
\begin{align}
 W_n(x_1, \cdots x_n) 
 = \frac{N\delta_{n,1}}{x_1} +
 \sum_{\mu_1, \cdots \mu_n=1}^\infty 
 \sum_{\Sigma \in \Sigma_{\mu_1, \cdots \mu_n}}
 \frac{N^{\chi(\Sigma)}}{\# \operatorname{Aut}(\Sigma)} \frac{\prod_k
 t_k^{\#k\text{-gons}}}{\prod_{i=1}^n x_i^{\mu_i+1}} \ .
 \label{eq:wns}
\end{align}
Here the sums run from $\mu_i=1$ rather than $\mu_i=0$, because the only contribution of $\mu_i=0$ terms to connected correlation functions is the term $\frac{N}{x_1}$ of $W_1$, which we wrote separately.

Let us draw an example of such a discrete surface, with genus $g=2$ and $n=3$ holes with $(\mu_1,\mu_2,\mu_3)=(5,6,4)$. 
Marked edges are denoted by arrows which follow the orientation of the surface, so that the hole is always on the left of the arrow:
% Discrete surface
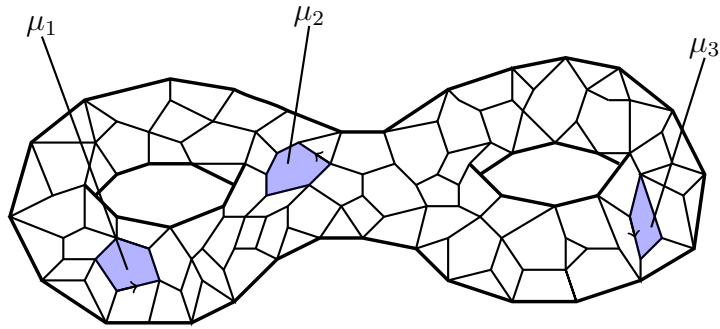
\begin{align}
\begin{tikzpicture}
 [y=0.80pt,x=0.80pt,yscale=-1, inner sep=0pt, outer sep=0pt, scale=.5,baseline=(current bounding box.center)]
\begin{scope}[shift={(-48.5,-291.86218)}]
  \path[draw=black,line join=miter,line cap=butt,miter limit=4.00,line
    width=1.300pt] (200.0000,342.3622) -- (120.0000,362.3622) --
    (70.0000,402.3622) -- (50.0000,472.3622) -- (80.0000,532.3622) --
    (140.0000,572.3622) -- (220.0000,572.3622) -- (260.0000,552.3622) --
    (300.0000,512.3622) -- (340.0000,492.3622) -- (380.0000,492.3622) --
    (430.0000,502.3622) -- (460.0000,532.3622) -- (520.0000,552.3622) --
    (580.0000,552.3622) -- (640.0000,532.3622) -- (690.0000,502.3622) --
    (700.0000,432.3622) -- (670.0000,372.3622) -- (620.0000,332.3622) --
    (570.0000,322.3622) -- (520.0000,332.3622) -- (460.0000,352.3622) --
    (400.0000,392.3622) -- (360.0000,392.3622) -- (310.0000,372.3622) .. controls
    (290.0000,362.3622) and (280.0000,362.3622) .. (260.0000,352.3622) -- cycle;
    \path[draw=black,line join=miter,line cap=butt,miter limit=4.00,line
      width=1.300pt] (120.0000,442.3622) -- (150.0000,472.3622) --
      (200.0000,482.3622) -- (250.0000,462.3622) -- (270.0000,422.3622);
    \path[draw=black,line join=miter,line cap=butt,miter limit=4.00,line
      width=1.300pt] (130.0000,452.3622) -- (150.0000,432.3622) --
      (180.0000,422.3622) -- (220.0000,422.3622) -- (260.0000,442.3622);
    \path[draw=black,line join=miter,line cap=butt,miter limit=4.00,line
      width=1.300pt] (480.0000,422.3622) -- (510.0000,452.3622) --
      (560.0000,462.3622) -- (600.0000,452.3622) -- (630.0000,412.3622);
    \path[draw=black,line join=miter,line cap=butt,miter limit=4.00,line
      width=1.300pt] (490.0000,432.3622) -- (520.0000,412.3622) --
      (560.0000,402.3622) -- (600.0000,412.3622) -- (620.0000,422.3622);
  \path[shift={(48.5,320.86218)},draw=black,line join=miter,line cap=butt,line
    width=0.800pt] (21.5000,81.5000) -- (41.5000,131.5000) -- (1.5000,151.5000) --
    (51.5000,171.5000) -- (51.5000,191.5000) -- (31.5000,211.5000) --
    (71.5000,211.5000) -- (91.5000,251.5000);
  \path[shift={(48.5,320.86218)},draw=black,line join=miter,line cap=butt,line
    width=0.800pt] (91.5000,251.5000) -- (101.5000,221.5000) --
    (141.5000,211.5000) -- (161.5000,221.5000) -- (171.5000,251.5000) --
    (181.5000,211.5000) -- (211.5000,231.5000) -- (201.5000,201.5000) --
    (211.5000,181.5000) -- (231.5000,181.5000) -- (251.5000,191.5000) --
    (261.5000,161.5000) -- (291.5000,171.5000) -- (311.5000,141.5000) --
    (331.5000,171.5000);
  \path[shift={(48.5,320.86218)},draw=black,line join=miter,line cap=butt,line
    width=0.800pt] (71.5000,211.5000) -- (81.5000,191.5000) -- (51.5000,191.5000);
  \path[shift={(48.5,320.86218)},draw=black,line join=miter,line cap=butt,line
    width=0.800pt] (101.5000,221.5000) -- (81.5000,191.5000) --
    (101.5000,171.5000) -- (131.5000,181.5000) -- (141.5000,211.5000);
  \path[shift={(48.5,320.86218)},draw=black,line join=miter,line cap=butt,line
    width=0.800pt] (101.5000,171.5000) -- (71.5000,161.5000) --
    (51.5000,171.5000);
  \path[shift={(48.5,320.86218)},draw=black,line join=miter,line cap=butt,line
    width=0.800pt] (101.5000,151.5000) -- (101.5000,171.5000) --
    (131.5000,181.5000) -- (151.5000,161.5000);
  \path[shift={(48.5,320.86218)},draw=black,line join=miter,line cap=butt,line
    width=0.800pt] (71.5000,121.5000) -- (71.5000,161.5000);
  \path[shift={(48.5,320.86218)},draw=black,line join=miter,line cap=butt,line
    width=0.800pt] (41.5000,131.5000) -- (71.5000,121.5000);
  \path[shift={(48.5,320.86218)},draw=black,line join=miter,line cap=butt,line
    width=0.800pt] (71.5000,41.5000) -- (81.5000,81.5000) -- (21.5000,81.5000);
  \path[shift={(48.5,320.86218)},draw=black,line join=miter,line cap=butt,line
    width=0.800pt] (71.5000,121.5000) -- (81.5000,81.5000);
  \path[shift={(48.5,320.86218)},draw=black,line join=miter,line cap=butt,line
    width=0.800pt] (101.5000,111.5000) -- (81.5000,81.5000);
  \path[shift={(48.5,320.86218)},draw=black,line join=miter,line cap=butt,line
    width=0.800pt] (131.5000,101.5000) -- (131.5000,71.5000) -- (101.5000,61.5000)
    -- (81.5000,81.5000);
  \path[shift={(48.5,320.86218)},draw=black,line join=miter,line cap=butt,line
    width=0.800pt] (161.5000,221.5000) -- (171.5000,191.5000) --
    (181.5000,211.5000);
  \path[shift={(48.5,320.86218)},draw=black,line join=miter,line cap=butt,line
    width=0.800pt] (201.5000,201.5000) -- (181.5000,181.5000) --
    (171.5000,191.5000) -- (151.5000,161.5000);
  \path[shift={(48.5,320.86218)},draw=black,line join=miter,line cap=butt,line
    width=0.800pt] (181.5000,181.5000) -- (201.5000,151.5000) --
    (201.5000,141.5000);
  \path[shift={(48.5,320.86218)},draw=black,line join=miter,line cap=butt,line
    width=0.800pt] (211.5000,181.5000) -- (221.5000,151.5000) --
    (201.5000,151.5000);
  \path[shift={(48.5,320.86218)},draw=black,line join=miter,line cap=butt,line
    width=0.800pt] (221.5000,151.5000) -- (241.5000,131.5000) --
    (241.5000,111.5000) -- (221.5000,101.5000);
  \path[shift={(48.5,320.86218)},draw=black,line join=miter,line cap=butt,line
    width=0.800pt] (231.5000,181.5000) -- (241.5000,151.5000) --
    (221.5000,151.5000);
  \path[shift={(48.5,320.86218)},draw=black,line join=miter,line cap=butt,line
    width=0.800pt] (241.5000,151.5000) -- (261.5000,141.5000) --
    (261.5000,161.5000);
  \path[shift={(48.5,320.86218)},draw=black,line join=miter,line cap=butt,line
    width=0.800pt] (261.5000,141.5000) -- (281.5000,121.5000) --
    (311.5000,141.5000);
  \path[shift={(48.5,320.86218)},draw=black,line join=miter,line cap=butt,line
    width=0.800pt] (281.5000,121.5000) -- (241.5000,131.5000);
  \path[shift={(48.5,320.86218)},draw=black,line join=miter,line cap=butt,line
    width=0.800pt] (241.5000,111.5000) -- (251.5000,91.5000) -- (271.5000,81.5000)
    -- (301.5000,101.5000) -- (281.5000,121.5000);
  \path[shift={(48.5,320.86218)},draw=black,line join=miter,line cap=butt,line
    width=0.800pt] (251.5000,91.5000) -- (231.5000,71.5000) --
    (221.5000,101.5000);
  \path[shift={(48.5,320.86218)},draw=black,line join=miter,line cap=butt,line
    width=0.800pt] (131.5000,251.5000) -- (141.5000,211.5000);
  \path[shift={(48.5,320.86218)},draw=black,line join=miter,line cap=butt,line
    width=0.800pt] (221.5000,101.5000) -- (171.5000,101.5000) --
    (161.5000,81.5000) -- (131.5000,71.5000);
  \path[shift={(48.5,320.86218)},draw=black,line join=miter,line cap=butt,line
    width=0.800pt] (161.5000,81.5000) -- (191.5000,61.5000) -- (231.5000,71.5000);
  \path[shift={(48.5,320.86218)},draw=black,line join=miter,line cap=butt,line
    width=0.800pt] (101.5000,61.5000) -- (71.5000,41.5000);
  \path[shift={(48.5,320.86218)},draw=black,line join=miter,line cap=butt,line
    width=0.800pt] (131.5000,71.5000) -- (151.5000,41.5000) -- (151.5000,21.5000);
  \path[shift={(48.5,320.86218)},draw=black,line join=miter,line cap=butt,line
    width=0.800pt] (211.5000,31.5000) -- (191.5000,61.5000) -- (151.5000,41.5000);
  \path[shift={(48.5,320.86218)},draw=black,line join=miter,line cap=butt,line
    width=0.800pt] (231.5000,71.5000) -- (261.5000,51.5000);
  \path[shift={(48.5,320.86218)},draw=black,line join=miter,line cap=butt,line
    width=0.800pt] (271.5000,81.5000) -- (311.5000,71.5000);
  \path[shift={(48.5,320.86218)},draw=black,line join=miter,line cap=butt,line
    width=0.800pt] (311.5000,71.5000) -- (301.5000,101.5000) --
    (331.5000,111.5000) -- (331.5000,131.5000) -- (311.5000,141.5000);
  \path[shift={(48.5,320.86218)},draw=black,line join=miter,line cap=butt,line
    width=0.800pt] (331.5000,171.5000) -- (351.5000,151.5000) --
    (391.5000,141.5000) -- (381.5000,181.5000);
  \path[shift={(48.5,320.86218)},draw=black,line join=miter,line cap=butt,line
    width=0.800pt] (381.5000,181.5000) -- (411.5000,171.5000) --
    (411.5000,211.5000) -- (451.5000,201.5000) -- (471.5000,231.5000);
  \path[shift={(48.5,320.86218)},draw=black,line join=miter,line cap=butt,line
    width=0.800pt] (471.5000,231.5000) -- (481.5000,201.5000) --
    (501.5000,201.5000) -- (521.5000,201.5000) -- (531.5000,231.5000) --
    (521.5000,201.5000) -- (541.5000,181.5000) -- (571.5000,181.5000) --
    (591.5000,211.5000) -- (591.5000,191.5000) -- (611.5000,171.5000) --
    (641.5000,181.5000) -- (631.5000,131.5000) -- (651.5000,111.5000);
  \path[shift={(48.5,320.86218)},draw=black,line join=miter,line cap=butt,line
    width=0.800pt] (651.5000,111.5000) -- (621.5000,101.5000) --
    (611.5000,71.5000) -- (621.5000,51.5000) -- (581.5000,41.5000) --
    (571.5000,11.5000) -- (541.5000,31.5000) -- (521.5000,1.5000);
  \path[shift={(48.5,320.86218)},draw=black,line join=miter,line cap=butt,line
    width=0.800pt] (521.5000,1.5000) -- (491.5000,31.5000) -- (471.5000,11.5000)
    -- (471.5000,41.5000) -- (441.5000,51.5000) -- (421.5000,41.5000) --
    (411.5000,31.5000) -- (401.5000,61.5000) -- (351.5000,71.5000);
  \path[shift={(48.5,320.86218)},draw=black,line join=miter,line cap=butt,line
    width=0.800pt] (351.5000,71.5000) -- (361.5000,101.5000) --
    (331.5000,111.5000);
  \path[shift={(48.5,320.86218)},draw=black,line join=miter,line cap=butt,line
    width=0.800pt] (331.5000,131.5000) -- (351.5000,151.5000);
  \path[shift={(48.5,320.86218)},draw=black,line join=miter,line cap=butt,line
    width=0.800pt] (391.5000,141.5000) -- (381.5000,121.5000) --
    (361.5000,101.5000);
  \path[shift={(48.5,320.86218)},draw=black,line join=miter,line cap=butt,line
    width=0.800pt] (401.5000,61.5000) -- (411.5000,91.5000) -- (401.5000,111.5000)
    -- (381.5000,121.5000);
  \path[shift={(48.5,320.86218)},draw=black,line join=miter,line cap=butt,line
    width=0.800pt] (431.5000,101.5000) -- (451.5000,81.5000) --
    (471.5000,91.5000);
  \path[shift={(48.5,320.86218)},draw=black,line join=miter,line cap=butt,line
    width=0.800pt] (561.5000,41.5000) -- (541.5000,31.5000) -- cycle;
  \path[shift={(48.5,320.86218)},draw=black,line join=miter,line cap=butt,line
    width=0.800pt] (581.5000,41.5000) -- (561.5000,41.5000);
  \path[shift={(48.5,320.86218)},draw=black,line join=miter,line cap=butt,line
    width=0.800pt] (401.5000,111.5000) -- (421.5000,121.5000) --
    (421.5000,141.5000) -- (391.5000,141.5000);
  \path[shift={(48.5,320.86218)},draw=black,line join=miter,line cap=butt,line
    width=0.800pt] (421.5000,121.5000) -- (431.5000,101.5000);
  \path[shift={(48.5,320.86218)},draw=black,line join=miter,line cap=butt,line
    width=0.800pt] (421.5000,141.5000) -- (441.5000,151.5000) --
    (461.5000,131.5000);
  \path[shift={(48.5,320.86218)},draw=black,line join=miter,line cap=butt,line
    width=0.800pt] (411.5000,171.5000) -- (441.5000,151.5000) --
    (461.5000,171.5000) -- (451.5000,201.5000);
  \path[shift={(48.5,320.86218)},draw=black,line join=miter,line cap=butt,line
    width=0.800pt] (481.5000,201.5000) -- (461.5000,171.5000);
  \path[shift={(48.5,320.86218)},draw=black,line join=miter,line cap=butt,line
    width=0.800pt] (511.5000,141.5000) -- (461.5000,171.5000);
  \path[shift={(48.5,320.86218)},draw=black,line join=miter,line cap=butt,line
    width=0.800pt] (521.5000,201.5000) -- (521.5000,171.5000);
  \path[shift={(48.5,320.86218)},draw=black,line join=miter,line cap=butt,line
    width=0.800pt] (521.5000,171.5000) -- (511.5000,141.5000);
  \path[shift={(48.5,320.86218)},draw=black,line join=miter,line cap=butt,line
    width=0.800pt] (581.5000,151.5000) -- (551.5000,131.5000);
  \path[shift={(48.5,320.86218)},draw=black,line join=miter,line cap=butt,line
    width=0.800pt] (581.5000,91.5000) -- (591.5000,111.5000) --
    (581.5000,151.5000);
  \path[shift={(48.5,320.86218)},draw=black,line join=miter,line cap=butt,line
    width=0.800pt] (611.5000,171.5000) -- (591.5000,111.5000);
  \path[shift={(48.5,320.86218)},draw=black,line join=miter,line cap=butt,line
    width=0.800pt] (621.5000,101.5000) -- (591.5000,111.5000) --
    (631.5000,131.5000);
  \path[shift={(48.5,320.86218)},draw=black,line join=miter,line cap=butt,line
    width=0.800pt] (471.5000,91.5000) -- (481.5000,51.5000) -- (511.5000,81.5000)
    -- (531.5000,61.5000) -- (551.5000,91.5000) -- (581.5000,91.5000);
  \path[shift={(48.5,320.86218)},draw=black,line join=miter,line cap=butt,line
    width=0.800pt] (451.5000,81.5000) -- (441.5000,51.5000);
  \path[shift={(48.5,320.86218)},draw=black,line join=miter,line cap=butt,line
    width=0.800pt] (411.5000,91.5000) -- (451.5000,81.5000);
  \path[shift={(48.5,320.86218)},draw=black,line join=miter,line cap=butt,line
    width=0.800pt] (481.5000,51.5000) -- (491.5000,31.5000);
  \path[shift={(48.5,320.86218)},draw=black,line join=miter,line cap=butt,line
    width=0.800pt] (531.5000,61.5000) -- (541.5000,31.5000);
  \path[shift={(48.5,320.86218)},draw=black,line join=miter,line cap=butt,line
    width=0.800pt] (581.5000,91.5000) -- (581.5000,41.5000);
  \path[shift={(48.5,320.86218)},draw=black,line join=miter,line cap=butt,line
    width=0.800pt] (481.5000,51.5000) -- (471.5000,41.5000);
  \path[shift={(48.5,320.86218)},draw=black,line join=miter,line cap=butt,line
    width=0.800pt] (611.5000,71.5000) -- (581.5000,91.5000);
  \path[shift={(48.5,320.86218)},draw=black,line join=miter,line cap=butt,line
    width=0.800pt] (581.5000,151.5000) -- (591.5000,191.5000);
  \path[shift={(48.5,320.86218)},draw=black,line join=miter,line cap=butt,line
    width=0.800pt] (571.5000,181.5000) -- (551.5000,131.5000);
  \path[shift={(48.5,320.86218)},draw=black,line join=miter,line cap=butt,line
    width=0.800pt] (541.5000,181.5000) -- (521.5000,171.5000);
  \path[shift={(48.5,320.86218)},draw=black,fill=blue,opacity=.3,line join=miter,line
    cap=butt,line width=0.800pt] (591.5000,111.5000) -- (581.5000,151.5000) --
    (591.5000,191.5000) -- (611.5000,171.5000) -- cycle;
  \path[shift={(48.5,320.86218)},draw=black,fill=blue,opacity=.3,line join=miter,line
    cap=butt,line width=0.800pt] (271.5000,81.5000) -- (251.5000,91.5000) --
    (241.5000,111.5000) -- (241.5000,131.5000) -- (281.5000,121.5000) --
    (301.5000,101.5000) -- cycle;
  \path[shift={(48.5,320.86218)},draw=black,fill=blue,opacity=.3,line join=miter,line
    cap=butt,line width=0.800pt] (101.5000,171.5000) -- (81.5000,191.5000) --
    (101.5000,221.5000) -- (141.5000,211.5000) -- (131.5000,181.5000) -- cycle;
  \path[shift={(48.5,320.86218)},draw=black,line join=miter,line cap=butt,miter
    limit=4.00,line width=.800pt,-latex] (301.5000,101.5000) -- (286.5000,91.5000);
  \path[shift={(48.5,320.86218)},draw=black,line join=miter,line cap=butt,miter
    limit=4.00,line width=.800pt,-latex] (101.5000,221.5000) -- (121.5000,216.5000);
  \path[shift={(48.5,320.86218)},draw=black,line join=miter,line cap=butt,miter
    limit=4.00,line width=.800pt,-latex] (581.5000,151.5000) -- (586.5000,171.5000);
  \path[shift={(48.5,320.86218)},draw=black,line join=miter,line cap=butt,line
    width=0.800pt] (261.5000,101.5000) -- (281.5000,-28.5000) node [above] {$\mu_2$};
  \path[shift={(48.5,320.86218)},draw=black,line join=miter,line cap=butt,line
    width=0.800pt] (111.5000,201.5000) -- (31.5000,-18.5000) node [above] {$\mu_1$};
  \path[draw=black,line join=miter,line cap=butt,line width=0.800pt]
    (650.0000,482.3622) -- (700.0000,322.3622) node [above] {$\mu_3$} ;
\end{scope}
\end{tikzpicture}
\end{align}
We insist that only the interior of a marked face is removed.
For example, two neighbouring marked faces count as two holes, as the edge which separates them is not removed.
Moreover, an edge between a marked face and itself is also not removed, and can suffice to keep our graph connected:
% Self-attached face
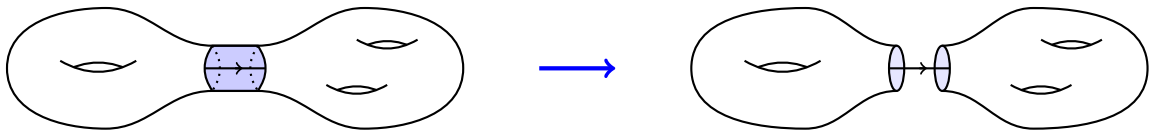
\begin{align}
 \begin{tikzpicture}[thick,baseline=(current bounding box.center)]
 %
   % shape
%
  \draw (-.3,.3) to [out=left,in=right] (-1.7,.8) to [out=left,in=up]
  (-3,0) to [out=down,in=left] (-1.7,-.8) to [out=right,in=left]
   (-.3,-.3) -- (.3,-.3) to [out=right,in=left] (1.7,-.8) to
   [out=right,in=down] (3,0) to [out=up,in=right] (1.7,.8) to
   [out=left,in=right] (.3,.3) -- cycle;
%
  % holes
%
  \begin{scope}[shift={(-1.8,0)}]
   \draw[clip] (-.5,.1) to [bend right] (.5,.1);
   \draw (-.5,-.07) to [bend left] (.5,-.07);
  \end{scope}
  \begin{scope}[shift={(1.6,-.3)},scale=.8]
   \draw[clip] (-.5,.1) to [bend right] (.5,.1);
   \draw (-.5,-.07) to [bend left] (.5,-.07);
  \end{scope}
  \begin{scope}[shift={(2,.3)},scale=.8]
   \draw[clip] (-.5,.1) to [bend right] (.5,.1);
   \draw (-.5,-.07) to [bend left] (.5,-.07);
  \end{scope}
%
  % face
%
  \filldraw [fill opacity=.2,fill=blue]
  (-.3,.3) to [out=180+60,in=up] (-.4,0) to [out=down,in=180-60] (-.3,-.3) --
  (.3,-.3) to [out=60,in=down] (.4,0) to [out=up,in=-60] (.3,.3) -- cycle;
  \draw [dotted] 
  (-.3,.3) to [out=-60,in=up] (-.2,0) to [out=down,in=60] (-.3,-.3);
  \draw [dotted] 
  (.3,.3) to [out=180+60,in=up] (.2,0) to [out=down,in=180-60] (.3,-.3);
  \draw[-latex] (-.4,0) -- (.1,0);
  \draw (.4,0) -- (0,0);
%
  % intermediate arrow
%  
  \draw [ultra thick,blue,-latex] (3.4,0) -- ++(.8,0);
  \begin{scope}[shift={(7.6,0)}]
   % shape
%
  \draw (-.3,.3) to [out=left,in=right] (-1.5,.8) to [out=left,in=up]
  (-3,0) to [out=down,in=left] (-1.5,-.8) to [out=right,in=left] (-.3,-.3);
  \draw (.3,.3) to [out=right,in=left] (1.5,.8) to [out=right,in=up]
  (3,0) to [out=down,in=right] (1.5,-.8) to [out=left,in=right] (.3,-.3);
%
  % holes
%
  \begin{scope}[shift={(-1.8,0)}]
   \draw[clip] (-.5,.1) to [bend right] (.5,.1);
   \draw (-.5,-.07) to [bend left] (.5,-.07);
  \end{scope}
  \begin{scope}[shift={(1.6,-.3)},scale=.8]
   \draw[clip] (-.5,.1) to [bend right] (.5,.1);
   \draw (-.5,-.07) to [bend left] (.5,-.07);
  \end{scope}
  \begin{scope}[shift={(2,.3)},scale=.8]
   \draw[clip] (-.5,.1) to [bend right] (.5,.1);
   \draw (-.5,-.07) to [bend left] (.5,-.07);
  \end{scope}
   \filldraw [fill opacity = .1,fill=blue]
   (-.3,0) circle [x radius=.1, y radius=.3];
%   (-.3,.3) to [out=180+60,in=up] (-.4,0) to [out=down,in=180-60]
%   (-.3,-.3) to [out=60,in=down] (-.2,0) to [out=up,in=-60] (-.3,.3);
%
   \filldraw [fill opacity = .1,fill=blue]
   (.3,0) circle [x radius=.1, y radius=.3];
%   (.3,.3) to [out=-60,in=up] (.4,0) to [out=down,in=60]
%   (.3,-.3) to [out=180-60,in=down] (.2,0) to [out=up,in=180+60] (.3,.3);
%
   % connected arrow
%
   \draw[-latex] (-.4,0) -- (.1,0);
   \draw (.4,0) -- (0,0);
  \end{scope}
 \end{tikzpicture}
\end{align}

\subsubsection{Difference between formal and convergent matrix integrals}

The formal matrix integral in general differs from the convergent matrix integral
\begin{align}
\int_{\mathrm{ formal}} dM \,
 e^{-N\operatorname{Tr} V(M)} 
\neq
 \int_{E} dM \,
 e^{-N\operatorname{Tr} V(M)} \, ,
\end{align}
where the ensemble $E$ and the values of the coefficients $t_k$ are chosen such that the integral absolutely converges.
(For example, $E=H_N$ and $t_k<0$.) 

The difference is due 
%neither to non-perturbative terms, nor to symmetry breaking as was initially thought~\cite{Brezin:1999PRE}, but 
to the fact that integration does not always commute with the Taylor expansion of the non-Gaussian factor $\exp\left({N \sum_{k=3}^\infty \frac{t_k}{k} \operatorname{Tr} M^k}\right)$ in powers of $t_k$. 
The formal series \eqref{eq:sumg} is actually divergent, since typically the number of graphs of genus $g$ grows like $(2g)! $. 
It can sometimes be Borel resummed in some domain of the $t_k$s, and analytically continued to a larger domain. 
On a domain where both exist, the Borel-resummed formal integral and the convergent integral may differ or agree, depending on the domain and on the ensemble $E$.

In general, a convergent matrix integral does not agree with the corresponding formal matrix integral, and its large $N$ behaviour is not directly described by a topological expansion, with its integer powers of $N$.  
Nevertheless, the enumeration of ribbon graphs and the associated topological expansions are useful for studying large $N$ asymptotics of convergent matrix integrals, as explained in \autoref{sec:non-perturbative}. 
Initially the converse was expected: that  asymptotic expansions of random matrix integrals would be useful for enumerating ribbon graphs.
But graph enumeration and formal series manipulations are purely algebraic, and the combinatorics of maps are much easier than large $N$ asymptotic analysis.

\section{Examples}

We have discussed the combinatorial expansion of the Hermitian random
matrix model.
Let us now consider other matrix ensembles.
See also Exercise \autoref{exo:kontsevich} for the example of the Kontsevich integral, and
\cite{Eynard:2006bs} for a short review.

\subsection{Multi-matrix models and colored graphs}\label{sec:multi-matrix}

\subsubsection{The Ising model}

Let us consider the formal two-matrix integral
\begin{align}
 \mathcal{Z}
 & =
 \int_{\text{formal}\,H_N\times H_N} dM_1 dM_2 \,
 e^{-\frac{N}{2} \operatorname{Tr}\left( M_1^2 + M_2^2 - 2c M_1 M_2 \right)} \,
 e^{\frac{g_1}{3} \operatorname{Tr} M_1^3} \,
 e^{\frac{g_2}{3} \operatorname{Tr} M_2^3} \, .
\end{align}
This involves two non--quadratic terms 
$\operatorname{Tr}M_1^3$ and $\operatorname{Tr}M_2^3$. 
In a graph, the two corresponding types of vertices can be distinguished by colors:
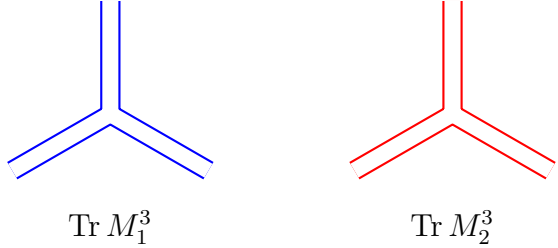
\begin{align}
 \begin{tikzpicture}[scale = 1.5,baseline=(current bounding box.center)]
  \draw [blue,thick,double,double distance=6pt]
  plot coordinates{(0,1) (0,0) (-.86, -.5)}
  plot coordinates{(0,0) (.86, -.5)};
  \draw [red,thick,double,double distance=6pt]
  plot coordinates{(3,1) (3,0) (3-.86, -.5)}
  plot coordinates{(3,0) (3.86, -.5)};
  \draw (0,-1) node {$\operatorname{Tr} M_1^3$};
  \draw (3,-1) node {$\operatorname{Tr} M_2^3$};
 \end{tikzpicture}
\end{align}
The color of a vertex can be interpreted as the state of a two-valued spin at this vertex.
The two-matrix model therefore describes the Ising model on a
random surface~\cite{Bershadsky:1986ws,Kazakov:1986hu}.

Let us write the quadratic terms in the action in terms of a rank two matrix,
\begin{align}
 M_1^2 + M_2^2 -2 c M_1 M_2
 & =
 \begin{pmatrix}
  M_1 \\ M_2
 \end{pmatrix}^\text{T}
  \begin{pmatrix}
   1 & -c \\ -c & 1
  \end{pmatrix}
  \begin{pmatrix}
   M_1 \\ M_2
  \end{pmatrix}
 \, .
\end{align}
The inverse of this matrix,
\begin{align}
 \begin{pmatrix}
   1 & -c \\ -c & 1
 \end{pmatrix}^{-1}
 & =
 \frac{1}{1-c^2}
  \begin{pmatrix}
   1 & c \\ c & 1
 \end{pmatrix}
 \, ,
\end{align} 
gives us the propagator
\begin{align}
 \Big<
  (M_\alpha)_{ij} (M_\beta)_{kl}
 \Big>
 & = \frac{1}{N}
 \frac{1}{1 - c^2} \,
 \delta_{il} \delta_{jk}
 \times
 \begin{cases}
  1 & (\alpha = \beta) \\
  c & (\alpha \neq \beta)
 \end{cases}
 \, .
\end{align}
For example, the two planar graphs with two vertices of different colors are 
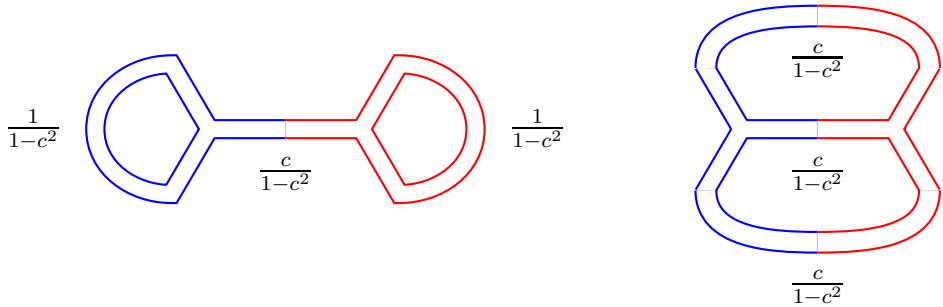
\begin{align}
 \begin{tikzpicture}[baseline=(current bounding box.center)]
  \draw [blue,thick,double,double distance=6pt]
  (0,0) to (-1,0) to (-1.5,.86) to [out=left,in=up] (-2.5,0) to
  [out=down,in=left] (-1.5,-.86) to (-1,0) ;
  \draw [red,thick,double,double distance=6pt]
   (0,0) to (1,0) to (1.5,.86) to [out=right,in=up] (2.5,0) to
   [out=down,in=right] (1.5,-.86) to (1,0) ;
  \draw (0,-.6) node {$\frac{c}{1-c^2}$};
  \draw (2.8,0) node [right] {$\frac{1}{1-c^2}$};
  \draw (-2.8,0) node [left] {$\frac{1}{1-c^2}$};
  \draw [blue,thick,double,double distance=6pt]
  plot coordinates{(7-1,0) (7-1.5,-.86)}
  plot coordinates{(7,0) (7-1,0) (7-1.5,.86)};
  \draw [blue,thick,double,double distance=7pt]
  (7-1.47,-.81) to [out=down,in=left] (7,-1.5);
  \draw [blue,thick,double,double distance=7pt]
  (7-1.47,.81) to [out=up,in=left] (7,1.5);
  \draw [red,thick,double,double distance=6pt]
  plot coordinates{(8,0) (8.5,-.86)}
  plot coordinates{(7,0) (8,0) (8.5,.86)};
  \draw [red,thick,double,double distance=7pt]
  (8.47,-.81) to [out=down,in=right] (7,-1.5);
  \draw [red,thick,double,double distance=7pt]
  (8.47,.81) to [out=up,in=right] (7,1.5);
  \draw (7,.9) node {$\frac{c}{1-c^2}$};
  \draw (7,-.6) node {$\frac{c}{1-c^2}$};
  \draw (7,-2.1) node {$\frac{c}{1-c^2}$};
 \end{tikzpicture}
\end{align}
The contributions of these two graphs would be equal in a one-matrix model, but here they differ due to the color dependence of the propagators.

Our formal two-matrix integral can be written as a sum over graphs:
\begin{align}
\frac{1}{2^N (\frac{\pi}{N})^{N^2} (1-c^2)^{\frac{N^2}{2}}}\mathcal Z 
= \sum_{\text{bi-colored triangulated surfaces}\,\Sigma}
\frac{N^{\chi(\Sigma)}}{\#{\mathrm{Aut}}(\Sigma)} \frac{g_1^{n_1} g_2^{n_2} c^{n_{12}}}{(1-c^2)^{\#\mathrm{edges}}} \ ,
\end{align}
where $n_1$ and $n_2$ are the numbers of triangles of each color, and $n_{12}$ is the number of edges separating triangles of different colors.

\subsubsection{The $O(n)$ matrix model}
 
Consider the following multi-matrix model,
\begin{align}\label{defOnmmgraphs}
 \mathcal{Z} = \int_{\text{formal}\,\,H_N^{1+n}} dM dA_1 \dots dA_n \,
 e^{-\frac{N}{2} \operatorname{Tr} M^2} \,
 e^{N \sum_{k \ge 3} \frac{t_k}{k} \operatorname{Tr} M^k} \,
 \prod_{i=1}^n \
 e^{-\frac{N}{2}  \operatorname{Tr} A_i^2} \,
 e^{N  \operatorname{Tr} MA_i^2} \, .
\end{align}
We will now color the links according to which quadratic term is used to build the propagator: $\operatorname{Tr} M^2$ in black, or $\operatorname{Tr} A_i^2$ in color, with $n$ possible colors $1,\dots, n$.
The specific feature of this model is the three-leg vertex with one $M$-link and two $A_i$-links,
\begin{align}
 \begin{tikzpicture}[baseline=(current bounding box.center)]
 \node at (0,0.1) [left] {$\displaystyle N\operatorname{Tr} M A_i^2$};
 \draw[very thick,blue,-latex] (0.3,0) -- (1.3,0);
%
  % vertex
% 
  \draw [red,thick] (3,0)++(120:.1) -- ++ (60:1);
  \draw [red,thick] (3,0)++(0:.1) -- ++ (60:1);
  \draw [red,thick] (3,0)++(-120:.1) -- ++ (-60:1);
  \draw [red,thick] (3,0)++(0:.1) -- ++ (-60:1);
  \draw [thick] (3,0)++(120:.1) -- ++ (180:1);
  \draw [thick] (3,0)++(-120:.1) -- ++ (180:1);
 \end{tikzpicture}
\end{align}
Since all terms of the action are quadratic in $A_i$, colored links must form
non-intersecting loops on the resulting graphs:
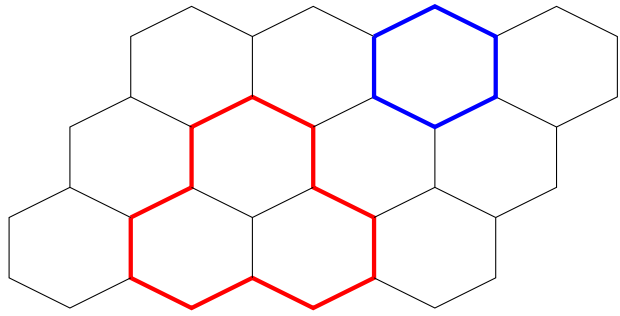
\begin{align}
\begin{tikzpicture}[scale = .4,baseline=(current bounding box.center)]
\draw [ultra thick, red]
(4, 0) -- (6,1) -- (6,3) -- (8,4) -- (10, 3) -- (10, 1) -- (12, 0) -- (12, -2) -- (10, -3) -- (8, -2) -- (6, -3) -- (4, -2) -- (4, 0);
\draw 
(6,1) -- (8, 0) -- (8, -2);
\draw 
(8, 0) -- (10, 1);
\draw 
(4, -2) -- (2, -3) -- (0, -2) -- (0, 0) -- (2, 1) -- (2, 3) -- (4, 4) -- (4, 6) -- (6, 7) -- (8, 6) -- (10, 7) -- (12, 6);
\draw 
(2, 1) -- (4, 0);
\draw 
(4, 4) -- (6, 3);
\draw 
(8, 6) -- (8, 4);
\draw 
(10, 3) -- (12, 4);
\draw 
(12, -2) -- (14, -3) -- (16, -2) -- (16, 0) -- (18, 1) -- (18, 3) -- (20, 4) -- (20, 6) -- (18, 7) -- (16, 6);
\draw 
(12, 0) -- (14, 1) -- (14, 3);
\draw 
(14, 1) -- (16, 0);
\draw 
(16, 4) -- (18, 3);
\draw [ultra thick, blue]
(12, 6) -- (14, 7) -- (16, 6) -- (16, 4) -- (14, 3) -- (12, 4) -- (12, 6);
\end{tikzpicture}
\end{align}
This model therefore describes a colored loop gas on a random surface
and is called the
 \textbf{\boldmath $O(n)$ matrix model}\index{On matrix@$O(n)$ matrix model}~\cite{Kostov:1988fy,Gaudin:1989vx,Eynard:1995nv}.

It is actually possible to continue $n$ to arbitrary non-integer values, by performing the Gaussian integrals over $A_i$, we recover the $M$-dependent factor that appears in the measure \eqref{eq:mon}:
\begin{equation}
\prod_{i=1}^n \int dA_i\,\,
 e^{-\frac{N}{2} \operatorname{Tr} A_i^2} \,
 e^{N \operatorname{Tr} MA_i^2} 
 \propto \det(\operatorname{Id}\otimes \operatorname{Id} - M\otimes \operatorname{Id}-\operatorname{Id}\otimes M )^{-n}\ .
\end{equation}
This is in particular useful for studying the replica limit $n \to 0$ and the Kosterlitz--Thouless transition $n\to 2$.
Moreover, the case $n \to 1$ is equivalent to
the Ising model.
(See \autoref{exo:ising}.)

\subsubsection{The four color problem}

Let us consider the following three-matrix
integral~\cite{Eynard:1998he,Kostov:2000vn},
\begin{equation}
\int_{\text{formal}\, H_N^3} dA\,dB\,dC\,\,e^{-\frac{N}{2}\operatorname{Tr}\left(A^2+B^2+C^2\right)}\,\,e^{\frac{N}{3} \operatorname{Tr}\left(ABC + \operatorname{Tr} ACB\right)}\ .
\end{equation}
In the resulting graphs, we assign a different color to each of the propagators $\operatorname{Tr}A^2$, $\operatorname{Tr}B^2$ and $\operatorname{Tr}C^2$. 
Then each vertex is trivalent, and must involve all three colors:
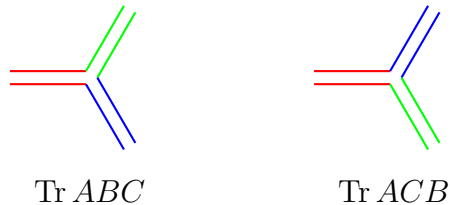
\begin{align}
 \begin{tikzpicture}[baseline=(current bounding box.center)]
  \draw [red,thick] (120:.1) -- ++(180:1);
  \draw [red,thick] (-120:.1) -- ++(180:1);
  \draw [green,thick] (120:.1) -- ++(60:1);
  \draw [green,thick] (0:.1) -- ++(60:1);
  \draw [blue,thick] (-120:.1) -- ++(-60:1);
  \draw [blue,thick] (0:.1) -- ++(-60:1);
  \draw [red,thick] (4,0)++(120:.1) -- ++(180:1);
  \draw [red,thick] (4,0)++(-120:.1) -- ++(180:1);
  \draw [blue,thick] (4,0)++(120:.1) -- ++(60:1);
  \draw [blue,thick] (4,0)++(0:.1) -- ++(60:1);
  \draw [green,thick] (4,0)++(-120:.1) -- ++(-60:1);
  \draw [green,thick] (4,0)++(0:.1) -- ++(-60:1);
  \draw (0,-1.5) node {$\operatorname{Tr} ABC$};
  \draw (4,-1.5) node {$\operatorname{Tr} ACB$};
 \end{tikzpicture}
\end{align}
The resulting tricolor graphs can actually be mapped to the \textbf{four
color problem}\index{four color problem}.
The idea is that each face should have one of four colors. 
The colors of two adjacent faces must differ, and determine the color of the edge which separates them by a two-to-one mapping:
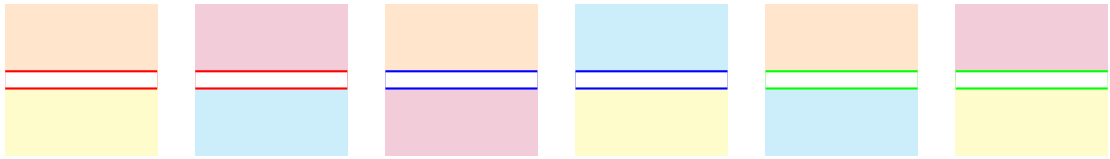
\begin{align}
 \begin{tikzpicture}[baseline=(current bounding box.center)]
\fill [orange, opacity = .2] (0,0) rectangle (2,1);
\fill [yellow, opacity = .2] (0,0) rectangle (2,-1);
\draw[red, thick, double, double distance = 2mm] (0,0) -- (2,0);
\fill [purple, opacity = .2] (2.5,0) rectangle (4.5,1);
\fill [cyan, opacity = .2] (2.5,0) rectangle (4.5,-1);
\draw[red, thick, double, double distance = 2mm] (2.5,0) -- (4.5,0);
\fill [orange, opacity = .2] (5,0) rectangle (7,1);
\fill [purple, opacity = .2] (5,0) rectangle (7,-1);
\draw[blue, thick, double, double distance = 2mm] (5,0) -- (7,0);
\fill [cyan, opacity = .2] (7.5,0) rectangle (9.5,1);
\fill [yellow, opacity = .2] (7.5,0) rectangle (9.5,-1);
\draw[blue, thick, double, double distance = 2mm] (7.5,0) -- (9.5,0);
\fill [orange, opacity = .2] (10,0) rectangle (12,1);
\fill [cyan, opacity = .2] (10,0) rectangle (12,-1);
\draw[green, thick, double, double distance = 2mm] (10,0) -- (12,0);
\fill [purple, opacity = .2] (12.5,0) rectangle (14.5,1);
\fill [yellow, opacity = .2] (12.5,0) rectangle (14.5,-1);
\draw[green, thick, double, double distance = 2mm] (12.5,0) -- (14.5,0);
 \end{tikzpicture}
\end{align}
Starting from a tricolor graph, the colors of all faces are determined once the color of one face is given:
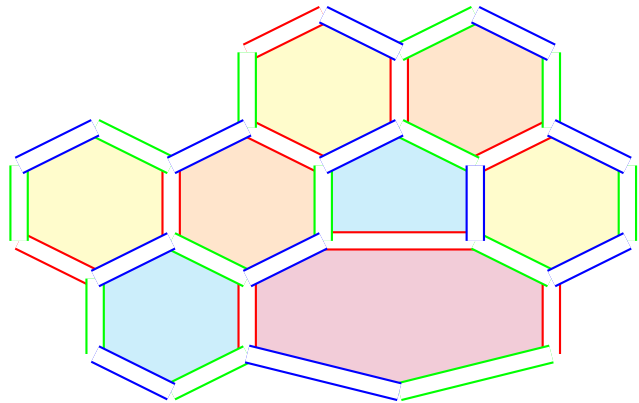
\begin{align}
 \begin{tikzpicture}[scale = .5,baseline=(current bounding box.center)]
\tikzset{style1/.style={thick, double, double distance = 2mm}}
\tikzset{style2/.style={opacity = .2}}
\fill[style2, cyan] (8,4)--(8,6)--(10,7)--(12,6)--(12,4)--(8,4);
\fill[style2, cyan] (4,4)--(6,3)--(6,1)--(4,0)--(2,1)--(2,3)--(4,4);
\fill[style2, yellow] (2,3)--(4,4)--(4,6)--(2,7)--(0,6)--(0,4)--(2,3);
\fill[style2, yellow] (6,7)--(8,6)--(10,7)--(10,9)--(8,10)--(6,9)--(6,7);
\fill[style2, yellow] (12,6)--(14,7)--(16,6)--(16,4)--(14,3)--(12,4)--(12,6);
\fill[style2, orange] (4,6)--(6,7)--(8,6)--(8,4)--(6,3)--(4,4)--(4,6);
\fill[style2, orange] (10,7)--(10,9)--(12,10)--(14,9)--(14,7)--(12,6)--(10,7);
\fill[style2, purple] (6,1)--(6,3)--(8,4)--(12,4)--(14,3)--(14,1)--(10,0)--(6,1);
\draw[red, style1] (0, 4) -- (2, 3);
\draw[red, style1] (4, 6) -- (4, 4);
\draw[red, style1] (6,3) -- (6,1);
\draw[red, style1] (14,3)--(14,1);
\draw[red, style1] (8, 4) -- (12, 4);
\draw[red, style1] (6, 7) -- (8, 6);
\draw[red, style1] (6, 9) -- (8, 10);
\draw[red, style1] (10, 7) -- (10, 9);
\draw[red, style1] (12, 6) -- (14, 7);
\draw[green, style1] (6,7) -- (6, 9);
\draw[green, style1] (10,9) -- (12, 10);
\draw[green, style1] (10,7) -- (12, 6);
\draw[green, style1](14, 7)--(14, 9);
\draw[green, style1](16, 6)--(16, 4);
\draw[green, style1](12, 4) --(14, 3);
\draw[green, style1](10, 0) -- (14, 1);
\draw[green, style1](4, 0)--(6,1);
\draw[green, style1](2,1)--(2,3);
\draw[green, style1](4, 4)--(6, 3);
\draw[green, style1](8, 4) --(8, 6);
\draw[green, style1](0, 4)--(0,6);
\draw[green, style1](2,7)--(4,6);
\draw[blue, style1](0, 6) -- (2, 7);
\draw[blue, style1](4, 6)--(6,7);
\draw[blue, style1](8, 10)--(10,9);
\draw[blue, style1](12,10)--(14,9);
\draw[blue, style1](14,7)--(16,6);
\draw[blue, style1](16,4)--(14,3);
\draw[blue, style1](6,1)--(10,0);
\draw[blue, style1](6,3)--(8,4);
\draw[blue, style1](8,6)--(10,7);
\draw[blue, style1](2,3)--(4,4);
\draw[blue, style1](2,1)--(4,0);
\draw[blue, style1](12,4)--(12,6);
 \end{tikzpicture}
\end{align}

\subsection{Non-Hermitian matrix model and oriented edges}\label{sec:non-Hermitian_oriented}

Let us consider the following formal integral over complex random matrices,
\begin{align}
 \mathcal{Z} & =
 \int_{\text{formal }M_N(\mathbb{C})} dM \,
 e^{-N \operatorname{Tr} MM^\dag} \,
 e^{t \operatorname{Tr} M^2 (M^\dag)^2} \,
 e^{\tilde{t} \operatorname{Tr} MM^\dag MM^\dag}
 \, .
\end{align}
We consider non-Hermitian matrices with $M\neq M^\dag$, and this can be captured graphically by orienting the propagators:
\begin{align}
 \begin{tikzpicture}[baseline=(current bounding box.center)]
  \draw (0,0) node [left]
  {$\displaystyle 
  \Big< M_{ij} M_{kl}^\dag \Big> = \frac{1}{N} \, \delta_{il} \delta_{jk}$};
  \draw [blue,very thick,-latex] (.5,0) -- ++(1,0);
  \draw (2.5,0) node {$\displaystyle M_{ij}$};
  \draw (6,0) node {$\displaystyle M_{kl}^\dag$};
  \draw [thick] 
  (3.5,.15) node [left] {$i$} --  ++(1.5,0) node [right] {$l$};
  \draw [thick]
  (3.5,-.15) node [left] {$j$} --  ++(1.5,0) node [right] {$k$};
  \draw [thick] (4.2,-.3) -- (4.4,.0) -- (4.2,.3);
 \end{tikzpicture}
\end{align}
The two types of vertices are therefore 
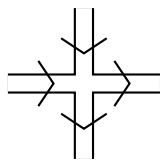
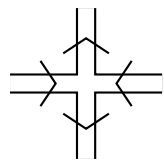
\begin{align}
 \begin{tikzpicture}[baseline=(current bounding box.center)]
  \draw [thick,double,double distance = 6pt]
  plot coordinates {(0,1)(0,0)(1,0)}
  plot coordinates {(-1,0)(0,0)(0,-1)};
  \draw [thick,double,double distance = 6pt]
  plot coordinates {(4,1)(4,0)(5,0)}
  plot coordinates {(3,0)(4,0)(4,-1)};
  \draw [thick] (-.3,.6) -- (0,.4) -- (.3,.6);
  \draw [thick] (-.3,-.4) -- (0,-.6) -- (.3,-.4);
  \draw [thick] (-.6,-.3) -- (-.4,0) -- (-.6,.3);
  \draw [thick] (.4,-.3) -- (.6,0) -- (.4,.3);
  \draw [thick] (4-.6,-.3) -- (4-.4,0) -- (4-.6,.3);
  \draw [thick] (4-.3,-.4) -- (4,-.6) -- (4.3,-.4);
  \draw [thick] (4-.3,.4) -- (4,.6) -- (4.3,.4);
  \draw [thick] (4.6,-.3) -- (4.4,0) -- (4.6,.3);
  \draw (0,-1.5) node [below] {$\operatorname{Tr} M^2(M^\dag)^2$};
  \draw (4,-1.5) node [below] {$\operatorname{Tr} MM^\dag MM^\dag$};
 \end{tikzpicture}
\end{align}
Putting these vertices on a square lattice and rotating them in all possible ways, we actually obtain six different vertices. Our matrix model therefore describes a six-vertex model on a random surface, whose continuum limit is
described by a $c=1$ conformal field theory coupled to two-dimensional gravity~\cite{Ginsparg:1991bi,Kostov:1999qx}.

\subsection{Real or quaternionic matrices and twisted ribbons}\label{sec:real_quaternionic_ribbons}

Let us consider the formal integral over real symmetric matrices,
\begin{align}
 \mathcal{Z} & =
 \int_{\text{formal}\,\, M_N^\text{sym}(\mathbb{R})} dM \,
 e^{-\frac{N}{2} \operatorname{Tr} M^2} \,
 e^{\frac{N}{4} t \operatorname{Tr} M^4} \, .
\end{align}
The quadratic term is
\begin{align}
 N\operatorname{Tr} M^2 
 & = N\sum_{i,j=1}^N M_{ij} M_{ji}
   = N\sum_{i,j=1}^N M_{ij} M_{ij}
 \, ,
\end{align}
from which we deduce the propagator
\begin{align}
 \Big< M_{ij} M_{kl} \Big>
  =
 \frac{1}{2N} 
 \left(
  \delta_{il} \delta_{jk} + \delta_{ik} \delta_{jl}
 \right)
 \, .
\end{align}
Graphically, we distinguish the two terms of the propagator by twisting (or not) the ribbon,
\begin{align}
 \begin{tikzpicture}[baseline=(current bounding box.center)]
  \draw (0,0) node [left] {$\displaystyle \Big< M_{ij} M_{kl} \Big>$};
  \draw [blue,very thick,-latex] (.5,0) -- ++(1,0);
  \draw [thick] 
  (2.5,.15) node [left] {$i$} -- ++(2,0) node [right] {$l$};
  \draw [thick]
  (2.5,-.15) node [left] {$j$} -- ++(2,0) node [right] {$k$};
  \draw (5.5,0) node {$+$};
  \begin{knot}[clip width=6]
   \strand [thick] 
   (6.5,.15) node [left] {$i$} -- ++(0.7,0) to [out=right,in=left]
   ++(.6,-.3) -- ++(.7,0) node [right] {$k$};
   \strand [thick]
   (6.5,-.15) node [left] {$j$} --  ++(.7,0) to [out=right,in=left]
   ++(.6,.3) -- ++(.7,0) node [right] {$l$}; 
  \end{knot}
 \end{tikzpicture}
\end{align}
Graphs with twisted ribbons live on non-orientable surfaces.
The surface associated to a graph is obtained by associating a square face to each four-valent vertex, and gluing squares together along edges, so that each edge is dual to a ribbon.
In the case of a graph with only one vertex, we obtain the 
Klein bottle and the real projective plane:
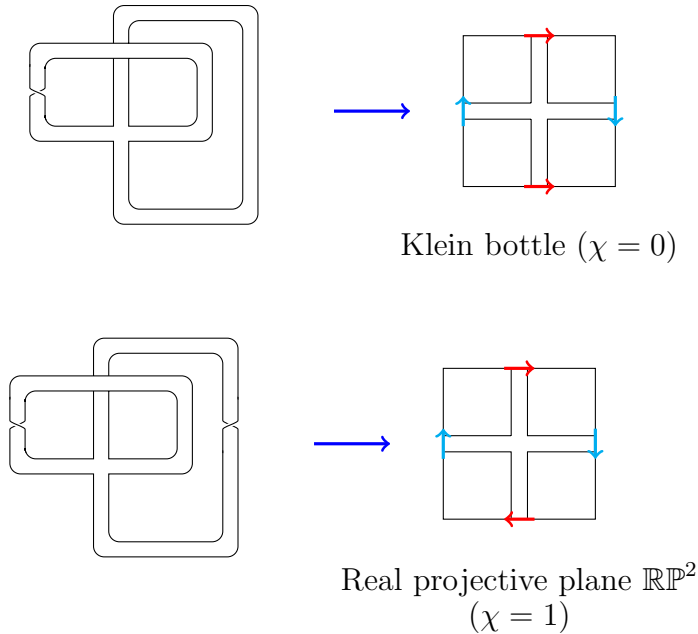
\begin{subequations}
\begin{align}
  \begin{tikzpicture}[baseline=(current bounding box.center)]
  \begin{knot}[clip width=2]
  \strand (.9,1) -- (.9,.1) [rounded corners] -- (0,.1) -- (0,.35) to
   [out=up,in=down] (-.2,.75) -- (-.2,1.2) -- (2.2,1.2) -- (2.2,-.1)
   [sharp corners] -- (1.1,-.1) [rounded corners] -- (1.1,-1) --
   (2.6,-1) -- (2.6,1.5) -- (1.1,1.5) -- (1.1,1.2);
  \strand (.9,1.2) [rounded corners] -- (.9,1.7) -- (2.8,1.7) --
   (2.8,-1.2) -- (.9,-1.2) [sharp corners] -- (.9,-.1) [rounded corners]
   -- (-.2,-.1) -- (-.2,.35) to [out=up,in=down] (0,.75) -- (0,1) --
   (2,1) -- (2,.1) [sharp corners] -- (1.1,.1) -- (1.1,1);
 \end{knot}
   \draw[blue,very thick,-latex] (3.8,.3) -- ++(1,0);
   \draw[double,double distance=.2cm]
   plot coordinates {(5.5,.3)(6.5,.3)(6.5,1.3)}
   plot coordinates {(7.5,.3)(6.5,.3)(6.5,-.7)};
   \draw (5.5,1.3) rectangle (7.5,-0.7);
   \draw[red,very thick,-latex] (6.3,1.3) -- ++(.4,0);
   \draw[red,very thick,-latex] (6.3,-0.7) -- ++(.4,0);
   \draw[cyan,very thick,-latex] (5.5,.1) -- ++(0,.4);
   \draw[cyan,very thick,-latex] (7.5,.5) -- ++(0,-.4);
   \draw (6.5,-1.5) node {Klein bottle ($\chi=0$)};
 \end{tikzpicture}
 \end{align}
% \vspace{1em}
 \begin{align}
  \begin{tikzpicture}[baseline=(current bounding box.center)]
  \begin{knot}[clip width=2]
  \strand (.9,1) -- (.9,.1) [rounded corners] -- (0,.1) -- (0,.35) to
   [out=up,in=down] (-.2,.75) -- (-.2,1.2) -- (2.2,1.2) -- (2.2,-.1)
   [sharp corners] -- (1.1,-.1) [rounded corners] -- (1.1,-1) --
   (2.6,-1) -- (2.6,.35) to [out=up,in=down] (2.8,.75) -- (2.8,1.7) --
   (.9,1.7) -- (.9,1.2);
  \strand (1.1,1) -- (1.1,.1) [rounded corners] -- (2,.1) -- (2,1) --
   (0,1) -- (0,.75) to [out=down,in=up] (-.2,.35) -- (-.2,-.1) [sharp
   corners] -- (.9,-.1) [rounded corners] -- (.9,-1.2) -- (2.8,-1.2) --
   (2.8,.35) to [out=up,in=down] (2.6,.75) -- (2.6,1.5) -- (1.1,1.5) --
   (1.1,1.2);
  \end{knot}	
   \draw[blue,very thick,-latex] (3.8,.3) -- ++(1,0);
   \draw[double,double distance=.2cm]
   plot coordinates {(5.5,.3)(6.5,.3)(6.5,1.3)}
   plot coordinates {(7.5,.3)(6.5,.3)(6.5,-.7)};
   \draw (5.5,1.3) rectangle (7.5,-0.7);
   \draw[red,very thick,-latex] (6.3,1.3) -- ++(.4,0);
   \draw[red,very thick,-latex] (6.7,-0.7) -- ++(-.4,0);
   \draw[cyan,very thick,-latex] (5.5,.1) -- ++(0,.4);
   \draw[cyan,very thick,-latex] (7.5,.5) -- ++(0,-.4);
   \draw (6.5,-1.5) node
   {Real projective plane $\mathbb{RP}^2$};
   \draw (6.5,-2) node {($\chi=1$)};
 \end{tikzpicture}
\end{align}
\end{subequations}
More generally, for $\beta\in\{1,2,4\}$, the formal integral over the Gaussian ensemble $E_N^\beta$
\begin{equation}
\int_{\text{formal }E_N^\beta} dM\,\,e^{-\frac{N\beta}{4}\,\operatorname{Tr} M^2} \,\,\, e^{\frac{N\beta}{2}\,\sum_k \frac{t_k}{k}\,\operatorname{Tr} M^k}\ ,
\end{equation}
yields the propagator
\begin{align}
 \Big< M_{ij} M_{kl} \Big>
 & =
 \epsilon_1   \delta_{il} \delta_{jk} -(\epsilon_1+\epsilon_2)\delta_{ik} \delta_{jl}
 \, ,
\end{align}
where we used the Nekrasov variables $\epsilon_1,\epsilon_2$ \eqref{Nekrasov_notation}.
Graphically, 
\begin{align}
 \begin{tikzpicture}[baseline=(current bounding box.center)]
  \draw (0,0) node [left] {$\displaystyle \Big< M_{ij} M_{kl} \Big>$};
  \draw [blue,very thick,-latex] (.3,0) -- ++(.7,0);
  \draw (2,0) node [left] {$\displaystyle \epsilon_1$};
  \draw [thick] 
  (2.5,.15) node [left] {$i$} -- ++(2,0) node [right] {$l$};
  \draw [thick]
  (2.5,-.15) node [left] {$j$} -- ++(2,0) node [right] {$k$};
  \draw (6.5,0) node {$- \quad (\epsilon_1+\epsilon_2)$};
  \begin{knot}[clip width=6]
   \strand [thick] 
   (8.5,.15) node [left] {$i$} -- ++(0.7,0) to [out=right,in=left]
   ++(.6,-.3) -- ++(.7,0) node [right] {$k$};
   \strand [thick]
   (8.5,-.15) node [left] {$j$} --  ++(.7,0) to [out=right,in=left]
   ++(.6,.3) -- ++(.7,0) node [right] {$l$}; 
  \end{knot}
 \end{tikzpicture}
\end{align}
This allows us to extend the formal integral to arbitrary values of $\beta\in \mathbb{C}$, as the generating function of weighted ribbon graphs, with weights $\epsilon_1$ and $-(\epsilon_1+\epsilon_2)$ associated to untwisted and twisted ribbons.
Twisted ribbons have a vanishing weight in the case $\beta=2$ of Hermitian matrices.

\section{Exercises}

\begin{exo}[Kontsevich integral]
~\label{exo:kontsevich}
 Consider the Gaussian matrix integral
 \begin{align}
  \mathcal Z_0(\Lambda,t) = \int_{H_N} dM\, e^{-Nt^{-\frac13}\operatorname{Tr} M^2 \Lambda }\ ,
 \end{align}
 where $\Lambda=\mathrm{ diag}(\lambda_1,\dots,\lambda_N)$ is a given diagonal matrix with distinct positive eigenvalues, and $t>0$.
Adding a cubic interaction, we define a formal matrix integral called the \textbf{Kontsevich integral}\index{Kontsevich integral}~\cite{Kontsevich:1992ti},
\begin{subequations}
\begin{align}
 \mathcal Z(\Lambda,t) 
&= \frac{1}{\mathcal Z_0(\Lambda,t)}\int_{\mathrm{ formal}\,H_N} dM\, e^{-N\operatorname{Tr} \left( t^{-\frac13} M^2 \Lambda - \frac{1}{3} M^3 \right) } \ ,
\label{eq:koni}
\\
&= \frac{1}{\mathcal Z_0(\Lambda,t)} \sum_{k=0}^\infty \frac{N^k}{3^k k!} \int_{H_N}dM\, (\operatorname{Tr} M^3)^k e^{-N t^{-\frac13}\operatorname{Tr} M^2 \Lambda }\ . \label{eq:koni_exp}
\end{align}
\end{subequations}

\begin{enumerate}

\item Compute the Gaussian integral $\mathcal Z_0(\Lambda,t)$.

\item Compute the propagator, and write $\mathcal Z(\Lambda,t) $ as a sum over ribbon graphs. Find the weights of edges and vertices.
Observe that the power of $t$ provides a grading of graphs by their numbers of edges. 
Explain why the small $t$ expansion is also a large $\Lambda$ expansion.

\item Find all graphs of genus $0$, with $3$ edges, and compute their weighted sum.

\item Find all graphs of genus $1$, with $3$ edges, and compute their weighted sum.

\item Compute the formal expectation value of the matrix element $M_{1,1}$ as a sum over graphs.
Compute the contribution of graphs of genus $1$ with $3$ edges to this expectation value.

\end{enumerate}

\end{exo}

\begin{exo}[Ising model on triangulated surfaces and $O(1)$ model]
~\label{exo:ising}
Let $\mathcal{T}$ be the set of triangulated surfaces, with two types of triangles called spin up and spin down. The Ising model on triangulated surfaces has the partition function
\begin{align}
 \mathcal{Z}^\text{Ising} = \sum_{T\in \mathcal{T}} g^{\#\mathrm{triangles}(T)} e^{h\#\mathrm{up}(T)-h\#\mathrm{down}(T)} a^{\#E(T)}N^{\chi(T)}\ ,
\end{align}
where $E(T)$ is the set of edges that separate triangles of different spins, and $\chi(T)$ is the Euler characteristic. The weights $g$, $h$, $a$ are called the cosmological constant, the magnetic field and the energy.
\begin{enumerate}
\item Show that  the partition function coincides with the formal matrix integral
\begin{equation}
\mathcal Z = \int_{\mathrm{ formal}} dM_1 dM_2\, e^{-N \operatorname{Tr}\left( \frac{b M_1^2}{2} + \frac{b M_2^2}{2} -c M_1 M_2  - \frac{g e^h}{3} M_1^3 - \frac{g e^{-h}}{3} M_2^3 \right)} \ ,
\end{equation}
where the parameters $b,c$ are defined by $\left(\begin{smallmatrix}
b & -c \\
-c & b
\end{smallmatrix}\right)=\left(
\begin{smallmatrix}
1 & a \\
a & 1
\end{smallmatrix}\right)^{-1}$.
\item Show that the Ising model on triangulated surfaces is equivalent to the $O(1) = \mathbb{Z}_2$ matrix model \eqref{defOnmmgraphs}.
\end{enumerate}
\end{exo}

\chapter{Coulomb gas method}\label{chap:SaddlePoint}

In this chapter we introduce the Coulomb gas method for studying the leading large $N$ asymptotics of matrix models.
This method is an infinite-dimensional version of the saddle-point approximation.
An integral has a \textbf{saddle-point approximation}\index{saddle-point approximation} in some asymptotic regime, if it is dominated by the largest value of its integrand. Then this largest value corresponds to a saddle point of the action. 
In probabilistic language, this is called the localization or concentration of the probability measure on the most probable configuration. 

In this chapter, our goal is to find and study the most probable configuration.
This provides an ansatz for the large $N$ behaviour of the distribution of eigenvalues.
We will admit without proof that the distribution of eigenvalues does have this behaviour.
Having the correct ansatz would actually be essential if we wanted to prove it.

Our saddle-point problem is analogous to 
finding an equilibrium configuration of electrostatic charges in a potential, or more generally an equilibrium density of charges.
The name of the Coulomb gas method comes from the Coulombian interaction between the charges.

We will find that the equilibrium density is an algebraic function, and that the equilibrium distribution of eigenvalues can be described in terms of an algebraic curve called the \textbf{spectral curve}\index{spectral curve}. 
In \autoref{chap:LoopEq} we will see that the spectral curve encodes not only the leading behaviour, but also the whole large $N$ asymptotic expansion of correlation functions.

\section{Saddle points and eigenvalue integrals}\label{sec:spei}

\subsection{Coulomb gas of eigenvalues}

\subsubsection{Why we focus on eigenvalues}

One could think of applying the saddle-point approximation directly to a matrix integral, rather than to the corresponding integral over eigenvalues. 
This would amount to finding the most probable matrix, rather than the most probable spectrum.
Let us explain why these two approaches are not equivalent, and why we do need the most probable spectrum.

Let us consider the one-matrix integral
\begin{equation}
 \mathcal{Z} = \frac{1}{N! \mathcal V_N^\beta} \int_{E_N^\beta} dM\ e^{-N\frac{\beta}{2}\operatorname{Tr} V(M)} \ .
\end{equation}
Due to the invariance under conjugations by unitary matrices, the extremums of $\operatorname{Tr} V(M)$ are hugely degenerate.
In general they are not saddle points, in the sense that they are not isolated, rather they form a submanifold of dimension $O(N^2)$, giving rise to Goldstone bosons.
As a result, they do not dominate the large $N$ asymptotics of $\mathcal{Z}$.
In order to find the correct large $N$ asymptotics, we should take invariance under conjugation into account, and reduce $\mathcal{Z}$ to an integral over eigenvalues.
Then the saddle points in the space of eigenvalue configurations do determine the large $N$ asymptotics.
However, since the number of eigenvalues still depends on $N$, we will have to introduce eigenvalue distributions, rather than keeping track of individual eigenvalues.

\subsubsection{Coulomb gas}

Reducing $\mathcal{Z}$ to an integral over eigenvalues  \eqref{eq:znormalbeta} and omitting the group volume prefactor, we have 
\begin{align}
 \mathcal{Z} = \frac{1}{N!}
 \left(\prod_{i=1}^N \int_{\gamma} d\lambda_i\right) e^{-N^2 \frac{\beta}{2}
 S(\lambda_1, \ldots, \lambda_N)}  \ ,
\label{eq:z}
\end{align}
where 
$\beta^{-1}$ is called the temperature in statistical physics, and 
the action is
\begin{align}
\boxed{
S(\lambda_1,\ldots,\lambda_N) 
= \frac{1}{N} \sum_{i=1}^N V(\lambda_i) -\frac{1}{N^2} \sum_{i<j}^N \log(\lambda_i-\lambda_j)^2 } \ .
\label{eq:action}
\end{align}
This action, and therefore also its extremums, are $\beta$-independent.
Moreover, assuming $V$ to be $N$-independent, both terms of the action are $O(1)$, and the action has a good chance of having a nontrivial large $N$ limit. 
This is because we followed the prescription of \autoref{sec:defuni} when writing our matrix integral $\mathcal{Z}$, and included a prefactor $N$ in front of the potential.

Our action can be identified with the energy of a \textbf{Coulomb gas}\index{Coulomb gas}: an ensemble of point particles in the one-dimensional space $\gamma$, subject to a potential $V$ and a repulsive logarithmic electrostatic Coulomb interaction. 
At equilibrium, the eigenvalues cannot all gather at the minimum of the potential, as their mutual repulsion keeps them at distances typically $O(\frac{1}{N})$. 
Rather, they tend to occupy finite segments near the bottoms of potential wells:
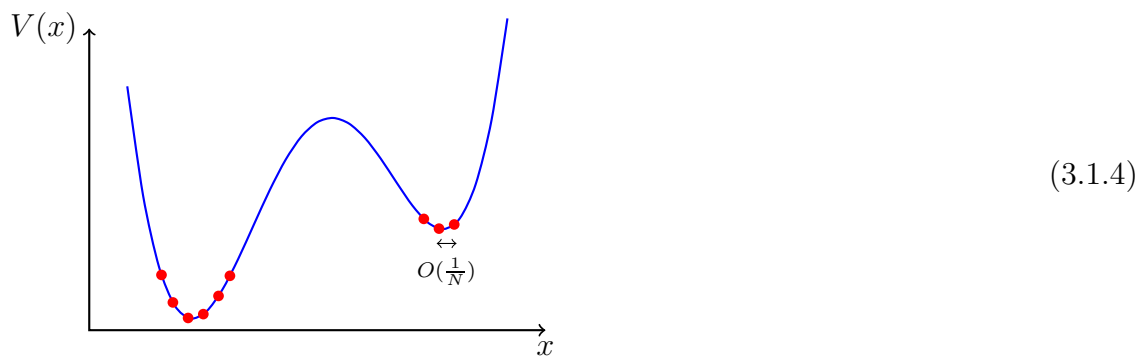
\begin{equation}
\begin{tikzpicture}[baseline=(current  bounding  box.center),
declare function = {potential(\x) = cosh(1.1*(\x - 3)) + 3*cos(1.2*(\x - 3.1) r) + .16*\x - 1.7;}
]
 \draw[thick, latex-latex] (0, 4) node[left]{$V(x)$} -- (0, 0) -- (12, 0) node[below]{$x$};
 \draw[domain=1:11,smooth,variable=\x,blue, thick] plot ({\x},{potential(\x/2)});
\foreach \x in {1.1, 1.3, .95, 1.5, 1.7, 1.85, 4.4, 4.6, 4.8}
 \fill [red] (2*\x, {potential(\x)}) circle (2pt);
 \draw[latex-latex] (9.14,1.15) -- (9.66,1.15);
 \node[below] at (9.4,1.1) {$\scriptstyle O(\frac{1}{N})$};
\end{tikzpicture} 
\label{fig:doublewell}
\end{equation}

\subsubsection{Saddle points of the action}

The saddle points are the solutions of the $N$ saddle-point equations $\frac{\partial S}{\partial \lambda_i} = 0$, 
explicitly
\begin{equation}
 V'(\lambda_i) = \frac{2}{N} \sum_{j(\neq i)} \frac{1}{\lambda_i - \lambda_j} 
  \quad \text{for} \quad i = 1, \ldots, N \, .
 \label{eq:saddlelambda}
\end{equation}
For rational potentials, these equations are algebraic, and have finitely many solutions. 
The number of solutions, up to permutations of the eigenvalues, is
\begin{equation}
d_N=\#\left\{\{\lambda_1,\ldots,\lambda_N\}\middle|\frac{\partial S}{\partial \lambda_i}=0\right\} \Big/ \mathfrak S_N  
=\binom{N+d-1}{N}\ ,
\end{equation}
where 
\begin{equation}
 d = \deg V'\ .
\end{equation}
(In the case of $d_2=\frac{d(d+1)}{2}$, we give the proof in \autoref{exo:nbsdpt}.)
This number coincides with the number $d_N$ \eqref{eq:dimh1} of homologically independent convergent integrals with the potential $V$, hinting at a correspondence between saddle points and integration contours.
In \autoref{sec:detpol}, we will explain which integration contours correspond to a given saddle point in the large $N$ limit.

Whether the potential is real or not, the solutions of the saddle-point equations are not always real. 
In order to account for complex saddle points, and the corresponding complex integration contours, we have to consider normal matrix ensembles. 
Even if we start with a Wigner ensemble $E_N^\beta$ with real eigenvalues, we may therefore have to homotopically deform the integration domain in the complex plane, so that it includes a saddle point.

\subsection{The large \texorpdfstring{$N$}{N} limit as a WKB approximation}

\subsubsection{Riccati equation for the resolvent}

Given a solution $\{\lambda_i\}$  of the saddle-point equations \eqref{eq:saddlelambda}, let us consider the Dirac comb eigenvalue distribution
\begin{align}
 \rho_\text{D}(x) = \frac{1}{N} \sum_{i=1}^N \delta(x-\lambda_i)\ ,
 \label{eq:rho_D}
\end{align}
and its Stieltjes transform, which is the resolvent of a matrix with spectrum $\{\lambda_i\}$,
\begin{equation}
 W(x) = \frac{1}{N} \sum_{i=1}^N \frac{1}{x-\lambda_i} \ .
 \label{eq:rho_D_Stieltjes}
\end{equation}
%If the solution $\{\lambda_i\}$ dominates the matrix integral, then the eigenvalue density coincides with $\rho_D(x)$, and its Stieltjes transform coincides with $W(x)$. By an abuse of terminology, we then call $W(x)$ the resolvent.
In order to find an equation for $W(x)$, we compute
\begin{align}
 W(x)^2 + \frac{1}{N} W'(x) 
 % &= \frac{1}{N^2}\left(\sum_{i,j=1}^N \frac{1}{x-\lambda_i} \frac{1}{x-\lambda_j} - \sum_{i=1}^N \frac{1}{(x-\lambda_i)^2}\right)\ , \nonumber  \\
 &= \frac{1}{N^2} \sum_{\substack{i,j=1 \\ (i \neq j)}}^N \frac{1}{x-\lambda_i} \frac{1}{x-\lambda_j}\ .
\end{align}
Using $\frac{1}{x-\lambda_i} \frac{1}{x-\lambda_j} = (\frac{1}{x-\lambda_i}-\frac{1}{x-\lambda_j}) \frac{1}{\lambda_i-\lambda_j}$, and Eq.~\eqref{eq:saddlelambda}, this leads to
\begin{equation}
 W(x)^2 + \frac{1}{N} W'(x) 
= \frac{2}{N^2} \sum_{i=1}^N \frac{1}{x-\lambda_i} \sum_{j(\neq i)} \frac{1}{\lambda_i -\lambda_j} \ 
= \frac{1}{N} \sum_{i=1}^N \frac{V'(\lambda_i)}{x-\lambda_i}\ .
\end{equation}
Introducing
\begin{align}
 P(x) = \frac{1}{N} \sum_{i=1}^N \frac{V'(x)-V'(\lambda_i)}{x-\lambda_i}  \ ,
\label{defPx1}
\end{align}
we obtain the \textbf{Riccati equation}\index{Riccati equation}
\begin{equation}
\boxed{ W(x)^2 + \frac{1}{N} W'(x) 
= V'(x) W(x) - P(x) }
\ .
 \label{eq:omegadiff}
\end{equation}
Assuming that $P(x)$ is known, this is a closed nonlinear differential equation for $W(x)$. 
From its definition, we do know that $P(x)$ is a rational function of degree $d-1$, whose leading behaviour  at $\infty$ is the same as that of $\frac{V'(x)}{x}$. 
This fully determines $P(x)=V''(x)$ if $d=1$, and leaves $d-1$ unknown coefficients in general. 
For the moment we will work with arbitrary values of these coefficients, and consider $P$ as known. 
In \autoref{sec:detpol}, we will explain how to determine $P$.

\subsubsection{Linear equation for the characteristic polynomial}

The Riccati equation for $W(x)$ can be rewritten as the linear second-order differential equation
\begin{align}\label{eq:BethetoBaxter}
 \frac{1}{N^2} \psi'' - \frac{1}{N} V'\psi' + P \psi = 0\ ,
\end{align}
for the function $\psi(x)$ defined by
\begin{align}
 W
 = \frac{1}{N} \frac{\psi'}{\psi}
 = \frac{1}{N} \left( \log \psi \right)'
 \ .
 \end{align}
Conversely, $\psi(x)$ can be written in terms of $W(x)$ as
\begin{equation}
 \psi(x) = x^N  \exp \left( {N \int_\infty^x (W(x')-\frac{1}{x'}) dx'} \right)
  \ ,
\label{eq:psiw}
\end{equation}
and in terms of the eigenvalues $\lambda_i$ as 
 \begin{align}
 \psi(x) = \prod_{i=1}^N (x-\lambda_i) \ .
\end{align}
So $\psi(x)$ is the characteristic polynomial of the diagonal matrix with eigenvalues $\lambda_i$.

\subsubsection{Analogy with Bethe ansatz and Baxter equations}

The relation between Riccati and linear equations, is analogous to the relation between Bethe ansatz and Baxter equations in integrable systems. 
Our saddle-point equations \eqref{eq:saddlelambda} are indeed identical to the Bethe ansatz equations for the Jaynes--Cummings--Gaudin model, if we identify our matrix eigenvalues $\lambda_i$ with Bethe roots. And our linear equation for $\psi(x)$ is identical to the linear Baxter equation for Baxter's $Q$-function, whose zeros are the Bethe roots.
This is a first hint of the deep link between random matrices and integrable systems.

\subsubsection{Large $N$ limit}

The large $N$ WKB formal approximation, amounts to 
neglecting the  term $\frac{1}{N} W'$ of the Riccati equation, which then reduces to the algebraic equation
\begin{align}
\boxed{
 \overline{W}^2 - V'\overline{W} + \overline{P} = 0 \quad \text{with} \quad 
 \left\{\begin{array}{l} 
 \overline{W} = \underset{N\to \infty}{\lim} W\ , \\
 \overline{P} = \underset{N\to \infty}{\lim} P\ .
\end{array}
\right. }
 \label{eq:ovomdef}
\end{align}
The solution is
\begin{align}
 \overline{W} = \frac12\left(V' - \sqrt{(V')^2 - 4\overline{P}}\right)\ .
 \label{eq:ovomsol}
\end{align}
According to Eq.~\eqref{eq:rewo}, the eigenvalues are found at the singularities of the resolvent $\overline W$. 
Therefore, in the WKB large $N$ limit, the eigenvalues coalesce into a one-dimensional curve: the cut of the square root in the expression for $\overline W$. We will determine this curve in \autoref{sec:supprho}, before determining the polynomial $\overline P$ in \autoref{sec:detpol}. Assuming the curve and the polynomial are known, the equilibrium density is deduced from the resolvent via Eq.~\eqref{eq:rewo}, 
\begin{align}
\boxed{
 \bar{\rho} = \frac{1}{2\pi} \sqrt{4\overline{P} - (V')^2} \times {\mathbf 1}_{\operatorname{supp}\bar\rho} } \ .
\label{eq:barrho}
\end{align}
In terms of the characteristic polynomial, the  WKB approximation amounts to
\begin{equation}
 \psi(x) \sim  \ C \ e^{\frac{N}{2} V(x)}   \  e^{-\frac{N}{2}\int^x \sqrt{V'(x')^2-4 \overline{P}(x')}dx'  }
.
\end{equation}

\subsubsection{Gaussian case}

In the case of the quadratic potential $V(x)=\frac12 x^2$, we have $\overline P = P=V'' =1$.
Then $\psi(x)$ satisfies the equation
\begin{align}
 \frac{1}{N^2} \psi''(x) - \frac{1}{N} x\psi'(x) + \psi(x) = 0\ .
\end{align}
Up to a rescaling of $x$, this is the differential equation for the Hermite polynomial $H_N(x)$, 
and the solution is 
\begin{equation}
\psi(x) = N^{-\frac{N}{2}}  H_N(x\sqrt{N})\ .
\end{equation}
Therefore, after a rescaling by $\frac{1}{\sqrt N}$, the equilibrium eigenvalues coincide with the zeros of $H_N(x)$.
Since these zeros lie in the interval $[-2\sqrt N,2\sqrt N]$, the equilibrium spectrum is in the interval $[-2,2]$, in agreement with Wigner's semi-circle law.

It is of course simpler to directly solve the large $N$ equations, than to solve the finite $N$ equations before taking the limit. 
In our Gaussian example, the solution of the large $N$ Riccati equation \eqref{eq:ovomsol} is 
\begin{align}
 \overline W = \frac12\left(x-\sqrt{x^2-4}\right)\ ,
 \label{eq:bwsc}
\end{align}
and the corresponding equilibrium density \eqref{eq:barrho} agrees with Wigner's semi-circle law \eqref{sc_law}.

\section{The equilibrium density}\label{sec:ded}

Instead of minimizing the action with respect to the $N$ eigenvalues before sending $N\to\infty$ as we did in \autoref{sec:spei},
we now want to send $N\to\infty$, and then minimize the action with respect to the eigenvalue density.
Given a potential $V$ and integration domain $\Gamma$, we will find a unique equilibrium density $\bar\rho$. 
The map $(V,\Gamma) \mapsto \bar{\rho}$ is however not invertible: an equilibrium density corresponds to a unique potential, but may correspond to many homologically inequivalent domains.

In order to find the equilibrium density, we will first minimize the real part of the action at fixed integration domain, and then maximize it with respect to the domain. 
As usual with the saddle-point approximation, this yields a saddle point of the action (not just of its real part) that dominates the large $N$ limit of the corresponding integral.

\subsection{Toy model: one-dimensional integrals}
\label{sec:sdp1d}

Before extremizing $N$-dimensional matrix integrals, let us deal with the case of 1-dimensional integrals, and discuss large deviations in that case.

For $\gamma$ a complex Jordan arc from $\infty$ to $\infty$, and $V(x)=tx^{d+1}+\cdots$ a polynomial of degree $d+1$, let us consider the partition function
\begin{equation}
\mathcal Z=\int_\gamma e^{-NV(x)}dx\ .
\end{equation}
Since the integrand is analytic, the integral depends only on the class of $\gamma$ in the homology space $H_1(e^{-NV(x)}dx)$, whose dimension is $d$.

\subsubsection{Extremization of $\Re V$}

Let us view $e^{-NV(x)}$ as a complex measure, and its modulus $e^{-N\Re V(x)}$ as a positive probability measure. 
We might think that in the large $N$ limit, the partition function is dominated by the point $X\in\gamma$ with the highest probability, i.e. the lowest value of $\Re V$. 
But this is not true if $V'(X)\neq 0$, because of presence of the pure imaginary subleading term in $V(x)\underset{x\to X}{=} V(X) + V'(X)(x-X) + \dots$, leading to oscillations of frequency $N$.
The partition function is actually dominated by points $X$ such that $V'(X)=0$, i.e. saddle points of the potential.
Since $V$ is a polynomial, it obviously has saddle points. 
For pedagogical reasons, we will now characterize saddle points in terms of $\Re V$, and infer their existence 
in a similar way as we will do for the equilibrium density in \autoref{sec:fspe}. 

Let $x_\gamma\in\gamma$ be a point such that $\Re V(x_\gamma)=\underset{x\in\gamma}{\min} \Re V(x)$. By definition of the homology space $H_1(e^{-NV(x)}dx)$, $\Re V$ is indeed bounded from below on $\gamma$, and reaches its lowest value. 
Moreover, if $\gamma$ does not vanish homologically, then it must intersect the set 
$A=(-t)^{\frac{1}{d+1}}\bigcup_{j=0}^d \left(e^{ \frac{2\pi i}{d+1}j}\mathbb{R}_+\right)$, on which $\Re V$ is bounded from above. 
It follows that $\Re V(x_\gamma) \leq \underset{x\in A}{\max} \Re V(x)$.
Therefore there exists
\begin{equation}
V_0 = \sup_{\gamma} \min_{x\in\gamma} \Re V(x)  < \infty\ .
\end{equation}
We admit that there are a Jordan arc $\gamma$ whose homology class  $\in H_1(e^{-NV(x)}dx)$ and a point $X=x_\gamma\in \gamma$ such that $V_0=\Re V(X)$.

By construction, $X$ extremizes $\Re V$ not only on $\gamma$, but also with respect to deformations of $\gamma$ itself.
Parametrizing our Jordan arc with a function $\gamma:\mathbb{R}\to \mathbb{C}$, the deformations of $x=\gamma(s)$ are 
\begin{align}
\delta x = \gamma'(s)\delta s + (\delta\gamma)(s)\ ,
\end{align}
which can take any complex value. 
That $X=\gamma(s) $ minimizes $\Re V$ for longitudinal deformations amounts to the condition
\begin{align}
\forall \delta s\in\mathbb{R}\ , \quad  \Re \Big( V'(X) \gamma'(s)\delta s \Big) \geq  0\ ,
\end{align}
and that $X$ maximizes $\Re V$ for transverse deformations amounts to the condition
\begin{align}
\forall \delta \gamma\in\mathbb{C}\ , \quad  \Re \Big( V'(X) \delta \gamma \Big) \leq  0\ .
\end{align}
This second condition is equivalent to $V'(X)=0$. (And $V'(X)=0$ implies the first condition.)
Then the saddle point with the lowest value of $\Re V$ dominates the partition function.

\subsubsection{Ascending paths}

The saddle point that dominates the partition function needs not belong to our original contour $\gamma$. Even if 
$\gamma=\mathbb{R}$ and $V$ is real, some saddle points can actually be complex. 
However, any Jordan arc from $\infty$ to $\infty$ has a homotopic deformation that contains a saddle point.
Conversely, let us associate an arc to a saddle point $X_i$.

We define the ascending path $\gamma_i$ as a Jordan arc  from $\infty$ to $\infty$ such that
\begin{itemize}
\item $X_i\in \gamma_i$,
\item $x\in \gamma_i\backslash\{X_i\}\implies \Re V(x)> \Re V(X_i) $, 
\item $\Re V(x)$ is never decreasing as $x$ moves away from $X_i$, in both directions.
\end{itemize}
From the Laplace saddle-point approximation theorem, the partition function associated to an ascending path behaves as 
\begin{equation}
\int_{\gamma_i} e^{-NV(x)}dx \underset{N\to\infty}{=} \frac{\sqrt{2\pi}}{\sqrt{N V''(X_i)}} e^{-NV(X_i)} \left(1+O(N^{-\frac12})\right) \ .
\end{equation}
Through any saddle point, there exists at least one ascending path. Choosing one ascending path for each one of the $d$ saddle points, we obtain a basis of $H_1(dx e^{-NV(x)})$.
Modulo deformations, any Jordan arc can be decomposed on this basis,
\begin{equation}
\gamma = \sum_{i=1}^d c_i \gamma_i \qquad \text{with} \qquad c_i \in \{-1,0,1\}\ .
\end{equation}
The 
corresponding partition function is then dominated by the saddle points with the smallest value $M=\min_{i, \ c_i\neq 0} \Re V(X_i)  $: 
\begin{equation}
\mathcal Z \sim \sum_{i|\Re V(X_i)=M}
 c_i \frac{\sqrt{2\pi}}{\sqrt{N V''(X_i)}} e^{-NV(X_i)} \left(1+O(N^{-\frac12})\right) \ .
\end{equation}

\subsubsection{Large deviations and concentration}

For $\gamma_i$ the ascending path through saddle point $X_i\in \gamma_i$, let us define the large deviation function
\begin{equation}
S_i(x) = \Re V(x)-\Re V(X_i) \ , \qquad (x\in\gamma_i)\ .
\end{equation}
This function is continuous on $\gamma_i$, with $S_i(x)\geq 0$.
For $M\in\mathbb{R}$ the set $\{x\in\gamma_i|S_i(x)\leq M\}$, called a \textbf{level set}\index{level set} is compact. 
Moreover, 
for any open subset $A$ of $\gamma_i$, 
\begin{equation}
\liminf_{N\to\infty} \frac{1}{N}\log \left| \frac{1}{\mathcal Z}\int_A e^{-NV(x)}dx \right| \geq - \inf_{x\in A} S_i(x) \ ,
\end{equation}
and for any closed subset set $A$ of $\gamma_i$,
\begin{equation}
\limsup_{N\to\infty} \frac{1}{N}\log \left| \frac{1}{\mathcal Z}\int_A e^{-NV(x)}dx \right| \leq - \inf_{x\in A} S_i(x)\ .
\end{equation}
The interpretation is that the probability of a closed subset $A$ of $\gamma_i$ tends to $1$ if $A$ contains $X_i$, and to $0$ exponentially fast if not.

One can show that $\lim_{N\to\infty}\left<x\right>= X_i$, with a variance that  tends to 0 as $O(\frac{1}{\sqrt{N}})$. This is called the concentration of the measure, and provides a precise sense in which the saddle point dominates the large $N$ limit of the partition function.
One can also show that a random sequence of points of $\gamma_i$  sampled with probability measure $\frac{1}{\mathcal Z}e^{-NV(x)}dx$, tends almost surely to the saddle point as $N\to \infty$.

\subsection{Functional saddle-point equation}\label{sec:fspe}

Now we come back to our $N$-dimensional probability measure.

\subsubsection{Functional action}

In the large $N$ limit, let us model the action $S(\lambda_1,\dots, \lambda_N)$ \eqref{eq:action} as a functional $S[\rho]$ of the eigenvalue density. 
We are looking for a functional such that $S[\rho_\text{D}]=S(\lambda_1,\dots, \lambda_N)$, where $\rho_\text{D}$ \eqref{eq:rho_D} is the Dirac comb density that describes $N$ eigenvalues. 
We first rewrite both terms of the action in terms of $\rho_\text{D}$,
\begin{subequations}
\begin{align}
N\sum_i V(\lambda_i) &= N^2 \int V(x) \rho_\text{D}(x) dx\ ,
\\
\sum_{i\neq j} \log{(\lambda_i-\lambda_j)}  &= N^2 \int \rho_\text{D}(x) dx \ \rho_\text{D}(x')dx' \ \log{(x-x')}\ ,
\end{align}
\end{subequations}
where the integral is performed outside the diagonal $\{x=x'\}$.
Moreover, in the large $N$ limit, we describe the integration domain $\Gamma=\vec\gamma^{\vec n}$ as a set of constraints
\begin{align}
 \int_{\gamma_i} \rho(x)dx = \epsilon_i\ ,
 \label{eq:igre}
\end{align}
where $\epsilon_i = \lim_{N\to\infty} \frac{n_i}{N}$ is a filling fraction, and $\sum_i\epsilon_i=1$.
We introduce Lagrange multipliers $\ell_i$ for these constraints, 
and we end up with the functional action
\begin{multline}\label{def:Srho}
 S[\rho] = \int_{\cup\gamma_i} V(x) \rho(x) dx - \int_{(\cup\gamma_i)^2} \log(x-x') {\rho}(x) dx \ {\rho}(x') dx'
\\
 + \sum_i\ell_i \left(\epsilon_i-\int_{\gamma_i}  \rho(x) dx \right)\ .
\end{multline}
After extremizing this action with respect to the density $\rho$, we will have the two equivalent options of either extremizing with respect to $\ell_i$, or directly imposing the constraints \eqref{eq:igre}. We will choose the latter option, and the constraints can therefore be viewed as the equations that determine $\ell_i$. In the case where we only impose one constraint $\sum_i \int_{\gamma_i} \rho(x)dx = 1$, we only have one Lagrange multiplier $\forall i,\, \ell_i=\ell$.

% Notice that when the constraints are satsfied, the last term in \eqref{def:Srho} vanishes.

\subsubsection{Space of probability measures}

We define a probability measure as an equivalence class of pairs $\rho=(\gamma, f)$ such that $\rho(x) dx = \rho(\gamma(s)) \gamma'(s) ds =  f(s)ds$, where:
\begin{itemize}
 \item $\gamma : \mathbb R \to \mathbb C$ parametrizes a finite union of piecewise $C^1$ Jordan arcs $\gamma=\cup_i\gamma_i$ that may end at infinity or at poles of $V'$, in directions such that
 \begin{align}
  \lim e^{-V(\gamma(s))} = 0\ , \qquad \text{or equivalently} \qquad \lim \Re V(\gamma(s)) = +\infty\ .
  \label{lsi}
 \end{align}
 \item The measure $f(s)ds$ is positive  and obeys $\int f(s)ds = 1$. This distribution is defined on test functions that are piecewise continuous and polynomially bounded: for $h:\mathbb{R}\to \mathbb{R}$ a piecewise continuous function such that $|h(s)|\underset{\gamma(s)\to\infty}{\leq} |\gamma(s)|^n$ with $n\geq 0$, we have $\left|\int_\mathbb{R} h(s)f(s)ds \right|<\infty$.
 \item Two measures are equivalent if and only if their Stieltjes transforms
  \begin{align}
  W[\rho](x) = \int_{\cup\gamma_i} \frac{\rho(x')dx'}{x-x'} = \int_{\mathbb{R}} \frac{f(s)ds}{x-\gamma(s)}
  \label{eq:strho}
 \end{align}
 coincide on a non empty open subset of $\mathbb C$, equivalently if they have the same polynomial moments $\int_\gamma x^n \rho(x)dx$ for all $n\geq 0$.

 Therefore, two measures related by a reparametrization $(\gamma,f)\to (\gamma\circ \phi,f\circ \phi)$ are equivalent. Two measures can be equivalent without being related by reparametrization, in particular if two arcs $\gamma_1,\gamma_2$ have a non-empty intersection.
\end{itemize}
Then we define the following spaces of measures:
\begin{itemize}
 \item Let $\mathcal{D}$ be the space of all probability measures, given a potential $V$.
 \item Given a domain $\gamma$, let $\mathcal{D}_{[\gamma]}$ be the space of measures that are supported on the corresponding Jordan arcs.
 \item Let $\mathcal{D}_{[\gamma],\epsilon}$ be the space of measures with a fixed domain and fixed filling fractions $\epsilon=(\epsilon_1,\epsilon_2,\dots)$.
\end{itemize}

\subsubsection{Minimization at fixed domain}

We will now show that the functional action $\Re S[\rho]$ \eqref{def:Srho}
admits a unique minimum $\bar \rho$ on $\mathcal D_{[\gamma],\epsilon}$. An important object will be the
\textbf{effective potential}\index{effective potential}
\begin{align}\label{def:Veff}
V_\text{eff}[\rho](x)
= \Re V(x) - 2 \int \rho(x')dx'\, \log|x-x'| \ .
\end{align}
The support of $\bar\rho$ is compact, and the values of the effective potential are given by the Lagrange multipliers:
\begin{align}
\left\{\begin{array}{lcl}
        x\in \gamma_i\cap \operatorname{supp}\bar{\rho} & \implies & V_\text{eff}[\bar\rho](x) =\ell_i \ ,
        \\
        x\in \gamma_i-\operatorname{supp}\bar{\rho} & \implies & V_\text{eff}[\bar\rho](x) \geq \ell_i\ .
       \end{array}
\right.
 \label{eq:saddlerho}
\end{align}
The proof of these results relies on the fact that $\Re S[\rho]$ is a \textbf{good rate functional}\index{good rate functional}, which means that $\Re S[\rho]$ is strictly convex, bounded from below, and lower semi-continous, with level sets that are closed, tight and compact.
Good rate functionals are at the heart of the large deviation theory in probabilities and are used to give a mathematical definition of functional integrals.
%These properties will imply the existence of a minimum.
%
%Then, from Cantor's intersection theorem, the intersection of all level sets is not empty, which shows that the minimum $\bar\rho$ exists. And strict convexity implies that $\bar\rho$ is unique. 
Let us prove the properties of $\Re S[\rho]$ one at a time:
\begin{itemize}
 \item \textbf{Strictly convex:} This is because the quadratic term $Q[\rho,\rho]$ of $\Re S[\rho]$ is positive definite on the space of signed distributions with zero charge $\int \rho(x)dx=0$. This can be seen by rewriting it as the $L^2$ norm of the Stieltjes transform $W[\rho]$, which is finite thanks to the zero charge condition:
 \begin{align}
  Q[\rho,\rho]=-\int \rho(x)dx\rho(x')dx'\log|x-x'| = \frac{1}{2\pi} \int_{\mathbb C} |W[\rho](x)dx|^2 \ .
 \end{align}

\item \textbf{Bounded from below:}
For $\rho_0\in \mathcal{D}_{[\gamma],\epsilon}$ an arbitrary reference measure, we have
\begin{align}
 \Re S[\rho] = -Q[\rho_0,\rho_0] + Q[\rho-\rho_0,\rho-\rho_0] + \int V_\text{eff}[\rho_0](x)\rho(x)dx
 \ .
 \label{rsr}
\end{align}
Since  $\rho-\rho_0$ is a signed measure with zero charge, we have $Q[\rho-\rho_0,\rho-\rho_0]\geq 0$.
Since $\Re V(x)\underset{x\to\infty}{\geq} 2\log |x|$, it follows that $V_\text{eff}[\rho_0]$ is bounded from below on $\gamma$, so that $\int V_\text{eff}[\rho_0](x)\rho(x)dx \geq \inf_\gamma V_\text{eff}[\rho_0]$, and thus $\Re S[\rho] \geq  -Q[\rho_0,\rho_0] + \int V_\text{eff}[\rho_0](x)\rho(x)dx > -\infty $.
\item \textbf{Lower semi-continuous:} The functional $\Re S[\rho]$ is not necessarily continuous, because the integration kernel $\log|x-x'|$ is unbounded. However, since $\rho$ is positive, we have
\begin{align}
 Q[\rho,\rho] = \limsup_{\Lambda\to +\infty} \int_{(\cup\gamma_i)^2}  {\rho}(x) dx \ {\rho}(x') dx' \ \inf\left(\Lambda,-\log|x-x'| \right)\ .
\end{align}
This shows that $Q[\rho,\rho]$ is the limit superior of functionals that are Lipschitz thus continuous. A limit superior of a family of continuous functions is necessarily lower semi-continuous, therefore, $\Re S[\rho]$ is lower semi-continuous.
\item \textbf{Closed level sets:} The level sets
\begin{align}
\mathcal{L}_M=\Big\{\rho\in \mathcal D_{[\gamma]}  \Big|  \Re S[\rho]\leq M\Big\}\ ,
\end{align}
are closed by lower semi-continuity.
\item \textbf{Tight level sets:}
Tightness means that $\lim_{n\to \infty} \int_{\gamma(\mathbb R-[-n,n])} \rho(x)dx  =0$, uniformly on the level set.
To prove it, let us introduce the sequence
\begin{align}
C_n = \inf_{\gamma(\mathbb R-[-n,n])} V_\text{eff}[\rho_0](x) \ ,
\end{align}
then by Eq.~\eqref{lsi} and because the potential is polynomial, we have $\lim_{n\to \infty} C_n = +\infty$. For any measure in the level set $\mathcal{L}_M$ we have, using Eq.~\eqref{rsr},
\begin{align}
 M \geq \Re S[\rho] \geq -Q[\rho_0,\rho_0] + C_0 + C_n \int_{\gamma(\mathbb R-[-n,n])} \rho(x)dx \ .
\end{align}
This implies
\begin{align}
\int_{\gamma(\mathbb R-[-n,n])} \rho(x)dx \leq \frac{ M + Q[\rho_0,\rho_0] - C_0}{C_n}
\end{align}
that tends to $0$ as $n\to\infty$, uniformly in $\rho$.
\item \textbf{Compact level sets:} Prohorov's theorem ensures that closed and tight level sets are then compact.
\end{itemize}
The minimum is at the intersection of all level sets, and since they are compact, Cantor's intersection theorem implies that the intersection is non-empty: a minimum exists. Strict convexity implies that  $\Re S[\rho]$ has a unique minimum $\bar\rho=(\gamma,\bar f)$.
For any infinitesimal fluctuation
$\delta\rho = (\gamma,\delta f)$ in the tangent space of $\mathcal{D}_{[\gamma]}$, we have $\delta\Re S = \Re S[\rho_0+\delta\rho]- \Re S[\rho_0]\geq 0$, with
\begin{align}
 \delta \Re S = \int_{\mathbb R}   \Big(V_{\text{eff}}(\gamma(s)) - \sum_i \ell_i \mathbf 1_{\gamma^{-1}(\gamma_i)}\Big)\delta f(s) ds\ .
\end{align}
For $s$ such that $\bar f(s)\neq 0$, the fluctuation $\delta f(s)$ can be positive or negative, which implies $V_\text{eff}(\gamma(s)) = \ell_i$. On the other hand, if $\bar f(s)=0$, then $\delta f(s)$ must be positive, and we can only conclude $V_\text{eff}(\gamma(s)) \geq \ell_i$.
This justifies Eq.~\eqref{eq:saddlerho}. The support of $\bar \rho$ is compact because of Eq.~\eqref{lsi}, which also applies to $V_\text{eff}[\bar\rho]$ because the logarithm is negligible compared to the polynomial potential.

\subsubsection{Maximization with respect to $\gamma$}

In order to find a saddle point of the functional action, let us now maximize $\Re S[\bar\rho]$ with respect to $\gamma$. We first have to show that it is bounded from above. Since $\bar\rho$ minimizes $\Re S$ at fixed $\gamma$, it is enough to find a density $\tilde{\rho}_\gamma$ for each $\gamma$, such that $\Re S[\tilde{\rho}_\gamma]$ is bounded from above with a $\gamma$-independent bound. The proof is proposed in \autoref{exo:ubrs}.

The space $\mathcal{D}$ of positive densities is not complete, as a limit of Jordan arcs needs not itself be a Jordan arc, and can even be fractal (think of Peano curve). Therefore, it is not obvious that there exists a positive density $\bar\rho$ that maximizes $\Re S[\bar\rho]$.
This can however be proved with the help of Stieltjes tranforms. The idea is to consider a sequence of densities $\bar\rho_n$ such that $\lim_{n\to\infty}\Re S[\bar\rho_n] = \sup_\gamma\Re S[\bar\rho]$, and to show that their Stieltjes transforms have a limit $\lim_{n\to\infty} W_n = \overline{W}\in L^2(\mathbb C)$.
Then, using the Euler--Lagrange equations, we show that $\overline{W}$ is in fact the Stieltjes transform of a positive density $\bar\rho$ on some domain $\bar \gamma$ -- actually, a steepest descent road.

We now admit the existence of a density $\bar\rho$ that maximizes $\Re S[\bar\rho]$ with respect to $\gamma$. Then the Euler--Lagrange equations
are obtained by writing that under an infinitesimal deformation $\delta\gamma$ of $\gamma$, we have 
\begin{align}\label{eq:drs}
0\geq  \delta \Re S[\bar\rho] = \Re \int_\mathbb{R} \delta\gamma(s)f(s)ds \Big(V'(\gamma(s)) - 2\overline W^\text{reg}(\gamma(s))\Big)\ ,
\end{align}
where we define the Stieltjes tranform $\overline W$ of $\bar \rho$ not only on $\mathbb{C} \backslash \operatorname{supp}\bar\rho$, but also on $\operatorname{supp}\bar\rho$, using the regularization
\begin{align}
 \overline W^{\text{reg}}(x) \underset{x\in \operatorname{supp}\bar\rho}{=} \frac12 \Big( \overline W(x+i0) +\overline W(x-i0) \Big)\ .
\end{align}
By definition of $\bar \rho$, we must have $\forall \delta\gamma(s),\ \delta \Re S[\bar \rho]\leq 0$, and the corresponding Euler--Lagrange equation takes the form of the \textbf{Riemann--Hilbert equation}\index{Riemann--Hilbert!---equation}
\begin{equation}
\boxed{
x\in\operatorname{supp}\bar{\rho}\ \ \implies \ \ 
 V'(x) = \overline{W}(x+i0) + \overline{W}(x-i0) } \ ,
 \label{eq:saddleder}
\end{equation}
which is the continuum limit of the saddle-point equation \eqref{eq:saddlelambda} for the matrix eigenvalues.

Let us give an alternative formulation of this Riemann--Hilbert equation.
Let us go back to Eq.~\eqref{eq:drs}, and rewrite $f(s)$ in terms of $\overline W$ using 
$f(s) = \gamma'(s)\bar\rho(\gamma(s))$ together with Eq.~\eqref{eq:rewo}. We then find
\begin{align}
 \delta \Re S[\bar \rho] = -\frac{1}{2\pi} \Im \int_\mathbb{R} \delta\gamma(s)\gamma'(s)ds\Big(\overline P(\gamma(s)+i0)-\overline P(\gamma(s)-i0)\Big)\ , 
\label{eq:saddlederP1}
\end{align}
where we introduced
\begin{align}
 \overline P(x) = V'(x) \overline W(x) - \overline W(x)^2\ .
 \label{eq:pvo}
\end{align}
Using again $\forall \delta\gamma(s),\ \delta \Re S[\bar \rho]\leq 0$, we find $\overline P(x+i0)=\overline P(x-i0)$ for $x\in \operatorname{supp}\bar\rho$. By definition, $\overline P(x)$ is analytic for $x\notin \operatorname{supp}\bar\rho$, so $\overline P(x)$ is an entire function on $\mathbb{C}$. This analyticity condition on $\overline P(x)$ is our alternative formulation of the Riemann--Hilbert equation.

The analyticity condition can be rewritten as $\bar\partial\, \overline{P}(x) =0$. This does not involve $\operatorname{supp} \bar\rho$, which need therefore not be a Jordan arc. A variational derivation of this equation is obtained by rewriting Eq.~\eqref{eq:saddlederP1} as a two-dimensional integral using Stokes' theorem,
\begin{align}
 \delta \Re S[\bar \rho] = -\frac{1}{2\pi} \Im \int_{\mathbb{C}}  \delta\gamma(x) \ \bar\partial \,\overline{P}(x) \ d^2x \ ,
\end{align}
where $\delta\gamma$ is an arbitrary complex function.

\subsubsection{Solving the Riemann--Hilbert equation}

We have just found that $\overline{P}(x)$ is an entire function that behaves polynomially at infinity, $\overline{P}(x) \underset{x\to \infty}{\sim} \frac{V'(x)}{x}$. Therefore, $\overline{P}(x)$ must be a polynomial of degree $\deg \overline{P} =d-1$, which actually coincides with the previously defined $\overline{P}(x)$ \eqref{eq:ovomdef}. 
Assuming this polynomial is known, 
we can solve Eq.~\eqref{eq:pvo} for $\overline W(x)$, and recover the equilibrium density \eqref{eq:barrho} via
\begin{equation}\label{eq:omegajump}
\bar \rho(x) = -\frac{1}{2\pi i} \left( \overline W(x+i0)-\overline W(x-i0) \right),
\end{equation}
which is equivalent to Eq.~\eqref{eq:rewo}.
Let us factorize the polynomial $V'^2-4\overline P$ into its odd and even zeros, by introducing two polynomials $M(x)$ and $\sigma(x)$ such that all zeros of $\sigma(x)$ are distinct and
\begin{equation}
\boxed{
(V')^2-4\overline{P} = M^2\sigma } \ .
\label{eq:Msigmadef}
\end{equation}
We then rewrite equations \eqref{eq:barrho} and \eqref{eq:pvo} as 
\begin{equation}
\boxed{
 \bar{\rho}(x) = \frac{1}{2\pi}M(x)\sqrt{-\sigma(x)} \times {\mathbf 1}_{\operatorname{supp}\bar\rho}(x) \quad , \quad \overline{W}(x)=\frac12\left(V'(x)-M(x)\sqrt{\sigma(x)}\right) } \ ,
 \label{eq:rsms}
\end{equation}
Since $\deg \left((V')^2-4\overline{P} \right) = (\deg V')^2 $ is even, $\deg \sigma$ is also even, and we can write
\begin{equation}
\boxed{\sigma(x) = \prod_{j=1}^{2s} (x-a_i)}\ ,
\end{equation}
where $a_i$ is called a \textbf{spectral edge}\index{spectral edge}. 
Let $k_m$ be the order of a zero $m$ of $M^2\sigma$, then $\sum_m k_m = 2d$. We have $k_{a_i}\geq 1$ for spectral edges and $k_m\geq 2$ for other zeros, which implies that the number of zeros is at most $d+s$.

Therefore, $\overline W$ is an algebraic function, and is the Stieltjes transform of a smooth density $\bar\rho$, called the equilibrium density, which is algebraic  with compact support.

\subsubsection{Tricomi relation}

Actually, the resolvent $\overline{W}$ can be written in terms of the polynomial $\sigma$ alone (and not $M$) via the \textbf{Tricomi relation}\index{Tricomi relation},
\begin{align}
\boxed{
 \overline{W}(x) = \frac{1}{2\pi i} \int_{\operatorname{supp} \bar{\rho}} dx' \sqrt{\frac{\sigma(x)}{\sigma(x')}} \ \frac{V'(x')}{x-x'} } \ .
\end{align}
To prove this relation, we first use the behaviour of $\overline{W}(x)$ near $x=\infty$, and write
\begin{align}
 \frac{\overline{W}(x)}{\sqrt{\sigma(x)}}
 = \frac{1}{2\pi i}\oint_x \frac{dx'}{x'-x}\frac{\overline{W}(x')}{\sqrt{\sigma(x')}}
 = -\frac{1}{2\pi i}\oint_{\operatorname{supp}\overline{\rho}}\frac{dx'}{x'-x}\frac{\overline{W}(x')}{\sqrt{\sigma(x')}}\ .
 \end{align}
Now we insert the expression \eqref{eq:rsms} for $\overline{W}$ in terms of $\sigma$ and $M$, whose second term does not contribute to the integral around $\operatorname{supp}\overline{\rho}$. The first term is evaluated using the identity $\oint_{\operatorname{supp}\bar{\rho}} \frac{f(x')}{\sqrt{\sigma(x')}} = -2 \int_{\operatorname{supp}\bar{\rho}} \frac{f(x')}{\sqrt{\sigma(x')}}$, which holds for any function $f$ that is analytic near $\operatorname{supp}\bar{\rho}$.

In the case where $\operatorname{supp}\bar{\rho}=[a,b]$ reduces to one interval, we can actually determine its boundaries $a$ and $b$ by analyzing the large $x$ behaviour of the Tricomi relation,
\begin{align}
\overline{W}(x) \underset{x\to\infty}{=} 
\frac{1-\frac{a+b}{2x}}{2\pi i} \int_a^b dx'\frac{1+\frac{x'}{x}}{\sqrt{(x'-a)(x'-b)}} V'(x') + O\left(\frac{1}{x^2}\right)\ .
\end{align}
Comparing with the known behaviour of $\overline{W}(x)$ \eqref{eq:omegainf}, we obtain two equations for the two unknowns $a$ and $b$,
\begin{align}
 \int_a^b dx'\frac{V'(x')}{\sqrt{(x'-a)(x'-b)}} = 0 \quad , \quad \frac{1}{2\pi i}\int_a^b dx'\frac{x'V'(x')}{\sqrt{(x'-a)(x'-b)}} = 1\ .
\end{align}
In the more general case with $\deg \sigma=2s >2$, the constraint \eqref{eq:omegainf} gives $s+1$ relations on the $2s$ spectral edges $a_i$. These constraints could alternatively be obtained by requiring that $(V')^2 -4\overline P$ has $d-s$ double zeros, but the Tricomi relations are linear in $V'$ and therefore simpler. 
The determination of the remaining $s-1$ unknowns 
will be discussed in \autoref{sec:detpol}.

\subsection{Determining the support}
\label{sec:supprho}
 
The equilibrium density is given by Eq.~\eqref{eq:barrho} in terms of its support $\operatorname{supp}\bar\rho$, and the polynomial $\overline P$.
Let us now determine $\operatorname{supp}\bar\rho$, given $\overline{P}$. 
Since $\bar \rho$ is an algebraic function on its support, assuming that it is positive and integrates to one implies that its support is compact.

\subsubsection{Real case}

If the potential and integration contour are real, and if moreover the spectral edges $a_i$ are real, we can 
assume $a_i < a_{i+1}$, so that
\begin{equation}
\operatorname{supp}\bar{\rho} = \mathop{{\bigcup}}_{i=1}^s  [a_{2i-1},a_{2i}] \ .
\end{equation}
Let us show that the constraint $\bar{\rho}>0$ leads to a bound on the number $s$ of intervals. 
Since $\sqrt{-\sigma}$ changes sign from one interval to the next, $M$ should also change sign, and therefore have a zero in $(a_{2i},a_{2i+1})$. Therefore $\deg M \geq s-1$, which implies $s\leq \frac{d+1}{2}$. 

The heuristic picture is that the support is a union of compact intervals, corresponding to the wells of the potential where the eigenvalues gather. The zeros of $M$ are approximately at the top of the potential barriers, and there is one barrier between each well.
The number of intervals is sensitive to the precise shape of the potential. 
For example, in the case of a quartic potential $\deg V = 4$ with two wells, we can have two intervals if the wells are deep enough, or one single interval if they are not:
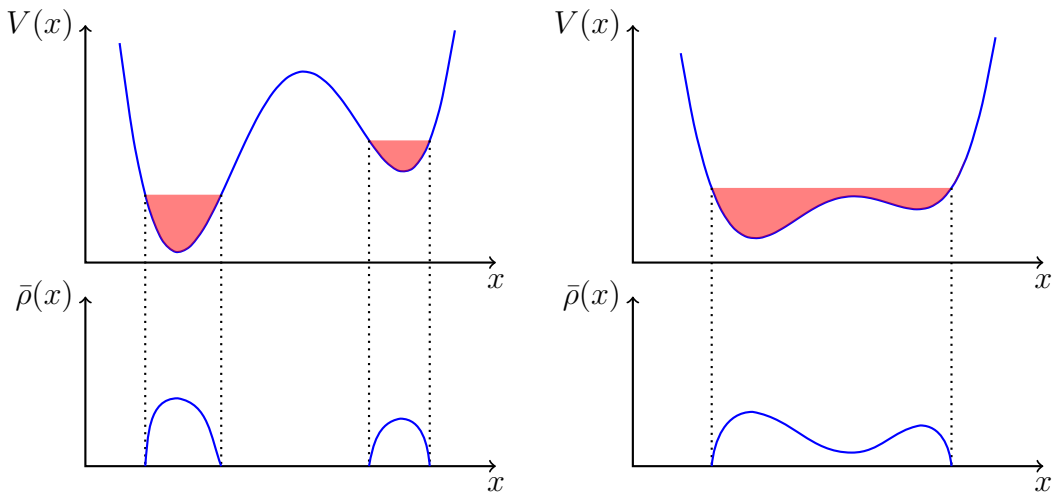
\begin{align}
\begin{tikzpicture}[scale = .9, baseline=(current  bounding  box.center),
declare function = {potential(\x) = cosh(1.1*(\x - 3)) + 3*cos(1.2*(\x - 3.1) r) + .16*\x - 1.7;},
declare function = {potential2(\x) = cosh(1.1*(\x - 3)) + 1.6*cos(1.2*(\x - 3.1) r) + .02*\x - 1.7;},
]
 \draw[thick, latex-latex] (0, 3.5) node[left]{$V(x)$} --
 	(0, 0) -- (6, 0) node[below]{$x$};
 \draw[domain=.5:5.4,smooth,variable=\x,blue, thick,name path=pot] plot (\x,{potential(\x)});
 \draw[opacity=0,name path=const1] (0,1) -- ++(3,0);
 \draw[opacity=0,name path=const2] (3,1.8) -- ++(3,0);
  \fill [red,opacity=.5,name intersections={of= pot and const1, by={a, b}}] 
  (a) -- plot [domain=1:2,smooth] ({\x},{potential(\x)}) -- (b) -- cycle;
  \fill [red,opacity=.5,name intersections={of= pot and const2, by={c, d}}] 
  (c) -- plot [domain=4.:4.9,smooth] ({\x},{potential(\x)}) -- (d) -- cycle;
  \draw [thick,latex-latex] (0,-.5) node[left]{$\bar{\rho}(x)$} -- (0,-3) -- ++(6,0) node [below] {$x$};
  \draw [dotted,thick] let \p1 = (a) in (a) -- (\x1,-3) coordinate (a2);
  \draw [dotted,thick] let \p1 = (b) in (b) -- (\x1,-3) coordinate (b2);
  \draw [dotted,thick] let \p1 = (c) in (c) -- (\x1,-3) coordinate (c2);
  \draw [dotted,thick] let \p1 = (d) in (d) -- (\x1,-3) coordinate (d2);
  \draw [thick,blue] (a2) to [out=85,in=left] (1.35,-2) to [out=-10,in=110] (b2);
  \draw [thick,blue] (c2) to [out=80,in=190] (4.6,-2.3) to [out=right,in=95] (d2);
  \begin{scope}[shift={(8,0)}]
   \draw[thick, latex-latex] (0, 3.5) node[left]{$V(x)$} --
 	(0, 0) -- (6, 0) node[below]{$x$};
 \draw[domain=.7:5.3,smooth,variable=\x,blue, thick,name path=pot] plot (\x,{potential2(\x)});
 \draw[opacity=0,name path=const3] (0,1.1) -- ++(6,0);
  \fill [red,opacity=.5,name intersections={of= pot and const3, by={a, b}}] 
  (a) -- plot [domain=1.1:4.9,smooth] ({\x},{potential2(\x)}) -- (b) -- cycle;
  \draw [thick,latex-latex] (0,-.5) node[left]{$\bar{\rho}(x)$} -- (0,-3) -- ++(6,0) node [below] {$x$};
  \draw [dotted,thick] let \p1 = (a) in (a) -- (\x1,-3) coordinate (a2);
  \draw [dotted,thick] let \p1 = (b) in (b) -- (\x1,-3) coordinate (b2);
  \draw [thick,blue] (a2) to [out=85,in=left] (1.75,-2.2) to [out=-10,in=left] (3.2,-2.8) to [out=right,in=190] (4.2,-2.4) to [out=right,in=95] (b2);
  \end{scope}  
\end{tikzpicture}
\end{align}
In this quartic example, since $\deg \left(4\overline{P}-(V')^2\right) = 6$, it may seem that the support could be made of three intervals, but this would be incompatible with the positivity of the equilibrium density.

\subsubsection{The spectral network}

The connected components of $\operatorname{supp}\bar{\rho}$ are arcs between the spectral edges $a_i$, on which $\bar\rho(x)dx$ is a positive measure.
This implies that $\operatorname{supp}\bar{\rho}$ must be a subset of the \textbf{spectral network}\index{spectral network}: the arcs between all the zeros of $M^2\sigma$, on which $M(x)\sqrt{-\sigma(x)}dx$ is a real measure. Let us first study the spectral network.

Given a zero $m$ of $M^2\sigma$, we define the function
\begin{align}
 g_m(x) = \pm \int_m^x M(x')\sqrt{\sigma(x')}dx'\ .
 \label{gmx}
\end{align}
Due to the square root, this function has a sign ambiguity, which does not affect the definition of the spectral network as
\begin{align}\label{def:spnetwork}
 \mathcal{S} = \bigcup_{\{m|M^2\sigma(m)=0\}}\left\{\Re g_m = 0\right\}_m\ ,
\end{align}
where $\left\{\Re g_m = 0\right\}_m$ is the connected component of $\left\{\Re g_m = 0\right\}$ that contains $m$. 
Let us describe the features of the spectral network, as a graph embedded in $\mathbb{C}$:
\begin{itemize}

 \item Vertices are the zeros of $M^2\sigma$.
 As a vertex of $\mathcal S$, a zero $m$ of order $k_m$ has the valency $k_m+2\geq 3$: spectral edges are generically trivalent, while zeros of $M$ are generically tetravalent. 

\item Since there is a finite number of vertices, there is a finite number of edges, $\mathcal S$ is a finite graph.

\item Edges end either at vertices, or at infinity.
Using $M\sqrt{\sigma}(x) \underset{x\to\infty}{\sim} V'(x)\underset{x\to\infty}{\sim} t_{d+1}x^{d}$, the edges of the spectral network go to infinity at angles in $  \frac{\pi}{d+1}(\mathbb{Z}+\frac12-\frac{\operatorname{Arg} t_{d+1}}{\pi})$.

\item Let \textbf{sectors}\index{sector} be the  connected components of $\mathbb C \backslash \mathcal S$. The sectors are simply connected and are never compact. This is because the spectral network has no univalent vertices and therefore no tree subgraphs, so a minimal compact sector $K$ would have to be simply connected. For a vertex $m\in\partial K$, the function $\Re g_m$ would be harmonic on $K$ and zero on $\partial K$, and therefore on $K$ too, which is impossible by definition of $g_m$. And without compact sectors, we cannot have non-simply connected sectors.

\item Given a sector $A$ and a vertex $m\in\partial A$, we arbitrarily choose the sign of $g_m(x)$ \eqref{gmx} in $A$. Since $A$ is simply connected, this unambiguously defines $g_m$ on $A$. Since $A$ contains no zero of $dg_m$, the function $g_m$ is a bijection from $A$ to $g_m(A)\subset \mathbb{C}$. The boundaries of $g_m(A)$ are vertical lines in $\mathbb{C}$, so $g_M(A)$ is a vertical strip, or the half-plane $\{\Re g_m<0\}$, or the half-plane $\{\Re g_m>0\}$.

Let us draw $g_m(A)$ in these three case, together with a conventional color and geographic terminology~\cite{Bertola:2002}, and conventional signs for each boundary ($+$ for left boundaries, $-$ for right boundaries) and half-plane:
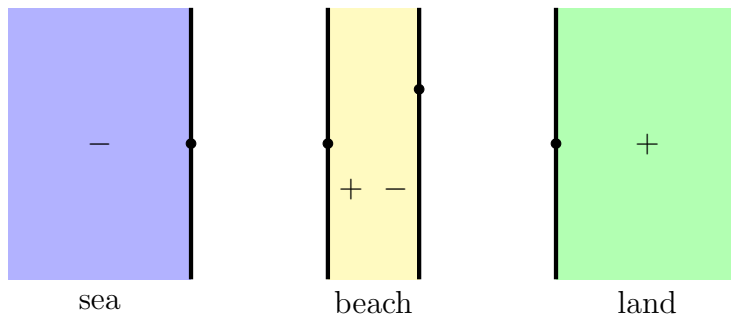
\begin{align}
\begin{tikzpicture}[scale = .6,baseline=(current bounding box.center)]
 \fill[blue!30] (-4, -3) rectangle (0, 3); 
 \draw[ultra thick] (0, -3) -- (0, 3);
 \filldraw (0, 0) circle [radius = 3pt]; 
 \node at (-2, 0) {$-$};
 \fill[yellow!30] (3, -3) rectangle (5, 3);
 \draw[ultra thick] (3, -3) -- (3, 3);
 \draw[ultra thick] (5, -3) -- (5, 3);
 \filldraw (3, 0) circle [radius = 3pt];
 \filldraw (5, 1.2) circle [radius = 3pt];
 %\node at (4, 0) {$0$};
 \node[right] at (3, -1) {$+$};
 \node[left] at (5, -1) {$-$};
 \fill[green!30] (8, -3) rectangle (12, 3);
 \draw[ultra thick] (8, -3) -- (8, 3);
 \filldraw (8, 0) circle [radius = 3pt];
 \node at (10, 0) {$+$};
 \node at (10, -3.5) {land};
 \node at (4, -3.5) {beach};
 \node at (-2, -3.5) {sea};
\end{tikzpicture}
\end{align}
We denoted vertices as dots, including $g_m(m)=0$. 
For $m'$ a vertex on the strip boundary that does not contain $m$, $|\Re g_m(m')|=|\Re g_{m'}(m)|\neq 0$ is the \textbf{strip width}\index{strip width}, while  $\Im g_m(m')$ is called the \textbf{strip shearing}\index{strip shearing}.

\item 
When mapped back to sectors, our signs for strip boundaries and half-planes indicate whether $dg_m$ is analytic across edges: $dg_m$ is analytic across edges that separate two different signs, and discontinuous across edges that separate two identical signs. According to Eq.~\eqref{eq:omegajump}, the latter edges therefore belong to $\operatorname{supp}\bar{\rho}$, and are called bridges. 
Since $g_m$ is locally an analytic function, Cauchy--Riemann relations relate the imaginary part of its derivative parallel to an oriented arc to the real part of its normal derivative to the arc, $ 2\pi \bar\rho = \Im d_\parallel g_m = -\Re d_\perp g_m $. Therefore, the equilibrium density is positive in the $(-, -)$ case, and negative in the $(+, +)$ case:
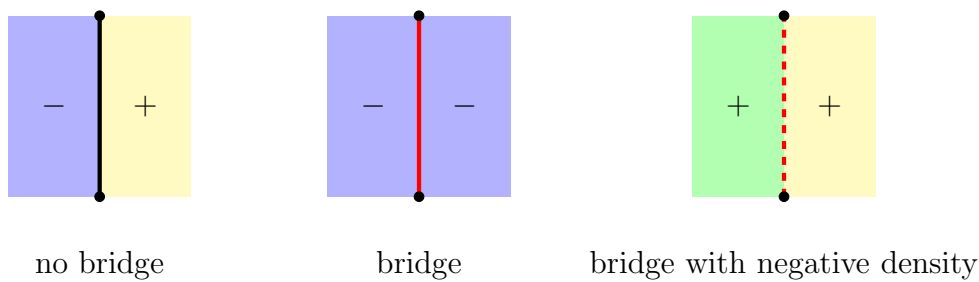
\begin{align}
\begin{tikzpicture}[scale = .6,baseline=(current bounding box.center)]
\begin{scope}[shift={(-7,0)}]
 \fill[blue!30] (-2, -2) rectangle (0, 2); 
 \fill[yellow!30] (0, -2) rectangle (2, 2); 
 \draw[ultra thick] (0, -2) -- (0, 2);
 \filldraw (0, -2) circle [radius = 3pt]; 
 \filldraw (0, 2) circle [radius = 3pt]; 
 \node at (-1, 0) {$-$};
 \node at (1, 0) {$+$};
 \node at (0, -3.5) {no bridge};
\end{scope} 
\begin{scope}
 \fill[blue!30] (-2, -2) rectangle (0, 2); 
 \fill[blue!30] (0, -2) rectangle (2, 2); 
 \draw[ultra thick,red] (0, -2) -- (0, 2);
 \filldraw (0, -2) circle [radius = 3pt]; 
 \filldraw (0, 2) circle [radius = 3pt]; 
 \node at (-1, 0) {$-$};
 \node at (1, 0) {$-$};
 \node at (0, -3.5) {bridge};
\end{scope} 
\begin{scope}[shift={(8,0)}]
 \fill[green!30] (-2, -2) rectangle (0, 2); 
 \fill[yellow!30] (0, -2) rectangle (2, 2); 
 \draw[ultra thick,red, dashed] (0, -2) -- (0, 2);
 \filldraw (0, -2) circle [radius = 3pt]; 
 \filldraw (0, 2) circle [radius = 3pt]; 
 \node at (-1, 0) {$+$};
 \node at (1, 0) {$+$};
 \node at (0, -3.5) {bridge with negative density};
 %\node at (0, -4.5) {not allowed};
\end{scope} 
\end{tikzpicture}
\end{align}

\end{itemize}
Let us conclude our discussion of the spectral network $\mathcal S$ with some combinatorial properties, in the case where $V'$ is a polynomial of degree $d$. If $c$ is the number of connected components of $\mathcal S$, then 
\begin{align}
 \#\text{lands} = \#\text{seas} = d+1 \qquad , \qquad \#\text{beaches} \leq c-1\ .
\end{align}
Let $e_c$ and $e_n$ be the numbers of compact and non-compact edges. Then $2e_c+e_n$ is the sum of vertex valencies, $2e_c+e_n=\sum_m(k_m+2) = 2d+2v$ where $v$ is the number of vertices. Moreover, each beach carries $4$ non-compact edges, and each land or sea carries $2$. This implies 
\begin{equation}
 \begin{split}
 \#\text{compact edges} &= v-1- \#\text{beaches}  \ , \\
 \#\text{non-compact edges} & = 2d+2+2 \#\text{beaches} \ .
 \end{split}
\end{equation}

\subsubsection{The support of the equilibrium density}

Since the equilibrium density $\bar \rho$ is positive, there exists a choice of signs of $g_m$ in each sector such that there is no $(+,+) $ edge.
Moreover, since $\bar \rho$ is normalizable, bridges are compact and never reach $\infty$ neither any pole of $V'$.
This restricts the set of possible $\overline P$.

Then $\operatorname{supp}\bar\rho$ is the union of all bridges. It is a disjoint union of compact, connected trees, which generically have trivalent or univalent vertices. A spectral edge $a_i$ always belongs to $\operatorname{supp}\bar{\rho}$, because $g_{a_i}(x)$ \eqref{gmx} is not continuous near $a_i$.

Now the one-form $dg_m(x) =M(x)\sqrt{\sigma(x)}dx$ is analytic in the connected domain $\mathbb C \backslash \operatorname{supp}\bar{\rho}$.
The function $g_m(x)$ cannot be globally defined, because it has non-trivial monodromies around components of $\operatorname{supp}\bar\rho$. 
These monodromies are however imaginary: for any loop in $\mathbb C \backslash \operatorname{supp}\bar{\rho}$,
\begin{align}
 \oint_{\text{loop}} dg_m = 2\pi i \int_{\text{inside the loop}} \bar \rho(x)dx\ \in i\mathbb R \ .
\end{align}
We notice that $d\Re g_m = d V_{\text{eff}} $, therefore
 $\Re g_m$ is a global harmonic function on $\mathbb C \backslash \operatorname{supp}\bar\rho$, which is related to the effective potential by a constant shift,
\begin{align}
 \Re g_m (x) = V_{\text{eff}}(x) + c_m\ .
\end{align}
The real constant $c_m$ can be computed by comparing the behaviour of $\Re g_m(x)$ near $x=\infty$, with the behaviour of $V_\text{eff}(x)$ that can be deduced from Eq.~\eqref{def:Veff},
\begin{align}
 V_{\text{eff}}(x) \underset{x\to\infty}{=}  \Re V(x) - 2 \log |x| + O\left(\frac{1}{x}\right)\ .
\end{align}

\subsubsection{Example of a quartic potential}

In the case $\gamma=\mathbb R$ and $V(x)=\frac{x^4}{4}-t\frac{x^2}{2}$ with $t\in \mathbb R_+>0$,  we find 
$\overline P(x) = x^2$ for $t>2$, and $\overline P(x) = x^2+\frac{1}{54}\left(t^3-36 t+(t^2+12)^{3/2}\right) $ for $t\leq 2$. (See \autoref{sec:cmu} for detailed computations.)
The edges of $\mathcal S $ go to infinity in the eight asymptotic directions $\arg x\in\frac{\pi}{8}+ \frac{\pi}{4}\mathbb{Z}$. 
Depending on the parameter $t$, we can have two or four spectral edges. For $t\leq 2$ we have two spectral edges, $M^2\sigma$ has two imaginary double zeros, and $\operatorname{supp}\bar{\rho}$ is one real segment (a bridge) between two beaches:

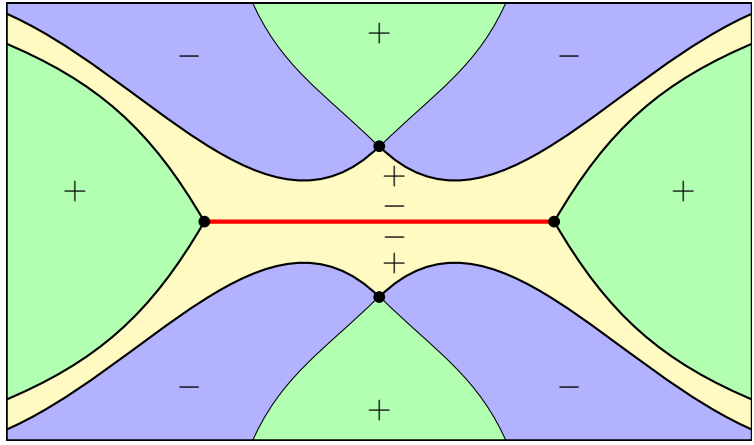
\begin{align}
 \begin{tikzpicture}[baseline=(current bounding box.center)]
\clip%[draw]
(-4.9, -2.9) -- (-4.9, 2.9) -- (4.9, 2.9) -- (4.9, -2.9) -- cycle;
\fill[yellow!30]
(-5,-3) -- (-5, 3) -- (5, 3) -- (5, -3) -- cycle;
\draw[ultra thick, red] (-2.3, 0) -- (2.3, 0); 
\foreach \xscale in {-1, 1}{
  \begin{scope}[xscale=\xscale]  
  \filldraw [fill=green!30, thick]
  (5,2.4) to [out=180+22.5, in=60] (2.3,0) to [out=-60,in=180-22.5]
  (5,-2.4) -- cycle;
  \node at (4,0.4) {$+$};
  \filldraw (2.3,0) circle [radius=2pt];
  \end{scope}
  }
  \foreach \yscale in {-1, 1}{
  \foreach \xscale in {-1, 1}{
  \begin{scope}[xscale=\xscale, yscale=\yscale]
   \filldraw [fill=blue!30, thick]
  (-5, 3) to (-5, 2.8) to [out=-22.5, in=180+45] (0,1) to [out=180-45,in=-67.5] (-1.7,3) -- cycle;
  \node at (-2.5, 2.2) {$-$};
  \end{scope}
  }}
  \foreach \yscale in {-1, 1}{
  \begin{scope}[yscale = \yscale]
   \fill [green!30]
   (-1.7, 3) to [out=-67.5, in=180-45] (0, 1) to [out=45, in=180+67.5] (1.7, 3) -- cycle;
   \node at (0, 2.5) {$+$};
   \filldraw (0, 1) circle [radius=2pt];
  \end{scope}
  }
%
%   \node at (-0.2, 0.3) {$0$};
%   \node at (-0.2, -0.3) {$0$};
   \node at (0.2, 0.6) {$+$};
   \node at (0.2, -0.55) {$+$};
   \node at (0.2, 0.2) {$-$};
   \node at (0.2, -0.2) {$-$};
   \draw [thick] (-4.9,-2.9) rectangle (4.9,2.9);
%
%  \draw [very thick,brown] (-2.3,0) -- (-5,0);
 % \draw [very thick,brown] (2.3,0) -- (5,0);
%
 \end{tikzpicture} 
\end{align}
If $t> 2$ we have four spectral edges, $M^2\sigma$ has one double zero $m=0$, and $\operatorname{supp}\bar{\rho}$ is made of two real segments, each of which is a bridge between two seas:
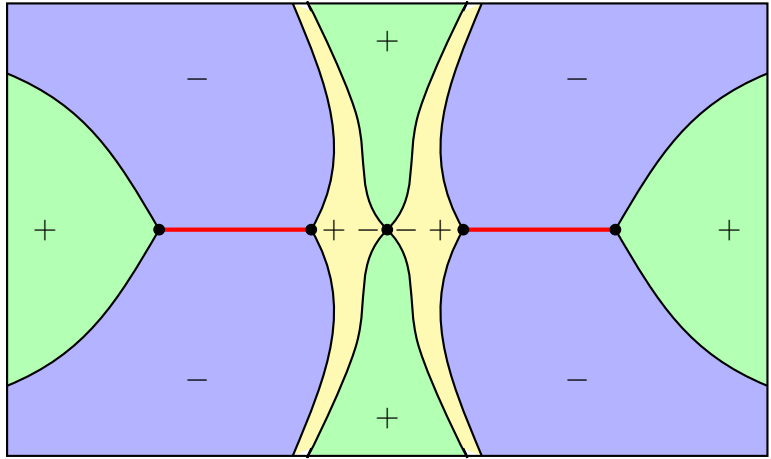
\begin{align}
 \begin{tikzpicture}[baseline=(current bounding box.center)]
%
  % domains
%
  \filldraw [fill=green,draw=none,opacity=.3]
  (5,2.071) to [out=180+22.5,in=60] (3,0) to [out=-60,in=180-22.5]
  (5,-2.071) -- cycle;
  \filldraw [fill=green,draw=none,opacity=.3]
  (-5,2.071) to [out=-22.5,in=180-60] (-3,0) to [out=180+60,in=22.5]
  (-5,-2.071) -- cycle;
  \filldraw [fill=blue,draw=none,opacity=.3]
  (5,3) -- (1.243,3) to [out=180+67.5,in=120] (1,0) to
  [out=240,in=180-67.5] (1.243,-3) -- (5,-3) -- (5,-2.071) to
  [out=180-22.5,in=-60] (3,0) to [out=60,in=180+22.5] (5,2.071) --
  cycle;
  \filldraw [fill=blue,draw=none,opacity=.3]
  (-5,3) -- (-1.243,3) to [out=-67.5,in=60] (-1,0) to
  [out=-60,in=67.5] (-1.243,-3) -- (-5,-3) -- (-5,-2.071) to
  [out=22.5,in=180+60] (-3,0) to [out=180-60,in=-22.5] (-5,2.071) -- cycle;
%
%  \filldraw [fill=green,draw=none,opacity=.3]
%  (-1.243,3) to [out=-67.5,in=60] (-1,0) to [out=-60,in=67.5] (-1.243,-3)
%  -- (1.243,-3) to [out=180-67.5,in=-120] (1,0) to [out=120,in=180+67.5]
%  (1.243,3) -- cycle;
%
  \filldraw [fill=green,draw=none,opacity=.3]
(0,0) to [out=45., in = 180+75] (0.4,1.5) to [out=75.,in=65] (1,2.9) to (1,3) to (-1,3) 
to (-1,2.9) to [out=-65, in=180-75] (-0.4,1.5) to [out= -75,in=180-45] (0,0) --cycle;
  \filldraw [fill=green,draw=none,opacity=.3]
(0,0) to [out=-45., in = 180-75] (0.4,-1.5) to [out=-75.,in=-65] (1,-2.9) to (1,-3) to (-1,-3) to (-1,-2.9) to [out=65, in=180+75] (-0.4,-1.5) to [out= 75,in=180+45] (0,0) --cycle;
  \filldraw [fill=yellow,draw=none,opacity=.3]
(0,0) to [out=45., in = 180+75] (0.4,1.5) to [out=75.,in=65] (1,2.9) 
to (1.243,3) to [out=180+67.5,in=120] (1,0) to   [out=240,in=180-67.5] (1.243,-3)   to (1,-2.9) to [out=180-65, in=-75] (0.4,-1.5) to [out=180-75 ,in=-45] (0,0)
--cycle;
  \filldraw [fill=yellow,draw=none,opacity=.3]
(0,0) to [out=180-45., in = -75] (-0.4,1.5) to [out=180-75.,in=180-65] (-1,2.9) 
to (-1.243,3) to [out=-67.5,in=180-120] (-1,0) to   [out=180-240,in=67.5] (-1.243,-3)   to (-1,-2.9) to [out=65, in=180+75] (-0.4,-1.5) to [out=75 ,in=180+45] (0,0)
--cycle;
%
  % frame
%
  \draw [thick] (-5,-3) rectangle (5,3);
%
  % borders
%
  \draw [thick] (5,2.071) to [out=180+22.5,in=60] (3,0);
  \draw [thick] (5,-2.071) to [out=180-22.5,in=-60] (3,0);
  \draw [thick] (-5,2.071) to [out=-22.5,in=180-60] (-3,0);
  \draw [thick] (-5,-2.071) to [out=22.5,in=180+60] (-3,0);
  \draw [thick] (1.243,3) to [out=180+67.5,in=120] (1,0);
  \draw [thick] (1.243,-3) to [out=180-67.5,in=-120] (1,0);
  \draw [thick] (-1.243,3) to [out=-67.5,in=180-120] (-1,0);
  \draw [thick] (-1.243,-3) to [out=67.5,in=180+120] (-1,0);
  \draw [thick] (0,0) to [out=45., in = 180+75] (0.4,1.5) to [out=75.,in=65] (1,2.9) ;
  \draw [thick] (0,0) to [out=180-45., in = -75] (-0.4,1.5) to [out=180-75.,in=180-65] (-1,2.9) ;
  \draw [thick] (0,0) to [out=-45., in = 180-75] (0.4,-1.5) to [out=-75.,in=-65] (1,-2.9) ;
  \draw [thick] (0,0) to [out=180+45., in = 75] (-0.4,-1.5) to [out=180+75.,in=180+65] (-1,-2.9) ;
  \draw [ultra thick,red] (1,0) -- (3,0);
  \draw [ultra thick,red] (-1,0) -- (-3,0);
%
  % end points
%
  \filldraw (0,0) circle [radius=2pt] ;
  \filldraw (1,0) circle [radius=2pt] ;
  \filldraw (3,0) circle [radius=2pt] ;
  \filldraw (-1,0) circle [radius=2pt] ;
  \filldraw (-3,0) circle [radius=2pt] ;
%
  % signs
%
  \node at (4.5,0) {$+$};
  \node at (-4.5,0) {$+$};
  \node at (0,2.5) {$+$};
  \node at (0,-2.5) {$+$};
  \node at (2.5,2) {$-$};
  \node at (2.5,-2) {$-$};
  \node at (-2.5,2) {$-$};
  \node at (-2.5,-2) {$-$};
%
%  \node at (0.55,1.3) {$0$};
%  \node at (-0.55,1.3) {$0$};
  \node at (0.7,0) {$+$};
  \node at (0.25,0) {$-$};
  \node at (-0.7,0) {$+$};
  \node at (-0.25,0) {$-$};
 \end{tikzpicture}
\end{align}

\subsubsection{Example with a branched support}

Consider a normal matrix model with the cubic potential $V(x)=\frac{4}{3}x^3$, and with eigenvalues on a path $\gamma$.
For the matrix integral to converge, $\gamma$ has to belong to the homology space associated to our potential, generated by three paths that go to infinity in the three directions $\arg x\in \frac{2\pi}{3} \mathbb{Z}$:
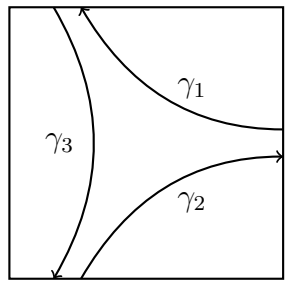
\begin{align}
 \begin{tikzpicture}[scale = .6,baseline=(current bounding box.center)]
%
% paths
%
 \draw [->,thick] (3,0.3) to [out=180,in=-60] (-1.73+0.3,3) ;
 \draw [latex-,thick] (3,-0.3) to [out=180,in=60] (-1.73+0.3,-3) ;
 \draw [->,thick] (-1.73-0.3,3) to [out=-60,in=60] (-1.73-0.3,-3) ;
\node at (1,1.2) {$\gamma_1$};
\node at (1,-1.3) {$\gamma_2$};
\node at (-1.9,0) {$\gamma_3$};
%
  % frame
%
  \draw [thick] (-3,-3) rectangle (3,3);
 \end{tikzpicture}
\end{align}
These three paths are not independent $\gamma_1+\gamma_2+\gamma_3=0$.
Let our integration path be the linear combination 
\begin{align}
 \gamma = \sum_{k=1}^3 e^{\frac{2\pi i}{3}k} \gamma_k\ ,
\end{align}
which is projectively invariant under rotations $x\mapsto  e^{\frac{2\pi i}{3}}x$. 
Then we expect the spectral network to be invariant as well. 
This requires that $(V')^2 - 4\overline{P}$ be projectively invariant. 
Since $V'(x) = 4x^2$ is homogeneous, this implies that the polynomial $\overline{P}(x)$ of degree 1 must also be homogeneous, therefore $\overline{P}(x)=4x$, and
\begin{align}
 V'(x)^2 - 4\overline{P}(x) = 16x(x^3-1)\quad \implies \quad \{a_i\} = \left\{0, 1, e^{\frac{2\pi i}{3}}, e^{\frac{4\pi i}{3}}\right\}\, .
\end{align}
The spectral network $\mathcal{S}$ turns out to be connected, and therefore determined by only one function
\begin{align}
 g_0(x) = 4\int_0^x dx' \sqrt{x'(x'^3-1)} = \frac13\left(\frac{1}{\zeta}-\zeta+2\log \zeta\right) \, , \ \ \text{with} \ \ 2x^\frac32 = \sqrt{\zeta} + \frac{1}{\sqrt{\zeta}}\, .
\end{align}
Then $\mathcal{S} = \{\Re g_0(x)=0\} = \{|\zeta| = 1\} \cup 
\left\{\cos{ \left(\operatorname{arg} \zeta\right)}=\frac{\log{|\zeta|}}{|\zeta|-|\zeta|^{-1}}\right\}$. 
In the complex $x$-plane, the circle $\{|\zeta| = 1\}$ corresponds to the three segments $[0,1]$, $[0,e^{\frac{2\pi i}{3}}]$ and
$[0,e^{\frac{4i\pi}{3}}]$, whose union $\operatorname{supp}\bar{\rho}$ is made of 3 bridges separating 3 seas:
\begin{ceqn}
\begin{equation}
 \begin{tikzpicture}[baseline=(current  bounding  box.center)]

  % domains

  \filldraw [fill=green,draw=none,opacity=.3]
  (3,1.6) to [out=180+22.5,in=60] (1,0) to [out=-60,in=180-22.5]
  (3,-1.6) -- cycle;

  \filldraw [fill=green,draw=none,opacity=.3]
  (-3,1.8) to [out=-30,in=180] (-0.5,0.86) to [out=60,in=270] 
  (-0.17,3) -- (-3,3) -- cycle;

  \filldraw [fill=green,draw=none,opacity=.3]
  (-3,-1.8) to [out=30,in=180] (-0.5,-0.86) to [out=-60,in=90] 
  (-0.17,-3) -- (-3,-3) -- cycle;

  \filldraw [fill=blue,draw=none,opacity=.3]
  (3,1.6) to [out=180+22.5,in=60] (1,0) -- (0,0) -- (-0.5,0.86) 
  to [out=60,in=270]   (-0.17,3) -- (3,3) -- cycle;

 \filldraw [fill=blue,draw=none,opacity=.3]
  (3,-1.6) to [out=180-22.5,in=-60] (1,0) -- (0,0) -- (-0.5,-0.86) 
  to [out=-60,in=90]   (-0.17,-3) -- (3,-3) -- cycle;

  \filldraw [fill=blue,draw=none,opacity=.3]
  (-3,1.8) to [out=-30,in=180] (-0.5,0.86)  -- (0,0) -- (-0.5,-0.86)    to [out=180,in=30]   (-3,-1.8)  -- cycle;

% borders

 \draw [thick]
  (3,1.6) to [out=180+22.5,in=60] (1,0) to [out=-60,in=180-22.5]
  (3,-1.6) ;

  \draw [thick]  (-3,1.8) to [out=-30,in=180] (-0.5,0.86) to [out=60,in=270] 
  (-0.17,3) ;

  \draw [thick]
  (-3,-1.8) to [out=30,in=180] (-0.5,-0.86) to [out=-60,in=90] 
  (-0.17,-3) -- (-3,-3) -- cycle;

  % cuts

  \draw [very thick,red] (0,0) -- (1,0);
  \draw [very thick,red] (0,0) -- (-0.5,0.86);
  \draw [very thick,red] (0,0) -- (-0.5,-0.86);

  % end points

  \filldraw (1,0) circle [radius=2pt];
  \filldraw (0,0) circle [radius=2pt];
  \filldraw (-0.5,0.86) circle [radius=2pt];
  \filldraw (-0.5,-0.86) circle [radius=2pt];

  % signs

  \node at (2.5,0) {$+$};
  \node at (-1.3,2) {$+$};
  \node at (-1.3,-2) {$+$};

  \node at (1.3,2) {$-$};
  \node at (1.3,-2) {$-$};
  \node at (-2.5,0) {$-$};

  % frame

  \draw [thick] (-3,-3) rectangle (3,3);

 \end{tikzpicture}
\label{fig:ebs}
 \end{equation}
 \end{ceqn}

\subsection{Determining the polynomial \texorpdfstring{$\overline{P}(x)$}{P(x)}} \label{sec:detpol}

In our determination of $\bar\rho$ and $\operatorname{supp}\bar\rho$, we have been requiring that the effective potential is constant on $\operatorname{supp}\bar\rho$. The Euler--Lagrange equation \eqref{eq:saddlerho} further requires that the effective potential be lower on $\operatorname{supp}\bar\rho$ than elsewhere on the Jordan arc that includes $\operatorname{supp}\bar\rho$. Let us now fulfill this further requirement.

\subsubsection{Descent-ascent roads}

Given a connected arc $I\subset \operatorname{supp}\bar\rho$,
let us find Jordan arcs $\gamma\in H_1(e^{-NV(x)}dx)$ that contain $I$ such that $\left. V_\mathrm{eff} \right|_{\gamma\backslash I} > \left. V_\mathrm{eff} \right|_{I}$. 
It is enough to build a representative of each homotopy class of such arcs, and we will build particular arcs called descent-ascent roads.

Let an ascending road be a path $\gamma$ originating at a vertex $m$ of the spectral network $\mathcal S$, 
such that $\gamma\subset \{\Im g_m=0\}$, and $V_\mathrm{eff}= \Re g_m- c_m$ strictly increases when moving away from $m$ on $\gamma$.
The image of an ascending road by $g_m$ is a horizontal segment or half-line originating at $g_m(m)=0$, and oriented to the right:
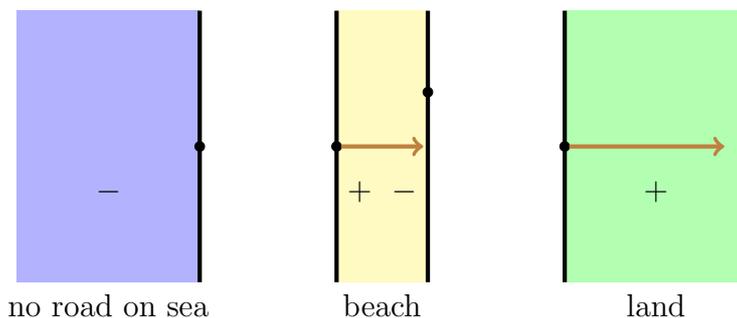
\begin{align}
\begin{tikzpicture}[scale = .6,baseline=(current bounding box.center)]
 \fill[blue!30] (-4, -3) rectangle (0, 3); 
 \draw[ultra thick] (0, -3) -- (0, 3);
 \filldraw (0, 0) circle [radius = 3pt]; 
 \node at (-2, -1) {$-$};
 \fill[yellow!30] (3, -3) rectangle (5, 3);
 \draw[ultra thick] (3, -3) -- (3, 3);
 \draw[ultra thick] (5, -3) -- (5, 3);
 \filldraw (3, 0) circle [radius = 3pt];
 \filldraw (5, 1.2) circle [radius = 3pt];
 %\node at (4, 0) {$0$};
 \node[right] at (3, -1) {$+$};
 \node[left] at (5, -1) {$-$};
  \draw [ultra thick,brown, -latex] (3.1,0) -- (4.9,0);
 \fill[green!30] (8, -3) rectangle (12, 3);
 \draw[ultra thick] (8, -3) -- (8, 3);
 \filldraw (8, 0) circle [radius = 3pt];
 \node at (10, -1) {$+$};
  \draw [ultra thick,brown, -latex] (8.1,0) -- (11.5,0);
 \node at (10, -3.5) {land};
 \node at (4, -3.5) {beach};
 \node at (-2, -3.5) {no road on sea};
\end{tikzpicture}
\end{align}
A land ascending road goes to $\infty$ in a sector where $\Re V \to +\infty$, and never crosses a bridge or another ascending road.
A beach ascending road must reach either a $(-,+)$ edge of the spectral network, beyond which it then continues, or a bridge, where it stops. 
We then define a full road by allowing the ascending road to follow the bridge to the left, until it reaches another vertex, from which it can then continue to follow an ascending road, until it eventually reaches $\infty$ in a land sector:
\begin{align}
\begin{tikzpicture}[scale = .6,baseline=(current bounding box.center)]
 \fill[yellow!30] (3, -3) rectangle (5, 3);
  \filldraw [fill=green,draw=none,opacity=.3]
  (5,1.2) to  (5,3) to (8,3) to (8,-1.5) to (5,1.2) -- cycle;
  \filldraw [fill=blue,draw=none,opacity=.3]
  (5,1.2) to  (8,-1.5) to (8,-3) to (5,-3) to (5,1.2) -- cycle;
 \draw[ultra thick] (3, -3) -- (3, 3);
 \draw[ultra thick] (5, 1.2) -- (8, -1.5);
 \draw[ultra thick] (5, 1.2) -- (5, 3);
 \draw[ultra thick, red] (5, -3) -- (5, 1.2);
  \draw [ultra thick,brown, -latex] (3,0) -- (4.8,0);
  \draw [ultra thick,brown, -latex] (4.8,0) -- (4.8,1.2);
  \draw [ultra thick,brown, -latex] (4.9,1.2) to [out=0, in =225] (7.9,2.6);
 %
 %\node at (4, 0) {$0$};
 \node[right] at (3, -1) {$+$};
 \node[left] at (5, -1) {$-$};
 \node[right] at (6.5, -2) {$-$};
 \node[right] at (6, 2.2) {$+$};
 \filldraw (3, 0) circle [radius = 3pt];
 \filldraw (5, 1.2) circle [radius = 3pt];
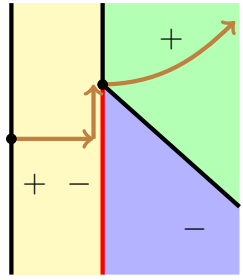
 \node at (5.5, -3.5) {road on a bridge};
\end{tikzpicture}
\end{align}
More generally,
let a full road be a path from a vertex $m$ to $\infty$, on which $V_\mathrm{eff}$ never decreases, and that is made of bridges and ascending roads.

If $k_m$ is the order of $m$ as a zero of $M^2\sigma$, there are $k_m+2$ paths starting at $m$ on which $\Im g_m=0$. But there can be roads only in $+$ sectors, and two such sectors cannot be neighbours. 
So the number of ascending roads starting at $m$ is at most $\frac12(k_m+2)$, and can be zero, as in the case of the central vertex in Figure~\eqref{fig:ebs}. The number of full roads starting at $m$ is $\frac12(k_m+2)+\frac12 \#\text{bridges}\geq 2$.

We then define a  \textbf{descent-ascent road}\index{descent-ascent road} as the union of two different full roads that start at the same vertex. 
The profile of the effective potential along a descent-ascent road looks as follows:
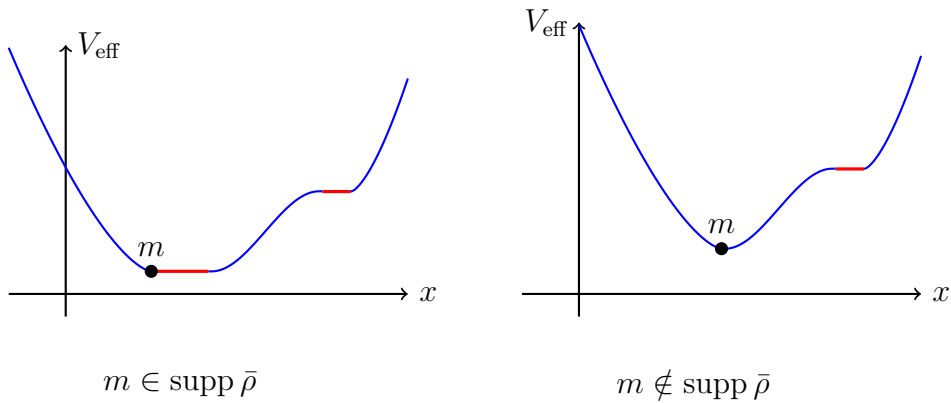
\begin{align}
\begin{tikzpicture}[scale = .75,baseline=(current bounding box.center)]
\begin{scope}
 \draw[thick,-latex] (-3,- 0.4) -- (4, -0.4) node [right] {$x$};
 \draw[thick,-latex] (-2,- 0.8) -- (-2, 4) node [right] {$V_{\text{eff}}$};
 \draw[red,very thick] (-0.5, 0) -- (0.5, 0);
\draw[domain=0.5:2.5,smooth,variable=\x,blue, thick] plot({\x},{
- 0.5* (sqrt(\x-0.5)^3 + 1.5*(2.5-\x)*sqrt(2) + 1.5*(\x-0.5)*(2.5-\x)^2/sqrt(8) - 1.5*2^(1.5)) 
+ 0.5* (sqrt(2.5-\x)^3 + 1.5*(\x-0.5)*sqrt(2) + 1.5*(\x-0.5)^2*(2.5-\x)/sqrt(8) -2^(1.5) ) 
 });
 \draw[red,very thick] (2.5, 1.41) -- (3, 1.41);
\draw[domain=-3:-0.5,smooth,variable=\x,blue, thick] plot({\x},{
 (sqrt(-0.5-\x)^3  )  });
\draw[domain=3:4,smooth,variable=\x,blue, thick] plot({\x},{
 1.41+2* (sqrt(\x-3)^3  )  });
 \filldraw (-0.5, 0) circle [radius = 3pt];
 \node at (-0.5,.4) {$m$};
 \node at (0, -2) {$m\in \operatorname{supp}\bar\rho $};
\end{scope}
\begin{scope}[shift={(9,0.4)}]
 \draw[thick,-latex] (-3,- 0.8) -- (4, -0.8) node [right] {$x$};
 \draw[thick,-latex] (-2,- 1.2) -- (-2, 4) node [left] {$V_{\text{eff}}$};
\draw[domain=0.5:2.5,smooth,variable=\x,blue, thick] plot({\x},{
- 0.5* (sqrt(\x-0.5)^3 + 1.5*(2.5-\x)*sqrt(2) + 1.5*(\x-0.5)*(2.5-\x)^2/sqrt(8) - 1.5*2^(1.5)) 
+ 0.5* (sqrt(2.5-\x)^3 + 1.5*(\x-0.5)*sqrt(2) + 1.5*(\x-0.5)^2*(2.5-\x)/sqrt(8) -2^(1.5) ) 
 });
 \draw[red,very thick] (2.5, 1.41) -- (3, 1.41);
\draw[domain=-2:0.5,smooth,variable=\x,blue, thick] plot({\x},{
 (sqrt(0.5-\x)^3  )  });
\draw[domain=3:4,smooth,variable=\x,blue, thick] plot({\x},{
 1.41+2* (sqrt(\x-3)^3  )  });
 \filldraw (0.5, 0) circle [radius = 3pt];
 \node at (0.5, .4) {$m$};
 \node at (0, -2.4) {$m\notin \operatorname{supp}\bar\rho $};
\end{scope}
\end{tikzpicture}
\end{align}
For a descent-ascent road $\gamma$ originating at a vertex $m$, we define 
\begin{itemize}

\item a level $\ell(\gamma) = V_\textrm{eff}(m)$,

\item a filling fraction $\bar{\epsilon}(\gamma) = \int_{\gamma \cap \{V_\textrm{eff}=\ell(\gamma)\}} \bar \rho (x)dx$. 
\end{itemize}
A descent-ascent road is called empty if its filling fraction is zero.
A road $\gamma$ that starts at a vertex $m$ is empty if and only if there is a neighbourhood $\Omega$ of $m$ such that $\Omega \cap \gamma \cap \operatorname{supp}\bar\rho  \in \{\emptyset, \{m\}\}$.

% NB: Is this topological characterization of empty roads correct?

\subsubsection{Example of the  quartic potential $V(x)=\frac{x^4}{4}-t\frac{x^2}{2} $}

Let us add the ascending roads to our diagrams of the spectral network from \autoref{sec:supprho}:
\begin{align}
 \begin{tikzpicture}[scale = .6,baseline=(current bounding box.center)]
%
  % domains
%
  \filldraw [fill=green,draw=none,opacity=.3]
  (5,2.071) to [out=180+22.5,in=60] (3,0) to [out=-60,in=180-22.5]
  (5,-2.071) -- cycle;
  \filldraw [fill=green,draw=none,opacity=.3]
  (-5,2.071) to [out=-22.5,in=180-60] (-3,0) to [out=180+60,in=22.5]
  (-5,-2.071) -- cycle;
  \filldraw [fill=blue,draw=none,opacity=.3]
  (5,3) -- (1.243,3) to [out=180+67.5,in=120] (1,0) to
  [out=240,in=180-67.5] (1.243,-3) -- (5,-3) -- (5,-2.071) to
  [out=180-22.5,in=-60] (3,0) to [out=60,in=180+22.5] (5,2.071) --
  cycle;
  \filldraw [fill=blue,draw=none,opacity=.3]
  (-5,3) -- (-1.243,3) to [out=-67.5,in=60] (-1,0) to
  [out=-60,in=67.5] (-1.243,-3) -- (-5,-3) -- (-5,-2.071) to
  [out=22.5,in=180+60] (-3,0) to [out=180-60,in=-22.5] (-5,2.071) -- cycle;
  \filldraw [fill=green,draw=none,opacity=.3]
(0,0) to [out=45., in = 180+75] (0.4,1.5) to [out=75.,in=65] (1,2.9) to (1,3) to (-1,3) 
to (-1,2.9) to [out=-65, in=180-75] (-0.4,1.5) to [out= -75,in=180-45] (0,0) --cycle;
  \filldraw [fill=green,draw=none,opacity=.3]
(0,0) to [out=-45., in = 180-75] (0.4,-1.5) to [out=-75.,in=-65] (1,-2.9) to (1,-3) to (-1,-3) to (-1,-2.9) to [out=65, in=180+75] (-0.4,-1.5) to [out= 75,in=180+45] (0,0) --cycle;
  \filldraw [fill=yellow,draw=none,opacity=.3]
(0,0) to [out=45., in = 180+75] (0.4,1.5) to [out=75.,in=65] (1,2.9) 
to (1.243,3) to [out=180+67.5,in=120] (1,0) to   [out=240,in=180-67.5] (1.243,-3)   to (1,-2.9) to [out=180-65, in=-75] (0.4,-1.5) to [out=180-75 ,in=-45] (0,0)
--cycle;
  \filldraw [fill=yellow,draw=none,opacity=.3]
(0,0) to [out=180-45., in = -75] (-0.4,1.5) to [out=180-75.,in=180-65] (-1,2.9) 
to (-1.243,3) to [out=-67.5,in=180-120] (-1,0) to   [out=180-240,in=67.5] (-1.243,-3)   to (-1,-2.9) to [out=65, in=180+75] (-0.4,-1.5) to [out=75 ,in=180+45] (0,0)
--cycle;
%
  % frame
%
  \draw [thick] (-5,-3) rectangle (5,3);
%
  % borders
%
  \draw [thick] (5,2.071) to [out=180+22.5,in=60] (3,0);
  \draw [thick] (5,-2.071) to [out=180-22.5,in=-60] (3,0);
  \draw [thick] (-5,2.071) to [out=-22.5,in=180-60] (-3,0);
  \draw [thick] (-5,-2.071) to [out=22.5,in=180+60] (-3,0);
  \draw [thick] (1.243,3) to [out=180+67.5,in=120] (1,0);
  \draw [thick] (1.243,-3) to [out=180-67.5,in=-120] (1,0);
  \draw [thick] (-1.243,3) to [out=-67.5,in=180-120] (-1,0);
  \draw [thick] (-1.243,-3) to [out=67.5,in=180+120] (-1,0);
  \draw [thick] (0,0) to [out=45., in = 180+75] (0.4,1.5) to [out=75.,in=65] (1,2.9) ;
  \draw [thick] (0,0) to [out=180-45., in = -75] (-0.4,1.5) to [out=180-75.,in=180-65] (-1,2.9) ;
  \draw [thick] (0,0) to [out=-45., in = 180-75] (0.4,-1.5) to [out=-75.,in=-65] (1,-2.9) ;
  \draw [thick] (0,0) to [out=180+45., in = 75] (-0.4,-1.5) to [out=180+75.,in=180+65] (-1,-2.9) ;
  \draw [ultra thick,red] (1,0) -- (3,0);
  \draw [ultra thick,red] (-1,0) -- (-3,0);
  \draw [ultra thick,brown,-latex] (-3.1,0) -- (-4.9,0);
  \draw [ultra thick,brown,-latex] (3.1,0) -- (4.9,0);
  \draw [ultra thick,brown,-latex] (0,0.1) -- (0,2.9);
  \draw [ultra thick,brown,-latex] (0,-0.1) -- (0,-2.9);
  \draw [ultra thick,brown,-latex] (-0.9,0) -- (-0.5,0);
  \draw [ultra thick,brown] (-0.5,0) -- (-0.1,0);
  \draw [ultra thick,brown,-latex] (0.9,0) -- (0.5,0);
  \draw [ultra thick,brown] (0.5,0) -- (0.1,0);
%
  % end points
%
  \filldraw (-3,0) circle [radius=3pt] ;
  \filldraw (-1,0) circle [radius=3pt] ;
  \filldraw (0,0) circle [radius=3pt] ;
  \filldraw (1,0) circle [radius=3pt] ;
  \filldraw (3,0) circle [radius=3pt] ;
%
  % signs
%
  \node at (4.5,0.3) {$+$};
  \node at (-4.5,0.3) {$+$};
  \node at (0.3,2.5) {$+$};
  \node at (0.3,-2.5) {$+$};
  \node at (2.5,2) {$-$};
  \node at (2.5,-2) {$-$};
  \node at (-2.5,2) {$-$};
  \node at (-2.5,-2) {$-$};
%
%  \node at (0.55,1.3) {$0$};
%  \node at (-0.55,1.3) {$0$};
%
  \node at (0, -4) {$t>2$};
 \end{tikzpicture}
 \hspace{1.cm}
 \begin{tikzpicture}[scale=.63,baseline=-8]
  \node at (0, -4) {$t\leq 2$};
\clip%[draw]
(-4.9, -2.9) -- (-4.9, 2.9) -- (4.9, 2.9) -- (4.9, -2.9) -- cycle;
\fill[yellow!30]
(-5,-3) -- (-5, 3) -- (5, 3) -- (5, -3) -- cycle;
\draw[ultra thick, red] (-2.3, 0) -- (2.3, 0); 
\foreach \xscale in {-1, 1}{
  \begin{scope}[xscale=\xscale]  
  \filldraw [fill=green!30, thick]
  (5,2.4) to [out=180+22.5, in=60] (2.3,0) to [out=-60,in=180-22.5]
  (5,-2.4) -- cycle;
  \node at (4,0.4) {$+$};
  \end{scope}
  }
  \foreach \yscale in {-1, 1}{
  \foreach \xscale in {-1, 1}{
  \begin{scope}[xscale=\xscale, yscale=\yscale]
   \filldraw [fill=blue!30, thick]
  (-5, 3) to (-5, 2.8) to [out=-22.5, in=180+45] (0,1) to [out=180-45,in=-67.5] (-1.7,3) -- cycle;
  \node at (-2.5, 2.2) {$-$};
  \end{scope}
  }}
  \foreach \yscale in {-1, 1}{
  \begin{scope}[yscale = \yscale]
   \fill [green!30]
   (-1.7, 3) to [out=-67.5, in=180-45] (0, 1) to [out=45, in=180+67.5] (1.7, 3) -- cycle;
  \end{scope}
  }
   \node at (0.4, 2.5) {$+$};
   \node at (0.4, -2.5) {$+$};
   \draw [thick] (-4.9,-2.9) rectangle (4.9,2.9);
  \draw [ultra thick,brown,-latex] (-2.4,0) -- (-4.85,0);
  \draw [ultra thick,brown,-latex] (2.4,0) -- (4.85,0);
  \draw [ultra thick,brown,-latex] (0,1.1) -- (0,2.9);
  \draw [ultra thick,brown,-latex] (0,-1.1) -- (0,-2.9);
  \draw [ultra thick,brown,-latex] (0,1.1) -- (0,0.1);
  \draw [ultra thick,brown,-latex] (0,-1.1) -- (0,-0.1);
   \filldraw (2.3,0) circle [radius=3pt];
    \filldraw (-2.3,0) circle [radius=3pt];
     \filldraw (0, 1) circle [radius=3pt];
      \filldraw (0, -1) circle [radius=3pt];
 \end{tikzpicture} 
\end{align}
In each case, we build a basis of descent-ascent roads using bridges and ascending roads:
\begin{align}
\renewcommand*{\arraystretch}{1.3}
 \begin{array}{ccccc}
  \text{Case} & \quad  &\text{Non-empty roads} & \quad & \text{Empty roads}
  \\
  \hline
  t \leq 2 & & \mathbb{R} & & \mathbb R_- + i\mathbb R_+, \mathbb R_- + i\mathbb R_- 
  \\
  t > 2 & & \mathbb R_- + i\mathbb R_+ , \mathbb R_+ - i\mathbb R_+ & & i\mathbb{R}
 \end{array}
\end{align}
In the case $t\leq 2$, there are two empty roads that nevertheless contain parts of $\operatorname{supp}\bar\rho $. On each one of these roads, the effective potential reaches its minimum at a vertex that does not belong to  $\operatorname{supp}\bar\rho $.

\subsubsection{Possible integration domains}

Since they reach $\infty$ only on land, descent-ascent roads belong to the homology space $H_1(e^{-NV(x)}dx)$. We admit that there is a basis of $H_1(e^{-NV(x)}dx)$ made of descent-ascent roads, and that the elements of that basis can be made disjoint by small deformations, namely pushing a road slightly off a bridge that already belongs to another road. Sketch of a proof:
\begin{itemize}
\item  Two different descent-ascent roads are never homotopic to each other. Indeed if they start from the same vertex but in different directions, it means that the 2 directions must be separated by at least 1 decreasing potential region, and it is possible to find an ascent-descent road (the opposite of a descent-ascent road) on which $V_{\text{eff}} \to -\infty$.
\item The descent-ascent roads generate $H_1(e^{-NV(x)}dx)$.
Indeed if we take all vertices and all possible directions, the consecutive descent-ascent roads must each go from a land to another land, and consecutive such roads can't cross. By a combinatorial argument, the number of lands that one can reach in this way must be at least equal to the total number of lands, i.e. all admissible paths of $H_1(e^{-NV(x)}dx)$ can be realized by these roads.
\end{itemize}
Let $(\hat \gamma_1,\dots,\hat \gamma_d)$ be a basis of descent-ascent roads, such that $\hat\gamma_i$ is empty if and only if $i>\hat s$. In generic situations, the number $\hat s$ of non-empty roads coincides with $s=\frac12 \deg \sigma$. Using this basis, we build an integration domain for our $N$-dimensional matrix integral, such that the large $N$ limit gives rise to the equilibrium density $\bar \rho$:
\begin{align}
 \Gamma = \prod_{i=1}^{\hat s} \hat \gamma_i^{\hat n_i} \quad \text{for} \quad \hat n_i = N\bar{\epsilon}(\hat \gamma_i) + o(N)\quad \text{with} \quad \sum_{i=1}^{\hat s} \hat n_i = N\ .
\end{align}
There are several choices of integers $\hat n_i$ that obey these constraints, and we can linearly combine domains $\Gamma$ of this type.

\subsubsection{Determining \texorpdfstring{$\overline{P}(x)$}{P(x)}} 

Given paths $\gamma_i$ and filling fractions $\epsilon_i$, let us now determine the polynomial $\overline P(x)$ that corresponds to the integration domain $\Gamma = \prod_i \gamma_i^{n_i} $ with $n_i=N\epsilon_i+o(N)$. 

Given the potential $V$, there are finitely many possibilities for the multiplicities for the zeros of $(V')^2-4\overline{P}=M^2 \sigma $, and finitely many possible topologies of the spectral network and ascending roads. We have to examine all these possible choices. 
Given a choice, we decompose the paths $\gamma_i$ in the basis of descent-ascent roads as
\begin{equation}
\gamma_i = \sum_{j=1}^{d} c_{i,j} \hat\gamma_j\ .
\end{equation}
The polynomial $\overline P$ is determined by the requirements that 
\begin{enumerate}
 \item $(V')^2-4\overline P$ has zeros of the given multiplicities,
 \item the descent-ascent roads that appear in the decomposition of a given path $\gamma_i$ all have the same level $\ell_i$,
 \item the filling fractions are respected, $\epsilon_i=\int_{\gamma_i}\bar \rho(x)dx$.
\end{enumerate}
For example, in the case where $\Gamma = \gamma^N$ and $\operatorname{supp}\bar\rho$ is made of $s$ intervals, we have
\begin{enumerate}
 \item $d-s$ equations from requiring that $(V')^2-4\overline{P}$ has $d-s$ double zeros,
 \item $s-1$ equations from requiring that the effective potential be the same on our $s$ intervals,
 \item no equation from filling fractions.
\end{enumerate}
The resulting $d-1$ equations determine the unknown coefficients of $\overline P(x)$.
In particular, in the 1-cut case $s=1$, requiring that $(V')^2-4\overline{P}$ has $d-1$ double zeros yields a finite set of solutions for $\overline P(x)$. The correct solution is chosen by minimizing the level of the effective potential.

\section{Algebraic geometry}
\label{sec:alggeomspcurve}

We will now introduce a geometric object called the spectral curve, which will allow us to determine which properties of our matrix integral are universal, and to efficiently compute universal quantities. 
Although we will construct it in the saddle-point approximation, the spectral curve can actually be used for doing perturbative and even non-perturbative computations of correlation functions, as we will see in \autoref{chap:LoopEq}.

\subsection{The spectral curve}\label{sec:spcurve}

\subsubsection{The Riemann surface}

The resolvent $\overline{W}$ is defined on the complex plane, minus the support of $\bar{\rho}$, where it has a discontinuity.
In other words, $\overline W$ differs from its analytic continuation across the support. An interpretation is that $\overline W$ is a multivalued function on the complex plane. 
However, we can also view $\overline W$ as a single-valued, holomorphic function, that lives on a two-sheeted cover of the complex plane. The two sheets correspond to the two solutions of the quadratic equation \eqref{eq:ovomdef}, and they are glued along $\operatorname{supp}\bar\rho$. Let us illustrate this in the 2-cut case:
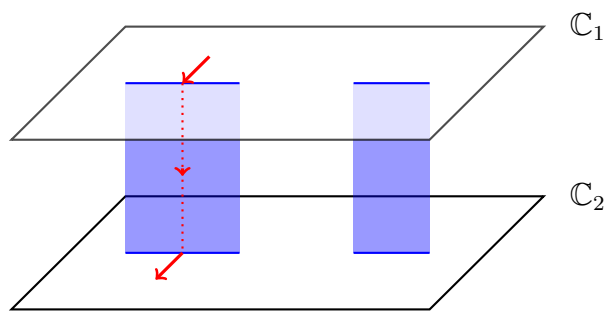
\begin{align}
 \begin{tikzpicture}[baseline=(current bounding box.center)]
%  
  % bottom sheet
  \filldraw [thick,fill=none,draw=black] 
  (-3.5,-3) -- (-2,-1.5) -- (3.5,-1.5) -- (2,-3) -- cycle;
%  
  % connecting cuts
  \filldraw [fill=blue,opacity=0.4,draw=none] (-2,-2.25) rectangle (-.5,0);
  \filldraw [fill=blue,opacity=0.4,draw=none] (1,-2.25) rectangle (2,0);  
%  
  % upper sheet
  \filldraw [thick,fill=white,opacity=0.7,draw=black] 
  (-3.5,-.75) -- (-2,.75) -- (3.5,.75) -- (2,-.75) -- cycle;
%  
  % cuts
  \draw [thick,blue] (-2.,0) -- (-.5,0);
  \draw [thick,blue] (1.,0) -- (2.,0);
  \draw [thick,blue] (-2.,-2.25) -- (-.5,-2.25);
  \draw [thick,blue] (1.,-2.25) -- (2.,-2.25);
%  
  % labels
  \draw (3.7,.75) node [right] {$\mathbb{C}_1$};
  \draw (3.7,-1.5) node [right] {$\mathbb{C}_2$};
%
  % arrows
  \draw [red,very thick,latex-] (-1.25,0) -- ++(45:.5);
  \draw [red, thick, dotted] (-1.25,0) -- (-1.25,-2.25);
  \draw [red,very thick,-latex] (-1.25,-1.13) -- ++(0,-.1);
  \draw [red,very thick,-latex] (-1.25,-2.25) -- ++(225:.5);
 \end{tikzpicture}
\end{align}
The intrinsic geometry of the resulting manifold is that of a \textbf{Riemann surface}\index{Riemann surface}.
That Riemann surface is actually compact, because $\overline W$ is regular at infinity, so each sheet can be completed into a Riemann sphere $\overline{\mathbb{C}} = \mathbb C \cup \{\infty\}$.
The cuts that connect the Riemann spheres can be viewed as contractions of tubes, and the whole surface as spheres with holes, glued along tubes:
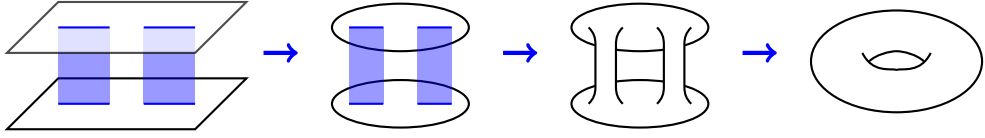
\begin{align}
 \begin{tikzpicture}[scale = .45,baseline=(current bounding box.center)]
  % Fig 1  
  % bottom sheet
  \filldraw [thick,fill=white,draw=black] 
  (-3.5,-3) -- (-2,-1.5) -- (3.5,-1.5) -- (2,-3) -- cycle;
%  
  % connecting cuts
  \filldraw [fill=blue,opacity=0.4,draw=none] (-2,-2.25) rectangle (-.5,0);
  \filldraw [fill=blue,opacity=0.4,draw=none] (.5,-2.25) rectangle (2,0);  
%  
  % upper sheet
  \filldraw [thick,fill=white,opacity=0.7,draw=black] 
  (-3.5,-.75) -- (-2,.75) -- (3.5,.75) -- (2,-.75) -- cycle;
%  
  % cuts
  \draw [thick,blue] (-2.,0) -- (-.5,0);
  \draw [thick,blue] (.5,0) -- (2.,0);
  \draw [thick,blue] (-2.,-2.25) -- (-.5,-2.25);
  \draw [thick,blue] (.5,-2.25) -- (2.,-2.25);
  \draw [blue,ultra thick,-latex] (4,-.75) -- ++(1,0);
%  
%
  % Fig 2    
  \draw [thick] (8,0) circle [x radius=2, y radius = .7];
  \draw [thick] (8,-2.25) circle [x radius=2, y radius = .7];
  \filldraw [fill=blue,opacity=0.4,draw=none] (6.5,-2.25) rectangle (7.5,0);
  \filldraw [fill=blue,opacity=0.4,draw=none] (8.5,-2.25) rectangle (9.5,0);  
  \draw [thick,blue] (6.5,0) -- (7.5,0);
  \draw [thick,blue] (8.5,0) -- (9.5,0);
  \draw [thick,blue] (6.5,-2.25) -- (7.5,-2.25);
  \draw [thick,blue] (8.5,-2.25) -- (9.5,-2.25);
%
%
 %  \draw [blue,ultra thick,-latex] (8,-3.4) -- ++(0,-1);
  \draw [blue,ultra thick,-latex] (11,-.75) -- ++(1,0);
%
%  
  % Fig 3
%  
  \begin{scope}[shift={(7,5.5)}]
  \draw [thick] (8,-5.5) circle [x radius=2, y radius = .7];
  \draw [thick] (8,-7.75) circle [x radius=2, y radius = .7];
  \filldraw [draw=none,fill=white] (6.7,-6) rectangle (7.3,-7.25);
  \filldraw [draw=none,fill=white] (8.7,-6) rectangle (9.3,-7.25);
  \draw [thick] (6.5,-5.5) to [out = -45,in = up] (6.7,-6) -- (6.7,-7.25) to [out=down,in=45] (6.5,-7.75);
  \draw [thick] (7.5,-5.5) to [out = 225,in = up] (7.3,-6) -- (7.3,-7.25) to [out=down,in=135] (7.5,-7.75);
  \draw [thick] (8.5,-5.5) to [out = -45,in = up] (8.7,-6) -- (8.7,-7.25) to [out=down,in=45] (8.5,-7.75);
  \draw [thick] (9.5,-5.5) to [out = 225,in = up] (9.3,-6) -- (9.3,-7.25) to [out=down,in=135] (9.5,-7.75);
  \end{scope}
%
%%  \draw [blue,ultra thick,-latex] (5,-6.25) -- ++(-1,0);
  \draw [blue,ultra thick,-latex] (18,-.75) -- ++(1,0);
% 
%
  % Fig 4
%  
  \begin{scope}[shift={(22.5,5.5)}]	
  \draw [thick] (0,-6.5) circle [x radius = 2.5, y radius = 1.5];
  \draw [thick] (-1,-6.25) .. controls ++(.3,-.6) and (-.2,-6.7) .. (0,-6.75) .. controls (.2,-6.7) and (.7,-6.85) .. (1,-6.25);
  \draw [thick] (-.8,-6.5) .. controls (-.6,-6.3) and (-.2,-6.2) .. (0,-6.2) .. controls (.2,-6.2) and (.6,-6.3) .. (.8,-6.5);
  \end{scope}
 \end{tikzpicture}
\end{align}
If we have one cut, the resulting surface is topologically a sphere. 
If we have two cuts, it is a torus. 
If we have $s$ cuts, and more generally if the support has $2s$ spectral edges, it is a surface whose number of holes, called the 
\textbf{genus}\index{genus}, is
\begin{align}
 \boxed{g = s - 1} \ .
\end{align}

\subsubsection{Lagrangian immersion into $\overline{\mathbb{C}}\times \mathbb{C}$}

Our Riemann surface $\Sigma$ comes with a natural holomorphic projection $x:\Sigma\to \overline{\mathbb{C}}$.
Moreover, the multivalued function $\overline W$ on $\overline{\mathbb{C}}$ becomes a single-valued function $\bar \omega:\Sigma \to \mathbb C$ on $\Sigma$,
\begin{equation}
\forall z \in \Sigma\ , \qquad  \overline W(x(z)) = \bar\omega(z).
\end{equation}
The function $\bar \omega$ is meromorphic, with poles at the two infinities, i.e. at the elements of $x^{-1}(\{\infty\})$.
If $V'$ were rational rather than polynomial, then $\bar \omega$ would have more poles, at the preimages by $x$ of poles of $V'$. If $V'$ had essential singularities, then $\Sigma$ would have to be non-compact for $\bar\omega$ to be meromorphic.

Renaming $y=\bar \omega$, $\Sigma$ has the immersion into $\overline{\mathbb{C}}\times {\mathbb C}$,
\begin{equation}
\begin{array}{ccl}
\Sigma & \overset{i}{\lhook\joinrel\longrightarrow} & \overline{\mathbb{C}}\times {\mathbb C} \ ,
\\
z & \longmapsto & (x(z),y(z)) \ ,
\end{array}
\end{equation}
whose image is the solution locus of the equation
\begin{equation}
\boxed{
 y^2 - V'(x) y + \overline{P}(x) = 0  } \ .
 \label{eq:yvp}
\end{equation}
At any point $x\in\overline{\mathbb{C}}$, we can identify
the complex plane $\mathbb{C}$ with the cotangent plane $T_x^*{\overline{\mathbb{C}}}$.
Then $\overline{\mathbb{C}}\times {\mathbb C}$ is identified with the cotangent fiber bundle $T^*\overline{\mathbb{C}} $ by $(x,y)\mapsto (x,ydx)$, and we have the commutative diagram
\begin{equation}
\begin{tikzcd}
\Sigma \arrow[r,hookrightarrow,"i"] \arrow[rd,"x"']& T^*\overline{\mathbb{C}} \arrow[d] \\
& \overline{\mathbb{C}}
\end{tikzcd}
\end{equation}
The one-form $ydx$ is the \textbf{tautological one-form}\index{tautological one-form} (or Liouville one-form) of $T^*\overline{\mathbb{C}} $, i.e. the one-form whose differential is the canonical symplectic form $dy\wedge dx = d(y dx)$.
That symplectic form vanishes on $i(\Sigma)$, because of Eq.~\eqref{eq:yvp}. So $i$ is a Lagrangian immersion of $\Sigma$ into $T^*\overline{\mathbb{C}} $.

This geometrical construction is a special case of a Hitchin system. To define a \textbf{Hitchin system}\index{Hitchin system}, we consider $x:\Sigma\to\Sigma_0$ for $\Sigma_0$ an arbitrary Riemann surface, and $ydx$ the tautological 1-form of $T^*\Sigma_0$, restricted to the spectrum $\det(ydx-\Phi(x))=0$. The Higgs field $\Phi(x)$ takes values in a Lie algebra tensored with 1-forms, i.e. in the adjoint bundle of some Lie group principal bundle.
In our case, the Riemann surface is $\Sigma_0 = \overline{\mathbb C}$, the Lie group is $GL(2,\mathbb C)$, and the Higgs field is
\begin{align}
 \Phi(x) = \begin{pmatrix}
V'(x) & - \overline{P}(x) \cr 1 & 0
\end{pmatrix}\ ,
\end{align}
so that $\Tr \Phi(x)=V'(x)$ and $\det\Phi(x) = \overline{P}(x)$. Then the equation $\det(ydx-\Phi(x))=0$ is equivalent to Eq.~\eqref{eq:yvp}.

\subsubsection{The spectral curve and universality}

We define the \textbf{spectral
curve}\index{spectral curve} of our matrix model as the Riemann surface $\Sigma$, together with its immersion $i=(x,y)$ into $\overline{\mathbb{C}}\times {\mathbb C}$, i.e. the triple $(\Sigma, x,y)$ with
\begin{equation}
\Sigma =\text{Riemann surface} \ \ ,\ \
x:\Sigma \to \overline{\mathbb{C}} \ \ ,\  \
y:\Sigma \to {\mathbb C} \ .
\end{equation}
Since the equation \eqref{eq:yvp} of $i(\Sigma)$ depends on the potential, the immersion  is not universal.
On the other hand, we consider the Riemann surface $\Sigma$ itself as universal, where 
by Riemann surface we mean an equivalence class under \textbf{biholomorphisms}\index{biholomorphism}, i.e. holomorphic bijections whose inverses are holomorphic.

Biholomorphisms preserve the topology and complex structure of $\Sigma$. 
The topology is characterized by the genus $g\in \mathbb{N}$, which is constant under small changes of the potential.
The space of possible complex structures has dimension $0$ if $g=0$, dimension $1$ if $g=1$, and dimension $3g-3$ if $g\geq  2$. 
This dimension does not depend on $d=\deg V'$, and there can be deformations of the potential that do not affect the complex structure.
This weak dependence of the complex structure on the potential is the reason why we consider the complex structure as universal.

\subsubsection{Cycles of $\Sigma$}

Assuming that the support of the equilibrium density is a union of intervals $\operatorname{supp}\bar\rho = \cup_{i=1}^s [a_i, b_i]$, let 
$A_i$ be a smooth closed loop that surrounds the cut $[a_i,b_i]$ in the first sheet:
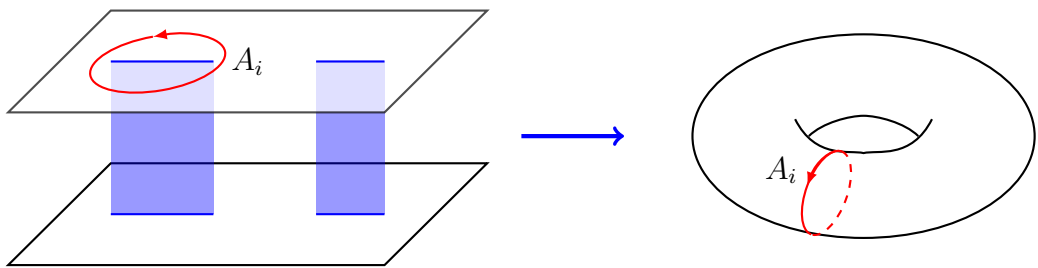
\begin{align}
 \begin{tikzpicture}[scale = .9,baseline=(current bounding box.center)]
%
 % bottom sheet
  \filldraw [thick,fill=white,draw=black] 
  (-3.5,-3) -- (-2,-1.5) -- (3.5,-1.5) -- (2,-3) -- cycle;
%    
  % connecting cuts
  \filldraw [fill=blue,opacity=0.4,draw=none] (-2,-2.25) rectangle (-.5,0);
  \filldraw [fill=blue,opacity=0.4,draw=none] (1,-2.25) rectangle (2,0);  
%  
  % upper sheet
  \filldraw [thick,fill=white,opacity=0.7,draw=black] 
  (-3.5,-.75) -- (-2,.75) -- (3.5,.75) -- (2,-.75) -- cycle;
%  
  % cuts
  \draw [thick,blue] (-2.,0) -- (-.5,0);
  \draw [thick,blue] (1.,0) -- (2.,0);
  \draw [thick,blue] (-2.,-2.25) -- (-.5,-2.25);
  \draw [thick,blue] (1.,-2.25) -- (2.,-2.25);
% 
  % label
  \node at (0, 0) {$A_i$};
%  
  % contour
  \draw[red, -latex, thick, rotate = 10] (-1.3, .6) arc [start angle = -270, end angle = 90, x radius = 1, y radius = .4];
%
  % arrow
  \draw[blue, ultra thick, -latex] (4, -1.1) -- (5.5, -1.1);
  \begin{scope}[shift={(9,5.4)}]	
  \draw [thick] (0,-6.5) circle [x radius = 2.5, y radius = 1.5];  
  \draw [thick] (-1,-6.25) .. controls ++(.3,-.6) and (-.2,-6.7) .. (0,-6.75) .. controls (.2,-6.7) and (.7,-6.85) .. (1,-6.25);
  \draw [thick] (-.8,-6.5) .. controls (-.6,-6.3) and (-.2,-6.2) .. (0,-6.2) .. controls (.2,-6.2) and (.6,-6.3) .. (.8,-6.5);
  \draw[red, thick, rotate = 70] (-6.43, -2) arc [start angle = 0, end angle = 180, x radius = .65, y radius = .3];
  \draw[red, thick, -latex, rotate = 70] (-6.43, -2) arc [start angle = 0, end angle = 90, x radius = .65, y radius = .3];
  \draw[red, thick, dashed, rotate = 70] (-6.43, -2) arc [start angle = 0, end angle = -180, x radius = .65, y radius = .3];
  \node at (-1.2,-7) {$A_i$};
  \end{scope}
 \end{tikzpicture}
\end{align}
Let $B_i$ be a closed loop that goes from $b_{i}$ to $a_{s}$ in the first sheet, and comes back to $b_{i}$ in the second sheet:
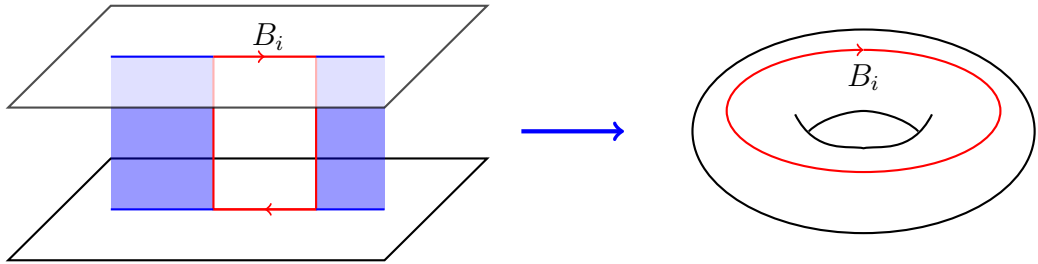
\begin{align}
 \begin{tikzpicture}[scale = .89,baseline=(current bounding box.center)]
%
 % bottom sheet  
  \filldraw [thick,fill=white,draw=black] 
  (-3.5,-3) -- (-2,-1.5) -- (3.5,-1.5) -- (2,-3) -- cycle;
%  
  % connecting cuts
  \filldraw [fill=blue,opacity=0.4,draw=none] (-2,-2.25) rectangle (-.5,0);
  \filldraw [fill=blue,opacity=0.4,draw=none] (1,-2.25) rectangle (2,0);  
%  
   % contour
  \draw[thick, red] (-.5, 0) -- (1,0) -- (1,-2.25) -- (-.5, -2.25) -- (-.5, 0); 
%  
  % upper sheet
  \filldraw [thick,fill=white,opacity=0.7,draw=black] 
  (-3.5,-.75) -- (-2,.75) -- (3.5,.75) -- (2,-.75) -- cycle;
%  
  % cuts
  \draw [thick,blue] (-2.,0) -- (-.5,0);
  \draw [thick,blue] (1.,0) -- (2.,0);
  \draw [thick,blue] (-2.,-2.25) -- (-.5,-2.25);
  \draw [thick,blue] (1.,-2.25) -- (2.,-2.25);
%  
%   \draw[thick, red, -latex] (-.5,0) -- (.3, 0);
%   \draw[thick, red] (-.5,0) -- (1, 0);
%
  % label
  \node at (.3, .3) {$B_i$};
%
  % contour arrow
%  
%  \draw[red, thick, -latex] (-.5, 0) -- (.3,0);
  \draw[thick,red,->-] (-.5,0) -- (1,0);
  \draw[thick,red,->-] (1,-2.25) -- (-.5,-2.25);
%  
  % arrow  
  \draw[blue, ultra thick, -latex] (4, -1.1) -- (5.5, -1.1);
  \begin{scope}[shift={(9,5.4)}]	
  \draw [thick] (0,-6.5) circle [x radius = 2.5, y radius = 1.5];
  \draw [thick] (-1,-6.25) .. controls ++(.3,-.6) and (-.2,-6.7) .. (0,-6.75) .. controls (.2,-6.7) and (.7,-6.85) .. (1,-6.25);
  \draw [thick] (-.8,-6.5) .. controls (-.6,-6.3) and (-.2,-6.2) .. (0,-6.2) .. controls (.2,-6.2) and (.6,-6.3) .. (.8,-6.5);
  \draw[red, thick, latex-] (0,-5.3) arc [start angle = 90, end angle = 450, x radius = 2, y radius = .9];
  \node at (0, -5.7) {$B_i$};
  \end{scope}
 \end{tikzpicture}
\end{align}
Then the $2g$ loops $(A_i,B_i)_{i=1,\dots, g}$ form a symplectic basis of cycles of the Riemann surface $\Sigma$, in the sense that
\begin{equation}\label{def:H1symplbasis}
A_i\cap B_j =\delta_{i,j}
 \, , \quad
A_i\cap A_j =\emptyset
 \, , \quad
B_i\cap B_j =\emptyset \, ,
\end{equation}
where intersections are counted algebraically according to the respective orientations of contours at crossing points.
Let us draw such a basis in the case $g=3$:
\begin{align}
 \begin{tikzpicture}[baseline=(current bounding box.center)]
%
  % shape
%  
  \draw [thick] 
  (0,1.5) to [out=left,in=right] (-2,1) to 
  [out=left,in=right] (-4,1.5) to [out=left,in=up] (-6,0) to 
  [out=down,in=left] (-4,-1.5) to [out=right,in=left] (-2,-1) to
  [out=right,in=left] (0,-1.5) to [out=right,in=left] (2,-1) to
  [out=right,in=left] (4,-1.5) to [out=right,in=down] (6,0) to
  [out=up,in=right] (4,1.5) to [out=left,in=right] (2,1) to
  [out=left,in=right] (0,1.5);
%
  % hole
%
  \draw [thick] 
  (-1,.3) to [out=-45,in=left] (0,-.2) to [out=right,in=225] (1,.3);
  \draw [thick]
  (-.7,.05) to [out=45,in=left] (0,.4) to [out=right,in=135] (.7,.05);
%
  % alpha cycle
%
  \draw [very thick,blue] 
  (0,.2) circle [x radius = 1.5,y radius = .7];
  \draw [very thick,blue,-latex]
  (-.05,.9) -- ++(-.05,0);
%  
  % beta cycle
%
  \draw [very thick,red]
  (0,-.2) to [out=-50,in=up] (.3,-.85) to [out=down,in=50] (0,-1.5);
  \draw [very thick,red,dotted]
  (0,-.2) to [out=230,in=up] (-.3,-.85) to [out=down,in=130] (0,-1.5);
  \draw [very thick,red,-latex]
  (.3,-.8) -- ++(0,.05);
  \node at (-0.9, -0.75) {$B_2$};
  \node at (0.7, -1.) {$A_2$};
  \begin{scope}[shift={(-4,0)}]
%
   % hole
%
   \draw [thick] 
   (-1,.3) to [out=-45,in=left] (0,-.2) to [out=right,in=225] (1,.3);
   \draw [thick]
   (-.7,.05) to [out=45,in=left] (0,.4) to [out=right,in=135] (.7,.05);
%   
   % alpha cycle
%
   \draw [very thick,blue] 
   (0,.2) circle [x radius = 1.5,y radius = .7];
   \draw [very thick,blue,-latex]
   (-.05,.9) -- ++(-.05,0);
%
   % beta cycle
%   
   \draw [very thick,red]
   (0,-.2) to [out=-50,in=up] (.3,-.85) to [out=down,in=50] (0,-1.5);
   \draw [very thick,red,dotted]
   (0,-.2) to [out=230,in=up] (-.3,-.85) to [out=down,in=130] (0,-1.5);
   \draw [very thick,red,-latex]
   (.3,-.8) -- ++(0,.05);
  \node at (-0.9, -0.75) {$B_1$};
  \node at (0.7, -1.) {$A_1$};
  \end{scope}
  \begin{scope}[shift={(4,0)}]
%
   % hole
%
   \draw [thick] 
   (-1,.3) to [out=-45,in=left] (0,-.2) to [out=right,in=225] (1,.3);
   \draw [thick]
   (-.7,.05) to [out=45,in=left] (0,.4) to [out=right,in=135] (.7,.05);
%   
   % alpha cycle
%
   \draw [very thick,blue] 
   (0,.2) circle [x radius = 1.5,y radius = .7];
   \draw [very thick,blue,-latex]
   (-.05,.9) -- ++(-.05,0);
%
   % beta cycle
%   
   \draw [very thick,red]
   (0,-.2) to [out=-50,in=up] (.3,-.85) to [out=down,in=50] (0,-1.5);
   \draw [very thick,red,dotted]
   (0,-.2) to [out=230,in=up] (-.3,-.85) to [out=down,in=130] (0,-1.5);
   \draw [very thick,red,-latex]
   (.3,-.8) -- ++(0,.05);
  \node at (-0.9, -0.75) {$B_3$};
  \node at (0.7, -1.) {$A_3$};
  \end{scope}
 \end{tikzpicture}
\end{align}
A complex homology cycle is a $\mathbb C$-linear combination of homotopy classes of Jordan loops on $\Sigma$, where
the concatenation of two loops is identified with the sum of the corresponding homology cycles.
The complex \textbf{homology space}\index{homology space!of a Riemann surface} $H_1(\Sigma)$ of $\Sigma$ can then be written as
\begin{equation}
H_1(\Sigma) = \bigoplus_{i=1}^{ g} \left(\mathbb{C} A_i \oplus \mathbb{C} B_i \right)\ .
\end{equation}

\subsubsection{Periods}

Thanks to the expression for $\bar\rho$ in terms of discontinuities of $\overline{W}$ \eqref{eq:omegajump}, the filling fraction of the cut $[a_j,b_j]$ can  be rewritten as a contour integral of the resolvent over the cycle $A_j$,
\begin{align}
\boxed{
 \bar{\epsilon}_j
= \int_{a_{j}}^{b_{j}}dx\, \bar{\rho}(x) 
= \frac{1}{2\pi i} \oint_{
A_j
}
dx\, \overline{W}(x) } \ 
= \frac{1}{2\pi i} \oint_{A_j} ydx \ .
\label{eq:aomega}
\end{align}
Moreover, the effective potential on each cut is constant according to Eq.~\eqref{eq:saddlerho}. 
If $\ell_j = \left. V_\mathrm{eff}\right|_{[a_j,b_j]}$, then 
$\int_{b_{j}}^{a_{s}} V'_{\text{eff}}(x)dx =\ell_{s}-\ell_j$, and 
\begin{equation}
\boxed{
\Re \oint_{B_j}dx\, \overline{W}(x) = \ell_{s}-\ell_j} \ ,
\label{eq:bomega}
\end{equation}
where we used that $V'_\text{eff}=M\sqrt\sigma = V'-2\overline W $ changes sign when we change sheets, together with  $\oint_{B_j} V' = 0$.
The integrals \eqref{eq:aomega} and \eqref{eq:bomega} of the one-form $\overline{W}(x)dx$ are called  the periods of the spectral curve.

The \textbf{periods of the Riemann surface}\index{period (of a Riemann surface)} $\Sigma$ are defined as
\begin{equation}
\eta_{k,j} = \frac{1}{2\pi i}\oint_{A_j} x^{k-1}\frac{dx}{\sqrt{\sigma(x)}}
\quad , \quad
\tilde \eta_{k,j} = \frac{1}{2\pi i}\oint_{B_j} x^{k-1}\frac{dx}{\sqrt{\sigma(x)}}
\quad , \quad
k=1, \dots, g \ .
\end{equation}
It is a classical theorem in algebraic geometry that the square matrix
$[\eta_{k,j}]_{j,k=1,\dots, g}$ is invertible \cite{Fay:1973}, and we define the
Riemann period matrix of $\Sigma$ as 
\begin{equation}
\tau = \eta^{-1} \tilde \eta \ .
\end{equation}
From the functional action~\eqref{def:Srho}, we obtain the relation $\ell_i = \frac{\partial S[\rho]}{\partial \epsilon_i}$, which leads to the \textbf{Seiberg--Witten equation}\index{Seiberg--Witten equation} for the periods,
\begin{align}
 \Re \oint_{B_j}dx\, \overline{W}(x)
 = \frac{\partial S[\rho]}{\partial \epsilon_{s}} - \frac{\partial S[\rho]}{\partial \epsilon_{j}}
 \ .
\end{align}
In case $g=1$, the complex number $\tau=\frac{\tilde \eta_1}{\eta_1}$ is the \textbf{modulus}\index{modulus (of a torus)}, i.e. the number such that $\Sigma$ is holomorphically equivalent to $\frac{\mathbb{C}}{\mathbb{Z} + \tau \mathbb{Z}}$. The modulus is defined modulo $\tau\equiv \tau+1 \equiv -\frac{1}{\tau}$, and obeys $\Im \tau>0$. 
More generally if $g\geq 1$, 
it can be proved that $\tau$ belongs to the \index{Siegel} \textbf{Siegel space} of genus $g$: the set of size $g$ symmetric complex matrices, whose imaginary parts are positive definite. 
The real dimension of that space is $ g(g+1)$. 
If $g\geq 2$, only a submanifold of real dimension $6g-6$ corresponds to period matrices of Riemann surfaces.
But the polynomial $\overline P$ has $g$ complex coefficients that are not fixed by the constraint that $(V')^2-4\overline{P}$ has $d-1-g$ double zeros. 
So the period matrices of Riemann surfaces that have equations of the type \eqref{eq:yvp} form a submanifold of real dimension $2g$.

\subsection{Construction of the spectral curve}

We will now explain how the spectral curve can be built from atlases of charts. 
A chart is an open domain of $\mathbb{C}$, and an atlas is a collection of charts, together with biholomorphic transition functions between overlapping charts.  Such an atlas then defines a Riemann surface, which is compact if it is covered by a finite number of compact charts. The spectral curve is obtained by moreover specifying an immersion $(x,y)$ in $\overline{\mathbb{C}}\times \mathbb{C}$, by defining it in each chart.

While this construction is not needed for computing correlation functions, it is useful for understanding the geometry of the spectral curve.
We will give two different constructions of the spectral curve: a direct construction that requires only the knowledge of $\operatorname{supp}\bar\rho$, and a construction based on the spectral network and its sectors. 
The latter construction involves a larger number of smaller charts, and provides a parametrization of the spectral curve in terms of simple moduli of the sectors. 

\subsubsection{Direct construction}

Let us construct the Riemann surface by gluing two copies of the complex plane along cuts.
Let our charts be:
\begin{itemize}
\item  two copies of $\mathbb{C} \backslash \operatorname{supp}\bar\rho$, called the upper and lower sheets;
\item for each edge in $\operatorname{supp}\bar\rho$, two rectangular strips called the left and right strips, that cover the edge but stop short of its vertices;
\item for each vertex in $\operatorname{supp}\bar\rho$ of valency $k$, one disc of $\mathbb C$ centered at 0 if $k$ is odd, or two discs if $k$ is even;
\item two discs, centered at 0, called the upper and lower infinities.
\end{itemize}
The transition functions that glue these charts are:
\begin{itemize}
\item for each edge in $\operatorname{supp}\bar\rho$, each one of the two strips is glued to one sheet on one side, and to the other sheet on the other side, using as transition function the map $x\mapsto x$;
\item for each vertex $a$ of odd valency, the corresponding disc is glued to different sheets and strips,
using the map $z\mapsto x=a+z^2$, whose inverse map is   $x\mapsto z= \pm\sqrt{x-a}$;
\item for each vertex $a$ of even valency, the corresponding discs are glued to different sheets and strips,
using the map $z\mapsto x=a+z$;
\item each infinity disc is glued to its corresponding sheet using the function $x\mapsto \frac{1}{x}$.
\end{itemize}
These charts and transition functions define a smooth compact Riemann surface $\Sigma$. 
We define the projection $x:\Sigma \to \overline{\mathbb{C}}$ as the identity function in the upper and lower sheets, and $z\mapsto a+z^2$ in a vertex disc, and $x\mapsto \frac{1}{x}$ in the infinity discs.
Each vertex has one preimage by the projection, while any other point of $\overline{\mathbb{C}}$ has two preimages.
And we define the function $y:\Sigma \to \mathbb{C}$ by solving Eq.~\eqref{eq:yvp} in the upper and lower sheets, and composing the solution with transition functions in other charts.

\subsubsection{Construction from the spectral network}

We will now construct the spectral curve from the topology of the spectral network, together with its real moduli.
The topology of the spectral network is given by a graph, together with signs associated to the different sectors:
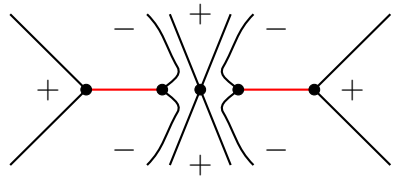
\begin{align}
 \begin{tikzpicture}[baseline=(current bounding box.center)]
\foreach \xscale in {1,-1}{
\begin{scope}[xscale=\xscale]
\draw[thick] (1.5,0) -> (2.5,1) ;
\draw[thick] (1.5,0) -> (2.5,-1) ;
\draw[thick] (0,0) -> (0.4,1) ;
\draw[thick] (0,0) -> (0.4,-1) ;
 \draw[thick] (0.5,0) to [out=90+45, in=180+60] (0.3,0.3) to  [out=60, in=180+45] (0.7,1) ;
\draw[thick] (0.5,0) to [out=-90-45, in=180-60] (0.3,-0.3) to  [out=-60, in=180-45] (0.7,-1) ;
\draw[thick, red ] (0.5,0) -> (1.5,0) ;
   \filldraw (0.5,0) circle [radius=2pt];
   \filldraw (1.5,0) circle [radius=2pt];
   \node at (1,0.8) {$-$};
   \node at (1,-0.8) {$-$};
   \node at (2,0) {$+$};
\end{scope}
}
   \node at (0,1) {$+$};
   \node at (0,-1) {$+$};
   \filldraw (0,0) circle [radius=2pt];
%
   % \node at (0,-1.5) {Graph of a spectral network.};
 \end{tikzpicture} 
\end{align}  
This determines a number of sectors, with vertices on their boundaries. 
To these sectors, we associate the corresponding strips (for beaches) and half-planes (for lands and seas). 
The real moduli are strip widths $w_i$, edge lengths $l_i$, and strip shearings $s_i$:
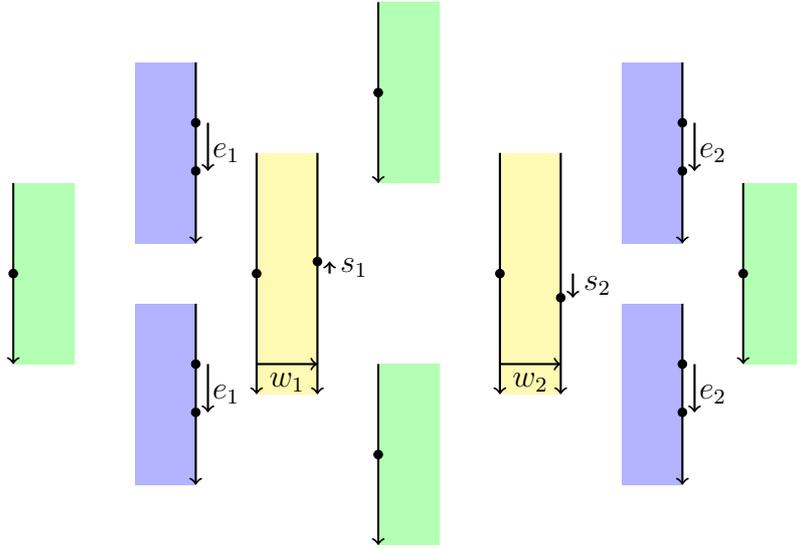
\begin{align}
 \begin{tikzpicture}[baseline=(current bounding box.center)]
\begin{scope}[scale=.8]
\foreach \x in {1,2}{
\foreach \y in {-2, 2}{
\begin{scope}[shift={(-12+8*\x,\y)}, rotate=-90] 
  \filldraw [fill=blue,opacity=0.3,draw=none] (-1.5,0) rectangle (1.5,1);  
  \draw[thick, -latex] (-1.5,1) -> (1.5,1);
   \filldraw (-0.5, 1) circle [radius=2pt];
   \filldraw (0.3, 1) circle [radius=2pt];
   \draw[thick, -latex] (-0.5,1.2) -> (0.3,1.2);
   \node at (0, 1.5) {$l_{\x}$};
   \end{scope}
   }
}
\foreach \x in {-6, 6}{
\begin{scope}[shift={(\x,0)}, rotate=-90] 
  \filldraw [fill=green,opacity=0.3,draw=none] (-1.5,0) rectangle (1.5,1);  
  \draw[thick, -latex] (-1.5,0) -> (1.5,0);
   \filldraw (0, 0) circle [radius=2pt];
   \end{scope}
   }
\foreach \y in {-3, 3}{
\begin{scope}[shift={(0,\y)}, rotate=-90] 
  \filldraw [fill=green,opacity=0.3,draw=none] (-1.5,0) rectangle (1.5,1);  
  \draw[thick, -latex] (-1.5,0) -> (1.5,0);
   \filldraw (0, 0) circle [radius=2pt];
   \end{scope}
   }
\begin{scope}[shift={(2.5,0)}, rotate=-90] 
  \filldraw [fill=yellow,opacity=0.3,draw=none] (-2,-0.5) rectangle (2,0.5);  
  \draw[thick, -latex] (-2,-0.5) -> (2,-0.5);
  \draw[thick, -latex] (-2,0.5) -> (2,0.5);
   \draw[thick, -latex] (1.5,-0.5) -> (1.5,0.5);
   \node at (1.8,0) {$w_2$};
   \draw[thick, -latex] (0,0.7) -> (0.4,0.7);
   \node at (0.2,1.1) {$s_2$};
   \filldraw (0,-0.5) circle [radius=2pt];
   \filldraw (0.4,0.5) circle [radius=2pt];
   \end{scope}
\begin{scope}[shift={(-1.5,0)}, rotate=-90] 
  \filldraw [fill=yellow,opacity=0.3,draw=none] (-2,-0.5) rectangle (2,0.5);  
  \draw[thick, -latex] (-2,-0.5) -> (2,-0.5);
  \draw[thick, -latex] (-2,0.5) -> (2,0.5);
   \draw[thick, -latex] (1.5,-0.5) -> (1.5,0.5);
   \node at (1.8,0) {$w_1$};
   \draw[thick, -latex] (0,0.7) -> (-0.8,0.7);
   \node at (-0.4,1.1) {$s_1$};
   \filldraw (0,-0.5) circle [radius=2pt];
   \filldraw (-0.8,0.5) circle [radius=2pt];
   \end{scope}
\end{scope}
%
  % \node at (0,-4) {8 half-planes and 2 strips. 6 real parameters.};
%
 \end{tikzpicture} 
\end{align}
Let our charts be:
\begin{itemize}
 \item two copies of each strip and half-plane, with the second copy obtained by a reflection with respect to the vertical axis thus reverting all signs -- for example the second copy of a land is a sea;
 \item one disc for each spectral edge;
 \item two discs for each vertex that is not a spectral edge;
 \item two discs called the upper and lower infinities.
\end{itemize}
These charts are glued as follows:
\begin{itemize}
 \item along each edge of the graph, glue the relevant strips and/or half-planes, with transition functions being translations in $\mathbb C$ that match the edge's vertices;
 \item for a spectral edge $v$ of valency $k_v+2$, glue all 
relevant $2(k_v+2)$ strips and/or half-planes to the disc, with a transition function $g \mapsto (g-g(v))^{\frac{1}{k_v+2} }$, with inverse map  $z\mapsto g(v)+z^{k_v+2}$, from each strip or half-plane to the disc;
\item for a vertex $v$ that is not a spectral edge, and therefore has even valency $k_v+2$, glue $k_v+2$ relevant strips and/or half-planes to each disc, with a transition function $g \mapsto (g-g(v))^{\frac{2}{k_v+2} }$, with inverse map  $z\mapsto g(v)+z^{\frac{k_v+2}{2}}$.
\item before gluing the two infinity discs, 
this defines a Riemann surface $\Sigma$, equipped with an involution $\sigma$ from our doubling or strips and half-planes.
\item the upper and lower infinity disc are each glued to half of the lands and seas, using transition functions of the type
$g \mapsto c_p+\xi^{-d-1}+2\log\xi$. Requiring that the transition functions from any two neighbouring sectors are compatible, determines the complex constants $c_p$ up to one overall constant for each infinity disc, leaving us with $2$ complex moduli, therefore $4$ real moduli.
\end{itemize}
We thus obtain a meromorphic one-form $dg$ on $\Sigma$, odd under the involution, and that has a pole of order $d+2$ at each copy of infinity.
There is also a natural projection $x:\Sigma\to\overline{\mathbb{C}}$ to the complex plane where the spectral network lives, which can be reconstructed similarly as $\Sigma$, but taking only one copy of each sector, plus one disc per vertex and one infinity disc. This projection is even under the involution.
Defining the function $y$ by $dg=ydx$ completes the construction of the spectral curve. 

\subsubsection{Recovering the potential}

The function $y^2$ is even under the involution, and has a pole of order $2d$ at infinity.
So there must exist a polynomial $U(x)$ of degee $2d$, such that $y^2 = U(x)$. 
Defining 
$V'(x) = (\sqrt{U(x)})_+$ and $\overline P(x) =  \frac14(V'(x)^2 - U(x))$, we 
recover the potential $V$, and the equation \eqref{eq:yvp} of $\Sigma$'s image under its immersion into $\overline{\mathbb{C}}\times \mathbb{C}$.

This allows us to relate the moduli of $\Sigma$, to the coefficients of the potential $V$. 
There is one length per compact edge, a strip width and a strip shearing for each beach,
and $4$ moduli from the gluing of charts at infinity, for a total number of moduli
\begin{equation}
\mathcal N = \#\{\text{compact edges}\}+ 2 \#\text{beaches}+4 \ .
\end{equation}
Let $E=\#\{\text{edges in }\operatorname{supp}\bar\rho\}$ and $V=\#\{\text{vertices}\}$, we have
$E \leq  \#\{\text{compact edges}\} = V-1-\#\text{beaches} $, which implies that
$\#\text{beaches}\leq V-E-1 $, and
$\mathcal N- E  = 3+V-E+\#\text{beaches} \leq 2+2V-2E $.
Odd vertices can only appear by pairs in connected components, which implies
$V-E \leq \#\text{even vertices} + \frac12  \#\text{odd vertices}$, and the total valency is
$2d = \sum_v k_v \geq 2  \#\text{even vertices} +  \#\text{odd vertices}$.
This implies $V-E \leq d$, thus
\begin{equation}
\mathcal N \leq 2d+2+\#\{\text{edges in }\operatorname{supp}\bar\rho\}\ ,
\end{equation}
which is saturated if and only if all connected components are segments and are separated by beaches.
In this case, $\mathcal N$ coincides with the number of real parameters from $V'$, plus the number of filling fractions.
So choosing $V'$ and filling fractions generically determines a zero-dimensional space of spectral curves, i.e. a finite number of choices of $\overline P$. 
%And we know that exactly one choice is a positive density.
%And it can be proved that this number is nonzero \cite{Bertola:2007}.

\subsection{Computing resolvents using uniformization}\label{sec:cmu}

Every Riemann surface can be \textbf{uniformized}\index{uniformization}, i.e. mapped to a standard Riemann surface with the same genus and complex structure. Applying this to the spectral curve will simplify calculations, and lead to universal results.

\subsubsection{Genus zero: the Joukowsky map}

Let us assume here that our spectral curve has genus zero, and thus one cut with two spectral edges $[a,b]$.
Since the Riemann sphere is the unique Riemann surface of genus zero, there exists a biholomorphism between the Riemann sphere, and our two-sheeted cover of the complex plane with one cut.
In terms of coordinates $z$ on the Riemann sphere and $x$ on the complex plane, this biholomorphism is the \textbf{Joukowsky map}\index{Joukowsky map}:
\begin{align}
\boxed{
z\mapsto x = \frac{a+b}{2} + \gamma \left(z + \frac{1}{z}\right) \quad \text{with} \quad \gamma = \frac{a-b}{4} }\ .
 \label{eq:xz}
\end{align}
The upper (resp. lower) sheet is mapped to the exterior (resp. interior) of the unit disc, the cut is mapped to the unit circle $|z|=1$, and changing sheets at $x$ fixed corresponds to the involution $z\to\frac{1}{z}$:
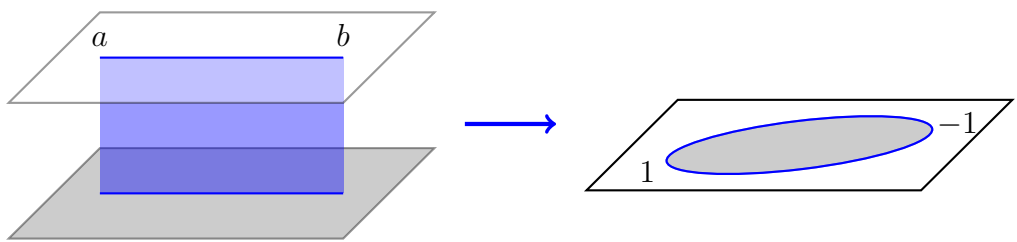
\begin{align}
 \begin{tikzpicture}[scale = .8,baseline=(current bounding box.center)]
%  
  % bottom sheet
  \filldraw [thick,fill= gray,opacity=0.4,draw=black] 
  (-3.5,-3) -- (-2,-1.5) -- (3.5,-1.5) -- (2,-3) -- cycle;
%  
  % connecting cut
  \filldraw [fill=blue,opacity=0.4,draw=none] (-2,-2.25) rectangle (2,0);
%  
  % upper sheet
  \filldraw [thick,fill= white,opacity=0.4,draw=black] 
  (-3.5,-.75) -- (-2,.75) -- (3.5,.75) -- (2,-.75) -- cycle;
%  
  % cuts
  \draw [thick,blue] (-2.,0) node[black, above]{$a$} -- (2,0) node[black,above]{$b$};  
  \draw [thick,blue] (-2.,-2.25) -- (2,-2.25);
%
   % arrow  
  \draw[blue, ultra thick, -latex] (4, -1.1) -- (5.5, -1.1);
%  
  % next sheet
  \draw [thick]
  (6, -2.2) -- (7.5, -.7) -- (13, -.7) -- (11.5, -2.2) -- cycle;
%  
  % ellipse
  \filldraw [thick, fill = gray, opacity = .4, draw = blue] (9.5, -1.45) circle [x radius = 2.2, y radius = .4, rotate = 7];
  \draw [thick, draw = blue] (9.5, -1.45) circle [x radius = 2.2, y radius = .4, rotate = 7];
%  
  % points
  \node at (7, -1.9){$1$};
  \node at (12.1, -1.1){$-1$};
 \end{tikzpicture}
\end{align}
The inverse map is
\begin{equation}
z = \frac1{2\gamma} \left(  x - \frac{a+b}{2} \pm \sqrt{\left(x-\frac{a+b}{2}\right)^2 - 4 \gamma^2} \right)\ ,
\end{equation}
where the sign is $+$ in the upper sheet and $-$ in the lower sheet.
The Joukowsky map was originally introduced for simplifying the study of the air flow around an airplane's thin wing, by holomorphically mapping the wing's section to a circle.

Let us write the resolvent $\overline{W}(x)$ \eqref{eq:rsms} in terms of the Joukowsky coordinate $z$, i.e. 
compute the single-valued function $\bar\omega(z) = \overline W(x(z))$.
The non-analytic contribution $\sqrt{\sigma(x)}$ becomes a rational function of $z$,
\begin{align}
 \sqrt{\sigma(x)} = \sqrt{(x-a)(x-b)} = \gamma\left(z-\frac{1}{z}\right)\ ,
\end{align}
where the sign of the square root is such that near $x=\infty$ we have $\sqrt{\sigma(x)}\sim x \sim \gamma z$ in the upper sheet, and $\sqrt{\sigma(x)}\sim -x \sim - \frac{\gamma}z$ in the lower sheet.
Thus $\bar\omega(z)$ is a rational function of $z$ with poles at $z=0$ and $z=\infty$, that is $\bar \omega(z)\in \mathbb C[z,\frac{1}{z}]$. 
Moreover, the behaviour \eqref{eq:omegainf} of $\overline{W}(x)$ near $x=\infty$ in the first sheet, determines the behaviour of  $\bar \omega(z)$ near $z=\infty$, and implies $\bar \omega(z)\in \mathbb C[\frac{1}{z}]$. Therefore,
\begin{align}
 \bar{\omega}(z) = \sum_{k=0}^d v_k z^{-k} \quad \text{with} \quad 
 \left\{\begin{array}{l} v_0 = 0 \ , \\ v_1 = \frac{1}{\gamma}\ . \end{array}\right.
 \label{eq:v0v1z}
\end{align}
Let us insert this in the Riemann--Hilbert equation \eqref{eq:saddleder}, after rewriting it as 
\begin{align}
 V'(x(z)) = \bar{\omega}(z) + \bar{\omega}\left(\frac{1}{z}\right)\ .
 \label{eq:RHJoukowsky}
\end{align}
We obtain 
\begin{align}
 V'\left(\tfrac{a+b}{2}+\gamma(z+\tfrac{1}{z})\right) = \sum_{k=0}^{d} v_k (z^k+z^{-k})\ .
 \label{eq:vkVz}
\end{align}
This determines $v_k$ in terms of $a,b$ and $V'$, then the equations \eqref{eq:v0v1z} for $v_0$ and $v_1$  determine $a$ and $b$, as we will now see in the case of a quartic potential.
For an application of the Joukovsky map to the combinatorics of partitions, see \autoref{exo:plancherel}.

\subsubsection{Example of a quartic potential}

Let us consider the quartic, even potential
\begin{align}\label{ex:potquartic}
 V(x) = \frac{x^4}{4} - t \frac{x^2}{2} \ .
\end{align}
We assume that the eigenvalue integration path is also even so that  $\operatorname{supp}\bar\rho$ is symmetric under $x\to -x$.

Let us first assume we have only one cut, which must therefore be of the type $(a,b)=(2\gamma,-2\gamma)$. 
Then, Eq.~\eqref{eq:vkVz} gives $(v_0, v_1, v_2, v_3) = \left(0, 3\gamma^3-t\gamma, 0, \gamma^3\right)$. 
The condition $v_1 = \frac{1}{\gamma}$ implies $3\gamma^3-t\gamma = \frac{1}{\gamma} $ whose solution is $\gamma^2 = \frac{1}{6}(t+\sqrt{t^2+12})$.
Using the inverse of the Joukowsky map $\frac{1}{z} = \frac{x-\sqrt{x^2-4\gamma^2}}{2\gamma}$, we then compute 
\begin{subequations}
\begin{align}
\overline{\omega}(z)  
&= y = \frac{1}{\gamma z} + \frac{\gamma^3}{z^3} \ , \\ 
\overline W(x) 
&= \frac{1}{2}\Big( x^3-t x - (x^2-t+2\gamma^2)\sqrt{x^2-4\gamma^2}\Big)\ ,
\\
\bar\rho(x) &= \frac{1}{2\pi } (x^2-t+2\gamma^2)\sqrt{4\gamma^2-x^2}\ ,
\end{align}
\end{subequations}
where $\bar{\rho}$ is deduced from $\overline{W}$ via Eq.~\eqref{eq:omegajump}.

This is only valid provided $\bar{\rho}$ is positive on its support $(2\gamma,-2\gamma)$, i.e. if $x^2-t+2\gamma^2$ has no zero on $(2\gamma,-2\gamma)$, which is true only for $t\leq 2$.
For $t> 2$ we must consider at least two cuts. 
Let us consider the 2-cut case $\operatorname{supp}\bar\rho = (-a,-b)\cup (b, a)$. The polynomial $M(x)$ is then of degree one, and it must be odd so that $M(x)^2$ is even, thus $M(x)=\lambda x$.
Using Eq.~\eqref{eq:rsms}, we have 
\begin{equation}
\overline W(x) = \frac{1}{2}\Big(x^3-t x - \lambda x\sqrt{(x^2-a^2)(x^2-b^2)}\Big)\ .
\end{equation}
The values of $\lambda$ and of the spectral edges $a,b$ can be deduced from the asymptotic behaviour of $\overline{W}(x)$ \eqref{eq:omegainf},
\begin{align}
\lambda = 1\quad , \quad \left\{\begin{array}{l}
 a^2 = t+2 \ , \\ b^2 = t-2\ . \end{array}\right.
\end{align}
For any $t>2$ we have $a^2>b^2>0$, and $\bar{\rho}$ is positive on its support.

\subsubsection{Genus one: elliptic functions}

The double-covered complex plane with two cuts $(a,b)$ and $(c,d)$ is equivalent to a torus, which we realize as the parallelogram $\frac{\mathbb{C}}{\mathbb{Z} + \tau \mathbb{Z}}$. 
The uniformizing coordinate $u$ on the torus is given in terms of the coordinate $x$ on the complex plane by 
\begin{equation}
u(x) = \frac{1}{2\int_a^b \frac{dx'}{\sqrt{\sigma(x')}}}
\int_a^x \frac{dx'}{\sqrt{\sigma(x')}} \quad \text{with} \quad \sigma(x) = (x-a)(x-b)(x-c)(x-d)\ ,
\label{eq:ux}
\end{equation}
where different choices of integration contours lead to different representations of $u\equiv u+1\equiv u+\tau$.
We have $u((a,b,c,d)) = (0, \frac12, \frac12+\frac{\tau}{2}, \frac{\tau}{2}$), 
and the modulus of the torus is 
\begin{equation}
 \tau = 2u(d) = i \frac{_2F_1(\frac12,\frac12;1;1-r)}{_2F_1(\frac12,\frac12; 1;r)} \quad \text{with} \quad r = \frac{(a-b)(c-d)}{(a-d)(c-b)}\ ,
\end{equation}
where $r$ is called the cross-ratio of the four points $(a,b,c,d)$, and $_2 F_1$ is the hypergeometric function.
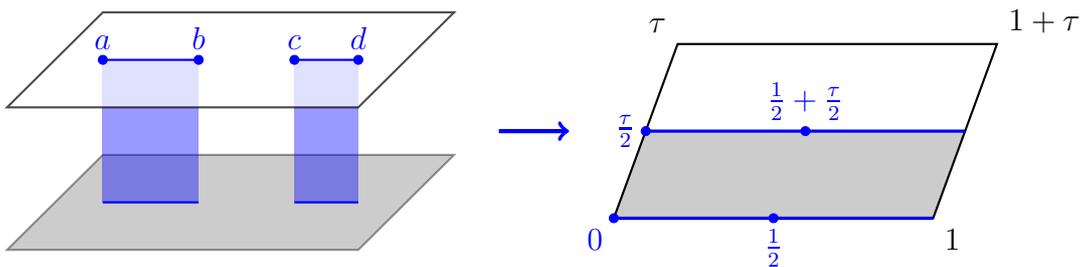
\begin{align}
 \begin{tikzpicture}[scale = .84,baseline=(current bounding box.center)]
  \filldraw [thick,fill= gray,opacity=0.4,draw=black] 
  (-3.5,-3) -- (-2,-1.5) -- (3.5,-1.5) -- (2,-3) -- cycle;
  \filldraw [fill=blue,opacity=0.4,draw=none] (-2,-2.25) rectangle (-.5,0);
  \filldraw [fill=blue,opacity=0.4,draw=none] (1,-2.25) rectangle (2,0);  
  \filldraw [thick,fill=white,opacity=0.7,draw=black] 
  (-3.5,-.75) -- (-2,.75) -- (3.5,.75) -- (2,-.75) -- cycle;
  \draw [thick,blue] (-2.,0) -- (-.5,0);
  \draw [thick,blue] (1.,0) -- (2.,0);
  \draw [thick,blue] (-2.,-2.25) -- (-.5,-2.25);
  \draw [thick,blue] (1.,-2.25) -- (2.,-2.25);
  \filldraw [blue] (-2,0) circle [radius = 2pt] node [above] {$a$};
  \filldraw [blue] (-.5,0) circle [radius = 2pt] node [above] {$b$};
  \filldraw [blue] (1,0) circle [radius = 2pt] node [above] {$c$};
  \filldraw [blue] (2,0) circle [radius = 2pt] node [above] {$d$};
  \draw [fill = gray, opacity = .4] (6, -2.5) -- (11, -2.5) -- (11.5, -1.125) -- (6.5, -1.125) -- cycle;
  \draw [thick] (6, -2.5) -- (7, 0.25) node [above left] {$\tau$} -- (12, 0.25) node [above right] {$1+\tau$} -- (11, -2.5) node [below right] {$1$} -- cycle;
  \draw [very thick, blue] (6, -2.5) -- (11, -2.5);
  \draw [very thick, blue] (6.5, -1.125) -- (11.5, -1.125);
  \filldraw [blue] (6.5, -1.125) circle [radius = 2pt] node [left] {$\frac{\tau}{2}$};
  \filldraw [blue] (9, -1.125) circle [radius = 2pt] node [above] {$\frac12 + \frac{\tau}{2}$};
  \filldraw [blue] (8.5, -2.5) circle [radius = 2pt] node [below] {$\frac12$};
  \filldraw [blue] (6, -2.5) circle [radius = 2pt] node [below left] {$0$};
  \draw [blue, ultra thick, -latex] (4.2, -1.125) -- (5.3, -1.125);
 \end{tikzpicture}
\end{align}
Written in terms of the variable $u$, the resolvent $\overline W(x) = \bar \omega(u)$ is then not only a meromorphic, but also an \textbf{elliptic function}\index{elliptic function} of $u$, that is a doubly periodic function such that 
\begin{equation}
\bar\omega(u+1)=\bar\omega(u+\tau)=\bar\omega(u) \ . 
\end{equation}
A meromorphic elliptic function is entirely characterized by its behaviour at its poles. 

The function $u(x)$ \eqref{eq:ux} can be rewritten in terms of standard special functions, provided the spectral edges $(a,b,c,d)$ have special positions in the complex plane. 
Such special positions can always be reached from generic positions via a M\"obius transformation, 
at the expense of changing the potential $V'(x)$ from a polynomial to a rational function of $x$. 
A first type of special positions is $(a,b,c,d) = (-\frac{1}{k}, -1, 1, \frac{1}{k})$, in which case $x$ and $\sigma(x)$ can be written in terms of the \textbf{Jacobi elliptic functions}\index{elliptic function!Jacobi---},
\begin{equation}
 x = \operatorname{sn}(u,k) \quad , \quad \sqrt{\sigma(x)} = \operatorname{cn}(u,k) \operatorname{dn}(u,k)\ .
\end{equation}
Another type of special positions is $d=\infty$ with $a+b+c=0$, in which
case $x$ and $\sigma(x)=(x-a)(x-b)(x-c) $ can be written in terms of the \textbf{Weierstrass
elliptic $\wp$-function}\index{elliptic function!Weierstrass---},
\begin{equation}
 x = \wp(u) \quad , \quad \sqrt{\sigma(x)}=\frac{1}{2}\wp'(u)\ .
\end{equation}
This elliptic function obeys the differential equation
\begin{equation}
 (\wp'(u))^2 = 4\sigma(\wp(u)) = 4 \wp(u)^3 - g_2 \wp(u) -g_3\ ,
\end{equation}
where the parameters $g_2,g_3$ are given in terms of the roots $a,b,c$ of $\sigma(x)$ by
\begin{equation}
 g_2 = -4 (ab+ac+bc) \quad ,\quad g_3 = 4 abc\ .
\end{equation}

\subsubsection{Higher genus}

Poincar\'e's uniformization theorem asserts that a Riemann surface $\Sigma$ of genus ${g}\geq 2$ can be holomorphically mapped to a right angles polygon with $4 g$ sides in the Poincar\'e hyperbolic disc $\mathcal{H}$, whose sides are glued pairwise. 
That polygon is a fundamental domain of the quotient of $\mathcal{H}$ by a discrete group $\Gamma$ of hyperbolic isometries, so that 
$\Sigma \sim \frac{\mathcal{H}}{\Gamma}$. 
Meromorphic functions on $\frac{\mathcal{H}}{\Gamma}$ can be written in terms  of Theta functions -- higher genus generalizations of elliptic functions. 
(See \cite{Fay:1973} for more details.)
For example, a Riemann surface of genus $2$ can be mapped to an octagon in the Poincar\'e disc:
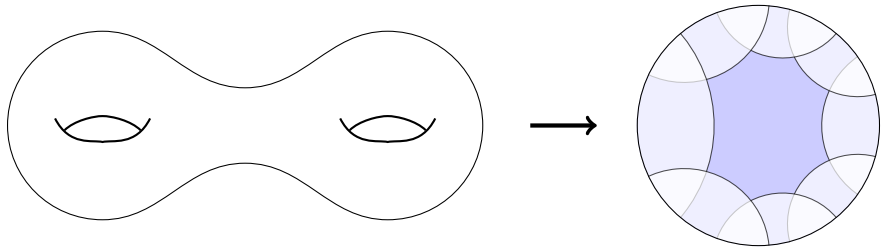
\begin{align}
\begin{tikzpicture}[scale = .5,baseline=(current bounding box.center)]
\begin{scope}[scale = .8] % the Poincare disc
  \fill [blue,opacity=0.2] (0,0) circle [radius = 4] ;
\clip (0,0) circle [radius = 4] ;
\newcommand {\geod}[2]{   % 1st argument = angle for the center, 2nd argument = angle of the arc
\begin{scope}[rotate=#1]
\newcommand{\ex}{{4*cos(#2)}}
\newcommand{\ey}{{4*sin(#2)}}
\newcommand{\er}{4*tan(#2)}
\filldraw[fill = white, opacity = .7] (\ex,\ey) arc ({90+#2}:{270+#2}:{\er});
\end{scope}
}
\geod{0}{35}
\geod{45}{30}
\geod{80}{30}
\geod{120}{35}
\geod{180}{50}
\geod{-45}{30}
\geod{-80}{30}
\geod{-120}{35}
\draw [thick] (0,0) circle [radius = 4] ;
\end{scope}
\begin{scope}[shift={(-13.5,0)},scale=2.5] % the surface
\draw (0,0.4) to [out=left,in=right] (-1.5,1) to [out=left,in=up] (-2.5,0) to [out=down,in=left] (-1.5,-1) to [out=right,in=left] (0,-0.4) to [out=right,in=left] (1.5,-1) to [out=right,in=down] (2.5,0) to [out=up,in=right] (1.5,1) to [out=left,in=right] (0,0.4) --cycle;
\foreach \x in {-1.5,1.5}{
\begin{scope}[shift={(\x,3.2)},scale=0.5]
  \draw [thick] (-1,-6.25) .. controls ++(.3,-.6) and (-.2,-6.7) .. (0,-6.75) .. controls (.2,-6.7) and (.7,-6.85) .. (1,-6.25);
  \draw [thick] (-.8,-6.5) .. controls (-.6,-6.3) and (-.2,-6.2) .. (0,-6.2) .. controls (.2,-6.2) and (.6,-6.3) .. (.8,-6.5);
\end{scope}
}
\draw[ultra thick, -latex] (3, 0) -- (3.7, 0);
\end{scope}
\end{tikzpicture}
\end{align}

\subsection{The two-point function}\label{sec:two-point_func}

In the large $N$ limit, the disconnected two-point function $\rho_2(x_1,x_2) + N^2 \rho(x_1)\rho(x_2)$ is dominated by the trivial term $N^2 \bar{\rho}(x_1)\bar{\rho}(x_2)$. 
The large $N$ limit (when it exists) of the connected two-point function $\bar{\rho}_2(x_1,x_2) = \lim_{N\to\infty}\rho_2(x_1,x_2)$ thus appears as a subleading contribution.
While we do not a priori expect the saddle-point approximation to be useful for computing such subleading terms, we can still obtain some information.

Let us renounce the large $N$ limit for a moment, and come back to our original eigenvalue integral \eqref{eq:z}.
Using the functional derivative $\frac{\delta}{\delta V(x)}$ with respect to the potential $V(x)$, which in particular obeys $\frac{\delta V(x')}{\delta V(x)} = \delta(x-x')$, we can write the eigenvalue density as 
\begin{align}\label{eq:rhodetalogZ}
 \rho(x) = -\frac{1}{N^2}\frac{\delta \log \mathcal{Z}}{\delta V(x)}\ .
\end{align}
Taking an extra functional derivative, we obtain the connected two-point function
\begin{align}\label{eq:drhodVrho2}
 \frac{\delta \rho(x)}{\delta V(x')} = - \rho_2(x,x')\ .
\end{align}
We now want to take the large $N$ limit of this exact relation. 
The problem is that in general limits do not commute with derivatives: for instance $f(x)=\frac{1}{N}\sin Nx=O(\frac{1}{N})$, while $f'(x)=\cos Nx = O(1)$. 
A similar phenomenon happens with $\rho_2(x,x')$, whose limit can receive contributions from  subleading terms of $\rho(x)$, which have oscillations of frequency $O(N)$.
As we will explain in \autoref{chap:LoopEq}, oscillations occur only for spectral curves of genus $g\geq 1$. In the 1-cut case there are no oscillations, so  that the limit and the derivative commute.

\subsubsection{Riemann--Hilbert equation}

In the heuristic spirit of the  Coulomb gas method, let us do as if we could  exchange the large $N$ limit and the functional derivative.
Taking the large $N$ limit of Eq.~\eqref{eq:drhodVrho2}, we obtain
\begin{align}
 \int_{\operatorname{supp}\bar\rho} \frac{dx''}{x'-x''} \frac{\delta \overline{W}(x)}{\delta V(x'')} = - \overline{W}_2(x, x')\ ,
\end{align}
where $\overline{W}_2$ is the double Stieltjes transform of $\bar{\rho}_2$,
\begin{equation}\label{eq:W2Stieljes}
\overline W_2(x,x') = \int_{\operatorname{supp}\bar\rho}\int_{\operatorname{supp}\bar\rho} \frac{d\lambda}{x-\lambda}\frac{d\lambda'}{x'-\lambda'}\bar\rho_2(\lambda,\lambda')\ .
\end{equation}
We can now obtain a Riemann--Hilbert equation for $\overline{W}_2$ by applying the functional derivative to the Riemann--Hilbert equation \eqref{eq:saddleder} for $\overline{W}$,
\begin{align}
\boxed{
 \overline{W}_2(x+i0, x') +\overline{W}_2(x-i0, x') = -\frac{1}{(x-x')^2} } \ .
 \label{eq:rhot}
\end{align}
This means that $\overline{W}_2(x,x') + \frac{1}{2(x-x')^2}$ changes sign when $x$ goes through a cut, and therefore behaves just like $\sqrt{\sigma(x)}$ in this respect. 
This implies
\begin{align}
 \overline{W}_2(x,x') = - \frac{1}{2(x-x')^2}\left(1 - \frac{Q_2(x, x')}{\sqrt{\sigma(x)} \sqrt{\sigma(x')}}\right)\ ,
\label{saddle2pt}
\end{align}
where $Q_2(x, x')=Q_2(x',x)$ is meromorphic in its two arguments. 
Let us analyze how $Q_2(x, x')$ behaves near its possible singularities:
\begin{itemize}
 \item Near $x=x'$ in the same sheet, $\overline{W}_2(x,x')$ is manifestly regular by Eq.~\eqref{eq:W2Stieljes}, which implies $\underset{x\to x'}{\lim} Q_2(x,x')= \sigma(x)$ as well as $\underset{x\to x'}{\lim} \partial_x Q_2(x,x')= \frac{1}{2}\sigma'(x)$.
 The neighbourhood of $x=x'$ in different sheets is then obtained by  $\sqrt{\sigma(x)}\to -\sqrt{\sigma(x)}$, and 
 $\overline W_2(x,x')$ has a double pole there.
\item $x\mapsto \overline W_2(x,x')$ is analytic outside the cut, and therefore $Q_2$ can have no other singularity than $x=\infty $ and the spectral edges. 
\item Near a spectral edge $x=a$, the worst possible singularity of $\overline{W}(x)$ is $O(\sqrt{x-a})$.
Applying $\frac{\delta}{\delta V(x')} \sim \frac{\delta a}{\delta V(x')}\ \frac{\partial}{\partial a}$, we obtain at worst an $O(\frac{1}{\sqrt{x-a}})$ singularity for $\overline{W}_2(x,x')$. 
This shows that $Q_2(x, x')$ must be regular at $x= a$. 
So $Q_2(x, x')$ can have a pole only at $x=\infty$, and is a symmetric polynomial of its two variables.
\item Near $x=\infty$, we must have by definition $\overline{W}_2(x, x') \underset{x\to \infty}{=} O(\frac{1}{x^2})$, which implies $\deg_x Q_2 \leq \frac12 \deg \sigma = s$. 
Using again the behaviour near $x=x'$, we actually find $\deg_x Q_2 =s$. 
\end{itemize}

\subsubsection{One-cut case}

If  $\operatorname{supp}\bar\rho= (a, b)$, our constraints fully determine $Q_2(x,x')$, and we find the only possible symmetric polynomial of degree $s=1$ such that $Q_2(x,x)=(x-a)(x-b)$:
\begin{equation}
Q_2(x,x') = x x' - \frac{a+b}{2}(x+x') + ab\ .
\end{equation}
Using the Joukowsky variables $z_1,z_2$ which correspond to $x_1, x_2$ \eqref{eq:xz}, this leads to 
$
\overline W_2(x_1,x_2)
= \frac{d}{dx_1}  \frac{d}{dx_2} \log\left(\frac{z_1-z_2}{x_1-x_2}\right)
$, equivalently 
\begin{equation}\label{eq:W2barB}
\overline W_2(x_1,x_2)dx_1dx_2 = \frac{dz_1dz_2}{(z_1-z_2)^2} - \frac{dx_1dx_2}{(x_1-x_2)^2} 
\ .
\end{equation}
We can then compute the connected two-point function $\bar{\rho}_2$ as the discontinuity of $\overline{W}_2$ on the cut. 
The term $- \frac{dx_1dx_2}{(x_1-x_2)^2}$ is a rational function of $x_1,x_2$, and does not contribute to $\bar \rho_2$, although it is essential for $\overline W_2(x_1,x_2)$ to be regular at $x_1=x_2$.
All the useful information is in the term
\begin{equation}\label{def:BRsphere}
\boxed{
B(z_1,z_2) = \frac{dz_1dz_2}{(z_1-z_2)^2} = d_{1}d_{2}\log (z_1-z_2) } \ .
\end{equation}
This symmetric bi-differential form is called the \textbf{fundamental second kind differential}\index{fundamental second kind differential!---of the Riemann sphere} of the Riemann sphere.
This is a meromorphic version of the Green's function $\log |z_1-z_2|$ of the Laplacian $4\partial\bar{\partial}$.
We can actually rewrite the two-point function in terms of the fundamental second kind differential as 
\begin{align}\label{eq:W2barBJ}
\boxed{
 \overline{W}_2(x_1,x_2)dx_1dx_2 = - B\left(z_1, \frac{1}{z_2}\right) } \ ,
\end{align}
using the identity
\begin{align}
 \frac{dxdx'}{(x-x')^2} = B(z, z') + B\left(z, \frac{1}{z'}\right)\ ,
\end{align}
which itself follows from the Joukowsky map. 

\subsubsection{Case of a general spectral curve}

Given an arbitrary Riemann surface $\Sigma$,
a \textbf{fundamental second kind differential}\index{fundamental second kind differential} is a meromorphic symmetric bilinear differential form $B$, whose only singularity is a double pole on the diagonal, with bi-residue one:
\begin{equation}
B(z_1,z_2) \mathop{\sim}_{z_1\to z_2} \frac{dz_1  dz_2}{(z_1-z_2)^2} + O(1)\ .
\end{equation}
We have $B:\Sigma\times\Sigma \to K_\Sigma \overset{\text{sym}}{\otimes} K_\Sigma(2\operatorname{diag})$, where $K_\Sigma$ is the canonical bundle of $\Sigma$, i.e. the bundle of locally holomorphic differential forms on $\Sigma$.
It has the property that for any locally analytic function $f:\Sigma\to \mathbb{C}$, we have
\begin{equation}
df(z) = \underset{z'\to z}{\operatorname{Res}} \, B(z,z') f(z')\ .
\end{equation}
The fundamental second kind differential is unique up to holomorphic terms and a choice of normalization. On a compact Riemann surface of genus $g$, the affine space of fundamental second kind differentials has the dimension $\frac12 g(g+1)$ \cite{Fay:1973}. (On a non-compact Riemann surface, it would be infinite-dimensional.) There is a unique differential that is normalized on $A$-cycles, which means $\forall z_2$ $\int_{A_i} B(.,z_2)=0$:
\begin{align}
B(u_1,u_2) =   d_1 d_2 \log{\Theta\left(\mathfrak a(u_1)-\mathfrak a(u_2) + \chi \right) }\ ,
\end{align}
where $\mathfrak a:\Sigma \to \mathbb C^{g}$ is the Abel map, $\chi \in \mathbb C^{g}$ is a non-singular odd half-integer characteristics and $\Theta:\mathbb C^{g} \to \mathbb C$ is the Riemann Theta function.

The Riemann sphere has a unique fundamental second kind differential \eqref{def:BRsphere}.
On the torus, with its canonical coordinate $u\in \frac{\mathbb C}{\mathbb Z+\tau\mathbb Z}$, related to $x$ by the map $x\mapsto u(x)$ \eqref{eq:ux}, the Abel map is $\mathfrak a(u)=u$, and the unique odd half-integer characteristic in genus $1$ is $\chi = \frac12(1+\tau)\in\mathbb{C}$.
Alternatively, this torus fundamental second kind differential can be expressed in terms of the Weierstrass elliptic function $\wp$ by
\begin{equation}
B(u_1,u_2) \ \underset{g=1}{=}\ \left( \wp(u_1-u_2;\tau) + G_2(\tau) \right)  du_1  du_2\ ,
\end{equation}
where $G_2$ is the second Eisenstein series.

For general algebraic curves, $B$ can be written as a rational function of $(x_1,y_1),(x_2,y_2)$, see \cite{Eynard:2018gor}. The spectral curves that we constructed from 2-sheeted covers of the complex plane are called curves of degree 2: for such curves, we have
\begin{multline}
 B(x_1,x_2)  =  dx_1 dx_2 \frac{\left[\sqrt{\sigma(x_1)}+\sqrt{\sigma(x_2)}\right]^2 - \left[\left(\sqrt{\sigma(x_1)}\right)_+-\left(\sqrt{\sigma(x_2)}\right)_+\right]^2}{4\sqrt{\sigma(x_1)\sigma(x_2)} (x_1-x_2)^2}
 \\
 +dx_1 dx_2\frac{S(x_1,x_2) }{4\sqrt{\sigma(x_1)\sigma(x_2)} }\ ,
\end{multline}
%\begin{multline}
% B(x_1,x_2)  =  \frac{dx_1 dx_2}{4\sqrt{\sigma(x_1)\sigma(x_2)} } \frac{2 \sqrt{\sigma(x_1)\sigma(x_2)} + 2 A(x_1)A(x_2)+B(x_1)+B(x_2)}{(x_1-x_2)^2} \\
%+ \frac{dx_1 dx_2}{4\sqrt{\sigma(x_1)} \sqrt{\sigma(x_2)} } S(x_1,x_2)
%\end{multline}
%where $A(x) = \left(\sqrt{\sigma(x)}\right)_+$  and $B(x) = \sigma(x)-A(x)^2$,
where $S(x_1,x_2)=S(x_2,x_1)$ is a symmetric polynomial of degree $\leq s-2$ in each variable, fixed by normalizing $B$ on $A$-cycles.

In the multi-cut case, the large $N$ two-point function is not given solely by the fundamental second kind differential, but may also involve oscillatory terms. (See \autoref{sec:non-perturbative}.)
The resulting expression is more complicated than in the 1-cut case, but it is still universal.
If there are no oscillatory terms, the two-point function is
\begin{align}
\overline W_2(x_1,x_2)dx_1dx_2 = B(u_1,u_2) - \frac{dx_1dx_2}{(x_1-x_2)^2}\ .
\end{align}

\subsection{The free energy}
\label{sec:F0}

Let us admit that the limit $F_0=-\underset{N\to\infty}{\lim}\frac{1}{N^2}\log \mathcal{Z}$ of the free energy can be calculated using the saddle-point approximation. 
Then $F_0$ is the value of the functional action \eqref{def:Srho} at the equilibrium density,
\begin{equation}\label{eq:F0Sbarrho}
F_0 = S[\bar\rho] = \int_{\operatorname{supp}\bar\rho} dx\, {\bar\rho}(x)V(x) - \int_{(\operatorname{supp}\bar\rho)^2} dxdx'\, {\bar\rho}(x){\bar\rho}(x')\log|x-x'|\ .
\end{equation}
This implies 
$
\frac{\delta F_0}{\delta V(x)} = \bar\rho(x)
$,
in agreement with Eq.~\eqref{eq:rhodetalogZ}.
For $V(x) = \sum_k \frac{t_k}{k} x^k$, this amounts to
\begin{align}
 k  \frac{\partial F_0}{\partial t_k} = \int_{\operatorname{supp}\bar\rho} x^k\bar \rho(x)dx 
 = -\underset{x=\infty}{\operatorname{Res}}\, x^k\overline W(x) dx = \underset{z=0}{\operatorname{Res}}\, x(z)^k\overline W(x(z))dx(z)\ .
 \label{eq:kpfz}
\end{align}
This relation is the large $N$ limit of the relation between the free energy and the moments $k\frac{\partial F}{\partial t_k} = N \left<  \operatorname{Tr} M^k \right>$, and can be rewritten as 
\begin{equation}
\overline W(x) = \frac{1}{x}+\sum_{k=1}^\infty \frac{k}{x^{k+1}}  \frac{\partial F_0}{\partial t_k} \ .
\label{eq:dkfz}
\end{equation}
This allows us to determine $F_0$ once we know the resolvent $\overline W(x)$.

\subsubsection{One-cut case}

Let us write a solution $F_0$ of Eq.~\eqref{eq:kpfz} as an integral of the one-form $\overline W(x)dx = \bar\omega(z)x'(z)dz$, where $x(z)$ is the Joukovsky map. The behaviour of this one-form at its poles $z=0,\infty$ is determined by Eqs. \eqref{eq:v0v1z} and \eqref{eq:RHJoukowsky}, allowing us to define the regularized integral
\begin{multline}
 \int^{\mathrm{reg}}_{z\in(0,\infty)} \overline W(x)dx
 = V(x(p)) -2\log x(p)
 \\
 +\int_{z\in(0,p)} \left( \overline W(x)dx -dV(x) + d\log x\right)
 + \int_{z\in(p,\infty)} \left(\overline W(x)dx -d\log x\right)
 \ ,
\end{multline}
which does not depend on the regulator $p\in \mathbb{C}^*$.
Then our solution for the free energy is
\begin{align}\label{eq:F0resom}
 \boxed{ F_0 = \frac12 \left( \int^{\mathrm{reg}}_{z\in(0,\infty)} \overline W(x)dx + \underset{z=0}{\operatorname{Res}}\, V(x)\overline W(x)dx \right) }\ .
\end{align}
In terms of the coefficients $(v_1,\dots, v_d)$ of the decomposition \eqref{eq:v0v1z} of $\bar \omega(z)$ in powers of $z$, this amounts to
\begin{equation}
F_0 = -\log\gamma -\frac{\gamma^2}{2} \sum_{j=1}^{\infty} \frac{1}{j}(v_{j+1}-v_{j-1})^2
\ ,
\end{equation}
where by convention $v_0 = v_{j>d}=0$.

Let us prove that our expression for $F_0$ does solve Eq.~\eqref{eq:dkfz}.
We denote fixed $x$ derivatives of the resolvent as $\partial_{t_k} \overline W(x)$. 
These differ from fixed $z$ derivatives, because the Joukowsky map $x(z)$ depends on $t_k$. 
By taking a fixed $z$ derivative of Eq.~\eqref{eq:v0v1z}, we learn that $\partial_{t_k}\overline W(x(z)) x'(z)$ is a polynomial of $\frac{1}{z}$.
We denote this as $\left(\partial_{t_k}\overline W(x) dx\right)_+=0$, where for a Laurent series $f(z) = \sum_{j\in\mathbb{Z}} c_jz^j$ we write 
\begin{align}
 f(z) = f(z)_-+f(z)_0 + f(z)_+ = \sum_{j<0}c_jz^j + c_0 +\sum_{j>0}c_jz^j\ .
\end{align}
Moreover, the asymptotic behaviour \eqref{eq:omegainf} of $\overline W(x)$ implies that $\partial_{t_k}\overline W(x(z)) x'(z) \underset{z\to\infty}{=} O(\frac{1}{z^2})$. This implies $ \left(\partial_{t_k}\overline W(x(\frac{1}{z})) x'(z) \right)_-=0$.
Now the Riemann--Hilbert equation \eqref{eq:saddleder} leads to
\begin{align}
 \partial_{t_k}\overline W(x+i0) + \partial_{t_k}\overline W(x-i0) = x^{k-1}\ .
 % OK
\end{align}
Multiplying with $dx=x'(z)dz$ and taking the negative part, we obtain
\begin{align}
 \partial_{t_k}\overline W(x)dx = \frac{1}{k} d(x^k)_-\ .
\end{align}
This is just what we need in order to compute the $t_k$-derivative of our expression \eqref{eq:F0resom} for $F_0$. 
In the $t_k$-derivative of the regularized integral $\int^{\mathrm{reg}}_{z\in(0,\infty)} \overline W(x)dx $, the only nonvanishing contribution is $\left(\partial_{t_k} V(x)-\frac{1}{k}(x^k)_-\right)\big|_{z=0}$, i.e.
\begin{align}
 k\partial_{t_k} \int^{\mathrm{reg}}_{z\in(0,\infty)} \overline W(x)dx = (x^k)_0\ .
 %OK 
\end{align}
Then let us compute the $t_k$-derivative of the remaining term of $F_0$:
\begin{align}
 k\partial_{t_k} \underset{z=0}{\operatorname{Res}}\, V(x)\overline W(x)dx 
 &= \underset{z=0}{\operatorname{Res}} \left( x^k\overline W(x)dx + V(x)d(x^k)_- \right)\ ,
 \cr
 &= \underset{z=0}{\operatorname{Res}} \left(x^k\overline W(x)-(x^k)_- V'(x)\right)dx\ .
\end{align}
We now use the Riemann--Hilbert equation \eqref{eq:RHJoukowsky} in the form $V'(x(z)) = \overline W(x(z))+\overline W(x(\frac{1}{z}))$. Using $x(z)=x(\frac{1}{z})$, we compute the term
\begin{align}
 \underset{z=0}{\operatorname{Res}}\, (x(z)^k)_- \overline W(x(\tfrac{1}{z})) dx(z) 
 = \underset{z=\infty}{\operatorname{Res}}\, (x^k)_+ \overline W(x) dx 
 = - \underset{z=0}{\operatorname{Res}}\, (x^k)_+ \overline W(x) dx\ .
 %OK
\end{align}
Using $\overline W(x)_+=0$ and $ \underset{z=0}{\operatorname{Res}}\, \overline W(x)dx = 1$, this leads to 
\begin{align}
 k\partial_{t_k} \underset{z=0}{\operatorname{Res}}\, V(x)\overline W(x)dx 
 &=  \underset{z=0}{\operatorname{Res}} \left(x^k\overline W(x) - (x^k)_- \overline W(x) + (x^k)_+ \overline W(x)\right)dx\ ,
 \cr
 &= - (x^k)_0 + 2\, \underset{z=0}{\operatorname{Res}}\, x^k\overline W(x)dx \ .
 %OK
\end{align}
This allows us to check that the expression 
\eqref{eq:F0resom} for $F_0$ provides a solution of Eq.~\eqref{eq:kpfz}.

%OK

\subsubsection{Arbitrary spectral curve}

Generalizing \eqref{eq:F0resom}, we will now define $F_0$ for an arbitrary spectral curve $(\Sigma, x,y)$.
For $\alpha$ a pole of the one-form $ydx$, let $d_\alpha $ be the order of $x$ at $\alpha$, and $\xi_\alpha(z)$ the canonical local coordinate that vanishes at $z=\alpha$, such that  
\begin{align}
 x(z) \underset{z\to\alpha}{\sim}  \left\{\begin{array}{ll} x(\alpha)+\xi_\alpha(z)^{d_\alpha}\ \ & (d_\alpha> 0)\ , 
                 \\  \xi_\alpha(z)^{d_\alpha} & (d_\alpha<0)\ .
                \end{array}\right.
\end{align}
We define 
\begin{align}
 t_\alpha = \underset{z=\alpha}{\operatorname{Res}}\, y(z)dx(z)\ ,
\end{align}
and 
\begin{align}
 V_\alpha(z) = \underset{z'=\alpha}{\operatorname{Res}}\, \log\left( 1-\frac{\xi_\alpha(z')}{\xi_\alpha(z)} \right) y(z')dx(z') 
 = -\sum_{k=1}^{s_\alpha} \frac{t_{\alpha,k}}{k}  \xi_\alpha(z)^{-k}\ .
\end{align}
In the 1-cut case, the poles of $ydx$ are $\alpha=0,\infty$, and we have:
\begin{align}
\renewcommand{\arraystretch}{1.3}
 \begin{array}{ccccl}
  \alpha & d_\alpha & \xi_\alpha & t_\alpha & V_\alpha(z) 
  \\
  \hline
  0 & -1 & \frac{1}{x(z)} & 1 & V(x(z))- V(0)
  \\
  \infty & -1 & \frac{1}{x(z)} & -1 & 0
 \end{array}
\end{align}
Near one of its poles, the one-form $ydx$ behaves as 
\begin{align}
 y(z)dx(z) \underset{z\to\alpha}{=} dV_\alpha(z)  + t_\alpha d\log \xi_\alpha(z) + \text{holomorphic}\ ,
\end{align}
and this allows us to define the regularized integral
\begin{align}
 \int^{\mathrm{reg}}_{(p,\alpha)} ydx = -V_\alpha(p)-t_\alpha \log\xi_\alpha(p) 
 + \int_{(p,\alpha)} \left( ydx - dV_\alpha-t_\alpha d\log\xi_\alpha\right)\ .
\end{align}
The generalization of the formula \eqref{eq:F0resom} for $F_0$ in the 1-cut case is
\begin{equation}
\boxed{
F_0 = \frac12 \sum_\alpha  \left(t_\alpha \int^{\mathrm{reg}}_{(\alpha,p)} ydx + \underset{\alpha}{\operatorname{Res}}\, V_{\alpha}  y dx  
- \sum_{i=1}^{g} \frac{1}{2\pi i}  \oint_{A_i} ydx \oint_{B_i} ydx
\right)
}
\ 
\label{eq:F0gen}
%OK
\end{equation}
which is independent of the choice of $p$.
We admit that this formula computes $F_0=-\lim_{N\to\infty} \frac{1}{N^2} \log \mathcal Z$ for 
matrix models with several cuts and/or rational potentials, and also in multi-matrix models.
The proof is done like for the genus zero case, by showing that this formula satisfies \eqref{eq:kpfz}.

\section{Examples and generalizations}

\subsection{Rational potentials}

For simplicity, we have been restricting to the case where $V(x)$ is polynomial. However, most results extend to the case where $V'(x)\in \mathbb C(x)$ is a rational function of $x$.
This allows $V(x)$ to also have logarithmic singularities:
\begin{align}
V'(x) = \sum_{p=\text{finite poles}} \sum_{k=0}^{d_p-1} t_{p,k} (x-p)^{-k-1} + \sum_{k=1}^{d_\infty+1} t_{\infty,k} x^{k-1}.
\end{align}
\begin{align}
V(x) = \sum_{p=\text{finite poles}} \left( t_{p,0} \log{(x-p)} - \sum_{k=1}^{d_p-1} \frac{t_{p,k}}{k} (x-p)^{-k} \right) + \sum_{k=1}^{d_\infty+1} \frac{t_{\infty,k}}{k} x^{k}.
\end{align}

\subsubsection{Differences with the case of polynomial potentials}

\begin{itemize}

\item \textbf{Homology space:} By definition, the homology space $H_1(e^{-NV(x)}dx)$ is generated by paths $\gamma$ where the measure $e^{-NV(x)}dx$ is integrable, and does not generate boundary terms under integration by parts. Such paths can only end at poles of $V'$, in directions where $e^{-NV(x)}$ vanishes. This leads to
\begin{align}
 \dim H_1(e^{-NV(x)}dx)= \deg V' = d_\infty+\sum_p d_p\ .
\end{align}
Let us discuss in more detail how $\gamma$ behaves near a pole $p$ of $V'$, depending on its order $d_p$:
\begin{itemize}
\item If $d_p>1$, then $\gamma$ must approach $p$ in a sector where $\arg(x-p) \in (\theta_{2n-\frac12},\theta_{2n+\frac12})$ with $n\in\mathbb{Z}$, where
\begin{align}\label{defthetaanglescaserational}
\theta_n=  \frac{1}{d_p-1}\left(\pi(n+1)+\arg t_{p,d_p-1}\right)\ .
\end{align}
\item If $d_p=1$, and $-Nt_{p,0}\in\mathbb{N}$, then $\gamma$ may end at $p$ in any direction.
\item If $d_p=1$, and $-Nt_{p,0}\notin \mathbb{N}$, then $\gamma$ may not end at $p$, it must go around $p$. If $Nt_{p,0}\in\mathbb{N}$, then $\gamma$ may be a small circle around $p$. If $-Nt_{p,0}\notin\mathbb{Z}$, then $e^{-NV(x)}$ is not analytic around $p$, and there is a cut ending at $p$, which $\gamma$ must avoid.
\end{itemize}

\item \textbf{Functional action:}
At fixed $\gamma$, the functional $\Re S[\rho]$ is still a good rate functional and still has a unique minimum.
Maximizing with respect to $\gamma$ no longer provides a unique $\gamma$, but the non-uniqueness is not very relevant. We will see this in the example of $V(x)=\log{x}$.

\item \textbf{Resolvent:}
The function $\overline{P}(x)$ defined by Eq. \eqref{eq:pvo} is no longer polynomial. it is now a rational function, whose poles can be only those of $V'(x)$.
At $\infty$ it behaves as $P(x)\sim t_{\infty,d_\infty+1} x^{d_\infty-1}$, and at a finite pole $p$ of $V'(x)$, it has a pole of order at most $d_p$. The resolvent $\overline{W}(x)$ is still the Stieltjes transform of a density $\bar{\rho}(x)$. It still has an expression of the type of Eq. \eqref{eq:rsms}, where the function $M(x)$ is now rational. At $\infty$ it must behave as $t_{\infty,d_\infty+1} x^{d_\infty-1-2s}$ where $2s=\deg\sigma$ is the number of branch points. At a finite pole $p$ of $V'(x)$, it can have a pole of order at most $d_p$, such that $M(x)\sqrt{\sigma(x)}\sim V'(x)$.

The support of $\bar\rho(x)$ is made of compact arcs of the spectral network, i.e. vertical trajectories of the 1-form $M(x)\sqrt{\sigma(x)}dx$.

\item \textbf{Spectral network:}
Arcs of the spectral network may go to $\infty$ or to poles of $V'$ of order $d_p>1$, at angles $\theta_{n}$ \eqref{defthetaanglescaserational} for some $n\in \mathbb Z$.
Near a pole $p$ of $V'$ of order $d_p=1$, we have $g_m \sim t_{p,0} \log{(x-x_p)}$, thus the spectral network can't approach the pole if $t_{p,0}$ is real, it can approach it as a straight line (at any angle) if $t_{p,0}$ is imaginary, and otherwise as a logarithmic spiral $\arg (x-p) = \theta_0+ \frac{\Re t_{p,0}}{\Im t_{p,0}} \log{|x-x_p|}$.
However, in almost all applications of random matrices the poles of $V'$ have real residues, so the spectral network can't approach the simple poles.

\item \textbf{Sectors:} The image $g_m(A)\subset \mathbb{C}$ of a sector $A$ under $g_m$ (with $dg_m=M\sqrt{\sigma}$) is still bounded by vertical lines, but now the $g_m$ may be defined modulo an imaginary constant. In addition to lands, seas and beaches, we can therefore one more types of sectors: a half-cylinder, i.e. a half-plane modulo an imaginary constant. This occurs around a simple pole with a real residue $t_{p,0}\in \mathbb R$. Then the perimeter is $2\pi |t_{p,0}|$.

We could in principle also have a cylinder, i.e. a strip modulo an imaginary constant. But one can prove that if $\overline{P}$ is the extremum of $\Re S$, then there is actually no cylinder.

\item \textbf{Support of the equilibrium density:}
This support is made of compact arcs of the spectral network, not ending at the poles. The arcs that can occur are:
\begin{itemize}
\item Open vertical arcs going from branch point to branch point, such that $\Re g_m$ is decreasing on both sides of the arc.
\item Closed vertical arcs winding around a half-cylinder, such that $\Re g_m$ is decreasing on both sides of the arc. Such arcs are compact even if they don't start from a branch point.
\end{itemize}
\end{itemize}

\subsubsection{Example of the logarithmic potential $V(x) = \log x$}

This corresponds to the ensemble $U_N$ of unitary matrices with the Haar measure. (See Eq.~\eqref{eq:circular_Haar}.)
The number of independent integration contours is $\deg V'=1$.
Up to homotopy, the  unique contour is a loop around $x=0$.
Since $V'$ has a simple pole at $x=0$, we must have $\overline{P}(x) = \frac{a}{x}$ for some coefficient $a$, and the asymptotic condition \eqref{eq:omegainf} actually implies $\overline P(x)=0$.
The resolvent is thus
\begin{equation}
\overline W(x) = \frac12 \left(\frac{1}{x} \pm  \sqrt{\frac{1}{x^2}} \right)
\end{equation}
where the sign must be $+$ near $x=\infty$ due to Eq.~\eqref{eq:omegainf}, and $-$ near $x=0$ by analyticity of $\overline W$ outside  $\operatorname{supp}\bar\rho$.
The equilibrium density is the discontinuity of $\overline W$,
\begin{align}
 \bar{\rho}(x) dx = \frac{1}{2\pi i} \frac{dx}{x} = \frac{d\theta}{2\pi} + \frac{1}{2\pi i}\frac{dr}{r} \quad \text{with} \quad x = re^{i\theta}\ .
\end{align}
Requiring the density to be real implies $\frac{dr}{r}=0$, so that $\operatorname{supp}\bar\rho$ must be a circle with constant radius, and $\bar{\rho}(x) dx =  \frac{d\theta}{2\pi}$ is the uniform measure on that circle.
For $m\in \operatorname{supp}\bar\rho$
we have $\Re g_m(x)= -\frac12 \left|\log \frac{|x|}{r} \right|<0 $, so that we have seas on both sides, in agreement with the density being positive.
\begin{align}
 \begin{tikzpicture}[scale = .6,baseline=(current bounding box.center)]
  \filldraw [fill=blue,draw=none,opacity=.3]
  (3,3) to (-3,3) to (-3,-3) to (3,-3) to (3,3);
  \filldraw [fill=white,draw=none,opacity=1.]  (0,0) circle [radius=1.];
  \filldraw [fill=blue,draw=none,opacity=.3]  (0,0) circle [radius=1.];
  \draw [very thick,red] (0,0) circle [radius=1.];
  \draw [thick] (-3,-3) rectangle (3,3);
  \node at (0,0) {$-$};
  \node at (2.3,0) {$-$};
 \end{tikzpicture}
\end{align}
We see that there is an ambiguity for choosing the radius $r$ of the circle, which is related to the non-uniqueness of the min-max problem of finding $\gamma$.
However, all choices of $r$ give $\gamma$ that are homotopic, and for all of them, $\overline{W}(x)$ is the same, and all the moments of $\bar \rho(x)$ are the same.
The canonical choice is $r=1$.

This shows that eigenvalues of a random unitary matrix with the Haar measure are uniformly distributed on the unit  circle. This result can of course be obtained as a direct consequence of the $U_N$ symmetry.

\subsection{Generalized measures} \label{sec:gvi}

After the rescaling $V\to NV$, the measure \eqref{eq:mon} of the $O(n)$ matrix model leads to the Riemann--Hilbert equation 
\begin{equation}
V'(x) = \overline W(x+i0)+\overline W(x-i0) + n \overline W(-x)   \ .
\end{equation}
This equation can in general be solved using Theta functions.
If moreover $n\in 2\cos\pi\mathbb{Q}$, then the solution $\overline W(x)$ is an algebraic function, see \autoref{exo:sheets} or \cite{Eynard:2006bs}.
More generally, let us consider a measure of the type of \eqref{eq:defRmeasures}, and the integral
\begin{equation}
\mathcal{Z} \propto \int_{\gamma^N} \prod_i e^{-N V(\lambda_i)}  d\lambda_i \,\prod_{i<j}  R(\lambda_i,\lambda_j)
\quad \text{with} \quad 
 R(\lambda,\tilde \lambda) = (\lambda-\tilde \lambda)^\beta \tilde R(\lambda,\tilde \lambda) \ ,
\label{eq:zr}
\end{equation}
where $\tilde R(\lambda,\lambda)$ is finite and nonzero.
The corresponding Riemann--Hilbert equation is 
\begin{equation}
V'(x) = \frac\beta{2} \left(\overline W(x+i0)+\overline W(x-i0) \right) + \frac{1}{2\pi i}\oint_\mathcal{C} \left(\partial_x\log \tilde R(x,x') \right)  \overline W(x')dx' \ ,
\end{equation}
where the contour $\mathcal{C}$ surrounds $\operatorname{supp}\bar\rho$, and can be deformed to surround the zeros and poles of $\tilde R$.

While this leads to more complicated spectral curves than our Hermitian matrix model, the universal results continue to hold. In particular, the solutions of the Riemann--Hilbert equation are still parametrized by periods of a spectral curve. 
The two-point function is still the fundamental second kind
differential of the spectral curve, plus possible oscillatory terms.

\subsubsection{An application to knot theory}

An interesting case is the matrix model with the measure
\eqref{HOMFLY}, whose correlation functions include the HOMFLY polynomials of
the torus knot $\mathfrak K_{P,Q}$.
This corresponds to $\beta=2$, together with
\begin{equation}
V(x) = \frac{1}{2PQ t} (\log x)^2 - \frac{1}{N}\log x \quad ,\quad 
\tilde R(x,x') = \frac{(x^P-x'^P)(x^Q-x'^Q)}{(x-x')^2}\ .
\end{equation}
In the saddle-point approximation, we neglect the $O(\frac{1}{N})$ term in the potential, and we obtain the Riemann--Hilbert equation
\begin{equation}
\frac{1}{PQt}\frac{\log x}{x} = \overline W(x+i0)+\overline W(x-i0) + \sum_{k=1}^{P-1} e^{2\pi i\frac{k}{P}}\overline W(e^{2\pi i\frac{k}{P}}x)+ \sum_{k=1}^{Q-1} e^{2\pi i\frac{k}{Q}}\overline W(e^{2\pi i\frac{k}{Q}}x) \ .
\label{eq:rhknot}
\end{equation}
This equation specifies how $\overline W$ behaves across a cut $\Gamma$. 
The appearance of $\overline W(e^{2\pi i\frac{k}{P}}x)$ and $\overline W(e^{2\pi i\frac{k}{Q}}x)$ then implies the existence of rotated cuts $e^{2\pi i (\frac{k}{P}+\frac{j}{Q})} \Gamma$.
Under analytic continuation across these cuts, the resolvent $\overline W(x)$ has an infinite orbit.
However, it is possible to build a function that has a finite orbit, and will therefore obey an algebraic equation,
\begin{equation}\label{eq:torusknotsfalg}
f(x) = \prod_{k=0}^{P-1} Y(e^{2\pi i\frac{k}{P}}x) \qquad \text{with} \qquad Y(x) = -x  e^{-\frac{P+Q}{PQ}(x\overline W(x)+\frac{t}{2})}\ .
\end{equation}
The function $Y(x)$ indeed obeys the Riemann--Hilbert equation 
\begin{equation}
Y(x+i0)Y(x-i0)\prod_{k=1}^{P-1} Y(e^{2\pi i\frac{k}{P}}x)\prod_{k=1}^{Q-1} Y(e^{2\pi i\frac{k}{Q}}x) =1\ ,
\end{equation}
which implies that the orbit of $f(x)$ is made of the $P+Q$ functions 
\begin{subequations}
\begin{align}
f_j(x) &= \prod_{k=0}^{P-1} Y(e^{2\pi i\frac{k}{P}+2\pi i\frac{ j}{Q}}x)\ , \qquad (j=0,\dots,Q-1)\ , 
\\
f_{Q+j}(x) &= \prod_{j=0}^{Q-1} \frac{1}{Y(e^{2\pi i\frac{k}{Q}+2\pi i\frac{ j}{P}}x)}\ , \qquad (j=0,\dots,P-1) \ .
\end{align}
\end{subequations}
Crossing the cut $e^{2\pi i (\frac{k}{P}+\frac{j}{Q})} \Gamma$ indeed permutes $f_j$ with $f_{Q+k}$, while $f_{j'\neq j}$ and $f_{Q+k'\neq Q+k}$ are smooth across the cut.  
It follows that $f(x)$ satisfies a polynomial equation of degree $P+Q$,
\begin{equation}
0 = \prod_{j=0}^{P+Q-1} (f-f_j(x))  = \sum_{j=0}^{P+Q}  e_j(x) f^{P+Q-j} \ ,
\end{equation}
whose coefficients $e_j(x)$ are smooth across the cuts, and invariant under rotations by multiples of $\frac{2\pi}{PQ}$. 
Therefore $e_j(x) \in \mathbb C[x^{PQ},x^{-PQ}]$. The asymptotic behaviours near $x=0$ and $x=\infty$ determine the $x$-dependence of $e_j(x)$, and the polynomial equation must be of the form
 \cite{Brini:2011}
\begin{equation}\label{eq:knotsfPQ}
0=f^{P+Q}+e^{\frac{P+Q}{2}t}\Big((-x)^{-PQ} (-f)^Q +(-x)^{PQ}(-f)^P\Big) + 1  + \sum_{k=1}^{P+Q-1} s_k f^k 
 \, ,
\end{equation}
where the $P+Q-1$ $x$-independent coefficients $s_k$ play the same role as the
coefficients of the polynomial $\overline{P}(x)$ in the case of our Hermitian matrix model.
In particular, the assumption that we have only one cut fixes all these coefficients, 
and leads to a rational parametrization of $f$ that generalizes the Joukowsky map,
\begin{equation}
x^{PQ}  = e^{t\frac{P+Q}{2}}z^{-P}\left(\frac{1-e^{-t} z}{1-z}\right)^Q\qquad , \qquad 
f = - z \frac{1-e^{-t} z}{1-z}\ .
\end{equation}
It is then possible to compute the equilibrium density, and to find the leading large $N$ asymptotic behaviour of the HOMFLY polynomial of the torus knot $\mathfrak K_{P,Q}$  \cite{Brini:2011}.

\subsubsection{Sheet-changing operators}

The notion of the orbit of the resolvent $\overline W(x)$ under analytic continuation across the cuts, can be formalized in terms of sheet-changing operators. In the case of the Riemann--Hilbert equation \eqref{eq:rhknot} for a cut $\Gamma$, the relevant operator is 
\begin{equation}
( T_\Gamma\cdot w )(x) = -w(x) - \sum_{k=1}^{P-1} e^{2\pi i\frac{k}{P}} w(e^{2\pi i\frac{k}{P}}x)- \sum_{k=1}^{Q-1} e^{2\pi i\frac{k}{Q}} w(e^{2\pi i\frac{k}{Q}}x)\ .
\end{equation}
In more general cases, sheet-changing operators can take the form 
\begin{equation}
( T_\Gamma\cdot w )(x) = - w(x) - \sum_{s\in G} a(s) w(s(x))\ .
\end{equation}
In our knot example, the group $G$ was made of rotations by multiples of $\frac{2\pi}{PQ}$, and the coefficients $a(s)$ were either zero or roots of unity. 
Sheet-changing operators associated to different cuts form a group. 
A function whose orbit under this group has $m\geq 2$ elements is multivalued on $\overline{\mathbb{C}}$, and becomes meromorphic on an $m$-sheeted cover of $\overline{\mathbb{C}}$.
Such a function therefore obeys an algebraic equation of degree  $m$.

The problem of finding algebraic solutions of the Riemann--Hilbert equation, is equivalent to finding finite orbits of the group of sheet-changing operators.
The groups $G$ and coefficients $a(s)$ that lead to finite orbits can be classified in terms of Dynkin diagrams, see \autoref{exo:sheets}.
Then the algebraic solutions are related to knot polynomials in Seifert spheres \cite{Borot:2014kda}.
A similar technique has been used in the context of quiver gauge theory~\cite{Nekrasov:2012xe}.

\subsection{Multi-matrix models}\label{sec:multi_saddle}

Let us consider the matrix chain with the partition function
\begin{align}\label{def:chainmatrix}
\mathcal{Z} = \prod_{j=1}^k \int_{\Gamma} dM_j\ e^{-N \operatorname{Tr} V_j(M_j)} e^{N\operatorname{Tr} M_j M_{j+1}}\ ,
\end{align}
In order for a saddle-point approximation to exist, we need not only have prefactors of $N$ for the potentials and quadratic terms $ \operatorname{Tr} M_j M_{j+1}$, but also assume that the external field $M_{k+1}$ has an $N$-independent number of distinct eigenvalues $m$, with the multiplicity $m_i$ of the eigenvalue $\lambda_i$ behaving as 
\begin{align}
m_i\underset{N\to\infty}{=} N\nu_i + o(N)\ ,
\end{align}
where $\nu_i$ is $N$-independent and $\sum_{i=1}^m \nu_i = 1$.  
Assuming that the potentials $V_j$ are polynomial, the Stieltjes transform $\overline{W}$ of the equilibrium density of eigenvalues of the first matrix $M_1$ is then an algebraic function \cite{Eynard:2009zz}, which we will now describe.
We will start with two special cases before dealing with the general case. 

\subsubsection{One-matrix model with an external field}

In the case $k=1$, the function $\overline{W}$ obeys a generalization of the saddle-point equation \eqref{eq:ovomdef} for the one-matrix model without an external field.
A prominent role in that equation was played by the polynomial $\overline{P}(x)$, and it turns out that the correct generalization of that polynomial is a function of two variables,
\begin{equation}\label{eq:sdptdefP1Mext}
\overline P(x,y) = \lim_{N\to \infty} \frac{1}{N}\left< \operatorname{Tr} \frac{V_1'(x)-V_1'(M_1)}{x-M_1}\ \frac{1}{y- M_2}\right>
\ .
\end{equation}
This is a polynomial of $x$ of degree $\deg V'_1-1$, and a rational function of $y$ of degree $m$ with simple poles at $y=\lambda_i$, such that 
\begin{equation}
 \overline P(x,y) \underset{x\to\infty}{\sim} \frac{V_1'(x)}{x} \sum_{i=1}^m \frac{\nu_i}{y -\lambda_i}  \ \left(1+O\left(\frac{1}{x}\right)\right) \ .
\end{equation}
Defining moreover
\begin{equation}
 Y_2(x) = V'_1(x) - \overline{W}(x)\ ,
 \label{eq:y2def}
\end{equation}
the saddle-point equation is 
\begin{equation}\label{eq:sdpt1Matrixext}
 V'_1(x) - Y_2(x) = \overline{P}(x, Y_2(x))\ ,
\end{equation}
which indeed reduces to Eq.~\eqref{eq:ovomdef} if $M_2=0$.
So $Y_2$ and therefore also $\overline{W}$ obey rational equations, equivalently polynomial equations, and $\overline{W}$ is 
an algebraic function of $x$. (See \autoref{exo:Kontsevich3} for the example of the Kontsevich integral.)

For example, let us consider the case of the Kontsevich integral, which is obtained for a cubic potential $V_1(x) = \frac13 x^3$. 
The polynomial $\overline P(x,y)$ is of the type 
\begin{equation}
\renewcommand{\arraystretch}{1.3}
\overline P(x,y) =  2x R(y) + Q(y) \quad \text{with} \quad 
 \begin{cases}
  \displaystyle
  R(y) = \frac12 \sum_{i=1}^m \frac{\nu_i}{y-\lambda_i} \ , \\
  \displaystyle  
  Q(y) = \sum_{i=1}^m \frac{\nu_i c_i}{y-\lambda_i}\ , 
 \end{cases}
\end{equation}
with $m$ unknown coefficients $c_i$.
The saddle-point equation is then
\begin{equation}
x^2 - 2R(Y_2)x - Q(Y_2)-Y_2 = 0\ .
\label{eq:xryt}
\end{equation}
This amounts to an algebraic equation of degree $m+1$ in $Y_2$ and of degree $2$ in $x$. 

In the case where the corresponding algebraic curve has genus zero, let us look for a rational parametrization of that curve. 
We want to express both $x$ and $Y_2$ as functions of a variable $z$ on the Riemann sphere $\overline{\mathbb{C}}$.
Assuming that $z=\infty$ is the point where both $x=\infty$ and  $Y_2=\infty$, and calling $z=\zeta_i$ the positions of the poles, this leads to 
\begin{align}\label{eq:Joukowsky1Mextfield}
 x(z) = z +\sum_{i=1}^m \frac{\nu_i}{2\zeta_i (z-\zeta_i)}\qquad , \qquad 
 Y_2(z) = z^2 +\sum_{i=1}^m \frac{\nu_i}{\zeta_i}\ .
\end{align}
Up to M\"obius transformations of $z$,
this is the unique rational parametrization of our algebraic curve that has the correct asymptotic behaviours. 
The parameters $\zeta_i$ are determined by $Y_2(\zeta_i)=\lambda_i$.
Moreover, we also have $Y_2(-\zeta_i)=\lambda_i$, and in the limit $z\to -\zeta_i$ Eq.~\eqref{eq:xryt} implies
\begin{align}
 c_i = -x(-\zeta_i) = \zeta_i +\sum_{j=1}^m \frac{\nu_j}{2\zeta_j(\zeta_i+\zeta_j)}\ .
\end{align}
So the assumption that our spectral curve has genus zero completely determines the coefficients $c_i$. 

\subsubsection{Two-matrix model}

Let us consider the case $k=2$ in the absence of an external field, equivalently with $M_3=0$. 
We now define 
\begin{equation}\label{eq:sdptdefP2mm}
\overline P_2(x,y) = \lim_{N\to \infty}\frac{1}{N}\left< \operatorname{Tr} \frac{V_1'(x)-V_1'(M_1)}{x-M_1}\ \frac{V_2'(y)-V_2'(M_2)}{y-M_2}\right> -1 \ ,
\end{equation}
which is a polynomial of $x$ of degree $\deg V_1'-1$, and a polynomial of $y$ of degree $\deg V'_2-1$.
The saddle-point equation for the resolvent $\overline{W}$ turns out to be 
\begin{equation}\label{eq:sdpt2matrixmm}
 (V_1'(x)-Y_2(x))(V'_2(Y_2(x))-x) = \overline P_2(x,Y_2(x)) \ ,
\end{equation}
where $Y_2$ is still given in terms of $\overline W$ by Eq.~\eqref{eq:y2def}. This is a polynomial equation of degree $\deg V_1$ in $x$ and $\deg V_2$ in $Y_2$, whose solution $\overline{W}$ is an algebraic function of $x$. 

In the case where the corresponding algebraic curve has genus zero, that curve has a rational parametrization of the type
\begin{align}\label{eq:Joukowsky2MM}
x(z) = \gamma z + \sum_{k=0}^{\deg V'_2} \alpha_k z^{-k} \qquad , \qquad
Y_2(z) = \frac{\gamma}{z} + \sum_{k=0}^{\deg V'_1} \beta_k z^{k}  \ .
\end{align}
The coefficients $\gamma$, $\alpha_k$, $\beta_k$, can be deduced from $z\to 0,\infty$ asymptotics
\begin{equation}
 V'_1(x(z)) - Y_2(z) \underset{z\to \infty}{=} \frac{1}{\gamma z} + O\left(\frac{1}{z^2}\right) \quad , \quad 
 V'_2(Y_2(z)) - x(z) \underset{z\to 0}{=} \frac{z}{\gamma} + O(z^2)\ ,
\end{equation}
which themselves follow from $\overline P_2(x, y) \underset{x,y\to \infty}{\sim} \frac{V'_1(x)}{x} \frac{V'_2(y)}{y}$ via Eq.~\eqref{eq:sdpt2matrixmm}.

\subsubsection{Matrix chain with an external field}

We now consider the most general matrix chain.
Let us generalize our definition \eqref{eq:y2def} of $Y_2(x)$, and define functions $Y_0(x), \dots, Y_{k+1}(x)$ by
\begin{equation}
Y_0(x)=\overline W(x) \quad , \quad  
Y_1(x)=x   \quad ,  \quad  
Y_{j+1}(x) = V'_j(Y_j(x))-Y_{j-1}(x)\ ,
\label{ywyx}
\end{equation}
where $\overline W(x)$ is the resolvent with respect to the first matrix in the chain. (See \autoref{exo:burgers} for the limit $k\to\infty$.)
All these functions are polynomials of $x$ and $\overline W(x)$. 
We now define 
\begin{equation}
\overline P(y_1,\dots,y_{k+1}) =   \lim_{N\to \infty}\frac{1}{N}\left< \operatorname{Tr} \operatorname{Pol}_{y_1,\dots, y_k} f_{1,k}(y_1,\dots,y_k)\frac{1}{y_1-M_1} \dots \frac{1}{y_{k+1}-M_{k+1}} \right> \ ,
\end{equation}
where 
\begin{equation}
f_{i,j}(y_i,y_2,\dots,y_j)
= \det \begin{pmatrix}
V'_i(y_i) & -y_{i+1} & & & \cr
-y_{i+1} & V'_2(y_{i+1}) & -y_{i+2} & & \cr
& &  \ddots &  &  \cr
& &  -y_{j-2} & V'_{j-1}(y_{j-1}) & -y_j\cr
& & & -y_{j-1} & V'_j(y_j)\cr
\end{pmatrix}\ ,
\label{fdet}
\end{equation}
and $\operatorname{Pol}_x g(x)$ is the polynomial part of the Taylor expansion of $g(x)$ near $x=\infty$. 
In particular $\operatorname{Pol}_x \frac{V'(x)}{x-M_1} = \frac{V'(x)-V'(M_1)}{x-M_1}$, and we recover the
polynomial $\overline{P}(x,y)$ \eqref{eq:sdptdefP1Mext} of the one-matrix model with an external field as a special case.
In the case of the two-matrix model, we recover the appropriate polynomial $\overline{P}_2(x, y)$ up to an overall factor,  $\overline{P}(y_1,y_2,y_3) = \frac{1}{y_3} \overline{P}_2(y_1,y_2)$.

The saddle-point equations then reduce to a system of algebraic equations, 
\begin{align}\label{eq:sdptchainmatrices}
Y_0 = \overline P(Y_1,\dots,Y_{k+1}) \quad , \quad  
Y_{j-1}+Y_{j+1} = V'_j(Y_j) \ \ , \ \ (1\leq j\leq k)\ .
\end{align}
(See \autoref{exo:loop_multi_matrix} for the proof.)
In the case $k=2$ this reduces to the equations
\begin{align}
 Y_0Y_3 = \overline{P}(Y_1,Y_2) \quad , \quad Y_0+Y_2 = V'_1(Y_1) \quad , \quad Y_1+Y_3 = V'_2(Y_2)\ ,
\end{align}
which are equivalent to \eqref{eq:sdpt2matrixmm}.

The function $\overline P(y_1,\dots, y_{k+1})$ is a polynomial of $y_1,\dots , y_k$ of degree at most $\deg V'_j-1$ in $y_j$, and a rational function of $y_{k+1}$, with $m$ simple poles at the eigenvalues of the external field $M_{k+1}$.
The total number of unknown coefficients of $\overline P$ is therefore 
\begin{equation}
d = m\prod_{j=1}^k \deg V'_j \ .
\end{equation}
Actually, one of these coefficients is fixed by the asymptotic behaviour \eqref{eq:omegainf} of the resolvent $\overline{W}(x)$. 
The remaining $d-1$ coefficients are in one-to-one correspondence with the periods
of the spectral curve, and also with the choices of integration contours for the
random matrices.
In contrast to the case of the one-matrix model with no external field, the spectral curve can have more than two sheets, in other words the degree of its equation can be larger than two. 

In the case where the spectral curve has genus zero, there exists a rational parametrization of that curve in terms of a variable $z\in\overline{\mathbb{C}}$, i.e. rational functions $Y_j(z)$  that obey the algebraic equations \eqref{eq:sdptchainmatrices}.
Assuming that $z=\infty$ corresponds to $x(z)=Y_1(z)=\infty$ in the first sheet, 
the asymptotic behaviour of $\overline{W}(x) = Y_0(z)$ near $x=\infty$ implies
\begin{align}
 Y_0(z) \underset{z\to \infty}{=} \frac{1}{Y_1(z)} + O\left(\frac{1}{Y_1(z)^2}\right)\ .
\end{align}
Let us moreover determine the behaviour of $Y_j(z)$ near a point $\zeta_i$ such that $Y_{k+1}(\zeta_i)= \lambda_i$, 
under the assumption $\deg V'_j \geq 2$. We have $\lim_{z\to \zeta_i} Y_j(z) = \infty$ for $0\leq j \leq k$ with $Y_0 \gg Y_1 \gg \dots \gg Y_{k+1}$, and the leading behaviour of the saddle-point equations \eqref{eq:sdptchainmatrices} is
\begin{align}
Y_0 \sim \prod_{j=1}^k \frac{V'_j(Y_j)}{Y_j}\cdot \frac{\nu_i}{Y_{k+1}-\lambda_i} \quad , \quad 
 Y_{j-1} \sim V'_j(Y_j)  \ \ , \ \ (1\leq j\leq k)\ .
\end{align}
Inserting the asymptotics of $Y_{j-1}$ into the asymptotics of $Y_0$ leads to the cancellation of all factors that involve $Y_{j<k}$, and we are left with
\begin{equation}
Y_k(z) \underset{z\to \zeta_i}{=} \frac{\nu_i}{Y_{k+1}(z)- \lambda_{i}}  + O(1) \ .
\label{eq:ykzetai}
\end{equation}
These asymptotic behaviours, together with the saddle-point equations themselves, completely determine the functions $Y_j(z)$ up to affine transformations of $z$. 
These functions are of the type
\begin{align}
Y_j(z) = \sum_{l=0}^{r_j} \alpha_{j,l} z^l + \sum_{i=1}^m  \nu_i \sum_{l=1}^{s_j} \frac{\beta_{j,i,l}}{(z-\zeta_i)^l} \ ,
\end{align}
where 
\begin{align}
r_0=-1\ ,\quad  r_{j\geq 1} =\prod_{l=1}^{j-1} \deg V'_l \ , \quad 
s_{j\leq k}=\prod_{j+1}^{k} \deg V'_l   \ , \quad s_{k+1} = 0\ ,
\end{align}
In particular, Eq.~\eqref{eq:ykzetai} amounts to $\beta_{k,i,1} = \frac{1}{Y_{k+1}'(\zeta_i)}$.
The resulting rational parametrization generalizes not only the Joukowsky map, but also the parametrizations that appeared in the two special cases that we considered.

\subsection{Complex matrix model with a harmonic potential}\label{sec:cmplx_harmonic}

Let us consider a Gaussian complex matrix model, and add a harmonic contribution to the potential. The partition function is 
\begin{align}
 {\mathcal Z} 
 = \int_{M_N(\mathbb C)} dM \, e^{-N \mathcal{V}(M,M^\dagger) }   \qquad \text{with} \qquad \mathcal{V}(x,\bar{x}) = Rx\bar{x} +V(x) + \overline{V}(\bar{x})\ .
\end{align}
The total potential
$\mathcal{V}(x,\bar{x})$
includes a Gaussian term with a constant coefficient $R$, and a harmonic term $V(x)+\overline{V}(\bar x)$, so that 
\begin{align}
 \partial \bar\partial \mathcal{V}(x,\bar x) = \frac14 \Delta \mathcal{V}(x,\bar x) = R \, .
 \label{eq:laplacianv}
\end{align}

\subsubsection{Functional saddle-point equation}

In terms of the eigenvalues $\lambda_i$ of the matrix $M$, the partition function is
\begin{align}
 {\mathcal Z} 
 \propto \int_{\mathbb C^N} \prod_{i<j} |\lambda_i-\lambda_j|^2 \prod_i e^{-N \left(R|\lambda_i|^2 + V(\lambda_i) + \overline V(\bar\lambda_i)\right)}d^2\lambda_i
 \, .
\end{align}
The corresponding saddle-point equation for the equilibrium density $\bar{\rho}$ is 
\begin{equation}
\forall x\in \operatorname{supp}\bar{\rho} \, ,
\qquad 
 \mathcal{V}(x,\bar x) - 2 \int d^2x'\ \bar\rho(x') \log{|x-x'|} = \ell 
 \, .
\end{equation}
(Compare with the saddle-point equation \eqref{eq:saddlerho} of our Hermitian matrix model.)
Taking the Laplacian of this saddle-point equation, and using Eq.~\eqref{eq:laplacianv}, we obtain 
\begin{equation}
\forall x\in \operatorname{supp}\bar{\rho} \ ,
\qquad
\bar\rho(x) = \frac{R}{\pi}\, .
\end{equation}
This means that the equilibrium density is constant on its support.

What remains to be found is the support itself. 
Since $\int_{\operatorname{supp}\bar\rho}\bar \rho = 1$, the support
must be a domain of area $\frac{\pi}{R}$. 
Let us parametrize this domain by the shape of its boundary curve $\partial\operatorname{supp}\bar{\rho}$, whose equation we write as 
\begin{equation}
\bar x = S(x) \, .
\end{equation}
The analytic function $S(x)$, which obeys 
\begin{align}
 \overline S \circ S = \operatorname{Id}\ ,
 \label{eq:scsi}
\end{align}
is called the \textbf{Schwarz function}\index{Schwarz function} of the curve $\partial\operatorname{supp}\bar{\rho}$.
Then the Stieltjes transform of the equilibrium density,
\begin{equation}
\overline{W}(x) = \int_{\mathbb{C}}\frac{\bar\rho(x')}{x-x'}d^2x' 
\, , 
\end{equation}
can be rewritten in terms of the function $S(x)$, with the help of Stokes' theorem,
\begin{equation}
\overline{W}(x) 
= \frac{R}{\pi} \int_{\operatorname{supp}\bar{\rho}}\frac{1}{x-x'}d^2x' 
= \frac{R}{2\pi i} \int_{\partial\operatorname{supp}\bar{\rho}}\frac{S(x')}{x-x'}dx' 
 \, ,
\end{equation}
where we used $\frac{1}{x-x'} d^2x'= \frac{1}{2i} \frac{1}{x-x'} dx' \wedge d\bar{x}' = \frac{1}{2i} d\left(\frac{\bar{x}'}{x-x'} dx'\right)$.
On the other hand, taking one derivative of the saddle-point equation gives
\begin{equation}
\forall x\in \operatorname{supp}\bar{\rho} \, ,
\qquad
R \bar x + V'(x) =  \overline{W}(x)\ , 
\end{equation}
which implies in particular 
\begin{equation}
\forall x\in \partial\operatorname{supp}\bar{\rho} \, ,
\qquad
R S(x) + V'(x)
=  \frac{R}{2\pi i} \int_{\partial\operatorname{supp}\bar{\rho}}\frac{S(x')dx' }{x-x'}
 \, .
 \label{eq:rsv}
\end{equation}
This is an affine integral equation for the Schwarz function $S(x)$. 
While originally derived for $x\in \partial\operatorname{supp}\bar{\rho}$, this equation must be valid outside $\operatorname{supp}\bar{\rho}$ by analyticity.

We now want to solve this equation, together with the similar equation for the function $\overline{S}$, under the constraint \eqref{eq:scsi}. 
Solving such an integral equation is in general difficult. 
As we will show in examples, in the case of polynomial or rational potentials, it is fruitful to look for solutions that obey a polynomial equation. 
The analysis of the behaviour of $S(x)$ as $x\to \infty$, knowing the behaviour \eqref{eq:omegainf} of $\overline{W}(x)$, then provides a lower bound on the degree of that polynomial, and the simplest assumption is that this bound is saturated. 
The coefficients of the polynomial that are not determined by the affine integral equation correspond to filling fractions.

\subsubsection{Gaussian complex matrix model}

We first consider the case $V(x)=\overline V(x)=0$.
From the equation \eqref{eq:rsv} and the behaviour of $\overline{W}(x)$, we must have $S(x) \underset{x\to\infty}{=} \frac{1}{Rx} + O(\frac{1}{x^2})$. 
Repeating the analysis for $\overline{S}(x)$, and using the relation \eqref{eq:scsi}, leads to the same asymptotic behaviour for $\overline S(x)$. 
Simple functions $S(x)$ and $\overline S(x)$ that have this asymptotic behaviour are 
\begin{align}
 S(x) = \overline S(x) = \frac{1}{Rx}\ ,
\end{align}
and they provide a solution of our affine integral equation. 
The boundary of the support of the equilibrium density then has the equation $\bar x=\frac{1}{Rx}$, and
is the circle of radius $R^{-\frac12}$. 
Therefore, in the large $N$ limit, the eigenvalues of the Gaussian matrix model fill a
disk of radius $R^{-\frac12}$. This is called the circular law:
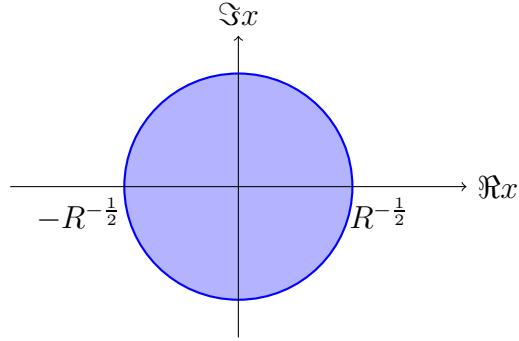
\begin{align}
 \begin{tikzpicture}[baseline=(current bounding box.center)]
  \filldraw [fill=blue,fill opacity=.3,thick,draw=blue] 
  (0,0) circle [radius=1.5]; 
  \draw [-latex] (-3,0) -- (3,0) node [right] {$\Re x$};
  \draw [-latex] (0,-2) -- (0,2) node [above] {$\Im x$};
  \node at (1.85,0) [below] {$R^{-\frac12}$};
%  \node at (-2.1,0) [below] {$-R^{-\frac12}$};
%
 \end{tikzpicture}
\end{align}

\subsubsection{Quadratic potentials}

We consider the case $V(x)= \frac{t_2}{2}x^2$, $\overline V(x)=\frac{\bar{t}_2}{2}x^2$.
We must have
\begin{equation}
S(x) \underset{x\to\infty}{=} -\frac{t_2}{R} x  + \frac{1}{Rx} + O\left(\frac{1}{x^2}\right)
\, , \quad
\overline S(x) \underset{x\to\infty}{=} -\frac{\bar{t}_2}{R} x  + \frac{1}{Rx} + O\left(\frac{1}{x^2}\right) \ .
\end{equation}
The inverse functions therefore behave as
\begin{equation}
S^{-1}(x) \underset{x\to\infty}{=} -\frac{R}{t_2} x  - \frac{1}{Rx} + O\left(\frac{1}{x^2}\right)
\, , \quad
\overline S^{-1}(x) \underset{x\to\infty}{=} -\frac{R}{\bar{t}_2} x  - \frac{1}{Rx} + O\left(\frac{1}{x^2}\right)
\, .
\end{equation}
This seems at odds with the equation \eqref{eq:scsi}, which says $S = \overline{S}^{-1}$. 
Actually, this discrepancy means that $S(x)$ lives on a cover of the complex $x$-plane with at least two sheets labelled $\pm$, and the asymptotic behaviours
\begin{align}
 S_+(x) \underset{x\to\infty}{=} -\frac{t_2}{R} x  + \frac{1}{Rx} + O\left(\frac{1}{x^2}\right)
\, , \quad S_-(x) \underset{x\to\infty}{=} -\frac{R}{\bar{t}_2} x  - \frac{1}{Rx} + O\left(\frac{1}{x^2}\right)
\, .
\end{align}
Let us assume that there are exactly two sheets.
Then $S(x)$ is a solution of an algebraic equation of degree two
$S^2 - S(S_++S_-)+S_+S_-=0$,
whose coefficients must be single-valued, and analytic outside $\operatorname{supp}\bar{\rho}$. 
Assuming that these coefficients are polynomials, we find
\begin{subequations}
\begin{align}
 S_+(x) S_-(x) 
 & = \frac{t_2}{\bar{t}_2}x^2 + \frac{t_2}{R^2}-\frac{1}{\bar{t}_2} \,,\\
 S_+(x) +  S_-(x) 
 & = -\left(\frac{t_2}{R} + \frac{R}{\bar{t}_2} \right)x  \, .
\end{align}
\end{subequations}
Then the support of the equilibrium density is the ellipse with the equation
\begin{equation}
t_2 x^2 + \bar{t}_2 \bar x^2 + \left(\frac{t_2 \bar{t}_2 }{R} + R \right)x \bar x =1 - \frac{t_2\bar{t}_2}{R^2} \ .
\end{equation}
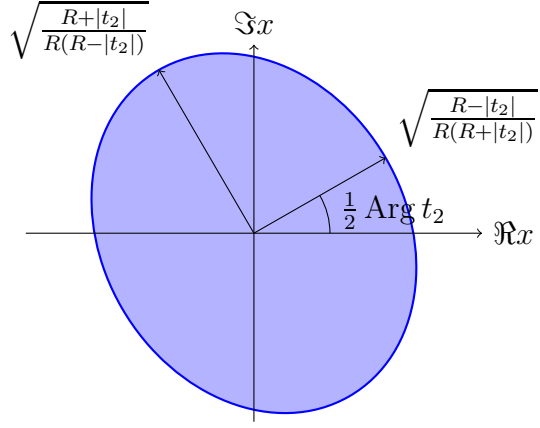
\begin{align}
 \begin{tikzpicture}[baseline=(current bounding box.center)]
  \filldraw [fill=blue,fill opacity=.3,thick,draw=blue,rotate=30] 
  (0,0) circle [x radius=2., y radius = 2.5]; 
  \draw [-latex] (-3.,0) -- (3.,0) node [right] {$\Re x$};
  \draw [-latex] (0,-2.5) -- (0,2.5) node [above] {$\Im x$};
  \draw [->, rotate=30] (0,0) -- (2.,0) node [above right]{$\sqrt{\frac{R-|t_2|}{R(R+|t_2|)}}$};
  \draw [->, rotate=30] (0,0) -- (0,2.5) node [above left]{$\sqrt{\frac{R+|t_2|}{R(R-|t_2|)}}$};
  \draw (1,0) arc [start angle = 0, end angle = 30, x radius = 1, y radius = 1]; 
  \node [right] at (1, .3) {$\frac12 \operatorname{Arg} t_2$};
 \end{tikzpicture}
\end{align}

\subsubsection{Polynomial potentials}\label{sec:cmmh}

Let us assume that $V$ is polynomial of degree $d+1$ with $V(x) = \sum_{k=1}^{d+1} \frac{t_k}{k}x^k$.
There must be one sheet where $S(x)$ behaves as 
\begin{align}
 S_1(x) \underset{x\to\infty}{=}
-\frac{t_{d+1}}{R}x^d + \cdots \ .
\end{align}
On the other hand, using the analogous behaviour of $\overline{S}$, together with the equation \eqref{eq:scsi}, produces $d$ sheets where 
\begin{align}
 S_k(x) \underset{x\to\infty}{=} e^{\frac{2\pi i}{d}k}\left(-\frac{Rx}{\bar
t_{d+1}}\right)^\frac{1}{d}+\cdots \ , \qquad k\in\{2,\dots, d+1\}\ .
\end{align}
Assuming that $S(x)$ is a solution of a polynomial equation of degree $d+1$,
the support of the equilibrium density is the interior of an algebraic curve of degree $d+1$ of the form $P(x,\bar x)=0$.
This involves a polynomial $P(x,y)$ of the type
\begin{equation}
P(x,y) =  (V'(x)+R y)({\overline V}'(y)+R x) -1 +\overline P(x,y)\ ,
\end{equation}
where $\overline P(x,y)$ is a polynomial of degree $d-1$ in each variable, whose leading term is $\overline P(x,y)\underset{x,y\to\infty}{\sim} \frac{t_{d+1} \bar t_{d+1}}{R} (xy)^{d-1}$, and which plays the same role as the polynomial  $\overline P(x)$ in \autoref{sec:ded}.
The polynomial equation $P(x,y)=0$ then defines the spectral curve of the model.

For example, for $d=3$, the support of the equilibrium density may look as follows:
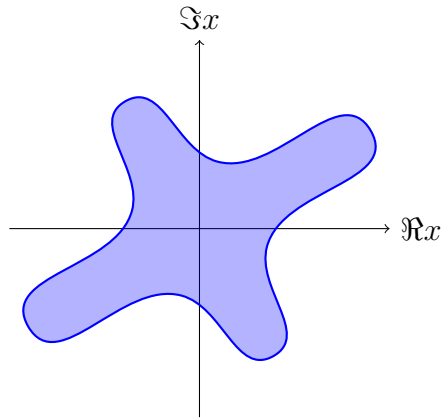
\begin{align}
 \begin{tikzpicture}[baseline=(current bounding box.center)]
%  
  % axes
  \draw [-latex] (-2.5,0) -- (2.5,0) node [right] {$\Re x$};
  \draw [-latex] (0,-2.5) -- (0,2.5) node [above] {$\Im x$};
%  
  % domain
  \begin{scope}[scale=1.3]
  \filldraw [fill=blue,fill opacity=.3,thick,draw=blue,rotate=30] 
  (2,0) to [out=up,in=-45] (.5,.5) to [out=90+45,in=right] (0,1.5)
  to [out=left,in=90-45] (-.5,.5) to [out=180+45,in=up] (-2,0) to
  [out=down,in=90+45] (-.5,-.5) to [out=-45,in=left] (0,-1.5) to
  [out=right,in=180+45] (.5,-.5) to [out=45,in=down] (2,0);
  \end{scope}
 \end{tikzpicture}
\end{align}
The large $N$ eigenvalue density of complex matrix models, provides the solution of the Hele--Shaw problem: the moments of the pocket of viscous liquid are given by the times $t_k$ and $\bar t_k$, and its shape by the support of the equilibrium density.

\section{Exercises}

\begin{exo}[Number of solutions of saddle-point equations]
\label{exo:nbsdpt}
Let us compute the number of solutions of the saddle-point equations \eqref{eq:saddlelambda} for matrix size $N=2$.
 \begin{enumerate}
\item
Consider the cubic potential $V(x)= \frac{t_3}{3}x^3+\frac{t_2}{2}x^2$. Show that the number of solutions up to permutations of the $N=2$ eigenvalues is $d_2=3$.

\item
For a general polynomial $V'$ of degree $d$, show that $d_2=\frac{d(d+1)}{2}$.
\end{enumerate}
\end{exo}

\begin{exo}[Upper bound of the functional action after minimizing at fixed domain]
~\label{exo:ubrs}
Let $V=tx^{d+1}+\cdots $ be a polynomial potential of degree $d+1$, and let 
\begin{equation}
\mathcal V(x) = \Re V(x) + 2 d \log(1+|x|)\ .
\end{equation}
We want to find an upper bound on the functional action $\Re S[\rho_{[\gamma]}]$ \eqref{def:Srho}, where the density $\rho_{[\gamma]}$ minimizes that action at fixed domain $\gamma$. For simplicity we set the Lagrange multipliers $\ell_i$ to zero.

\begin{enumerate}
 \item Show that there is a number $\mathcal{V}_0\in\mathbb{R}$ such that $\mathcal{V}(x)$ takes the value $\mathcal{V}_0$ at least once on any domain $\gamma\in H_1(e^{-V(x)}dx)$  that is not homologically trivial. 
 To do this, show that any nontrivial domain must intersect the set $A=(-t)^{\frac{1}{d+1}}\bigcup_{j=0}^d \left(e^{ \frac{2\pi i}{d+1}j}\mathbb{R}_+\right)$, and show that $\mathcal V_0=\sup_A \mathcal V$ is finite.
 \item Show that $K<\infty$, where we define
 \begin{align}
  K = \sup_{(x,x')\in \mathbb C^2} 
\frac{|\mathcal V(x)-\mathcal V(x')|}{|x-x'|} \frac{1}{(1+|x|)^d (1+|x'|)^d} \ .
 \end{align}
\item For any density $\rho=(\gamma,f)$, show that
\begin{equation}
\Re S[\rho] \leq \log K + \int_{\gamma} \mathcal V(x) \rho(x)dx
- \int_\gamma\int_\gamma \log{|\mathcal V(x)-\mathcal V(x')|} \  \rho(x)dx \ \rho(x')dx'\ .
\label{eq:resrho}
\end{equation}
%\item For any domain $\gamma$, consider the density
\item Let $\gamma$ a Jordan arc $\gamma:\mathbb R \to \mathbb C$, parametrized by a variable $s\in \mathbb R$.
For any $v\geq \mathcal V_0$, define
\begin{align}
R(v) = \inf \{s\in \mathbb R  \ | \ \mathcal V(\gamma(s))=v \} \in \mathbb{R}\ .
\end{align}
Show that the function  $v\mapsto R(v)$ is well-defined on $[\mathcal V_0,+\infty[$ and strictly increasing thus invertible.
Define
\begin{align}
F(s) =
\begin{cases}
 0 & \text{if} \quad  \mathcal V(\gamma(s))<\mathcal V_0\ , \\
 1-e^{\mathcal V_0-R^{-1}(s)} & \text{if} \quad  \mathcal V(\gamma(s))\geq \mathcal V_0\ , \\ %\text{where} \quad v=\inf \{v' \in [\mathcal V_0,+\infty[ \ | \ s=R(v') \}
\end{cases}
\end{align}
Show that $F(s)$ is the cumulative distribution function of a positive measure on $\mathbb R$, and induces a probability measure $\tilde\rho_\gamma(x)dx$ on $\gamma$.
Compute the last 2 terms of the right-hand side of Eq.~\eqref{eq:resrho} for $\rho = \tilde{\rho}_{\gamma}$.
\item
Conclude that $\Re S[\tilde{\rho}_\gamma]$ is bounded from above.
\end{enumerate}
\end{exo}

\begin{exo}[Plancherel measure and Logan--Shepp--Vershik--Kerov profile]
~\label{exo:plancherel}
Consider the integral
\begin{equation}
\mathcal Z  = \frac{1}{N!\,(2\pi i)^N }\int_{\gamma^N} \Delta(\{x_i\})^2 \prod_{i=1}^N \varphi\left(\frac{x_i}{\hbar}\right)dx_i \quad \text{with} \quad \varphi(x) = \frac{\Gamma(-x)}{\Gamma(x+1)} e^{i\pi x}\hbar^{-2x-1}\ ,
\end{equation}
where $\Gamma$ is the Gamma function, and 
the integration contour $\gamma$ surrounds the positive real axis and therefore the set of poles $x_i\in \hbar\mathbb{N}$ of the integrand.
\begin{enumerate}
 \item Evaluate the integral as a sum of residues at the poles, and show
 \begin{align}
\mathcal Z  = \hbar^{N^2-N} \sum_{h_1>\dots >h_N\geq 0}
\left(\frac{\prod_{i<j} (h_i-h_j)}{\prod_i h_i!}\right)^2  \hbar^{-2\sum_i h_i} \ . 
\label{eq:Plancherel_measure_discrete_sum}
\end{align}
\end{enumerate}
Writing $\lambda_i = h_i+i-N$, we have $\lambda_1\geq \lambda_2 \geq \dots \geq \lambda_N\geq 0$, and $\lambda=(\lambda_1,\dots,\lambda_N)$ is a partition of length $\ell(\lambda) \le N$. We define the dimension of the irreducible representation of the symmetric group $\mathfrak{S}_{|\lambda|}$ parametrized by a partition $\lambda$ as 
\begin{equation}
\dim\lambda = |\lambda|! \ \frac{\prod_{i<j} (h_i-h_j)}{\prod_i h_i!}\ .
\end{equation}
For example, the partition $\lambda=(8,5,4,4,2,1,0,0)$ with $N=8$ has weight $|\lambda|=24$, and $(h_1,\dots,h_N) = (15,11,9,8,5,3,1,0)$, and dimension $\dim\lambda=52718561280$. The Russian representation of this partition is
\newcommand{\squarebox}[1]{\begin{scope}[shift = {(#1)}] \draw (0, 0) -- (1, 0) -- (1, 1) -- (0, 1) -- cycle; \end{scope}}
\newcommand{\linemark}[1]{\begin{scope}[shift = {(#1)}] \draw (0, 0) -- (0, 1); \draw (-.1, 1) -- (.1, 1); \end{scope}}
\newcommand{\lambdas}{8,5,4,4,2,1,0,0}
\newcounter{ga}\setcounter{ga}{0}
\begin{align}
 \begin{tikzpicture}[scale = .35, rotate = 45,baseline=(current bounding box.center)]
  \foreach \x in \lambdas 
 {\ifthenelse{\equal{\x}{0}}{\linemark{1, \thega}}{
 \foreach \y in {1, ..., \x}
 {\squarebox{\y, \thega}
 }}
 \stepcounter{ga}
 }
 \draw[latex-latex] (11, 0) node[below right]{$\lambda_i$} -- (1, 0) -- (1, 10) node[below left]{$i$};
 \end{tikzpicture}
\end{align}
We finally introduce the Plancherel measure on partitions of a given weight, whose definition and normalization property are 
\begin{align}
 \mathcal P_\mathrm{ Plancherel}(\lambda) = \frac{(\dim\lambda)^2}{|\lambda|!}  \quad , \quad \sum_{|\lambda|=k} \mathcal P_\mathrm{ Plancherel}(\lambda) = 1\ .
\end{align}

\begin{enumerate}
\setcounter{enumi}{1}
 \item Show that our integral $\mathcal Z$ can be expressed in terms of the Plancherel measure as 
 \begin{align}
  \mathcal Z  = %\hbar^{N^2-N} q^{\frac{N(N-1)}{2}} 
\sum_{k=0}^\infty \frac{\hbar^{-2k}}{k!}\sum_{|\lambda|=k,\, \ell(\lambda)\leq N} \mathcal P_\mathrm{ Plancherel}(\lambda)\ ,
 \end{align}
and deduce its large $N$ limit.

\item Now we are interested in the $\hbar\to 0$ limit with $N \hbar =O(1)$.
Write the saddle-point equation for the $x_i$s.

\item Using Stirling's formula for the Gamma function, write the leading order of the saddle-point equation, and deduce
\begin{equation}
\overline W(x+i0)+\overline W(x-i0) = 2\log{x} + O(\hbar)\quad , \quad \overline W(x) \underset{x\to\infty}{\sim} \frac{N\hbar}{x} + O\left(\frac{1}{x^2}\right)\ .
\label{owxi}
\end{equation}

\item Assuming a 1-cut situation with $N\hbar =2$, use the Joukowsky map and show that the saddle-point equation has the solution
\begin{align}
\overline W(x(z)) = 2\log{\left(1+z^{-1}\right)}\quad \text{with} \quad
x(z) = \gamma(z+z^{-1})+\gamma^2+1 \ \  , \ \ \gamma^2 = \frac{N\hbar}{2}\ .
\label{owxz}
\end{align}

\item Consider the case $N\hbar=2$. 
Compute the corresponding density $\bar \rho(x)$ and its support.
Deduce that the asymptotic shape
of a large partition in the Russian representation $f(x-2) = x-2-2\int_4^x \bar\rho(x')dx' $ is given by the Logan--Shepp--Vershik--Kerov function,
\begin{equation}
f(x) = \frac{2}{\pi} \left( \sqrt{4 - x^2}  + x\arcsin\frac{x}{2} \right)\ . \quad
 \begin{tikzpicture}[baseline = (current  bounding  box.center), scale = 1.3]
\draw[very thick, latex-] (0, 2) node[left]{$f(x)$} -- (0, 0);
\draw[very thick, -latex] (-2,0)  -- (-1,0) node[below]{$-2$} -- (0, 0) node[below]{$0$} -- (1,0) node[below]{$2$}-- (2, 0) node[below]{$x$};
\draw[domain=-.98:.98,smooth,variable=\x,blue, thick] plot({\x},{2*sqrt(1-\x*\x)/pi+\x*asin(\x)/90)}) ;
%\draw[domain=-1.:1.,smooth,variable=\x,blue, thick] plot({\x},{2*sqrt(1-\x*\x)/pi-\x*acos(\x)/90+\x)}) ;
\draw (-1.5,1.5) -- (0,0) -- (1.5,1.5);
\draw[dotted] (1, 0) -- (1, 1);
\draw[dotted] (-1,0) -- (-1, 1);
 \end{tikzpicture}
\end{equation}

\end{enumerate}

\end{exo}

\begin{exo}[Kontsevich integral]
~\label{exo:Kontsevich3}
The Kontsevich integral \eqref{eq:koni} belongs to the space $\mathbb C[[t]]$ of formal power series of $t$. This exercise is about quantities and equations in that space.
\begin{enumerate}
\item For any matrix $\Lambda$ of size $N$, show that there exists a unique $\hat{t}_1 \in \mathbb C[[t]]$ such that
\begin{equation}
\hat{t}_1 = \frac{1}{N}\operatorname{Tr} %\Lambda^{-1} \ (1 + t \hat t_1 \Lambda^{-2})^{\frac{-1}{2}}.
 \left( \Lambda \, \sqrt{1 + t \hat{t}_1 \Lambda^{-2}} \right)^{-1}
 \, .
\end{equation}
Show that the formal power series
\begin{equation}\label{exo:K:defhatLambda}
\hat \Lambda = \Lambda  \sqrt{1+ t \hat{t}_1 \Lambda^{-2}} 
\qquad \text{and} \qquad
\hat{t}_k = \frac{1}{N}\operatorname{Tr} \hat{\Lambda}^{-k} \ ,
\end{equation}
are such that
\begin{equation}
\hat{\Lambda} = \Lambda + O(t)
\ , \qquad
\hat{t}_k = \frac{1}{N} \operatorname{Tr} \Lambda^{-k} + O(t) \ ,
\end{equation}
where the $O(t)$ terms vanish if $\operatorname{Tr} \Lambda^{-1}=0$.

\item Show that the Kontsevich integral is proportional to the partition function of a one-matrix model, with the external field $-\frac{1}{t}\Lambda^2$ and cubic potential $V_1(x)=-\frac{1}{3t}x^3$.

\item Show that the saddle-point equation \eqref{eq:sdpt1Matrixext} has the solution
\begin{align}
x(z) = z - \frac{t}{N} \operatorname{Tr}   \frac{1}{2 \hat{\Lambda} (z-\hat{\Lambda})} \qquad , \qquad 
 Y_2(z) =  -\frac{1}{t}z^2 + \hat{t}_1 \ .
 \label{xzy2z}
\end{align}
Show that saddle-point resolvent $\overline W = -\frac{1}{t}x^2-Y_2$ has the leading behaviour
\begin{equation}
\overline W(x) = \frac{1}{N} \operatorname{Tr}   \frac{1}{ x-\Lambda} + O(t)\ .
\label{wxot}
\end{equation}
Since it obeys the same equation \eqref{eq:sdpt1Matrixext}
 and same leading behaviour, conclude that $\overline W$ coincides with the large $N$ limit of the resolvent.
\item
Show that the small $z$ expansion of $x(z)$ is
\begin{align}
x(z) = z + \frac t 2 \sum_{k=0}^\infty \hat{t}_{k+2}  z^k \ .
\end{align}
In particular compute $x(z)$ in the case  $\operatorname{Tr} \Lambda^{-1}=0$.
\end{enumerate}
\end{exo}

\begin{exo}[Sheet-changing operators]
~\label{exo:sheets}
Consider the generalized matrix integral \eqref{eq:zr}, with a potential $V(x)$ such that $V'(x)$ is rational, and
a symmetric rational function $R(x,x')=R(x',x)$.
As a function of $x'$, this function has finitely many poles and zeros, and we write
\begin{equation}
R(x,x') = C(x) \prod_{s\in G} (x'-s(x))^{a(s)}\ .
\end{equation}
Here $G$ is the group of algebraic functions generated by the poles and zeros by composition, with $a(s)\in\mathbb{Z}$ and $a(s)=0$ if $s(x)$ is not a pole or zero. We assume $a(\mathrm{Id}) = 2$.

We look for a 1-cut solution of the saddle-point equation for the resolvent. This means that the resolvent has one cut $\Gamma$ in the physical sheet, such that $s\neq \mathrm{Id} \implies s(\Gamma)\cap \Gamma= \emptyset$, and $V'$ has no poles in $\cup_{s\in G}s(\Gamma)$.

\begin{enumerate}

\item On a cut $\Gamma$, show that the functional saddle-point equation for the resolvent is
\begin{equation}\label{eq:RHsG}
\overline W(x+i0)+\overline W(x-i0) + \sum_{s\in G \backslash \{\mathrm{Id}\}} a(s) \overline W(s(x))s'(x) = V'(x) \ .
\end{equation}
Show that
\begin{equation}
\overline W_0 = \sum_{s\in G} b(s) (V\circ s)'\ ,
\end{equation}
is a solution provided the coefficients $b(s)$ satisfy
\begin{equation}
\forall s_0 \in G\ , \qquad \sum_{s_1,s_2\in G|s_2\circ s_1 = s_0} a(s_1) b(s_2) = \delta_{s_0,\mathrm{ Id}}\ .
\label{fsig}
\end{equation}

\item Consider the group algebra $\hat{\mathcal G}=\text{Span}_\mathbb{C}(\{e_g\}_{g\in\mathbb{G}})$, with the multiplication $e_g.e_{g'} = e_{g\circ g'}$. Let
$(\ell_g)_{g\in G}$ be the dual basis of $\hat{\mathcal G}^*$, i.e. $
\ell_g(e_{g'}) = \delta_{g,g'}$.
For $g\in G$ let $\hat{T}_g \in \operatorname{End}(\mathcal G)$ be defined by its action on $v\in \hat{\mathcal G}$:
\begin{equation}
\hat{T}_g(v) = v  - \ell_g(v) \,\hat\alpha.e_g
\qquad \text{where} \qquad
 \hat\alpha = \sum_{g\in G} a(g) e_g\ .
\end{equation}
Show that $\hat{T}_g$ is a pseudo-reflection, i.e. an involution such that $\operatorname{Ker}(\mathrm{ Id}+\hat{T}_g)= \operatorname{Span}(\hat\alpha.e_g)$ has dimension $1$.

Let $\hat{G}$ be the group generated by $\hat{T}_g$ for $g\in G$. It is in general a non-commutative infinite discrete subgroup of $\operatorname{Aut}(\mathcal G)$.

\item
Let the differential form $w(x) =(\overline W(x) - \overline W_0(x)) dx $  be a solution of the homogeneous equation
\begin{equation}
w(x+i0)+w(x-i0) + \sum_{s\in G \backslash \{\mathrm{Id}\}} a(s)  w(s(x)) = 0\ .
\end{equation}
For $v \in \hat{\mathcal G}$, let us define the function
\begin{equation}
\phi_v(x) = \sum_{g\in G} \ell_g(v) w(g(x))\ .
\end{equation}
On the cut $g^{-1}(\Gamma)$, show that
\begin{equation}
\phi_v(x-i0) = \phi_{\hat T_g(v)}(x+i0)\ .
\label{fvftv}
\end{equation}
Conclude that $\phi_v(x)$ is algebraic if and only if the orbit of $v $ by the group  $\hat G$ is finite.

\item Let us relate the question of finite orbits to the root system associated to the matrix $\left<e_g,e_h\right> = \frac12 a(g\circ h^{-1})$ \cite{Borot:2014kda}. To do this, show that this matrix is symmetric, and corresponds to the quadratic form $\left<v,v'\right> = \frac12 \ell_{v'}(\hat \alpha\cdot v)$ on $\hat{\mathcal G}$.
Show that
\begin{equation}
\hat T_g(\hat\alpha.v) = \hat\alpha.T_g(v) \qquad \text{for} \qquad T_g(v) =  v-2\left<v,e_g\right> e_g\ .
\end{equation}
Assuming the quadratic form is non-degenerate, show that $\hat G$ has finite orbits if and only if the reflection group $\hat G_{\text{reduced}} $ generated by $(T_g)_{g\in G}$ has finite orbits.

\item In the example of the $O(n)$ model, show that $G=\mathbb Z_2$. Writing
$(e_+,e_-)$ the basis of $\hat{\mathcal G}$, show that $\hat\alpha = 2 e_+ - n e_-$, and that
the symmetric bilinear form is non-degenerate if $n\neq \pm 2$.
Compute the reflections $T_+,T_-$, as well as
\begin{equation}
R \equiv T_+ T_- = \begin{pmatrix}
n^2-1 & -n \cr  n  & -1
\end{pmatrix} \ .
\end{equation}
Deduce that the spectral curve is algebraic if and only if $n\in 2\cos \pi \mathbb{Q}$.

\item In the example of the torus knot $\mathfrak K_{P,Q}$, we have $G=\frac{\mathbb Z}{PQ \mathbb Z} $. Let  $e_0, e_1,\dots, e_{PQ-1} $ be the canonical basis of $\hat{\mathcal G}$.
Show that $\hat\alpha = 2  e_0 + \sum_{k=1}^{P-1}  e_{Qk} + \sum_{k=1}^{Q-1}  e_{Pk}$.
Show that $v_0=\sum_{k=0}^{P-1}  e_{Qk}$ has a finite orbit, and conclude that the function $f(x)$ \eqref{eq:torusknotsfalg} is algebraic.

\item 
For $(p_1,\dots,p_r)$ a tuple of $r$ integers $p_i\geq 2$, let
\begin{equation}
R(x,x') = \frac{\prod_{j=1}^r (x^{p_j'}-x'^{p_j'})}{(x^p-x'^p)^{(r-2)}} \quad \text{with} \quad p =\prod p_j \quad , \ p'_j=\frac{p}{p_j}.
\end{equation}
Which case corresponds to the torus knot $\mathfrak K_{P,Q}$?
Show that $G= \frac{\mathbb Z}{p \mathbb Z}$. Writing $e_0, e_1,\dots, e_{p-1} $ the canonical basis of $\hat{\mathcal G}$, show that
\begin{align}
\hat\alpha = 2  e_0 + \sum_{j=1}^r \sum_{k=1}^{p'_j-1}  e_{k p_j } \ .
\end{align}
Let $\chi = 2-r+ \sum_{j=1}^r \frac{1}{p_j}$.
Show that if $ \chi<0$, then $n(v)=\sum_{k=0}^{p-1}\ell_k(v)$ takes infinitely many values on any orbit.
Classify the families $( p_1,\dots,p_r)$ such that $\chi\geq 0$.

Bonus question: show that if $\chi\geq 0$, then there is a finite orbit.
(See \cite{Borot:2014kda} for the solution.)

\end{enumerate}

\end{exo}

\begin{exo}[Matrix chains and Burgers' equation]
~\label{exo:burgers}
Consider a matrix chain with matrices $(M_j)_{j=1,\dots k}$ and partition function \eqref{def:chainmatrix}. We are interested in the limit $k\to\infty$ where the number of matrices becomes infinite.

\begin{enumerate}

\item Find a redefinition of the potentials, and appropriate rescalings of matrices and potentials, such that the partition function has the finite limit
\begin{equation}
 \mathcal Z = \int [dM(t)]\ \exp
  \left(
   - N \operatorname{Tr} \int_0^T \Big[ \tfrac12 M'(t)^2 + V(M(t),t) \Big] dt
  \right)
  \ .
  \label{eq:matrix_chain_partition_fn}
\end{equation}
This is the partition function of a continuous chain of matrices, also called Matrix quantum mechanics.

\item Calling $Y(z,t) = \lim Y_j(z)$, show that the limit of the saddle-point equation \eqref{eq:sdptchainmatrices} is
\begin{equation}
\frac{\partial^2 Y(z,t)}{\partial t^2} = \left.\frac{\partial V(y,t)}{\partial y}\right|_{y= Y(z,t)}\ .
\label{ptyzt}
\end{equation}

\item Let us define the resolvent $\overline W(x,t)= \left. \partial_t Y(z,t) \right|_{z=z(x,t)}$, where $z(x,t)$ is such that $Y(z(x,t),t)=x$.
Show that $-\overline W(x,0)$ is the resolvent of the first matrix $M_1$, up to adding an analytic function.

\item
Show that $\overline W(x,t)$ satisfies the non-homogeneous inviscid Burgers' equation,
\begin{equation}
\frac{\partial \overline W(x,t) }{\partial t}  + \overline W(x,t) \frac{\partial \overline W(x,t) }{\partial x}  = \frac{\partial}{\partial x}V(x,t)\ .
\end{equation}

\end{enumerate}
\end{exo}

\chapter{Loop equations}\label{chap:LoopEq}

\textbf{Loop equations}\index{loop equations} are relations between correlation functions, obtained by integrating by parts
in the matrix integral. 
Equivalently, loop equations follow from the invariance of the matrix integral under changes of integration variables: in this sense, loop equations are Schwinger--Dyson
equations.
Loop equations are difficult to solve at finite $N$, but they can be used for efficiently computing correlation functions order by order in a large $N$ expansion.

In this chapter we will first derive the exact loop equations. We will then introduce the topological expansion of the loop equations, and
solve them 
recursively using the topological recursion equation. 
The resulting solutions are formal series in powers of $\frac{1}{N}$.  Whether these series converge, and how they are related to convergent matrix integrals, will be discussed in \autoref{sec:non-perturbative}.

\section{Exact loop equations}

\subsection{Loop equations}\label{sec:loop}

\subsubsection{Direct derivation}

Consider the matrix integral
\begin{align}
 \mathcal{Z} = \int_\Gamma dM \, e^{-N\operatorname{Tr}V(M)} \ ,
 \label{mat_int}
\end{align}
where $N$ is the size of the matrix $M$, which for the moment belongs to a unitary ensemble ($\beta=2$). 
In order to derive the loop equations, the ensemble $\Gamma$ and potential $V(M)$ need only be such that the integrand vanishes at the boundaries of $\Gamma$. For simplicity, we further assume that the potential is polynomial, and write
\begin{align}
 d  = \deg V'\ .
\end{align}
We are interested in correlation functions, such as the moments
\begin{align}
 \Big< \operatorname{Tr} M^k \Big>
 =
 \frac{1}{\mathcal{Z}} \int_\Gamma dM \left(\operatorname{Tr}M^k\right) 
 e^{-N\operatorname{Tr}V(M)}  \, .
\end{align}
Assuming that the integrand vanishes at
the boundaries of $\Gamma$, the integral of the following total derivative vanishes:
\begin{align}
 \sum_{i,j} \int_\Gamma dM \, \frac{\partial}{\partial M_{ij}}
 \left(
  \left(M^k\right)_{ij} \, e^{-N\operatorname{Tr}V(M)}
 \right) = 0\ .
\end{align} 
This apparently trivial equation leads to our first example of a loop equation.
Computing the derivative yields
\begin{align}
\sum_{i,j} \int_\Gamma dM \,
 \left(
 \sum_{l=0}^{k-1} \left(M^l\right)_{ii} \left(M^{k-l-1}\right)_{jj}
 - N \left(M^k\right)_{ij} \left(V'(M)\right)_{ji}
 \right) \, e^{-N\operatorname{Tr}V(M)}  = 0\ .
\end{align}
Rewriting this equation as a relation for correlation functions, we obtain the loop equation
\begin{align}
 \sum_{l=0}^{k-1}
 \Big<
  \operatorname{Tr}M^l \operatorname{Tr}M^{k-l-1}
 \Big>
 - N 
 \Big<
  \operatorname{Tr}M^k V'(M)
 \Big> = 0 \ .
\label{eq:1loop}
\end{align}
This equation can alternatively be obtained by performing the infinitesimal change of variable $M \to M +
\epsilon M^k$  in the integral (\ref{mat_int}) at the order $O(\epsilon)$.
The first term is the contribution of the Jacobian
\begin{align}
\frac{d(M+\epsilon M^k)}{dM} 
&= 1 + \epsilon \sum_{i,j} \frac{\partial }{\partial M_{i,j}} (M^k)_{i,j} +O(\epsilon^2)\ ,
\\
&= 1+\epsilon  \sum_{l=0}^{k-1}   \operatorname{Tr}M^l \operatorname{Tr}M^{k-l-1}
 +O(\epsilon^2)\ ,
\end{align}
and the second term comes from
the variation of the integrand $e^{-N\operatorname{Tr} V(M)}$. 
More general loop equations can be derived from
\begin{align}
 \sum_{i,j}
 \int_\Gamma dM \, \frac{\partial}{\partial M_{ij}}
 \left(
  \left(M^{\mu_1}\right)_{ij}
  \operatorname{Tr}M^{\mu_2} \cdots \operatorname{Tr}M^{\mu_n} \,
  e^{-N\operatorname{Tr}V(M)}
 \right) = 0\ .
\end{align}
Computing the action of the derivative, and rewriting the resulting equation in terms of correlation functions, we obtain the loop equation
\begin{empheq}[marginbox = \fbox]{multline}
 \sum_{l=0}^{\mu_1-1}
 \Big<
  \operatorname{Tr}M^l \operatorname{Tr}M^{\mu_1-l-1}
  \prod_{i=2}^n \operatorname{Tr}M^{\mu_i}
 \Big>
 \\
 + \sum_{j=2}^n \mu_j
 \Big<
  \operatorname{Tr}M^{\mu_1+\mu_j-1}
  \prod_{\substack{i=2\\ i\neq j}}^n \operatorname{Tr}M^{\mu_i}
 \Big>
 =  N 
 \Big<
  \operatorname{Tr} V'(M) M^{\mu_1}
  \prod_{i=2}^n \operatorname{Tr}M^{\mu_i}
 \Big> \, .
 \label{loopeq_npt}
\end{empheq}
(See \autoref{exo:loop_multi_matrix} for the case of multi-matrix models.)

\subsubsection{Eigenvalue derivation}

An alternative way to derive loop equations, which has the advantage of being valid for any $\beta\in\mathbb{C}$, is to start from the eigenvalue integral representation
\begin{align}
 \mathcal{Z} = \int_\Gamma \Delta(\lambda)^\beta d\lambda_1 \dots d\lambda_N\   e^{-N\frac\beta{2} \sum_{i=1}^N V(\lambda_i)} \ ,
 \label{mat_int_ev}
\end{align}
where $\Gamma$ is such that the integral absolutely converges, and the integrand vanishes on $\partial \Gamma$.
In this representation, the moments are 
\begin{equation}
\left\langle \operatorname{Tr} M^k \right\rangle  
= \left\langle \sum_{i=1}^N \lambda_i^k \right\rangle
= \frac{1}{\mathcal Z} \int_\Gamma \Delta(\lambda)^\beta d\lambda_1 \dots d\lambda_N\   e^{-N\frac\beta{2} \sum_{i=1}^N V(\lambda_i)}  \sum_{i=1}^N \lambda_i^k \ .
\end{equation}
Loop equations follow from the vanishing of integrals of total derivatives, such as
\begin{multline}
0 =  \sum_{i=1}^N \frac{1}{\mathcal{Z}}\int_\Gamma d\lambda_1 \dots d\lambda_N
\frac{\partial}{\partial \lambda_i} \Big( \Delta(\lambda)^\beta e^{-N\frac\beta{2} \sum_{j=1}^N V(\lambda_j)}  \lambda_i^k \Big)\ ,
\\
= \sum_{i=1}^N \frac{\beta}{2} \left<  \sum_{j(\neq i)} \frac{2\lambda_i^k}{\lambda_i-\lambda_j}- N V'(\lambda_i)  \lambda_i^k + \frac{2k}{\beta} \lambda_i^{k-1} \right> \ .
\end{multline}
The term $\sum_i \sum_{j(\neq i)} \frac{2\lambda_i^k}{\lambda_i-\lambda_j} 
 =\sum_{i\neq j} \frac{\lambda_i^k-\lambda_j^k}{\lambda_i-\lambda_j}$ is actually polynomial in the eigenvalues, and can be written as 
\begin{align}
 \sum_i \sum_{j(\neq i)} \frac{2\lambda_i^k}{\lambda_i-\lambda_j} 
  = - \sum_i k \lambda_i^{k-1}+\sum_{l=0}^{k-1}  \left( \sum_i \lambda_i^l\right) \left(\sum_j \lambda_j^{k-l-1}\right)\ .
\end{align}
This leads to 
\begin{equation}
\left\langle\sum_{l=0}^{k-1}  \left( \sum_i \lambda_i^l\right) \left(\sum_j \lambda_j^{k-l-1}\right) \right\rangle 
+\left(\frac{2}{\beta}-1\right) \left\langle  \sum_i k\lambda_i^{k-1} \right\rangle 
= N  \left\langle  \sum_i V'(\lambda_i)\lambda_i^k \right\rangle\ .
\end{equation}
Rewriting this in terms of a matrix $M$ with eigenvalues $\lambda_i$, 
we obtain a generalization of the loop equation \eqref{eq:1loop} for arbitrary values of $\beta$,
\begin{equation}
\left\langle\sum_{l=0}^{k-1} \operatorname{Tr} M^l  \operatorname{Tr} M^{k-l-1} \right\rangle\
+\left(\frac{2}{\beta}-1\right) k \left\langle\operatorname{Tr} M^{k-1} \right\rangle\
= N  \left\langle  \operatorname{Tr} V'(M) M^k \right\rangle\ .
\label{eq:1loopbeta}
\end{equation}
The loop equation \eqref{loopeq_npt} can similarly be rederived and generalized,
\begin{multline}
 \sum_{l=0}^{\mu_1-1}
 \Big<
  \operatorname{Tr}M^l \operatorname{Tr}M^{\mu_1-l-1}
  \prod_{i=2}^n \operatorname{Tr}M^{\mu_i}
 \Big> 
 + \left( \frac{2}{\beta}-1\right) \mu_1 \Big<
 \operatorname{Tr}M^{\mu_1-1}
  \prod_{i=2}^n \operatorname{Tr}M^{\mu_i}
 \Big> \\
 + \frac{2}{\beta}\sum_{j=2}^n \mu_j
 \Big<
  \operatorname{Tr}M^{\mu_1+\mu_j-1}
  \prod_{\substack{i=2\\ i\neq j}}^n \operatorname{Tr}M^{\mu_i}
 \Big>
 =   N 
 \Big<
  \operatorname{Tr} V'(M) M^{\mu_1}
  \prod_{i=2}^n \operatorname{Tr}M^{\mu_i}
 \Big> \, .
 \label{loopeqbeta_npt}
\end{multline}

\subsubsection{Loop equations for correlation functions}

Let us write the loop equations in terms of the connected and disconnected correlation functions $W_n$ \eqref{wn_conn} and $\widehat{W}_n$ \eqref{wn_disconn}.
Due to the appearance of $\operatorname{Tr}V'(M) M^k $ in
the loop equations, it is useful to introduce 
\begin{align}
 \boxed{ P_n(x;x_1, \ldots, x_n)  = \left(\frac{\beta}{2}\right)^{\frac{n+1}{2}}
 \left< 
  \operatorname{Tr} \frac{V'(x) - V'(M)}{x-M} 
  \prod_{i=1}^n \operatorname{Tr}\frac{1}{x_i - M}
 \right>_\text{c} }
 \, ,
 \label{eq:pndef}
\end{align}
which is a polynomial in its first variable $x$ of degree $d-1$.
Summing over the integer $k$ with a factor $x^{-k-1}$ in the loop equation \eqref{eq:1loopbeta} yields
\begin{align}
 \left< \operatorname{Tr}\frac{1}{x-M} \operatorname{Tr}\frac{1}{x-M} \right> +\left(\tfrac{2}{\beta}-1\right) \left<\operatorname{Tr}\frac{1}{(x-M)^2}\right> - N \left< \operatorname{Tr} \frac{V'(M)}{x-M} \right> = 0\ .
\end{align}
Let us rewrite this equation 
in terms of the connected correlation functions $W_n$ and $P_n$.
In the case $\beta=2$, we find
\begin{align}
W_2(x,x) + W_1(x)^2  = N\left(V'(x)W_1(x) - P_0(x)\right)\, ,
 \label{eq:firstloop}
\end{align}
whose generalization to arbitrary values of $\beta $ is
\begin{align}
 W_2(x,x) + W_1(x)^2 + \left[\sqrt{\tfrac{\beta}{2}}-\sqrt{\tfrac{2}{\beta}}\right]W_1'(x) = N\sqrt{\tfrac{\beta}{2}}\left(V'(x)W_1(x) - P_0(x)\right)\, .
\end{align}
Higher loop equations 
lead to linear equations for disconnected correlation functions $\widehat{W}_n$.
In the case $\beta=2$, the loop equation \eqref{loopeq_npt} leads to
\begin{empheq}[marginbox = \fbox]{multline}
 \widehat{W}_{n+2}(x,x,I) + \sum_{i=1}^n\frac{\partial}{\partial x_i}\frac{\widehat{W}_{n}(x,I \backslash \{x_i\})-\widehat{W}_{n}(I)}{x-x_i} 
 \\
 = N\Big( V'(x)\widehat{W}_{n+1}(x, I) - \widehat{P}_{n}(x;I)\Big)\ , 
 \label{eq:loopdisc}
\end{empheq}
where $I=\{x_1,\cdots x_n\}$, and $\widehat{P}_n$ is the disconnected version of the polynomial $P_n$ \eqref{eq:pndef}.
Rewriting this in terms of the connected correlation functions $W_n$ yields the equations
\begin{empheq}[marginbox = \fbox]{multline}
 W_{n+2}(x,x,I) + \sum_{J\subset I} W_{1+|J|}(x,J)\,W_{1+n-|J|}(x,I \backslash J)
\\
+ \sum_{i=1}^n \frac{\partial }{\partial x_i}\frac{W_n(x,I \backslash \{x_i\})- W_n(I)}{x-x_i} = N \Big( V'(x) W_{n+1}(x,I) - P_n(x;I) \Big)\ ,
\label{eq:loopconn}
\end{empheq}
whose generalization to arbitrary values of $\beta $ is
\begin{multline}
W_{n+2}(x,x,I) + \sum_{J\subset I} W_{1+|I|}(x,J) W_{1+n-|J|}(x,I \backslash J) 
+ \left[ \sqrt{\tfrac\beta{2}}-\sqrt{\tfrac{2}{\beta}}\right] \frac{\partial}{\partial x} W_{n+1}(x,I) \\
+ \sum_{i=1}^n \frac{\partial }{\partial x_i}\frac{W_n(x,I \backslash \{x_i\})- W_n(I)}{x-x_i} 
= N\sqrt{\tfrac{\beta}{2}} \Big( V'(x) W_{n+1}(x,I) - P_n(x;I) \Big)\ .
\end{multline}
This only depends on $N$ and $\beta$ through the prefactors $\sqrt{\frac\beta{2}}-\sqrt{\frac{2}{\beta}}$ and $N\sqrt{\frac{\beta}{2}}$, which are invariant under the duality \eqref{beta_duality}. Therefore, the loop equations are self-dual.

\subsection{Space of solutions}
\label{sec:solsloopeq}

Since loop equations are linear equations for expectation values of symmetric polynomials of the eigenvalues, their solutions form a vector space.
In the case of a polynomial (or rational) potential, and for $\beta=2$, we will show that this space is finite-dimensional, and compute its dimension. 

\subsubsection{Relation between solutions and integration domains}

Given a potential $V$ and integration domain $\Gamma\in H_1(dM\, e^{-N\operatorname{Tr}V(M)})$, let us define the following linear form on the space $\mathcal V_N$ of symmetric polynomials of $N$ variables
\begin{align}\label{def:MGamma}
\begin{array}{cccc}
\mathcal{M}_{V,\Gamma}: \, & 
\mathcal{V}_N  & \longrightarrow & \mathbb C
 \\
& p  &\longmapsto  & \displaystyle \int_\Gamma p(M) \ dM\, e^{-N\operatorname{Tr}V(M)} \ ,
\end{array}
\end{align}
where we used the notation $p(M)= p(\lambda_1,\dots,\lambda_N)$.
As a function of $\Gamma$, this linear form is a linear map from integration domains to linear forms on $\mathcal{V}_N$,
\begin{align}\label{def:MGammapol}
\begin{array}{cccc}
 \mathcal{M}_V: \, &  H_1(dM e^{-N \operatorname{Tr} V(M)}) & \longrightarrow & \mathcal V_N^*
 \\
 & \Gamma & \longmapsto & \mathcal M_{V,\Gamma}\ .
\end{array}
\end{align}
The linear map $\mathcal{M}_V$ is injective. This is proved by showing that if $\Gamma\neq 0$, then we can find a polynomial $p$ such that $\mathcal{M}_{V,\Gamma}( p ) \neq 0$.

Loop equations \eqref{loopeq_npt} can now be interpreted as the subspace $\mathcal L_{V,N} \subset \mathcal V_N$ of symmetric polynomials whose expectation values vanish. If
$ V(M) = \sum_{k=1}^{d+1} \frac{t_k}{k} M^k$, this subspace is spanned by the following combinations of power sum polynomials,
\begin{align}
 \sum_{\ell=0}^{\mu_1-1}p_{(\ell,\mu_1-\ell-1,\mu_2,\dots,\mu_n)} +\sum_{j=2}^n \mu_jp_{(\mu_1+\mu_j-1,\mu_2,\dots,\mu_n)} -N\sum_{k=1}^{d+1} t_k p_{(\mu_1+k-1,\mu_2,\dots,\mu_n)} \ ,
\label{eq:pmloop}
 \end{align}
for any $\mu_1,\dots,\mu_n\geq 0$. 
The space of solutions of the loop equations is
\begin{equation}
\mathcal L_{V,N}^\perp = \big\{ f\in \mathcal V_N^* \,\big|\, \forall p\in \mathcal L_{V,N}, \, f(p)=0\big\} \ .
\end{equation}
Since all matrix integrals satisfy loop equations, we must have
$
\mathrm{ Im}\, \mathcal M_V \subset \mathcal L_{V, N}^\perp
$.
In fact, $\mathcal M_V$ is a bijection between the space of solutions of the loop equations, and the space of integration domains, which have the same finite dimension $d_N$ \eqref{eq:dimh1}:
\begin{align}
\boxed{
\mathcal L_{V, N}^\perp \stackrel{\mathcal M_V}{\sim} H_1(dM e^{-N \operatorname{Tr} V(M)})
 }\ .
 \label{eq:lmh}
\end{align}
To prove this,
consider the space of power sum polynomials of $N$ variables generated by partitions that fit in a box of  at most $N$ rows and $d-1$ columns:
\begin{align}
\mathcal V_{N, d} = \operatorname{Span}\left(\{p_\nu \mid \ell(\nu)\leq N,\nu_i\leq d-1\}\right) \subset \mathcal V_N\ .
\end{align}
Let us show that modulo loop equations, any symmetric polynomial belongs to $\mathcal V_{N, d}$, i.e. $\mathcal V_N = \mathcal L_{V, N} \oplus \mathcal V_{N,d} $.
More specifically, let us show by recursion on $k$ that 
\begin{quote}
any power sum polynomial $p_\mu$ with $|\mu|\leq k$ is a linear combination of loop equations, and elements of $\mathcal V_{N, d}$. 
\end{quote}
This clearly holds for $k=0,1$. 
Assuming this holds for all weights $\leq k-1$, let $p_\nu$ be a power sum polynomial with $|\nu|=k$. 
We can rewrite $p_\nu$ in terms of polynomials with the same weight and length at most $N$, and therefore assume that $\ell(\nu)\leq N$.
If $\nu$ has a row of length $d$ or more, say the first row, then we can rewrite $\nu= (\mu_1+d,\mu_2,\dots,\mu_n)$.
Now the loop equation polynomial \eqref{eq:pmloop} is a combination of $p_{\nu}$, and polynomials of strictly lower weight, that therefore belong to $\mathcal L_{V, N} \oplus \mathcal V_{N,d} $ by the recursion hypothesis. 

So $\mathcal V_N = \mathcal L_{V, N} \oplus \mathcal V_{N,d} $, which implies $\dim \mathcal{L}_{V, N}^\perp \leq \dim \mathcal V_{N,d}$. Our previous inclusion $
\mathrm{ Im}\, \mathcal M_V \subset \mathcal L_{V, N}^\perp
$ implied $\dim \mathrm{ Im}\, \mathcal M\leq \dim \mathcal{L}_{V, N}^\perp $.
But, according to Eq.~\eqref{eq:dimh1} $\dim \mathrm{ Im}\, \mathcal M = \dim H_1(dM e^{-N\operatorname{Tr}V(M)})=d_N=\binom{N+d-1}{N}$. 
This coincides with $\dim \mathcal V_{N,d}$. 
It follows that all our dimensions are in fact equal, and that $
\mathrm{ Im}\, \mathcal M_V = \mathcal L_{V, N}^\perp
$. This proves Eq.~\eqref{eq:lmh}.

\subsection{Split and merge rules}

\subsubsection{Split and merge rules}

The loop equation \eqref{loopeq_npt} can be interpreted in terms of splitting and merging traces: $\operatorname{Tr}M^{\mu_1} \rightarrow \operatorname{Tr}M^l\,\times \, \operatorname{Tr}M^{\mu_1-l-1}$ in the first term, 
$\operatorname{Tr}M^{\mu_1}\,\times \, \operatorname{Tr}M^{\mu_j} \rightarrow \operatorname{Tr}M^{\mu_1+\mu_j-1}$ in the second term. This can be formalized into rules for deriving loop equations in more general matrix models, including multi-matrix models, and matrix integrals that cannot be reduced to eigenvalue integrals. 
These rules specify the Jacobians of certain changes of variables in matrix integrals.
These changes of variables depend on constant matrices $A$ and $B$ that were set to $A=B=\mathrm{ Id}_N$ in the derivation of Eq.~\eqref{loopeq_npt}.
Elementary changes of variables need not keep $M$ Hermitian, as they can be combined into more complicated changes of variables that do.
\begin{itemize}
\item \textbf{Split rule:}\index{split rule}
\begin{equation}
\frac{d(M+\epsilon A M^k B)}{dM} = 1+\epsilon \sum_{l=0}^{k-1} 
\operatorname{Tr} A M^l \operatorname{Tr}M^{k-l-1}B +O(\epsilon^2) \ .
\label{eq:split}
\end{equation}

\item \textbf{Merge rule:}\index{merge rule}
\begin{equation}
\frac{d(M+\epsilon A \operatorname{Tr}B M^k)}{dM} = 
1+\epsilon \sum_{l=0}^{k-1} 
\operatorname{Tr}  A M^l B M^{k-l-1} +O(\epsilon^2)\ .
\label{eq:merge}
\end{equation}

\item \textbf{Chain rule:}\index{chain rule}
When computing the derivative of a product, the $O(\epsilon)$ contribution to the Jacobian is a sum of terms, obtained by applying the split and merge rules to each factor.
For example, if the matrix $A$ in the merge rule is itself a function of $M$, then additional $O(\epsilon)$ terms are obtained by applying the split rule to $A$, while treating $M^k$ and $B$ as constants. 

\end{itemize}

\subsubsection{Generalizations:}

It is useful to sum over $k$ and write split and merge rules for changes of variables that involve $\frac{1}{x-M}$ instead of $M^k$. 
Moreover, split and merge rules can be generalized to the Gaussian ensembles with $\beta=1,4$, and furthermore to arbitrary values of $\beta$. 
(Whenever these rules involve terms that are not functions of the eigenvalues of $M$, they extend the definition of $\beta$-matrix models beyond eigenvalue integrals.)
These generalizations obviously do not modify the chain rule.

\begin{itemize}
\item Split rule at arbitrary $\beta$:
\begin{multline}
\label{eq:splitbeta}
\frac{d(M+\frac\epsilon{2} (A M^k B+ B^\dagger M^k A^\dagger) )}{dM} 
= 1 +\epsilon \tfrac{\beta}{2} \sum_{l=0}^{k-1} 
\operatorname{Tr} A M^l \operatorname{Tr}M^{k-l-1}B  
\\
 + \epsilon \left(1-\tfrac{\beta}{2}\right) \sum_{l=0}^{k-1} 
\operatorname{Tr} A M^l B^\dagger  M^{k-l-1}  
 +O(\epsilon^2) \ .
\end{multline}

\item Split rule in terms of $\frac{1}{x-M}$:
\begin{multline}
\label{eq:splitbetaW}
\frac{d(M+\frac\epsilon{2} (A \frac{1}{x-M} B+ B^\dagger \frac{1}{x-M} A^\dagger)\ )}{dM} 
= 1 +\epsilon \tfrac{\beta}{2} 
\operatorname{Tr} A \frac{1}{x-M} \operatorname{Tr} \frac{1}{x-M} B 
\\
 + \epsilon \left(1-\tfrac{\beta}{2}\right)
\operatorname{Tr} A \frac{1}{x-M} B^\dagger  \frac{1}{x-M}  
 +O(\epsilon^2) \ .
\end{multline}

\item Merge rule at arbitrary $\beta$: Eq.~\eqref{eq:merge} is unchanged.

\item Merge rule in terms of $\frac{1}{x-M}$:

\begin{equation}
\frac{d(M+\epsilon A \operatorname{Tr}B \frac{1}{x-M})}{dM} = 
1+\epsilon 
\operatorname{Tr}  A \frac{1}{x-M} B \frac{1}{x-M} +O(\epsilon^2)\ .
\label{eq:mergebetaW}
\end{equation}

\end{itemize}

\subsection{Loop equations for formal matrices: Tutte's recursion}

Since \autoref{chap:FeynmanGraph}, we know that formal matrix integrals can be interpreted as sums over discrete polygonal surfaces. We will now show that loop equations for formal matrix integrals then amount to splitting and merging faces in such surfaces.

Using the formula \eqref{eq:vom} for the polynomial potential, 
the loop equation \eqref{loopeq_npt} becomes
\begin{multline}
 N
 \Big<
  \operatorname{Tr} M^{\mu_1+1} 
  \operatorname{Tr}M^{\mu_2} \cdots \operatorname{Tr}M^{\mu_n}
 \Big>
 =
 \sum_{j=0}^{\mu_1-1}
 \Big<
  \operatorname{Tr}M^j \operatorname{Tr}M^{\mu_1-j-1}
  \operatorname{Tr}M^{\mu_2} \cdots \operatorname{Tr}M^{\mu_n}
 \Big>
  \\
  +
 \sum_{i=2}^n \mu_i
 \Big<
  \operatorname{Tr}M^{\mu_1+\mu_i-1} 
  \prod_{\substack{j=2 \\ j\neq i}}^n \operatorname{Tr}M^{\mu_j}
 \Big>
 +
 \sum_{k=3}^{\infty} t_k
 \Big<
  \operatorname{Tr}M^{\mu_1+k-1}
  \operatorname{Tr}M^{\mu_2} \cdots \operatorname{Tr}M^{\mu_n}
 \Big> \, .
\end{multline}
We will now interpret each term of this equation as a sum over surfaces, using Eq.~\eqref{eq:nsd}. 
On the left-hand side, we have all discrete surfaces with $n$ marked faces of respective lengths $\mu_1+1,\mu_2,\dots,\mu_n$.
On the right-hand side, we have three terms, corresponding to the three situations that can occur if we erase the marked edge of the first marked face:
\begin{enumerate}
 \item If the erased marked edge is between the first marked face and itself, then we disconnect its boundary into two boundaries of lengths $j$ and $\mu_1-1-j$, and obtain a surface with $n+1$ boundaries. 
To get the correct combinatorics of surfaces, the two ``daughter'' marked faces must have well-defined marked edges, which we choose as the edges preceding and following the erased marked edge. 
So the first term of the right-hand side corresponds to splitting a marked face.
This can either disconnect our surface, or decrease its genus by one:
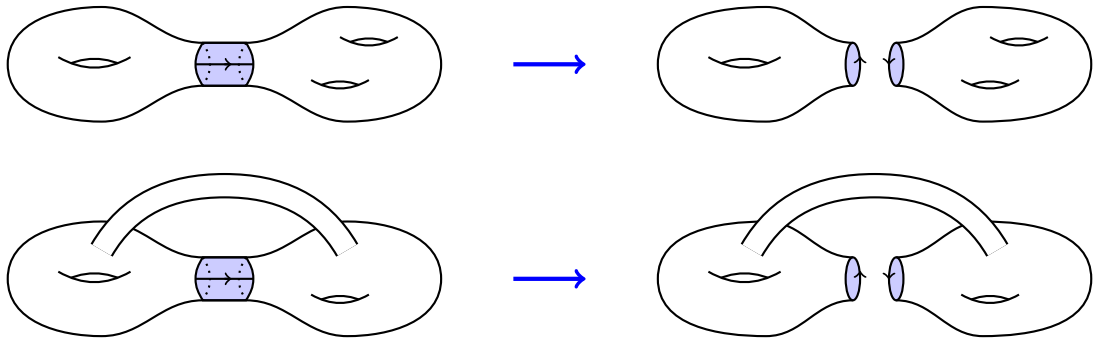
\begin{subequations}
% Splitting face 1
\begin{align}
 \begin{tikzpicture}[thick, scale = .85,baseline=(current bounding box.center)]
   % shape
  \draw (-.3,.3) to [out=left,in=right] (-1.7,.8) to [out=left,in=up]
  (-3,0) to [out=down,in=left] (-1.7,-.8) to [out=right,in=left]
   (-.3,-.3) -- (.3,-.3) to [out=right,in=left] (1.7,-.8) to
   [out=right,in=down] (3,0) to [out=up,in=right] (1.7,.8) to
   [out=left,in=right] (.3,.3) -- cycle;
  % holes
  \begin{scope}[shift={(-1.8,0)}]
   \draw[clip] (-.5,.1) to [bend right] (.5,.1);
   \draw (-.5,-.07) to [bend left] (.5,-.07);
  \end{scope}
  \begin{scope}[shift={(1.6,-.3)},scale=.8]
   \draw[clip] (-.5,.1) to [bend right] (.5,.1);
   \draw (-.5,-.07) to [bend left] (.5,-.07);
  \end{scope}
  \begin{scope}[shift={(2,.3)},scale=.8]
   \draw[clip] (-.5,.1) to [bend right] (.5,.1);
   \draw (-.5,-.07) to [bend left] (.5,-.07);
  \end{scope}
  % face
  \filldraw [fill opacity=.2,fill=blue]
  (-.3,.3) to [out=180+60,in=up] (-.4,0) to [out=down,in=180-60] (-.3,-.3) --
  (.3,-.3) to [out=60,in=down] (.4,0) to [out=up,in=-60] (.3,.3) -- cycle;
  \draw [dotted] 
  (-.3,.3) to [out=-60,in=up] (-.2,0) to [out=down,in=60] (-.3,-.3);
  \draw [dotted] 
  (.3,.3) to [out=180+60,in=up] (.2,0) to [out=down,in=180-60] (.3,-.3);
  \draw[-latex] (-.4,0) -- (.1,0);
  \draw (.4,0) -- (0,0);
  % intermediate arrow
%
  \draw [ultra thick,blue,-latex] (4,0) -- ++(1,0);
  \begin{scope}[shift={(9,0)}]
   % shape
  \draw (-.3,.3) to [out=left,in=right] (-1.5,.8) to [out=left,in=up]
  (-3,0) to [out=down,in=left] (-1.5,-.8) to [out=right,in=left] (-.3,-.3);
  \draw (.3,.3) to [out=right,in=left] (1.5,.8) to [out=right,in=up]
  (3,0) to [out=down,in=right] (1.5,-.8) to [out=left,in=right] (.3,-.3);
  % holes
  \begin{scope}[shift={(-1.8,0)}]
   \draw[clip] (-.5,.1) to [bend right] (.5,.1);
   \draw (-.5,-.07) to [bend left] (.5,-.07);
  \end{scope}
  \begin{scope}[shift={(1.6,-.3)},scale=.8]
   \draw[clip] (-.5,.1) to [bend right] (.5,.1);
   \draw (-.5,-.07) to [bend left] (.5,-.07);
  \end{scope}
  \begin{scope}[shift={(2,.3)},scale=.8]
   \draw[clip] (-.5,.1) to [bend right] (.5,.1);
   \draw (-.5,-.07) to [bend left] (.5,-.07);
  \end{scope}   
   % marked faces
   \filldraw [fill opacity = .2,fill=blue]
   (-.3,0) circle [x radius=.1, y radius=.3];
   \filldraw [fill opacity = .2,fill=blue]
   (.3,0) circle [x radius=.1, y radius=.3];
   % marked arrows
%   \draw[-latex] (-.4,0) -- (.1,0);
%   \draw (.4,0) -- (0,0);
%
   \draw[-latex] (-.2,.1) -- ++(0,.01);
   \draw[-latex] (.2,0) -- ++(0,-.01);
  \end{scope}
 \end{tikzpicture}
\end{align}

% Splitting face 2
\begin{align}
 \begin{tikzpicture}[thick, scale = .85,baseline=(current bounding box.center)]
   % shape
  \draw (-.3,.3) to [out=left,in=right] (-1.7,.8) to [out=left,in=up]
  (-3,0) to [out=down,in=left] (-1.7,-.8) to [out=right,in=left]
   (-.3,-.3) -- (.3,-.3) to [out=right,in=left] (1.7,-.8) to
   [out=right,in=down] (3,0) to [out=up,in=right] (1.7,.8) to
   [out=left,in=right] (.3,.3) -- cycle;
  % holes
  \begin{scope}[shift={(-1.8,0)}]
   \draw[clip] (-.5,.1) to [bend right] (.5,.1);
   \draw (-.5,-.07) to [bend left] (.5,-.07);
  \end{scope}
  \begin{scope}[shift={(1.6,-.3)},scale=.8]
   \draw[clip] (-.5,.1) to [bend right] (.5,.1);
   \draw (-.5,-.07) to [bend left] (.5,-.07);
  \end{scope}
%
%  \begin{scope}[shift={(2,.3)},scale=.8]
%   \draw[clip] (-.5,.1) to [bend right] (.5,.1);
%   \draw (-.5,-.07) to [bend left] (.5,-.07);
%  \end{scope}
%
  % handle
  \draw[double,double distance=8]
  (-1.7,.4) to [out=60,in=left] (0,1.3) to [out=right,in=180-60] (1.7,.4);
  % face
  \filldraw [fill opacity=.2,fill=blue]
  (-.3,.3) to [out=180+60,in=up] (-.4,0) to [out=down,in=180-60] (-.3,-.3) --
  (.3,-.3) to [out=60,in=down] (.4,0) to [out=up,in=-60] (.3,.3) -- cycle;
  \draw [dotted] 
  (-.3,.3) to [out=-60,in=up] (-.2,0) to [out=down,in=60] (-.3,-.3);
  \draw [dotted] 
  (.3,.3) to [out=180+60,in=up] (.2,0) to [out=down,in=180-60] (.3,-.3);
  \draw[-latex] (-.4,0) -- (.1,0);
  \draw (.4,0) -- (0,0);
  % intermediate arrow
  \draw [ultra thick,blue,-latex] (4,0) -- ++(1,0);
  \begin{scope}[shift={(9,0)}]
   % shape
  \draw (-.3,.3) to [out=left,in=right] (-1.5,.8) to [out=left,in=up]
  (-3,0) to [out=down,in=left] (-1.5,-.8) to [out=right,in=left] (-.3,-.3);
  \draw (.3,.3) to [out=right,in=left] (1.5,.8) to [out=right,in=up]
  (3,0) to [out=down,in=right] (1.5,-.8) to [out=left,in=right] (.3,-.3);
  % holes
  \begin{scope}[shift={(-1.8,0)}]
   \draw[clip] (-.5,.1) to [bend right] (.5,.1);
   \draw (-.5,-.07) to [bend left] (.5,-.07);
  \end{scope}
  \begin{scope}[shift={(1.6,-.3)},scale=.8]
   \draw[clip] (-.5,.1) to [bend right] (.5,.1);
   \draw (-.5,-.07) to [bend left] (.5,-.07);
  \end{scope}
%
%  \begin{scope}[shift={(2,.3)},scale=.8]
%   \draw[clip] (-.5,.1) to [bend right] (.5,.1);
%   \draw (-.5,-.07) to [bend left] (.5,-.07);
%  \end{scope}
%
  % handle
  \draw[double,double distance=8]
  (-1.7,.4) to [out=60,in=left] (0,1.3) to [out=right,in=180-60] (1.7,.4);
   % marked faces
   \filldraw [fill opacity = .2,fill=blue]
   (-.3,0) circle [x radius=.1, y radius=.3];
   \filldraw [fill opacity = .2,fill=blue]
   (.3,0) circle [x radius=.1, y radius=.3];
   % marked arrows
%
%   \draw[-latex] (-.4,0) -- (.1,0);
%   \draw (.4,0) -- (0,0);
%
   \draw[-latex] (-.2,.1) -- ++(0,.01);
   \draw[-latex] (.2,0) -- ++(0,-.01);
  \end{scope}
 \end{tikzpicture}
\end{align}
\end{subequations}
\item If the erased marked edge separates two different marked faces, then erasing it amounts to merging them, and we obtain a marked face with $\mu_1+\mu_i-1$ edges:
% Merging faces 1
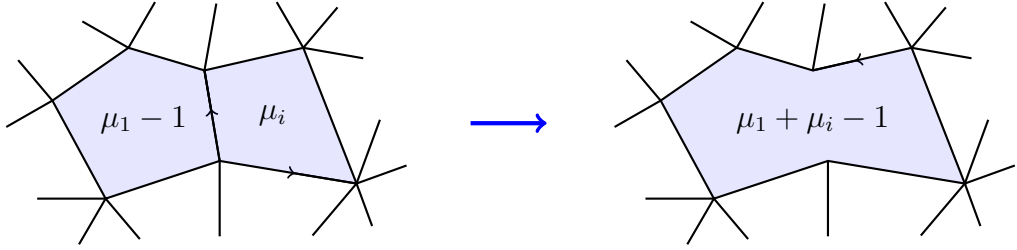
\begin{align}
 \begin{tikzpicture}[baseline=(current bounding box.center)]
  % arrows
  \draw[thick,latex-] (.6,.1) -- (.5,.7);
  \draw[thick,latex-] (1.6,-.65) -- (2.5,-.8);
  % marked face
  \filldraw [thick,fill opacity=.1,fill=blue]
  (-.5,1) -- (-1.5,.3) -- (-.8,-1) -- (.7,-.5) -- (.5,.7) -- cycle;
  \node at (-.3,.05) {$\mu_1-1$};
  % neighboring face
  \filldraw [thick,fill opacity=.1,fill=blue]
  (.5,.7) -- (.7,-.5) -- (2.5,-.8) -- (1.8,1) -- cycle;
  \node at (1.4,0.1) {$\mu_i$};
  % edges going out
  \draw [thick] (-.5,1) -- ++(60:.7);
  \draw [thick] (-.5,1) -- ++(150:.7);
  \draw [thick] (-1.5,.3) -- ++(130:.7);
  \draw [thick] (-1.5,.3) -- ++(210:.7);
  \draw [thick] (-.8,-1) -- ++(180:.9);
  \draw [thick] (-.8,-1) -- ++(240:.7);
  \draw [thick] (-.8,-1) -- ++(310:.7);
  \draw [thick] (.7,-.5) -- ++(270:1);
  \draw [thick] (2.5,-.8) -- ++(230:.9);
  \draw [thick] (2.5,-.8) -- ++(290:.6);
  \draw [thick] (2.5,-.8) -- ++(20:.7);
  \draw [thick] (2.5,-.8) -- ++(70:.9);
  \draw [thick] (1.8,1) -- ++(-10:.8);
  \draw [thick] (1.8,1) -- ++(50:.7);
  \draw [thick] (1.8,1) -- ++(120:.7);
  \draw [thick] (.5,.7) -- ++(80:.9);
  % intermediate arrow
  \draw [ultra thick,blue,-latex] (4,0) -- ++(1,0);
  \begin{scope}[shift={(8,0)}]
   % arrow
   \draw [thick,latex-] (1.15,.85) -- (.5,.7);
  % merged face
  \filldraw [thick,fill opacity=.1,fill=blue]
  (-.5,1) -- (-1.5,.3) -- (-.8,-1) -- (.7,-.5) -- (2.5,-.8) -- (1.8,1)
   -- (.5,.7) -- cycle;
  \node at (0.5,.05) {$\mu_1+\mu_i-1$};
  % edges going out
  \draw [thick] (-.5,1) -- ++(60:.7);
  \draw [thick] (-.5,1) -- ++(150:.7);
  \draw [thick] (-1.5,.3) -- ++(130:.7);
  \draw [thick] (-1.5,.3) -- ++(210:.7);
  \draw [thick] (-.8,-1) -- ++(180:.9);
  \draw [thick] (-.8,-1) -- ++(240:.7);
  \draw [thick] (-.8,-1) -- ++(310:.7);
  \draw [thick] (.7,-.5) -- ++(270:1);
  \draw [thick] (2.5,-.8) -- ++(230:.9);
  \draw [thick] (2.5,-.8) -- ++(290:.6);
  \draw [thick] (2.5,-.8) -- ++(20:.7);
  \draw [thick] (2.5,-.8) -- ++(70:.9);
  \draw [thick] (1.8,1) -- ++(-10:.8);
  \draw [thick] (1.8,1) -- ++(50:.7);
  \draw [thick] (1.8,1) -- ++(120:.7);
  \draw [thick] (.5,.7) -- ++(80:.9);
  \end{scope}
\end{tikzpicture}
\end{align}
We choose the edge following the erased edge as the marked edge of the new marked face.
By doing that, we forget the combinatorial information which was encoded in the marked edge of the $i$-th marked face.
So we must restore a factor $\mu_i$. 
This explains the second term of the right-hand side.
\item
If there is an unmarked $k$-gon on the other side of the marked edge, then the number of marked faces does not change, but the first marked face becomes a $\mu_1+k-1$-gon:
% Merging faces 2
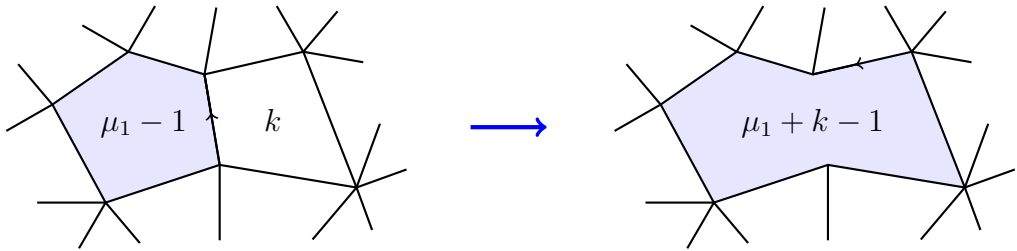
\begin{align}
 \begin{tikzpicture}[baseline=(current bounding box.center)]
  % arrows
  \draw[thick,latex-] (.6,.1) -- (.5,.7);
  % marked face
  \filldraw [thick,fill opacity=.1,fill=blue]
  (-.5,1) -- (-1.5,.3) -- (-.8,-1) -- (.7,-.5) -- (.5,.7) -- cycle;
  \node at (-.3,.05) {$\mu_1-1$};
  % neighboring face
  \draw [thick]
  (.5,.7) -- (.7,-.5) -- (2.5,-.8) -- (1.8,1) -- cycle;
  \node at (1.4,0.1) {$k$};
  % edges going out
  \draw [thick] (-.5,1) -- ++(60:.7);
  \draw [thick] (-.5,1) -- ++(150:.7);
  \draw [thick] (-1.5,.3) -- ++(130:.7);
  \draw [thick] (-1.5,.3) -- ++(210:.7);
  \draw [thick] (-.8,-1) -- ++(180:.9);
  \draw [thick] (-.8,-1) -- ++(240:.7);
  \draw [thick] (-.8,-1) -- ++(310:.7);
  \draw [thick] (.7,-.5) -- ++(270:1);
  \draw [thick] (2.5,-.8) -- ++(230:.9);
  \draw [thick] (2.5,-.8) -- ++(290:.6);
  \draw [thick] (2.5,-.8) -- ++(20:.7);
  \draw [thick] (2.5,-.8) -- ++(70:.9);
  \draw [thick] (1.8,1) -- ++(-10:.8);
  \draw [thick] (1.8,1) -- ++(50:.7);
  \draw [thick] (1.8,1) -- ++(120:.7);
  \draw [thick] (.5,.7) -- ++(80:.9);
  % intermediate arrow
  \draw [ultra thick,blue,-latex] (4,0) -- ++(1,0);
  \begin{scope}[shift={(8,0)}]
   % arrow
   \draw [thick,latex-] (1.15,.85) -- (.5,.7);
  % merged face
  \filldraw [thick,fill opacity=.1,fill=blue]
  (-.5,1) -- (-1.5,.3) -- (-.8,-1) -- (.7,-.5) -- (2.5,-.8) -- (1.8,1)
   -- (.5,.7) -- cycle;
  \node at (0.5,.05) {$\mu_1+k-1$};
  % edges going out
  \draw [thick] (-.5,1) -- ++(60:.7);
  \draw [thick] (-.5,1) -- ++(150:.7);
  \draw [thick] (-1.5,.3) -- ++(130:.7);
  \draw [thick] (-1.5,.3) -- ++(210:.7);
  \draw [thick] (-.8,-1) -- ++(180:.9);
  \draw [thick] (-.8,-1) -- ++(240:.7);
  \draw [thick] (-.8,-1) -- ++(310:.7);
  \draw [thick] (.7,-.5) -- ++(270:1);
  \draw [thick] (2.5,-.8) -- ++(230:.9);
  \draw [thick] (2.5,-.8) -- ++(290:.6);
  \draw [thick] (2.5,-.8) -- ++(20:.7);
  \draw [thick] (2.5,-.8) -- ++(70:.9);
  \draw [thick] (1.8,1) -- ++(-10:.8);
  \draw [thick] (1.8,1) -- ++(50:.7);
  \draw [thick] (1.8,1) -- ++(120:.7);
  \draw [thick] (.5,.7) -- ++(80:.9);
  \end{scope}
 \end{tikzpicture}
\end{align}
We choose the edge that follows the erased edge as our new marked edge. 
We must now include a factor $t_k$ -- the weight of the unmarked $k$-gon. 
\end{enumerate}
Applying this edge-erasing procedure recursively, all the discrete surfaces can be
reduced to surfaces with very few edges. 
Counting discrete surfaces by recursively erasing edges is called \textbf{Tutte's recursion}\index{Tutte's recursion}. And we have just seen that Tutte's recursion is equivalent to repeatedly using loop equations in formal matrix integrals.

This equivalence, and the view of polygon boundaries as loops, explain why the Schwinger--Dyson equations for random matrices were called loop equations by Migdal in 1984 \cite{Migdal:1984gj}.

\section{Topological expansion and perturbative solutions}\label{sec:teps}

We define the \textbf{topological expansion}\index{topological expansion} of a connected correlation function as the formal asymptotic expansion in powers of $N^2$,
\begin{align}
 \boxed{ W_n = \sum_{g=0}^\infty N^{2-2g-n}W_{g,n} } \ . 
 \label{eq:wgnexp}
\end{align}
In the case $n=0$, we identify $W_0$ with the free energy $F$. With the notation $F_g = W_{g,0}$, the topological expansion reads
\begin{align}
\boxed{ F = \sum_{g=0}^\infty N^{2-2g} F_g } \ .
 \label{eq:logzexp}
\end{align}
Correlation functions of random matrix models do not always have topological expansions.
However, for the moment, let us look for solutions of loop equations that do have topological expansions.
Such solutions will be called \textbf{perturbative solutions}\index{perturbative solution (of loop equations)}.

\subsection{Structure of the topological expansion}\label{sec:ste}

\subsubsection{Topological structure}

The name topological expansion comes from the case of formal matrix integrals. 
In this case, $W_n$ is a formal series that has a topological expansion \eqref{eq:wns}, where $W_{g,n}$ is the generating function counting discrete surfaces of genus $g$ with $n$ boundaries. Let us adopt this topological interpretation of the indices $(g,n)$ not only in the name of the expansion, but also as a graphical notation of coefficients as surfaces. For example, we will represent $F_1$ as a torus, $W_{0,1}$ as a disc, $W_{0,2}$ as a cylinder, and $W_{1,1}$ as a torus minus a disc:
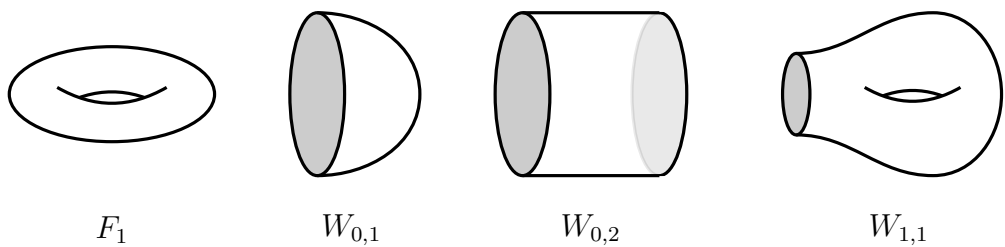
\begin{align}
\begin{tikzpicture}[very thick, scale = .9,baseline=(current bounding box.center)]
\draw (0,0) circle [x radius = 1.5, y radius = .7];
\begin{scope}
\draw[clip] (-.8, .1) to [bend right] (.8, .1);
\draw (-.8, -.2) to [bend left] (.8, -.2);
\end{scope}
\node at (0, -2) {$F_{1}$};
\begin{scope}[shift={(3,0)}]
\filldraw[fill opacity = .2] (0,0) circle [x radius = .4, y radius = 1.2];
\draw (0, -1.2) to [out = 0, in = -90] (1.5,0) to [out = 90, in = 0] (0, 1.2); 
\node at (.5, -2) {$W_{0, 1}$}; 
\end{scope}
\begin{scope}[shift= {(6,0)}]
\filldraw[fill opacity = .2] (0,0) circle [x radius = .4, y radius = 1.2];
\filldraw[opacity = .09] (2,0) circle [x radius = .4, y radius = 1.2];
\draw (0, 1.2) -- (2, 1.2);
\draw (0, -1.2) -- (2, -1.2);
\node at (1, -2) {$W_{0, 2}$};
\clip (2, 1.2) -- (2, -1.2) -- (3, -1.2) -- (3, 1.2) -- cycle;
\draw (2, 0) circle [x radius = .4, y radius = 1.2];
\end{scope}
\begin{scope}[shift = {(10, 0)}]
 \filldraw[fill opacity = .2] (0,0) circle [x radius = .2, y radius = .6];
 \draw (0, .6) to [out = 0, in = 180] (2, 1.2) to [out = 0, in = 90] (3, 0) to [out = -90, in = 0] (2, -1.2) to [out = 180, in = 0] (0, -.6);
\node at (1.5, -2) {$W_{1,1}$};
\draw[clip] (1, .1) to [bend right] (2.4, .1);
\draw (1, -.15) to [bend left] (2.4, -.15);
\end{scope}
\end{tikzpicture}
\end{align}
We will call a surface stable or unstable depending on 
the sign of its
Euler characteristic $\chi=2-2g-n$, and extend this terminology to the corresponding pairs $(g,n)$ and coefficients $W_{g,n}$:
\begin{itemize}
\item $(g,n)$ is \textbf{unstable}\index{unstable (surface)} if $\chi\geq 0$. The four unstable surfaces are the sphere $(0,0)$, disc $(0,1)$, cylinder $(0,2)$ and torus $(1,0)$.
\item $(g,n)$ is \textbf{stable}\index{stable (surface)} if $\chi<0$.
\end{itemize}
As we will see in \autoref{sec:toprec}, loop equations will allow us to recursively determine all the stable coefficients $W_{g,n}$ from the disc and cylinder only.

\subsubsection{Recursive structure}
 
The loop equations \eqref{eq:loopconn} are not closed equations: to compute $W_1$ we need $W_2$, for $W_2$ we need $W_3$, and so on. 
However, inserting the topological expansion, we will obtain a triangular system of equations for the coefficients $W_{g,n}$.
We will demonstrate this for a few low values of $g$ and $n$ in the present  \autoref{sec:teps}, before deriving the general solution of these closed equations using topological recursion in \autoref{sec:toprec}.

To solve the equations for $W_{g,n}$, we need to know the polynomials $P_n$ that appear in the loop equations.
These polynomials have the topological expansions 
\begin{align}\label{defPgn}
 \boxed{ P_n = \sum_{g=0}^\infty N^{1-2g-n} P_{g,n} } \ .
\end{align}
The polynomial $P_{g,n}$ appears in the equation for $W_{g,n+1}$, and in particular $W_{0,1}$ depends on $P_{0,0}$. 
Each choice of polynomials $P_{g, n}$ such that  the resulting $W_{g, n}(x_1,\dots x_n)$ are symmetric under permutations of $x_i$, yields a perturbative solution of the loop equations.

\subsubsection{Linear structure}

Loop equations are linear equations for the disconnected correlation functions $\widehat{W}_n$, whose space of solutions has the dimension $d_N$ \eqref{eq:dimh1} for any given $N$. 
In order to generate all solutions, we therefore only need to find at least $d_N$ independent solutions, and
we will look for solutions that have topological expansions.
To do this, we introduce particular choices of polynomials $P_{g,n}$:
\begin{itemize}

\item $P_{0,0}$ is an arbitrary polynomial of degree $d-1$ such that $P_{0,0}(x)\underset{x\to \infty}{\sim} \frac{V'(x)}{x}$. 
Choosing $P_{0,0}$ is equivalent to choosing the set of \textbf{filling fractions}\index{filling fraction}
\begin{align}
\boxed{ \epsilon_i = \frac{1}{2\pi i}\oint_{{\mathcal A}_i} W_{0,1}(x)dx } \quad i=1,\dots,d \ .
\label{eq:ecff}
\end{align}
Here $(\mathcal{A}_i)$ are cycles that surround the cuts of $W_{0,1}$, whose 
sum is homologically equivalent to a countour around $\infty$. 
We have 
$\operatorname{Res}_\infty W_{0, 1} = -1$ according to  \eqref{eq:wninf}, and this leads to the constraint $\sum_i \epsilon_i = 1$.

\item Given a choice of $P_{0,0}$ and $(\mathcal A_i)$, for $(g,n)\neq (0,0)$ the polynomials $P_{g,n}(x_1;x_2,\dots,x_n)$ are the unique polynomials of $x_1$ such that
\begin{align}
\renewcommand{\arraystretch}{1.5}
\boxed{ \begin{array}{l}
         \deg P_{g,n}\leq d-2 \, ,
         \\ \displaystyle
         \forall\,(g,n)\neq (0,1)\,, \quad \oint_{{\mathcal A}_i} W_{g,n}(x_1,\dots,x_n)dx_1 = 0\, .
        \end{array}}
        \label{eq:iad}
\end{align}
\end{itemize}
These assumptions on $P_{g,n}$ are obeyed in the case of topological expansions of formal matrix integrals. The uniqueness of these polynomials will follow from the uniqueness of the solution of the loop equations under these assumptions, as demonstrated in \autoref{sec:higherWgnJoukowsky} in the 1-cut case and beyond.

The existence of a topological expansion is a linear condition on the connected correlation functions $W_n$, which are nonlinear combinations of $\widehat{W}_n$. So a linear combination of solutions that have a topological expansion, does not in general have a topological expansion.
(See \autoref{sec:non-perturbative}.)

\subsection{First few orders}\label{sec:W11}

\subsubsection{Determining \texorpdfstring{$W_{0,1}$}{W01}} 

The first loop equation \eqref{eq:firstloop} is not a closed equation for $W_1(x)$, as it also involves $W_2(x,x)$. 
If however we perform the topological expansion, the coefficient of $N^{2-2g}$ reads 
\begin{align}
 \sum_{h_1+h_2=g} W_{h_1,1}(x) W_{h_2,1}(x) + W_{g-1,2}(x,x) = V'(x) W_{g,1}(x) -P_{g,0}(x)\ .
 \label{loop1g}
\end{align}
In the case $g=0$, the term $W_{g-1,2}(x,x)$ is actually absent, and we obtain
\begin{align}
 W_{0,1}(x)^2 = V'(x)W_{0,1}(x) - P_{0,0}(x)\ .
 \label{eq:w01eq}
\end{align}
This is  a closed equation for $W_{0,1}$, for any given choice of $P_{0,0}$. 
This equation is formally identical to the saddle-point equation \eqref{eq:pvo} for the resolvent $\overline{W}$, and we reproduce the solution of \autoref{chap:SaddlePoint} with $\overline P = P_{0,0} $
\begin{align}
\boxed{ W_{0,1}
= \frac{1}{2}\left( V' - M\sqrt{\sigma}\right) \qquad \text{where} \qquad  (V')^2-4P_{0,0}=M^2\sigma} \ ,
\label{eq:w01x}
\end{align}
where $M(x)$ and $\sigma(x)$ are polynomials, with $\sigma(x)$ having only simple roots. 
So $W_{0, 1}$ is an algebraic function of $x$, with a finite number of cuts in the complex $x$-plane. (See \autoref{exo:loop_On_matrix} for the case of the $O(n)$ matrix model, and \autoref{exooopeqtrv} for matrix integrals on contours with boundaries.)

\subsubsection{Case of formal matrix integrals}

In this case, correlation functions are formal series in the coefficients $t_k$ of the potential $V(x)=\frac12 x^2+O(t_k)$.
The case $t_k=0$ is the Gaussian matrix model, and therefore
\begin{equation}\label{eq:W01formalleading}
W_{0,1}(x) = \frac{1}{2}\left( x - \sqrt{x^2-4} \right) + O(t_k)\ .
\end{equation}
(This agrees with Eq.~\eqref{eq:wns}, after preforming the sum over the number $\mu_1$ of edges of the marked face.)
According to Tutte's recursion, there is a unique solution of the loop equations that is a formal series in $t_k$, such that $W_{0,1}$ obeys Eq.~\eqref{eq:W01formalleading}. 
This solution can be determined by assuming that $W_{0,1}(x)$ still has one cut beyond the Gaussian limit. 
This amounts to letting the parameters of the Joukowsky map $x(z)=\alpha +\gamma(z+\frac{1}{z})$ be formal series $\alpha=O(t_k)$ and $\gamma=1+O(t_k)$. Then the coefficients $v_j$ of $W_{0,1}(x(z))=\sum_{j=1}^d v_jz^{-j}$ are formal series too, and are determined by Eq.~\eqref{eq:vkVz}. The corresponding polynomial $P_{0,0}(x)$ is given in terms of the Chebyshev polynomials $T_k(2\cosh x)=2\cosh(k x)$  by
\begin{equation}
P_{0,0}(x) = \sum_{k=0}^{d-1} \left(\sum_{j=1}^{d-k} v_j v_{j+k}\right) \left( T_k\big(\tfrac{x-\alpha}{\gamma}\big) -\delta_{k,0} \right)\ .
\end{equation}

\subsubsection{Determining \texorpdfstring{$W_{0, 2}$}{W02}}

In the topological expansion, the 
leading term of the $n=1$ loop equation \eqref{eq:loopconn} is 
\begin{align}
 2W_{0,2}(x,x')W_{0,1}(x) +\frac{\partial}{\partial x'} \frac{W_{0,1}(x) - W_{0,1}(x')}{x-x'} = V'(x) W_{0,2}(x,x') - P_{0,1}(x;x')\ .
\end{align}
Now that we know $W_{0,1}$, this is a closed equation for $W_{0,2}$, assuming we also know $P_{0,1}$. The solution is
\begin{align}
W_{0,2}(x,x')
 = -\frac{1}{2(x-x')^2} +
\frac{\displaystyle
\frac{\partial}{\partial x'} \frac{ M(x')\sqrt{\sigma(x')}}{2(x-x')} + \frac{\partial}{\partial x'} \frac{V'(x) - V'(x')}{2(x-x')} + P_{0,1}(x;x')}{M(x)\sqrt{\sigma(x)}} 
\ .
\label{eq:w02sol}
\end{align}
Since $P_{0,1}(x;x')$ is a polynomial function of $x$, $W_{0,2}(x, x')$ is an
algebraic function of $x$, and by symmetry also of $x'$. 
Let us choose $P_{0, 1}$ according to our assumptions \eqref{eq:iad}. 
To begin with, $W_{0, 2}(x, x')$ must have no poles at the zeros of $M(x)$, and therefore be singular only at the spectral edges -- the zeros of $\sigma(x)$. Since $M(x)$ has $d-s$ zeros where $s$ is the number of cuts, and $P_{0, 1}$ has $d-1$ unknown coefficients,
this determines $P_{0, 1}$ if $s=1$. If $s>1$, we need to further impose that $W_{0,2}$ has vanishing $\mathcal A$-cycle integrals.
In every case, the resulting $W_{0, 2}$ obeys the same Riemann--Hilbert equation \eqref{eq:rhot} as the saddle-point limit of the two-point function $\overline{W}_2$, and therefore must have an expression of the same form \eqref{saddle2pt}. 

All this shows that $(W_{0,2}(x,x')+1/(x-x')^2)dx dx'$ is a fundamental second kind differential normalized on $A$-cycles described in \autoref{sec:two-point_func},
\begin{align}
W_{0,2}(x,x') dx dx ' + \frac{dx dx'}{(x-x')^2} = B.
\end{align}

\subsubsection{Determining \texorpdfstring{$W_{1,1}$}{W11}}

Once we know $W_{0,1}$ and $W_{0,2}$, we can find $W_{1,1}$ using the case $g=1$ of the first loop equation \eqref{loop1g},
\begin{align}
 2W_{0,1}(x) W_{1,1}(x) + W_{0,2}(x,x) = V'(x) W_{1,1}(x) - P_{1,0}(x)\ ,
 \label{eq:w11}
\end{align}
whose solution is 
\begin{align}
 W_{1,1}(x) = \frac{W_{0,2}(x, x) + P_{1,0}(x)}{M(x)\sqrt{\sigma(x)}}
 \ .
 \label{eq:w11e}
\end{align}
This proves that $W_{1,1}(x)$ is an algebraic function, whose sign flips when we cross a cut. 
In order to determine it, we should choose $P_{1,0}$. 
As in the case of $W_{0, 2}$, this polynomial is uniquely determined by our assumptions \eqref{eq:iad}.

\subsection{One-cut case}\label{sec:higherWgnJoukowsky}

In the 1-cut case, we can use the Joukowsky map $x(z)$ from the Riemann sphere to the complex $x$-plane.
Let us rewrite correlation functions as differential forms on the Riemann sphere,
\begin{equation}\label{eq:wdsc}
\boxed{ \omega_{g,n}(z_1,\dots,z_n) = W_{g,n}(x_1,\dots,x_n) dx_1 \dots dx_n + \delta_{g,0}\delta_{n,2} \frac{dx_1 dx_2}{(x_1-x_2)^2} }\ ,
\end{equation}
where $x_i=x(z_i)$.
We are looking for a solution of the loop equations that obeys our assumptions \eqref{eq:iad}, so that 
\begin{itemize}
\item $\omega_{g,n}(z_1,\dots,z_n)$ is a rational, symmetric function of $z_1,\dots,z_n$,

\item $\omega_{0,1}(z) = \frac12 \left( V'(x(z)) - M(x(z)) \gamma(z-1/z) \right) x'(z)dz$,

\item $\omega_{0,2}(z_1,z_2) = B(z_1,z_2) = \frac{dz_1 dz_2}{(z_1-z_2)^2}$,

\item for $2g-2+n>0$, there are poles at $z_i=\pm 1$ only,

\item we have the antisymmetry property
\begin{equation}\label{eq:omgnantisym}
2g-2+n>0\quad \implies \quad \omega_{g,n}(\tfrac{1}{z_1},\dots,z_n) = -\omega_{g,n}(z_1,\dots,z_n)\ .
\end{equation}
\end{itemize}

\subsubsection{Determining \texorpdfstring{$W_{1,1}$}{W11}}

%Since $W_{0,2}$ coincides with the saddle-point two-point function $\overline W_2$, it has a universal expression \eqref{eq:W2barBJ} in terms of the fundamental second kind differential of the Riemann sphere
%\begin{align}
% \omega_{0,2}=  B\left(z, z'\right) = \frac{dz \ dz'}{(z-z')^2} \ .
%\label{eq:w02}
% \end{align}
Inserting  $\omega_{0,2}=  B\left(z, z'\right) = \frac{dz \ dz'}{(z-z')^2} $  into  \eqref{eq:w11e} for $W_{1,1}$, we find 
\begin{align}
\omega_{1,1}(z) = \frac{-B(z,\frac{1}{z}) + P_{1,0}(x(z)) dx(z)^2 }{-w(z)dx(z)}\ ,
 \label{eq:w11z}
\end{align}
where we introduced the function
\begin{align}
w(z) = - M(x(z))\sqrt{\sigma(x(z))} 
\ .
\label{eq:woz}
\end{align}
Knowing that $\omega_{1,1}(z)$ has poles at $z=\pm 1$ only and obeys $\omega_{1,1}(\frac{1}{z})=-\omega_{1,1}(z)$, we use the Cauchy residue formula and compute
\begin{align}
\omega_{1,1}(z) &= \frac{dz}{2\pi i}\oint_z \frac{1}{z'-z} \omega_{1,1}(z') = \sum_\pm \frac{dz}{2\pi i}\oint_{\pm 1}\frac{1}{z-z'}  \omega_{1,1}(z')\ ,
 \\
 & = \frac12\sum_\pm \frac{dz}{2\pi i}\oint_{\pm 1} \left(\frac{1}{z-z'}-\frac{1}{z-\frac{1}{z'}}\right)\omega_{1,1}(z')\ . 
 \label{oooz}
\end{align}
Inserting Eq.~\eqref{eq:w11z} for $\omega_{1,1}(z')$, the contribution of $P_{1,0}$ now vanishes, because $x(z')$ and polynomials thereof have no poles at $z'=\pm 1$. (So we do not need to compute $P_{1, 0}$.)
Introducing the recursion kernel
\begin{align}
K(z,z') = \frac{1}{2 w(z')dx(z')}\int_{\frac{1}{z'}}^{z'} B(z, \cdot)= \frac{1}{2 w(z')dx(z')} \left(\frac{1}{z-z'} - \frac{1}{z-\frac{1}{z'}}\right) dz \ ,
\label{r_kernel1}
\end{align}
we finally find 
\begin{align}
\omega_{1,1}(z) =  \sum_\pm \frac{1}{2\pi i}\oint_{\pm 1} K(z,z')B\left(z',\frac{1}{z'}\right)\ .
\label{eq:W11KB}
\end{align}
Explicitely evaluating the residues at $z'=\pm 1$, we get
\begin{equation}
\omega_{1,1}(z)
= -\frac{1}{32\gamma} \sum_{\eta=\pm 1} \frac{\eta}{w'(\eta)} \left(
\frac{1}{(z-\eta)^4} +\frac{\eta}{(z-\eta)^3} 
- \frac{1+\eta \frac{w''(\eta)}{w'(\eta)} + \frac{w'''(\eta)}{3 w'(\eta)} }{2 (z-\eta)^2} \right) dz\ .
\end{equation}

\subsubsection{Diagrammatic representation}

Let us introduce a diagrammatic representation of Eq.~\eqref{eq:W11KB}. We need two kinds of edges, associated to $B$ and $K$ respectively. The edge for $B(z,z')=B(z',z)$ is depicted as an unoriented edge, while the edge for $K(z, z')=K(z,\frac{1}{z'})$ is depicted as an oriented edge from $z$ to $z'$, ending on a trivalent vertex:
\begin{subequations}
\begin{align}
B(z, z') &= 
\begin{tikzpicture}[baseline=(current  bounding  box.center), very thick]
   \draw (0,0) node[below]{$z$} -- (2,0) node[below]{$z'$};
 \end{tikzpicture}
 \label{eq:brep}
\\
K(z,z') &= 
 \begin{tikzpicture}[baseline=(current  bounding  box.center), very thick]
   \draw[->-] (0,0) node[below]{$z$} -- (2,0) node[below left]{$z'$} -- +(60:.3);
  \draw (2,0) -- +(-60:.3);
 \end{tikzpicture}
= 
\begin{tikzpicture}[baseline=(current  bounding  box.center), very thick]
   \draw[->-] (0,0) node[below]{$z$} -- (2,0) -- +(60:.3) node[right]{$z'$};
  \draw (2,0) -- +(-60:.3) node[right]{$\frac{1}{z'}$};
 \end{tikzpicture}
 \label{rec_kernel1}
\end{align}
\end{subequations}
Then Eq.~\eqref{eq:W11KB} can be depicted by the following diagram, where the vertex signifies an integration around contours that surround $\pm 1$:
\begin{align}
 W_{1,1} = \begin{tikzpicture}[baseline=(current  bounding  box.center), very thick]
 \draw[->-] (0,0) -- (2,0) -- +(-60: .3);
  \draw (2,0) -- +(-60:.3) to [out = -60, in = -90] (3.5, 0) to [out = 90, in = 60] (2,0);
 \end{tikzpicture}
\end{align}

\subsubsection{Higher correlation functions}

The formula \eqref{oooz} for $\omega_{1,1}$ can be generalized to higher correlation functions $\omega_{g,n}$:
\begin{align}
\omega_{g,n}(z_1,\dots,z_n)
 = \frac12 \sum_{\eta = \pm 1} \underset{z = \eta}{\operatorname{Res}} \left(\frac{dz_1}{z_1-z}-\frac{dz_1}{z_1-\frac{1}{z}}\right) \omega_{g,n}(z,\dots,z_n)\ .
\end{align}
Inserting the relevant loop equation, and noticing that the contribution of $P_{g,n}$ has no residue at $z=\pm 1$ and therefore disappears, we eventually obtain
\begin{empheq}[marginbox = \fbox]{multline}
\label{eq:TRJoukowsky}
 \omega_{g,n}(z_1,\cdots z_n)
  =
 \sum_{\eta = \pm 1} \underset{z = \eta}{\operatorname{Res}}\,  K(z_1,z)
 \Bigg[
  \omega_{g-1,n+1} (z,\tfrac{1}{z},z_2,\cdots z_n)
  \\
 + \sum_{\substack{h+h'=g \\ I \sqcup I' = \{z_2,\cdots z_n\}}}'
  \omega_{h,1+|I|}(z,I) \omega_{h',1+|I'|}(\tfrac{1}{z},I')
 \Bigg]
 \ ,
\end{empheq}
where the modified summation $\sum'$ excludes the terms $(h,I) =
(0,\emptyset)$ and $(h,I) =(g,\{z_2,\cdots z_n\})$. 

We have thus shown that the correlation functions are completely determined by the loop equations in the 1-cut case, and therefore that the polynomials $P_{g,n}$ are completely determined too. (See \autoref{exogauss} for the case of the Gaussian matrix model.)
This can be generalized beyond the 1-cut case, using the assumptions \eqref{eq:iad} for eliminating the contributions of $P_{g,n}$ by contour manipulations.

\section{Topological recursion}\label{sec:toprec}

Let us generalize the recursive solution \eqref{eq:TRJoukowsky} of the topologically expanded loop equations, that we have sketched at the first few orders and in the 1-cut case.
We will see that the loop equations determine $W_{g,n\geq 1}$, once $W_{0, 1}$ and $W_{0, 2}$ are given.
We will therefore define the spectral curve as a geometric object that encodes $W_{0, 1}$ and $W_{0, 2}$, and consider the correlation functions as invariants of that spectral curve. 
We will introduce the topological recursion formula for recursively solving the loop equations, and computing these invariants. 
This defines invariants for spectral curves that do not necessarily come from matrix models \cite{Eynard:2007kz}.

\subsection{The spectral curve and its geometry}

\subsubsection{Immersed Riemann surface}

Let $\Sigma$ be the Riemann surface whose immersion $i(\Sigma)\subset \overline{\mathbb{C}}\times \mathbb{C}$ is given by the quadratic polynomial equation \eqref{eq:w01eq} for $y=W_{0,1}(x)$,
\begin{equation}
i(\Sigma) =  
 \Big\{(x,y)\in \overline{\mathbb{C}}\times \mathbb C \mid %\Big|
 y^2 - V'(x) y + P_{0,0}(x)=0\Big\} 
\ .
\end{equation}
The compact Riemann surface $\Sigma$, together with its meromorphic immersion $(x,y)$, form a spectral curve in the sense of \autoref{sec:spcurve}, provided we identify $W_{0,1}$ with $\overline W$ and $P_{0,0}$ with $\overline P$.

In matrix models, the polynomial $\overline P$ is strongly constrained by the requirement that the equilibrium density \eqref{eq:barrho} be positive and normalized; we now relax this requirement and allow $V$ and $P_{0,0}$ to be arbitrary rational functions.
We also allow spectral curves with polynomial equations of degrees higher than two: in \autoref{sec:multi_saddle} we saw that such curves appear in multi-matrix models. 
For a matrix model interpretation to exist, it is necessary that the polynomial equation has a root that behaves as $y\underset{x\to\infty}{\sim} \frac{1}{x} $; in the quadratic case, this amounts to $P_{0,0}(x)\underset{x\to\infty}{\sim} \frac{V'(x)}{x}$.
Possible further generalizations include non-compact spectral curves with non-polynomial equations (see \autoref{exo:plancherel} for an example), and non-commutative spectral curves for $\beta$-matrix models \cite{Chekhov:2010zg}.

\subsubsection{Branch points}

The analytic map $x:\Sigma\to \mathbb C $ is locally invertible everywhere, except at $\infty$ and at the points where $x'(z)=0$, that is the points where the differential $dx = x'(z) dz$ vanishes. 
These points are called the \textbf{branch points}\index{branch point} of the spectral curve, and their images by the map $x$ are the spectral edges. 

Generically the differential $dx$ has a simple zero at a branch point $a$. 
If $z$ is a local complex coordinate on $\Sigma$, we have 
\begin{equation}
x(z) = x(a) + \frac{x''(a)}{2}(z-a)^2  + \frac{x'''(a)}{6} (z-a)^3 + O\left((z-a)^4\right)\  ,
\end{equation}
whose inverse map behaves locally like a square root
\begin{equation}
z(x) \sim a+ \sqrt{\frac{2}{x''(a)}}\ \sqrt{x-x(a)} - \frac{x'''(a)}{3 x''(a)^2} (x-x(a)) + O((x-x(a))^{\frac32}).
\end{equation}
Since $x$ depends quadratically on $z$ at the leading order, there must exist a nontrivial local map $\sigma_a(z)$ such that $x(z) = x(\sigma_a(z))$ that exchanges the sign of the square root, in a vicinity of $a$. 
This map is called the \textbf{local Galois
involution}\index{Galois involution!local---} near the branch point $a$, and it reads
\begin{equation}
\sigma_a(z) = 2a-z - \frac{x'''(a)}{3x''(a)}(z-a)^2 + O\left((z-a)^3\right)\ .
\end{equation}
This involution exchanges the two sheets of $\Sigma$ that meet at the branch point $a$. 

For example, in the 1-cut case, the differential of the Joukowsky map $x(z)$ is $dx=\gamma(1-\frac{1}{z^2})dz$, and the two branch points are $z=\pm 1$. Their images by the Joukowsky map are the spectral edges $(x(1), x(-1))=(a, b)$.
In this case, the local Galois involutions can actually be extended into a global analytic involution $\sigma(z) = \frac{1}{z}$ such that $x(\sigma(z))=x(z)$.

Near a non-generic branch point $a$, we have $x-x(a) \propto (z-a)^{d_a}$ with $d_a>2$.
The $d_a$ local maps $z-a \to e^{2\pi i \frac{k}{d_a}}\ (z-a) + \cdots$ where $k=1,\dots,d_a$ form a local Galois group $\frac{\mathbb{Z}}{d_a\mathbb{Z}}$. Non-generic branch points can be obtained by smoothly merging generic branch points, allowing topological recursion to be used for curves with non-generic branch points \cite{Bouchard:2012yg}.

\subsubsection{Fundamental second kind differentials}

Our constraints \eqref{eq:iad} on $W_{g,n}$ determine a unique fundamental second kind differential $B=\omega_{0,2}$.
However, let us relax those constraints, and allow $\omega_{0,2}$ to be given by any fundamental second kind differential $B$. Then we include
include the choice of $B$ in the definition of the \textbf{spectral curve}\index{spectral curve}
\begin{equation}
\boxed{
\mathcal S = (\Sigma, x,y,B)
}\ ,
\end{equation}
where $\Sigma$ is a Riemann surface, and $(x,y)$ a meromorphic immersion of $\Sigma$ into $\overline{\mathbb{C}}\times \mathbb{C}$.

\subsection{Invariants of the spectral curve}

Given a spectral curve $(\Sigma, x,y,B)$, let us define its \textbf{topological recursion invariants}\index{topological recursion!---invariant} $\omega_{g,n}$ with $g,n\in\mathbb{N}$.

\subsubsection{Definition}

We first define $\omega_{0,1}$ as the tautological one-form of the immersion, and $\omega_{0,2}$ as the fundamental second kind differential $B$:
\begin{align}
 \boxed{\omega_{0, 1} = y dx} \quad , \quad \boxed{\omega_{0,2}=B} \ .
\end{align}
For each branch point $a$ with its local Galois involution $\sigma_a$, we define the \textbf{recursion kernel}\index{kernel!recursion---}
\begin{align}
 \boxed{ K_a(z_1,z)  =
 \frac{1}{2}
 \frac{\int_{\sigma_a(z)}^z \omega_{0,2}(z_1,\cdot)}
      {\omega_{0,1}(z) - \omega_{0,1}(\sigma_a(z))}
 }  \quad \text{so that}\quad
 K_a(z_1,z)=  K_a(z_1,\sigma_a(z))\ .
 \label{eq:kasym}
 \end{align}
We then define the stable invariants $\omega_{g,n\geq 1}$ recursively from $\omega_{0,1}$ and $\omega_{0,2}$ by the \textbf{topological recursion}\index{topological recursion} formula
\begin{empheq}[marginbox = \fbox]{multline}
 \omega_{g,n}(z_1,\cdots z_n)
  =
 \sum_{a} \underset{z = a}{\operatorname{Res}}\,  K_a(z_1,z)
 \Bigg[
  \omega_{g-1,n+1} (z,\sigma_a(z),z_2,\cdots z_n)
  \\
 + \sum_{\substack{h+h'=g \\ I \sqcup I' = \{z_2,\cdots z_n\}}}'
  \omega_{h,1+|I|}(z,I) \omega_{h',1+|I'|}(\sigma_a(z),I')
 \Bigg]
 \ ,
 \label{TRecursion}
\end{empheq}
where the truncated summation $\sum'$ excludes the terms $(h,I) =
(0,\emptyset)$ and $(h,I) =(g,\{z_2,\cdots z_n\})$.
This formula expresses $\omega_{g,n}$ in terms of invariants with larger Euler characteristic $\chi = 2-2g-n$, hence the name topological recursion. 
The recursion stops after $|\chi|$ steps, when it reaches $\chi=0$ and therefore $\omega_{0,2}$. 

In the notations of algebraic geometry, our invariants are sections of vector bundles,
\begin{equation}
\omega_{g,n} \in H^0(\Sigma^n, K_\Sigma(*\{a\})^{\boxtimes n})^{\mathfrak{S}_n} \qquad \text{if} \quad (g,n)\neq (0, 1), (0, 2)\ .
\label{eq:omh}
\end{equation}
Here $K_\Sigma$ is the canonical bundle over $\Sigma$, i.e. the bundle whose fibers are the cotangent planes. A section $\omega\in H^0(\Sigma,K_\Sigma)$ of $K_\Sigma$ is a holomorphic one-form on $\Sigma$. Then $K_\Sigma(*\{a\})$ is the canonical bundle with arbitrary twists at the branch points, and its sections are meromorphic one-forms on $\Sigma$ with poles of arbitrary orders at the branch points $a$.
(In contrast, $\omega_{0, 2}(z_1,z_2)$ has a double pole at $z_1=z_2$, and $\omega_{0,1}$ can have poles that are not branch points.)
The boxed tensor product $\boxtimes n$ means that $\omega_{g,n}$ is a one-forms in each of its variables, and the symmetric group $\mathfrak{S}_n$ indicates that $\omega_{g,n}$ is symmetric under permutations of its $n$ arguments.

The stable \textbf{free energy invariants}\index{free energy!---invariant} $F_g=-\omega_{g,0}$ are defined from $\omega_{g,1}$ by
\begin{align}
 \boxed{ F_{g\geq 2}
 = 
 \frac{1}{2g-2} \sum_a \underset{z = a}{\operatorname{Res}} \,
 \left(\int^z\omega_{0,1}\right) \omega_{g,1}(z) }\ .
 \label{eq:frio}
\end{align}
This does not depend on the choice of a local primitive $\int^z\omega_{0,1}$ of $\omega_{0,1}$ near each branch point, because 
$\underset{z=a}{\operatorname{Res}} \, \omega_{g,1}(z) =0$, a property that we admit.
The expression of $F_1$ can be found in \cite{Eynard:2007kz}. If $\Sigma$ is the Riemann sphere it reduces to 
\begin{equation}
\boxed{
F_1 = \frac{1}{24}  \log\left( \gamma^2  w'(1) w'(-1) \right)
}\ ,
\end{equation}
where $\gamma$ is the parameter of the Joukowsky map, and $w(z)$ was defined in Eq.~\eqref{eq:woz}.
The expression of $F_0$ has been given in Eq.~\eqref{eq:F0gen}.

\subsubsection{Some properties}

Let us give a few essential properties of the topological recursion invariants, which can be proved recursively using the topological recursion formula.
For more properties, see \cite{Eynard:2007kz}.
\begin{itemize}
\item
$\omega_{g,n}$ is invariant under permutations of its $n$ variables, although this is not manifest in the topological recursion formula. 
Stable invariants are meromorphic, with poles at the branch points only. 
At a generic branch point, the order of the pole of the stable invariant $\omega_{g,n}$ does not exceed $2n+6g-4$. And the residues of stable invariants at branch points vanish.
\item
Stable invariants satisfy the relation
\begin{equation}
\sum_{a} \underset{z=a}{\operatorname{Res}} \, \left(\int^z\omega_{0,1}\right) \omega_{g,n+1}(z_1,\dots,z_n,z)  = (2-2g-n) \omega_{g,n}(z_1,\dots,z_n)\ ,
\end{equation}
which we already encountered in the case $n=0$ as the definition \eqref{eq:frio} of $F_{g\geq 2}$.
This relation is called the dilaton relation in string theory applications \cite{Eynard:2007kz}.
\item 
Let us reveal how the invariants behave under deformations of the spectral curve. 
Let us consider a one-parameter analytic family of spectral curves $(\Sigma_t, x_t,y_t,B_t)$.
Let $\gamma_t$ be a family of closed Jordan curves such that the homology class of $x_t(\gamma_t)$ is $t$-independent.
We admit that there exists an analytic family of functions $f_t$ that are analytic on a vicinity of $\gamma_t$, such that $f_t\circ x_t^{-1}$ is analytic on a vicinity of $x_t(\gamma_t)$, and such that
\begin{equation}
\left. \frac{\partial}{\partial t} y_tdx_t(z) \right|_{x_t(z) \, \text{fixed}}
= \oint_{\gamma_t} f_t(z') B(z,z') \ ,
\end{equation}
and such that $B_t$ satisfies the Rauch formula
\begin{equation}
\left. \frac{\partial}{\partial t} B_t(z_1,z_2) \right|_{x_t(z_1), x_t(z_2) \, \text{fixed}}
= \oint_{\gamma_t} f_t(z') \omega_{0,3}(z',z_1,z_2)\ .
\end{equation}
For all stable pairs of indices $(g,n)$, we then have 
\begin{equation}
\left. \frac{\partial}{\partial t}  \omega_{g,n}(z_1,\dots,z_n)\right|_{x_t(z_1),\dots,x_t(z_n) \, \text{fixed}}
= \oint_{\gamma_t} f_t(z') \omega_{g,n+1}(z',z_1,\dots,z_n)\ .
\end{equation}
In the case $n=0$, this gives the behaviour of the free energy invariants under deformations, and thereby justifies our definition \eqref{eq:frio}.
\end{itemize}

\subsubsection{Graphical representation}

One of the reasons that make the topological recursion easy to use, is that it can be represented graphically.
Inspired by the interpretation of formal integrals in terms of discrete surfaces, we depict  $\omega_{g,n}(z_1,\dots,z_n)$ as a genus $g$ surface with $n$ marked points:
\begin{align}
\begin{tikzpicture}[baseline=(current  bounding  box.center)]
  \begin{scope}
    \node at (0,0) [left]
   {\quad $\omega_{g,n}(z_1,\cdots z_n) = $};   
   % shape
   \draw [thick] (3,0) circle [x radius = 2., y radius = 1];
   % holes
   \begin{scope}[thick,shift={(2,-.4)}]
    \draw [clip] (-.5,.3) to [bend right] (.5,.3);
    \draw (-.5,0.1) to [bend left] (.5,0.1);
   \end{scope}
   \begin{scope}[thick,shift={(3,.2)}]
    \draw [clip] (-.5,.3) to [bend right] (.5,.3);
    \draw (-.5,0.1) to [bend left] (.5,0.1);
   \end{scope}
   \begin{scope}[thick,shift={(4,-.2)}]
    \draw [clip] (-.5,.3) to [bend right] (.5,.3);
    \draw (-.5,0.1) to [bend left] (.5,0.1);
   \end{scope}
   \node at (5.5,0) [right] {($g$ handles)};
   % vertices
   \draw (1.5,.3) -- ++(120:1) node [above] {$z_1$};
   \draw (2,.6) -- ++(100:1) node [above] {$z_2$};
   \draw (4.2,.5) -- ++(70:1) node [above] {$z_n$};
   \draw [thick,dotted] (2.5,1.5) to [bend left] (4,1.4);
  \end{scope}
 \end{tikzpicture}
\end{align}
The fundamental second kind differential $\omega_{0,2}=B$ is depicted as an unoriented line \eqref{eq:brep}, and the recursion kernel is depicted as 
\begin{align}
 \begin{tikzpicture}[baseline=(current  bounding  box.center)]
  \node at (-.5,0) [left] {$K_a(z_1,z) \ =$};
  \draw[->-,thick]
  (0,0) node[below]{$z_1$} -- (2,0) -- +(60:.3) node[right]{$z$};
  \draw[thick] (2,0) -- +(-60:.3) node[right]{$\sigma_a(z)$};
\end{tikzpicture}
\end{align}
Then the recursion relation  \eqref{TRecursion} is depicted as
\begin{align}\label{eq:TRgraphical}
 \begin{tikzpicture}[baseline=(current  bounding  box.center), scale=.58]
  % frame 1
  \draw [thick] (0,0) circle [x radius = 1.5, y radius = 2];
  % holes 1
  \draw [thick] (0-0.3,0.15+1) arc (225:315:.7);
  \draw [thick] (0.19-0.3,0+1) arc (130:50:.45);
  \draw [thick] (0-0.8,0.15) arc (225:315:.7);
  \draw [thick] (0.19-0.8,0) arc (130:50:.45);
  \draw [thick] (0-0.3,0.15-1) arc (225:315:.7);
  \draw [thick] (0.19-0.3,0-1) arc (130:50:.45);
  % vertices 1
  \draw (-2.3,0) node [left] {$z_1$} -- (-1.2,0);
  \draw (2.,1.5) node [right] {$z_2$} 
  -- (1.,1);
  \draw (2.3,.5) node [right] {$z_3$} 
  -- (1.1,.3);
  \draw (2.3,-.5) node [right] {$z_4$} 
  -- (1.1,-.2);
  \draw (2.,-1.5) node [right] {$z_5$} 
  -- (1.,-1);
  \node at (4.2,0) {$=$};
  \node at (12.2,0) {$+ \sum'$};
 \begin{scope}[shift={(10,0)}]
  \draw [thick] (6.5,2) circle [radius = .8];
  \draw [thick] (6.5,-2) circle [x radius = 1.5, y radius = 1];
  % holes 2
  \draw [thick] (0+6.,0.15+2) arc (225:315:.7);
  \draw [thick] (0.19+6.,0+2) arc (130:50:.45);
  \draw [thick] (0+5.3,0.15-2) arc (225:315:.7);
  \draw [thick] (0.19+5.3,0-2) arc (130:50:.45);
  \draw [thick] (0+6.7,0.15-2) arc (225:315:.7);
  \draw [thick] (0.19+6.7,0-2) arc (130:50:.45);
  % vertices 2
  \draw [->-] (4.,0) node [left] {$z_1$}-- (5.,0);
  \draw (5.,0) -- (6.,1.7);
  \draw (5.,0) -- (5.8,-1.5);
  \node at (5.4,1.) [left] {$z$};
  \node at (5.4,-1.) [left] {$\sigma_a(z)$};
  \draw (8.,2.3) node [right] {$z_2$} 
  -- (7.2,2.1);
  \draw (7.,-0.3) node [above] {$z_3$} 
  -- (6.6,-1.4);
  \draw (8.,-0.7) node [right] {$z_4$} 
  -- (7.2,-1.5);
  \draw (7.8,-3.2) node [right] {$z_5$} 
  -- (6.9,-2.5);
 \end{scope}
 \begin{scope}[shift={(-3.5,0)}]
  \draw [->-] (9.5,0) node [left] {$z_1$}-- (10.5,0);
  \draw (10.5,0) .. controls (10.7,.3) and (11.1,.5) .. 
  (11.4,0.5);
  \draw (10.5,0) .. controls (10.7,-.3) and (11.1,-.7) .. 
  (11.4,-0.7);
  \node at (10.4,.6) {$z$};
  \node at (10.1,-.8) {$\sigma_a(z)$};
  \draw (11.5+2.,1.5) node [right] {$z_2$} 
  -- (11.5+1.,1);
  \draw (11.5+2.3,.5) node [right] {$z_3$} 
  -- (11.7+1.1,.3);
  \draw (11.5+2.3,-.5) node [right] {$z_4$} 
  -- (11.7+1.1,-.2);
  \draw (11.5+2.,-1.5) node [right] {$z_5$} 
  -- (11.5+1.,-1);
  % frame 3
  \draw [thick] (12.,0) circle [x radius = 1., y radius = 1.5];
  % holes 3
  \draw [thick] (0+11.6,0.15+.5) arc (225:315:.7);
  \draw [thick] (0.19+11.6,0+.5) arc (130:50:.45);
  \draw [thick] (0+11.4,0.15-0.5) arc (225:315:.7);
  \draw [thick] (0.19+11.4,-0.5) arc (130:50:.45);
 \end{scope} 
 \end{tikzpicture}
\end{align}
In this graphical representation, $\sum'$ is the summation over 
all possible distributions of handles and marked points in the last term, truncated so that there is no disc factor.
The vertex means an integration, here a sum of residues at all branch points.

We will use this graphical representation in three examples. We will see that some graphs come with symmetry factors of $2$, because of the invariance of the kernel under the local Galois involution \eqref{eq:kasym}.

\subsubsection{Examples of $\omega_{1,1}$, $\omega_{0,3}$ and $\omega_{1,2}$}

\begin{subequations}
\begin{align}
 \omega_{1,1}(z_1) & =
 \begin{tikzpicture}[baseline=(current  bounding  box.center), thick]  
  % 1st figure
  \draw (0,0) circle [x radius = 1, y radius = .7];
  \begin{scope}
   \draw [clip] (-.5,.1) to [bend right] (.5,.1);
   \draw (-.5,-.1) to [bend left] (.5,-.1);
  \end{scope}
  \draw (-0.8,0) -- ++(-1,0) node [left] {$z_1$};
  % 2nd figure
  \node at (1.5,0) {$=$};
  \draw [->-] (2.5,0) node [left] {$z_1$} -- (3.5,0);
  \draw (3.5,0) to [out=60,in=up] (5,0) to [out=down,in=-60] (3.5,0);
  \node at (3.5,.5) {$z$};
  \node at (3.4,-.5) {$\sigma_a(z)$};
 \end{tikzpicture}
   \\
 & = 
 \sum_a \underset{z = a}{\operatorname{Res}} \, 
 K_a(z_1,z) B(z,\sigma_a(z)) \ .
 \label{eq:omega11_TR}
\end{align}
\end{subequations}

\begin{subequations}
\begin{align}
 \omega_{0,3}(z_1,z_2,z_3) & =
 \begin{tikzpicture}[baseline=(current  bounding  box.center), thick]
  \draw (0,0) circle [x radius = 1, y radius = .7];
  \draw (-.8,0) -- ++(-1,0) node [left] {$z_1$};
  \draw (.6,.3) -- ++(40:.7) node [right] {$z_2$};
  \draw (.6,-.3) -- ++(-40:.7) node [right] {$z_3$};
 \end{tikzpicture}
   \\
 & =
 \begin{tikzpicture}[baseline=(current  bounding  box.center), thick]
  \draw [->-] (-1,0) node [left] {$z_1$} -- (0,0);
  \draw (0,0) -- ++(30:1) node [right] {$z_2$};
  \draw (0,0) -- ++(-30:1) node [right] {$z_3$};
  \node at (0,.3) {$z$};
  \node at (-.3, -.4) {$\sigma_a(z)$};
 \end{tikzpicture}
 +
 \begin{tikzpicture}[baseline=(current  bounding  box.center), thick]
  \draw [->-] (-1,0) node [left] {$z_1$} -- (0,0);
  \draw (0,0) -- ++(30:1) node [right] {$z_3$};
  \draw (0,0) -- ++(-30:1) node [right] {$z_2$};
  \node at (0,.3) {$z$};
  \node at (-.3, -.4) {$\sigma_a(z)$};
 \end{tikzpicture}
   \\
 &=  2\times \left(
  \begin{tikzpicture}[baseline=(current  bounding  box.center), thick]
  \draw [->-] (-1,0) node [left] {$z_1$} -- (0,0);
  \draw (0,0) -- ++(30:1) node [right] {$z_2$};
  \draw (0,0) -- ++(-30:1) node [right] {$z_3$};
  \node at (0,.3) {$z$};
  \node at (-.3, -.4) {$\sigma_a(z)$};
 \end{tikzpicture}
  \right)
   \\
 & =
2 \sum_a \underset{z = a}{\operatorname{Res}} \,
 K_a(z_1,z) B(z,z_2) B(\sigma_a(z),z_3) \ .
 \label{eq:omega03_TR}
\end{align}
\end{subequations}
\begin{subequations}
\begin{align}
 \omega_{1,2}
 & =
 \begin{tikzpicture}[baseline=(current  bounding  box.center), thick]
  \draw (0,0) circle [x radius = 1, y radius = .7];
  \begin{scope}
   \draw [clip] (-.5,.1) to [bend right] (.5,.1);
   \draw (-.5,-.1) to [bend left] (.5,-.1);
  \end{scope}
  \draw (-0.8,0) -- ++(-1,0) ;
  \draw (0.8,0) -- ++(1,0);
 \end{tikzpicture}
   \\
 & =
 \begin{tikzpicture}[baseline=(current  bounding  box.center), thick]
  \draw [->-] (-1,0)  -- (0,0);
  \draw (0,0) -- ++(30:.5) to [out=30,in=180-60] (1.5,.3);
  \draw (0,0) -- ++(-30:.5) to [out=-30,in=180+60] (1.5,-.3);
  \draw (2,0) circle [x radius = 1, y radius = .7];  
  \draw (2.8,0) -- ++(1,0);
 \end{tikzpicture}
  + 2\times \left(
 \begin{tikzpicture}[baseline=(current  bounding  box.center), thick]
  \draw [->-] (-1,0)  -- (0,0);
  \draw (0,0) -- ++(30:1.5);
  \draw (0,0) -- ++(-30:.5) to [out=-30,in=left] (1.3,-.6);
  \draw (2,-.6) circle [x radius=1,y radius=.5];
  \begin{scope}[shift={(2,-.6)}]
   \draw [clip] (-.5,.1) to [bend right] (.5,.1);
   \draw (-.5,-.1) to [bend left] (.5,-.1);
  \end{scope}
 \end{tikzpicture} 
 \right)
   \\
 & =
 2 \times 
 \left(
 \begin{tikzpicture}[baseline=(current  bounding  box.center), thick]
  \draw [->-] (-1,0)  -- (0,0);
  \draw [->-] (0,0) to [out=60,in=180-60] (2,0);
  \draw (0,0) to [out=-60,in=180+60] (2,0);
  \draw (2,0) -- (3,0);
 \end{tikzpicture} 
 \right)
 + 2 \times
 \left(
 \begin{tikzpicture}[baseline=(current  bounding  box.center), thick]
  \draw [->-] (-1,0)  -- (0,0);
  \draw (0,0) -- ++(30:1.5);
  \draw (0,0) -- ++(-30:.5);
  \draw [->-] (-30:.5) to [out=-30,in=left] (1.3,-.6);
  \draw (1.3,-.6) to [out=40,in=up] (2.5,-.6);
  \draw (1.3,-.6) to [out=-40,in=down] (2.5,-.6);
 \end{tikzpicture}
 \right)
   \\
 & =
 2  \sum_{a,b} \underset{z = a}{\operatorname{Res}}\, \underset{z' = b}{\operatorname{Res}}\, 
 K_a(z_1,z) K_b(z,z')
 \nonumber \\
 & \qquad \quad \times 
 \Big(
   B (\sigma_a(z),\sigma_a(z')) B (z',z_2) 
 + B (\sigma_a(z),z_2) B (z',\sigma_a(z')) 
 \Big)\ .
\end{align}
\end{subequations}

\subsubsection{General solution of the recursion}

Any invariant $\omega_{g,n}$ can be computed by recursively applying the recursion relation  \eqref{eq:TRgraphical}, leading to a representation of $\omega_{g,n}$ as a sum of graphs with oriented and unoriented edges,
\begin{equation}
\boxed{
\omega_{g,n}
= \sum_{G\in \mathcal G_{g,n}} m_G \prod_{\text{vertices}}
\underset{\text{all branch points}}{\operatorname{Res}} \prod_{\text{oriented edges}} K
\prod_{\text{unoriented edges}} B\ .
}
\end{equation}
In this formula:
\begin{itemize}
 \item $\mathcal G_{g,n}$ is the set of graphs with $2g-2+n$ trivalent vertices, $n$ external edges labelled $1,2,\dots,n$, and $3g-3+n$ internal edges forming $g$ loops.
 Edges can be oriented or unoriented, such that 
 \begin{itemize}
 \item oriented edges form an oriented tree covering the graph (i.e. reaching every vertex), rooted at the external edge $1$,
 \item each loop carries exactly one unoriented edge,
 \item each internal unoriented edge joins a vertex to an ancestor along the oriented tree (i.e. does not join vertices belonging to different branches). This non-local constraint is not of Wick's theorem type, so that there are fewer graphs in $\mathcal G_{g,n}$ than in a comparable set of Feynman graphs.
 \end{itemize}
 \item The symmetry factor $m_G$ of a graph is a power of two that comes from Eq.~\eqref{eq:kasym}. This symmetry factor could be set to $1$, by distinguishing the left and right branches at each vertex.
 \item The variable $z_1$ is assigned to the root of the oriented tree, and the variables $z_2,\dots,z_n$ to the corresponding  leaves.
 \item A variable $z'_v$ is assigned to each vertex $v$, with $z'_v$ and $\sigma_a(z'_v)$ being assigned to the outgoing legs.
 Then one evaluates the sum of the residues in $z'_v$ at all branch points, $\displaystyle \prod_{\text{vertices}}
\underset{\text{branch points}}{\operatorname{Res}} = \prod_v \sum_a\underset{z'_v = a}{\operatorname{Res}}$. Residues are computed from the leaves towards the root, i.e. following the arrows backwards.
\end{itemize}

\subsection{Loop equations}

We will now justify the interest of
topological recursion invariants in matrix models, by showing that these invariants satisfy loop equations. 
Actually, loop equations are only the $k=2$ case of a family of generalized loop equations satisfied by topological recursion invariants, parametrized by $k\in\mathbb{N}^*$.

\subsubsection{Generalized loop equations}

Under the map $x:\Sigma \to \mathbb{C}$, a generic point $x_0\in \mathbb C$ generically has $\deg x$ preimages. 
For $k\geq 1 $ let $K_1 \sqcup \dots \sqcup K_\ell\vdash_k x^{-1}(x_0)$ be a partition of a subset of size $k$ of $x^{-1}(x_0)$ into $\ell$ non-empty parts. 
Then, for $I=\{z_1,\dots,z_n\}$ and $x_0\in \mathbb C$ let us define 
\begin{multline}
 Q^{(k)}_{g,n}(x_0;I) = \sum_{K_1 \sqcup \dots \sqcup K_\ell\vdash_k x^{-1}(x_0) }\ \ \sum_{J_1 \sqcup \dots \sqcup J_\ell=I}
 \\
  \sum_{(g_i) \, | \, \sum_{i=1}^\ell(2g_i-2+|K_i|+|J_i|)= 2g-2+n} \
 \prod_{i=1}^\ell \omega_{g_i,|K_i|+|J_i|}(K_i\cup J_i)\ .
\end{multline}
In words, we first sum over subsets of size $k$ of $x^{-1}(x_0)$ (this sum is empty if $k>\deg x$), then over partitions of such subsets into non-empty parts, then over ordered partitions of $I$ into $\ell$ possibly empty subsets $J_i$, and finally over positive integers $g_1,\dots, g_\ell$  subject to a constraint on their sum. 
For example,
\begin{subequations}
\begin{align}
Q^{(1)}_{g,n}(x_0;I)
= \sum_{z \, | \, x(z)=x_0} \omega_{g,n+1}(z,I)\ ,
\end{align}
\begin{multline}
Q^{(2)}_{g,n}(x_0;I)
=  \frac12 \sum_{z\neq z' \, | \, x(z)=x(z')=x_0} 
\\
\Bigg( \omega_{g-1,n+2}(z,z',I) 
+\sum_{J_1\sqcup J_2=I}\sum_{g_1+g_2=g} \omega_{g_1,1+|J_1|}(z,J_1)\omega_{g_2,1+|J_2|}(z',J_2)\Bigg)\ 
\end{multline}
\end{subequations}
where the factor $\frac12$ comes from the fact that the pair $(z,z')$ and $(z',z)$ are actually the same partition.

$Q^{(k)}_{g,n}(x_0;I)$ is a meromorphic differential form of order $k$ in the variable $x_0$, because it is a symmetric function of the preimages of $x_0$ under the map $x$, and $x^{-1}$ is locally analytic.
A priori, this form could have poles at the spectral edges $x_0=x(z_i)$ (from $\omega_{0,2}(z,z_i)$), and at the poles of $\omega_{0,1}$.
However, a theorem \cite{Eynard:2007kz, Bouchard:2012yg} called \textbf{generalized loop equations}\index{loop equations!generalized---} states that
\begin{quote}
\fbox{
$Q^{(k)}_{g,n}(x;I)$
has no poles at the spectral edges.
}
\end{quote}

\subsubsection{Correlation functions of matrix models and topological recursion invariants}

Let us show that the one-matrix model correlation functions obey these generalized loop equations in the 1-cut case using Joukowsky variables. 
Since the spectral curve is a two-sheeted cover of the complex plane, only the cases $k=1,2$ of the generalized loop equations are nontrivial.
The first equation involves
\begin{align}\label{eq:Q1gnJ}
Q^{(1)}_{g,n}(x;I)
= \omega_{g,n+1}(z,I) + \omega_{g,n+1}(\tfrac{1}{z},I) \ .
\end{align}
This vanishes  due to Eq.~\eqref{eq:omgnantisym}, except if $g=0$ and $n\in\{0,1\}$. The Riemann--Hilbert equation \eqref{eq:saddleder} for $W_{0, 1}$ implies 
$
 Q^{(1)}_{0, 0}(x) =  V'(x)dx
$,
which has no poles at the spectral edges. And $Q^{(2)}_{0, 1}(x;z_1)= \frac{dx dx(z_1)}{(x-x(z_1))^2}$ also has no poles at the spectral edges. 
To check the second generalized loop equation, we use the matrix model loop equations 
\eqref{eq:loopconn} and compute 
\begin{align}
Q^{(2)}_{g,n}(x;I)
= -P_{g,n}(x;I) dx^2 
 +dx^2 \sum_{i=1}^n \frac{dz_i}{x'(z_i)}\ \frac{d}{dz_i} \frac{\omega_{g,n}(I)}{(x(z)-x(z_i))dx(z_i)} \ ,
\end{align}
which indeed has no pole at spectral edges.

Then it  can be proved \cite{Eynard:2007kz}  that the solutions $W_{g,n}$ of the loop equations that obey our assumptions \eqref{eq:iad}, are related to the topological recursion invariants $\omega_{g,n}$ of the spectral curve by Eq.~\eqref{eq:wdsc}.

\section{Non-perturbative asymptotics and oscillations}
\label{sec:non-perturbative}

For each choice of $P_{0,0}$, topological recursion provides a formal solution of the loop equations, in the form of a topological expansion. Let us now describe actual asymptotics of convergent matrix integrals. For the example of the Airy function, see \autoref{exo:airy}.

%\sr{Could we give a reference to a more detailed work on this subject?}

\subsection{Asymptotics of convergent matrix integrals}

If $\mathcal{Z}$ is a convergent matrix integral, it can be proved that $ \underset{N\to \infty}{\lim} \frac{1}{N^2} \log \mathcal{Z}$ always exists. 
In some cases, including the 1-cut case of the GUE with a real convex potential \cite{Borot:2013}, $\log \mathcal{Z}$ and $W_n$ furthermore have  asymptotic series expansions in powers of $\frac{1}{N}$.
These expansions are typically divergent, in particular $\log \mathcal Z \sim \sum_g N^{2-2g} F_g $ with $ F_g \underset{g\to\infty}{\simeq} (2g)!$, but a Borel summation can be performed. 
Then, due to loop equations, these asymptotic expansions must coincide with the corresponding topological expansions, whose coefficients can be computed using topological recursion.

However, for generic potentials and integration domains, $\log \mathcal{Z}$ and $W_n$ have terms that depend exponentially on $N$, see Eq.~\eqref{eq:fene}, including oscillatory terms of the type $\cos(A N)$ with $A\in \mathbb{R}$.
These oscillations typically appear in the multi-cut case,  when the support of the equilibrium density is disconnected \cite{Borot:2013}. 

Nevertheless, topological recursion can be used for computing asymptotics of matrix integrals that have no large $N$ series expansions, and oscillatory asymptotics can be deduced from topological recursion asymptotics. 
The basic reason is that loop equations are linear in the disconnected correlation functions and partition functions.
And we will now show that a linear combination of partition functions that have topological expansions, typically has oscillatory asymptotics.

For a given potential $V$, let $\Gamma^{(i)}$ be integration domains such that the corresponding partition functions $\mathcal{Z}^{(i)}$ and correlation functions $\widehat{W}^{(i)}_n$ have topological expansions.
An integration domain $\Gamma = \sum_i a_i \Gamma^{(i)}$ leads to the partition function and disconnected correlation functions
\begin{align}
 \mathcal{Z}=\sum_i a_i \mathcal{Z}^{(i)} \quad , \quad \widehat{W}_n = \sum_i a_i \frac{\mathcal{Z}^{(i)}}{\mathcal{Z}} \widehat{W}_n^{(i)}\ .
\end{align}
Assuming $\Re F_0^{(1)}= \operatorname{min}\{\Re F_0^{(i)} \}$, the asymptotic behaviour of $\log\mathcal{Z}$ is dominated by $\log \mathcal{Z}^{(1)}$, but the terms $i\neq 1$ lead to exponential corrections: 
\begin{multline}
\log\mathcal{Z} 
=  -N^2 F_0^{(1)} -  F_1^{(1)}  + \log c_1 - \sum_{g=2}^\infty N^{2-2g} F_g^{(1)}
\\
 + \log\left(1+\sum_{i\neq 1} \frac{c_i}{c_1} e^{-N^2(F_0^{(i)}-F_0^{(1)})} e^{F_1^{(1)}-F_1^{(i)}} \left(1+O\left(\tfrac{1}{N^2}\right)\right)\right)  \ .
\end{multline}
Exponential corrections also appear in correlation functions, because the expression for $\widehat{W}_n$ in terms of $\widehat{W}_n^{(i)}$ involves the corresponding partition functions.

We will now study this asymptotic behaviour in detail, in the case where the indices $i$ correspond to $N$-dependent filling fractions. Then $F^{(i)}_g$ becomes $N$-dependent, and typically $F_0^{(i)}-F_0^{(1)} = O(\frac{1}{N})$, so that the oscillatory terms of $\log \mathcal{Z}$ have frequencies of order $N$, as in Eq.~\eqref{eq:fene}.

\subsection{Sum over filling fractions}\label{sec:sff}

We focus on the case of the normal matrix integral $\mathcal{Z}(\gamma^N)$, for some path $\gamma$.
We will first perform a formal computation of the large $N$ asymptotics, and then discuss the conditions for the asymptotic series to converge.

\subsubsection{Formal computation}

Let us write our path $\gamma$ in terms of $s$ independent contours $\gamma_i$,
\begin{equation}
\gamma = \sum_{i=1}^s c_i \gamma_i = \sum_{i=1}^s e^{2\pi i \mu_i}  \gamma_i\ ,
\end{equation}
where $c_i=e^{2\pi i\mu_i}\neq 0$ is assumed to be $N$-independent for simplicity. 
Redefining the partition functions $\mathcal{Z}(\vec\gamma^{\vec n})\to  \frac{1}{\prod_i n_i!} \mathcal{Z}(\vec\gamma^{\vec n})$ to eliminate factorial normalization factors, our matrix integral $\mathcal{Z}(\gamma^N)$ is given by the truncated discrete Fourier transform
\begin{align}
\boxed{
 \mathcal{Z}(\gamma^N) = \sum_{n_1+\dots+n_s=N} e^{2\pi i \sum_{i=1}^s n_i \mu_i}  \mathcal{Z}(\vec\gamma^{\vec n}) }  \ .
 \label{eq:zgns}
\end{align}
We now assume that $\mathcal{Z}(\vec\gamma^{\vec n}) $ have topological expansions in terms of invariants $F_g(\vec \epsilon)$, associated to spectral curves of genus $g=s-1$ with filling fractions $\vec\epsilon = (\frac{n_1}{N},\dots,\frac{n_s}{N})$, 
\begin{align}
 \mathcal{Z}(\vec\gamma^{\vec n}) =\exp\left( -\sum_{g=0}^\infty N^{2-2g} F_g(\vec\epsilon)  \right) \ .
\end{align}
According to Eqs.~\eqref{eq:ecff} and \eqref{eq:iad}, the genus and filling fractions uniquely determine the spectral curves, and the resulting invariants depend analytically on the filling fractions.

%\sr{Could we say why it is necessary (or advantageous) to expand around a reference point? How is this related to an expansion in $N$?}
%\nbbe{In fact this is answered by the fact that we obtain beautifful formulas below. Also we want to make variations of the filling fractions later, small variations of order 1/N, so it is useful to understand how these variations change the perturbative partition function }
Let us see how $ \mathcal{Z}(\vec\gamma^{\vec n})$ depends on changing the filling fractions, by performing the Taylor expansion of the invariants near an arbitrary reference point $\vec\epsilon^*$ called the \textbf{background}\index{background}, 
\begin{multline}
 F_g(\vec{\epsilon})
 = F_g(\vec \epsilon^*)
 + \sum_{m=1}^\infty \frac{1}{m!}
 (\vec{\epsilon}-\vec \epsilon^*)^{\vec m} F_g^{(\vec m)}(\vec\epsilon^*)\ ,
 \\
 = F_g(\vec \epsilon^*)
 + \sum_{m=1}^\infty \frac{1}{m!} \sum_{1\leq i_1 \cdots i_m\leq s} 
 (\epsilon_{i_1}-\epsilon^*_{i_1}) \cdots
 (\epsilon_{i_m}-\epsilon^*_{i_m})\,
 \partial_{i_1} \cdots \partial_{i_m} F_g(\vec \epsilon^*)\ . 
\end{multline}
(In this formula the first line is a condensed tensorial notation, where the sum over the vector of indices $\vec m=(i_1,\cdots i_m)$ is kept implicit, and where $m=\# \vec m$.)
We now want to insert this Taylor expansion into the topological expansion of $\mathcal{Z}(\vec\gamma^{\vec n})$. For the two expansions to be compatible, the background must be close enough to the vector $\vec \epsilon_\text{min}$ that minimizes $\Re F_0(\vec \epsilon)$. More specifically, we need $\Re F_0(\vec\epsilon^*) -\Re F_0(\vec \epsilon_\text{min}) = O(\frac{1}{N^2})$, and therefore
\begin{align}
 \vec \epsilon^* -\vec \epsilon_\text{min} = O\left(\frac{1}{N}\right) \ .
 \label{eq:esem}
\end{align}
We then obtain 
\begin{multline}\label{eq:ZNoscil1}
\mathcal{Z}(\vec\gamma^{\vec n})
 = e^{-\sum_{g=0}^\infty N^{2-2g} F_g(\vec\epsilon^*)} 
 \, e^{-N (\vec n-N\vec\epsilon^*) F_0^{(\vec 1)}(\vec\epsilon^*)}
 \, e^{-\frac12 (\vec n-N\vec\epsilon^*)^2   F_0^{(\vec 2)}(\vec\epsilon^*) } \\
\times 
 \left( 
 1 + \sum_{k=1}^\infty \frac{(-1)^k}{k!} \sum_{\substack{m_i>0,\, g_i\\ 2g_i+m_i-2>0}} 
 \prod_{i=1}^k \frac{N^{2-2g_i-m_i}}{m_i!} 
 \prod_{i=1}^k (\vec n-N\vec\epsilon^*)^{\vec m_i} F_{g_i}^{(\vec m_i)} (\vec\epsilon^*)
 \right)
 \ .
\end{multline}
In this expression, the $O(e^{N})$ factors are actually oscillatory. In particular, the factor $e^{-N \vec n F_0^{(\vec 1)}(\vec\epsilon^*)}$  is oscillatory because $\Re F_0^{(\vec 1)}(\vec\epsilon^*) = O(\frac{1}{N})$, as follows from  the
condition \eqref{eq:esem}.

We will now sum over $\vec n$ in order to compute $\mathcal{Z}(\gamma^N)$ \eqref{eq:zgns}. 
Due to the quadratic exponential in $\vec n$, the sum over the integers $n_j$ is dominated by the region $\vec n = N\vec{\epsilon}^*+O(1)$. Contributions beyond this region are damped by $O(e^{-N^2})$ factors, and are negligible with respect to the $O(e^{-N})$ oscillatory terms that we want to compute. 
We will therefore approximate the finite sum over integers $n_j$, with an infinite sum over the lattice $\vec n \in \mathbb Z^{s-1}$.
And this infinite sum will produce Theta functions.

\subsubsection{Theta functions}

Let ${g}>0$ be a positive integer, called the genus, and $\tau$ a \textbf{Siegel matrix}\index{Siegel matrix}, i.e. a ${g}\times {g}$ symmetric complex matrix whose imaginary
part is positive definite,
\begin{equation}
\tau^\text{T}=\tau \quad 
 , \quad
\Im \tau > 0 \ .
\label{eq:itsz}
\end{equation}
The associated genus ${g}$ \textbf{Theta
function}\index{Theta function} is an entire function of a complex vector $\vec u \in \mathbb C^{{g}} $, 
which is defined by the absolutely convergent series
\begin{equation}
\boxed{
\theta(\vec u;\tau) 
 = \sum_{\vec n\in \mathbb{Z}^{{g}}} 
 e^{i\pi(\vec n, \tau \vec n)}\,e^{2\pi i(\vec n,\vec u)} } \ .
 \label{eq:ts}
\end{equation}
The Theta function is even, $\mathbb Z^{{g}} $-periodic, and quasi-periodic in the direction
$\tau \mathbb{Z}^{{g}}$:
\begin{subequations}
\begin{align}
\theta(-\vec u) &= \theta(\vec u)
\ ,
\\
\forall \vec{m}, \vec{n}\in\mathbb{Z}^g\ , \quad 
\theta(\vec u+\vec n + \tau\vec m) 
&= e^{-2\pi i (\vec m,\vec u) - i\pi (\vec m,\tau\vec m)}\theta(\vec u)\ 
 .
\end{align}
\end{subequations}
Let us now approximate $\mathcal{Z}(\gamma^N)$ \eqref{eq:zgns} in terms of Theta functions, using Eq.~\eqref{eq:ZNoscil1}.
For the leading term, we find,
\begin{align}
\left. \mathcal{Z}(\gamma^N) \right|_\text{leading}
\sim e^{-N^2 F_0(\vec\epsilon^*) - F_1(\vec\epsilon^*)}
 e^{-\frac{N^2}{2} (\vec\epsilon^*,F_0^{(\vec 2)}(\vec\epsilon^*) \vec\epsilon^*)} 
 e^{N^2 \vec\epsilon^* F_0^{(\vec 1)}(\vec\epsilon^*)}
 \theta
 \left(\frac{\vec u_0}{2\pi i}; \tau
 \right)\ ,
\end{align}
where the genus of the Theta function is $g=s-1$, and 
the relevant symmetric matrix and complex vector are 
\begin{align}
\boxed{
 \tau = \frac{i}{2\pi } F_0^{(\vec 2)}(\vec\epsilon^*) \quad , \quad \vec u_0 = -N \left(F^{(\vec 1)}_0(\vec\epsilon^*) + 2\pi i \tau \vec\epsilon^*\right)+ 2\pi i \vec \mu} \ .
 \label{eq:tuz}
\end{align}
(We admit that $\tau$ is a Siegel matrix.)
By the same method, the subleading terms of $\mathcal{Z}(\gamma^N)$ can be approximated by derivatives of the Theta function, and we obtain
\begin{equation}\label{eq:oscilZN}
\boxed{
  \mathcal{Z}(\gamma^N) 
\sim
 \left.\left( 
  1 + \sum_{k=1}^\infty \frac{(-1)^k}{k!} \sum_{\substack{m_i\geq 0, g_i\\ 2g_i+m_i-2>0}}  
  \prod_{i=1}^k
  \frac{N^{2-2g_i-m_i}}{m_i!}  F_{g_i}^{(\vec m_i)}(\vec\epsilon^*)
  \vec\partial_{\vec u}^{ \vec m_i}  
 \right)
 \Theta
 \left(
  \vec u ; \tau
 \right)\right|_{\vec u = \vec u_0}
 } \ ,
\end{equation}
where we introduced the twisted Theta function
\begin{align}\label{eq:defThetatwisted}
\boxed{
\Theta(\vec u;\tau ) =  e^{-N^2 F_0(\vec\epsilon^*)-F_1(\vec\epsilon^*)}
e^{-\pi i N^2 (\vec\epsilon^*,\tau \vec\epsilon^*) } 
 e^{2\pi i N (\vec\epsilon^*, \vec \mu)}
 e^{-N (\vec\epsilon^*,\vec u)}
 \theta\left(\frac{1}{2\pi i}\vec u;\tau\right)
 } \ .
\end{align} 
Schematically, we can write the first few terms as 
\begin{align}\label{eq:oscilZNorder2}
  \mathcal{Z}(\gamma^N) 
  \sim
  \Theta - \frac{1}{N} \left( F_1^{(\vec 1)} \Theta^{(\vec 1)} +
  \frac16 F_0^{(\vec 3)}\Theta^{(\vec 3)}\right)
+ \frac{1}{N^2} \mathcal{Z}_2(\gamma^N) + O\left(\frac{\Theta}{N^3}\right) \ ,
\end{align}
where we introduced the quantity
\begin{multline}
\mathcal{Z}_2(\gamma^N) = 
-F_2 \Theta
\\
- \frac{F_1^{(\vec{2})}}{2}\,\Theta^{(\vec 2)}
+ \frac{\left(F_1^{(\vec{1})}\right)^2}{2}\,\Theta^{(\vec 2)}
- \frac{F_0^{(\vec{4})}}{24}\,\Theta^{(\vec 4)}
+ \frac{F_0^{(\vec{3})} F_1^{(\vec{1})}}{6}\,\Theta^{(\vec 4)}
+ \frac{\left(F_0^{(\vec{3})}\right)^2}{72}\,\Theta^{(\vec 6)}
 \ .
\end{multline}
Expression \eqref{eq:oscilZN} is the formal non-perturbative asymptotic formula that was announced in \eqref{eq:fene}.
Provided $\vec \epsilon^*$ obeys the condition \eqref{eq:esem}, it does not depend on the choice of the background $\vec \epsilon^*$: see \autoref{exobackground} for a check that its $\vec\epsilon^*$-derivatives vanish order by order in $\frac{1}{N}$. This property is called \textbf{background independence}\index{background!---independence}.
%\sr{Our Theta functions are the result of an approximation: what are we neglecting? Not only when we introduced the Theta functions, but also in the final result: neglecting terms of the type $e^{-N}$ at large $N$?}
%\nbbe{This is answered above, just after eq 4.4.8.  Theta function is a sum over filling fractions from $-\infty$ to $+\infty$, whereas the matrix model should have filling fractions $0\leq n/N \leq 1$. The approximation is to neglect the filling fractions outside interval $[0,1]$. See above.}

\subsubsection{Boutroux spectral curves}

Let us provide a geometrical interpretation of our condition \eqref{eq:esem}, by studying the implications of minimizing $\Re F_0$. 
At its minimum, $\Re F_0$ has vanishing derivatives, and we admit that
$
\frac{\partial \Re F_0}{\partial \epsilon_i}
 = \Re \oint_{{\mathcal B}_i} ydx =0
$.
This comes in addition to $\Re \oint_{{\mathcal A}_i} ydx\ = \Re 2\pi i \epsilon_i = 0$, which holds because the filling fractions are real. 
So the real part of the integral of $ydx$ on any contour vanishes, 
\begin{equation}
\forall\,\mathcal C , \qquad  \Re \oint_{{\mathcal C}} ydx\ = 0  \ .
\end{equation}
This condition on the spectral curve is called the \textbf{Boutroux property}\index{Boutroux property (algebraic curve)}.
Our expression \eqref{eq:oscilZN} for the partition function holds provided the spectral curve that corresponds to the reference filling fractions $\vec \epsilon^*$ is in a neighborhood of size $O(\frac{1}{N})$ of a Boutroux algebraic curve. 

Actually, minimizing $\Re F_0$ with respect to filling fractions is a consequence of minimizing the functional action with respect to the density of eigenvalues (without fixing filling fractions), as we did in \autoref{sec:ded}. So the spectral curve that corresponds to the equilibrium density is a Boutroux spectral curve.

\iftrue
The functional $\vec\epsilon \mapsto F_0(\vec\epsilon)$ is analytic (see Section \ref{sec:F0}), in particular it satisfies
\begin{equation}
d F_0 = \frac{1}{2\pi i} \ \sum_{i=1}^s  \oint_{\mathcal B_i} ydx  \, d\left( \oint_{\mathcal A_i} ydx \right),
\end{equation}
thus in terms of the real $\epsilon_i$s
\begin{equation}
d \Re F_0 =  \sum_{i=1}^s  \Re \oint_{\mathcal B_i} ydx  \, d\epsilon_i .
\end{equation}
Therefore at $\vec\epsilon^*$ we have $d \Re F_0=0$.
The second derivative is
\begin{equation}
d^2 \Re F_0 =  \frac{1}{2\pi} \sum_{i,j=1}^s  \Im \tau_{i,j}  \, d\epsilon_i  d\epsilon_j
\end{equation}
which is a positive definite quadratic form.
Therefore the Boutroux curve is the minimum of $\Re F_0$.
\fi
% NB: The above paragraph starts with equations that are not obvious from our text, and reaches a conclusion that is apparently obvious on the relation between second derivatives of F_0, and the matrix \tau. (This is how we defined \tau!)
%BE: indeed this needs a rewriting. These equations, and in particular the fact that the Hessian is the matrix $\Im\tau$ is a useful equation to have in a textbook. We need to better formulate.

\subsection{Graphical representation}

\subsubsection{Partition function}

The expansion \eqref{eq:oscilZN} of the partition function has 
an efficient graphical representation, where

\begin{itemize}

\item $-F_{g}^{(\vec m)}$ corresponds to a surface
      of genus $g$ with $m$ outgoing legs, which we call a vertex, 
\begin{align}
\begin{tikzpicture}[baseline=(current bounding box.center)]
  \begin{scope}
   \node at (0,.3) [left]
   { $-F_{g}^{(\vec m)} \ =$};   
   % shape
   \draw [thick] (3,0) circle [x radius = 2., y radius = 1];
   % holes
   \begin{scope}[thick,shift={(2,-.4)}]
    \draw [clip] (-.5,.3) to [bend right] (.5,.3);
    \draw (-.5,0.1) to [bend left] (.5,0.1);
   \end{scope}
   \begin{scope}[thick,shift={(3,.2)}]
    \draw [clip] (-.5,.3) to [bend right] (.5,.3);
    \draw (-.5,0.1) to [bend left] (.5,0.1);
   \end{scope}
   \begin{scope}[thick,shift={(4,-.2)}]
    \draw [clip] (-.5,.3) to [bend right] (.5,.3);
    \draw (-.5,0.1) to [bend left] (.5,0.1);
   \end{scope}
   \node at (5.5,0) [right] {$g$ handles};
   % vertices
   \draw [->-,thick] (1.5,.3) -- ++(120:1) ;
   \draw [->-,thick] (2,.6) -- ++(100:1) ;
   \draw [->-,thick] (4.2,.5) -- ++(70:1) ;
   \draw [thick,dotted] (2.5,1.5) to [bend left] (4,1.4);
    \node at (3,1.6) [above] {$m$ legs};
  \end{scope}
 \end{tikzpicture}
\end{align}

\item $\Theta^{(\vec m)}$ corresponds to a black dot with $m$ incoming legs,
\begin{align}
 \begin{tikzpicture}[baseline=(current bounding box.center)]
  \node at (0,0) [left]
  { $\Theta^{(\vec m)} \ =$};   
  \filldraw (2,0) circle [radius =.1] node (dot) {};
  \foreach \x in {0,1,2,3,4}
  {
  \draw [thick] (2,0) --++(40+\x*360/5:.7) node (m\x) {} --++(40+\x*360/5:.3);
  \draw [thick,latex-] (m\x) -- (dot);
  }
  \node at (4,0) {$m$ legs};
 \end{tikzpicture}
\end{align}

\end{itemize}
The partition function can then be written as a sum over all possible graphs with one black dot and arbitrarily many vertices,
\begin{equation}
\boxed{
  \mathcal{Z}(\gamma^N) 
 \sim 
 \left( 
  \Theta + \sum_{G} 
 \frac{N^{\chi(G)}}{\#\text{Aut}(G)} 
 \left[\prod_{\text{vertices}
 \,v} (-F^{(\vec m_v)}_{g_v} )\right] \Theta^{(\sum_v \vec m_v)}
 \right)
}\ ,
\end{equation}
where $\text{Aut}(G)$ is the group of automorphisms of the graph $G$, and $\chi(G)=\sum_v (2-2g_v-m_v)$ is its Euler characteristic. 
For example, the first two orders are:
\begin{subequations}\label{eq:oscilgraphicorder2}
\begin{align}
 O\left(\tfrac{1}{N}\right) \ : \quad &
 \begin{tikzpicture}[baseline=(current  bounding  box.center),thick,scale=.7]
  \draw (0,0) circle [x radius = 1, y radius = .7];
  \begin{scope}
   \draw [clip] (-.5,.1) to [bend right] (.5,.1);
   \draw (-.5,-.1) to [bend left] (.5,-.1);
  \end{scope}
  \draw [->-] (0.8,0) -- ++(1,0) node (dot1) {};
  \filldraw (dot1) circle [radius=.1];
 \end{tikzpicture}
 \ + \ \frac{1}{6} \
 \begin{tikzpicture}[baseline=(current  bounding  box.center),thick,scale=.7]
  \draw (0,0) circle [x radius = 1, y radius = .7]; 
  \draw [->-] (.7,0) -- ++(1,0) node (dot2) {};
  \draw [->-] (.7,.3) to [out=30,in=180-45] (1.7,0);
  \draw [->-] (.7,-.3) to [out=-30,in=180+45] (1.7,0);
  \filldraw (dot2) circle [radius=.1];
 \end{tikzpicture}
  \\
 O\left(\tfrac{1}{N^2}\right) \ : \quad &
 \begin{tikzpicture}[baseline=(current  bounding  box.center),thick,scale=.7]
   \draw (0,0) circle [x radius = 1, y radius = .7];   
   \begin{scope}[scale=.6,shift={(-.8,0)}]
    \draw [clip] (-.5,.1) to [bend right] (.5,.1);
    \draw (-.5,-.1) to [bend left] (.5,-.1);
   \end{scope}
   \begin{scope}[scale=.6,shift={(.8,0)}]
    \draw [clip] (-.5,.1) to [bend right] (.5,.1);
    \draw (-.5,-.1) to [bend left] (.5,-.1);
   \end{scope}
   \filldraw (1.8,0) circle [radius=.1];
 \end{tikzpicture}
 \ + \ \frac{1}{2} \
 \begin{tikzpicture}[baseline=(current  bounding  box.center),thick,scale=.7]
  \draw (0,0) circle [x radius = 1, y radius = .7]; 
   \begin{scope}
    \draw [clip] (-.5,.1) to [bend right] (.5,.1);
    \draw (-.5,-.1) to [bend left] (.5,-.1);
   \end{scope}
   \node (dot2) at (1.7,0) {};
   \draw [->-] (.7,.3) to [out=30,in=180-45] (1.7,0);
   \draw [->-] (.7,-.3) to [out=-30,in=180+45] (1.7,0);
   \filldraw (dot2) circle [radius=.1];
  \end{tikzpicture}
 \ + \ \frac{1}{24} \
 \begin{tikzpicture}[baseline=(current  bounding  box.center),thick,scale=.7]
  \draw (0,0) circle [x radius = 1, y radius = .7]; 
  \node (dot2) at (1.7,0) {};
  \draw [->-] (.7,.1) to [out=20,in=180-30] (1.7,0);
  \draw [->-] (.7,.3) to [out=45,in=180-60] (1.7,0);
  \draw [->-] (.7,-.1) to [out=-20,in=180+30] (1.7,0);
  \draw [->-] (.7,-.3) to [out=-45,in=180+60] (1.7,0);
  \filldraw (dot2) circle [radius=.1];
 \end{tikzpicture}
 \nonumber
 \\
 & + \ \frac{1}{2} \
 \begin{tikzpicture}[baseline=(current  bounding  box.center),thick,scale=.7]
   \draw (0,0) circle [x radius = 1, y radius = .7];   
   \begin{scope}
    \draw [clip] (-.5,.1) to [bend right] (.5,.1);
    \draw (-.5,-.1) to [bend left] (.5,-.1);
   \end{scope}
   \begin{scope}[shift={(.7,-1.7)}]
    \draw (0,0) circle [x radius = 1, y radius = .7];
    \begin{scope}
     \draw [clip] (-.5,.1) to [bend right] (.5,.1);
     \draw (-.5,-.1) to [bend left] (.5,-.1);
    \end{scope}
   \end{scope}
   \draw [->-] (0.8,0) -- ++(1,0) node (dot1) {};
   \draw [->-] (1.2,-1.3) -- (1.8,0);
   \filldraw (1.8,0) circle [radius=.1];
 \end{tikzpicture}
 \ + \ \frac{1}{6} \
 \begin{tikzpicture}[baseline=(current  bounding  box.center),thick,scale=.7]
   \draw (0,0) circle [x radius = 1, y radius = .7];
   \begin{scope}
    \draw [clip] (-.5,.1) to [bend right] (.5,.1);
    \draw (-.5,-.1) to [bend left] (.5,-.1);
   \end{scope}
   \begin{scope}[shift={(.7,-1.7)}]
    \draw (0,0) circle [x radius = 1, y radius = .7];
   \end{scope}
   \draw [->-] (0.8,0) -- ++(1,0) node (dot1) {};
   \draw [->-] (1.2,-1.4) -- (1.8,0);
   \draw [->-] (1.1,-1.3) to [out=up,in=180+30] (1.8,0);
   \draw [->-] (1.3,-1.5) to [out=30,in=down] (1.8,0);
   \filldraw (1.8,0) circle [radius=.1];
 \end{tikzpicture}
 \ + \ \frac{1}{72} \
 \begin{tikzpicture}[baseline=(current  bounding  box.center),thick,scale=.7]
   \draw (0,0) circle [x radius = 1, y radius = .7];
   \begin{scope}[shift={(.7,-1.7)}]
    \draw (0,0) circle [x radius = 1, y radius = .7];
   \end{scope}
   \draw [->-] (.7,0) -- ++(1.1,0) node (dot1) {};
   \draw [->-] (.7,.3) to [out=30,in=180-45] (1.8,0);
   \draw [->-] (.7,-.3) to [out=-30,in=180+45] (1.8,0);
   \draw [->-] (1.2,-1.4) -- (1.8,0);
   \draw [->-] (1.1,-1.3) to [out=up,in=180+30] (1.8,0);
   \draw [->-] (1.3,-1.5) to [out=30,in=down] (1.8,0);
   \filldraw (dot1) circle [radius=.1];
 \end{tikzpicture}
\end{align}
\end{subequations}

\subsubsection{Free energy}

Let us define the derivative of $\log \Theta$,
\begin{equation}
T^{(\vec m)}(\vec u) = \partial_{\vec u}^{\vec m} \log\Theta(\vec u)\ ,
\end{equation}
which we represent as a white dot with $m$ ingoing legs:
\begin{align}
 \begin{tikzpicture}[baseline=(current bounding box.center)]
  \node at (0,0) [left]
  { $T^{(\vec m)} \ =$};   
  \node (dot) at (2,0) {};
  \foreach \x in {0,1,2,3,4}
  {
  \draw [thick] (2,0) --++(40+\x*360/5:.7) node (m\x) {} --++(40+\x*360/5:.3);
  \draw [thick,latex-] (m\x) -- (dot);
  }
  \filldraw[fill=white,thick] (dot) circle [radius =.1];
  \node at (4,0) {$m$ legs};
 \end{tikzpicture}
\end{align}
The relation between the derivatives of $\Theta$ and $\log \Theta$,
\begin{align}
\frac{\Theta^{(\vec m)}}{\Theta} 
 = \sum_{\sqcup \vec m_i = \vec m} \prod_i T^{(\vec m_i)} \ ,
\end{align}
amounts to decomposing the black dot into white dots, for instance,
% black node to white nodes
\begin{align}
 \begin{tikzpicture}[baseline=(current bounding box.center)]  
  % black
  \filldraw (0,0) circle [radius =.1] node (dot) {};
  \foreach \x in {0,1,2}
  {
  \draw [thick] (0,0) --++(50+\x*360/3:.7) node (m\x) {} --++(50+\x*360/3:.3);
  \draw [thick,latex-] (m\x) -- (dot);
  }  
  \node at (1.5,0) {$=$};
  \node at (5.,0) {$+$ \ $3$};
  \node at (8.5,0) {$+$};
  \begin{scope}[shift={(3.5,0)}]
  \node (dot) at (0,0) {};
  \foreach \x in {0,1,2}
  {
  \draw [thick] (0,0) --++(50+\x*360/3:.7) node (m\x) {} --++(50+\x*360/3:.3);
  \draw [thick,latex-] (m\x) -- (dot);
  }
  \filldraw[fill=white,thick] (dot) circle [radius =.1];
%
%   \node at (0,-1.5) [below] {$\yng(3)$};
%
  \end{scope}
  \begin{scope}[shift={(7.,0)}]
  \node (dot1) at (-.1,.2) {};
  \node (dot2) at (.1,-.2) {};
  \draw [thick] (-.1,.2) --++(50:.7) node (m1) {} --++(50:.3);
  \draw [thick] (-.1,.2) --++(50+120:.7) node (m2) {} --++(50+120:.3);
  \draw [thick] (.1,-.2) --++(50+240:.7) node (m3) {} --++(50+240:.3);
  \draw [thick,latex-] (m1) -- (dot1);
  \draw [thick,latex-] (m2) -- (dot1);
  \draw [thick,latex-] (m3) -- (dot2);
  \filldraw[fill=white,thick] (dot1) circle [radius =.1];
  \filldraw[fill=white,thick] (dot2) circle [radius =.1];
%
%   \node at (0,-1.5) [below] {$\yng(2,1)$};
%
  \end{scope}
  \begin{scope}[shift={(10.5,0)}]
   \foreach \x in {0,1,2} {
   \node (dot\x) at (50+\x*120:.2) {};
   \draw [thick] 
   (50+\x*120:.2) --++(50+\x*120:.7) node (m\x) {} --++(50+\x*120:.3);
  \draw [thick,latex-] (m\x) -- (dot\x);
  \filldraw[fill=white,thick] (dot\x) circle [radius =.1];
   }
%
%   \node at (0,-1.5) [below] {$\yng(1,1,1)$};
%
  \end{scope}
 \end{tikzpicture}
\end{align}
Using this representation, $\mathcal Z_N(\gamma)$ is then a sum of graphs made of surfaces and white dots, and taking the log amounts to keeping only the connected graphs.
The free energy $\log \mathcal Z_N(\gamma)$ can thus be written as a sum over all connected graphs with arbitrarily many white dots and vertices,
\begin{equation}\label{eq:oscillnZN}
  \log \mathcal{Z}(\gamma^N) 
 \sim 
 \log\Theta+ \sum_{G\, \text{connected}} 
 \frac{N^{\chi(G)}}{\#\text{Aut}(G)} 
 \prod_{\text{vertices} \,v} F^{(\vec m_v)}_{g_v} 
 \prod_{\text{white dots} \,v'} T^{(\vec m_{v'})}\ .
\end{equation}
In particular, the leading nontrivial contribution $F_1(\gamma^N)$ such that $\log \mathcal{Z}(\gamma^N) 
  \sim \log\Theta  + \frac{1}{N} F_1(\gamma^N) + O(\frac{1}{N^2})$, is 
\begin{align}
 F_1(\gamma^N) & = 
 \begin{tikzpicture}[baseline=(current  bounding box.center),thick,scale=.7]
  \draw (0,0) circle [x radius = 1, y radius = .7];
  \begin{scope}
   \draw [clip] (-.5,.1) to [bend right] (.5,.1);
   \draw (-.5,-.1) to [bend left] (.5,-.1);
  \end{scope}
  \draw [->-] (0.8,0) -- ++(1,0) node (dot1) {};
  \filldraw[fill=white] (dot1) circle [radius=.1];
 \end{tikzpicture}
 \ + \ \frac{1}{6} \
 \begin{tikzpicture}[baseline=(current  bounding box.center),thick,scale=.7]
  \draw (0,0) circle [x radius = 1, y radius = .7]; 
  \draw [->-] (.7,0) -- ++(1,0) node (dot2) {};
  \draw [->-] (.7,.3) to [out=30,in=180-45] (1.7,0);
  \draw [->-] (.7,-.3) to [out=-30,in=180+45] (1.7,0);
  \filldraw [fill=white] (dot2) circle [radius=.1];
 \end{tikzpicture}
 \ + \ \frac{1}{2} \
 \begin{tikzpicture}[baseline=(current  bounding box.center),thick,scale=.7]
  \draw (0,0) circle [x radius = 1, y radius = .7]; 
  \draw [->-] (.7,.3) to [out=20,in=180-30] (1.7,.15) node (dot1) {};
  \draw [->-] (.7,0) to [out=-20,in=180+30] (1.7,.15);
  \draw [->-] (.7,-.3) -- (1.7,-.3) node (dot2) {};
  \filldraw [fill=white] (dot1) circle [radius=.1];
  \filldraw [fill=white] (dot2) circle [radius=.1];
 \end{tikzpicture}
 \ + \ \frac16 \
 \begin{tikzpicture}[baseline=(current  bounding box.center),thick,scale=.7]
  \draw (0,0) circle [x radius = 1, y radius = .7]; 
  \draw [->-] (.7,.3) --++ (1,0) node (dot1) {};
  \draw [->-] (.7,-.3) --++ (1,0) node (dot2) {};
  \draw [->-] (.7,0) --++ (1,0) node (dot3) {};
  \filldraw [fill=white] (dot1) circle [radius=.1];
  \filldraw [fill=white] (dot2) circle [radius=.1];
  \filldraw [fill=white] (dot3) circle [radius=.1];
 \end{tikzpicture}
\nonumber \\
& = F_1^{(\vec 1)}T^{(\vec 1)} + \frac16 F_0^{(\vec 3)}T^{(\vec 3)}
+ \frac12 F_0^{(\vec 3)}T^{(\vec 2)}T^{(\vec 1)}+ \frac16 F_0^{(\vec 3)}T^{(\vec 1)}T^{(\vec 1)}T^{(\vec 1)}\ .
\end{align}

\subsubsection{Wave functions}

For $D$ a divisor of degree zero, let us give a graphical representation of the wave function $\psi(D)$ \eqref{eq:psid}. Before generating oscillatory terms by summing over filling fractions, we consider the case when correlation functions have topological expansions. We then have 
\begin{align}
 \log\psi(D) = \sum_{n=1}^\infty \sum_{g=0}^\infty \frac{N^{2-2g-n}}{n!} \int_{D} \dots \int_{D} \left(\omega_{g,n}-\delta_{g.0}\delta_{n,2}\frac{dx_1dx_2}{(x_1-x_2)^2} \right)\ .
 \label{lpd}
\end{align}
Let us pay particular attention to the $g=0,n=2$ term. We recall that $\omega_{0,2}(z_1,z_2)=B(z_1,z_2)$ is the fundamental second kind differential on the Riemann surface $\Sigma$, where $z$ is a coordinate on $\Sigma$ and $x_i=x(z_i)$. Let us introduce its primitive, called the \textbf{prime form}\index{prime form} \cite{Fay:1973}, a $(-\frac12,-\frac12)$ form $E(z_1,z_2)$ such that 
\begin{align}
 B(z_1,z_2)= dz_1 dz_2 \frac{d}{dz_1} \frac{d}{dz_2}  \log{E(z_1,z_2)}\ .
\end{align}
(In the 1-cut case, $\Sigma$ is the Riemann sphere, and $ E(z_1,z_2) = \frac{z_1-z_2}{\sqrt{dz_1 dz_2}}$.)
Writing $D=\sum_i\alpha_i\cdot [x_i]$, we then find  
\begin{align}
 \int_D \int_D \left(\omega_{0,2}-\frac{dx_1dx_2}{(x_1-x_2)^2} \right) = - \sum_i \frac{\alpha_i^2}{2} \log{dx(z_i)} +  \sum_{i<j} \alpha_i\alpha_j \log\frac{E(z_i,z_j)}{x(z_i)-x(z_j)}\ .
\end{align}
Then, for $2g-2+n>0$, we introduce the graphical representation
\begin{align}
 \begin{tikzpicture}[baseline=(current  bounding box.center),thick,scale=.7]
  \begin{scope}%[shift={(1.1,0)}]
   \node at (0,.3) [left]
   { $\displaystyle \int_D \cdots \int_D \omega_{g,n} \ =$};   
   % shape
   \draw [thick] (3,0) circle [x radius = 2., y radius = 1];
   % holes
   \begin{scope}[thick,shift={(2,-.4)}]
    \draw [clip] (-.5,.3) to [bend right] (.5,.3);
    \draw (-.5,0.1) to [bend left] (.5,0.1);
   \end{scope}
   \begin{scope}[thick,shift={(3,.2)}]
    \draw [clip] (-.5,.3) to [bend right] (.5,.3);
    \draw (-.5,0.1) to [bend left] (.5,0.1);
   \end{scope}
   \begin{scope}[thick,shift={(4,-.2)}]
    \draw [clip] (-.5,.3) to [bend right] (.5,.3);
    \draw (-.5,0.1) to [bend left] (.5,0.1);
   \end{scope}
   \node at (5.5,0) [right] {$g$ holes};
   % vertices
   \draw [-triangle 60 reversed, thick] (1.5,.3) -- ++(120:1) node [above] {$D$};
   \draw [-triangle 60 reversed, thick] (2,.6) -- ++(100:1) node [above] {$D$};
   \draw [-triangle 60 reversed, thick] (4.2,.5) -- ++(70:1) node [above] {$D$};
   \draw [thick,dotted] (2.5,1.5) to [bend left] (4,1.4);
    %\node at (3,1.6) [above] {$m$ legs};
  \end{scope}
 \end{tikzpicture}
\end{align}
We define the Euler characteristic of such a graph $G$ as $\chi(G)=2-2g-n$.
Finally, summing over filling fractions amounts to introducing legs that end on white dot vertices $T^{(\vec m)}$, in such a way that the whole graph is connected. The resulting diagrammatic representation of the wave function is 
\begin{multline}
\log \psi(D) 
\sim  N\int_D ydx  - \sum_i \frac{\alpha_i^2}{2} \log{dx(z_i)} +  \sum_{i<j} \alpha_i\alpha_j \log\frac{E(z_i,z_j)}{x(z_i)-x(z_j)} 
\\
+ \sum_{G\text{ connected}} \frac{N^{\chi(G)}}{\#\mathrm{ Aut}(G)}
\begin{tikzpicture}[baseline=(current  bounding box.center),thick,scale=.7]
  \begin{scope}%[shift={(1.1,0)}]
   %\node at (0,.3) [left]   { $F_{g}^{(m)} \ =$};   
   % shape
   \draw [thick] (3,0) circle [x radius = 2., y radius = 1];
   % holes
   \begin{scope}[thick,shift={(2,-.4)}]
    \draw [clip] (-.5,.3) to [bend right] (.5,.3);
    \draw (-.5,0.1) to [bend left] (.5,0.1);
   \end{scope}
   \begin{scope}[thick,shift={(3,.2)}]
    \draw [clip] (-.5,.3) to [bend right] (.5,.3);
    \draw (-.5,0.1) to [bend left] (.5,0.1);
   \end{scope}
   \begin{scope}[thick,shift={(4,-.2)}]
    \draw [clip] (-.5,.3) to [bend right] (.5,.3);
    \draw (-.5,0.1) to [bend left] (.5,0.1);
   \end{scope}
   % vertices
   \draw [-triangle 60 reversed, thick] (1.5,.3) -- ++(120:1) node [above] {$D$};
   \draw [-triangle 60 reversed, thick] (2,.6) -- ++(100:1) node [above] {$D$};
   \draw [-triangle 60 reversed, thick] (4.2,.5) -- ++(70:1) node [above] {$D$};
   \draw [thick,dotted] (2.5,1.5) to [bend left] (4,1.4);
    %\node at (3,1.6) [above] {$m$ legs};
   \draw [->-, thick] (4.2, -.5) to [bend right] (5.4, -1.8) node[label={-90:$T$}] (dot1) {};
   \draw [->-, thick] (4.6, -.3) to [bend left] (5.4, -1.8);
   \draw [->-, thick] (6.3, -.5) to [bend left] (5.4, -1.8);
   \filldraw[fill=white] (dot1) circle [radius=.1];   
   \draw [->-, thick] (2.2, -.7) to [bend right] (2.4, -1.9) node[label={-90:$T$}] (dot1) {};
   \draw [->-, thick] (2.7, -.8) to [bend left] (2.4, -1.9);
   \filldraw[fill=white] (dot1) circle [radius=.1];
  \end{scope}  
  \begin{scope}[thick, shift = {(6.5, 0)}]
  \draw (0,0) circle [x radius = 1, y radius = .7];
  \draw [-triangle 60 reversed, thick] (0,.4) -- ++(90:1) node [above] {$D$};
  \begin{scope}
   \draw [clip] (-.5,.1) to [bend right] (.5,.1);
   \draw (-.5,-.1) to [bend left] (.5,-.1);
  \end{scope}
  \end{scope}
 \end{tikzpicture}
%\text{SURFACEs WITH n LEGS ENDING ON D,D,...,D, and legs ending on white dots}
\end{multline}

\section{Exercises}

\begin{exo}[Loop equations of the multi-matrix model]
~\label{exo:loop_multi_matrix}
Consider the chain of matrices whose partition function is given in Eq.~\eqref{def:chainmatrix}.
\begin{enumerate}
\item 
For the one-matrix model with an external field, derive the loop equation \eqref{eq:sdpt1Matrixext} using the infinitesimal change of variables
\begin{equation}
M_1 \to M_1 + \epsilon \frac{1}{x-M_1} \frac{1}{y-M_2} \ .
\end{equation}
\item 
For the two-matrix model, derive the loop equations \eqref{eq:sdpt2matrixmm} using the
infinitesimal changes of variables
\begin{equation}
M_1 \to M_1 + \epsilon \frac{1}{x-M_1} \frac{V'_2(y)-V'_2(M_2)}{y-M_2} \qquad \text{then}\qquad
M_2 \to M_2 + \epsilon \frac{1}{x-M_1} \ .
\end{equation}
\item For the full chain, which changes of variables lead to the loop equations \eqref{eq:sdptchainmatrices}?
\end{enumerate}

\end{exo}

\begin{exo}[Loop equations of the $O(n)$ matrix model]
~\label{exo:loop_On_matrix}
In the partition function \eqref{defOnmmgraphs} of the $O(n)$ model, let $V(x) = \frac{x^2}{2} - \sum_k \frac{t_k}{k} x^k$ be the potential such that the integrand has a factor $e^{-N\operatorname{Tr}V(M)}$. We define
\begin{equation}
W_1(x) = \left<\operatorname{Tr} \frac{1}{x-M}\right> \  , \
G(x) \underset{\forall i}{=} \left<  \operatorname{Tr} \frac{1}{x-M}  A_i^2\right>\  , \
P_0(x) = \left< \operatorname{Tr} \frac{V'(x)-V'(M)}{x-M} \right>\, .
\end{equation}
\begin{enumerate}

\item Perform the infinitesimal change of variable
\begin{equation}
A_i \to A_i + \epsilon \frac{1}{x-M}  A_i \frac{1}{\tilde x+M}\ ,
\end{equation}
and choose $\tilde x = 1-x$ to derive the loop equation
\begin{equation}
W_1(x)W_1(1-x) + W_2(x,1-x)  = N (G(x)+G(1- x) ) \ . 
\end{equation}

\item Prove the loop equation
\begin{equation}
W_1(x)^2 +W_2(x,x)  = N\left( V'(x) W_1(x) - P_0(x) - n G(x) \right) \ .
\end{equation}

\item Assuming a topological expansion, deduce the equation for $W_{0,1}(x)$:
    \begin{multline}
W_{0,1}(x)^2 + W_{0,1}(1-x)^2 + n W_{0,1}(x)W_{0,1}(1-x)  \\
   = V'(x) W_{0,1}(x) +V'(1-x) W_{0,1}(1-x) - P_{0,1}(x)- P_{0,1}(1-x)\ .
    \end{multline}
These non-local functional equations can be solved in some cases. For example, there are 1-cut solutions in terms of  elliptic
functions \cite{Eynard:2006bs}.
\end{enumerate}

\end{exo}

\begin{exo}[Loop equations in the presence of boundaries]
~\label{exooopeqtrv}
Let $\gamma\subset \mathbb C$ be a contour, and let $V$ be a polynomial of degree $d+1$.
Consider the integral
\begin{equation}
\mathcal Z = \int_{\gamma^N} \prod_{i<j} (\lambda_i-\lambda_j)^2 \prod_{i=1}^N e^{-N V(\lambda_i)} d\lambda_i \ .
\end{equation}

\begin{enumerate}

\item Assuming that $\gamma$ goes from $\infty$ to $\infty$, prove the loop equations
\begin{equation}
\langle \operatorname{Tr} V'(M)\rangle = \left\langle \sum_i V'(\lambda_i) \right\rangle = 0
\ , \quad
\langle \operatorname{Tr} MV'(M)\rangle = \left\langle \sum_i \lambda_i V'(\lambda_i) \right\rangle = N\ .
\end{equation}
Justify the interpretation of these equations as translation and dilation invariance.

\item Assume that $\gamma=[a,\infty)$ has one boundary $a$, also called a hard edge. Find changes of variables that vanish at the boundary. 
Deduce the loop equation
\begin{equation}
\widehat{W}_2(x,x) - N V'(x) W_1(x) + N P_0(x) - \frac{N}{x-a} \left< \operatorname{Tr} V'(M)\right> =0\ .
\end{equation}
Observe that the pole at $x=a$ is present if and only if translation invariance is broken, while dilation invariance is preserved as $
\langle \operatorname{Tr} (M-a)V'(M)\rangle = N
$.
In the formal large $N$ limit, derive the leading order resolvent
\begin{align}
W_{0,1}(x) = \frac12 \left( V'(x) - \sqrt{V'(x)^2 - 4 P_{0,1}(x) + \frac{4v}{x-a}} \right)  \ ,
\end{align}
where $v=\lim_{N\to\infty} \frac{1}{N} \left<\Tr V'(M)\right>$.
How does the spectral density behave as $x\to a$?
Show that such a singularity is integrable (i.e. the integration of the integral of the density is convergent). 

\item Assuming that $\gamma=[a,b]$ has two boundaries, prove the loop equation
\begin{multline}
\widehat{W}_2(x,x) - N V'(x) W_1(x) + N P_0(x) + \frac{N^2}{(x-a)(x-b)} \\
- \frac{N}{(x-a)(x-b)} \Big( (x-a-b)\langle \operatorname{Tr} V'(M)\rangle + \langle \operatorname{Tr} MV'(M)\rangle \Big) = 0\ .
\end{multline}
Conclude that both translation and dilation invariances can be broken, leading to poles at the boundaries. Compute the leading order resolvent.

\item Assuming that $\gamma$ is a union of intervals with boundaries at points $a_i$, let
\begin{equation}
 \sigma(x)=\prod_i (x-a_i)=\sum_k \sigma_k x^k\ .
\end{equation}
Find changes of variables that vanish at all boundaries, and prove the loop equation
\begin{multline}
\widehat{W}_2(x,x) - N V'(x) W_1(x) + N P_0(x) 
- \frac{N}{\sigma(x)} \left\langle \operatorname{Tr} \frac{\sigma(x)-\sigma(M)}{x-M} V'(M)\right\rangle 
\\
+ \frac{1}{\sigma(x)} \sum_k \sigma_k \sum_{i+j\leq k-2} x^{k-2-i-j} \langle \operatorname{Tr} M^i \operatorname{Tr} M^j \rangle 
= 0\ .
\end{multline}
Show that in the formal large $N$ limit, the solution is 
\begin{align}
W_{0,1}(x) = \frac12 \left( V'(x) - \sqrt{V'(x)^2 - 4 P_{0,1}(x) + \frac{4 Q_{0,1}(x)}{\sigma(x)}   } \right) \ ,
\end{align}
with some polynomial $Q_{0,1}(x)$ to be determined. Conlude that $W_{0,1}(x)$ and the limit spectral density may have $(x-a_i)^{-\frac12}$ singularities at the hard edges.

\end{enumerate}

\end{exo}

\begin{exo}[Gaussian matrix model]
~\label{exogauss}
Consider the matrix integral \eqref{mat_int} with the potential $V(x)=\frac{x^2}{2}$.

\begin{enumerate}

\item
Compute the polynomials $P_n$ \eqref{eq:pndef}, and deduce
\begin{equation}
P_{g,n}(x;x_1,\dots,x_n) = \delta_{g,0}\delta_{n,0}\ .
\end{equation}

\item 
Show that the leading order resolvent is
\begin{equation}
W_{0,1}(x) = \frac{1}{2}\left( x-\sqrt{x^2-4}\right) = z^{-1}\ ,
\end{equation}
where $x$ and $z$ are related by the Joukowsky map as $x=z+z^{-1}$.

\item
Compute $W_{0,2}$, $W_{1,1}$ and $W_{0,3}$ in the $x$ and $z$ variables, and check in particular
\begin{subequations}
\begin{align}
 W_{1,1}(x) &= \frac{1}{(x^2-4)^{5/2}}\ ,
 \\
 W_{0,3}(x_1,x_2,x_3) dx_1 dx_2 dx_3 
&=  \sum_{a = \pm 1} \frac{a}{2}   \frac{dz_1 dz_2 dz_3}{(z_1 - a)^2 (z_2 - a)^2 (z_3 - a)^2}  \ .
\end{align}
\end{subequations}

\end{enumerate}
\end{exo}

\begin{exo}[From Airy curve to Airy function with Topological Recursion]
~\label{exo:airy}
Consider the spectral curve of equation $y^2-x=0$, with the rational parametrization $x(z)=z^2, y(z)=z$, where $z$ is a coordinate on the Riemann sphere. 
We recursively define differential forms $\omega_{g,n}$ by the topological recursion equation.

\begin{enumerate}

\item Discuss the cuts of $y(x)$, and the eigenvalue density. Explain why our spectral curve cannot occur as the spectral density in a matrix model.

\item Find the branch points and the local Galois involution.

\item 
Show that if $2g-2+n>0$, then the differential $\omega_{g,n}(z_1,\dots,z_n)$ is $dz_1\dots dz_n$ times an even polynomial of $z_i^{-1}$, which is homogeneous of degree $6g+4n-6$.
Deduce that the numbers $\langle \tau_{d_1} \dots \tau_{d_n}\rangle_g$ defined by
\begin{align}
\omega_{g,n}(z_1,\dots,z_n) = (-1)^n 2^{2-2g-n} \sum_{d_1,\dots,d_n} \langle \tau_{d_1} \dots \tau_{d_n}\rangle_g \prod_{i=1}^n \frac{(2d_i+1)!!\, dz_i}{z_i^{2d_i+2}} \ ,
\label{ognz}
\end{align}
can only be non-vanishing if $\sum_i d_i = 3g-3+n$.
These numbers are called Kontsevich--Witten intersection numbers, and they are integrals of Chern classes over the moduli spaces of Riemann surfaces~\cite{Kontsevich:1992ti}.

\item Compute the recursion kernel, and then compute $\omega_{0,3}$, $\omega_{0,4}$ and $\omega_{1,1}$, and 
deduce
\begin{equation}
\langle\tau_0^3\rangle_{0} = 1
\quad , \quad
\langle\tau_0^3\tau_1\rangle_{0} = 1
\quad , \quad
\langle\tau_1\rangle_{1} = \frac{1}{24}\ .
\label{eq:Airy_curve_tau_av}
\end{equation}

\item Let us define $ W_{0,1}(z) =  z $ and $W_{0,2}(z_1,z_2) =  \frac{z_1^2+z_2^2}{4z_1 z_2 (z_1^2-z_2^2)^2}$, while for $(g,n)\notin\{(0,1),(0,2)\}$ we define $W_{g,n}$ from $\omega_{g,n}$ \eqref{eq:wdsc}.
Then, for $(g,n)\neq (1,0)$, we define $P_{g,n}$ from $W_{g,n}$ using the topological expansion of the loop equation \eqref{eq:loopconn}:
\begin{multline}
 P_{g,n}(z;z_1,\dots,z_n)
 =  W_{g-1,n+2}(z,z,z_1,\dots,z_n) - \sum_{j=1}^n  \frac{d}{dx_j}    \frac{  W_{g,n}(z_1,\dots,z_n)  }{x-x_j}
  \\+ \sum_{g_1+g_2=g}\sum_{I_1\sqcup I_2 = \{z_1,\dots,z_n\} } W_{g_1,1+|I_1|}(z,I_1) W_{g_2,1+|I_2|}(z,I_2)   \ .
  \label{eq:AiryLoopeqPgn}
\end{multline}
We also define $P_{1,0}(z)=2W_{0,1}(z)W_{1,1}(z) - \frac{1}{16z^4}$.
Compute $P_{g,n}$ using the topological recursion equation \eqref{TRecursion} for $\omega_{g,n}$, and show that
\begin{equation}
P_{g,n}(z_;z_1,\dots,z_n) = \delta_{g,0}\delta_{n,0} z^2\ .
\end{equation}

\item Let us define $F_{g,n}(z_1,\dots,z_n)$ as 
\begin{equation}
F_{0,1}(z) = \frac{2}{3} z^{3}
\quad , \quad
F_{0,2}(z_1,z_2) = -\log{(z_1+z_2)}\ ,
\end{equation}
and, if $2g-2+n>0$, 
\begin{equation}
F_{g,n}(z_1,\dots,z_n) = \int_{z'_1=\infty}^{z_1}\dots \int_{z'_n=\infty}^{z_n} \omega_{g,n}(z'_1,\dots,z'_n)\ .
\end{equation}
Write the wave function $\psi(z)$ \eqref{lpd} in terms of $F_{g,n}$, and show
\begin{align}
\psi(z) 
= \frac{e^{N \frac23 z^3}}{\sqrt{2z}} \left[1 +  \frac{1}{N} \frac{5 }{48 z^3} +O\left(\frac{1}{N^2}\right)\right]\ . \label{eq:Airy_wf_WKB}
\end{align}
Deduce that $u(z) = \frac{1}{N} \frac{d\log\psi(z) }{dx(z)}$ obeys
\begin{equation}
u(z)^2 + \frac{1}{N} \frac{d u(z) }{dx(z)} 
= z^2 + O\left(\frac{1}{N^3}\right)\ .
\label{eq:uqdu}
\end{equation}

\item 
Let us
define the following rational functions of $z$,
\begin{subequations}
\begin{align}
\Phi_{g,n}(x(z)) = & \frac{1}{n!} F_{g,n}(z,\cdots,z)\ , \\
\Phi^{(1)}_{g,n+1}(z_1;x(z)) = & \frac{1}{n!} \frac{d}{dx(z_1)}  F_{g,n+1}(z_1,z,\cdots,z) \ ,  \\
\Phi^{(1,1)}_{g,n+2}(z_1,z_2;x(z)) = & \frac{1}{n!} \frac{d}{dx(z_1)}\frac{d}{dx(z_2)}   F_{g,n+2}(z_1,z_2,z,\cdots,z) \ , \\
\Phi^{(2)}_{g,n+1}(z_1;x(z)) = & \frac{1}{n!} \frac{d^2}{dx(z_1)^2}  F_{g,n+1}(z_1,z,\cdots,z)\  .
\end{align}
\end{subequations}
Compute $\Phi_{g,n}(x)$ and $\Phi^{(1)}_{g,n}(z_1,x),\Phi^{(2)}_{g,n}(z_1,x),\Phi^{(1,1)}_{g,n}(z_1,z_2,x)$ for $(g,n)=(0,1),(0,2)$.
For $2g-2+n>0$, write $\Phi_{g,n}$ in terms of intersection numbers.

\item
Using the loop equations \eqref{eq:AiryLoopeqPgn}, show that
\begin{multline}
 \Phi''_{g-1,n+2}(x)+  \Phi^{(2)}_{g,n}(z,x)  - \Phi^{(2)}_{ g-1,n+2}(z,x)
 \\
+ \sum_{g_1+g_2=g, n_1+n_2=n} \Phi'_{g_1,n_1+1}(x) \Phi'_{g_2,n_2+1}(x)
= x \delta_{g,0}\delta_{n,0}\ ,
\label{eq:Phi_recursion}
\end{multline}
where primes denote $x$-derivatives.

\item Deduce that the $O\left(\frac{1}{N^3}\right)$ term and all higher order terms vanish in Eq.~\eqref{eq:uqdu}, so that the wave function $\psi$ satisfies the Airy equation to all orders
\begin{equation}
\frac{1}{N^2} \frac{d^2\psi}{dx^2} = x \psi \ .
\label{eq:Airy_curve_quantization}
\end{equation}
Solving the Airy equation, we can write the wave function in terms of the Airy function $\psi(z) =\operatorname{Ai}(N^\frac23 z^2)$. This is why our spectral curve is called the Airy curve. The Airy equation can actually be considered as the quantization $y\to \frac{1}{N}\frac{d}{d x}$ of the spectral curve's equation.

\end{enumerate}

\end{exo}

\begin{exo}[Background independence]
~\label{exobackground}
Let us consider a spectral curve of genus $1$, so that the filling fraction $\vec \epsilon$ and vector $\vec u$ are actually single complex numbers $\epsilon, u \in \mathbb{C}$. We want to study how the partition function $\mathcal{Z}(\gamma^N)$ \eqref{eq:oscilZN} depends on the reference filling fraction $\epsilon^*$. 
\begin{enumerate}
\item Compute the $\epsilon^*$-derivative of the twisted Theta function $\Theta$ \eqref{eq:defThetatwisted}, and show that 
its multiple $u$-derivatives $\Theta^{(k)}$, evaluated at $\vec u = \vec u_0$ (which depends on $\epsilon^*$), obey
\begin{equation}
\partial_{\epsilon^*} \Theta^{(k)} = - F'_1 \Theta^{(k)} - \frac12  {F_0'''} \Theta^{({k+2})} - N  k   \Theta^{({k-1})} \ .
\label{petk}
\end{equation}
This is illustrated graphically in the case $k=3$ as
\begin{align}
-\partial_{{\epsilon}^*} \ 
 \begin{tikzpicture}[baseline=(current  bounding  box.center),thick,scale=.7]
  \draw  (0.8,0) -- ++(1,0) node (dot1) {};
  \draw (1.1,0.5) -> (1.8,0);
  \draw (1.1,-0.5) -> (1.8,0);
  \filldraw (dot1) circle [radius=.1];
 \end{tikzpicture}
= \quad &
 \begin{tikzpicture}[baseline=(current  bounding  box.center),thick,scale=.7]
  \draw (0.8,0) -- ++(1,0) node (dot1) {};
  \draw (1.1,0.5) -> (1.8,0);
  \draw (1.1,-0.5) -> (1.8,0);
  \filldraw (dot1) circle [radius=.1];
  \draw (1.5,-1.) circle [x radius = .5, y radius = .4];
  \begin{scope}
   \draw (1.2,-1.) to [bend right] (1.8,-1.);
   \draw (1.35,-1.) to [bend left] (1.65,-1.);
  \end{scope}
  \draw (1.9,-1.) -> (2.4,-1.) node (dot2) {};
  \draw (dot2) circle [radius=.1];
 \end{tikzpicture}
 \ + \ \frac{1}{2} \
 \begin{tikzpicture}[baseline=(current  bounding  box.center),thick,scale=.7]
  \draw (0.8,0) -- ++(1,0) node (dot1) {};
  \draw (1.1,0.5) -> (1.8,0);
  \draw (1.1,-0.5) -> (1.8,0);
  \filldraw (dot1) circle [radius=.1];
  \draw (1.8,-1.) circle [x radius = .5, y radius = .4];
  \draw (2.,-1.) to (2.6,-1.) node (dot2) {};
   \draw (dot2) circle [radius=.1];
  \draw (1.6,-.85) to  [out=90+30, in=90+30] (1.8,-.1);
  \draw (2.0,-.85) to [out=90-30, in=90-30] (1.8,-.1);
 \end{tikzpicture}
 \ + N \ 
 \begin{tikzpicture}[baseline=(current  bounding  box.center),thick,scale=.7]
  \draw  (0.8,0) -- ++(1,0) node (dot1) {};
  \draw (1.1,0.5) -> (1.8,0);
  \filldraw (dot1) circle [radius=.1];
  \draw (0.8,-1.) to  (2.4,-1.0) node (dot2) {};
   \draw (dot2) circle [radius=.1];
 \end{tikzpicture}
\end{align}
where $\Theta^{(k)}$ is a vertex, $F_0$ is a sphere, and $F_1$ is a torus, while legs stand for derivatives.

\item Compute the ${\epsilon}^*$-derivatives of the first three terms of the partition function~\eqref{eq:oscilZNorder2}
\begin{align}
 \mathcal{Z}(\gamma^N) \sim \mathcal{Z}_0(\gamma^N) + N^{-1} \mathcal{Z}_1(\gamma^N) + N^{-2} \mathcal{Z}_2(\gamma^N) + O(N^{-3}) \ .
\end{align}
Redo these computations graphically.

\item Prove graphically that $\mathcal{Z}(\gamma^N)$ does not depend on $\epsilon^*$.

\end{enumerate}

\end{exo}

\chapter{Orthogonal polynomials and integrable systems}\label{chap:ortho}

In this chapter we introduce the orthogonal polynomials method for matrix integrals.
This method leads to exact results, in contrast to the large $N$ limits of \autoref{chap:SaddlePoint} and the asymptotic expansions of \autoref{chap:LoopEq}.
This method also provides the link between matrix integrals and integrable systems.

\section{Determinantal formulas}

We consider a normal matrix integral \eqref{eq:znormal}, written as an integral over eigenvalues,
\begin{align}
 \mathcal{Z}_N = \frac{1}{N!} \int_{\gamma^N} d\lambda_0\cdots d\lambda_{N-1}\ \Delta(\lambda_0, \cdots, \lambda_{N-1})^2 \prod_{i=0}^{N-1} e^{-V(\lambda_i)}\ ,
 \label{eq:normaleig}
\end{align}
where $\Delta(\lambda_0, \cdots \lambda_{N-1})$ is the Vandermonde determinant~\eqref{eq:vandermonde}. 
The integration path $\gamma$ can be arbitrary.
%Our analysis can be generalized to the cases $\beta = 1, 4$ of the Orthogonal and Symplectic ensembles.
For later convenience we have introduced the overall normalization factor $\frac{1}{N!}$, and labelled the eigenvalues from $0$ to $N-1$. 

\subsection{Matrix integrals}

Our first aim is to separate the variables in the matrix integral $\mathcal{Z}_N$. 
To do this, we rewrite the Vandermonde determinant as a linear combination of monomials,
\begin{align}
 \Delta(\lambda_0, \cdots ,\lambda_{N-1}) = \underset{i,j=0,\cdots N-1}{\det}\lambda_i^j = \sum_{\sigma\in \mathfrak S_N} (-1)^\sigma \prod_{i=0}^{N-1} \lambda_i^{\sigma(i)}\ ,
 \label{eq:Vandermonde_det}
\end{align}
where $\mathfrak S_N$ is the symmetric group of permutations of $N$ elements $\{0,\dots,N-1\}$, and $(-1)^\sigma$ is the signature of $\sigma\in
\mathfrak S_N$.
This leads to 
\begin{align}
 \mathcal{Z}_N = \frac{1}{N!} \sum_{\sigma, \sigma'\in \mathfrak S_N} (-1)^{\sigma\sigma'} \prod_{i=0}^{N-1} \int_\gamma d\lambda_i \ \lambda_i^{\sigma(i)} \lambda_i^{\sigma'(i)} e^{-V(\lambda_i)}\ ,
\end{align}
where we have indeed separated the variables $\lambda_i$.
We notice that the resulting one-dimensional integrals are examples of the scalar product for the measure $e^{-V(\lambda)}d\lambda$, 
\begin{align} 
 \left< f\middle| g \right> = \int_\gamma d\lambda\ e^{-V(\lambda)}f(\lambda)g(\lambda)\ ,
\end{align}
where in our case the functions $f$ and $g$ are monomials of the type
\begin{align}
 m_i(\lambda) = \lambda^i\ .
\end{align}
We thus compute
\begin{align}
 \mathcal{Z}_N &= \frac{1}{N!} \sum_{\sigma, \sigma'\in \mathfrak S_N} (-1)^{\sigma\sigma'} \prod_{i=0}^{N-1} \left< m_{\sigma(i)}\middle|m_{\sigma'(i)}\right> = \sum_{\sigma \in \mathfrak S_N} (-1)^\sigma \prod_{i=0}^{N-1} \left<m_{\sigma(i)}\middle| m_i\right> \ ,
\end{align}
and we end up with a \textbf{determinantal formula} \index{determinantal formulas!for matrix integrals} for our matrix integral,
\begin{align}\label{eq:ZNdetmimj}
\mathcal{Z}_N &=  \underset{i,j=0,\cdots N-1}{\det} \left< m_i\middle| m_j\right>\ .
\end{align}
(See \autoref{exo:HOMFLY_Jones_unknot} for an application.)
This is a \textbf{Hankel determinant}\index{Hankel determinant}, i.e. a determinant of a matrix whose entries depend only on $i+j$
\begin{equation}
\mathcal{Z}_N = \det_{i,j=0,\cdots N-1} M_{i+j}
\qquad , \quad M_k = \int_\gamma d\lambda\ \lambda^k e^{-V(\lambda)}\ .
\label{eq:Hankel_rep}
\end{equation}
By permuting the columns $j\to N-1-j$, this can be rewritten in terms of a \textbf{Toeplitz determinant}\index{Toeplitz determinant} whose entries are constant on diagonals, i.e. depend only on $i-j$:
\begin{equation}
\mathcal{Z}_N = (-1)^{\frac{N(N-1)}{2}} \det_{i,j=0,\cdots N-1} M_{N-1+i-j}\,\, .
\end{equation}
(See \autoref{exo:volun} for an application.)
By linearly combining lines and columns of the matrix $\left< m_i\middle| m_j\right>$, the monomials $m_i$ can be replaced with arbitrary monic polynomials $p_i$ of the same degrees. 
Introducing two families of such polynomials $p_i(\lambda)= \lambda^i + \cdots $ and $\tilde{p}_i(\lambda)= \lambda^i + \cdots $, and their scalar products 
\begin{align}
 \boxed{ H_{i,j} = \left< \tilde{p}_i\middle| p_j\right> } \ ,
\label{eq:hij}
\end{align}
we obtain
\begin{align}
  \boxed{ \mathcal{Z}_N = \underset{i,j=0,\cdots N-1}{\det} H_{i,j} } \ .
\label{eq:zdeth}
\end{align}

\subsection{Joint eigenvalue distributions}

Let us define the \textbf{joint eigenvalue distributions}\index{joint eigenvalue distribution} of $k$ eigenvalues by integrating out the $N-k$ other eigenvalues,
\begin{align}
 \boxed{ R_k(\lambda_0, \cdots \lambda_{k-1}) = \frac{1}{\mathcal{Z}_N} \frac{1}{N!}\int_{\gamma^{N-k}} d\lambda_k \cdots d\lambda_{N-1}\ \Delta(\lambda_0, \cdots \lambda_{N-1})^2 \prod_{i=0}^{N-1} e^{-V(\lambda_i)} } \ .
\end{align}
In particular, 
\begin{itemize}
 \item $R_0 = 1$,
 \item $R_1(\lambda) =  \rho(\lambda)$ is the density of eigenvalues,
 \item $R_2$ is related to the connected two-point function $\rho_2$ by
 \begin{align}
  \rho_2(\lambda_1,\lambda_2) = N(N-1)R_2(\lambda_1,\lambda_2) +N\delta(\lambda_1-\lambda_2)R_1(\lambda_1) -N^2 R_1(\lambda_1)R_1(\lambda_2)\ ,
 \end{align}
 \item $R_N$ is (up to a normalization) the integrand of our matrix integral $\mathcal{Z}_N$:
 \begin{align}
 R_N(\lambda_0,\ldots,\lambda_{N-1}) = \frac{1}{\mathcal{Z}_N} \frac{1}{N!} \Delta(\lambda_0,\ldots,\lambda_{N-1})^2 \prod_{i=0}^{N-1} e^{-V(\lambda_i)} . 
 \label{eq:R_N}
 \end{align}
\end{itemize}
We first remark that $R_N$ can be expressed in terms of three determinant factors: one from the determinantal formula \eqref{eq:zdeth} for $\mathcal{Z}_N$, and two from the Vandermonde determinants,
\begin{align}
 R_N(\lambda_0,\cdots \lambda_{N-1}) = \frac{1}{N!}\det H^{-1} \det_{i,j=0,\cdots N-1} p_j(\lambda_i) \det_{i',j'=0,\cdots N-1} \tilde{p}_{j'}(\lambda_{i'}) \prod_{i=0}^{N-1} e^{-V(\lambda_i)}\ .
\end{align}
The product of these three determinants can be rewritten as the determinant of a product matrix, and we find 
\begin{align}
 R_N(\lambda_0,\cdots \lambda_{N-1}) = \frac{1}{N!}\det_{i,j=0,\cdots N-1} K(\lambda_i,\lambda_j)\ , 
 \label{eq:R_N_det}
\end{align}
where we introduce 
\begin{align}
 \boxed{ K(\lambda, \tilde{\lambda}) = 
 \sum_{k=0}^{N-1} \sum_{k'=0}^{N-1} p_k(\lambda) \left(H^{-1}\right)_{k,k'} 
 \tilde{p}_{k'}(\tilde{\lambda}) e^{-\frac12 V(\lambda)} e^{-\frac12 V(\tilde{\lambda})} } \ .
\label{eq:kll}
\end{align}
The kernel $K$ enjoys the following properties:
\begin{itemize}
 \item $K$ is independent of the choice of the families of polynomials $p_k$ and $\tilde{p}_k$. 
  Choosing another family of polynomials $p_k$ indeed amounts to multiplying the vector $p=(p_0, \cdots p_{N-1})$ and the matrix $H$ on the right with the same matrix, so that $\sum_{k=0}^{N-1} p_k \left(H^{-1}\right)_{k,k'}$ is unchanged. 
  \item The trace-class condition,
\begin{align}
\int_\gamma d\lambda\ K(\lambda,\lambda) 
  = \sum_{k=0}^{N-1} \sum_{k'=0}^{N-1}\left(H^{-1}\right)_{k,k'} H_{k',k} = N \, .
\end{align}
  \item $K$ is a \textbf{self-reproducing kernel}\index{kernel!self-reproducing---}, which obeys
\begin{align}
 K\star K = K\ ,
\end{align}
where we define the star product of two functions $f(\lambda^1, \cdots \lambda^p)$ and $g(\mu^1,\cdots \mu^q)$ of one or more variables as 
\begin{align}
 (f \star g)(\lambda^1, \cdots \lambda^{p-1},\mu^2, \cdots \mu^q) = \int_\gamma d\lambda\ f(\lambda^1, \cdots \lambda^{p-1},\lambda) g(\lambda,\mu^2, \cdots \mu^q)\ .
\end{align}
\end{itemize}
Let us now compute the next eigenvalue distribution
\begin{align}
 R_{N-1}
 = \int_\gamma d\lambda_{N-1}\ R_N(\lambda_0,\cdots \lambda_{N-1})
 = \frac{1}{N!} \int_\gamma d\lambda_{N-1}\ \sum_{\sigma\in \mathfrak S_N} (-1)^\sigma K(\lambda_i, \lambda_{\sigma(i)}) \ .
\end{align}
Let us rewrite the permutations $\sigma\in \mathfrak S_N$ in terms of  permutations $\sigma'\in \mathfrak{S}_{N-1}$ of the first $N-1$ elements.
If $\sigma(N-1) = N-1$, then we identify $\sigma$ with an element $\sigma'\in \mathfrak{S}_{N-1}$. 
And if $\sigma(N-1) = i\neq N-1$, then we write $\sigma = \sigma'\cdot (N-1,i)$ as the composition of $\sigma'\in \mathfrak{S}_{N-1}$ with a transposition. 
We thus find 
\begin{subequations}
\begin{align}
 N!R_{N-1}
 &= \sum_{\sigma'\in \mathfrak{S}_{N-1}} (-1)^{\sigma'} 
 \prod_{i=0}^{N-2}K(\lambda_i,\lambda_{\sigma(i)}) 
 \int_\gamma d\lambda_{N-1}\, K(\lambda_{N-1},\lambda_{N-1}) 
 \nonumber
 \\
 & \hspace{-8mm} - \sum_{i=0}^{N-2}\sum_{\sigma'\in \mathfrak{S}_{N-1}} (-1)^{\sigma'} \prod_{\substack{j=0 \\ j\neq i}}^{N-2} K(\lambda_j,\lambda_{\sigma'(j)})
 \int_\gamma d\lambda_{N-1}\, K(\lambda_i,\lambda_{N-1}) K(\lambda_{N-1},\lambda_{\sigma'(i)}) \ ,
 \\
 &= N \det_{i,j=0,\cdots N-2} K(\lambda_i,\lambda_j) - (N-1)\det_{i,j=0,\cdots N-2} K(\lambda_i,\lambda_j)\ ,
 \\
 &= \det_{i,j=0,\cdots N-2} K(\lambda_i,\lambda_j)\, .
\end{align}
\end{subequations}
So $R_{N-1}$ is also given by a determinantal formula.
Similarly, we can find $R_{N-2}$ from $R_{N-1}$ by integrating over $\lambda_{N-2}$. 
Iterating, we obtain \textbf{Dyson's theorem}\index{Dyson's theorem} \cite{Mehta:2004RMT} --
determinantal formulas for all the joint eigenvalue distributions,  
\begin{align}
 \boxed{ R_k(\lambda_{0},\cdots \lambda_{k-1}) = \frac{(N-k)!}{N!} \det_{i,j=0,\cdots k-1} K(\lambda_i,\lambda_j) } \ .
\end{align}
In particular, the eigenvalue density and the connected two-point function are
\begin{subequations}
\begin{align}
 \rho(\lambda) &= \frac{1}{N} K(\lambda,\lambda)\ ,
 \\
 \rho_2(\lambda_1,\lambda_2) &= -K(\lambda_2,\lambda_1)K(\lambda_1,\lambda_2) + \delta(\lambda_1-\lambda_2)K(\lambda_1,\lambda_1)\ .
\end{align}
\end{subequations}
More generally, any disconnected $k$-point correlation function is given by a determinant $\det_{i,j=0,\cdots k-1} K(\lambda_i,\lambda_j)$, plus delta-function terms.

\subsection{Spacing distributions} \label{sec:spacing}

Given a subset $I\subset \gamma$, let $\mathcal{E}(I)$ be the probability that there is no eigenvalue in $I$, which is called the \textbf{gap pobability}.
From this probability we can in particular deduce the \textbf{spacing distribution}, i.e. the probability density that $[a, b]$ is an interval between two consecutive eigenvalues $a$ and $b$,
\begin{equation}
\boxed{ p_{\text{spacing}}(a,b) = -\frac{1}{N(N-1)}\frac{\partial^2}{\partial a\partial b} \mathcal{E}([a,b]) } \ .
\end{equation}
In the case $I=[a,\infty)$, we find the probability density that $a$ is the largest eigenvalue:
\begin{equation}
\boxed{ p_{\text{largest}}(a) = -\frac{1}{N}\frac{\partial}{\partial a} \mathcal{E}([a,\infty)) } \ .
\end{equation}
The probability $\mathcal{E}(I)$ is given by the matrix integral 
\begin{align}
 \mathcal{E}(I) = \frac{1}{\mathcal{Z}_N} \frac{1}{N!} \int_{(\gamma-I)^N} d\lambda_0\cdots d\lambda_{N-1}\ \Delta(\lambda_0, \cdots, \lambda_{N-1})^2 \prod_{i=0}^{N-1} e^{-V(\lambda_i)}\ .
\end{align}
This is very similar to the matrix integral $\mathcal{Z}_N$ itself, with the integration contour $\gamma$ replaced with $\gamma-I$. 
Explicitly writing the $\gamma$-dependences of the scalar product and of the $N\times N$ matrix $(H_\gamma)_{i,j} = \left<\tilde{p}_i|p_j\right>_\gamma$ \eqref{eq:hij}, we find 
\begin{align}
 \mathcal{E}(I) 
 = \frac{\det H_{\gamma - I}}{\det H_\gamma} 
 = 
 \det (\operatorname{Id} - A) \qquad \text{with} \qquad 
 A = H^{-1}_\gamma H_{ I}\ .
\end{align}
Now let us show that the matrix $A$ can be interpreted in terms of an operator $K_I$ that acts on functions on $I$,
\begin{align}
 \boxed{ K_I(f)(\lambda) = (K \star_I f)(\lambda) = \int_I d\tilde{\lambda}\ K(\lambda,\tilde{\lambda}) f(\tilde{\lambda}) } \ .
\end{align}
The action of $K_I$ on functions of the type $p_j e^{-\frac12 V}$ is
\begin{multline}
 K_I(p_j e^{-\frac12 V})(\lambda) 
= \int_I d\tilde{\lambda} \sum_{k=0}^{N-1} \sum_{k'=0}^{N-1} p_k(\lambda) \left(H^{-1}_\gamma\right)_{k,k'}
 \tilde{p}_{k'}(\tilde{\lambda}) p_j(\tilde{\lambda}) e^{-\frac12 V(\lambda)} e^{-V(\tilde{\lambda})}\ ,
\\
= \sum_{k=0}^{N-1} \sum_{k'=0}^{N-1} p_k(\lambda)e^{-\frac12 V(\lambda)}\left(H^{-1}_\gamma\right)_{k,k'} \left(H_I\right)_{k',j} = \left( \sum_{k=0}^{N-1} p_k e^{-\frac12 V} A_{k,j}\right)(\lambda)\ .
\end{multline}
This shows that the matrix of $K_I$ in the basis $(p_j e^{-\frac12 V})_j$ of functions on $I$ is $A^\text{T}$.
Therefore, $\mathcal{E}(I)$ can be written as the $N\times N$ determinant of the operator $K_I$ in the vector space spanned by $p_0, \dots p_{N-1}$
\begin{equation}
\mathcal{E}(I) = \det_{N\times N}\left(\operatorname{Id}-K_I\right)  \ .
\end{equation}
In the large $N$ limit, this becomes a Fredholm determinant:
\begin{equation}
\boxed{ \lim_{N\to\infty} \mathcal{E}(I) = \det_{\text{Fredholm}}\left(\operatorname{Id}-K_I\right) } \ .
\end{equation}
This shows that in the large $N$ limit, the spacing disctribution can be expressed in terms of Fredholm determinants, as we announced in \autoref{sec:micro} in the context of microscopic limits.

\section{Orthogonal polynomials}\label{sec:ortho}

Our formulas for matrix integrals and eigenvalue distributions, which involve scalar products of polynomials, would simplify if the polynomials were orthogonal. 
Let us assume that $p_k = \tilde{p}_k$ is a family of \textbf{orthogonal polynomials}\index{orthogonal polynomials}, satisfying
\begin{align}
 \boxed{ H_{k, k'} = \left<p_k\middle|p_{k'}\right> = h_k \delta_{k,k'} } \ ,
\end{align}
with normalizations $\langle p_k | p_k \rangle=h_k\in\mathbb C$.
(We cannot set $h_k=1$ because our polynomials are assumed to be monic, and cannot be arbitrarily normalized.)
The partition function is then 
\begin{align}
 \boxed{ \mathcal{Z}_N = \prod_{k=0}^{N-1} h_k } \ .
 \label{eq:zph}
\end{align}
Moreover, the self-reproducing kernel can then be written as 
\begin{align}
 \boxed{ K_N(\lambda,\tilde{\lambda}) = \sum_{k=0}^{N-1} \psi_k(\lambda)\psi_k(\tilde{\lambda}) } \ ,
\label{eq:kpp}
 \end{align}
where we  define 
\begin{align}
 \boxed{ \psi_k(\lambda) = \frac{p_k(\lambda)}{\sqrt{h_k}} e^{-\frac12 V(\lambda)} \qquad \text{so that} \qquad \int_\gamma d\lambda\ \psi_k(\lambda)\psi_{k'}(\lambda) = \delta_{k,k'} } \ .
 \label{eq:psik}
\end{align}

The existence and uniqueness of an orthogonal family of monic polynomials is equivalent to the non-degeneracy of the scalar product, as follows from the Gram--Schmidt construction. 
Due to the determinantal formula, this is also equivalent to $\mathcal{Z}_N \neq 0$ for all $N$.
This condition is fulfilled if 
$\gamma = \mathbb{R}$ and the potential is real, in which case $\mathcal{Z}_N > 0$ for all $N$, and the existence of the orthogonal family  implies that the density of eigenvalues is strictly positive,
\begin{align}
 N\rho(\lambda) = K_N(\lambda,\lambda) = \sum_{k=0}^{N-1} \psi_k(\lambda)^2 \geq \psi_0(\lambda)^2 = \frac{e^{-V(\lambda)}}{h_0} > 0\ .
\end{align}
Given a potential,
the scalar product depends polynomially on the coefficients $c_i$ of the contour $\gamma=\sum_{i=1}^d c_i \gamma_i$, and is therefore degenerate on an enumerable union of algebraic submanifolds. So orthogonal polynomials exist on a dense open subset of the space $\mathbb{C}^d$ of contours.

\subsection{How to compute orthogonal polynomials}

In the particularly simple case of a Gaussian potential $V(\lambda) = \frac12\lambda^2$, an orthogonal family is provided by the \textbf{Hermite polynomials}\index{Hermite polynomial},
\begin{align}
 p_k(\lambda) = (-1)^k e^{\frac{\lambda^2}{2}} \left(\frac{\partial}{\partial\lambda}\right)^k e^{-\frac{\lambda^2}{2}}\ .
\end{align}
Hermite polynomials satisfy the recursion relations
\begin{equation}
p_{k+1}(\lambda) = \lambda p_k(\lambda) - k p_{k-1}(\lambda) 
\quad , \quad 
p'_{k}(\lambda) = k p_{k-1}(\lambda) \ .
\end{equation}
Combining these two relations leads to the second-order differential equation
\begin{equation}
p''_{k}(\lambda) = \lambda p'_k(\lambda) - k p_k(\lambda)\ .
\end{equation}
We will study orthogonal polynomial in more general cases, and show that they satisfy recursion relations and differential equations.

\subsubsection{Heine's formula}

Expressions for orthogonal polynomials in terms of matrix integrals are given by \textbf{Heine's formula}\index{Heine's formula}, 
\begin{align}
 \boxed{ p_k(\lambda) = \Big<\det(\lambda- M)\Big>_{k\times k \text{-matrix}} } \ ,
 \label{Heine_formula}
 \end{align}
 or more explicitly 
 \begin{align}
 p_k(\lambda) &= \frac{\int_{\gamma^k} d\lambda_0 \cdots d\lambda_{k-1}\ \Delta(\lambda_0, \cdots \lambda_{k-1})^2 \prod_{i=0}^{k-1} (\lambda -\lambda_i)e^{-V(\lambda_i)}}
 {\int_{\gamma^k} d\lambda_0 \cdots d\lambda_{k-1}\ \Delta(\lambda_0, \cdots \lambda_{k-1})^2 \prod_{i=0}^{k-1} e^{-V(\lambda_i)}} \ .
\end{align}
Indeed this $p_k$ is manifestly a monic polynomial of degree $k$, let us show that $p_k$ is orthogonal to $p_j$ for $j<k$. We compute
\begin{equation}
\mathcal Z_k \left<p_k\middle|p_j\right> 
= \int_{\gamma^{k+1}} d\lambda d\lambda_0 \cdots d\lambda_{k-1}\ \Delta(\lambda_0, \cdots \lambda_{k-1})^2 p_j(\lambda) \prod_{i=0}^{k-1} (\lambda -\lambda_i)e^{-V(\lambda_i)}\ .
\end{equation}
The presence of the factor $\Delta(\lambda_0, \cdots \lambda_{k-1}) \prod_{i=0}^{k-1} (\lambda-\lambda_i) = \Delta(\lambda_0, \cdots \lambda_{k-1}, \lambda)$ allows us to antisymmetrize the rest of the integrand.
Starting with 
\begin{align}
 \Delta(\lambda_0, \cdots \lambda_{k-1}) p_j(\lambda)  =
     \det\begin{pmatrix}
      1 & 1 & \cdots & 1 & 1
      \\ 
      \lambda_0 & \lambda_1 & \cdots & \lambda_{k-1} & \lambda
      \\
       \vdots & \vdots & \ddots & \vdots & \vdots 
       \\
       \lambda_0^{k-1} & \lambda_1^{k-1} & \cdots & \lambda_{k-1}^{k-1} & \lambda^{k-1}
       \\
       0 & 0 & \cdots & 0 & p_j(\lambda)
     \end{pmatrix} \ ,
\end{align}
and antisymmetrizing with respect to permutations of $\{\lambda_0,\cdots,\lambda_{k-1},\lambda\}$, we obtain 
\begin{align}
\frac{1}{k+1}
\det\begin{pmatrix}
      1 & 1 & \cdots & 1 & 1
      \\ 
      \lambda_0 & \lambda_1 & \cdots & \lambda_{k-1} & \lambda
      \\
       \vdots & \vdots & \ddots & \vdots & \vdots 
       \\
       \lambda_0^{k-1} & \lambda_1^{k-1} & \cdots & \lambda_{k-1}^{k-1} & \lambda^{k-1}
       \\
       p_j(\lambda_0) & p_j(\lambda_1) & \cdots & p_j(\lambda_{k-1}) & p_j(\lambda)
     \end{pmatrix}\ ,
\end{align}
which vanishes if $j<k$.
This implies $\left<p_k\middle|p_j\right>=0$ for $j<k$, which proves Heine's formula.

\subsubsection{Orthogonal polynomials as Hankel determinants}

Another formula for orthogonal polynomials is
\begin{equation}
p_k(\lambda) = \frac{\det\begin{pmatrix}
 M_{0} & M_1 & M_2 & \dots & M_{k-1} & 1 \cr
 M_1 & M_2 & & & M_{k} & \lambda \cr
 M_2 &  & & & M_{k+1} & \lambda^2 \cr
 \vdots & & & & \vdots & \vdots \cr
 M_{k} & & & & M_{2k-1} & \lambda^k
\end{pmatrix}}{\det\begin{pmatrix}
 M_{0} & M_1 & M_2 & \dots & M_{k-1}  \cr
 M_1 & M_2 & & & M_{k}  \cr
 M_2 &  & & & M_{k+1}  \cr
 \vdots & & & & \vdots  \cr
 M_{k-1} & & & & M_{2k-2} 
\end{pmatrix}}\ ,
\label{eq:pkhd}
\end{equation}
where $M_j=\int_\gamma \lambda^j e^{-V(\lambda)}d\lambda $ are the moments.
Indeed this is manifestly a monic polynomial of degree $k$. Then $\left<p_k\middle|p_j\right> = 0$ if $j<k$ follows from
\begin{equation}
\left<p_k\middle|m_j\right>
\propto \det\begin{pmatrix}
 M_{0} & M_1 & M_2 & \dots & M_{k-1} & M_j \cr
 M_1 & M_2 & & & M_{k} & M_{j+1} \cr
 M_2 &  & & & M_{k+1} & M_{j+2} \cr
 \vdots & & & & \vdots & \vdots \cr
 M_{k} & & & & M_{2k-1} & M_{j+k}
\end{pmatrix}\ .
\end{equation}

\subsubsection{Three-term recursion relation}

With $m_1(x)=x$ the monomial of degree one, notice that $m_1p_k$ is a monic polynomial of degree $k+1$, and therefore can be decomposed on the basis of orthogonal polynomials as
\begin{align}
 m_1 p_k = \sum_{j=0}^{k+1} \hat{Q}_{k,j}p_j\quad \text{with} \quad \hat{Q}_{k, k+1}=1\ ,
\end{align}
which can be seen as a recursion relation.
Inserting this recursion relation in the symmetric scalar product $\left< m_1p_k\middle|p_j\right> = \left< p_k\middle|m_1p_j\right>$, we obtain the following identity for the unknown coefficients $\hat{Q}_{k,j}$,
\begin{align}
 h_j\hat{Q}_{k,j} = h_k \hat{Q}_{j,k}\ .
\end{align}
In particular, if $j<k-1$ then $\hat{Q}_{k,j}=0$, so the right-hand side of the recursion relation has only three non-vanishing terms. 
Using the functions $\psi_k$ \eqref{eq:psik} instead of the polynomials $p_k$, the three-term recursion relation becomes
\begin{align}
 \boxed{ x \psi_k(x) = Q_{k, k+1}\psi_{k+1}(x) + Q_{k,k} \psi_k(x) + Q_{k, k-1}\psi_{k-1}(x) } \ ,
 \label{eq:mpsi}
\end{align}
where we defined
\begin{align}
 Q_{j,k} = Q_{k,j} = \sqrt{\frac{h_j}{h_k}}\hat{Q}_{k, j}\ .
\end{align}
The infinite-dimensional, symmetric, \textbf{tridiagonal} matrix $Q$  is called a \textbf{Jacobi matrix}\index{Jacobi matrix}, and can be written as  
\begin{align}
Q = 
 \begin{pmatrix}
  S_0 & \gamma_1 & 0 & 0 & \cdots 
  \\
  \gamma_1 & S_1 & \gamma_2 & 0 & \ddots
  \\
  0 & \gamma_2 & S_2 & \gamma_3 & \ddots 
  \\
  0 & 0 & \gamma_3 & S_3 & \ddots 
  \\
  \vdots & \ddots & \ddots & \ddots & \ddots  
 \end{pmatrix}
 \ \  \text{with} \ \  \boxed{ \gamma_k = Q_{k,k-1}= \sqrt{\frac{h_k}{h_{k-1}}} } \ \  \text{and} \ \  \boxed{ S_k = Q_{k,k} } \ .
 \label{eq:qel}
\end{align}
The Jacobi matrix $Q$ is useful for computing expectation values of single-trace operators $\operatorname{Tr}f(M)$.
For any polynomial function $f$, we have 
\begin{align}
 \boxed{ \Big<\operatorname{Tr}f(M)\Big>_{N\text{-matrix}} = \operatorname{Tr}\Pi_N f(Q) } \ ,
\label{eq:trf}
 \end{align}
 where $\Pi_N= \begin{pmatrix} 
           \operatorname{Id}_{N\times N} & 0_{\infty \times N} 
           \\
           0_{N\times \infty} & 0_{\infty \times\infty}
        \end{pmatrix}
$ is the projector on the first $N$ components. 
To prove this, we will directly compute the relevant integral over eigenvalues.
Writing the Vandermonde determinant as 
$
 \Delta(\lambda_0,\dots,\lambda_{N-1}) = \sum_{\sigma\in\mathfrak{S}_N} (-1)^\sigma \prod_{i=0}^{N-1}p_{\sigma(i)}(\lambda_i)
$, we find 
\begin{subequations}
\begin{align}
 \Big<\operatorname{Tr}f(M)\Big>_{N\text{-matrix}}
 &= \frac{1}{N!\prod_{k=0}^{N-1}h_k} \sum_{i=0}^{N-1} \sum_{\sigma,\sigma'\in \mathfrak S_N} 
 \left< fp_{\sigma(i)}\middle| p_{\sigma'(i)}\right>
 \prod_{j\neq i} \left< p_{\sigma(j)}\middle| p_{\sigma'(j)} \right>\ ,
 \\
 &= \frac{1}{N!} \sum_{\sigma\in \mathfrak{S}_N} \sum_{i=0}^{N-1}\frac{1}{h_{\sigma(i)}} \left< fp_{\sigma(i)}\middle| p_{\sigma(i)}\right> = \sum_{k=0}^{N-1} \frac{1}{h_k}\left< fp_k\middle|p_k\right>\ .
\end{align}
\end{subequations}
Replacing the polynomials $p_k$ with the functions $\psi_k$, and using the recursion relation $f\psi_k = \sum_j f(Q)_{k, j}\psi_j$, we obtain Eq.~\eqref{eq:trf}. 

\subsubsection{Determinantal formulas}

The orthogonal polynomials $p_k$ can  be expressed in terms of the upper left submatrix of size $k$ of $Q$ by the \textbf{determinantal formula} \index{determinantal formulas!for orthogonal polynomials}
\begin{align}
 \boxed{ p_k(\lambda) = \underset{k\times k \text{ submatrix}}{\det} (\lambda  - Q) } \ .
\label{pkdet}
\end{align}
To prove this relation between two monic polynomials of degree $k$, it suffices to show that they have the same zeros.
To do this, notice that 
if $p_k(\lambda)= 0$, then the vector $(p_0(\lambda), \cdots p_{k-1}(\lambda))$ is annihilated by the matrix $(\lambda - Q)_{k\times k \text{ submatrix}}$, due to the recursion relation. This implies $\underset{k\times k \text{ submatrix}}{\det} (\lambda - Q) = 0$.

Combining the determinantal formula with Heine's formula yields
\begin{align}
 \Big<\det(\lambda-M)\Big>_{k\text{-matrix}} = \underset{k\times k \text{ submatrix}}{\det} (\lambda  - Q)\ .
\end{align}
The matrix $Q$ also provides  a determinantal formula for the self-reproducing kernel,
\begin{equation}
\boxed{ K_{k+1}(\lambda,\mu) = \frac{e^{-\frac12 V(\lambda)} e^{-\frac12 V(\mu)}}{h_k}\underset{k\times k \text{ submatrix}}{\det} (\lambda  - Q)(\mu-Q) } \ .
\label{eq:kds}
\end{equation}
Let us prove this formula by induction on $k$.
The case $k=0$ is obvious.
Denoting $M_k$ the upper left $k\times k$ block of a matrix $M$, we have 
\begin{equation}
\Big((\lambda-Q)(\mu-Q)\Big)_k = (\lambda-Q)_k (\mu-Q)_k + \gamma_k^2 \operatorname{diag}_{k\times k}(0,0,\dots,0,1)\ .
\end{equation}
So, assuming the formula holds for $K_k$, we find
\begin{subequations}
\begin{align}
\underset{k\times k}{\det} (\lambda  - Q)(\mu-Q)
& = \det(\lambda-Q)_k(\mu-Q)_k + \gamma_k^2\underset{k-1\times k-1}{\det} (\lambda  - Q)_k(\mu-Q)_k \\
& = p_k(\lambda) p_k(\mu)  + \gamma_k^2 h_{k-1}e^{\frac12 V(\lambda)} e^{\frac12 V(\mu)}  K_{k}(\lambda,\mu) \\
& = h_{k}e^{\frac12 V(\lambda)} e^{\frac12 V(\mu)} K_{k+1}(\lambda,\mu) \ , 
\end{align}
\end{subequations}
where we used the expression \eqref{eq:kpp} of the kernel in terms of the orthogonal polynomials. This shows that the formula holds for $K_{k+1}$.

\subsubsection{Derivatives of orthogonal polynomials}

In the same way as we wrote a recursion relation for the multiplication of our polynomials with the monomial $m_1$, we can decompose the derivatives of our polynomials $p_k$ on the basis of orthogonal polynomials.
Let us do this for the orthonormal functions $\psi_k$. Derivating the exponential factor of $\psi_k$ produces a polynomial $V'$ of degree $d$, so that
\begin{align}
 \boxed{ \psi_k ' = \sum_{j=0}^{k+d} P_{k,j}\psi_j } \ ,
 \label{eq:psiprime}
\end{align}
with some coefficients $P_{k,j}$ to be determined below. 
Integration by parts $\int_\gamma \psi_k' \psi_j = -\int_\gamma \psi_k \psi_j'$, implies that the matrix $P$ is antisymmetric, and its nonvanishing coefficients therefore reside on $2d$ bands:
\begin{align}
 P = \begin{pmatrix}
       0 & P_{0,1} & \cdots & P_{0,d} & 0 & 0 & \cdots
       \\
       -P_{0,1} & 0 & P_{1,2} & \cdots & P_{1, d+1} & 0 & \cdots 
       \\
       \vdots & -P_{1,2} & 0 & P_{2,3} & \cdots & P_{2,d+2} & \ddots 
       \\
       - P_{0,d} & \vdots & -P_{2,3} & 0 & P_{3,4} & \ddots & \ddots 
       \\
        0 & -P_{1,d+1} & \vdots & -P_{3,4} & 0 & P_{4,5} & \ddots 
        \\
        0 & 0 & -P_{2,d+2} & \ddots & -P_{4,5} & 0 & \ddots 
        \\
        \vdots & \ddots & \ddots & \ddots & \ddots & \ddots & \ddots 
     \end{pmatrix} \ .
\end{align}
Now, using the definition of $\psi_k$, we have 
\begin{align}
 \frac{1}{\sqrt{h_k}} p'_k e^{-\frac12 V} 
 = \psi'_k + \frac12 V' \psi_k 
 = \sum_{j=0}^{k+d} \left(P + \frac12 V'(Q)\right)_{k,j} \psi_j \ .
\end{align}
However, since $\deg p'_k = k-1$, the left-hand side of this equation must be a linear combination of only $\psi_0, \cdots \psi_{k-1}$, showing that $P+\frac12 V'(Q)$ must be strictly lower triangular matrix.
Denoting $M_+, M_0$ and $M_-$ the upper triangular, diagonal and lower triangular parts of a matrix $M$, 
this implies
\begin{align}
 P_+ = -\frac12 V'(Q)_+\qquad , \qquad
 P_0 = 0 = -\frac12 V'(Q)_0\ .
\end{align}
By antisymmetry of $P$ and symmetry of $Q$, this is also implies 
\begin{align}
 P_- = \frac12 V'(Q)_-\ ,
\end{align}
thus
\begin{align}\label{eq:PVQpm}
 \boxed{ P = -\frac12 \Big( V'(Q)_+ - V'(Q)_- \Big) } \ .
\end{align}
Moreover, from $p'_k = k p_{k-1} + \dots$ we can determine the diagonal and subdiagonal coefficients of $P+\frac12 V'(Q)$, therefore
\begin{align}
 \forall\,k\, , \qquad
 \begin{cases}
  \displaystyle
  \left(P + \tfrac12 V'(Q)\right)_{k,k-1} = \frac{k}{\gamma_k}\ , \\[.5em]
  \left(P + \tfrac12 V'(Q)\right)_{k,k} = 0\ . 
 \end{cases}
% \left\{\begin{array}{l}
%\left(P + \tfrac12 V'(Q)\right)_{k,k-1} = \frac{k}{\gamma_k}\ , \\
%\left(P + \tfrac12 V'(Q)\right)_{k,k} = 0\ . \\
%	\end{array}\right.
\end{align}
Using Eq.~\eqref{eq:PVQpm}, this can be rewritten in terms of $Q$ alone as
\begin{equation}\label{eq:recurPQ}
 \boxed{ \forall\,k\, , \qquad
  \begin{cases}
   \displaystyle
   \Big( V'(Q)\Big)_{k,k-1} = \frac{k}{\gamma_k}\ , \\[.5em]
   \Big( V'(Q)\Big)_{k,k} = 0\ . \\
  \end{cases}
%\left\{\begin{array}{l}
%\Big( V'(Q)\Big)_{k,k-1} = \frac{k}{\gamma_k}\ , \\
%\Big( V'(Q)\Big)_{k,k} = 0\ . \\
  %\end{array}\right.
  }
\end{equation}
These two equations lead to recursion relations for $S_k=Q_{k,k}$ and $\gamma_k = Q_{k,k-1}$, which determine them.
Let us give two examples:
\begin{itemize}
 \item Gaussian potential $V'(Q) = Q$: $\left\{\begin{array}{l} \gamma_k = \frac{k}{\gamma_k} \\ S_k = 0 \end{array}\right.$ leads to $\gamma_k = \sqrt{k}$.
\item Quartic potential $V'(Q) = \frac{1}{t}(Q-Q^3)$: 
\begin{equation}
\renewcommand{\arraystretch}{1.4}
\left\{\begin{array}{l}
\gamma_k(1-\gamma_{k-1}^2-\gamma_{k}^2-\gamma_{k+1}^2 -  S_k^2-S_kS_{k-1}-S_{k-1}^2) =  t\frac{k}{\gamma_k}\ , \\
S_k - S_k^3 - 2 S_k(\gamma_k^2+\gamma_{k+1}^2)-S_{k+1}\gamma_{k+1}^2-S_{k-1}\gamma_k^2 = 0\ . \\
\end{array}\right. 
\end{equation}
If we moreover assume that the contour $\gamma$ is symmetric under $x\mapsto -x$, then we must have $p_k(-x)=(-1)^k p_k(x)$, and therefore $S_k=0$. This leads to a recursion relation for $R_k=\gamma_k^2$,
\begin{equation}
R_k(1-R_{k-1}-R_k-R_{k+1}) =  tk \ ,
\end{equation}
which is known as the discrete Painlev\'e I equation.
This relation can be also obtained from the compatibility condition of the Lax matrices~\eqref{eq:HR_conpatibility}.
\end{itemize}
Finally, let us mention an alternative derivation of the equations \eqref{eq:recurPQ}. 
Instead of using  the relation $p'_k = kp_{k-1} + \cdots$, we could use the \textbf{string equation}\index{string equation}
\begin{align}
 \boxed{ [Q,P] = \operatorname{Id} } \ ,
\end{align}
which follows from the definitions of $P$ and $Q$ as the matrices of the derivative $\frac{d}{d \lambda}$ and multiplication by $\lambda$ respectively. 
Here we will check that the results \eqref{eq:PVQpm} - \eqref{eq:recurPQ} for $P$ and $Q$ imply the string equation. 
Since $P+\frac12 V'(Q)$ is strictly lower triangular and
$Q$ has at most one band above the diagonal, $[Q,P]=[Q,P+\frac12
V'(Q)]$ must be lower triangular.
The same reasoning with $P-\frac12 V'(Q)$ shows that $[Q,P]$ is also upper triangular, and therefore diagonal.  
And its diagonal coefficients are 
\begin{multline}
\left[Q,P+\tfrac12 V'(Q)\right]_{k,k}
= Q_{k,k+1} \left(P+\tfrac12 V'(Q)\right)_{k+1,k} - \left(P+\tfrac12 V'(Q)\right)_{k,k-1}Q_{k-1,k}\ ,
\\
 = \gamma_{k+1}\frac{k+1}{\gamma_{k+1}} - \frac{k}{\gamma_k}\gamma_k = 1\ .
\end{multline}

\subsection{Hilbert transforms of orthogonal polynomials}

We will shortly see that $\psi_k$ obeys a second-order differential equation. 
In order to construct another linearly independent solution, it is useful to introduce the \textbf{Hilbert transform}\index{Hilbert transform} of $\psi_k$,
\begin{align}
 \boxed{ \varphi_k(x) = e^{\frac12 V(x)}\int_\gamma dx' \frac{\psi_k(x')}{x-x'} e^{-\frac12 V(x')} } \ .
 \label{defvarphi}
\end{align}
Let us study the properties of $\varphi_k$, starting with its behaviour at infinity. 
Expanding $\frac{1}{x-x'} = \sum_{j=0}^\infty \frac{x'^j}{x^{j+1}}$, and noticing that $\int_\gamma dx' \psi_k(x')x'^j e^{-\frac12 V(x')} =0$ if $j<k$, we find
\begin{equation}
\boxed{ \varphi_k(x) \underset{x\to\infty}{=} \frac{\sqrt{h_k}}{x^{k+1}}\,e^{\frac12\,V(x)} \left(1+O\left(\frac{1}{x}\right)\right) } \ .
\label{varphi_inf}
\end{equation}
As proved in \autoref{exoHeinephi}, the functions $\varphi_k$ obey the Heine-type formulas
\begin{equation}\label{eq:Heinephi}
 \varphi_{k-1}(x) = \sqrt{h_{k-1}} e^{\frac12 V(x)} \left< \frac{1}{\det(x-M)} \right>_{k\text{-matrix}}\ .
\end{equation}

\subsubsection{Recursion and derivation relations}

We leave it to the reader to show that the functions 
$\varphi_k$ obey recursion and derivation relations, which only differ from the corresponding relations for $\psi_k$ for the first few values of $k$:
\begin{subequations}
\begin{align}
x \varphi_k(x) &= \sum_{j=k-1}^{k+1} Q_{k,j} \varphi_j(x)
 +\frac{\delta_{k,0}}{\psi_0(x)} \underset{k\geq 1}{=} \sum_{j=k-1}^{k+1} Q_{k,j} \varphi_j(x) \ , \label{eq:mphi}
\end{align}
\begin{multline}
 \varphi_k'(x) = \sum_{j=k-d}^{k+d} P_{k,j} \varphi_j(x)  +   \frac{e^{\frac12 V(x)}}{2}\int_\gamma dx' \psi_k(x')\frac{V'(x)-V'(x')}{x-x'} e^{-\frac12 V(x')}
 \ , \\
 \underset{k\geq d}{=} \sum_{j=k-d}^{k+d} P_{k,j} \varphi_j(x) \ .
\end{multline}
\end{subequations}

\subsubsection{Discontinuities}

$\varphi_k(x)$ is discontinuous at the integration domain $\gamma$: if $\gamma = \sum_i c_i \gamma_i$ is a linear combination of paths with complex coefficients $c_i$, then
\begin{align}
\boxed{ x\in \gamma_i\ : \qquad  \varphi_k(x+i0) - \varphi_k(x-i0) = - 2\pi i\, c_i \psi_k(x) } \ ,
\label{eq:phi_jump}
\end{align}
The difference of the two deformations of $\gamma_i$ that are needed for accommodating the points $x\pm i0$ is indeed a small circle around $x$, so that computing $\varphi_k(x+i0) - \varphi_k(x-i0)$ from Eq.~\eqref{defvarphi} amounts to taking a residue at $x$:
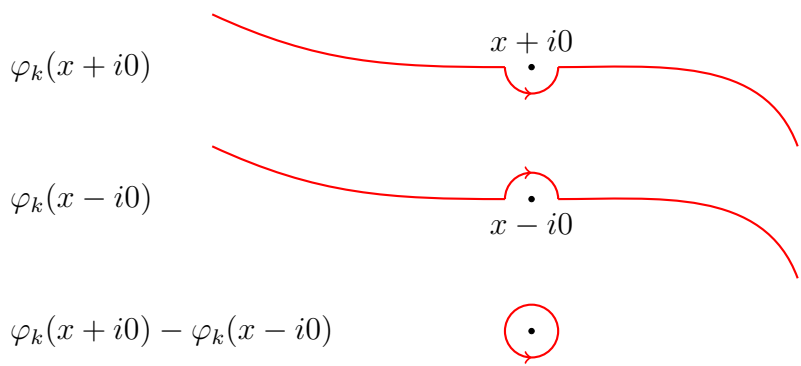
\begin{align}
 \begin{tikzpicture}[scale = .7,baseline=(current bounding box.center)]
  \draw[thick, red] (-6,1) to[out = -25, in = 180] (-.5, 0);
  \draw[thick, red] (.5,0) to[out = 0, in = 110] (5, -1.5);
  \draw[thick, red, ->-] (-.5,0) arc[radius = .5, start angle = -180, end angle =0];
  \node[right] at (-10,0) {$\varphi_k(x+i0)$};
  \node [label=above:$x+i0$,draw,fill=black,circle,inner sep=0pt,minimum size=2pt] at (0,0) {};
  \begin{scope}[shift = {(0,-2.5)}]
  \draw[thick, red] (-6,1) to[out = -25, in = 180] (-.5, 0);
  \draw[thick, red] (.5,0) to[out = 0, in = 110] (5, -1.5);
  \draw[thick, red, ->-] (-.5,0) arc[radius = .5, start angle = 180, end angle =0];
  \node[right] at (-10,0) {$\varphi_k(x-i0)$};
  \node [label=below:$x-i0$,draw,fill=black,circle,inner sep=0pt,minimum size=2pt] at (0,0) {};
  \end{scope}
  \draw[thick, red, ->-] (0,-4.5) arc[radius = .5, start angle =90, end angle = 450];
 \node [draw,fill=black,circle,inner sep=0pt,minimum size=2pt] at (0,-5)
 {};
  \node[right] at (-10,-5) {$\varphi_k(x+i0) - \varphi_k(x-i0)$};
 \end{tikzpicture}
\end{align}

\subsubsection{Determinantal formula}

$\varphi_k$ is given by the determinantal formula, 
\begin{equation}
\boxed{ \varphi_{k-1}(x) = \sqrt{h_{k-1}} e^{\frac12 V(x)}\det_{k\times k\text{ submatrix}} \left(\frac{1}{x-Q} \right) } \ .
\label{phi_det}
\end{equation}
To prove this, let us study matrix inverses of $x-Q$, starting with 
\begin{align}
 \left(\frac{1}{x-Q}\right)_{j,k} = \int_\gamma dx' \frac{\psi_j(x')\psi_k(x')}{x-x'}\ ,
\end{align}
which is a bilateral inverse of $x-Q$ as follows from the recursion relation \eqref{eq:mpsi}.
This can be rewritten as 
\begin{align}
 \left(\frac{1}{x-Q}\right)_{j,k}  = \psi_j(x)\varphi_k(x) + R_{j,k}(x)\ ,
\end{align}
where we introduce
\begin{align}
 R_{j,k}(x) = -\frac{1}{\sqrt{h_jh_k}}\int_\gamma dx' \frac{p_j(x)-p_j(x')}{x-x'} p_k(x')e^{- V(x')}\ .
\end{align}
The matrix $R(x)$ is still a right inverse of $x-Q$, but is simpler than $\frac{1}{x-Q}$, as it is strictly lower triangular, and polynomial in $x$.
More specifically, 
$R_{j,k}(x)$ is a polynomial of degree $j-k-1$, with
\begin{equation}
R_{j,k}(x) \underset{x\to \infty}{=} -\sqrt{\frac{h_k}{h_j}}x^{j-k-1}\left(1+O\left( \frac{1}{x}\right)\right)\ .
\end{equation}
The equation $(x-Q)R(x)=\text{Id}$ for $R(x)$ is a lower triangular linear system, that can be solved with Cramer's rule and yields
\begin{equation}\label{rightinvQdetQ}
\boxed{ R_{n,m}(x) = \left\{
\begin{array}{ll}
 \displaystyle
  -\frac{1}{\gamma_{m+1} \dots \gamma_n}\det_{m+1\leq i,j\leq n-1} (x-Q)_{i,j}   & \text{if } m+1<n\ , \\[1em]
 \displaystyle
-\frac{1}{\gamma_n} & \text{if } m+1=n\ ,   \\[1em]
0 & \text{if } m\geq n \ .
\end{array}
\right. }
\end{equation}
Similarly, the transpose $L(x)=R(x)^\text{T}$ is an upper triangular left inverse of $x-Q$.
Introducing 
the infinite vectors $\vec{\psi} = (\psi_0, \psi_1, \cdots )^\text{T}$ and $\vec{\varphi} = (\varphi_0, \varphi_1, \cdots )^\text{T}$, we can write
\begin{align}
\boxed{ \frac{1}{x-Q} = \vec{\psi}(x) \vec{\varphi}(x)^\text{T} + R(x) = \vec{\varphi}(x) \vec{\psi}(x)^\text{T} + L(x) } \ .
 \label{eq:invrl}
\end{align}
The existence of these three different inverses illustrates the exotic properties of infinite matrices, which in particular do not form an associative algebra.
For example, 
\begin{align}
 0 = \frac{1}{x-Q}\left((x-Q)\vec{\psi}\right) \neq \left(\frac{1}{x-Q}(x-Q)\right)\vec{\psi} = \vec{\psi}\ .
\end{align}
The determinantal formula \eqref{phi_det} for $\varphi_{k-1}$ can now be proved by starting with the trivial relation
\begin{align}
\varphi_{k-1}(x)  = \sqrt{h_{k-1}} e^{\frac12 V(x)} (-1)^{k-1}\psi_0(x)\varphi_{k-1}(x) \prod_{j=1}^{k-1} R_{j,j-1}\ ,
\end{align}
and rewriting the right-hand side as the desired minor of $\frac{1}{x-Q}$, with the help of Eq.~\eqref{eq:invrl}.

\subsection{Motzkin paths and continuous fractions}\label{sec:motzkin}

We have found that the orthogonal polynomials \eqref{pkdet} and the moments of our matrix model \eqref{eq:trf}
can be expressed in terms of an infinite matrix $Q$, which is tridiagonal: $|i-j|>1\implies Q_{i,j}=0$. 
So a sum of the type $\sum_i Q_{i,j} \cdots $ only has three non-vanishing terms, which suggests that it can have a combinatorial interpretation in terms of paths with three possible directions, called Motzkin paths. 

\subsubsection{Motzkin paths}

A \textbf{Motzkin path}\index{Motzkin path} is a path in the lattice $\mathbb{N}\times \mathbb{N}$, made of edges with three possible directions:
 \begin{tikzpicture}[scale = .35]     
 \draw [thick, blue] (0.,0.) -- (1,1) ;   
 \draw [thick, blue] (1.5,1) -- (2.5,0.) ; 
 \draw [thick, blue] (3.,0.5) -- (4,0.5) ; 
 \end{tikzpicture}.
  For example,
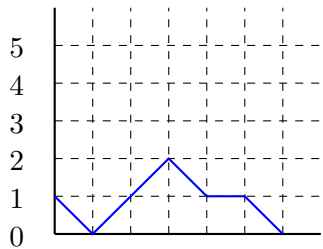
\begin{center}
 \begin{tikzpicture}
  
  \draw [thick] (0.,0.) -- (3.5,0.) ;

  \draw [thick] (0.,0.) -- (0.,3.) ;

  \draw [dashed] (0.,0.5) -- (3.5,0.5) ;

  \draw [dashed] (0.,1.) -- (3.5,1.) ;

  \draw [dashed] (0.,1.5) -- (3.5,1.5) ;

  \draw [dashed] (0.,2.) -- (3.5,2.) ;

  \draw [dashed] (0.,2.5) -- (3.5,2.5) ;

  \draw [thin, dashed] (0.5,0.) -- (0.5,3.) ;

  \draw [thin, dashed] (1.,0.) -- (1.,3.) ;

  \draw [thin, dashed] (1.5,0.) -- (1.5,3.) ;

  \draw [thin, dashed] (2.,0.) -- (2.,3.) ;

  \draw [thin, dashed] (2.5,0.) -- (2.5,3.) ;

  \draw [thin, dashed] (3.,0.) -- (3.,3.) ;

  \draw (-0.5,-0.3) node [above] {\small $0$};

  \draw (-0.5,0.2) node [above] {\small $1$};

  \draw (-0.5,0.7) node [above] {\small $2$};

  \draw (-0.5,1.2) node [above] {\small $3$};

  \draw (-0.5,1.7) node [above] {\small $4$};

  \draw (-0.5,2.2) node [above] {\small $5$};

% a path:
  \draw [thick,blue] (0.,0.5) -- (0.5,0.) -- (1.,0.5) -- (1.5,1.) -- (2.,0.5) -- (2.5,0.5) -- (3.,0.) ;

 \end{tikzpicture}
\end{center}
We assign a weight to each edge depending on its height and direction, 
and the weight of a path is the product of its edge weights,
\begin{equation}
w(\mathfrak m) = \prod_{e\in\{\text{edges}(\mathfrak m)\}} w(e)\ .
\end{equation}
In particular, the weights that correspond to the 3--band matrix $Q$ are 
\begin{align}
 \begin{tikzpicture}[baseline=(current bounding box.center), scale = 1.2]
 %
 % weights
  \draw [thick,blue] (0.2,0.) -- (0.7,0.5) ;
  \draw [dashed] (0.,0.0) -- (1.,0.0) ;
  \draw [dashed] (0.,0.5) -- (1.,0.5) ;
  \draw (-0.8,-0.3) node [above] {\small $k-1$};
  \draw (-0.5,0.2) node [above] {\small $k$};
  \draw (0.2,-0.4) node [below] { $w_Q(e)= Q_{k-1,k}=\gamma_k$};
  \draw [thick,blue] (4.2,0.5) -- (4.7,0.0) ;
  \draw [dashed] (4.,0.0) -- (5.,0.0) ;
  \draw [dashed] (4.,0.5) -- (5.,0.5) ;
  \draw (3.2,-0.3) node [above] {\small $k-1$};
  \draw (3.5,0.2) node [above] {\small $k$};
  \draw (4.2,-0.4) node [below] { $w_Q(e)= Q_{k,k-1}=\gamma_k$};
  \draw [thick,blue] (8.2,0.5) -- (8.7,0.5) ;
  \draw [thick,blue] (8.2,0.51) -- (8.7,0.51) ;
  \draw [dashed] (8.,0.0) -- (9.,0.0) ;
  \draw [dashed] (8.,0.5) -- (9.,0.5) ;
  \draw (7.2,-0.3) node [above] {\small $k-1$};
  \draw (7.5,0.2) node [above] {\small $k$};
  \draw (8.2,-0.4) node [below] { $w_Q(e)= Q_{k,k}=S_k$};
 \end{tikzpicture}
\end{align}

\subsubsection{Combinatorial expressions}

Let us write a moment in terms of the matrix $Q$, using Eq.~\eqref{eq:trf}:
\begin{equation}
\Big< \operatorname{Tr} M^m\Big>_{n\text{-matrix}} = \operatorname{Tr} Q^m\Pi_n
= \sum_{i_m=0}^{n-1}\sum_{i_1,\dots,i_{m-1}=0}^\infty Q_{i_m,i_1} Q_{i_1,i_2} Q_{i_2,i_3} \ldots Q_{i_{m-1},i_m}\ .
\end{equation}
We now interpret each nonzero matrix element as the weight of an edge, 
and each product of $m$ matrix elements as the weight of a Motzkin path of length $m$. 
If $\mathcal{M}^{(m)}_{k,l}$ is the set of Motzkin paths of length $m$ from height $k$ to height $l$, our moment has the combinatorial expression 
\begin{equation}
\Big< \operatorname{Tr} M^m\Big>_{n\text{-matrix}} 
= \sum_{k=0}^{n-1} \sum_{\mathfrak m\in \mathcal{M}^{(n)}_{k,k}}  w_Q(\mathfrak m) \ .
\end{equation}
The matrix elements of $Q$ are given in Eq.~\eqref{eq:qel}, and we find for example:
\begin{align}
 \begin{tikzpicture}[baseline=(current bounding box.center)]
  \draw [thick,blue] (0.,0.) -- (0.4,0.) -- (0.8,0.0) -- (1.2,0.) ;
  \draw [thick,blue] (2.,0.) -- (2.4,0.4) -- (2.8,0.) -- (3.2,0.) ;
  \draw [thick,blue] (4.,0.) -- (4.4,0.) -- (4.8,0.4) -- (5.2,0.) ;
  \draw [thick,blue] (6.,0.) -- (6.4,0.4) -- (6.8,0.4) -- (7.2,0.) ;
  \draw (-1,0) node [above] {$(Q^3)_{0,0}\, =$};
  \draw (1.6,0.) node [above] {$+$};
  \draw (3.6,0.) node [above] {$+$};
  \draw (5.6,0.) node [above] {$+$};
  \draw (0.4,0.5) node [above] {$S_0^3$};
  \draw (2.4,0.5) node [above] {$\gamma_1^2 S_0$};
  \draw (4.4,0.5) node [above] {$S_0 \gamma_1^2$};
  \draw (6.4,0.5) node [above] {$\gamma_1 S_0 \gamma_1$};
 \end{tikzpicture}
\end{align}
The orthogonal polynomial $p_n(x)$ can be expressed as the determinant of a size $n$ submatrix of $x-Q$.
According to the Lindstr\"om--Gessel--Viennot lemma, this determinant is a weighted sum over a set of non-intersecting paths:
\begin{equation}
 p_n(x) = \sum_{\mathfrak{m}\in \mathcal{M}^{(1,n)}_{\{0,\cdots,n-1\}}}  w_{x-Q}(\mathfrak{m})\ ,
\end{equation}
where $\mathcal{M}^{(1,n)}_{\{0,\cdots,n-1\}}$ is the set
of families of $n$ Motzkin paths of length 1, starting from  $\{0,\dots, n-1\}$ and ending at $\{0,\dots, n-1\}$, such that two paths do not intersect on the lattice. For example:
\begin{align}
 \begin{tikzpicture}[baseline=(current bounding box.center), scale = 1.4]
  \draw [thick,blue] (0.,0.) -- (0.5,0.) ;
  \draw [thick,blue] (0.,0.5) -- (0.5,0.5) ;
  \draw [thick,blue] (0.,1.) -- (0.5,1.) ;
  \draw [thick,blue] (2.,0.) -- (2.5,0.) ;
  \draw [thick,blue] (2.,0.5) -- (2.5,1.) ;
  \draw [thick,blue] (2.,1.) -- (2.5,0.5) ;
  \draw [thick,blue] (4.,0.) -- (4.5,0.5) ;
  \draw [thick,blue] (4.,0.5) -- (4.5,0.) ;
  \draw [thick,blue] (4.,1.) -- (4.5,1.) ;
  \draw (-1,0) node [above] {$p_3(x)\, =$};
  \draw (1.25,0.) node [above] {$+$};
  \draw (3.25,0.) node [above] {$+$};
  \draw (0.25,-0.2) node [below] {$(x-S_2)$};
  \draw (0.25,-0.7) node [below] {$(x-S_1)$};
  \draw (0.25,-1.2) node [below] {$(x-S_0)$};
  \draw (2.25,-0.2) node [below] {$(x-S_0) \gamma_2^2$};
  \draw (4.25,-0.2) node [below] {$ (x-S_2) \gamma_1^2$};
 \end{tikzpicture}
\end{align}
The determinant that appears in the expression \eqref{eq:kds} of the self-reproducing kernel $K_{n+1}(x,y)$ is a sum over the set $\mathcal{M}^{(2,n)}_{\{0,\cdots,n-1\}}$ of $n$ non-intersecting paths of length $2$ starting and ending at $\{0,\dots, n-1\}$
\begin{align}
 \mathop{\det}_{n\times n\text{ submatrix}} (x-Q)(y-Q) = \sum_{\mathfrak{m}\in \mathcal{M}^{(2,n)}_{\{0,\cdots,n-1\}}}  w_{x-Q,y-Q}(\mathfrak{m})\ ,
\end{align}
where the notation $w_{x-Q,y-Q}$ means that the first edge of a path has weight $w_{x-Q}(e)$, and the second edge has weight $w_{y-Q}(e)$. 
For example:
\begin{align}
 \begin{tikzpicture}[baseline=(current bounding box.center), scale = 1.1]
  \draw [thick,blue] (0.,0.) -- (0.5,0.) -- (1.,0.) ;
  \draw [thick,blue] (0.,0.5) -- (0.5,0.5) -- (1.,0.5) ;
  \draw [thick,blue] (2.,0.) -- (2.5,0.5) -- (3.,0.5) ;
  \draw [thick,blue] (2.,0.5) -- (2.5,0.) -- (3.,0.) ;
  \draw [thick,blue] (4.,0.) -- (4.5,0.0) -- (5.,0.5) ;
  \draw [thick,blue] (4.,0.5) -- (4.5,0.5) -- (5.,0.) ;
  \draw [thick,blue] (6.,0.) -- (6.5,0.5) -- (7.,0.0) ;
  \draw [thick,blue] (6.,0.5) -- (6.5,0.0) -- (7.,0.5) ;
  \draw [thick,blue] (8.,0.) -- (8.5,0.0) -- (9.,0.0) ;
  \draw [thick,blue] (8.,0.5) -- (8.5,1.0) -- (9.,0.5) ;
  \draw [thick,blue] (10.,0.) -- (10.5,0.5) -- (11.,0.0) ;
  \draw [thick,blue] (10.,0.5) -- (10.5,1.0) -- (11.,0.5) ;
  \draw (-1.2,-0.2) node [above] {$K_3(x,y)\, \propto$};
  \draw (1.6,0.) node [above] {$+$};
  \draw (3.6,0.) node [above] {$+$};
  \draw (5.6,0.) node [above] {$+$};
  \draw (7.6,0.) node [above] {$+$};
  \draw (9.6,0.) node [above] {$+$};
  \draw (0.4,-0.25) node [below] {\tiny $(x-S_1)(y-S_1)$};
  \draw (0.4,-0.7) node [below] {\tiny $(y-S_0)(y-S_0)$};
  \draw (2.5,-0.2) node [below] {\tiny $(y-S_0)(y-S_1) \gamma_1^2$};
  \draw (4.8,-0.2) node [below] {\tiny $ (x-S_0)(x-S_1) \gamma_1^2$};
  \draw (6.5,-0.2) node [below] {\tiny $  \gamma_1^4$};
  \draw (8.8,-0.2) node [below] {\tiny $ (x-S_0)(y-S_0) \gamma_1^2$};
  \draw (10.8,-0.2) node [below] {\tiny $  \gamma_1^2 \gamma_2^2$};
 \end{tikzpicture}
\end{align}
Other functions of the matrix $Q$ can also be expressed as sums of Motzkin paths.
This holds in particular for the elements of the right inverse of $x-Q$, using Eq.~\eqref{rightinvQdetQ}.

\subsubsection{Continuous fractions}

Matrix elements of $\frac{1}{x-Q}$
are generating functions of Motzkin paths, and can be expressed as continuous fractions. Recursively using the identity
\begin{equation}
 M = \left(\begin{array}{c|c} a & \begin{array}{cccc} b & 0 & \cdots & 0 \end{array} \\ \hline
           \begin{array}{c}
           b \\ 0 \\ \vdots \\ 0 
           \end{array}
           & \tilde M \end{array}\right)
 \quad \implies \quad (M^{-1})_{0,0} = \frac{1}{a -b^2 ({\tilde M}^{-1})_{0,0}}\ ,
\end{equation}
starting with $M=x-Q$, we indeed obtain the continuous fraction representation
\begin{equation}
\left(\frac{1}{x-Q}\right)_{0,0} =
\frac{1}{x-S_0 - \frac{\gamma_1^2}{x-S_1-\frac{\gamma_2^2}{x-S_2-\frac{\gamma_3^2}{x-\dots}}}}\ .
\end{equation}
Continuous fractions are limits of rational functions, and the rational functions that are relevant in our case are minors of $x-Q$, 
\begin{align}
 \left(\frac{1}{x-Q}\right)_{0,0} =\lim_{n\to\infty} \frac{\det (x-Q)_{[1,n-1]^2}}{\det (x-Q)_{[0,n-1]^2}} =  \lim_{n\to\infty} \frac{\tilde p_n(x)}{p_n(x)}\ ,
\end{align}
where  $\det (x-Q)_{I\times J}$ is the minor of $x-Q$ with rows in $I$ and columns in $J$, and the polynomials $\tilde{p}_n(x)=\det (x-Q)_{[1,n-1]^2}$ that we just introduced obey the same recursion relation as the orthogonal polynomials $p_n(x)=\det (x-Q)_{[0,n-1]^2}$, with however the initial conditions 
\begin{align}
(\tilde p_0,  \tilde p_{1}) = (0, 1)\ ,
\end{align}
instead of $(p_{-1}, p_0) = (0, 1)$.
Let us also derive an expression of $\left(\frac{1}{x-Q}\right)_{n,n}$ with $n>0$ as a continuous fraction, starting with $\left(\frac{1}{x-Q}\right)_{n,n} = \psi_n \varphi_n$ from Eq.~\eqref{eq:invrl}.
The Hilbert transform $\varphi_n$ of $\psi_n$ obeys the same recursion relation as $\psi_n$, and therefore as $\frac{p_n}{\sqrt{h_n}}$. 
So there must exist two functions $\alpha(x),\beta(x)$ such that 
$\sqrt{h_n}\varphi_n = \alpha p_n +\beta\tilde{p}_n$. 
Let us determine these functions by considering the cases $n=0,1$. 
In the case $n=0$ we simply have $\alpha(x) = \sqrt{h_0}\varphi_0(x) = h_0 e^{\frac12 V(x)} \left(\frac{1}{x-Q}\right)_{0,0}$. 
In the case $n=1$ we have 
$
 \beta = \sqrt{h_1}\varphi_1 -\sqrt{h_0}\varphi_0 p_1 
$.
From the definitions of $\varphi_0,\varphi_1$ as Hilbert transforms, we then deduce 
\begin{align}
 \beta(x) 
 = e^{\frac12 V(x)} \int_\gamma \frac{p_1(x')-p_1(x)}{x-x'}e^{-V(x')}dx' 
 = -e^{\frac12 V(x)} \int_\gamma e^{-V(x')}dx' = -h_0 e^{\frac12 V(x)}\ .
\end{align}
This leads to the  expression of $\left(\frac{1}{x-Q}\right)_{n,n}$ as a continuous fraction,
\begin{align}
 \left(\frac{1}{x-Q}\right)_{n,n} = \frac{h_0}{h_n}p_n(x)^2\left( \left(\frac{1}{x-Q}\right)_{0,0} - \frac{\tilde{p}_n(x)}{p_n(x)}\right)\ .
\end{align}

\subsection{Christoffel--Darboux formulas}

Using
the projector $\Pi_N$ on the first $N$ components,
the self-reproducing kernel \eqref{eq:kpp} can be written as 
\begin{align}
 K_N(x,y) = \vec{\psi}(x)^\text{T} \Pi_N \vec{\psi}(y)\ .
\end{align}
We thus have 
\begin{align}
 (y-x)K_N(x,y) = \vec{\psi}(x)^\text{T} [\Pi_N, Q] \vec{\psi}(y)\ .
\end{align}
This involves the \textbf{Christoffel--Darboux matrix}\index{Christoffel--Darboux!---matrix}
\begin{align}
 \boxed{ A^N = [\Pi_N, Q] } \ ,
 \label{anpnq}
\end{align}
whose only nonzero coefficients are $A^N_{N-1,N} = -A^N_{N,N-1} = \gamma_N$. %OK
And we obtain the \textbf{Christoffel--Darboux formula}\index{Christoffel--Darboux!---formula}
for the self-reproducing kernel,
\begin{align}
 \boxed{ K_N(x,y) = \frac{\sum_{j,k = N-1}^N A^N_{j,k}\psi_j(x)\psi_k(y)}{y-x} 
 = \gamma_N\frac{\psi_{N-1}(x)\psi_N(y) - \psi_N(x)\psi_{N-1}(y)}{y-x} } \ ,
 %OK
 \label{CD_kernel}
\end{align}
This formula is more convenient than \eqref{eq:kpp} at large $N$, since it only involves two terms instead of $N$ terms. 
Moreover, this formula shows that the self-reproducing kernel is an integrable kernel.

\subsubsection{Matrix kernel}

Our formulas for the kernel $K_N$ suggest other kernels, which are built
using the Hilbert transforms $\vec{\varphi}$ in addition to the original
orthogonal functions $\vec{\psi}$.
We thus define a matrix kernel
\begin{equation}
\hat K_N = \begin{pmatrix}
\hat{K}^N_{1,1} & \hat{K}^N_{1,2} \cr
\hat{K}^N_{2,1} & \hat{K}^N_{2,2}
\end{pmatrix}
\ ,
\end{equation}
whose elements are 
\begin{subequations}
\begin{align}
{\hat{K}^N}_{2,1}(x,y) &= K_N(x,y) =  \vec{\psi}(x)^\text{T} \Pi_N \vec{\psi}(y) \ ,
 \\
{\hat{K}^N}_{2,2}(x,y) &=   \vec{\psi}(x)^\text{T} \Pi_N \vec{\varphi}(y) + \frac{e^{\frac12(V(y)-V(x))}}{x-y} \ ,
 \\
{\hat{K}^N}_{1,1}(x,y) &=   - \vec{\varphi}(x)^\text{T} \Pi_N \vec{\psi}(y)  + \frac{e^{\frac12(V(x)-V(y))}}{x-y}  = - \hat K^N_{2,2}(y,x)\ ,
\label{eq:knoo}
 \\
{\hat{K}^N}_{1,2}(x,y) &= -  \vec{\varphi}(x)^\text{T} \Pi_N \vec{\varphi}(y)  + \int_\gamma \frac{e^{\frac12 V(x)}e^{\frac12 V(y)}}{(x-x')(y-x')}e^{-V(x')}dx'\ .
\end{align}
\end{subequations}
We introduce the matrix 
\begin{equation}
\boxed{ \Psi_N = \begin{pmatrix}
\psi_{N-1} & \varphi_{N-1} \cr
\psi_{N} & \varphi_{N} \cr
\end{pmatrix} } \ ,
\label{eq:psindef}
\end{equation}
which obeys $\det \Psi_N(x) = -\frac{1}{\gamma_N}$, and
using the recursion relation for $\psi$ and $\varphi$, we obtain the matrix Christoffel--Darboux formula
\begin{equation}
\boxed{ (x-y) \hat K_N(x,y) = \Psi_N(x)^\text{T} A^N \Psi_N(y) =\Psi_N(x)^{-1}\Psi_N(y) } \ ,
\label{eq:khpp}
\end{equation}
This formula can be generalized to
multi-matrix models, which give rise to matrices $\hat K_N$ and $\Psi_N$ of rank higher than two \cite{Bergere:2006}.

\subsubsection{Heine-type formulas for the matrix kernel}

The entries of the matrix kernel obey the Heine-type formulas 
\begin{subequations}
\begin{align}
 \hat K^N_{2,1} (x,y) &= \frac{e^{-\frac{V(x)}{2}}e^{-\frac{V(y)}{2}}}{h_{N-1}} \Big\langle \det(x-M)\det(y-M)\Big\rangle_{N-1\times N-1}\ ,
\\
\hat K^N_{2,2}(x,y) &= -\hat K^N_{1,1}(y,x) = \frac{e^{-\frac{V(x)}{2}}e^{\frac{V(y)}{2}}}{(x-y)} \left\langle \frac{\det(x-M)}{\det(y-M)}\right\rangle_{N\times N} \ ,
\label{eq:Khatratiodet}
\\
\hat K^N_{1,2} (x,y) &= h_{N-1} e^{\frac{V(x)}{2}} e^{\frac{V(y)}{2}} \left\langle \frac{1}{\det(x-M)\det(y-M)}\right\rangle_{N-1\times N-1} \ .
\end{align}
\end{subequations}
To prove these formulas, the idea is to write their right-hand sides in terms of the functions $\psi_k$ and/or $\varphi_k$, and to show that the resulting expressions agree with the Christoffel--Darboux formulas for the kernels.
Let us do this in the case of the first formula. 
Since $\det(x-M)$ is a polynomial function of $x$ of degree $N-1$ for $M\in H_{N-1}(\gamma)$, we can write $(y-x)\Big\langle \det(x-M)\det(y-M)\Big\rangle_{N-1\times N-1}$ as a linear combination of the orthogonal polynomials $p_0(x),\dots , p_{N}(x)$, and the coefficient of $p_{N}(x)$ is simply $-\Big\langle \det(y-M)\Big\rangle_{N-1\times N-1}$. 
This implies that there must exist $x$-independent functions $c_j(y)$ such that 
\begin{multline}
 (y-x) \frac{e^{-\frac{V(x)}{2}}e^{-\frac{V(y)}{2}}}{h_{N-1}} \Big\langle \det(x-M)\det(y-M)\Big\rangle_{N-1\times N-1} 
\\  = 
 -\gamma_N\psi_{N}(x)\psi_{N-1}(y) + \sum_{j=0}^{N-1} c_j(y)\psi_j(x) \ .
\end{multline}
Let us compute $c_j(y)$ as the scalar product of the left-hand side of this formula with $\psi_j(x)$. 
Writing $\Big\langle \det(x-M)\det(y-M)\Big\rangle_{N-1\times N-1} $ as an integral over the eigenvalues $\lambda_1,\dots,\lambda_{N-1}$ of $M$ and renaming $x$ as $\lambda_N$, we find 
\begin{align}
\sqrt{h_j} \mathcal{Z}_{N}e^{\frac{V(y)}{2}} c_j(y) = \int_{\gamma^{N}} d\lambda_1\dots d\lambda_{N} 
\Delta_{N}(\lambda) p_j(\lambda_N)  \Delta_{N-1}(\lambda) \prod_{i=1}^{N} (y-\lambda_i)e^{-V(\lambda_i)} \ .
\end{align}
Symmetrizing the integrand over the $N$ integration variables (exactly as in our proof of Heine's formula), we find $c_{j<N-1}=0$ and $c_{N-1} = \gamma_N \psi_N(y)$. 
This completes the proof of the Heine-type formula for $K_N(x,y)$. 
We leave it to the reader to prove the Heine-type formulas for the three other kernels. 

\section{Integrable systems}\label{sec:ais}

\subsection{Differential system and Riemann--Hilbert problem}

As a consequence of their recursion and derivation relations, the functions $\psi_N$ must obey a matrix differential equation of the type
\begin{align}
 \begin{pmatrix} \psi_{N-1}' \\ \psi_N' \end{pmatrix}  
 = D_N \begin{pmatrix} \psi_{N-1} \\ \psi_N \end{pmatrix}\ ,
\end{align}
where the size two matrix $D_N$ is called the \textbf{Lax matrix}\index{Lax matrix}.
The system is of size two because the recursion relation \eqref{eq:mpsi} involves three consecutive functions $\psi_{k+1}, \psi_k, \psi_{k-1}$, which implies that every $\psi_j$ can be expressed as a linear combination of $\psi_N$ and $\psi_{N-1}$, with polynomial coefficients of degrees at most $\max(j-N,N-1-j)$. 
The derivation relation \eqref{eq:psiprime} then gives $\psi_k'$ as a combination of $\psi_j$ with $|j-k|\leq d$, and this leads to a Lax matrix whose coefficients are polynomials of degrees at most $d$. 
We will now determine this matrix more explicitly.

\subsubsection{Folding}

From the three-term recursion relation \eqref{eq:mpsi}, we can deduce a polynomial \textbf{folding matrix}\index{folding matrix} $F^N(x)$ of size $\infty\times 2$, such that 
\begin{equation}
\forall\,j, \quad \psi_{j}(x) = F^N_{j,N-1}(x)\psi_{N-1}(x)+F^N_{j,N}(x)\psi_{N}(x)\ ,
\end{equation}
or in vector notation
\begin{subequations}
\begin{align}
 \vec \psi (x) &= F^N(x) \vec \psi(x)\ ,
 \label{eq:pefp}
 \\
 k\notin\{N-1,N\} & \implies F^N_{j,k}(x) = 0\ .
 \label{eq:knin}
\end{align}
\end{subequations}
The folding matrix can be expressed in terms of the right and left inverses of $(x-Q)$, computed in \eqref{rightinvQdetQ}, \eqref{eq:invrl}, and of the Christoffel--Darboux matrix $A^N$, as
\begin{equation}
F^N(x) = (R(x)-L(x)) A^N \ .
\label{fnrla}
\end{equation}
To prove this, notice that $(R(x)-L(x)) A^N$ is a polynomial of $x$ that satisfies Eq.~\eqref{eq:knin}. Then Eq.~\eqref{eq:pefp} can be checked with the help of Eq.~\eqref{eq:invrl}:
\begin{multline}
(R(x)-L(x)) A^N \vec\psi
 = (\vec \psi \vec\varphi^\text{T} - \vec\varphi\vec \psi^\text{T})
\Big( (x-Q)\Pi_N -\Pi_N(x-Q) \Big) \vec \psi \ , \\
 = \vec \psi \vec\varphi^\text{T} (x-Q)\Pi_N \vec \psi= \vec \psi \begin{pmatrix} \frac{1}{\psi_0}, 0, 0,\cdots \end{pmatrix} \Pi_N \vec \psi = \vec \psi\ .
\end{multline}
% NB: Sign wrong? Should be R - L = - (\vec \psi \vec\varphi^\text{T} - \vec\varphi\vec \psi^\text{T}) ?

\subsubsection{Determination of the Lax matrix}

The folding matrix can be used to fold the $d$-terms derivation relation $\vec \psi'=P\vec \psi$, onto the vector space generated by $\psi_{N},\psi_{N-1}$, giving an expression for the Lax matrix as a $2\times 2$ block of the infinite matrix, 
\begin{align}
D_N(x) = \left( PF^N(x)\right)_{[N-1,N]^2}\ .
\end{align}
Let us insert our expression for $F^N(x)$, and the formula \eqref{eq:PVQpm} for $P$. 
Since $V'(Q)_+$ and $L(x)$ are both strictly upper triangular, we have $\left(V'(Q)_+L(x)A^N\right)_{[N-1,N]^2}=0$, and similarly $\left(V'(Q)_-R(x)A^N\right)_{[N-1,N]^2}=0$. 
This allows us to compute
\begin{subequations}
\begin{align}
D_N(x) 
&= -\frac{1}{2}\Big( (V'(Q)_+-V'(Q)_-)(R(x)-L(x))A^N \Big)_{[N-1,N]^2} \ ,  \\
&= -\frac{1}{2}\Big( (V'(Q)_++V'(Q)_-)(R(x)+L(x))A^N \Big)_{[N-1,N]^2} \ , \\
&= -\frac{1}{2}\Big( V'(Q)(R(x)+L(x))A^N \Big)_{[N-1,N]^2}\ .
\end{align}
\end{subequations}
Using Eq.~\eqref{rightinvQdetQ}, we can compute $\left(R(x)A^N\right)_{[N-1,N]^2}$ and $\left(L(x)A^N\right)_{[N-1,N]^2}$, and we find $\left((R(x)+L(x))A^N\right)_{[N-1,N]^2} = \sigma_3$ where $\sigma_3=\mathrm{ diag}(1,-1)$ is a Pauli matrix. 
Moreover, in 
$R(x)+L(x) = 2(x-Q)^{-1} + \vec \psi\vec\varphi^\text{T} - \vec\varphi \vec \psi^\text{T}$, 
% NB:  Should be - \vec \psi\vec\varphi^\text{T} ?
the last two terms do not contribute to $\Big( (V'(x)-V'(Q))(R(x)+L(x))A^N \Big)_{[N-1,N]^2}$.
We then find 
%\cite{Eynard:2001}
\begin{align}
 \boxed{ D_N(x) = -\frac12 V'(x) \sigma_3  + \left( \frac{V'(x) - V'(Q)}{x-Q} A^N \right)_{[N-1,N]^2} }
\ .
\label{formulaDN}
 \end{align}
Defining the polynomial  
 \begin{align}
 \boxed{ P_N(x) = \operatorname{Tr}\left(\Pi_N \frac{V'(x)-V'(Q)}{x-Q}\right) =\left< \operatorname{Tr}\frac{V'(x)-V'(M)}{x-M} \right>_{N\times N \text{ matrices}} } \ ,
\label{eq:P_N_traceform}
\end{align}
the trace and determinant of the Lax matrix are then 
\begin{align}
 \boxed{ \operatorname{Tr} D_N = 0 } \quad , \quad \boxed{ \det D_N = P_N - \frac14(V')^2 } \ ,
\label{dn_inv}
\end{align}
The calculation of the trace is immediate, while the calculation of the determinant is done in \autoref{exodetD}.

We now define the \textbf{spectral curve}\index{spectral curve} of the system as the eigenvalue locus of the Lax matrix, and its equation is
\begin{align}
\boxed{ \det(y-D_N(x)) = 0 } \quad \iff \quad y^2 = \frac14(V'(x))^2  -P_N(x)\ .
\end{align}
Our polynomial $P_N(x)$ coincides with the polynomial $P_0(x)$ \eqref{eq:pndef} that we defined in the context of the loop equations. Therefore, in an appropriate large $N$ limit, our spectral curve tends to the spectral curve \eqref{eq:w01x} that we defined in that context.

\subsubsection{Isomonodromy}

For $N>d$, since $\varphi_N$ satisfies the same three-term recursion and derivation relations as $\psi_N$, 
 the whole matrix $\Psi_N$ \eqref{eq:psindef} obeys the differential equation
\begin{align}
 \boxed{ \Psi_N' = D_N \Psi_N } \ .
 \label{Lax_eq}
\end{align}
Since $\varphi_N$ is discontinuous \eqref{eq:phi_jump} across the path $\gamma_i$, the matrix $\Psi_N$ is discontinuous too, and its jump is described by a \textbf{monodromy matrix}\index{monodromy matrix} $S_{\gamma_i}$, 
\begin{align}
 \Psi_N (x+i0) = \Psi_N (x-i0)\, S_{\gamma_i} \quad \text{with} \quad 
 S_{\gamma_i} = \begin{pmatrix} 1 & -2\pi i c_i \\ 0 & 1 \end{pmatrix}\ .
 \label{eq:psi_path}
\end{align}
Since the monodromies $S_{\gamma_i}$ are independent of $x$, 
our differential system is called \textbf{isomonodromic}\index{isomonodromic!---equation}.
This is equivalent to the Lax matrix $D_N=\Psi_N'\Psi_N^{-1}$ being analytic across the path $\gamma_i$,
\begin{equation}
\forall x\in\gamma_i\ \quad  D_N(x+i0)=D_N(x-i0) \qquad \iff \qquad S'_{\gamma_i}=0 \ .
\label{cst_mono}
\end{equation}
The monodromies $S_{\gamma_i}$ are independent not only of $x$, but also of the potential $V(x)=\sum_{k=1}^{d+1} \frac{t_k}{k} x^k$, whose coefficients $t_k$  are called \textbf{isomonodromic times}\index{isomonodromic!---time}. The monodromies only depend on the coefficients $c_i$ of the path $\gamma=\sum_i c_i \gamma_i$. (See \autoref{exo:cft1} for an example from conformal field theory.)

\subsubsection{Riemann--Hilbert problem and large $N$ asymptotics}

Instead of solving the matrix differential equation \eqref{Lax_eq}, whose coefficients are not necessarily easy to determine, it is possible to determine $\Psi_N$ by solving a \textbf{Riemann--Hilbert problem}\index{Riemann--Hilbert!---problem}, i.e. requiring that it has particular analyticity, monodromy and asymptotic properties: 
\begin{itemize}
 \item $\Psi_N$ is analytic on $\mathbb{C} \backslash \cup_i \gamma_i$, and is everywhere invertible $\det \Psi_N(x)\neq 0$, 
 \item the behaviour of $\Psi_N$ across the path $\gamma_i$ is given by Eq.~\eqref{eq:psi_path},
 \item the asymptotic behaviour of $\Psi_N$ near infinity is given by
 \begin{align}
  \boxed{ \Psi_N(x) = \hat{\Psi}_N(x) e^{T_N(x)} \quad \text{with} \quad T_N(x) = \left(N\log x-  \frac{V(x)}{2} \right)\sigma_3 } \ ,
 \label{eq:psi_infty}
 \end{align}
 where $\hat{\Psi}_N(x)$ is analytic at $x=\infty$:
 \begin{equation}
 \hat\Psi_N(x) \sim \begin{pmatrix}
 0 & \sqrt{h_{N-1}} \cr
 \frac{1}{\sqrt{h_N}} & 0
 \end{pmatrix} + O\left(\frac{1}{x}\right)\ .
 \end{equation}
\end{itemize}
When it exists, the solution of the Riemann--Hilbert is unique.
Indeed if $\Psi_N$ and $\tilde \Psi_N$ were two solutions, then $\Psi_N \tilde \Psi^{-1}_N$ would be an entire function that tends to identity at infinity, therefore the identity everywhere.

The formulation of the conditions on $\Psi_N$ as a Riemann--Hilbert problem is particularly convenient for studying its large $N$ limit, because $N$ appears only in the $x\to\infty$ asymptotic behaviour \eqref{eq:psi_infty}. 
The matrix \textbf{steepest-descent method}\index{steepest-descent method} of Bleher--Its, Deift--Zhou--Mac Laughlin--Venakides~\cite{Bleher:2001random,Deift:1993steepest}, also called the Deift--Zhou steepest descent method~\cite{Deift:2000orthogonal}, was the first complete mathematical proof of the large $N$ asymptotic expansion for $\Psi_N$, and actually for all correlations functions. 
The idea is that if $\Psi_N$  and $\tilde \Psi_N$ solve two Riemann--Hilbert  problems with monodromies and $x\to\infty$ asymptotics that agree at large $N$ up to $(1+O(\frac{1}{N}))$ factors, then $\Psi_N(x) \sim \tilde \Psi_N(x) (1+O(\frac{1}{N}))$ uniformly in $x$.
This method can be used to prove that a matrix $\tilde \Psi_N$ constructed using the approximate Coulomb gas and topological recursion methods of Chapters~\ref{chap:SaddlePoint} and~\ref{chap:LoopEq}, correctly gives the asymptotics of $\Psi_N$.

\subsection{Deformations and flat connections}

We will now study how the matrix $\Psi_N$ depends on the 
coefficients $t_k$ in the potential $V(x) = \sum_{k} \frac{t_k}{k}
x^k$, and on the matrix size $N$.
We will see that $t_k$ and $N$ play the roles of time variables in an integrable system -- a discrete time variable in the case of $N$.

\subsubsection{Dependence on the times $t_k$}

Let us show that  $\Psi_N(x)$ obeys a matrix differential equation of the type 
\begin{align}
 \boxed{ \frac{\partial}{\partial t_k} \Psi_N
  = H_{N,k} \Psi_N  }\ ,
\label{eq:dtk}
\end{align}
where the dependence of $H_{N, k}$ on $x$ is polynomial of degree $k$. Due to $\partial_{t_k} V = \frac{1}{k} x^k$ and the Heine formula \eqref{Heine_formula}, $\partial_{t_k} p_N$ is a polynomial of degree at most $N-1$. Decomposing $\partial_{t_k} p_N$ on the basis of orthogonal polynomials, we obtain
\begin{equation}
\frac{\partial \psi_N(x)}{\partial t_k} = \sum_{j=0}^{N+k} \left(\mathcal{H}_{k}\right)_{N,j}\psi_j(x)\ ,
\end{equation}
where $\mathcal{H}_k$ is an infinite, $x$-independent matrix such that
and $\mathcal{H}_k + \frac{Q^k}{2k}$ is strictly lower triangular. Since $\partial_{t_k} \int_\gamma \psi_i\psi_j =0$, $\mathcal{H}_k$ must be antisymmetric, which implies
\begin{equation}
\mathcal{H}_k = -\frac{1}{2k}\left((Q^k)_+ - (Q^k)_-\right)\ .
\end{equation}
As in our derivation of the Lax matrix \eqref{formulaDN}, we obtain $H_{N,k}$ by multiplying with the folding matrix,
$
H_{N,k}(x) = \left(\mathcal{H}_k F^N(x)\right)_{[N-1,N]^2} 
$,
and we find
\begin{equation}
\boxed{ H_{N,k}(x) =  -\frac{x^k}{2k} \sigma_3 + 
  \left( \frac{x^k-Q^k}{x-Q} A^N \right)_{[N-1,N]^2} }
\ . 
\end{equation}
So $H_{N, k}(x)$ is a polynomial of $x$ of degree $k$. In particular, $H_{N, k}(x)$ is analytic across $\gamma_i$, which confirms that the time $t_k$ is isomonodromic.

Now the invertible matrix $\Psi_N$ is annihilated by the matrix differential operators $\frac{\partial}{\partial x} - D_N$ and $\frac{\partial}{\partial t_k} - H_{N,k}$, which must therefore commute with one another:
\begin{subequations}
\begin{align}
  \left[
  \frac{\partial}{\partial t_k} - H_{N,k}  ,
  \frac{\partial}{\partial t_j} - H_{N,j} 
 \right] & = \frac{\partial}{\partial t_j} H_{N,k}
 - \frac{\partial}{\partial t_k} H_{N,j}
 -
 \left[ H_{N,j} , H_{N,k} \right] = 0 \ ,
 \label{flatness}
 \\
 \left[
  \frac{\partial}{\partial x} - D_N ,
  \frac{\partial}{\partial t_k} - H_{N,k}
 \right]
 &=
 \frac{\partial}{\partial t_k} D_N
 - \frac{\partial}{\partial x} H_{N,k} 
 +  \left[ D_N, H_{N,k} \right] = 0\ .
 \label{eq:dhk}
\end{align}
\end{subequations}
In other words, $\Psi_N$ is a flat section of the following flat
connection on a $GL(2,\mathbb C)$ bundle over the space $\mathbb C^{d+1}$ with coordinates $x$ and $t_k$, 
\begin{align}
 \nabla = d - D_N dx - \sum_{k} H_{N,k} dt_k
 \ ,
\end{align}
where $d$ is the exterior derivative.
Eq.~\eqref{eq:dhk} can alternatively be reformulated as a Lax equation 
\begin{align}
 \frac{\partial}{\partial t_k} L
 = [H_{N,k}, L]  \qquad \text{with} \qquad L = \frac{\partial}{\partial x} - D_N\ ,
\end{align}
which justifies the name of the Lax matrix $D_N$.

\subsubsection{Dependence on the discrete time $N$}

The three-term recursion relations \eqref{eq:mpsi} and \eqref{eq:mphi} for $\psi_N$ and $\varphi_N$, can be rewritten as a matrix recursion relation for $\Psi_N$,  
\begin{align}
 \boxed{ \Psi_{N+1} = \mathcal R_N \Psi_N 
 \quad \text{with} \quad
\mathcal  R_N =
 \left(
  \begin{array}{cc}
   0 & 1 \\ - \frac{\gamma_N}{\gamma_{N+1}} & \frac{x-S_N}{\gamma_{N+1}}
  \end{array}
 \right) }
 \ .
\label{recPsiRN}
\end{align}
The matrix $\mathcal R_N$ is a polynomial of $x$ of degree one, and plays the same role for the discrete time $N$ as the matrices $D_N$ and $H_{N,k}$ for the variables $x$ and $t_k$ respectively. 
The compatibility of the discrete time evolution with the dependences on $x$ and $t_k$ leads to difference equations for $D_N$ and $H_{N,k}$,
\begin{subequations} 
\begin{align}
 D_{N+1}
 & = 
  \mathcal R_N \left( D_N + \mathcal R_N^{-1}\partial_x \mathcal R_N  \right) \mathcal R_N^{-1} 
 \ , 
 \label{eq:prp}
 \\
 H_{N+1,k}
 & =
 \mathcal R_N \left( H_{N,k} + \mathcal R_{N}^{-1} \partial_{t_k} \mathcal R_N\right) \mathcal R_{N}^{-1}
 \ .
 \label{eq:HR_conpatibility}
\end{align}
\end{subequations}
These equations can be interpreted as flatness conditions involving differences
in addition to differentials.
Due to the presence of the terms $\mathcal R_N^{-1}\partial_x \mathcal R_N$ and $\mathcal R_N^{-1}\partial_{t_k} \mathcal R_N$, $D_{N+1}$ and $H_{N+1,k}$ are gauge transformations of $D_N$ and $H_{N,k}$ respectively.

\subsection{Tau function and Baker--Akhiezer function}\label{sec:tau-func}

\subsubsection{Tau function}

Given an isomonodromic matrix differential equation \eqref{Lax_eq} with an analytic Lax matrix $D_N$, and a solution $\Psi_N$ whose asymptotic behaviour at infinity is governed by a function $T_N$ as in Eq.~\eqref{eq:psi_infty}, there exists a \textbf{Tau function}\index{Tau function} $\tau_N(t_1,\dots,t_{d+1})$ of the isomonodromic times $t_k$ such that
\begin{align}
 \boxed{ \frac{\partial}{\partial t_k} \log \tau_N
  =
-\underset{x=\infty}{\operatorname{Res}} \, dx \operatorname{Tr}
 \Psi_N^{-1}(x) \Psi_N'(x) \frac{\partial}{\partial t_k} T_N(x) }
 \ .
 \label{tau_rel}
\end{align}
The existence of the Tau function was established by Jimbo--Miwa~\cite{Jimbo:1983if} and
Ueno--Takasaki~\cite{Ueno:1984zz}, by showing that its definition is compatible with $\partial_{t_k}\partial_{t_j} \log\tau_N = \partial_{t_j}\partial_{t_k} \log\tau_N$, using the flatness conditions \eqref{flatness}.

Tau functions are fundamental objects in integrable systems, analogous to partition functions in statistical physics \cite{Babelon:2003CIS}. We will now prove that the Tau function of our matrix differential equation actually coincides with the partition function of our matrix model, 
\begin{align}
\boxed{  \tau_N \propto \mathcal{Z}_N 
= \int_{H_N(\gamma)} dM\, e^{-\operatorname{Tr} V(M)} } \ ,
\end{align}
where the proportionality coefficient is independent of the isomonodromic times $t_k$, but may depend on the non-isomonodromic variables: the matrix size $N$ and the path $\gamma=\sum_j c_j \gamma_j$. We first use $\Psi_N =\hat\Psi_N e^{T_N}$ \eqref{eq:psi_infty}. Using $[T_N,\frac{\partial}{\partial t_k} T_N]=0$, this leads to
\begin{align}
\frac{\partial}{\partial t_k} \log \tau_N
  =
-\underset{x=\infty}{\operatorname{Res}} \, dx \operatorname{Tr}\left[
\hat \Psi_N^{-1}(x) \hat\Psi_N'(x)   \frac{\partial}{\partial t_k} T_N(x) -
  T'_N(x)  \frac{\partial}{\partial t_k} T_N(x)\right] \ ,
\end{align}
where the contribution of the last term vanishes because $T'_N(x)  \frac{\partial}{\partial t_k} T_N(x) $ is a polynomial.
Then, using the identity $\frac{\partial}{\partial t_k} T_N(x) = \frac{x^k}{kV'(x)} \left(T'_N(x)-\frac{N}{x}\sigma_3\right)$,
we rewrite
\begin{align}
 \frac{\partial}{\partial t_k} \log \tau_N = - \underset{x=\infty}{\operatorname{Res}} \, dx \frac{x^k}{kV'(x)} \operatorname{Tr}  \hat\Psi_N^{-1}(x) \hat\Psi'_N(x) \left(T'_N(x)-\frac{N}{x}\sigma_3\right) \ .
\end{align}
In order to compute this, we will use the identity
\begin{align}
 \operatorname{Tr} D_N^2 = \operatorname{Tr} \left(\Psi_N^{-1}\Psi'_N\right)^2
= \operatorname{Tr}\left( \left(\hat\Psi_N^{-1}\hat\Psi_N'\right)^2 + (T'_N)^2 + 2 \hat\Psi_N^{-1} \hat\Psi'_N T_N' \right)\ .
\end{align}
Moreover, using respectively $\hat\Psi_N^{-1} \hat\Psi'_N(x) \underset{x\to\infty}{=} O(\frac{1}{x^2})$ and $T'_N(x)=\left(\frac{N}{x}-\frac12 V'(x)\right)\sigma_3$, we obtain
\begin{subequations}
\begin{align}
 \underset{x=\infty}{\operatorname{Res}} \, dx \frac{x^k}{kV'(x)}  \left(\hat\Psi_N^{-1} \hat\Psi'_N(x)\right)^2 &=0\ ,
 \\
 \underset{x=\infty}{\operatorname{Res}} \, dx \frac{x^k}{kV'(x)} \operatorname{Tr} (T'_N(x))^2
 &= \underset{x=\infty}{\operatorname{Res}} \, dx \frac{x^k}{kV'(x)}  \frac{2N^2}{x^2}\ .
\end{align}
\end{subequations}
This leads to 
\begin{align}
 \frac{\partial}{\partial t_k} \log \tau_N =  \underset{x=\infty}{\operatorname{Res}} \, dx \frac{x^k}{kV'(x)}  \left(-\frac12\operatorname{Tr} D_N(x)^2 + \frac{N^2}{x^2} \right)\ .
\end{align}
According to Eq.~\eqref{dn_inv}, we have 
\begin{align}
 -\frac12 \operatorname{Tr} D_N(x)^2 = \det D_N(x) = \left<
     \operatorname{Tr} \frac{V'(x) - V'(M)}{x-M}
    \right>_{N\times N} -\frac14 V'(x)^2\ .
\end{align}
The term $-\frac14 V'(x)^2$ does not contribute to our residue.
The term $\left<\operatorname{Tr} \frac{V'(x)}{x-M}\right>_{N\times N}$ yields the contribution 
\begin{align}
 \underset{x=\infty}{\operatorname{Res}} \, dx \frac{x^k}{k} 
\left<\operatorname{Tr} \frac{1}{x-M} \right>_{N\times N} = -\frac{1}{k} \left<\operatorname{Tr} M^k \right>_{N\times N} 
= \frac{\partial}{\partial t_k} \mathcal\log \mathcal{Z}_N   \ . 
\end{align}
Moreover, using the equations \eqref{eq:trf} and \eqref{eq:recurPQ}, we have
\begin{subequations}
 \begin{align}
 \Big< \operatorname{Tr} V'(M)\Big>_{N\times N} &=  \operatorname{Tr} V'(Q)\Pi_N = 0\ ,
 \\
\Big< \operatorname{Tr} MV'(M)\Big>_{N\times N} &= \operatorname{Tr} QV'(Q)\Pi_N =  N^2\ ,
\end{align}
\end{subequations}
which implies $\left<\operatorname{Tr}\frac{V'(M)}{x-M}\right>_{N\times N} = \frac{N^2}{x^2} + O\left(\frac{1}{x^3}\right)$.
This leads to the announced result $ 
\frac{\partial}{\partial t_k} \log \tau_N = \frac{\partial}{\partial t_k} \log \mathcal{Z}_N
$.

\subsubsection{Baker--Akhiezer functions} 

Let us express the orthogonal polynomials in terms of the Tau function, in order to find their interpretation in the associated isomonodromic integrable system. 
According to Heine's formula, we have 
\begin{multline}
 p_N(x)
= \Big< \det (x-M) \Big>_{N\times N} 
=
 \frac{1}{\mathcal{Z}_N}
 \int_{H_N(\gamma)} dM \,
 e^{\operatorname{Tr}\log(x-M) - \operatorname{Tr}V(M)}
 \ ,
\\
=
\frac{1}{\mathcal{Z}_N}
 \int_{H_N(\gamma)} dM \,
 e^{N\log x - \operatorname{Tr}\sum_{k=1}^\infty \left(t_k + \frac{1}{x^k}\right) \frac{M^k}{k} } \ .
\end{multline}
This matrix integral can be interpreted as the value of our partition function, after performing an $x$-dependent shift of the coefficients $t_k$ of the potential. 
In terms of the Tau function $\tau_N(\vec{t} = (t_1, t_2,\cdots))$, we therefore have 
\begin{align}
 \boxed{ p_N(x)  =
 x^N 
 \frac{\tau_N(\vec{t}+[x])}{\tau_N(\vec{t})} }
 \ ,
\end{align}
where we introduced Sato's notation for the infinite vector of times
\begin{align}
 [x] = \left(  \frac{1}{x},  \frac{1}{x^2}, \frac{1}{x^3}, \ldots \right)
 \ .
\end{align}
This formula for $p_N(x)$ coincides with the 
 \textbf{Sato formula}\index{Sato formula} which expresses the 
\textbf{Baker--Akhiezer function}\index{Baker--Akhiezer function} of an integrable system in terms of its Tau function.  
Therefore, the orthogonal polynomial $p_N(x)$ is the Baker--Akhiezer function of the associated integrable system.
Similarly, the Heine-type formula \eqref{eq:Heinephi} for the Hilbert transform $\varphi_{N-1} $ leads to 
\begin{align}
\varphi_{N-1}(x)\frac{e^{-\frac12 V(x)}}{\sqrt{h_{N-1}}} 
= x^{-N}  
 \frac{\tau_N(\vec{t}-[x])}{\tau_N(\vec{t})}
 \ ,
\end{align} 
which coincides with the Sato formula for the \textbf{dual
Baker--Akhiezer function}\index{Baker--Akhiezer function!dual---}.
And Heine-type formulas for the matrix kernel allow it to be expressed in terms of the Tau function. 
For example, Eq.~\eqref{eq:Khatratiodet} for the matrix kernel element $\hat K^N_{1,1}$ leads to
\begin{align}
e^{\frac12 V(x)} e^{-\frac12 V(y)}\hat K^N_{1,1}(y,x)
  = 
 \frac{1}{y-x} 
 \frac{\tau_N(\vec{t}+[x]-[y])}{\tau_N(\vec{t})}
 \ .
 \label{ch_poly_ratio}
\end{align}

\subsubsection{Hirota equation}

Orthogonality of orthogonal polynomials leads to bilinear equations for Tau functions, that we will now derive. 
Let us consider two families of times $(\vec t - \vec u , \vec t + \vec u)$, the associated potentials $(V^-, V^+)$, and the corresponding families of orthogonal polynomials $(p_n^-, p_n^+)$. We find the following equalities of formal series in $\vec u$,
\begin{subequations}
\begin{align}
 \forall m\leq n, \quad h_m^-\delta_{n,m} 
 &= \int_\gamma dx'\, p_n^-(x') p_m^+(x') e^{-V^-(x')} \ ,
 \\
 &= \int_\gamma dx'\, \mathop{\operatorname{Res}}_{x= x'} \frac{dx}{x-x'}   
e^{V^+(x) -  V^-(x)} p_n^-(x) p_m^+(x')  e^{- V^+(x')} \ ,
\\
&= -h_m^+ \mathop{\operatorname{Res}}_{x= \infty}\, dx e^{V^+(x) -  V^-(x)} p_n^-(x) \varphi_m^+(x) \frac{e^{-\frac12 V^+(x)}}{\sqrt{h_m^+}}\ .
\end{align}
\end{subequations}
Using the expressions of $p_n^-$ and $\varphi^+_m$ in terms of Tau functions, we obtain
the \textbf{Hirota equation}\index{Hirota equation}
\begin{align}
 \hspace{-8mm} \boxed{\delta_{n,m}\tau_{n}(\vec t+\vec u)\tau_{m+1}(\vec t-\vec u)
\underset{m\leq n}{=}
 -\mathop{\operatorname{Res}}_{x= \infty} \frac{dx}{x^{m+1-n}}
 e^{2\sum_k \frac{u_k x^k}{k}}
 %e^{\sum_k x^{-k} \frac{\partial}{\partial u_k}}
 \tau_n(\vec t-\vec u+[x]) \tau_{m+1}(\vec t+\vec u-[x])}\, .
\end{align}
These are equalities of formal series in powers of $\vec u$, whose coefficients form an infinite hierarchy of quadratic partial differential equations for the Tau functions, called the \textbf{KP hierarchy}\index{KP hierarchy}.
In order to facilitate the extraction of these coefficients, let us rewrite the Hirota equation in terms of the 
Hirota operator $D_{t_k} = (\overset{\leftarrow}{\partial}_{t_k}- \overset{\rightarrow}{\partial}_{t_k})$, 
\begin{multline}
\delta_{n,m}
\tau_{m+1}(\vec t) 
e^{-\sum_k u_k D_{t_k}}
\tau_{n}(\vec t)
\\ \underset{m\leq n}{=} \tau_{m+1}(\vec t)
 \left( -\mathop{\operatorname{Res}}_{x= \infty} \frac{dx}{x^{m+1-n}}
 e^{\sum_k u_k (D_{t_k}+\frac{2x^k}{k})} e^{-\sum_k x^{-k} D_{t_k}} \right)
 \tau_n(\vec t) \ .
\end{multline}
Then the coefficient of the monomial $\prod_k \frac{u_k^{\mu_k}}{\mu_k!}$ provides the equation 
\begin{multline}
\tau_{m+1}(\vec t) 
\left(\prod_k (-D_{t_k})^{\mu_k}\right)
\tau_{n}(\vec t)
\\ \underset{m\leq n}{=} \tau_{m+1}(\vec t) \left(-\mathop{\operatorname{Res}}_{x=\infty} \frac{dx}{x^{m+1-n}}\ e^{-\sum_k x^{-k} D_{t_k}}  \ \prod_k \left(D_{t_k}+\frac{2 x^k}{k}\right)^{\mu_k}\right)
 \tau_n(\vec t)\ .
\end{multline}
Let us focus on the case $n=m+1$. Then not only the left-hand side vanishes, but also products of odd numbers of Hirota operators vanish, $ \tau_n \left(\prod_{i=1}^{2m+1} D_{t_{k_i}}\right) \tau_n=0$.
As an example, the coefficient of $u_1^3$ is the equation
\begin{equation}\label{eq:KPhirota}
0 = \tau_n  \left( \frac{1}{3} D_{t_1}^4 + 4 D_{t_2}^2 -4  D_{t_1} D_{t_3} \right)  \tau_n\ .
\end{equation}
(We would find the same equation from the coefficients of $u_1u_2$ and of $u_3$.) 
\iffalse
\begin{equation}
0 = \tau_n  \left( \frac{1}{3} D_{t_1}^4 + 2 D_{t_1}^2 D_{t_2} + 4 D_{t_2}^2 -4  D_{t_1} D_{t_3} - 8 D_{t_4}\right)  \tau_n
\end{equation}
\fi
This can be rewritten as the KP equation for $u=2\partial_{t_1}^2 \log \tau$,
\begin{equation}
0 =  \partial_{t_2}^2 u + \partial_{t_1} \left(\frac{1}{12} \partial_{t_1}^3 u - \frac12 u \partial_{t_1} u -  \partial_{t_3} u\right)\ .
\end{equation}
Hirota equations, sometimes called bilinear equations, are ubiquitous in integrable systems, to the extent that integrability can be defined by their existence. Our Hirota equation comes from a one-matrix model, with a Lax matrix of size two: as a result, it involves residues at $x=\infty$ only, and one family of times $\vec t$. In more general integrable systems, Hirota equations can involve several families of times, and residues at several singular points \cite{Babelon:2003CIS}.

\subsubsection{Determinantal formulas}

In the language of Tau functions, the agreement between the Christoffel--Darboux formula \eqref{CD_kernel} and the Heine-type formula \eqref{ch_poly_ratio} for the matrix kernel becomes
\begin{equation}
\tau_N(\vec t)\tau_N(\vec t+[x]-[y])
= \tau_{N-1}(\vec t+[x])\tau_{N+1}(\vec t-[y])-\tau_N(\vec t+[x])\tau_N(\vec t-[y]) \ .
\end{equation}
Moreover, the identity $\hat K^{N+1}_{1,1}(x,y)-\hat K^{N}_{1,1}(x,y)=\psi_N(x)\varphi_N(y)$ (which follows from the definition \eqref{eq:knoo}) becomes
\begin{equation}
\tau_N(\vec t)\tau_{N+1}(\vec t+[x]-[y])
- \tau_{N+1}(\vec t)\tau_{N}(\vec t+[x]-[y])
= (y-x)\tau_N(\vec t+[x])\tau_{N+1}(\vec t-[y]) \ .
\end{equation}
Furthermore, as shown in \autoref{exoproofdetformula}, the Tau function $\tau_N$ obeys the quadratic equation
\begin{multline}
\frac{(x_1-x_2)(y_1-y_2)}{(x_1-y_1)(x_1-y_2)(x_2-y_1)(x_2-y_2)}\tau_N(\vec t)\tau_N(\vec t +[x_1]+[x_2]-[y_1]-[y_2]) \\
= \frac{\tau_N(\vec t+[x_2]-[y_1])\tau_N(\vec t+[x_1]-[y_2])}{(x_2-y_1)(x_1-y_2)} - \frac{\tau_N(\vec t+[x_1]-[y_1])\tau_N(\vec t+[x_2]-[y_2])}{(x_1-y_1)(x_2-y_2)} \ .
\label{eq:qtau}
\end{multline}
We call this a determinantal formula, as the right-hand side is the determinant of a matrix of size two. 
Actually, iterating this formula leads to the more general \textbf{determinantal formula} \index{determinantal formulas!for Tau functions}
\begin{align}
 \boxed{ \frac{\tau_N(\vec t+\sum_{i=1}^k [x_i]-\sum_{i=1}^k [y_i])}{\tau_N(\vec t)} 
 = \frac{ \mathop{{\det}}_{1 \le i, j \le k} \left(\frac{\tau_N(\vec t+[x_i]-[y_j])}{(x_i-y_j) \ \tau_N(\vec t)} \right) }{ \det_{1 \le i, j \le k} \left( \frac{1}{x_i - y_j} \right) } } \ ,
 \label{eq:tau_det}
\end{align}
whose right-hand side involves the \textbf{Cauchy
determinant}\index{Cauchy determinant},
\begin{align}
 \det_{1 \le i, j \le k} \left( \frac{1}{x_i - y_j} \right)
 & =
 \prod_{i<j}^k (x_i - x_j) (y_j - y_i)
 \prod_{i,j}^k (x_i - y_j)^{-1}
 \, .
 \label{Cauchy_det}
\end{align}
The determinantal formula for the Tau function is a generalization of Fay's trisecant identity for Theta functions. 
Theta functions are indeed particular Tau functions of certain integrable systems that are called isospectral or finite gap integrable systems \cite{Babelon:2003CIS}.

\subsection{Correlation functions}

\subsubsection{Determinantal formulas}

Let us see how correlation functions can be computed in terms of our integrable system. 
Using Heine-type formulas, our orthogonal polynomials and matrix kernels can be written in terms of insertions of $\det(x-M)$ in matrix integrals, and therefore in terms of wave functions. We already know the expression of wave functions in terms of correlation functions \eqref{eq:psid}. Conversely, correlation functions involve insertions of $\operatorname{Tr} \frac{1}{x-M}$, and can be expressed in terms of wave functions using 
\begin{equation}
\operatorname{Tr} \frac{1}{x-M} = \lim_{y\to x} \frac{1}{x-y}\left( \frac{\det (x-M)}{\det (y-M)} -1 \right)
 \, .
\end{equation}
In the case of the correlation function $W_1$, using the Heine-type formula \eqref{eq:Khatratiodet}, we obtain
\begin{align}
W_1(x) 
&= \lim_{y\to x}\left( e^{\frac12 V(x)-\frac12 V(y)} \hat K^N_{1,1}(y,x) - \frac{1}{x-y} \right)
\ .
\end{align}
Using the expression \eqref{eq:khpp} for the matrix kernel, we compute the limit and find
$
W_1(x) 
= \frac{V'(x)}{2}+\left(\Psi_N^{-1}(x)\Psi_N'(x)\right)_{1,1}
$.
Using moreover the matrix differential equation \eqref{Lax_eq} for $\Psi_N$, this leads to 
\begin{align}
\boxed{ W_1(x) = 
 \frac{V'(x)}{2}+\operatorname{Tr} D_N(x) M_N(x) } \ ,
\end{align}
where we introduce the projector
\begin{equation}
\boxed{ M_N  = \Psi_N \begin{pmatrix} 1 & 0 \\ 0 & 0 \end{pmatrix} \Psi_N^{-1} }\ ,
\end{equation}
which satisfies
\begin{align} 
\left\{\begin{array}{l} M_N^2 = M_N\ , \\ \operatorname{Tr} M_N = 1 \ , \\ \det M_N = 0 \ , \end{array} \right. \quad \text{and} \quad \frac{d}{dx} M_N = [D_N, M_N]\ .
\end{align}
In geometric terms, the projector $M_N(x)$ is a flat section of the adjoint bundle of the principal bundle $\mathbb{C}\times GL(2,\mathbb C)$, with the connection $d - [D_N(x)dx , \cdot \,]$.

Then let us compute the disconnected two-point function, starting with its expression in terms of insertions of determinants,
\begin{align}
\widehat{W}_2(x,x')  
& = \lim_{\substack{y\to x \\  y'\to x'}} \frac{1}{(x-y)(x'-y')}\left< \left(\frac{\det(x-M)}{\det(y-M)}-1\right)\left(\frac{\det(x'-M)}{\det(y'-M)}-1\right) \right>  
\ .
\end{align}
Writing this in terms of Tau functions, and using the quadratic equation \eqref{eq:qtau}, we obtain 
\begin{multline}
\widehat{W}_2(x,x')  
=  \lim_{\substack{y\to x \\  y'\to x'}} \Bigg\{
 \frac{(x-y')(x'-y)}{(x-x')(y'-y)} e^{\frac{V(x)+V(x')}{2}-\frac{V(y)+V(y')}{2}}
 \\ \hspace{3.5cm}
 \times \Big( \hat K^N_{1,1}(x,y)  \hat K^N_{1,1}(x',y') - \hat
 K^N_{1,1}(x,y') \hat K^N_{1,1}(x',y) \Big)
 \\
 - e^{\frac12 V(x)-\frac12 V(y)}\frac{\hat K^N_{1,1}(x,y)}{(x'-y')} - e^{\frac12 V(x')-\frac12 V(y')}\frac{\hat K^N_{1,1}(x',y')}{(x-y)} + \frac{1}{(x-y)(x'-y')} \Bigg\}\ .
 \end{multline}
 Performing the limits, we obtain
 \begin{align}
 \widehat{W}_2(x,x')  = W_1(x)W_1(x') - \hat K^N_{1,1}(x,x') \hat K^N_{1,1}(x',x)  -\frac{1}{(x-x')^2}\ .
\end{align}
Equivalently, the connected two-point function is $W_2(x,x') = - \hat K^N_{1,1}(x,x') \hat K^N_{1,1}(x',x)  -\frac{1}{(x-x')^2}$, which can be rewritten in terms of the projector $M_N$ as 
\begin{equation}
\boxed{ W_2(x,x') =  \frac{\operatorname{Tr} M_N(x) M_N(x')}{(x-x')^2} -\frac{1}{(x-x')^2} } \ .
\label{eq:w2mn}
\end{equation}
Similarly, writing the correlation function $W_{n\geq 3}$ in terms of insertions of determinants, using the determinantal formula \eqref{eq:tau_det} for the corresponding Tau functions, and writing the result in terms of the projector $M_N$, we obtain the \textbf{determinantal formula} \index{determinantal formulas!for correlation functions}
\begin{align}
 \boxed{ W_{n\geq 3}(x_1,\cdots x_n)
  = \frac{1}{n}
 \sum_{\sigma \in \mathfrak{S}_n} (-1)^\sigma
 \frac{\operatorname{Tr} \prod_{i=1}^n M_N(x_{\sigma(i)})}
      {\prod_{i=1}^n (x_{\sigma(i)} - x_{\sigma(i+1)})} } \ .
\end{align}

\subsubsection{Loop equations}

Let us check that our formulas for the correlation functions are compatible with the loop equations. 
From Eq.~\eqref{eq:w2mn} and the properties of $M_N$, we see that $W_2(x,x')$ has the finite limit
\begin{align}
W_2(x,x) &=  \frac{1}{2} \operatorname{Tr} M_N(x)M_N''(x) =   \operatorname{Tr} \Big( M_N^2D_N^2 - (M_ND_N)^2 \Big)(x) \ .
\end{align}
Now, using the characteristic equation of size two matrices $A^2 = A\operatorname{Tr} A - \operatorname{Id} \det A$, we find $\operatorname{Tr}M_N^2 D_N^2 = -\det D_N$ and $\operatorname{Tr} (M_ND_N)^2 = \left(\operatorname{Tr} M_ND_N\right)^2$, therefore
\begin{align}
W_2(x,x) =  - \det D_N(x)  -\left(W_1(x)-\frac{V'(x)}{2}\right)^2 \ .
\end{align}
Using the formula \eqref{dn_inv} for $\det D_N(x)$, this can be rewritten as
\begin{equation}
W_1(x)^2+W_2(x,x) =  V'(x) W_1(x) - P_N(x)\ .
\end{equation}
This agrees with the first loop equation \eqref{eq:firstloop}, up to the change of normalization $V\to NV$ of the potential. Higher loop equations can similarly be recovered by this method.

\section{Multi-matrix models}\label{sec:multi}

As we will see in \autoref{sec:multi_angular}, the partition function of a multi-matrix model has an eigenvalue representation of the type
\begin{empheq}[marginbox = \fbox]{multline}
\label{eq:Z2matrixDelta}
 \mathcal{Z}_N= \frac{1}{(N!)^2} \int_{\gamma^N\times \tilde\gamma^N} d\lambda_0\cdots d\lambda_{N-1}d\tilde{\lambda}_0\cdots d\tilde{\lambda}_{N-1}
\\
 \Delta(\lambda_0, \cdots \lambda_{N-1})\Delta(\tilde{\lambda}_0, \cdots \tilde{\lambda}_{N-1}) 
 \underset{i,j=0,\cdots N-1}{\det} \omega(\tilde{\lambda}_j,\lambda_i)\ .
\end{empheq}
In the case of a two-matrix model, the function $\omega(\tilde\lambda,\lambda)$ 
is given in terms of the potentials $V_1,V_2$ by 
\begin{align}
 \omega(\tilde{\lambda},\lambda) = e^{-V_1(\lambda)}e^{-V_2(\tilde{\lambda})} e^{\lambda\tilde{\lambda}}\ .
 \label{eq:omega2mm}
\end{align}
In the more general case of a matrix chain of length $L$, this function is
\begin{align}
 \omega(\mu_L,\mu_1) =  \int_{\gamma^{L-2}} d\mu_2  d\mu_3 \cdots d\mu_{L-1}  \prod_{i=1}^L e^{-V_i(\mu_i)} \prod_{i=1}^{L-1} e^{\mu_i \mu_{i+1} } \ .
 \label{eq:omegamulti}
\end{align}

\subsection{Determinantal formulas}\label{sec:multi_det}

The determinantal formula \eqref{eq:zdeth} for the partition function still holds in multi-matrix models, provided the matrix elements $H_{i,j}=\left<\tilde p_i\middle|p_j\right>$ are computed from the scalar product
\begin{align}
 \boxed{ \left< f\middle| g \right> 
 = \int_{\gamma\times \tilde\gamma} d\lambda d\tilde{\lambda}\ f(\tilde\lambda)g({\lambda})\omega(\tilde{\lambda},\lambda) 
 = f\operatorname{\tilde{\star}} \omega \star g } \ .
\end{align}
Let us now compute the joint eigenvalue distributions
\begin{multline}
 R_{k,k'}(\lambda_0, \cdots \lambda_{k-1},\tilde \lambda_0, \cdots \tilde \lambda_{k'-1}) = 
  \frac{1}{\mathcal{Z}_N} \frac{1}{N!^2}\int_{\gamma^{N-k}\times {\tilde\gamma}^{N-k'}} d\lambda_k \cdots d\lambda_{N-1}d{\tilde\lambda}_{k'} \cdots d{\tilde\lambda}_{N-1}
\\  
 \Delta(\lambda_0, \cdots \lambda_{N-1})
 \Delta(\tilde\lambda_0, \cdots \tilde\lambda_{N-1}) 
 \det_{i,j=0,\ldots N-1}\omega(\lambda_i,\tilde\lambda_j)\ .
\end{multline}
For this purpose we introduce the polynomial kernel 
\begin{align}
 \boxed{ K(\lambda, \tilde{\lambda}) = 
 \sum_{k=0}^{N-1} \sum_{k'=0}^{N-1} p_k(\lambda) \left(H^{-1}\right)_{k,k'} 
 \tilde{p}_{k'}(\tilde{\lambda}) } \ ,
\label{eq:kllbio}
\end{align}
and we construct three other kernels by convoluting with $\omega$,
\begin{align}
 \boxed{H = K \operatorname{\tilde\star} \omega \quad , \quad \tilde{H} = \omega \star K \quad , \quad J = - \omega + \omega \star K \operatorname{\tilde\star} \omega} \ .
\end{align}
These four kernels form a self-reproducing family, in the sense that 
\begin{subequations}
\begin{align}
 & H \star H = H \ \ , \ \
 H \star K = K \ \ , \ \
 \tilde{H} \operatorname{\tilde\star} \tilde{H} = \tilde{H}\ \ , \ \
 K \operatorname{\tilde\star} \tilde{H} = K \ ,
 \\ 
 & J \star H = J \star K = \tilde{H} \operatorname{\tilde\star} J = K  \operatorname{\tilde\star} J = 0\ .
\end{align}
\end{subequations}
In addition, we have 
\begin{equation}
\int d\lambda \ H(\lambda,\lambda)
= \int d\tilde\lambda\ \tilde H(\tilde\lambda,\tilde\lambda)
= N \ .
\end{equation}
Dyson's formula for the joint eigenvalue distributions, and its proof, can be generalized, and we obtain the \textbf{Eynard--Mehta theorem}\index{Eynard--Mehta theorem} \cite{Eynard:1998JPA,Mehta:2004RMT}, 
\begin{equation}
\boxed{ R_{k,k'} = \frac{(N-k)!(N-k')!}{N!^2}
\det_{\substack{i,j=0,\cdots k-1 \\ i',j'=0, \cdots k'-1}}
\begin{pmatrix}
H(\lambda_i,\lambda_j) & K(\lambda_i,\tilde\lambda_{j'}) \\
J(\tilde\lambda_{i'},\lambda_j) & \tilde H(\tilde \lambda_{i'},\tilde\lambda_{j'})
\end{pmatrix} } \,  ,
\end{equation}
which involves the determinant of a matrix of size $k+k'$.
As special cases, we have the total joint  distribution
\begin{subequations}
\begin{align}
N!^2 R_{N,N} 
&= \det
\begin{pmatrix}
H & K \\
J & \tilde H
\end{pmatrix} = \det
\begin{pmatrix}
K\operatorname{\tilde\star} \omega & K \\
- \omega + \tilde H \operatorname{\tilde\star}  \omega  & \tilde H
\end{pmatrix}
= \det
\begin{pmatrix}
0 & K \\
- \omega  & \tilde H
\end{pmatrix} \ , \\
&= 
\det_{i,j=0, \cdots N-1} K(\lambda_i,\tilde \lambda_j)\det_{i,j=0, \cdots N-1} \omega(\lambda_i,\tilde\lambda_j)\ ,
\end{align}
\end{subequations}
and the joint distributions of $\lambda$ and $\tilde\lambda$, respectively,
\begin{align}
R_{k,0} = \frac{(N-k)!}{N!}
\det_{i,j=0, \cdots k-1} H(\lambda_i,\lambda_j)\ \ ,  \ \
R_{0,k} = \frac{(N-k)!}{N!}
\det_{i,j=0, \cdots k-1} \tilde H(\tilde\lambda_i,\tilde\lambda_j)\, .
\end{align}
In particular, the densities of eigenvalues are
\begin{align}
R_{1,0}(\lambda) = \frac{1}{N} H(\lambda,\lambda)\ \ , \ \
R_{0,1}(\tilde \lambda) = \frac{1}{N} \tilde H(\tilde \lambda,\tilde \lambda)\ . 
\end{align}

\subsection{Biorthogonal polynomials}

Let \textbf{biorthogonal polynomials}\index{biorthogonal polynomials} be monic polynomials $p_k,\tilde{p}_k$ such that $\deg p_k = \deg\tilde{p}_k=k$ and 
\begin{align}
 \boxed{ \left<\tilde{p}_k\middle| p_{k'} \right> = h_k \delta_{k,k'} } \ .
 \label{tpkpk}
\end{align}
We also introduce the biorthonormal polynomials
\begin{align}
 \psi_k = \frac{p_k}{\sqrt{h_k}} \quad , \quad \tilde\psi_k = \frac{\tilde{p}_k}{\sqrt{h_k}}\ .
\end{align}
As in the case of orthogonal polynomials, the partition function is then a product of the coefficients $h_k$ \eqref{eq:zph}. 
Biorthogonal polynomials exist if and only if $\forall N \ \mathcal{Z}_N\neq 0$, and for a given function $\omega(\tilde\lambda,\lambda)$ this condition is obeyed almost everywhere in the homology space of integration contours.
Let us now see how the biorthogonal polynomials can be computed. 

\subsubsection{Generalized Heine's formula}

In our multi-matrix model, let us denote the expectation value of a function of $\lambda = (\lambda_0, \cdots \lambda_{N-1})$ and $\tilde{\lambda} = (\tilde{\lambda}_0,\cdots \tilde{\lambda}_{N-1})$ as
\begin{multline}
 \Big< f(\lambda, \tilde{\lambda}) \Big>_{N\times N} =  \frac{1}{\mathcal{Z}_N}\frac{1}{(N!)^2} \int_{\gamma^N\times \tilde\gamma^N} d\lambda d\tilde{\lambda}\
 \Delta(\lambda)\Delta(\tilde{\lambda}) 
 \underset{i,j=0,\cdots N-1}{\det} \omega(\tilde{\lambda}_j,\lambda_i) \times f(\lambda, \tilde{\lambda})\ .
\end{multline}
The reasoning that lead to Heine's formula can be generalized to multi-matrix models, and leads to the \textbf{generalized Heine's formula}\index{Heine's formula!generalized---},
\begin{align}
p_N(x) = \left< \prod_{i=0}^{N-1} (x-\lambda_i)\right>_{N\times N}\qquad , \qquad 
\tilde p_N(y) = \left< \prod_{i=0}^{N-1} (y-\tilde{\lambda}_i)\right>_{N\times N}\ .
\end{align}
Similarly, as proved in \autoref{exproofKdetbio}, there is a Heine-type formula for the polynomial kernel
(which we now call $K_N$ rather than $K$),
%\sr{ATTENTION: problem of indices} 
\begin{equation}\label{eq:KbioHeine}
K_{N+1}(x,y) = \frac{1}{h_N} \left< \prod_{i=0}^{N-1} (x-\lambda_i) (y-\tilde \lambda_i)\right>_{N\times N}\ .
\end{equation}
And we have determinantal formulas for biorthogonal polynomials in terms of the moments $M_{k,l} = \big< \tilde\lambda^l \big| \lambda^k  \big>$,
\begin{equation}
p_k(x) = \frac{\det\begin{pmatrix}
 M_{0,0} & M_{0,1} & M_{0,2} & \dots & M_{0,k-1} & 1 \cr
 M_{1,0} & M_{1,1} & & & M_{1,k-1} & x \cr
 M_{2,0} &  & & & M_{2,k-1} & x^2 \cr
 \vdots & & & & \vdots & \vdots \cr
 M_{k,0} & & & & M_{k,k-1} & x^k
\end{pmatrix}}{\det\begin{pmatrix}
 M_{0,0} & M_{0,1} & M_{0,2} & \dots & M_{0,k-1}  \cr
 M_{1,0} & M_{1,1} & & & M_{1,k-1}  \cr
 M_{2,0} &  & & & M_{2,k-1}  \cr
 \vdots & & & & \vdots  \cr
 M_{k-1,0} & & & & M_{k-1,k-1} 
\end{pmatrix}}
\ ,
\end{equation}
and a similar formula for $\tilde p_k$ with $M^t$. %\sr{Does this mean same formula with transposed $M$?}
In contrast to the one-matrix model, the moments $M_{k,l}$ are in general not functions of $k+l$, and our determinantal formulas do not involve Hankel determinants.

\subsubsection{Recursion relations}

Since biorthogonal polynomials provide two bases of polynomials, there exist coefficients $Q_{n,m}$ and $\tilde{Q}_{n,m}$ such that 
\begin{align}\label{defQbio}
x\psi_n(x) = \sum_{m=0}^{n+1} Q_{n,m}\psi_m(x)
\qquad , \qquad 
y\tilde \psi_n(y) = \sum_{m=0}^{n+1} \tilde Q_{n,m}\tilde \psi_m(y) \ .
\end{align}
Then $Q$ and $\tilde Q$ are two semi-infinite matrices, with at most one band above the diagonal, and
\begin{equation}\label{eq:Qsupdiag}
Q_{n,n+1} = \tilde Q_{n,n+1} = \gamma_{n+1} = \sqrt{\frac{h_{n+1}}{h_n}}\ .
\end{equation}
In contrast to the matrix that appears in the recursion relation for orthogonal polynomials, the matrices $Q$ and $\tilde{Q}$ are however not necessarily symmetric. 
Moreover, derivatives of biorthogonal polynomials are again linear combinations of biorthogonal polynomials, 
\begin{align}
\psi_n'(x) = \sum_{m=0}^{n-1} P_{n,m}\psi_m(x)
\qquad , \qquad 
\tilde \psi'_n(y) = \sum_{m=0}^{n-1} \tilde P_{n,m}\tilde \psi_m(y) \ .
\end{align}
Then $P$ and $\tilde P$ are strictly lower triangular matrices, with
\begin{equation}\label{eq:Pbiosubdiag}
P_{n,n-1} = \tilde P_{n,n-1} = \frac{n}{\gamma_n}\ .
\end{equation}
By definition of biorthogonal polynomials, the coefficients $P_{n,m}$ can be computed as scalar products, explicitly
\begin{align}
 P_{n,m} = \int_{\gamma\times \tilde\gamma} dxdy\ \omega(x,y) \psi_n'(x)\tilde{\psi}_m(y)\ .
\end{align}
Integrating by parts in the variable $x$ leads to the expression of $P$ in terms of $Q$ and $\tilde{Q}$, and we find 
\begin{align}
 \boxed{ P =  - :\partial_x\log\omega(x,y)\big|_{(x,y)=(Q, \tilde Q^\text{T})}: \quad , \quad \tilde{P}^\text{T} = -  :\partial_y\log\omega(x,y)\big|_{(x,y)=(Q, \tilde Q^\text{T})}: } \ ,
 \label{eq:PfromQomega}
\end{align}
where $:f(x,y):$ is the function of the non-commuting variables $x$ and $y$ where $y$ is inserted from the left and $x$ from the right, for example $:(x+y)^2: = x^2 + 2yx + y^2$.
In many cases, for example chains of matrices, 
the requirements that $P$ and $\tilde P$ are strictly lower triangular and obey Eq.~\eqref{eq:Pbiosubdiag} gives enough constraints on $Q$ and $\tilde Q$ to determine all their coefficients.
Moreover, it follows immediately from their definitions that the matrices $Q, \tilde Q,P,\tilde P$ obey the string equations
\begin{equation}
[Q,P]=\text{Id} 
\qquad , \qquad [\tilde Q,\tilde P]=\text{Id}\ .
\end{equation}
Together with the expressions of $P, \tilde{P}$ as functions of $Q, \tilde{Q}$, this is in principle enough for determining these four infinite matrices.
% NB: Now this works in all cases, not just in many cases?
% BE: let us leave it like that. 

\subsubsection{Determinantal formulas}

As in the case of orthogonal polynomials, there are determinantal formulas for biorthogonal polynomials,
\begin{equation}
p_N(x) =\underset{N\times N\,\mathrm{ submatrix}}{\det} \left(x - Q\right)
\qquad , \qquad 
\tilde p_N(y) = \underset{N\times N\,\mathrm{ submatrix}}{\det} \left(y - \tilde Q\right)\ .
\end{equation}
and for our polynomial kernel
\begin{equation}
K_{N+1}(x,y) = \frac{1}{h_{N}} \underset{N\times N\,\mathrm{ submatrix}}{\det} (x-Q)(y-\tilde Q^\text{T})\ .
\end{equation}

\subsubsection{Examples}

\begin{itemize}

\item 
In the case of a two-matrix model \eqref{eq:omega2mm}, equations \eqref{eq:PfromQomega} reduce to
\begin{align}
P = V'(Q) - \tilde Q^\text{T} \qquad , 
\qquad
\tilde P^\text{T} = \tilde V'(\tilde Q^\text{T}) - Q \ .
\end{align}
This not only reduces the string equations to
\begin{equation}
[\tilde Q^\text{T}, Q]=\text{Id}\ ,
\end{equation}
but also implies, since $P$ is lower triangular, 
\begin{align}
Q_- &= \tilde V'(\tilde Q^\text{T})_- \qquad , \qquad \tilde Q_- = V'(Q^\text{T})_-\ .
\end{align}
We know that $Q$ and $\tilde{Q}$ have one non-vanishing band above the diagonal, and these equations now tell us what happens below the diagonal.
In particular, if $V$ is a polynomial of degree $d+1$, then $\tilde Q$ has at most $d$ bands below the diagonal
Therefore, if $V$ and $\tilde V$ are polynomials, then all the matrices $Q,\tilde Q,P,\tilde P$ are \textbf{band matrices}\index{band matrix} -- matrices with finitely many non-vanishing bands.

\item In the rational case, which we define by the condition that $\partial_x \log\omega(x,y)$ and $\partial_y \log\omega(x,y)$ are rational functions of their arguments,
the string equations and the expressions for $P, \tilde{P}$ in terms of $Q,\tilde{Q}$ amount to recursion relations for the elements of these matrices.
The matrices $Q,\tilde Q,P,\tilde P$ are not band matrices, but they can be written as algebraic combinations of  band matrices. 

\item Consider the radial case
\begin{align}
 \omega(x,y) = e^{-f(xy)}\ .
\end{align}
The naming and interest of this case become clear if the integration domain is the complex plane, written as
$\{(x,y)\in \mathbb{C}^2| y=\bar x\}$, rather than a factorized domain
$\gamma\times \tilde\gamma \subset \mathbb{C}^2$. 
Then $\omega(z,\bar z)=e^{-f(|z|^2)}$ depends only on the modulus.
Now, using the symmetry $\omega(x,y)=\omega(y,x)$, we have $p_n=\tilde p_n$, $Q=\tilde Q$, and Eq.~\eqref{eq:PfromQomega} leads to
\begin{equation}
P=\tilde P = Q^\text{T} :  f'(QQ^\text{T}) :   \ .
\end{equation}
It follows that $PQ$ is a symmetric matrix. But we also know that $PQ$ is lower triangular, because $P$ is strictly lower triangular while $Q$ has at most one band above the diagonal. It follows that $PQ$ is in fact diagonal, and from Eq.~\eqref{eq:Qsupdiag} and Eq.~\eqref{eq:Pbiosubdiag} we find $(PQ)_{n,n}=n$.

A solution is to choose a matrix $Q$ whose only nonvanishing coefficients are on the band just above the diagonal. Then the only nonvanishing coefficients of $P$ are on the band just below the diagonal, so that 
\begin{equation}
x p_n(x) = p_{n+1}(x) \qquad \text{thus} \qquad p_n(x)=x^n\ .
\end{equation}
We can indeed directly check that the monomials $p_n(x)=x^n$ are orthogonal with respect to any radial measure,
\begin{equation}
\frac{1}{2i} \int_\mathbb{C} dxd\bar x\, x^n\bar x^m e^{-f(x\bar x)} =
\pi \delta_{m,n} \int_{0}^\infty dr\  r^{n} e^{-f(r)} \ .
\end{equation}
For example, in the case of the Gaussian complex matrix model $f'(x)=1$, we have $P=Q^\text{T}$ and $h_n  = \pi n!$ and $\gamma_n  = \sqrt{n}$.

\item Consider a matrix chain \eqref{eq:omegamulti}, where the potentials $V_i$ are polynomials of degrees $\deg V'_i=d_i$ \cite{Bertola:2002, Bergere:2006}.
It can be proved that $Q,\tilde Q,P,\tilde P$ are band matrices, with $Q$ having one band above and $d_2d_3\dots d_L$ bands below the diagonal, while $\tilde Q$ has one band above and $d_1d_2\dots d_{L-1}$ bands below the diagonal.
In order to determine these matrices, it is useful to define the partial integrals
\begin{equation}
\psi_n^{(k)}(x_k) = \frac{1}{\sqrt{h_n}} \int dx_1dx_2\dots dx_{k-1} \,
p_n(x_1)  
\prod_{i=1}^{k} e^{-V_i(x_i)} \prod_{i=1}^{k-1} e^{x_i x_{i+1}}\ , 
\end{equation}
with in particular 
\begin{equation}
\psi_n^{(1)}(x) = \psi_n(x) e^{-V_1(x)}\ .
\label{eq:psino}
\end{equation}
Consider the two families of semi-infinite matrices $Q^{(k)}$ and $P^{(k)}$ such that
\begin{align}
x \psi^{(k)}_n(x) &= \sum_{m} Q^{(k)}_{n,m} \psi^{(k)}_m(x)\quad , \quad
\frac{d}{dx} \psi^{(k)}_n(x) = \sum_{m} P^{(k)}_{n,m} \psi^{(k)}_m(x)\ .
\end{align}
Of course, these matrices obey the string equations
\begin{align}
 [Q^{(k)}, P^{(k)}] = \text{Id}\ . 
\end{align}
Moreover, 
together with the analogously defined $\tilde{Q}^{(k)}$ and $\tilde{P}^{(k)}$,
these matrices obey the relations 
\begin{subequations}
\begin{align}
Q^{(k)} = \tilde Q^{(k)}{}^\text{T}
\qquad  & , \qquad
P^{(k)} + \tilde P^{(k)}{}^\text{T} =  V'_k(Q^{(k)})\ ,
\\
P^{(k)} =   \tilde Q^{(k+1)}{}^\text{T}
\qquad  & , \qquad
\tilde P^{(k)} =   Q^{(k-1)}{}^\text{T}\ .
\end{align}
\end{subequations}
These equations in principle determine $Q^{(k)}$, $\tilde Q^{(k)}$, $P^{(k)}$, and $\tilde P^{(k)}$, which are band matrices. 
From the matrices $Q=Q^{(1)}$ and $\tilde{Q} =\tilde{Q}^{(1)}$, we can then compute the biorthogonal polynomials.
\end{itemize}

\subsection{Integrable systems}

\subsubsection{Differential system}

From the matrices $Q$ and $P$ that appear in the recursion and derivative relations of the biorthogonal polynomials, let us deduce the
matrix differential equation that these polynomials obey. 
We assume that $Q$ is a band matrix, with one band above and $d$ bands below the diagonal, as happens in the case of a matrix chain with polynomial potentials. 
Then the recursion relations imply folding relations of the type
\begin{equation}
\psi_n(x) = \sum_{m=N-d}^N  \left(F^{N}(x)\right)_{n,m} \psi_n(x) \ ,
\end{equation}
where the folding matrix $F^{N}(x)$ depends polynomially on $x$. The folding matrix is still given by Eq.~\eqref{fnrla} 
in terms of the Christoffel--Darboux matrix $A^N$, and the upper triangular left inverse $L(x)$ and lower triangular right inverse $R(x)$ of $(x-Q)$.
Now the Christoffel--Darboux matrix is nonzero only in a square block of size $d+1$,
\begin{equation}
A^N_{i,j} \neq 0
\qquad \implies \qquad \left\{
\begin{array}{l}
 N- d \leq i\leq N \ , 
 \\
 N-1\leq j\leq N-1+ d\ ,
\end{array}\right.
\end{equation}
and the matrix $R(x)$ is
\begin{equation}
R(x)_{n,m} = - \int_\Gamma \frac{\psi_n(x)-\psi_n(x')}{x-x'}\ \omega(x',y')\ \tilde \psi_m(y') \ ,
\end{equation}
with $R(x)_{n,n-1}=-\frac{1}{\gamma_n}$ and $R(x)_{n,m}=0$ if $m\geq n$.

The folding and derivative relations of biorthogonal polynomials, imply that 
the vector $\vec \psi_N = (\psi_{N-d}, \dots, \psi_N)^\text{T}$ satisfies 
a differential system of order $ d+1$,
\begin{equation}
\vec\psi_N'(x) = D_N(x)\vec\psi_N(x)\ ,
\end{equation}
where the Lax matrix $D_N(x)$ is the folding of the semi-infinite matrix $P$,
\begin{equation}
D_N(x) = (\Pi_{N+1}-\Pi_{N- d+1})  P F^{(N)}(x)\ .
\end{equation}
The coefficients of $D_N(x)$ are polynomial functions of $x$, whose degrees are bounded by the number of bands of $P$.
Therefore, our differential system is isomonodromic.

\subsubsection{Riemann--Hilbert problem}

The vector $\vec \psi_N$ of biorthonormal polynomials provides only one solution of our differential system. Let us construct the remaining $d$ independent solutions. We assume that not only $Q$, but also $\tilde Q$, is a finite band matrix. We accept that the band sizes $d$ and $\tilde d$ of $Q$ and $\tilde Q$ coincide with the numbers of independent contours for the eigenvalues $\tilde \lambda$ and $\lambda$ respectively.
Given bases of contours $(\gamma_i)_{i=1,\dots \tilde d}$ and $(\tilde \gamma_j)_{j=1,\dots d}$, we introduce a dual basis $(\tilde\gamma^*_j)_{j=1,\dots d}$ such that ${\tilde\gamma}^*_j\cap {{\tilde\gamma}}_k=\delta_{j,k}$, and write our integration domain as $\Gamma= \sum_{i=1}^{\tilde d}\sum_{j=1}^d c_{i,j} \gamma_i \times \tilde \gamma_j$.
We then introduce the functions
\begin{equation}
{\psi}_n^{(j)}(x)
= \int_{\tilde\gamma^*_j} dy \int_\Gamma \frac{\omega(x',y')dx'dy'}{\omega(x,y)}
\ \frac{\partial_y \log\omega(x,y)- \partial_{y'}\log\omega(x,y')}{(x-x')(y-y')} \ \psi_n(x') \ ,
\end{equation}
which are discontinuous across $\gamma_i$, with the discontinuities
\begin{equation}
{\psi}_n^{(j)}(x+i0)
- {\psi}_n^{(j)}(x-i0) = 2\pi i c_{i,j}  \psi_n(x)\ .
\end{equation}
Then the matrix of size $d+1$
\begin{equation}
\Psi_N(x) = \begin{pmatrix}
\psi_{N-d}(x) & {\psi}_{N-d}^{(1)}(x) & \dots & {\psi}_{N-d}^{(d)}(x) \cr
\vdots & \vdots & & \vdots \cr
\psi_{N-1}(x) & {\psi}_{N-1}^{(1)}(x) & \dots & {\psi}_{N-1}^{(d)}(x) \cr
\psi_{N}(x) & {\psi}_{N}^{(1)}(x) & \dots & {\psi}_{N}^{(d)}(x) \cr
\end{pmatrix}\ , 
\end{equation}
satisfies an isomonodromic matrix differential equation
\begin{equation}
\frac{d}{dx} \Psi_N(x) = D_N(x) \Psi_N(x)\ .
\end{equation}
Its monodromy matrix across $\gamma_i$ is 
\begin{equation}
S_{\gamma_i}  = \mathrm{ Id}_{d+1} +
2\pi i \begin{pmatrix}
1 & 0 & \dots & 0
\end{pmatrix}^\text{T} 
\begin{pmatrix}
0 & c_{i,1} & \dots & c_{i,d}
\end{pmatrix}\ .
\end{equation}
Near $x=\infty$, we still have $\psi_n(x) \sim \frac{x^n}{\sqrt{h_n}}$. 
The asymptotic behaviour of $\psi^{(j)}_n(x)$ depends on the asymptotic behaviour of $\omega(x,y)$ near $y=\infty$, which we write as
\begin{equation}
\partial_y \log \omega(x,y) \mathop{\sim}_{y\to \infty} \sum_{k=1}^{r} R_k(x) y^{k} +O(1) \ .
\end{equation}
We then have 
\begin{equation}
\psi^{(j)}_n(x) \underset{x\to\infty}{\sim} \sum_{0\leq l,m\leq r-1} \sum_{\substack{k\in\mathbb{Z} \\ k\geq \frac{n-m}{d}}}C_{n,m,k}\frac{R_{1+l+m}(x)}{x^{k+1}}  \int_{\tilde\gamma^*_j} \frac{y^l dy}{\omega(x,y)} \ ,
\end{equation}
% NB: Strange formula: is it asymptotic or exact? What about the x dependence in the last factor?
where we introduced 
\begin{equation}
C_{n,m,k} = \sqrt{h_0}  \left< \tilde \psi_0 \middle| \tilde Q^{mT} Q^k \middle| \psi_n \right> .
\end{equation}
%\sr{Did we define the bra and ket states?}
If $\omega(x,y)$ has finite poles, then the asymptotics of $\Psi_N(x)$ near these poles can similarly be deduced from the asymptotics of $\partial_y \log\omega(x,y) $.

In the example of the two-matrix model $\partial_y\log\omega(x,y) = N(x-V'_2(y))$,
we have $r=d=\deg V'_2$, and $R_k(x) = -N \tilde t_{k+1} + N x \delta_{k,0}$, and 
$C_{n,m,k}$ vanishes if $k<\frac{n-m}{d}$. 
% NB: So what? we never need such low values of k.
Let $n=qd+r$ be the Euclidian division of $n$ by $d$, then
\begin{equation}
\psi^{(j)}_n(x) \mathop{\sim }_{x\to \infty}
- \sqrt{\frac{\pi N}{d}}C_{n,r,q}  e^{NV_1(x)}
\frac{ \left(e^{2\pi ij}\frac{x}{\tilde t_{d+1}}\right)^{-\frac{r+1}{d}} }{x^{q-\frac12}} 
e^{-\frac{N d}{d+1} x  \left(e^{2\pi ij}\frac{x}{\tilde t_{d+1}}\right)^{\frac{1}{d} } }
\ .
\end{equation}
% NB: Here we do need a reference

\subsubsection{Tau function}

Among the many properties of orthogonal polynomials that have generalizations to biorthogonal polynomials, let us mention that in two-matrix models, the partition function and biorthogonal polynomial $p_n(x)$ respectively coincide with the Tau function and Baker--Akhiezer function of the associated integrable system \cite{Bergere:2006}.

\section{Exercises}

\begin{exo}[HOMFLY and Jones polynomials of the unknot]
~\label{exo:HOMFLY_Jones_unknot}
Following \autoref{exo:jones}, we know that the coloured HOMFLY polynomial of the unknot $\mathfrak K_{1,1}$ is written in terms of the integral
\begin{align}
\mathcal Z_{\mathfrak K_{1,1}, \lambda}(q)
= \int_{\mathbb R^n} du_1 \dots du_n\ e^{- \sum_i \frac{u_i^2}{2g_s}} \,
\frac{\det_{1\leq i,j\leq n} e^{(\lambda_i-i+n)u_j}}{\det_{1\leq i,j\leq n} e^{(n-i)u_j}}
 \left(\prod_{i<j} 2\sinh\tfrac{1}{2}(u_i-u_j)\right)^2\, .
\end{align}
We now want to compute this integral.

\begin{enumerate}

\item We introduce the Weyl vector $\rho=(\frac{n+1}{2}-i)_{i=1,\dots, n}$ of the root system $A_{n-1}$.
Prove the Weyl denominator formula,
\begin{equation}
\prod_{1 \le i<j \le n} 2\sinh\frac{u_i-u_j}{2}
= \det_{1 \le i, j \le n}\left( e^{\rho_i u_j}\right)\ .
\end{equation}

\item
Using the determinantal formula \eqref{eq:ZNdetmimj}, show that
the coloured HOMFLY polynomial of the unknot is
\begin{equation}
\mathrm{HOMFLY}_{\mathfrak K_{1,1},\lambda}(q)
 = \frac{\mathcal Z_{\mathfrak K_{1,1},\lambda}(q)}{\mathcal Z_{\mathfrak K_{1,1},\emptyset}(q)}
 = q^{-\frac{1}{2} \sum_i \lambda_i(\lambda_i-2i+n+1)}
 \frac{\det_{1\leq i,j\leq n}\left( q^{{-(\lambda_i+\rho_i) \rho_j}}  \right)}{\det_{1\leq i,j\leq n}\left( q^{-{\rho_i \rho_j}}  \right)}\ ,
\end{equation}
and is therefore a Laurent polynomial of $\sqrt{q}=e^{-\frac12 g_s}$.

\item For $n=2$ and $\lambda=(N-1,0)$, deduce the coloured Jones polynomial of the unknot,
\begin{equation}
J_{\mathfrak K_{1,1}, N}(q)
=
q^{-\frac{N(N-1)}{2}}\,\frac{ q^{\frac{N}{2}}-q^{-\frac{N}{2}}}{ q^{\frac{1}{2}}-q^{-\frac{1}{2}}}\ .
\end{equation}

\end{enumerate}

\end{exo}

\begin{exo}[Integrals on $U_N$ and Toeplitz determinants]
~\label{exo:volun}
Let us consider $U_N$ as the normal matrix ensemble $U_N=H_N(S^1)$, where $S^1$ is the unit circle.
\begin{enumerate}
\item Using the formula \eqref{eq:dmhaar} for the Haar measure, show that
\begin{align}
\frac{\operatorname{Vol}(U_N)}{\operatorname{Vol}(U_N/U_1^N)} = \frac{1}{N!} \int_{(S^1)^N} \prod_{i<j} |z_i-z_j|^2 \prod_{i=1}^N \frac{dz_i}{i z_i}\ .
\end{align}
Compute the integral using
the determinantal formula \eqref{eq:ZNdetmimj}, and check that you obtain the expected result for $\operatorname{Vol}((U_1)^N)$.

\item
Let $\left<\det f(U)\right>_{U_N}$ be the Haar measure expectation value of the determinant of a matrix function $f(U)$ on $U_N$.
For $f(z)=\sum_{k\in\mathbb{Z}} f_k z^k$ a convergent series in the unit disc, show that this expectation value is a Toeplitz determinant,
\begin{align}
 \left<\det f(U)\right>_{U_N} = \det_{0\leq i,j\leq N-1} f_{i-j}   \ .
\end{align}

\item
For $0<|q|<1$ let us define the Theta function $\Theta(z;q)$ and coefficient $C_N(q)$ as
\begin{align}
 \Theta(z;q)
 = \sum_{n\in \mathbb Z} q^{n^2} z^n
% = \prod_{n=1}^\infty (1+q^{2n-1} z) (1+q^{2n-1}/z)(1-q^{2n}),
\qquad , \qquad
C_N(q) = \prod_{j=1}^{N-1} (1-q^{2j})\ .
\end{align}
Show that
\begin{align}
 \left<\det \frac{\Theta(U;q)}{C_N(q)}\right>_{U_N} = \prod_{j=1}^N(1-q^{2j})^{-j}\ .
\end{align}
In the large $N$ limit, show that this coincides with the MacMahon function that counts plane partitions (see \cite{Okuda:2004mb} for details):
\begin{align}
 \sum_{\pi} q^{2|\pi|}= 1 + q^2 + 3q^4 + 6q^6+13q^8+ \cdots\ .
\end{align}
\end{enumerate}

\end{exo}

\begin{exo}[Proof of the Heine-type formula for $\varphi_{k-1}$]
~\label{exoHeinephi}
Inserting Heine's formula for $\psi_{k-1}$ into the definition \eqref{defvarphi} of $\varphi_{k-1}$, prove the Heine-type formula \eqref{eq:Heinephi} for $\varphi_{k-1}$.
Notice in particular how the formulas for $\varphi_{k-1}$ and $\psi_{k-1}$ involve matrices of sizes $k$ and $k-1$ respectively, with the integration variable in \eqref{defvarphi} providing the $k$-th eigenvalue.
One may use the identity
\begin{equation}
\frac{1}{\prod_{i=1}^k (x-\lambda_i)} = \sum_{i=1}^k \frac{1}{x-\lambda_i} \prod_{j(\neq i)}\frac{1}{\lambda_i-\lambda_j}\ ,
\label{oopxl}
\end{equation}
which follows from an analysis of poles and residues.
\end{exo}

\begin{exo}[Computation of the determinant of the Lax matrix]
~\label{exodetD}
Compute the determinant $\det D_N = -\frac12 \operatorname{Tr} D_N^2$ of the Lax matrix \eqref{formulaDN}, using the recursion relation \eqref{eq:prp}. Check that the result agrees with Eq.~\eqref{dn_inv}.
\end{exo}

\begin{exo}[Proof of a determinantal formula for Tau functions]
~\label{exoproofdetformula}
We want to prove the determinantal formula \eqref{eq:qtau} for the Tau function $\tau_N(\vec t)$. Let $H(x_1,x_2;y_1,y_2)$ be the difference of the two sides of this formula.
\begin{enumerate}
\item Write the relevant Tau functions as eigenvalue integrals, and
show that $\prod_{i,j=1}^2 (x_i-y_j)\, H(x_1,x_2;y_1,y_2)$ is a polynomial of $x_1,x_2$ of degree at most $N+1$ in each variable $x_i$.
\item Show that this polynomial has zeros at $x_i=y_j$, and that $H(x_1,x_2;y_1,y_2)$ can be written as a bilinear combination of orthogonal polynomials of $x_1,x_2$ with degrees at most $N-1$.
\item Write the $y_1,y_2$-dependent coefficients of that bilinear combination as eigenvalue integrals. By symmetrizing the integrands as in the proof of Heine's formula, show that these coefficients vanish.
\end{enumerate}

\end{exo}

\begin{exo}[Proof of a Heine-type formula in multi-matrix models]
~\label{exproofKdetbio}
In the multi-matrix model \eqref{eq:Z2matrixDelta} of size $N$, let us compute the polynomial expectation value $h_NK_{N+1}(x,y)=\left< \det(x-M)\det(y-\tilde M)\right>$ Eq.~\eqref{eq:KbioHeine}.
\begin{enumerate}
\item Show that $h_NK_{N+1}(x,y)$ is a linear combination of biorthogonal polynomials of $x$ of the type
$
\sum_{k=0}^N c_k(y) p_k(x)
$, where the coefficients are
\begin{align}
 h_k c_k(y) = \int d\lambda_N d\tilde\lambda_N \left< \det(\lambda_{N}-M)\det(y-\tilde M)\right> \omega(\lambda_N,\tilde\lambda_N) \tilde p_k (\tilde\lambda_N)\, .
 \label{hkcky}
\end{align}

\item Compute $c_{k}(y)$ as an integral on $2(N+1)$ variables. 
Decomposing the Vandermonde as determinants of polynomials, show that $h_kc_k(y) = \tilde p_k(y)$, and deduce the Heine-type formula
\begin{align}
\left< \det(x-M)\det(y-\tilde M)\right> = \sum_{k=0}^N \frac{1}{h_k} p_k(x) \tilde p_k(y)\ .
\end{align}

\end{enumerate}

\end{exo}

\begin{exo}[Conformal field theory at central charge one and Schlesinger system]
~\label{exo:cft1}
Consider a normal matrix integral of the type \eqref{eq:normaleig}, with the potential
\begin{align}
 V'(x) = \sum_{j=1}^n \frac{2i\alpha_j}{x-z_j}\quad \text{with} \quad \sum_{j=1}^n i\alpha_j = N\ .
\end{align}
This is a holomorphic version of the Dotsenko--Fateev integral of \autoref{exo:dotsenko}, with $\beta=2$ and therefore the central charge $c=1$. We want to study the corresponding isomonodromic differential equation for the matrix $\Psi_N(x)$.

\begin{enumerate}

\item
Show that the Lax matrix is of the type
\begin{align}
 D_N(x) = \sum_{j=1}^n \frac{A_j}{x-z_j}\ ,
\end{align}
where the size two, $x$-independent matrices $A_j$ have eigenvalues $(i\alpha_j,-i\alpha_j)$ and obey $\sum_{j=1}^n A_j = 0$.

\item
Show that the  monodromy of $\Psi_N(x)$ around $x=z_j$ has the eigenvalues $e^{\pm 2\pi \alpha_j}$. Study the matrix
$\frac{\partial}{\partial z_j} \Psi_N(x)\Psi_N(x)^{-1}$, and show that
\begin{equation}
\frac{\partial}{\partial z_j} \Psi_N(x) = \frac{A_j}{z_j-x} \Psi_N(x)\ .
\end{equation}

\item
Prove Schlesinger's isomonodromy equations for $i\neq j$:
\begin{equation}
\frac{\partial A_i}{\partial z_j} = \frac{[A_j,A_i]}{z_j-z_i}\ .
\end{equation}
\end{enumerate}
\end{exo}

\chapter{Angular integrals}\label{chap:AngularIntegral}

In this chapter, we consider matrix integrals that are not invariant under conjugation, and can therefore not be reduced to integrals over eigenvalues.
For $Y$ a constant matrix and $f_\text{invariant}$ a function that is invariant under conjugation, 
we consider the integral 
\begin{align}
 \mathcal{Z} = \int_{E_N^\beta} dM\ e^{-\operatorname{Tr}MY} f_\text{invariant}(M)\ ,
 \label{eq:zdm}
\end{align}
where invariance under conjugation is broken by the factor $ e^{-\operatorname{Tr}MY}$.
Using the behaviour \eqref{eq:dM} of the  measure $dM$ under the angular-radial decomposition $M=U\Lambda U^{-1}$, we find
\begin{align}
 \mathcal{Z} = \int_{\mathbb{R}^N} d\Lambda\, \left|\Delta(\Lambda)\right|^\beta f_\text{invariant}(\Lambda) \mathcal{Z}(\Lambda, Y)\ ,
\end{align}
where we introduce the \textbf{angular integral}\index{angular integral} over the corresponding circular ensemble
\begin{align}
 \boxed{ \mathcal{Z}(X,Y) = \int_{U_N^\beta} dU\, e^{-\operatorname{Tr} UXU^{-1}Y} } \ .
\end{align}
There are two particularly interesting cases:
\begin{itemize}
\item If $X$ and $Y$ belong to the Gaussian ensemble $E_N^\beta$, since the Haar measure $dU$ is invariant under the left and right actions of the group $U_N^\beta$ on itself, we can assume that $X$ and $Y$ are diagonal matrices. Then our angular integral is called an \textbf{Itzykson--Zuber integral}\index{Itzykson--Zuber integral}~\cite{Itzykson:1979fi}.

\item If $X$ and $Y$ belong to the Lie algebra $\operatorname{Lie} U_N^\beta$, we obtain \textbf{Harish-Chandra integrals}\index{Harish-Chandra!---integral}~\cite{HarishChandra:1957AJM}.

\end{itemize}
For $\beta=2$, the Gaussian ensemble $H_N$ and Lie algebra $\operatorname{Lie} U_N$ are the sets of Hermitian and anti-Hermitian matrices respectively. They are related by multiplication with $i = \sqrt{-1}$, so that Itzykson--Zuber and Harish-Chandra integrals coincide up to a $-1$ sign in the formulas. 
For $\beta=1$, the sets of real symmetric and antisymmetric matrices are genuinely different, in particular their respective dimensions $\frac{N(N+1)}{2}$ and $\frac{N(N-1)}{2}$ differ.

In angular integrals, integrands are not invariant under the left and right actions of the full group $U_N^\beta$ on itself: there is only a residual invariance under the action of a subgroup.
This residual invariance leads to constraints on the 
moments
\begin{align}
 \left< \prod_{k=1}^m U_{i_k,j_k} \prod_{l=1}^n U^{-1}_{j'_l, i'_l} \right> 
 = \frac{1}{\mathcal{Z}(X, Y)} \int_{U_N^\beta} dU\, e^{-\operatorname{Tr} UXU^{-1}Y}\prod_{k=1}^m U_{i_k,j_k} \prod_{l=1}^n U^{-1}_{j'_l, i'_l}\ .
\end{align}
In the $U_N$ case with $X$ and $Y$ diagonal, the residual invariance includes the left and right actions of the diagonal subgroup $(U_1)^N$, which act on matrix elements as $U_{i,j} \to e^{\varphi_i} U_{i,j} e^{\psi_j}$.
This implies that the moments
vanish unless there are permutations $\pi, \rho \in \mathfrak{S}_n$ such that $i_k = i'_{\pi(k)}$ and $j_l = j'_{\rho(l)}$. 
This characterization of non-vanishing moments can be generalized to $\beta = 1, 4$, by replacing permutations with elements of the Weyl group of $U_k^{\frac{4}{\beta}}$ (the Langlands dual of $U_k^{\beta}$).

\section{Harish-Chandra integrals}

Let us first give a more general formulation of Harish-Chandra integrals, valid not only for our circular ensembles $U_N^\beta$, but for arbitrary semi-simple compact Lie groups $G$. 
The case $G=U_N$ will be dealt with more explicitly in \autoref{sec:HCUN}.

\subsection{Harish-Chandra formula}

Define the Harish-Chandra integral as
\begin{align}
 \boxed{ \mathcal{Z}(X,Y) = \int_G dU\ e^{-\left<\operatorname{Ad}_U(X), Y \right>} } \ ,
\end{align}
where $X$ and $Y$ are elements of a Cartan subalgebra $\mathfrak{h}\subset \mathfrak{g} = \operatorname{Lie}G$, on which $G$ acts by the adjoint group action, and $\langle \cdot , \cdot \rangle$ is the Killing form -- a non-degenerate symmetric bilinear form that satisfies in particular $\left<X,[Y,Z]\right> = \left<[X,Y],Z\right>$ and $\left<\operatorname{Ad}_U(X),\operatorname{Ad}_U(Y)\right> = \left<X,Y\right>$ with $U \in G$.

Assuming that $G$ is the maximal compact group in $\exp \mathfrak{g}$,  
the value of the integral is then given by the \textbf{Harish-Chandra formula}\index{Harish-Chandra!---formula}~\cite{HarishChandra:1957AJM},
\begin{align}
 \boxed{ \mathcal{Z}(X,Y) = \frac{C(G)}{\Delta(X)\Delta(Y)} \sum_{\sigma \in W} (-1)^\sigma e^{-\left<\operatorname{Ad}_\sigma(X),Y\right>} } \ ,
\end{align}
where we introduce the following notations:
\begin{itemize}
 \item $W = \text{Weyl}(G)$ is the Weyl group, and the signature $(-1)^\sigma\in \{1,-1\}$ of $\sigma\in W$ depends on the parity of the length of $\sigma$ when written as a product of simple reflections.
 \item The \textbf{generalized Vandermonde determinant}\index{Vandermonde determinant!generalized---} is 
 \begin{align}
  \Delta(X) = \prod_{\alpha > 0}\alpha(X)\ ,
 \end{align}
where the product runs over the positive roots of the Lie algebra $\mathfrak{g}$.
A root $\alpha\in \mathfrak h^*$ is such that $\exists u\neq 0\in\mathfrak g \ , \ \forall h\in\mathfrak h \  :  [h,u]=\alpha(h) u$.
The total number of roots is $\dim \mathfrak g-\dim\mathfrak h$.
If $\alpha$ is a root, then $-\alpha$ is also a root. One arbitrarily splits the set of roots into two disjoint subsets of positive and negative roots, related by $\alpha\to -\alpha$.

\item The constant $C(G)$ is
\begin{align}
 C(G) = \prod_{j=1}^{\dim \mathfrak h} m_j! \prod_{\alpha>0} \frac{\left<\alpha,\alpha \right>}{2}\ .
\end{align}
In this formula, the quadratic form $\langle \cdot , \cdot\rangle$ on roots is induced by restriction of the Killing form to $\mathfrak{h}$. 
 The $\{m_j\}$ are the exponents of the Lie algebra $\mathfrak{g}$, i.e. the integers such that the eigenvalues of a Coxeter element of $W$ when acting on $\mathfrak{h}$ are $e^{2\pi i \frac{m_j}{h}}$, where $h$ is the Coxeter number of $\mathfrak{g}$.
\end{itemize}
The proof of the Harish-Chandra formula relies on the invariance of the exponent $S(U)=-\left<\operatorname{Ad}_U(X), Y \right>$ under the right action of the maximal torus $T = e^{\mathfrak h}$ on $U$. This allows us to reduce the integral to a sum over its critical points,
using the Duistermaat--Heckman localization theorem~\cite{Duistermaat:1982vw}:
\begin{equation}
\int_{G/T} dU e^{S(U)}
 = \sum_{\sigma\in\{\text{critical points}\}} \frac{e^{S(\sigma)}}{\sqrt{\det S''(\sigma)}}\ .
\end{equation}
Let us determine the critical points, i.e. the elements $U\in G/T$ such that $dS(U)=0$. We compute
\begin{equation}
d S(U) = -\left<d U U^{-1},[\operatorname{Ad}_U(X), Y ]\right> \ .
\end{equation}
Since the Killing form is non-degenerate, the critical points must obey $[\operatorname{Ad}_U(X),Y]=0$. For generic $X,Y\in\mathfrak h$, this implies that
$U$ belongs to the stabilizer of $\mathfrak h$, thus $U$ is an element $\sigma$ of the Weyl group $W$. 
The eigenvalues of the Hessian matrix $S''(\sigma)$ are then $\alpha(\operatorname{Ad}_\sigma(X))\alpha(Y)$ for $\alpha$ a root. The roots $\alpha$ and $-\alpha$ give the same eigenvalue, so that
\begin{equation}
\sqrt{\det S''(\sigma)} \propto \Delta(\operatorname{Ad}_\sigma(X)) \Delta(Y) = (-1)^\sigma \Delta(X) \Delta(Y)\ .
\end{equation}
The sum over the critical points is therefore proportional to the right-hand side of the Harish-Chandra formula. We admit that the correct $X,Y$-independent coefficient of proportionality is $C(G)$.

\subsubsection{Moments of the Harish-Chandra formula}

We admit that for $f(X,Y)$ a function such that $f(\operatorname{Ad}_U(X) , \operatorname{Ad}_U(Y) ) = f(X, Y)$,
the expectation value of $f(\operatorname{Ad}_U(X), Y)$ is given by the \textbf{generalized Harish-Chandra formula}\index{Harish-Chandra!---formula (generalized)},
\begin{multline}
 \int_G dU\ e^{-\left<\operatorname{Ad}_U(X), Y \right>} f(\operatorname{Ad}_U(X), Y) 
 = \frac{\pi^{-\dim \mathfrak n_+(G) }\prod_{j=1}^{\dim \mathfrak h} m_j!}{\Delta(X)\Delta(Y)} 
 \\ \times 
 \sum_{\sigma \in W} (-1)^\sigma e^{-\left<\operatorname{Ad}_\sigma(X),Y\right>} 
 \int_{\mathfrak n_+(G)} dT\ e^{- \left< T, T^\dagger\right>} f(\operatorname{Ad}_\sigma(X)+T, Y+T^\dagger)\ ,
\end{multline}
where the positive Borel subalgebra $\mathfrak n_+(G)$ is the complex nilpotent subalgebra of $\mathfrak g$ that is spanned by the positive root vectors.  
The generalized Harish-Chandra formula therefore converts an integral over the Lie group $G$ with its curved Haar measure, into a much simpler Gaussian integral over the vector space $\mathfrak n_+(G)$ with its flat Lebesgue measure.
If $f$ is polynomial, the Gaussian integral can then be efficiently computed using Wick's theorem \cite{Bertola:2008}, as we will shortly demonstrate in the case $G=U_N$.

\subsection{Case of \texorpdfstring{$U_N$}{UN}}
\label{sec:HCUN}

In the case $G=U_N$:
\begin{itemize}

 \item $\mathfrak g=i H_N$ is the set of anti-Hermitian matrices, with the Killing form $\left<A,B\right> = \operatorname{Tr} AB$.

 \item A maximal torus is $(U_1)^N$, and the Cartan subalgebra $\mathfrak h\simeq (i\mathbb{R})^N$ is the set of diagonal imaginary matrices, whose elements can be parametrized by their imaginary eigenvalues $X=\mathrm{ diag}(X_i)$.
 
 \item The adjoint action is $\operatorname{Ad}_U(X) = UXU^{-1}$. The Weyl group is the group of permutations $W = \mathfrak{S}_N$, and $\operatorname{Ad}_\sigma(X) = \mathrm{diag}(X_{\sigma(i)})$.
 
 \item The positive roots $\alpha$ are the linear forms $e_i-e_j$ with $i>j$, where $e_i\in\mathfrak h^* $ is the linear form defined by extracting the $i^{\text{th}}$ diagonal element $e_i(X)=X_i$. The generalized Vandermonde determinant therefore coincides with the Vandermonde determinant.
 
 \item $\mathfrak n_+(G)=\mathfrak n_+$ is the set of strictly upper triangular complex matrices, and $T^\dagger$ is the Hermitian conjugate of $T$.

\item The Coxeter number is $h=N$, and the Coxeter exponents are $m_j=j-1$.
All positive roots have norm $\left<\alpha,\alpha\right>=2$, and $C(U_N)=\prod_{j=0}^{N-1} j!$.
 
\end{itemize}
In the Harish-Chandra formula, the sum over the Weyl group amounts to the determinant 
\begin{align}
 \mathcal{Z}(X,Y) = \frac{C(U_N)}{\Delta(X)\Delta(Y)} \det_{i,j} e^{-X_i Y_j}\ .
 \label{eq:hcu}
\end{align}
We will now compute the moments, using the generalized Harish-Chandra formula.

\subsubsection{Morozov's formula for quadratic moments}

We start with the moments $\left<|U_{i,j}|^2\right>$, which we write as residues of a generating function,
\begin{align}
 \left<|U_{i,j}|^2\right> = \frac{1}{(2\pi i)^2}\oint_{X_i}dx \oint_{Y_j}dy\ 
 \left< \operatorname{Tr} U \frac{1}{x-X} U^{-1} \frac{1}{y-Y} \right>  \ .
 \label{eq:moment_residue}
\end{align}
Using the generalized Harish-Chandra formula, we find 
\begin{multline}
 \left< \operatorname{Tr} U \frac{1}{x-X} U^{-1} \frac{1}{y-Y} \right> 
 = \frac{\pi^{-\frac{N(N-1)}{2}}} {\det_{i,j} e^{-X_i Y_j}}
 \\
 \times
 \sum_{\sigma \in \mathfrak{S}_N} (-1)^\sigma e^{-\left<\sigma(X),Y\right>} 
 \int_{\mathfrak n_+} dT\ e^{- \left< T, T^\dagger\right>} 
 \operatorname{Tr} \frac{1}{x-\sigma(X)-T}\, \frac{1}{y-Y-T^\dagger}\ .
\end{multline}
Now $\frac{1}{x-X-T}$ is polynomial in $T$:
\begin{align}
 \frac{1}{x-X-T} = \frac{1}{x-X} \sum_{k=0}^{N-1} \left(T\frac{1}{x-X}\right)^k\ ,
\end{align}
which follows from
the formula $\frac{1}{1 - M} = \sum_{k=0}^\infty M^k$, applied to the strictly upper triangular matrix $M = T\frac{1}{x-X}$, whose $N$-th power therefore vanishes.
In terms of matrix elements, we have  
\begin{align}
\left(\frac{1}{x-X-T} \right)_{i,j} 
= \frac{1}{x-X_i} \sum_{k=0}^{N-1}\, \sum_{i=j_0<j_1<\dots<j_{k-1}<j_k=j}\, \prod_{l=1}^k T_{j_{l-1},j_l} \frac{1}{x-X_{j_l}}\ .
\end{align}
We can now use Wick's theorem for computing the integral over $T$, using the two-point function
\begin{equation}
 \pi^{-\frac{N(N-1)}{2}}\int_{\mathfrak n_+} dT\ e^{- \left< T, T^\dagger\right>} T_{i,j} T^\dagger_{k,l}  
 = \delta_{i,l}\delta_{j,k}\ .
\end{equation}
Using Wick's theorem is particularly simple because we are integrating  $T^k = \prod_{l=1}^k T_{j_{l-1},j_l}$ with ordered indices $j_0<j_1<\dots <j_k$.
The integral of $T^k (T^\dagger)^k$ therefore involves at most one pairing, and we find
\begin{multline}\label{eq:morozov1appear}
  \pi^{-\frac{N(N-1)}{2}} \int_{\mathfrak n_+} dT\ e^{- \left< T, T^\dagger\right>}
 \operatorname{Tr} \frac{1}{x-\sigma(X)-T}\frac{1}{y-Y-T^\dagger}
 \\
  =
 \sum_{k=0}^{N-1} \sum_{j_0<j_1<\dots<j_k} \prod_{l=0}^k \frac{1}{(x-X_{\sigma(j_l)})(y-Y_{j_l})}  =
 -1 + \prod_{j=1}^N \left[1+\frac{1}{(x-X_{\sigma(j)})(y-Y_j)}\right]\, .
\end{multline}
We therefore have 
\begin{align}
 1+\left< \operatorname{Tr} U \frac{1}{x-X} U^{-1} \frac{1}{y-Y} \right> 
 =  \frac{\det_{i,j} e^{-X_i Y_j}\left(1+\frac{1}{(x-X_i)(y-Y_j)}\right)} {\det_{i,j} e^{-X_i Y_j}} \ .
\end{align}
We can now extract our moments as residues \eqref{eq:moment_residue}, and we find \textbf{Morozov's formula}\index{Morozov's formula}~\cite{Morozov:1992zb}
\begin{equation}
 \boxed{ \left< |U_{i,j}|^2 \right>= \frac{\text{Minor}_{i,j} \left( E + \frac{1}{X_i-X}E\frac{1}{Y_j-Y}\right)}{\det E}\ ,  \qquad \text{with} \qquad E_{i,j} = e^{-X_i Y_j}}\ .
\end{equation}

\subsubsection{Higher moments}

Let us similarly compute higher moments, following \cite{Eynard:2005wg}. (See \cite{Bertola:2008} for the generalization to arbitrary Lie groups.) 
Nonvanishing moments
$\left< U_{i_1,j_{\pi(1)}}\dots U_{i_n,j_{\pi(n)}} U^{-1}_{j_{\rho(1)},i_1}\dots U^{-1}_{j_{\rho(n)},i_n} \right>$ depend on
two sets of indices $\{i_1,\dots i_n\}\subset\{1,\dots N\}$ and $\{j_1,\cdots j_n\}\subset \{1,\dots N\}$, and two permutations $\pi,\rho \in \mathfrak S_n$.
We will however trade the indices $\{i_1,\dots i_n\}$ and $\{j_1,\cdots j_n\}$ for complex variables $\vec x= (x_1,\dots x_n)$, $\vec y=(y_1,\dots y_n)$, using the transformation
\begin{multline}
U_{i_1,j_{\pi(1)}}\dots U_{i_n,j_{\pi(n)}} U^{-1}_{j_{\rho(1)},i_1}\dots U^{-1}_{j_{\rho(n)},i_n}\\
= \frac{1}{(2\pi i)^{2n}} \oint_{X_{i_1}} dx_1 \dots \oint_{X_{i_n}} dx_n \oint_{Y_{i_1}} dy_1 \dots \oint_{Y_{i_n}} dy_n
\ H(\vec x,\vec y;UXU^{-1},Y)_{\pi,\rho}\ ,
\end{multline}
where we introduced the generating function
\begin{equation}
\boxed{ H(\vec x,\vec y;X,Y)_{\pi,\rho}
= \prod_{c = \text{cycle of }\pi\circ \rho^{-1}}
\left( \delta_{\text{length}(c),1} + \operatorname{Tr} \prod_{i\in c} \frac{1}{x_{\rho(i)}-X}\,\frac{1}{y_i-Y}
\right) } \ .
\end{equation}
Here the term $\delta_{\text{length}(c),1}$ does not contribute to the moments, because it has no pole at $X_i$. Including this term however leads to simpler formulas.

We view $H$ as an $n!\times n!$ matrix, whose rows and columns are indexed by permutations $\pi,\rho$.
For example, in the case $n=2$, we have
\begin{multline}
 H(\vec{x},\vec{y};X,Y)  \\
 \hspace{-5mm} = %\left[]
 \scriptstyle{
 \begin{pmatrix}
  \left[1+\operatorname{Tr}\frac{1}{x_1-X}\frac{1}{y_1-Y}\right]
  \left[1+\operatorname{Tr}\frac{1}{x_2-X}\frac{1}{y_2-Y}\right]
 &
  \operatorname{Tr}\frac{1}{x_1-X}\frac{1}{y_1-Y}\frac{1}{x_2-X}\frac{1}{y_2-Y}
  \\
  \operatorname{Tr}\frac{1}{x_1-X}\frac{1}{y_2-Y}\frac{1}{x_2-X}\frac{1}{y_1-Y}
  &
  \left[1+\operatorname{Tr}\frac{1}{x_1-X}\frac{1}{y_2-Y}\right]
  \left[1+\operatorname{Tr}\frac{1}{x_2-X}\frac{1}{y_1-Y}\right]
 \end{pmatrix}} %\right]
 \, .
\end{multline}
Using Wick's theorem, it is possible to compute
\begin{align}\label{eq:HpipiM}
\boxed{ \Big< H(\vec x,\vec y;U X U^{-1},Y) \Big>
= \frac{1}{\det E}\sum_{\sigma\in\mathfrak{S}_N} (-1)^\sigma \prod_{i=1}^N M(\vec x,\vec y;X_{\sigma(i)},Y_i) } \ ,
\end{align}
where we introduced the $n!\times n!$ matrix $M(\vec x,\vec y;x, y)$: a rational function of $\vec x,\vec y$ and of the auxiliary complex variables $x,y$, whose elements are
\begin{equation}
\boxed{ M(\vec x,\vec y;x, y)_{\pi,\rho} 
= e^{-xy}\prod_{i=1}^n \left( \delta_{\pi(i),\,\rho(i)} + \frac{1}{(x-x_{\rho(i)})(y-y_{i})}\right) } \ .
\end{equation}
For example, in the case $n=2$,
\begin{multline}
  M(\vec x,\vec y;x, y)
  \\
 \hspace{-2mm} = e^{-xy} %\left[
\scriptstyle{ \begin{pmatrix}
  \left[1+ \frac{1}{(x-x_1)(y-y_1)}\right]\left[1 + \frac{1}{(x-x_2)(y-y_2)}\right]
  &
  \frac{1}{(x-x_1)(x-x_2)(y-y_1)(y-y_2)}
  \\
  \frac{1}{(x-x_1)(x-x_2)(y-y_1)(y-y_2)}
  &
  \left[1+ \frac{1}{(x-x_1)(y-y_2)}\right]\left[1 + \frac{1}{(x-x_2)(y-y_1)}\right]
 \end{pmatrix}}¨ %\right]
 \, .
\end{multline}
The matrix $M(\vec x,\vec y;x, y)$ enjoys beautiful properties:
\begin{subequations}
\begin{align}
\text{Symmetry:} & \qquad M(\vec x,\vec y;x,y)= M(\vec x,\vec y;x,y)^\text{T} 
\ ,
\\
\text{Commutativity:} & \qquad \big[ M(\vec x,\vec y;x,y),M(\vec x,\vec y;x',y')\big] = 0\ .
\end{align}
\end{subequations}
In particular, commutativity implies that the matrix product in Eq.~\eqref{eq:HpipiM} does not depend on the order.
In the case $n=1$, this reproduces Morozov's formula.

\subsection{Application to matrix chains}\label{sec:multi_angular}

Let us reduce partition functions of matrix chains to eigenvalue integrals, using the 
Harish-Chandra formula.

\subsubsection{One-matrix model with an external field}

Let us write the partition function of the Hermitian one-matrix model with an external field as 
\begin{align}
 \mathcal{Z}(A) & =
 \int_{H_N(\gamma)} dM \, e^{-\operatorname{Tr}V(M) + \operatorname{Tr} MA} \, .
\end{align}
We assume that the external field is diagonal, $A= \text{diag}(a_1, \dots, a_N)$, without loss of generality.
Let us use the angular-radial decomposition $M= U\Lambda U^{-1}$ with $\Lambda = \text{diag}(\lambda_1,\dots ,\lambda_N)$, and perform the angular integral with the help of the Harish-Chandra formula \eqref{eq:hcu}. 
Ignoring $A$ and $V$-independent prefactors, we find
\begin{align}
 \mathcal{Z}(A) \propto
 \frac{1}{N! \Delta(A)} \int_{\gamma^N} d\Lambda \,
 e^{-\operatorname{Tr}V(\Lambda)} 
 \Delta(\Lambda)  \det_{1 \le i, j \le N} e^{\lambda_i a_j} \, ,
\end{align}
where one of the two $\Delta(\Lambda)$ factors from the Jacobian was cancelled by the denominator of the Harish-Chandra formula.
Using the expression \eqref{eq:Vandermonde_det} of the Vandermonde determinant $\Delta(\Lambda)$, and the identity, called \textbf{Andr{\'e}ief's formula}\index{Andr{\'e}ief's formula},
\begin{align}
 \frac{1}{N!} \int d^N \lambda 
 \left( \det_{1 \le i, j \le N} f_i(\lambda_j) \right)
 \left( \det_{1 \le i, j \le N} g_i(\lambda_j) \right)
 & =
 \det_{1 \le i, j \le N}
 \left( \int d\lambda \, f_i (\lambda) g_j(\lambda) \right)
  ,
 \label{eq:det_prod}
\end{align}
we deduce 
\begin{align}
 \mathcal{Z}(A) \propto
 \frac{1}{\Delta(A)}
 \det_{1 \le i, j \le N}
 \int_\gamma d\lambda \, p_{i-1}(\lambda)  e^{-V(\lambda) + \lambda a_j}\ ,
 \label{eq:ext_det}
\end{align}
for any family of monic polynomials $p_i(\lambda) = \lambda^i + \cdots$. 
In particular, in the case $p_i(\lambda)=\lambda^i$, we can take the limit $A\to 0$ and recover the expression \eqref{eq:Hankel_rep} of the partition function of the one-matrix model as a Hankel determinant. (The singularity of the $\frac{1}{\Delta(A)}$ prefactor is cancelled by the determinant, as can be seen using the Taylor expansion at $A=0$.)

\subsubsection{Two-matrix model}

In the Hermitian two-matrix model with the potentials $V_1$ and $V_2$, let us use the angular-radial decompositions of the two matrices, 
$M_1 = U_1 \Lambda U_1^{-1}$, $M_2 = U_2 \tilde{\Lambda} U_2^{-1}$
with 
$\Lambda=\text{diag}(\lambda_1,\ldots\lambda_N)$,
$\tilde\Lambda=\text{diag}(\tilde\lambda_1,\ldots\tilde\lambda_N)$.
The Harish-Chandra formula allows us to perform the angular integrals,
\begin{align}
 \int_{(U_N)^2} dU_1 dU_2 \, 
 e^{\operatorname{Tr} U_1 \Lambda U_1^{-1} U_2 \tilde{\Lambda} U_2^{-1}}
 & \propto
 \frac{1}{\Delta(\Lambda)\Delta(\tilde{\Lambda})}
 \det_{1 \le i, j \le N} e^{\lambda_i \tilde{\lambda}_j} \, ,
\end{align}
and the partition function can be written as the radial integral
\begin{align}
 \mathcal{Z} & \propto
 \int d\Lambda d\tilde{\Lambda} \,
 e^{-\operatorname{Tr}\left(V_1(\Lambda)+V_2(\tilde{\Lambda})\right)} \,
 \Delta(\Lambda) \Delta(\tilde{\Lambda})
 \det_{1 \le i, j \le N} e^{\lambda_i \tilde{\lambda}_j}
 \, ,
\end{align}
therefore proving Eq.~\eqref{eq:Z2matrixDelta}.

\subsubsection{Matrix chain}

In the Hermitian matrix chain of length $L$ with no external field, a similar calculation leads to the representation of the partition function as a radial integral,
\begin{align}
 \mathcal{Z} & \propto
 \int \prod_{i=1}^L d\Lambda_i \, e^{-\operatorname{Tr}V_i(\Lambda^{(i)})} 
 \Delta(\Lambda^{(1)}) \Delta(\Lambda^{(L)}) 
 \prod_{i=1}^{L-1}
 \left(
  \det_{1 \le j, k \le N} e^{\lambda_j^{(i)} \lambda_{k}^{(i+1)}}
 \right) \, .
\end{align}
Applying the formula \eqref{eq:det_prod} to the integrals over $\Lambda^{(2)},\ldots \Lambda^{(L-1)}$, we recover Eq.~\eqref{eq:Z2matrixDelta}.

\section{Itzykson--Zuber integrals}\label{sec:izi}

Although there is no simple expression for Itzykson--Zuber
integrals for $\beta\neq 2$, there are many known relations, which we will now review following~\cite{Bergere:2008da}.
These relations can be used for defining Itzykson--Zuber integrals for
arbitrary complex values of $\beta$, beyond their original definition as
matrix integrals for
$\beta\in\{1,2,4\}$.

\subsection{Calogero--Moser equation}

We will show that the Itzykson--Zuber integral $\mathcal{Z}(X, Y)$ obeys a second-order linear partial differential equation.
We start with first-order equations for the quadratic moments 
\begin{equation}
M_{i,j} = \mathcal{Z}(X, Y) \left< U_{i,j} U^\dagger_{j,i} \right> = \int_{U_N^\beta}dU\ e^{-\operatorname{Tr}UXU^\dagger Y} U_{i,j} U^\dagger_{j,i}\ ,
\label{eq:mij}
\end{equation}
which obey
\begin{equation}\label{eq:IZMijrels}
 \sum_{i=1}^N M_{i,j} = \sum_{j=1}^N M_{i,j} = \mathcal{Z}(X, Y)
  \, , \qquad 
 \frac{\partial \mathcal{Z}(X, Y)}{\partial X_j}  = -\sum_{i=1}^N M_{i,j}Y_i\ .
\end{equation}
Furthermore, as a consequence of loop equations, the quadratic moments obey the following system of first-order linear differential equations,
\begin{equation}
 \frac{\partial M_{i,j}}{\partial X_j }
 = -M_{i,j}Y_i - \frac{\beta}{2}\sum_{k\neq j} \frac{M_{i,j}-M_{i,k}}{X_j-X_k}
 \ .
 \label{eq:dunkl}
\end{equation}
(See \autoref{exodunkl} for the proof.)
Taking one more derivative and summing over $i$ we get
\begin{align}
 \sum_j \frac{\partial^2}{\partial X_j^2} \mathcal{Z}(X, Y) 
 = 
 \sum_{i,j} M_{i,j}Y_i^2 +\frac{\beta}{2} \sum_i \sum_{k\neq j} \frac{M_{i,j}Y_i - M_{i,k}Y_i}{X_i-X_k}\ ,
 \end{align}
which leads to 
 \begin{align}\label{eq:CalogeroMoser}
 \boxed{
 \left( \sum_j \frac{\partial^2}{\partial X_j^2} + \frac{\beta}{2} \sum_{k\neq j} \frac{1}{X_j-X_k}\left(\frac{\partial }{\partial X_j }-\frac{\partial }{\partial X_k }\right) \right) \mathcal{Z}(X, Y)
 = \left(\sum_i Y_i^2\right) \mathcal{Z}(X, Y)  } \ .
\end{align}
Therefore, as a function of $X$, the Itzykson--Zuber integral is an eigenfunction of a second-order differential operator, with the eigenvalue $\sum_j Y_j^2$. 
Actually, $\mathcal{Z}(X, Y)$ can be characterized as the eigenfunction that has the symmetries
\begin{align}
 \mathcal{Z}(X, Y) = \mathcal{Z}(Y, X)= \mathcal{Z}(\sigma(X), Y)\ , \qquad (\forall \sigma \in \mathfrak S_N)\ .
\end{align}
Since the eigenvalue equation coincides with the Schr\"odinger equation of the Calogero--Moser system, we call it the \textbf{Calogero--Moser equation}\index{Calogero--Moser equation}.

\subsection{Jack polynomials and duality equation}

Since $\mathcal Z(X,Y)$ is symmetric under permutations of $(X_i)$ and $(Y_j)$, it is natural to expand it on a basis of symmetric functions. The relevant symmetric functions are the Jack polynomials, which are known to appear in
solutions of the Calogero--Moser equation. Let us review their properties, following the textbook~\cite{Macdonald:1997}. (See also  \cite{Desrosiers:2008tp} or \cite{DUMITRIU2007587}.)

\subsubsection{Jack polynomials}

For any partition $\lambda = (\lambda_1, \dots , \lambda_N)$ of length $N$,
we define the symmetrized monomial of $N$ variables
\begin{equation}
 m_\lambda(X_1,\dots,X_N) = \frac{1}{z_\lambda}\sum_{\sigma\in\mathfrak{S}_N} \prod_{i=1}^N X_{\sigma(i)}^{\lambda_i}\ 
\ ,\quad
z_\lambda = \prod_k n_k(\lambda)!
\ ,\quad
  n_k(\lambda) = \#\{i \mid \lambda_i=k\} \, .
\end{equation}
We also define a partial ordering on partitions of length $N$,
\begin{equation}
 \lambda \leq \mu \qquad \iff \qquad \forall \ 1\leq k\leq N \ , \quad \sum_{i=1}^k \lambda_i \leq \sum_{i=1}^k\mu_i\ .
\end{equation}
For any $\alpha\in\mathbb R^*_+$ we define a scalar product on symmetric functions of $N$ variables,
\begin{equation}
 \left< f\middle| g\right>_\alpha = \frac{(2\pi i)^{-N}}{N!} \int_{(S^1)^N} \frac{dX}{\det X} \left[ \Delta(X)  \Delta(X^{-1})\right]^{\frac{1}{\alpha}} f(X)g(X^{-1})\ .
\end{equation}
(For $X\in (S^1)^N$, $\left[ \Delta(X)  \Delta(X^{-1})\right]^{\frac{1}{\alpha}} = |\Delta(X)|^{\frac{2}{\alpha}}$ is well-defined.)
For this scalar product, power sum polynomials are orthogonal:
\begin{equation}
\langle p_\lambda | p_\mu \rangle_\alpha = \delta_{\lambda,\mu} \alpha^{\ell(\lambda)} z_\lambda \prod_{i=1}^{\ell(\lambda)} \lambda_i \ .
\end{equation}
The Jack polynomials are then defined by orthogonalizing the symmetrized monomials with respect to this scalar product. 
Namely, the Jack polynomial $J^{(\alpha)}_\lambda$ is the unique homogeneous symmetric polynomial, whose degree is the weight $|\lambda|$, such that 
\begin{align}
 J_\lambda^{(\alpha)}(X) =  m_\lambda(X) + \sum_{\substack{\mu<\lambda \\ |\mu|=|\lambda|}} c_{\lambda,\mu}  m_\mu(X) \quad \text{and} \quad \left<J_\lambda^{(\alpha)}\middle| J_\mu^{(\alpha)} \right>_\alpha \neq 0 \implies \lambda = \mu \ .
\end{align}
For example, 
\begin{subequations}
\begin{align}
& J^{(\alpha)}_{\begin{tikzpicture}[scale=0.8]
\draw (0,0) rectangle (0.2,0.2);
\end{tikzpicture}}
 =m_{\begin{tikzpicture}[scale=0.8]
\draw (0,0) rectangle (0.2,0.2);
\end{tikzpicture}} 
&&= p_{\begin{tikzpicture}[scale=0.8]
\draw (0,0) rectangle (0.2,0.2);
\end{tikzpicture}} 
\ , && 
\left\langle 
J^{(\alpha)}_{\begin{tikzpicture}[scale=0.8]
\draw (0,0) rectangle (0.2,0.2);
\end{tikzpicture}}
\middle|
 J^{(\alpha)}_{\begin{tikzpicture}[scale=0.8]
\draw (0,0) rectangle (0.2,0.2);
\end{tikzpicture}}
\right\rangle
= \alpha \ ,
\\
& J^{(\alpha)}_{\begin{tikzpicture}[scale=0.8]
\draw (0,0) rectangle (0.2,0.2);
\draw (0,0.2) rectangle (0.2,0.4);
\end{tikzpicture}}
 =m_{\begin{tikzpicture}[scale=0.8]
\draw (0,0) rectangle (0.2,0.2);
\draw (0,0.2) rectangle (0.2,0.4);
\end{tikzpicture}} 
&&= \frac12 \left( p_{\begin{tikzpicture}[scale=0.8]
\draw (0,0) rectangle (0.2,0.2);
\draw (0,0.2) rectangle (0.2,0.4);
\end{tikzpicture}} - p_{\begin{tikzpicture}[scale=0.8]
\draw (0,0) rectangle (0.2,0.2);
\draw (0.2,0) rectangle (0.4,0.2);
\end{tikzpicture}}\right) 
\, , \ \ \ 
&& \left\langle 
J^{(\alpha)}_{\begin{tikzpicture}[scale=0.8]
\draw (0,0) rectangle (0.2,0.2);
\draw (0,0.2) rectangle (0.2,0.4);
\end{tikzpicture}}
\middle|
J^{(\alpha)}_{\begin{tikzpicture}[scale=0.8]
\draw (0,0) rectangle (0.2,0.2);
\draw (0,0.2) rectangle (0.2,0.4);
\end{tikzpicture}}
\right\rangle
= \frac{\alpha(1+\alpha)}{2} \ ,
\\
& J^{(\alpha)}_{\begin{tikzpicture}[scale=0.8]
\draw (0,0) rectangle (0.2,0.2);
\draw (0.2,0) rectangle (0.4,0.2);
\end{tikzpicture}} 
 = m_{\begin{tikzpicture}[scale=0.8]
\draw (0,0) rectangle (0.2,0.2);
\draw (0.2,0) rectangle (0.4,0.2);
\end{tikzpicture}} + \frac{2}{1+\alpha} m_{\begin{tikzpicture}[scale=0.8]
\draw (0,0) rectangle (0.2,0.2);
\draw (0,0.2) rectangle (0.2,0.4);
\end{tikzpicture}} 
&& = \frac{  p_{\begin{tikzpicture}[scale=0.8]
\draw (0,0) rectangle (0.2,0.2);
\draw (0,0.2) rectangle (0.2,0.4);
\end{tikzpicture}} +\alpha  p_{\begin{tikzpicture}[scale=0.8]
\draw (0,0) rectangle (0.2,0.2);
\draw (0.2,0) rectangle (0.4,0.2);
\end{tikzpicture}}  }{1+\alpha}  
\ ,
&& \left\langle 
J^{(\alpha)}_{\begin{tikzpicture}[scale=0.8]
\draw (0,0) rectangle (0.2,0.2);
\draw (0.2,0) rectangle (0.4,0.2);
\end{tikzpicture}}
\middle|
J^{(\alpha)}_{\begin{tikzpicture}[scale=0.8]
\draw (0,0) rectangle (0.2,0.2);
\draw (0.2,0) rectangle (0.4,0.2);
\end{tikzpicture}}
\right\rangle
= \frac{2\alpha^2}{1+\alpha} \ .
\end{align}
\end{subequations}
The squared norm of a Jack polynomial is \cite{Macdonald:1997}
\begin{align}
 \left<J_\lambda^{(\alpha)}\middle| J_\lambda^{(\alpha)} \right>_\alpha 
 = 
\prod_{\square\in\lambda} \frac{\text{leg}(\square) +\alpha (1+\text{arm}(\square))}{\text{leg}(\square) + 1 +\alpha\ \text{arm}(\square)}
%\prod_{j=1}^N 
% \frac{\Gamma((N-j+1)\alpha^{-1}+\lambda_j)\Gamma((N-j)\alpha^{-1}+1)}
%      {\Gamma((N-j+1)\alpha^{-1})\Gamma((N-j)\alpha^{-1}+\lambda_j+1)}
 \, ,
\end{align}
where arm$(\square)$ and leg$(\square)$ are respectively the numbers of boxes to the right and bottom of the box $\square$ in the Young diagram of $\lambda$.

Jack polynomials provide a simple decomposition of the product
\begin{align}
\frac{1}{\prod_{i,j=1}^N (1-X_i Y_j)^{\frac{1}{\alpha}}}
= e^{\frac1\alpha \sum_{k=1}^\infty \frac1 k \operatorname{Tr} X^k\operatorname{Tr} Y^k}
= \sum_\lambda \frac{p_\lambda(X) p_\lambda(Y)}{ \left<p_\lambda| p_\lambda \right>_\alpha}
= \sum_\lambda \frac{J_\lambda^{(\alpha)}(X) J_\lambda^{(\alpha)}(Y)}{ \left<J_\lambda^{(\alpha)}\middle| J_\lambda^{(\alpha)} \right>_\alpha} \,.
\label{eq:Jack_exp}
\end{align}
The second equality actually involves the variables $X$ and $Y$ only through their invariants $\operatorname{Tr} X^k,\operatorname{Tr} Y^l$.
It is possible to consider these invariants as independent variables if the number of variables $N$ is sufficiently large, and to set $\operatorname{Tr} Y^k =  -\alpha \delta_{k,1}$. Calling $J_\lambda^{\alpha}(\delta_{k,1})$ the value of $J_\lambda^{\alpha}(Y)$ if $\operatorname{Tr} Y^k =  \delta_{k,1}$, we obtain
\begin{equation}
e^{- \operatorname{Tr} X}
= \sum_\lambda (-\alpha)^{|\lambda|} \ \frac{J_\lambda^{(\alpha)}(X) J_\lambda^{\alpha}(\delta_{k,1})}{ \left<J_\lambda^{(\alpha)}\middle| J_\lambda^{(\alpha)} \right>_\alpha} \,.
\label{eq:Jack_exp1}
\end{equation}
Finally, for $\frac{2}{\alpha}\in\{1,2,4\}$,
the Jack polynomials obey
\begin{equation}\label{eq:JackintUN}
\int_{U_N^{\frac2\alpha}} J^{(\alpha)}_\lambda(U X U^\dagger Y) dU = \operatorname{Vol}(U_N^{\frac2\alpha})\,
 \frac{J_\lambda^{(\alpha)}(X) J_\lambda^{(\alpha)}(Y)}{ J_\lambda^{(\alpha)}(\mathrm{ Id}_N)}\ ,
\end{equation}
where $dU$ is the Haar measure on $U_N^{\frac2\alpha}$, and $\operatorname{Vol}(U_N^{\frac2\alpha})=\int_{U_N^{\frac2\alpha}} dU$ is its volume.

\subsubsection{Expansion of Itzykson--Zuber integrals in terms of Jack polynomials}

Combining \eqref{eq:Jack_exp1} with \eqref{eq:JackintUN}, we obtain an expression for
Itzykson--Zuber integrals in terms of Jack polynomials \cite{Desrosiers:2008tp},
\begin{align} \label{eq:IZJack}
 \boxed{\mathcal{Z}(X, Y)
  = 
\operatorname{Vol}(U_N^{\beta})
\sum_\lambda (-1)^{|\lambda|}
\left[\prod_{j=1}^N
 \frac{\Gamma((N-j+1)\frac{\beta}{2})}{\Gamma((N-j+1)\frac{\beta}{2}+\lambda_j)}\right]
 J_\lambda^{(\frac{2}{\beta})}(X) J_\lambda^{(\frac{2}{\beta})}(Y)
}\ .
\end{align}
Although we derived it for $\beta\in\{1,2,4\}$, this expression holds for $\beta\in \mathbb R_+$, provided $\mathcal{Z}(X,Y)$ is defined as the solution of the Calogero--Moser equation.
This expression should be understood as a Taylor expansion near $X=Y=0$. In contrast to the Harish-Chandra formula, this expression involves an infinite sum, which moreover does not converge fast. 

Combining Eq.~\eqref{eq:IZJack} with the orthogonality of Jack polynomials, we obtain
\begin{align}
 \mathcal Z(X,Y) = \sum_\lambda 
 \frac{\Big< J_\lambda^{(\frac{2}{\beta})} \Big| \mathcal{Z}(\cdot, Y) \Big>_{\frac{2}{\beta}} }
 { \Big< J_\lambda^{(\frac{2}{\beta})} \Big| J_\lambda^{(\frac{2}{\beta})} \Big>_{\frac{2}{\beta}} }
 \, J_\lambda^{(\frac{2}{\beta})}(X)\ .
\end{align}
Using Eq.~\eqref{eq:Jack_exp} and the definition of the scalar product, we deduce the duality equation
\begin{align}\label{eq:dualityIZJack}
\boxed{ \mathcal Z(X,Y)
=  \frac{(2\pi i)^{-N}}{N!} \int_{(S^1)^N} \frac{d\Lambda}{\det \Lambda} \ \frac{(\Delta(\Lambda) \Delta(\Lambda^{-1}))^{\frac \beta 2}}
{\prod_{i,j} (1-X_i\Lambda_j^{-1})^{\frac\beta 2}}
\mathcal Z(\Lambda,Y) } \ .
\end{align}
This duality equation holds as an expansion near $X=Y=0$. In particular, since the integrand has poles at $\Lambda_j=X_i$, $X_i$ should be inside the integration contour $S^1$. This contour can however be deformed, so long the poles at $\Lambda_j=0$ and $\Lambda_j=X_i$ stay inside.

\subsection{Lagrange multipliers and recursion relation}
\label{sec:LagrangeUNbeta}

As in \autoref{sec:gce}, let us realize our circular ensemble $U_N^\beta$ as a set of size $N$ matrices with real, complex or quaternionic coefficients depending on $\beta\in \{1,2,4\}$. Let $U_{i,j}^{(\alpha)}$ be the real components of these coefficients, and let $dU_{\mathrm{Lebesgue}} = \prod_{i,j=1}^N \prod_{\alpha=0}^{\beta-1}
d U^{(\alpha)}_{i,j}$ be the Lebesgue measure on the space $\mathbb{R}^{\beta N^2}$ of these components.
Since the Lebesgue measure is invariant under left and right multiplications by elements of $U_N^\beta$, the Haar measure $dU$ on $U_N^\beta$ can be rewritten as 
\begin{equation}
 dU = dU_{\mathrm{ Lebesgue}} \delta(\mathrm{ Id} - U U^\dagger )\ .
\end{equation}
Let us rewrite the multidimensional Dirac delta function through its Fourier transform, i.e. as an integral over a Lagrange multiplier. Since $\mathrm{ Id} - U U^\dagger$ is self-conjugate, the Lagrange multiplier is an element of $E_N^\beta$, and 
\begin{equation}
\delta(\mathrm{ Id}-U U^\dagger )  = \frac{1}{(2\pi)^{\dim E_N^\beta}} \int_{E_N^\beta} dS\
e^{i\operatorname{Tr} S(\mathrm{ Id}- U  U^\dagger  )}\ .
\end{equation}

\subsubsection{Another duality equation}

Using Lagrange multipliers,
we rewrite the Itzykson--Zuber integral as 
\begin{align}
 \mathcal{Z}(X,Y)  = \frac{1}{(2\pi)^{\dim E_N^\beta}} \int_{\mathbb{R}^{\beta N^2}} dU_{\mathrm{Lebesgue}} e^{-\operatorname{Tr} UXU^\dagger Y} \int_{E_N^\beta}dS\
e^{i\operatorname{Tr} S(\mathrm{ Id}- U  U^\dagger  )}\ .
\end{align}
In terms of the vectors $v_k = (U_{1,k},\dots,U_{N,k}) \in \mathbb R^{\beta N}$, this can be reformulated as 
\begin{align}
 \mathcal{Z}(X,Y)  = \frac{1}{(2\pi)^{\dim E_N^\beta}} \int_{E_N^\beta}dS\ e^{i\operatorname{Tr} S} \prod_{k=1}^N \int_{\mathbb{R}^{\beta N}} dv_k\ e^{- v_k^\dagger (iS + X_k Y) v_k }\ .
 \label{zxylag}
\end{align}
Performing the Gaussian integrals over the vectors $v_k$, and changing the integration variable $S\to  Y^{\frac12} S Y^{\frac12}$, we obtain
\begin{equation}
 \mathcal{Z}(X, Y) = \frac{(2\pi)^{N\left(\frac{\beta}{2}-1\right)} 2^{-N^2 \frac\beta 2}}{\det Y^{\frac{\beta}{2}-1}} \
 \int_{E_N^\beta} dS\ 
  \frac{e^{i\operatorname{Tr} SY}}{\prod_{k=1}^N\det (iS+X_k )^{\frac{\beta}{2}}}\ .
  \label{zxyie}
\end{equation}
(For $\beta=4$ the determinant should be interpreted as a product of singular values, i.e. a product where each eigenvalue is counted only once.)

In this formulation of the matrix integral, the Haar measure $dU$ has been replaced with the flat measure $dS$. 
The resulting integral is itself of the type of Eq.~\eqref{eq:zdm}, so that diagonalizing $S=U \Lambda U^\dagger$ leads to an angular integral which is again an Itzykson--Zuber integral. 
This leads to the duality equation
\begin{equation}
 \mathcal{Z}(X, Y) = \frac{(2\pi)^{N\left(\frac{\beta}{2}-1\right)}\ 2^{-N^2 \frac\beta 2}}{\det Y^{\frac{\beta}{2}-1}} \
 \int_{\mathbb{R}^N} d\Lambda\ 
\frac{|\Delta(\Lambda)|^{\beta} }{\prod_{k=1}^N\det (i\Lambda+X_k )^{\frac{\beta}{2}}}\ \mathcal{Z}(-i\Lambda, Y)  \ .
\label{eq:dualityIZLagrange}
\end{equation}
This is similar to our previous duality equation \eqref{eq:dualityIZJack}, which  was however valid near $X=Y=0$. The present equation was derived for $X_i,Y_i>0$, and is valid provided the integral converges, which requires $\Re X_i,\Re Y_i>0$ sufficiently large.
Relating the two duality equations would involve exchanging the integral with the expansion near $X=Y=0$, and is non-trivial.

\subsubsection{Recursion relation}

Let us use Lagrange multipliers in order to derive a recursion relation in the matrix size $N$. 
The constraint $UU^\dagger =\mathrm{Id}$ means that the vectors $v_k = (U_{1,k},\dots,U_{N,k})$ form an orthonormal basis $v_k^\dagger v_l=\delta_{k,l}$. Given $(v_1,\dots ,v_{N-1})$, this fixes $v_N$ up to a rotation in $U_1^\beta$. Instead of introducing a Lagrange multiplier for $v_N$, let us perform the integration over $U_1^\beta$.
Before doing that, we will eliminate the dependence of the Itzykson--Zuber integrand on $v_N$ by shifting $Y$ in order to make its $N$th component vanish. Renaming $ \mathcal{Z}(X, Y)  \to  \mathcal{Z}_N(X, Y)$ to make the $N$-dependence explicit, we thus have
\begin{multline}
 \mathcal{Z}_N(X, Y) 
 = e^{-Y_N\operatorname{Tr} X} \mathcal{Z}_N(X, Y-Y_N\textrm{Id}) \ ,
 \\
 = \mathrm{Vol}(U_1^\beta)  e^{-Y_N\operatorname{Tr} X}
 \left(\prod_{k=1}^{N-1}\int_{\mathbb{R}^{\beta N}} dv_k\  e^{-(Y_k-Y_N)v_k^\dagger Xv_k} \right) \prod_{k,l=1}^{N-1}\delta(\delta_{k,l}-v_k^\dagger v_l)\ .
\end{multline}
The integral is now over $N-1$ vectors $v_k$ of size $N$. The Lagrange multiplier $S$ however only depends on the number of vectors, and therefore belongs to $E^\beta_{N-1}$. Performing the Gaussian integration over the vectors $v_k$, and changing the integration variable $S\to (Y-Y_N\mathrm{Id})^{\frac12} S (Y-Y_N\mathrm{Id})^{\frac12}$, we find 
\begin{align}
 \mathcal{Z}_N(X, Y) 
 &=
\frac{\mathrm{Vol}(U_1^\beta)}{(2\pi)^N 2^{\frac{\beta N (N-1)}{2}}} 
 \frac{e^{-Y_N\operatorname{Tr} X} }{ \prod_{i=1}^{N-1}(Y_i-Y_N)^{\beta-1} }
 \int_{E_{N-1}^\beta} dS\   \frac{e^{i\operatorname{Tr}S(Y-Y_N\mathrm{Id})}}{\prod_{k=1}^N\det(iS+X_k)^\frac{\beta}{2}}\ ,
\end{align}
where $Y$ means either the original size $N$ diagonal matrix, or its size $N-1$ submatrix, depending on the context.
Diagonalizing $S=U \Lambda U^\dagger$ leads to an angular integral which is an Itzykson--Zuber integral over matrices of size $N-1$, 
and we end up with the recursion relation
\begin{align}\label{eq:recurIZ}
\boxed{
 \mathcal{Z}_N(X, Y) \propto 
  \frac{e^{-Y_N\operatorname{Tr} X} }{ \prod_{i=1}^{N-1}(Y_i-Y_N)^{\beta-1} }
 \int_{\mathbb{R}^{N-1}}d\Lambda\ 
 \frac{|\Delta(\Lambda)|^\beta e^{-iY_N\operatorname{Tr}\Lambda} }{\prod_{k=1}^N\det(i\Lambda+X_k)^\frac{\beta}{2}} \mathcal{Z}_{N-1}(-i\Lambda, Y) } \ ,
\end{align}
where we omitted $X,Y$-independent prefactors. (See \autoref{exo:izuto} for the case $N=2$.)

\subsubsection{Analytic properties of Itzykson--Zuber integrals}

Together with the initial condition $\mathcal Z_1(X,Y) = \mathrm{Vol}(U_1^\beta) e^{-X_1 Y_1}$, 
the recursion relation can be used for defining and computing Itzykson--Zuber integrals for real values of $\beta\in \mathbb R_+$.

If $\beta \in 2\mathbb{N}^*$,
the integral over $\Lambda_j$ can be reduced to a sum of residues at $\Lambda_j =  iX_{\sigma(j)}$, parametrized by maps $\sigma:\{1,\dots N-1\}\to \{1,\dots N\}$. 
Due to the factor $|\Delta(\Lambda)|^\beta$, only injective maps give nonzero contributions. The residues are at poles of order $\frac\beta 2$ and are thus combinations of derivatives of order up to $\frac\beta 2 -1$, so that the order of poles is at most $\beta-1$.
By an easy recursion, there must exist a  polynomial  $\check{\mathcal{Z}}_N(X, Y)\in \mathbb Z[X,Y,\frac\beta 2]$ of degree at most $\leq (N-1)(\frac\beta 2-1)$ in each variable, such that 
\begin{equation}
\mathcal{Z}_N(X, Y)  \underset{\beta \in 2\mathbb{N}^*}{=}
\frac{\mathrm{Vol}(U_1^\beta)^N}{(2\pi)^{N} 2^{\frac{\beta N (N-1)(N+1)}{6}}} 
\sum_{\sigma \in \mathfrak S_N} 
\frac{\prod_{i=1}^N e^{-\sum_i X_i Y_{\sigma(i)}}}{\left(\Delta(X)\Delta(\sigma(Y))\right)^{\beta-1}}
\,\check{\mathcal{Z}}_N(X, \sigma(Y))  \ .
\end{equation}
This implies that there exists a rational function $\hat{\mathcal{Z}}_N(X, Y)= \frac{\check{\mathcal{Z}}_N(X, Y)}{ (\Delta(X)\Delta(Y))^{\frac \beta 2-1}}$, with poles only at coinciding points, and bounded at large $X$ or $Y$, such that 
\begin{equation}
\boxed{
\mathcal{Z}_N(X, Y)  = 
\frac{\mathrm{Vol}(U_1^\beta)^N}{(2\pi)^N 2^{\frac{\beta N (N-1)(N+1)}{6}}}
\sum_{\sigma \in \mathfrak S_N} 
\frac{\prod_{i=1}^N e^{-\sum_i X_i Y_{\sigma(i)}}}{\left(\Delta(X)\Delta(\sigma(Y))\right)^{\frac\beta 2}}
\,\hat{\mathcal{Z}}_N(X, \sigma(Y)) } \ .
\label{eq:zzh}
\end{equation}
Our rational function obeys the recursion relation
%\begin{equation}
%\hat{\mathcal Z}_N(X,Y) 
%= \frac{\Delta(X)^{\frac{\beta}{2}}}{\prod_{j=1}^{N-1} (Y_j-Y_N)^{\frac{\beta}{2} -1}}
% \mathop{\operatorname{Res}}_{\Lambda_j\to X_j} \frac{\Delta(\Lambda)^{\frac\beta 2} e^{\sum_{j=1}^{N-1} (Y_j-Y_N)(X_j-\Lambda_j)}}{\prod_{i=1}^N\prod_{j=1}^{N-1} (\Lambda_j-X_i)^{\frac\beta 2}}   \hat{\mathcal Z}_{N-1}(\Lambda,Y) \ .
%\end{equation}
%
%\begin{equation}
%\hat{\mathcal Z}_N(X,Y) 
%=
%\sum_{0\leq k_1,\dots,k_{N-1}< \frac\beta 2}
% \prod_{j=1}^{N-1} \frac{ (Y_j-Y_N)^{-k_j} \Gamma(\frac\beta 2)}{ k_j! \Gamma(\frac \beta 2 - k_j)} \left. \partial_{\Lambda_j}^{k_j}
% \frac{ \Delta(X)^{\frac{\beta}{2}}\Delta(\Lambda)^{\frac\beta 2} \hat{\mathcal Z}_{N-1}(\Lambda,Y)}{\prod_{i=1}^N\prod_{j=1, \neq i}^{N-1} (\Lambda_j-X_i)^{\frac\beta 2}}   \right|_{\Lambda_j=X_j} \ .
%\end{equation}
%
\begin{multline}
\hat{\mathcal Z}_N(X,Y) 
=
 \prod_{j=1}^{N-1} \left(\sum_{k_j=0}^\infty\sum_{m_j=0}^\infty   \frac{ \tau_{N,j}^{k_j} \Gamma(\frac\beta 2+k_j)}{ 2^{k_j} k_j! m_j! \Gamma(\frac \beta 2 - k_j-m_j)} (Y_j-Y_N)^{-m_j}\partial_{\Lambda_j}^{m_j} \right) \\
 \left.  
 \frac{ \Delta_{N-1}(X)^{\frac{\beta}{2}}\Delta(\Lambda)^{\frac\beta 2} \hat{\mathcal Z}_{N-1}(\Lambda,Y)}{\prod_{i=1}^{N-1}\prod_{j=1, \neq i}^{N-1} (\Lambda_j-X_i)^{\frac\beta 2}}   \right|_{\Lambda_j=X_j} \ .
\end{multline}
In the case $\beta = 2$, that rational function is a constant, and Eq.~\eqref{eq:zzh} reduces to the Harish-Chandra formula. 
For any $\beta \in 2\mathbb{N}^*$, that rational function is a symmetric polynomial of degree at most $\frac{\beta}{2}$ in the variables 
\begin{equation}
 \tau_{i,j} = -\frac{2}{(X_i-X_j)(Y_i-Y_j)}\ .
\end{equation}
These $\frac12 N(N-1)$ variables are redundant for $N\geq 6$, since they are combinations of the $2N$ variables $X_i$, $Y_i$. 

Actually, even when $\beta$ is not an even integer, the singularities of $\mathcal{Z}_N(X, Y)$ are still described by the formula \eqref{eq:zzh}, where the function $\hat{\mathcal{Z}}_N(X, Y)$ is an analytic function of the variables $\tau_{i,j}$ instead of a polynomial. 
This function can be written explicitly for $N=1,2,3$. To do this, we will use the analytic function
\begin{align}
 y_\alpha(x) = \sum_{k=0}^\infty \frac{\Gamma(\alpha+k+1)}{k! \Gamma(\alpha-k+1)} \left(\frac{x}{2}\right)^k = \sqrt{\frac{2}{\pi x}} e^{\frac{1}{x}} K_{\alpha+\frac12}\left(\frac{1}{x}\right)\ ,
 \label{eq:IZ_Bessel_fn}
\end{align}
where $K_{\alpha+\frac12}$ is the modified Bessel function of the second kind. If $\alpha\in\mathbb{N}$, then $y_\alpha$ is the Bessel polynomial of degree $\alpha$ if $\alpha\in\mathbb{N}$. Using the  recursion relation, we find
\begin{subequations}
\begin{align}
 \hat{\mathcal{Z}}_1(X, Y) & = 1 \ ,
 \\
 \hat{\mathcal{Z}}_2(X, Y) & = y_{\frac{\beta}{2}-1}(\tau_{1,2})\ ,
 \\
 \hat{\mathcal{Z}}_3(X, Y) & = \sum_{k=0}^\infty \frac{\Gamma(\frac\beta 2 -k)}{2^{6k}k!\Gamma(\frac\beta 2+k)}
 \prod_{1\leq i<j\leq 3} y^{(k)}_{\frac\beta 2-1}(\tau_{i,j}) \ ,
\end{align}
\end{subequations}
where $y^{(k)}_{\frac\beta 2-1}$ is the $k$-th derivative of $y_{\frac\beta 2-1}$, which vanishes for $k\geq \frac{\beta}{2}$ if $\beta$ is an even integer.

\section{Exercises}

\begin{exo}[Itzykson--Zuber integral on $U_{N=2}^{\beta =1}$]
~\label{exo:izuto}
Let us study the Itzykson--Zuber integral $\mathcal{Z}(X, Y)$ on the group $U_2^1$. This group is one of the two connected components of the group $O_2$, its elements are rotation matrices 
$
U = \left(\begin{smallmatrix}
\cos\theta & -\sin\theta \\
\sin\theta & \cos\theta
\end{smallmatrix}\right)
$, and 
the Haar measure is $dU=d\theta$.

We assume that $X_1>X_2>0$ and $Y_1>Y_2>0$ and are large enough real numbers.

\begin{enumerate}

\item 
By a direct computation, show that 
\begin{align}
 \mathcal{Z}(X, Y) = e^{-\frac12(X_1+X_2)(Y_1+Y_2)}\hat{\mathcal{Z}}(z) 
 \quad \text{with} \ \left\{\begin{array}{l}
 \hat{\mathcal{Z}}(z)  = 2\pi I_0(z)\, ,
 \\[1em]
 z = \tfrac{1}{2}(X_1-X_2)(Y_1-Y_2)\, ,
 \label{eq:zis}
 \end{array}\right.
\end{align}
where $I_0(z)=\frac{1}{\pi}\int_{-1}^1 \frac{ds}{\sqrt{1-s^2}} e^{sz}$ is the modified Bessel function.
\item
By a change of integration variable, show that this formula agrees with the recursion relation \eqref{eq:recurIZ}, in the case of matrix size $N=2$.
\item 
Show that $\hat{\mathcal{Z}}(z)$ obeys the second-order differential equation
\begin{align}
 \hat{\mathcal{Z}}''(z)+\frac{\hat{\mathcal{Z}}'(z)}{z}-\hat{\mathcal{Z}}(z) = 0\ ,
\end{align}
and that $\mathcal{Z}(X, Y)$ obeys the Calogero--Moser equation.
\end{enumerate}

\end{exo}

\begin{exo}[Proof of the Calogero--Moser equation]
~\label{exodunkl}
We want to fill the gaps in the derivation of the Calogero--Moser equation, by further studying the quadratic moments $M_{i,j}$ \eqref{eq:mij}.

\begin{enumerate}

\item
Introducing Lagrange multipliers as in \autoref{sec:LagrangeUNbeta}, show that
\begin{equation}\label{eq:Mijintexo}
M_{i,j} \propto 
\frac{\beta }{2}
\int_{E_N^\beta} dS \frac{e^{i\operatorname{Tr} S }}{\prod_{k=1}^N \det(iS+X_kY)^{\frac\beta 2}}  \left( (iS+X_j Y)^{-1}\right)_{i,i}\ ,
\end{equation}
where we omit $X, Y$-independent coefficients.
Check that this expression satisfies the last relation in Eq.~\eqref{eq:IZMijrels}.

\item Check that this expression for $M_{i,j}$ satisfies the first relation in Eq.~\eqref{eq:IZMijrels}, by using the loop equation
\begin{equation}
 \int dS \frac{\partial}{\partial S_{j,j}} \left( \frac{e^{i\operatorname{Tr} S }}{\prod_{k=1}^N \det(iS+X_kY)^{\frac\beta 2}} \right) = 0\ .
\end{equation}

\item Using the split rule \eqref{eq:splitbetaW} with $x=x_j$, $A=\mathrm{Id}, B=\mathrm{ diag}(0,\dots,0,\overset{i}{1},0,\dots,0)$ to a matrix integral over  $ M=-iSY^{-1}$, show that the quadratic moments $M_{i,j}$ satisfy the first-order differential equation \eqref{eq:dunkl}.

\end{enumerate}

\end{exo}

\appendix

\chapter{Solutions of exercises}

%\numberwithin{equation}{chapter}

%============ section 1  Introduction =========

\section{Chapter 1}

\paragraph{Solution \ref{exo:gaussian}.}

\begin{enumerate}
\item
Using the identity $|\lambda_1-\lambda_2|^2 = \lambda_1^2 +\lambda_2^2 -2\lambda_1\lambda_2$, we decompose $\mathcal{Z}$ into sums of products of integrals over $\lambda_1$ and $\lambda_2$. The last term gives rise to integrals of odd functions, which therefore vanish. Therefore,
\begin{align}
\mathcal Z = 2 \int_{\mathbb R} \lambda_1^2 e^{-\frac12 \lambda_1^2} \ d\lambda_1 \int_{\mathbb R} e^{-\frac12 \lambda_2^2} \ d\lambda_2  = 4\pi\ .
\end{align}

\item
From its definition, and by invariance under the permutation $\lambda_1\leftrightarrow \lambda_2$, the eigenvalue density is
\begin{align}
 \rho(x) &= \left<\frac{1}{2} \left( \delta(x - \lambda_1) +  \delta(x - \lambda_2) \right) \right> = \left<\delta(x-\lambda_1)\right>\ ,
 \nonumber \\
 & = \frac{1}{\mathcal Z} \int_{\mathbb R}  (x-\lambda_2)^2 e^{-\frac12 (x^2+\lambda_2^2)} \,d\lambda_2 \ .
\end{align}
Expanding the square, we integrate and obtain the result
\begin{align}
 \rho(x) = \frac{e^{-\frac12 x^2} }{2\sqrt{2\pi}}  (x^2+1)\ .
\end{align}
We verify that this density function is properly normalized, $\int_\mathbb{R} \rho(x) dx = 1$.

\item
Viewed as an equation for $\lambda_2$, the constraint $s=|\lambda_1-\lambda_2|$ has the two solutions $\lambda_2=\lambda_1\pm s$. Therefore, $p(s)$ decomposes into two terms,
\begin{align}
p(s) = \frac{1}{\mathcal Z}\left(\int_{\mathbb R} s^2 e^{-\frac12 (\lambda_1^2+(\lambda_1+s)^2)} \,d\lambda_1 + \int_{\mathbb R} s^{2} e^{-\frac12 (\lambda_1^2+(\lambda_1-s)^2)} \,d\lambda_1\right) \ .
\end{align}
The two terms are equal due to the parity symmetry $\lambda_1\to -\lambda_1$, leading to
\begin{align}
p(s) = \frac{2s^2}{\mathcal Z} \int_{\mathbb R} e^{-\lambda_1^2-s\lambda_1-\frac12 s^2} \,d\lambda_1 = \frac{2s^2}{\mathcal Z} \int_{\mathbb R} e^{-(\lambda_1-\frac{s}{2})^2-\frac14 s^2} \,d\lambda_1 = \frac{s^2 e^{-\frac14 s^2}}{2\sqrt{\pi}}\  .
\end{align}
We can check that this probability density is properly normalized $\int_0^\infty p(s) ds = 1$.
Modulo a renormalization $s\to \lambda s$, the result for $p(s)$ coincides with the Wigner surmise for $\beta=2$, which is therefore exact in the case of Gaussian Hermitian matrices of size $2$.
\end{enumerate}

\paragraph{Solution \ref{exo:gaussian2}.}

\begin{enumerate}

\item
In the partition function, we trade the variable $\lambda_2$ for $s=\lambda_2-\lambda_1$, and perform the Gaussian integral over $\lambda_1$:
\begin{align}
\mathcal Z = \int_{\mathbb R^2} |s|^{\beta} e^{-\frac12 (\lambda_1^2+(\lambda_1+s)^2)} \,d\lambda_1 ds = \sqrt{\pi}  \int_{\mathbb R}  |s|^{\beta} e^{-\frac14 s^2}ds  \ .
\end{align}
We then trade the variable $s$ for $u=\frac14 s^2$, and find that the integral yields a Gamma function:
\begin{align}
 \mathcal Z =  2^{\beta+1}\sqrt{\pi}  \int_{0}^\infty du \ u^{\frac12(\beta-1)} e^{-u}  = 2^{\beta+1}\sqrt{\pi}  \Gamma \left(\tfrac{\beta+1}{2}\right)\ .
\end{align}
\item
The calculation is done in exactly the same way as in \autoref{exo:gaussian}, with the result
\begin{align}
p(s)
= \frac{2\sqrt{\pi}}{\mathcal Z} s^{\beta}  e^{-\frac14 s^2}
= \frac{s^{\beta}  e^{-\frac14 s^2}}{2^\beta \Gamma \left(\frac{\beta+1}{2}\right)} \ .
\end{align}
We can check the normalization condition $\int_0^\infty p(s) ds=1$, by trading $s$ for $u=\frac14 s^2$. Again, the Wigner surmise is exact for Gaussian matrices of size $2$.
\end{enumerate}

\paragraph{Solution \ref{exo:Haarvolu2}.}

\begin{enumerate}
 \item Writing $M = \left[\begin{smallmatrix} v & x+iy \\ x-iy & w\end{smallmatrix}\right]$ with $v,w,x,y\in\mathbb{R}$, we compute
 \begin{align}
  \mathcal{Z} = \int_{\mathbb R^4} e^{-\frac12\left(v^2+w^2+2x^2+2y^2\right)}dvdwdxdy = 2\pi^2 \ .
 \end{align}
\item Diagonalizing $M$, we reduce $\mathcal{Z}$ to an integral over the eigenvalues $\lambda_1,\lambda_2$:
\begin{align}
\mathcal{Z} = \frac{1}{2!} \mathrm{Vol}(U_2/U_1^2) \int_{\mathbb R^2} (\lambda_1-\lambda_2)^2 e^{-\frac12(\lambda_1^2+\lambda_2^2)} d\lambda_1 d\lambda_2\ .
\end{align}
Then $\mathrm{Vol}(U_2/U_1^2) = \frac{\mathrm{Vol}(U_2)}{(2\pi)^2}$. Expanding the
square $(\lambda_1-\lambda_2)^2$ reduces the two-dimensional integral over eigenvalues to a combination of one-dimensional integrals, leading to
$
 \mathcal{Z} =  \frac{\mathrm{Vol}(U_2)}{2\pi}
$.
It follows that
\begin{align}
\mathrm{Vol}(U_2) = 4 \pi^3, \quad
\mathrm{Vol}(U_2/U_1) = \mathrm{Vol}(SU_2) = 2 \pi^2, \quad
\mathrm{Vol}(U_2/U_1^2) = \pi\ .
\end{align}
The volume of $SU_2$ agrees with that of unit three-sphere $S^3$, due to their equivalence.
\end{enumerate}

\paragraph{Solution \ref{exo:volu2}.}

\begin{enumerate}
 \item Using the change of variables $\lambda_i=e^{i\theta_i}$, and its consequence $\bar\lambda_i=\frac{1}{\lambda_i}$, we compute
 \begin{align}
  |\lambda_1-\lambda_2|^2 d\theta_1d\theta_2 = \left(\frac{\lambda_1-\lambda_2}{\lambda_1\lambda_2}\right)^2 d\lambda_1d\lambda_2\ .
 \end{align}
 Comparing with the measure $dM = dU_\text{Haar} (\lambda_1-\lambda_2)^2 d\lambda_1d\lambda_2$ \eqref{eq:dMnormal}, we indeed recover the case $N=2$ of Eq.~\eqref{eq:circular_Haar}.
\item We compute
\begin{align}
 \frac{\int_{U_2} dM_\text{Haar}}{\int_{U_2/U_1^2} dU_\text{Haar}} = \int_{S^1\times S^1}\left(\frac{\lambda_1-\lambda_2}{\lambda_1\lambda_2}\right)^2 d\lambda_1d\lambda_2 = 8\pi^2 \ .
\end{align}
We have expanded the square $(\lambda_1-\lambda_2)^2 = \lambda_1^2 -2\lambda_1\lambda_2+\lambda_2^2$, where only the second term yielded a nonzero residue, and we have used $\int_{S^1}\frac{d\lambda}{\lambda}=2\pi i$. The result is $8\pi^2 = 2!\mathrm{Vol}(U_1^2)$.
\end{enumerate}

\paragraph{Solution \ref{exo:DumiEdel}.}

\begin{enumerate}

\item This measure is the product of two normal Gaussian distributions for $a$ and $b$, and one Chi-distribution for $u$.

\item These identities follow from $\operatorname{Tr}M=a+b=\lambda_1+\lambda_2$ and $\det M = ab-u^2=\lambda_1\lambda_2$.

\item We first write the measure in terms of $\lambda_1,\lambda_2,u$:
\begin{align}
 d\mu(M)
=  \frac{1}{\pi \Gamma(\frac{\beta}{2})} e^{-\frac12 (\lambda_1^2+\lambda_2^2) }  \frac{|\lambda_1-\lambda_2|}{\sqrt{(\lambda_1-\lambda_2)^2 - 4 u^2}}  u^{\beta-1} \ d\lambda_1 \wedge d\lambda_2 \wedge du\ .
\end{align}
Then we perform the change of variable from $u$ to $y$.
\item
The integral over $y$ is done using the formula
\begin{align}
\int_0^1 \frac{ y^{\beta-1}}{\sqrt{1-y^2}} dy
= \sqrt{\pi} \frac{\Gamma(\frac{\beta}{2})}{\Gamma(\frac{\beta+1}{2})}\ .
\end{align}
The result agrees with Eq.~\eqref{dml} by the Legendre duplication formula for the Gamma function.
\end{enumerate}

\paragraph{Solution \ref{exo:dotsenko}.}

\begin{enumerate}

\item For a M\"obius transformation
$
f(z)= \frac{Az+B}{Cz+D}
$ with $AD-BC=1$, we have
\begin{align}
f'(z) = \frac{1}{(Cz+D)^2}\ ,
\quad
f(z)-f(z') = \frac{z-z'}{(Cz+D)(Cz'+D)}\ .
\end{align}
Applying the change of variables $x_i\mapsto f(x_i)$, $z_i\mapsto f(z_i)$, the integral behaves as
\begin{align}
\mathcal Z \mapsto
&  \int_{\mathbb C^N} \prod_{i=1}^N |f'(x_i)|^2 dx_id\bar x_i \,\prod_{i<j}^N |f(x_i)-f(x_j)|^{2\beta} \prod_{i=1}^N\prod_{j=1}^n |f(x_i)-f(z_j)|^{-4b\alpha_j} \ ,
\nonumber \\
& =  \prod_{j=1}^n |Cz_j+D|^{4b\alpha_j}\int_{\mathbb C^N} \prod_{i=1}^N |Cx_i+D|^{4b\Xi} dx_id\bar x_i
\nonumber \\  & \hspace{4cm}
\prod_{i<j}^N \left|x_i-x_j\right|^{2\beta} \prod_{i=1}^N\prod_{j=1}^n \left|x_i-z_j\right|^{-4b\alpha_j} \ .
\end{align}
The factors $|Cx_i+D|^{4b\Xi}$ become trivial if $\Xi=0$, in which case $\mathcal{Z}$ is conformally covariant:
\begin{align}
\mathcal Z(f(z_1),\dots,f(z_n))  \prod_{i=1}^n |f'(z_i)|^{-2bN\alpha_i}
= \mathcal Z(z_1,\dots,z_n)\ .
\end{align}

\item Under the assumptions, $\mathcal{Z}_\gamma$ is the integral over $x_1$ of a rational function of $x_1$, with poles at $z_j$. The integral is nonzero only if the integration contour $\gamma$ surrounds one or more poles.

Let $\mathcal C_{z_j}$ be a small circle that surrounds $z_j$, and no other pole. Then
$\sum_{j=1}^n \mathcal C_{z_j}$ is contractible, therefore $\sum_{j=1}^n\mathcal{Z}_{\mathcal C_{z_j}}=0$, and only $n-1$ of the $n$ integrals $\mathcal{Z}_{C_{z_j}}$ are independent.
A basis of the relevant space of contours is therefore $\mathcal C_{z_1},\dots,\mathcal C_{z_{n-1}}$.

This implies that $\mathcal{Z}$ is a bilinear combination of residues at $x_i=z_k$ and their complex conjugates.

\item
Let us write $2b\alpha_i=k_i\in\mathbb{N}$ for $i=1,2$, then the conformal covariance condition is $k_1+k_2 = 2N$. We compute $\mathcal{Z}$ by integrating each variable $x_i$ over the contour $\mathcal C_{z_1}$, and find $\mathcal Z= |\mathcal{Z}_1|^2$ with
\begin{align}
\mathcal{Z}_1&= (2\pi i)^N \left(\prod_{i=1}^N \res_{x_i\to z_1} dx_i\right) \frac{\prod_{i<j}^N (x_i-x_j)^{2}}{ \prod_{i=1}^N (x_i-z_1)^{k_1}(x_i-z_2)^{k_2}}\ .
\end{align}
We can already determine the dependence on $z_1,z_2$ thanks to the change of variables $x_i=z_1+z_{21}y_i$:
\begin{align}
\mathcal{Z}_1
= (-2\pi i)^N z_{21}^{-N^2} A_{k_1,k_2} ,
\ \text{with} \quad
A_{k_1,k_2} = \left(\prod_{i=1}^N \res_{y_i\to 0} dy_i\right)  \frac{\prod_{i<j} (y_i-y_j)^{2}}{ \prod_{i=1}^N y_i^{k_1}(1-y_i)^{k_2}} .
\end{align}
Now
$\prod_{i<j}^N (y_i-y_j)^{2}$ is a symmetric polynomial of $N$ variables, of total homogeneous degree $N(N-1)$, and maximal degree in each variable $2(N-1)$.
Only the terms of degrees $d_i<k_1$ in the variable $y_i$ may yield a non-zero residue.
This implies that $N(N-1) = \sum_{i=1}^N d_i < k_1 N$, thus the result can be nonzero only if $k_1\geq N$.
By permutation symmetry we must also have $k_2\geq N$. Since $k_1+k_2=2N$, we deduce
\begin{align}
A_{k_1,k_2} = A_{N,N} \delta_{k_1,N} \delta_{k_2,N}.
\end{align}
In order to compute $A_{N,N}$, we replace the $N$-th order pole at $y_i=0$ with $N$ first-order poles $(a_j)_{j=1,\dots,N}$, before taking the limit $a_j\to 0$. Each $y_i$ tends to a pole, and in fact each $y_i$ must tend to a different pole, otherwise the result is zero due to the factor $\prod_{i<j} (y_i-y_j)^{2}$. Any permutation $\sigma \in \mathfrak{S}_N$ leads to a contribution of the type $\res_{y_i\to a_{\sigma(i)}}dy_i\cdots$, but all these $N!$ contributions are equal, so that
\begin{align}
    A_{N,N} &= N! \lim_{a_i \to 0} \left(\prod_{i=1}^N \res_{y_i \to a_i} dy_i\right)   \frac{\prod_{i<j} (y_i-y_j)^{2}}{\prod_{i,j = 1}^N (y_i - a_j) \prod_{i=1}^N(1-y_i)^{N}}\ , \nonumber \\
    & =  N! \lim_{a_i \to 0} \frac{\prod_{i<j} (a_i-a_j)^{2}}{\prod_{i \neq j}^N (a_i - a_j) \prod_{i=1}^N(1-a_i)^{N}}  = (-1)^{N(N-1)/2} N!\ ,
\end{align}
which leads to the announced result for $\mathcal{Z}$.
\end{enumerate}

\paragraph{Solution \ref{exo:duality}.}

The integral decomposes into a product of Gaussian integrals over the matrix elements $M_{i,j}$,
\begin{align}
\mathcal Z(N,\beta)
&= 2^{-\frac{N}{2}} \prod_{i=1}^N \int_\mathbb{R} dM_{i,i} e^{-N\sqrt{\frac \beta 2} M_{i,i}^2} \, \prod_{i<j}^N \prod_{\alpha=0}^{\beta-1} \int_\mathbb{R} dM_{i,j}^{(\alpha)} e^{-2N\sqrt{\frac \beta 2} (M_{i,j}^{(\alpha)})^2} \ , \nonumber \\
& = 2^{-\frac{N}{2}}   \left(\frac{N}{\pi}\sqrt{\frac \beta 2} \right)^{-N/2} \left(\frac{2N}{\pi}\sqrt{\frac \beta 2} \right)^{-N(N-1)\beta/4} \ .
\end{align}
This leads to the announced result, which is manifestly invariant under the duality $\epsilon_1\leftrightarrow \epsilon_2$, equivalently
$\mathcal Z(N,\beta) = \mathcal Z(-\frac{N\beta}{2},\frac{4}{\beta}) $.

\paragraph{Solution \ref{exo:jones}.}

\begin{enumerate}
\item
In the case $n=2$, let us rewrite the hyperbolic factors in the integrand of $\mathcal Z_{\mathfrak K_{P,Q},\emptyset}(q)$ in termes of exponentials:
\begin{multline}
 4\sinh\tfrac{P}{2}(u_1-u_2)\sinh\tfrac{Q}{2}(u_1-u_2) = e^{\frac{P+Q}{2} u_1} e^{-\frac{P+Q}{2} u_2} + e^{-\frac{P+Q}{2} u_1} e^{\frac{P+Q}{2} u_2}
 \\
	- e^{\frac{P-Q}{2} u_1} e^{-\frac{P-Q}{2} u_2} - e^{-\frac{P-Q}{2} u_1} e^{\frac{P-Q}{2} u_2}\ .
	\label{4shsh}
\end{multline}
In each one of the 4 terms, the dependences on $u_1$ and $u_2$ factorize. As a result, the integral can be written in terms of the Gaussian integral $G(a)$ as
\begin{align}
 \mathcal Z_{\mathfrak K_{P,Q},\emptyset}(q) = 2G(\tfrac{P+Q}{2})^2 - 2 G(\tfrac{P-Q}{2})^2=
 \frac{4\pi g_s}{PQ} q^{- \frac{1}{4} \left( \frac{P}{Q} + \frac{Q}{P} \right)} \left( q^{-\frac{1}{2}} - q^{\frac{1}{2}} \right)\ .
\end{align}
\item
In the case of the Jones polynomial, the relevant integral is
\begin{multline}
 \mathcal Z_{\mathfrak K_{P,Q},(1,0)}(q) = \int_{\mathbb{R}^2} du_1du_2\, e^{-\frac{PQ}{2g_s}(u_1^2+u_2^2)} \left(e^{PQu_1}+e^{PQu_2}\right)
 \\
\times 4\sinh\tfrac{P}{2}(u_1-u_2)\sinh\tfrac{Q}{2}(u_1-u_2) \ .
\end{multline}
Again, Eq.~\eqref{4shsh} allows us to write this integral in terms of the Gaussian integral $G(a)$,
\begin{align}
	\mathcal Z_{\mathfrak K_{P,Q},(1,0)}(q) & =
	2G\left(\tfrac{P+Q}{2} + PQ)G(\tfrac{P+Q}{2}\right)
	+ 2G\left(-\tfrac{P+Q}{2} + PQ\right)G\left(\tfrac{P+Q}{2}\right)
	\nonumber \\
  & \quad - 2G\left(\tfrac{P-Q}{2} + PQ\right)G\left(-\tfrac{P-Q}{2}\right)
	- 2G\left(-\tfrac{P-Q}{2} + PQ\right)G\left(\tfrac{P-Q}{2}\right)\ ,
	\nonumber \\
	& =
	\frac{4\pi g_s}{PQ} q^{- \frac{1}{4} \left( \frac{P}{Q} + \frac{Q}{P} \right)}
	q^{-\frac{1}{2} (P+1)(Q+1)}
	\left( q^{P+Q} - q^{P+1} - q^{Q+1} +1\right) .
\end{align}
This leads to the announced result for the Jones polynomial
\begin{align}
 J_{\mathfrak K_{P,Q}}(q) = \frac{q^{PQ}}{q}
	\frac{ \mathcal Z_{\mathfrak K_{P,Q},(1,0)}(q) }{ \mathcal Z_{\mathfrak K_{P,Q},\emptyset}(q) }
	\frac{ \mathcal Z_{\mathfrak K_{1,1},\emptyset}(q) }{ \mathcal Z_{\mathfrak K_{1,1},(1,0)}(q) }\ .
\end{align}
\item
In the case of the unknot $P=Q=1$, the hyperbolic factors are
\begin{align}
 \prod_{i<j} 4\sinh^2\tfrac12(u_i-u_j) = \prod_{i<j} (e^{u_i}-e^{u_j})(e^{-u_j}-e^{-u_i})\ ,
\end{align}
whereas the Vandermonde determinant in the denominator is
\begin{align}
 \det_{1\leq i,j\leq n} e^{(n-i)u_j} = \prod_{i<j} (e^{u_i}-e^{u_j})\ .
\end{align}
This leads to the simplified integral
\begin{align}
\mathcal Z_{\mathfrak K_{1,1},\lambda}(q)
= \int_{\mathbb R^n} du_1 \dots du_n\ e^{-\frac{1}{2g_s} \sum_i u_i^2} \,
\left(\det_{1\leq i,j\leq n} e^{(\lambda_i-i+n) u_j}\right)
\prod_{i<j} (e^{-u_j}-e^{-u_i})\ .
\label{eq:simint}
\end{align}
\item
In the case $n=2$ with $\lambda=(N-1,0)$, the exponential factors in the simplified integral reduce to $(e^{Nu_1}-e^{Nu_2})(e^{-u_2}-e^{-u_1})$. Expanding the product, we find
\begin{align}
 \mathcal Z_{\mathfrak K_{1,1},(N-1,0)}(q) = 2G(N)G(1)-2G(N-1)G(0)=4 \pi g_s q^{-\frac{(N-1)^2}{2}} (q^{-N} - 1)\ ,
\end{align}
and we deduce the coloured Jones polynomial of the unknot,
\begin{align}
	J_{\mathfrak K_{1,1},N} (q)
	& = \frac{ \mathcal Z_{\mathfrak K_{1,1},(N-1,0)}(q) }{ \mathcal Z_{\mathfrak K_{1,1},\emptyset}(q) }
	= q^{-\frac{(N-1)^2}{2}} \frac{q^{-N} - 1}{q^{-1} - 1}
        = q^{-\frac{N(N-1)}{2}} \frac{q^{-\frac{N}{2}} - q^{\frac{N}{2}}}{q^{-\frac{1}{2}} - q^{\frac{1}{2}}}\ .
\end{align}
This is a polynomial in $q^{-1}$ if $N$ is odd, or in $q^{-\frac12}$ if $N$ is even.
\item We need to compute the simplified integral \eqref{eq:simint} in the case $n=3$ and $\lambda=(N-1,0,0)$. By permutation symmetry, the $6$ terms of the size $3$ determinant all lead to the same integral, and we may replace the determinant with $6 e^{(N+1)u_1}e^{u_2}$. Expanding
\begin{multline}
 \prod_{i<j} (e^{-u_j}-e^{-u_i})
 \\
 = e^{-2u_1-u_3}+e^{-2u_2-u_1}+e^{-2u_3-u_2}-e^{-2u_1-u_2}-e^{-2u_2-u_3}-e^{-2u_3-u_1}\ ,
\end{multline}
we evaluate the contribution of each term to the simplified integral, and find
\begin{multline}
 \mathcal Z_{\mathfrak K_{1,1},(N-1,0,0)}(q)
= 6\Big(G(N-1)G(1)^2+G(N)G(1)G(0)+G(N+1)G(2)G(0)
\\
-G(N-1)G(0)^2-G(N+1)G(1)^2-G(N)G(2)G(1)\Big)\ .
\end{multline}
Explicitly, this is
\begin{align}
 \mathcal Z_{\mathfrak K_{1,1},(N-1,0,0)}(q)
= -6 (2\pi g_s)^\frac32 q^{-\frac{(N-1)^2}{2}}\left(1-q^{-1}\right)\left(1-q^{-N}\right)\left(1-q^{-N-1}\right)\ .
\end{align}
To obtain the HOMFLY polynomial, it remains to divide this integral by its value for $N=1$, and we find
\begin{align}
 \mathrm{HOMFLY}_{\mathfrak K_{1,1},(N-1,0,0)}(q)= q^{-\frac{N^2-1}{2}} \frac{q^{-\frac{N}{2}}-q^\frac{N}{2}}{q^{-\frac12}-q^\frac12}\cdot \frac{q^{-\frac{N+1}{2}}-q^{\frac{N+1}{2}}}{q^{-1}-q}\ .
\end{align}

\end{enumerate}

%============ section 2  Ribbon graphs =========

\section{Chapter 2}

\paragraph{Solution \ref{exo:kontsevich}.}

\begin{enumerate}
    \item
   In the exponent, we write the trace in terms of matrix elements:
    \begin{align}
        \operatorname{Tr} M^2 \Lambda
        = \sum_{1 \le i,j,k \le N} M_{ij} M_{jk} \lambda_{k} \delta_{ki}
        = \sum_{i=1}^N \lambda_i M_{ii} ^2 + \sum_{1 \le i < j \le N} (\lambda_i + \lambda_j) |M_{ij}|^2 .
        \label{eq:Kontsevich_model_quadratic_part}
    \end{align}
    Hence, diagonal elements lead to real Gaussian integrals, while non-diagonal elements to complex Gaussian integrals, and we find
    \begin{align}
        \mathcal Z_0(\Lambda,t) & = \left( \frac{t^{\frac{1}{3}} \pi}{N}  \right)^{N(N-\frac{1}{2})} \frac{1}{\sqrt{\det \Lambda}} \prod_{ 1 \le i < j \le N} \frac{1}{\lambda_i + \lambda_j} ,
     \label{eq:Kontsevich_model_Gaussian_part}
    \end{align}
 where $\det \Lambda = \prod_{i=1}^N \lambda_i$.

    \item
    We denote the expectation value with respect to the Gaussian integral $\mathcal{Z}_0(\Lambda,t)$ by $\left< \cdot \right>_0$.
%    \begin{align}
%     \left< \mathcal{O} \right>_0 = \frac{1}{\mathcal{Z}_0(\Lambda,t)} \int_{H_N} dM \, \mathcal{O} \, e^{-N t^{-\frac{1}{3}} \Tr M^2 \Lambda} . 
%    \end{align}
    By Wick's theorem, the propagator is
    \begin{align}
        \left< M_{ij} M_{kl} \right>_0 =
        \frac{1}{\lambda_i + \lambda_j} \frac{t^{\frac{1}{3}}}{N} \delta_{il} \delta_{jk} . \label{eq:KW_propagator}
    \end{align}    
    Then, the formal series expansion \eqref{eq:koni_exp} is written using the correlation functions,
    \begin{align}
     \mathcal{Z}(\Lambda,t) = \sum_{k=0}^\infty \frac{N^k}{3^k k!} \left< (\operatorname{Tr} M^3)^k \right>_0 .
    \end{align}
    Applying 't Hooft--Br\'{e}zin--Itzykson--Parisi--Zuber's theorem \eqref{top_exp01} to this expansion, the integral is a sum over ribbon graphs $G$ with trivalent vertices.
    Since the propagator \eqref{eq:KW_propagator} depends on the indices, we assign an index for each face of the ribbon graph, and we obtain
    \begin{align}
        \mathcal Z(\Lambda,t) & = \sum_{\text{trivalent graphs } G}  \frac{N^{\#\text{vertices}}}{\#\operatorname{Aut}(G)} \left( \frac{t^{\frac{1}{3}}}{N} \right)^{\#\text{edges}} \sum_{1 \le i_1,\ldots,i_F \le N} \prod_{(l,m) \in \text{edges}} \frac{1}{\lambda_{i_l} + \lambda_{i_m}} ,
    \end{align}
    where $F = \#\{\text{faces in } G\}$, and we denote the edge between two faces $l$ and $m$ by $(l,m)$.
    Now, the weights of edges and vertices are given by $t^{\frac{1}{3}}/N$ and $N$, respectively.
    The parameter $t$ plays a role of the counting parameter for the edges, and the small $t$ expansion is equivalent to the large $\Lambda$ expansion because they appear as the combination $\Lambda / t^{\frac{1}{3}}$ in the matrix integral.

    \item
    The dual graph of the ribbon graph of $k$ trivalent vertices is a triangulized surface of $k$ faces.
    There are 4 genus 0 surfaces with 3 edges, 3 vertices and 2 faces:
    \begin{align}
    \begin{tikzpicture}[baseline=(current bounding box.center),scale = .85]
     \draw (0,0) circle (1);
     \draw (-1,0) arc[start angle=180, end angle=360, x radius=1, y radius=.3]; 
     \draw[dashed] (1,0) arc[start angle=0, end angle=180, x radius=1, y radius=.3]; 
     \draw (0,1) arc[start angle=90, end angle=270, x radius=.3, y radius=1]; 
     \filldraw (-.05,1) circle (.05) node [above] {$1$};
     \filldraw (-.05,-1) circle (.05) node [below] {$2$};
     \filldraw (-.28,-.28) circle (.05) node [above right] {$3$};
     \begin{scope}[shift={(4,0)}]
     \draw (0,0) circle (1);
     \draw (-1,0) arc[start angle=180, end angle=360, x radius=1, y radius=.3]; 
     \draw[dashed] (1,0) arc[start angle=0, end angle=180, x radius=1, y radius=.3]; 
     \draw (0,1) arc[start angle=90, end angle=270, x radius=.3, y radius=1]; 
     \filldraw (-.05,1) circle (.05) node [above] {$2$};
     \filldraw (-.05,-1) circle (.05) node [below] {$3$};
     \filldraw (-.28,-.28) circle (.05) node [above right] {$1$};      
     \end{scope}
     \begin{scope}[shift={(8,0)}]
     \draw (0,0) circle (1);
     \draw (-1,0) arc[start angle=180, end angle=360, x radius=1, y radius=.3]; 
     \draw[dashed] (1,0) arc[start angle=0, end angle=180, x radius=1, y radius=.3]; 
     \draw (0,1) arc[start angle=90, end angle=270, x radius=.3, y radius=1]; 
     \filldraw (-.05,1) circle (.05) node [above] {$3$};
     \filldraw (-.05,-1) circle (.05) node [below] {$1$};
     \filldraw (-.28,-.28) circle (.05) node [above right] {$2$};      
     \end{scope}
     \begin{scope}[shift={(12,0)}]
     \draw (0,0) circle (1);
     \draw (-1,0) arc[start angle=180, end angle=360, x radius=1, y radius=.3]; 
     \draw[dashed] (1,0) arc[start angle=0, end angle=180, x radius=1, y radius=.3]; 
     \filldraw (-.28,-.28) circle (.05) node [below left] {$1$};      
     \filldraw (.28,-.28) circle (.05) node [below right] {$2$};      
     \filldraw (0,.3) circle (.05) node [above] {$3$};      
     \end{scope}
    \end{tikzpicture}
    \end{align}
    We assign labels to vertices of these surfaces, equivalently to faces of the dual ribbon graphs:
    \begin{subequations}
    \begin{align}
    \begin{tikzpicture}[baseline=(current bounding box.center),scale = .8]
     \draw (-1,0) node [left] {$3$} -- (1,0) node [right] {$3$};
     \draw (-1,0) -- node[midway, sloped] {\tiny $||$} (0,1) node [above] {$1$};
     \draw (-1,0) -- node[midway, sloped] {\tiny $|$} (0,-1) node [below] {$2$};
     \draw (1,0) -- node[midway, sloped] {\tiny $||$} (0,1);
     \draw (1,0) -- node[midway, sloped] {\tiny $|$} (0,-1);
     \begin{scope}[shift={(4,0)}]
     \draw (-1,0) node [left] {$1$} -- (1,0) node [right] {$1$};
     \draw (-1,0) -- node[midway, sloped] {\tiny $||$} (0,1) node [above] {$2$};
     \draw (-1,0) -- node[midway, sloped] {\tiny $|$} (0,-1) node [below] {$3$};
     \draw (1,0) -- node[midway, sloped] {\tiny $||$} (0,1);
     \draw (1,0) -- node[midway, sloped] {\tiny $|$} (0,-1);      
     \end{scope}
     \begin{scope}[shift={(8,0)}]
     \draw (-1,0) node [left] {$2$} -- (1,0) node [right] {$2$};
     \draw (-1,0) -- node[midway, sloped] {\tiny $||$} (0,1) node [above] {$3$};
     \draw (-1,0) -- node[midway, sloped] {\tiny $|$} (0,-1) node [below] {$1$};
     \draw (1,0) -- node[midway, sloped] {\tiny $||$} (0,1);
     \draw (1,0) -- node[midway, sloped] {\tiny $|$} (0,-1);      
     \end{scope}
     \begin{scope}[shift={(12,0)},rotate=90]
     \draw (-1,0) node [below] {$2$} -- (1,0) node [above] {$1$};
     \draw (-1,0) -- node[midway, sloped] {\tiny $-$} (0,1) node [left] {$3$};
     \draw (-1,0) -- node[midway, sloped] {\tiny $-$} (0,-1) node [right] {$3$};
     \draw (1,0) -- node[midway, sloped] {\tiny $=$} (0,1);
     \draw (1,0) -- node[midway, sloped] {\tiny $=$} (0,-1);      
     \end{scope}
    \end{tikzpicture}     
     \\
    \begin{tikzpicture}[baseline=(current bounding box.center),scale = .8]
    \begin{knot}[double distance=4pt,clip width=2]
        \strand [double] (0,.5) arc [start angle = -90,end angle = 270,radius = .5] -- ++ (0,-1) arc [start angle = 90,end angle = 450,radius = .5];
    \end{knot}
        \node at (0,1) {$1$};
        \node at (0,-1) {$2$};
        \node at (0.5,0) {$3$};
    \begin{scope}[shift={(4,0)}]
    \begin{knot}[double distance=4pt,clip width=2]
        \strand [double] (0,.5) arc [start angle = -90,end angle = 270,radius = .5] -- ++ (0,-1) arc [start angle = 90,end angle = 450,radius = .5];
    \end{knot}
        \node at (0,1) {$2$};
        \node at (0,-1) {$3$};
        \node at (0.5,0) {$1$};
    \end{scope}
    \begin{scope}[shift={(8,0)}]
    \begin{knot}[double distance=4pt,clip width=2]
        \strand [double] (0,.5) arc [start angle = -90,end angle = 270,radius = .5] -- ++ (0,-1) arc [start angle = 90,end angle = 450,radius = .5];
    \end{knot}
        \node at (0,1) {$3$};
        \node at (0,-1) {$1$};
        \node at (0.5,0) {$2$};
    \end{scope}
    \begin{scope}[shift={(12,0)}]
    \begin{knot}[double distance=4pt,clip width=2]
        \strand [double] (0,0) circle (.9);
        \strand [double] (-.82,0) -- (.82,0);
    \end{knot}
        \node at (0,.45) {$1$};
        \node at (0,-.45) {$2$};
        \node at (.95,0) [right] {$3$};
    \end{scope}
    \end{tikzpicture}
    \end{align}     
    \end{subequations}
    The first ribbon graph has 2 trivalent vertices, 3 faces $\{1,2,3\}$, and 3 edges $\{(1,3), (2,3), (3,3) \}$.
    Hence its weight is given by
    \begin{align}
     N^2 \left(\frac{t^{\frac{1}{3}}}{N}\right)^3 \frac{1}{\lambda_{i_1}+\lambda_{i_3}} \frac{1}{\lambda_{i_2}+\lambda_{i_3}} \frac{1}{2 \lambda_{i_3}} = \frac{t N^{-1}}{2(\lambda_{i_1} + \lambda_{i_3})(\lambda_{i_2} + \lambda_{i_3}) \lambda_{i_3}} .
    \end{align}
    Applying the same computation for other graphs and summing over the indices, we obtain the weighted sum of the graphs of genus 0 with 3 edges,
    \begin{align}
     & \frac{t}{N} \sum_{1 \le i_1, i_2, i_3 \le N} \Bigg( \frac{1}{\lambda_{i_1}+\lambda_{i_3}} \frac{1}{\lambda_{i_2}+\lambda_{i_3}} \frac{1}{2 \lambda_{i_3}} + \frac{1}{\lambda_{i_2}+\lambda_{i_1}} \frac{1}{\lambda_{i_3}+\lambda_{i_1}} \frac{1}{2 \lambda_{i_1}} 
     \cr
     & \hspace{7em} 
     + \frac{1}{\lambda_{i_3}+\lambda_{i_2}} \frac{1}{\lambda_{i_1}+\lambda_{i_2}} \frac{1}{2 \lambda_{i_2}} + \frac{1}{\lambda_{i_1} + \lambda_{i_2}} \frac{1}{\lambda_{i_2} + \lambda_{i_3}} \frac{1}{\lambda_{i_3} + \lambda_{i_1}} \Bigg)
     \cr
     & = \frac{t}{N} \sum_{1 \le i_1, i_2, i_3 \le N} \frac{1}{2 \lambda_{i_1} \lambda_{i_2} \lambda_{i_3}} = \frac{t}{2N} \left( \Tr \Lambda^{-1} \right)^3 .
    \end{align}
    This may be rewritten as $\frac{t}{2} \mathsf{t}_1^3 N^2$ in the notations of \autoref{exo:Kontsevich3}, which defines  $\mathsf{t}_k = N^{-1} \Tr \Lambda^{-k}$ and assumes $\mathsf{t}_k = O(N^0)$. This expression is then of order $N^2$, as expected for a genus $0$ contribution.
    
    \item
    The unique graph of genus 1 with 3 edges, 1 vertex and 2 faces, and its dual ribbon graph with 2 vertices, 3 edges and 1 face, are:
    \begin{align}
    \begin{tikzpicture}[baseline=(current bounding box.center),scale = .7]
      \draw (0,0) -- node[midway, sloped] {\tiny $|$} (3,0) -- node[midway, sloped] {\tiny $||$} (2,3) -- node[midway, sloped] {\tiny $|$} (-1,3) -- node[midway, sloped] {\tiny $||$} cycle;
      \draw (0,0) -- (2,3);
     \begin{scope}[shift={(8,1.5)},scale=2]
      \begin{knot}[double distance=4pt,clip width=2]
       \strand [double]
       (1,0) arc [x radius = 1, y radius = .3, start angle = 0, end angle = 360];
       \strand [double]
       (-.6,-.2) to [out=90,in=left] (-.3,.1) to [out=right,in=left] (.3,-.7) to [out=right,in=-90] (.6,-.28);
      \end{knot}      
     \end{scope}
    \end{tikzpicture}
     \label{eq:triangulization_g1e3}
    \end{align}
    The weighted sum is then given by
    \begin{align}
        N^2 \left(\frac{t^{\frac{1}{3}}}{N}\right)^3 \sum_{i=1}^N \left( \frac{1}{2 \lambda_i} \right)^3 = \frac{t}{8N} \Tr \Lambda^{-3} = \frac{t}{8} \mathsf{t}_3 .
    \end{align}

    \item
    In the diagramatic expansion of the expectation value $\left<M_{11}\right>$, the insertion of $M_{11}$ corresponds to a tadpole:
    \begin{align}
        \left< M_{11} \right> = \sum_{\substack{ \text{trivalent graphs } G \\ \text{with } 1 \text{ tadpole} }} \frac{N^{\#\text{vertices}}}{\#\operatorname{Aut}(G)} \left(\frac{t^{\frac{1}{3}}}{N}\right)^{\#\text{edges}} \sum_{1 \le i_1,\ldots,i_{F-1} \le N} \prod_{(l,m) \in \text{edges}} \frac{1}{\lambda_{i_l} + \lambda_{i_m}} ,
    \end{align}
    where we denote the number of faces in $G$ by $F$, and the face associated with the tadpole is fixed by $i_F = 1$.
    For the $k$-point function $\left< M_{i_1 i_1} \cdots M_{i_k i_k} \right>$, we would have $k$ tadpoles.

    We consider the surface of genus 1 wih 3 edges~\eqref{eq:triangulization_g1e3} and we add a tadpole in the dual ribbon graph.
    In this case, there is a unique face, which is attached to the tadpole.
    Hence, we fix the face index $i = 1$, and there is no summation of the indices in the weighted sum.
    Adding the tadpole, the graph has 5 edges and 3 vertices in total.
    All the edge factors are given by $1/(2\lambda_1)$ since there is the unique face index $i = 1$.
    The weight of the dual ribbon graph is then given by
    \begin{align}
    \begin{tikzpicture}[baseline=(current bounding box.center),scale = 1.5]
     \begin{scope}%[shift={(5,0)}]
      \begin{knot}[double distance=4pt,clip width=2]
       \strand [double]
       (1,0) arc [x radius = 1, y radius = .3, start angle = 0, end angle = 360];
       \strand [double]
       (-.6,-.2) to [out=90,in=left] (-.3,.1) to [out=right,in=left] (.3,-.7) to [out=right,in=-90] (.6,-.28);
       \strand [double,-{Circle[open,scale=.4]}] (1.02,0) -- ++ (.3,0);
      \end{knot}      
     \end{scope}
    \end{tikzpicture}
    \qquad 
    N^3 \left(\frac{t^{\frac{1}{3}}}{N}\right)^5 \left(\frac{1}{2 \lambda_1}\right)^5 = \frac{t^{\frac{5}{3}}}{32 N^2 \lambda_1^5} .
    \end{align}

\end{enumerate}

\paragraph{Solution \ref{exo:ising}.}

\begin{enumerate}
    \item
    We associate the matrix $M_1$ to spin up triangles, and $M_2$ to spin down triangles. The propagator is
    \begin{align}
        \Big< (M_\alpha)_{ij} (M_\beta)_{kl} \Big> = \frac{1}{N} \delta_{ik} \delta_{jl} \times
        \begin{cases}
            1 & (\alpha = \beta) \\ a & (\alpha \neq \beta)
        \end{cases}
    \end{align}
    therefore the weight $a$ is assigned to each edge between two triangles of different types.
    \item
    The formal matrix integrals are related by a change of variables between $M_1,M_2$ and the matrices $M,A_1$ that appear in the $O(1)$ model. In the case $h=0$, the change of variables is $M_1 = M + A$ and $M_2 = M - A$, so that
    \begin{multline}
        \frac{b}{2} (M_1^2 + M_2^2) - c M_1 M_2 - \frac{g}{3}( M_1^3 + M_2^3)\\ =
        (b-c) M^2 + (b+c) A^2 - \frac{2}{3} g M^3 - 2g M A^2\ ,
    \end{multline}
    which agrees with the $O(1)$ matrix model after rescaling the variables.
\end{enumerate}

%============ section 3  Coulomb gas method =========

\section{Chapter 3}

\paragraph{Solution \ref{exo:nbsdpt}.}

The idea is to write the saddle-point equations as polynomial equations, and to eliminate one of the two eigenvalues.
\begin{enumerate}
 \item In the case of a cubic potential, with the derivative $V'(x)=t_3x^2 +t_2x$, the saddle-point equations may be written as
 \begin{align}
  (\lambda_1-\lambda_2)(t_3\lambda_1^2+t_2\lambda_1)=1 \quad , \quad (\lambda_2-\lambda_1)(t_3\lambda_2^2+t_2\lambda_2)=1\ .
 \end{align}
 Using the first equation, we write $\lambda_2 = \lambda_1-\frac{1}{t_3\lambda_1^2+t_2\lambda_1}$. Replacing $\lambda_2$ with this expression in the second equation, we obtain
 \begin{align}
  (t_3\lambda_1^2 + t_2\lambda_1)^3 + t_3\left(\lambda_1(t_3\lambda_1^2+t_2\lambda_1)^2-1\right)^2 +t_2 \left(\lambda_1(t_3\lambda_1^2+t_2\lambda_1)^2-1\right)= 0\ .
 \end{align}
This is now a polynomial equation of degree $6$ in $\lambda_1$. We therefore have $6$ solutions. Modulo permutations of the $2$ eigenvalues, the number of solutions is $3$.
\item For a generic potential, the equations read
\begin{align}
 (\lambda_1-\lambda_2)V'(\lambda_1)=1 \quad , \quad (\lambda_2-\lambda_1)V'(\lambda_2)=1\ .
\end{align}
Using the first equation, we write $\lambda_2 = \lambda_1-\frac{1}{V'(\lambda_1)}$. Replacing $\lambda_2$ with this expression in the second equation, we obtain
\begin{align}
  V'(\lambda_1)+ V'\left(\lambda_1-\frac{1}{V'(\lambda_1)}\right) = 0\ .
\end{align}
Multiplying with $V'(\lambda_1)^d$, we obtain a polynomial equation of degree $d(d+1)$ for $\lambda_1$. Modulo permutations of the $2$ eigenvalues, the number of solutions is $\frac{d(d+1)}{2}$.

\end{enumerate}

\paragraph{Solution \ref{exo:ubrs}.}

 \begin{enumerate}
\item
The set
$A$ separates all allowed sectors, therefore any $\gamma\in H_1(e^{-V(x)}dx)$ going from one allowed sector to another must cross $A$.
On this set, $\mathcal V$ is decreasing near $\infty$, so it is bounded from above, i.e. $\mathcal{V}_0<\infty$.
On any $\gamma$, there must exist a point $x\in \gamma\cap A$, at which $\mathcal V(x)\leq \mathcal V_0$, and on $\gamma$, $\mathcal V$ tends to $+\infty$ at the boundaries.
Since $\mathcal V$ is continuous on $\gamma$, it must take the value $\mathcal V_0$.
\item
Let us bound the first factor as follows,
\begin{align}
\frac{|\mathcal V(x)-\mathcal V(x')|}{|x-x'|}
\leq \frac{|V(x)-V(x')|}{|x-x'|} + 2d\frac{\left|\log\frac{1+|x|}{1+|x'|}\right|}{|x-x'|}\ .
\end{align}
We write the potential of degree $d+1$ as $V(x)=\sum_{k=0}^{d+1} t_k x^k$, where $t_{d+1}=t\neq 0$.
Recalling that $(x^k - {x'}^k)/(x-x') = \sum_{j=0}^{k-1} x^{k-1-j} x'^j$, we have
\begin{align}
\frac{V(x)-V(x')}{x-x'} = \sum_{k=1}^{d+1} t_k \sum_{j=0}^{k-1} x^{k-1-j} x'^j\ .
\end{align}
On the right-hand side, we have a finite sum of monomials of degree $j\leq d$.
From the inequality $|x|^j \leq (1+|x|)^j\leq (1+|x|)^d$ for $j\leq d$, we obtain
\begin{align}
\left|\frac{V(x)-V(x')}{x-x'}\right|\frac{1}{(1+|x|)^d(1+|x'|)^d}
 \leq \sum_{k=1}^{d+1} k |t_k| \ .
\end{align}
For the second term, we use the inquality $\left|\frac{d}{dy}\log y\right| \leq 1$ for $y\geq 1$ to have
\begin{align}
 \left|\log\frac{1+|x|}{1+|x'|}\right| \leq |x-x'| \ .
\end{align}
We then deduce that $K \leq \sum_{k=1}^{d+1} k |t_k|  + 2d  < +\infty$, thus $K$ is finite.

\item
By the definition of the functional action $S[\rho]$, we compute
\begin{multline}
\Re S[\rho] - \int_{\gamma} \mathcal V(x) \rho(x)dx
+ \int_\gamma\int_\gamma \log{|\mathcal V(x)-\mathcal V(x')|} \  \rho(x)dx \ \rho(x')dx'
\\ = -2d\int_{\gamma} \log(1+|x|) \rho(x)dx\ ,
+ \int_\gamma\int_\gamma \log{\frac{|\mathcal V(x)-\mathcal V(x')|}{|x-x'|}} \  \rho(x)dx \ \rho(x')dx'\ .
\end{multline}
Since the density function is normalized $\int_\gamma \rho(x)dx = 1$, we may rewrite the first term of the right-hand side as the double integral as follows,
\begin{align}
-d\int_\gamma\int_\gamma \log(1+|x|)(1+|x'|)\  \rho(x)dx \ \rho(x')dx'\ .
\end{align}
Then, we obtain a double integral whose integrand is bounded by $\log K$ and conclude the desired inequality.

% #TODO: correct in the main body, definition of the density
\item 
Since the value $\mathcal V_0$ is reached on $\gamma$, and $\mathcal V$ tends to $+\infty$ at the ends of $\gamma$, and $\gamma$ and $\mathcal V$ are continuous, every value on $[\mathcal V_0,+\infty[$ is reached by $\mathcal V$. Therefore, $\{s \in \mathbb R \ | \ \mathcal V(\gamma(s))=v\}$ is not empty thus $R(v)$ is well-defined.
The function $v\mapsto R(v)$ is strictly increasing by construction, and is thus invertible.

The map $s\mapsto F(s)$ is thus well-defined, it has image in the interval $[0,1]$, and by construction it is an increasing function.
Since $\mathcal V_0$ is reached, the value $0$ is reached by $F$. And since $\mathcal V\to +\infty$ at the extremities of $\gamma$, we have $\sup F=1$.
An increasing function taking values in $[0,1]$ can be taken as the cumulative distribution function of a measure on $\mathbb R$, whose density we call
\begin{align}
d F(s) = \tilde{\rho}_{\gamma}(\gamma(s))d\gamma(s)= \tilde{\rho}_{\gamma}(x)dx.
\end{align}
By construction, we have
\begin{align}
\int_{\mathbb R} \mathcal V(\gamma(s))dF(s)
= \int_{\mathcal V_0}^\infty e^{\mathcal V_0-v} v dv = \int_{0}^\infty e^{-v} (\mathcal V_0+v) dv = \mathcal V_0+1 .
\end{align}
We also compute the double integral,
\begin{multline}
\int_\gamma\int_\gamma \log{|\mathcal V(x)-\mathcal V(x')|}\  \tilde{\rho}_\gamma(x)dx \tilde{\rho}_\gamma(x')dx'
\\
= \int_{\mathcal V_0}^\infty\int_{\mathcal V_0}^\infty  \log{|v-v'|}\  e^{\mathcal V_0-v}  e^{\mathcal V_0-v'}  dv  dv'  = \int_{0}^\infty\int_{0}^\infty  \log{|v-v'|}\  e^{-v-v'}  dv  dv'\ ,
\end{multline}
which is a real constant. To evaluate it, we use the change of variables $v'=v+w$. Our double integral factorizes into a trival integral on $v$ and the integral $\int_{0}^\infty  \log{w}\ e^{-w}  dw
= -\psi(1) = 0.57721...$, which is Euler's constant, and $\psi(z) = \Gamma'(z)/\Gamma(z)$ is the digamma function.

\item
In the end, we have proved that
\begin{align}
\Re S[\tilde{\rho}_\gamma]
\leq \log K + \mathcal V_0+1 +\psi(1)\ .
\end{align}
This implies that
\begin{align}
\sup_\gamma \left(\inf_{\rho\in \mathfrak M(\gamma)} \Re S[\rho] \right) \leq \log K + \mathcal V_0+1 +\psi(1)\ .
\end{align}
\end{enumerate}

\paragraph{Solution \ref{exo:plancherel}.}

\begin{enumerate}
\item
Let us first rescale the variable $x_i=\hbar y_i$ to obtain
 \begin{align}
\mathcal Z
 = \frac{1}{N!\,(2\pi i)^N } \hbar^{N^2-N} \int_{\hbar^{-1}\gamma^N} \Delta(\{y_i\})^2 \prod_{i=1}^N \frac{\Gamma(-y_i)}{\Gamma(y_i+1)}e^{i\pi y_i}  \hbar^{-2 y_i} dy_i .
\end{align}
The Gamma function $\Gamma(-y)$ has simple poles for $y\in\mathbb{N}$, with the residues
\begin{align}
\res_{y = n} \Gamma(-y) =  \frac{(-1)^n}{n!}\ .
\end{align}
The contour $\hbar^{-1}\gamma$ surrounds all the poles. Therefore, the integral is the sum of their residues:
\begin{align}
\mathcal Z
= \frac{\hbar^{N^2-N}}{N!} \sum_{h_1,h_2,\dots,h_N \in \mathbb N^N} \Delta(\{h_i\})^2 \prod_{i=1}^N \frac{\hbar^{-2 h_i}}{(h_i!)^2} \ .
\end{align}
Since the summand is symmetric under permutations of the $h_i$s, and vanishes if some $h_i$s are not distinct, this formula is equivalent to Eq.~\eqref{eq:Plancherel_measure_discrete_sum}.

\item
Writing $h_i = \lambda_i + N - i$, where $\lambda_1 \ge \lambda_2 \ge \cdots \lambda_N \ge 0$, we recognize
\begin{align}
\mathcal Z
= \sum_{\lambda_1 \ge \lambda_2 \ge \cdots \ge \lambda_N \ge 0 } \left[\frac{\dim \lambda}{|\lambda|!}\right]^2 \hbar^{-2 \sum_i \lambda_i}   =  \sum_{\lambda, \ell(\lambda) \le N} \frac{1}{|\lambda|!}   \mathcal P_\mathrm{ Plancherel}(\lambda)  \hbar^{-2 |\lambda|}\ ,
\end{align}
or equivalently
 \begin{align}
  \mathcal Z  = %\hbar^{N^2-N} q^{\frac{N(N-1)}{2}}
\sum_{k=0}^\infty \frac{\hbar^{-2k}}{k!}\sum_{|\lambda|=k,\, \ell(\lambda)\leq N} \mathcal P_\mathrm{ Plancherel}(\lambda)\ .
 \end{align}
Since the dependence on $N$ is only through the constraint $\ell(\lambda)\leq N$, we can take the large $N$ limit by removing this constraint, thus we obtain the Poissonized-Plancherel measure,
 \begin{align}
\lim_{N\to \infty}  \mathcal Z  =
\sum_{k=0}^\infty \frac{\hbar^{-2k}}{k!}\sum_{|\lambda|=k} \mathcal P_\mathrm{ Plancherel}(\lambda)\ .
 \end{align}
\item
From the integral formula, the saddle-point equation in the limit $\hbar\to 0$ with $N\hbar=O(1)$ reads
\begin{align}
2\hbar\sum_{j\neq i} \frac{1}{x_i-x_j}
= -\frac{\varphi'(\frac{x_i}{\hbar})}{\varphi(\frac{x_i}{\hbar})}  = -i\pi + \psi(-\tfrac{x_i}{\hbar})+\psi(\tfrac{x_i}{\hbar}) +\frac{\hbar}{x_i} + 2\hbar\log{\hbar}\ ,
\end{align}
where we denote the digamma function by $\psi(z) = \Gamma'(z)/\Gamma(z)$.
\item
Using Stirling's formula for the Gamma function, we have the asymptotic behavior of $\psi(z)$,
\begin{align}
 \psi(z) \underset{z \to \infty}{=} \log z + \frac{1}{2z}+O\left(\frac{1}{z^2}\right)\ ,
\end{align}
which yields the leading order contribution of the saddle-point equation,
\begin{align}
2\hbar\sum_{j\neq i} \frac{1}{x_i-x_j}
\underset{\hbar\to 0}{=} 2\log{x_i} + O(\hbar)\ .
\end{align}
In the continuous limit, we may write this equation in terms of the resolvent, and we obtain Eq.~\eqref{owxi}.
\item
Let us look for a solution parametrized by
\begin{align}
 x(z) = \alpha+\gamma(z+z^{-1})
 \quad , \quad
\overline W(x(z)) = w(z)\ .
\end{align}
The saddle-point equation becomes
\begin{align}
w(z)+w(z^{-1}) =  2\log(x(z)) = 2\log(\alpha+\gamma(z+z^{-1}))\ .
\end{align}
Let us try the ansatz
\begin{align}
w(z) = 2\log(1+uz^{-1})\ ,
\end{align}
where the aysmptotic behaviour $\overline W(x) \underset{x\to\infty}{\sim} \frac{N\hbar}{x} $ requires $u = \frac{N\hbar}{2\gamma}$.
The saddle-point equation is satisfied provided
$u=\gamma$ and $\alpha=1+\gamma^2$, leading to Eq.~\eqref{owxz}. Notice that the
two branch points $z=\pm 1$ correspond to $x=(1\pm\gamma)^2$.

\item Now $\gamma=u=1$ and $\alpha=2$, leading to branch points at $x=0$ and $x=4$.
The discontinuity of the resolvent is related to the density function by
\begin{align}
2i\pi\overline{\rho}(x(z)) =  w(z) - w(z^{-1}) =  - 2 \log z\ ,
\end{align}
which yields  
\begin{align}
 \bar{\rho}(x) =
 \begin{cases}
  1 & (x < 0)\ , \\
  \displaystyle
  \frac{1}{\pi} \operatorname{arccos} \left( \frac{x-2}{2} \right) & ( 0 \le x < 4)\ , \\
  0 & (4 < x)\ .
 \end{cases}
\end{align}
Using the identities
\begin{subequations}
\begin{align}
 \operatorname{arccos}(x) &= - i \log \left( x + i \sqrt{1 - x^2} \right) = \frac{\pi}{2} - \operatorname{arcsin}(x)\ ,
\\
 \int^x dx' \, \operatorname{arccos}(x') &= x \operatorname{arccos}(x) - \sqrt{1 - x^2} + \text{constant} \ ,
\end{align}
\end{subequations}
we obtain the limit shape function
\begin{align}
 f(x)  = x - 2\int_4^{x} \bar\rho(x'+2)dx'
 = \frac{2}{\pi} \left( \sqrt{4 - x^2}  + x\arcsin\frac{x}{2} \right)\ .
\end{align}
\end{enumerate}

\paragraph{Solution \ref{exo:Kontsevich3}.}

\begin{enumerate}
\item
In terms of $
t_k =\frac{1}{N}\Tr \Lambda^{-k}
$,
the equation that defines  $\hat{t}_1$ may be rewritten as $t_1 = f(\hat{t}_1)$, where
\begin{align}
f(\hat{t}_1) = \hat{t}_1 - \sum_{k=1}^\infty \binom{-\frac{1}{2}}{k} t^k \hat t_1^k t_{2k+1}\ .
\end{align}
This formal power series is invertible, so that
$
\hat{t}_1 = f^{-1}(t_1) = t_1 + O(t)
$.
Furthermore,
\begin{align}
\hat t_k = \sum_{j=0}^\infty \binom{-\frac{k}{2}}{j} t^j \hat{t}_1^j t_{k+2j} = t_k +O(t)\ .
\end{align}
In the case $t_1=0$, we have $\hat t_1=t_1=0$, so that $\hat t_k=t_k$ for all $k$.
\item
Applying the change of variable $M \to t^{-\frac13}(M+\Lambda)$, the Kontsevich integral becomes
\begin{align}
 \mathcal Z(\Lambda,t)
&= \frac{t^{-\frac{N^2}{3}} e^{-\frac{2N}{3t}\operatorname{Tr} \Lambda^3}}{\mathcal Z_0(\Lambda,t)}\int_{\mathrm{ formal}\,H_N} dM\, e^{-\frac{N}{t}\operatorname{Tr} \left(  M \Lambda^2 - \frac{1}{3} M^3 \right) } \ .
\end{align}
\item
This corresponds to a one-matrix model with external field with potential $V(x) = -\frac1{3t} x^3$ and external field $-\frac{1}{t}\Lambda^2$. In our case,
the saddle-point equation \eqref{eq:sdpt1Matrixext} takes the form:
\begin{align}
V'(x)-Y_2(x) = \frac{1}{N}\left< \operatorname{Tr} \frac{V'(x)-V'(M)}{x-M} \frac{1}{Y_2(x)+\Lambda^2/t}\right>\ .
\end{align}
Redefining $\tilde{Y}(x) = t Y_2(x)$, this becomes
\begin{align}
 x^2+\tilde{Y}(x)- x \ \frac{t}{N}\operatorname{Tr} \frac{1}{\tilde{Y}(x)+\Lambda^2} - \frac{t}{N}\left< \operatorname{Tr} M \frac{1}{\tilde{Y}(x)+\Lambda^2}\right>=0\ .
\end{align}
This expression is polynomial in $x$ of degree 2, and a rational function of $\tilde Y$, therefore $\tilde Y(x)$ is an algebraic function of $x$. There must exist a parametric solution with $x$ and $\tilde Y$ meromorphic functions on a certain Riemann surface. The degree of $\tilde Y$ is $2$, and the degree of $x$ is $N+1$.
Eq.~\eqref{xzy2z} is a rational solution, i.e. the case of genus zero, parametrized by $z\in \mathbb C P^1$.

When $\tilde Y$ is large, $x$ must be large, and must behave like $\tilde Y\sim -x^2$, and since $\tilde Y$ is of degree 2, it can have at most a double pole, so $x$ can have at most a simple pole. Let us choose the pole at $z=\infty$, and up to a rescaling, we can choose $x=z(1+O(1/z))$ at $z\to\infty$, thus $\tilde Y = -z^2(1+O(1/z))$.
Since $\tilde Y$ is of degree 2, it has no other pole, and must be a polynomial of degree 2, which we write
\begin{align}
\tilde Y(z) = -z^2 + bz + c\ .
\end{align}
Moreover, $x$ must have a pole for $\tilde Y = - \Lambda_i^2$, with
$ x\sim \frac{t}{N} \frac{1}{\tilde Y+\Lambda_i^2}$. Introducing $\zeta_i$ such that $\tilde Y(\zeta_i)=-\Lambda_i^2$, $x$ must have a pole of the form $\frac{t}{N} \frac{1}{ \tilde Y'(\zeta_i)(z-\zeta_i)}$.
Therefore,
\begin{align}
x(z) = z + \frac{t}{N} \sum_{i=1}^N \frac{1}{ (-2\zeta_i+b)(z-\zeta_i)} = z + \frac{t}{N} \Tr \frac{1}{ (-2\zeta+b)(z-\zeta)}\ ,
\end{align}
where $\zeta_i^2-b\zeta_i-c = \Lambda_i^2$ and  $\zeta = \operatorname{diag}(\zeta_1,\dots,\zeta_N)$.

As $\tilde Y\to \infty$ i.e. $z\to\infty$, we must have $\tilde Y \sim -x^2 + O(1/x)$, and since $x(z)\sim z - t c_1/z  + O(z^{-2})$ with $c_1=\frac{1}{N}\Tr \frac{1}{2\zeta-b}$, this implies $x^2 \sim z^2-2c_1+O(1/z)$ thus
$-z^2+bz+c = - (z^2-c_1) +O(1/z)$. This implies $b=0$ and $c=2c_1$, leading to
\begin{align}
x(z)  = z - \frac{t}{N} \Tr \frac{1}{ 2\zeta (z-\zeta)} \  , \ \
\tilde Y(z)  = -z^2 + \frac{t}{N}  \Tr \frac{1}{\zeta} \  , \ \
\zeta_i^2 - \frac{t}{N}  \Tr \frac{1}{\zeta}  = \Lambda_i^2\ .
\end{align}
We thus recognize that $\zeta=\hat\Lambda$ defined in \eqref{exo:K:defhatLambda}, leading to Eq.~\eqref{xzy2z}.

Let us compute the leading behaviour of the resolvent as a power series in $t$, using the known behaviour of $\hat\Lambda$:
\begin{align}
 \overline{W}  =  \frac{1}{N} \left( \operatorname{Tr} \frac{1}{z-\hat{\Lambda}} + \operatorname{Tr} \frac{1}{\hat{\Lambda} }  \right)  - \hat{t}_1 +O(t)\ .
 \end{align}
 Using $z = x + O(t)$, and the cancellation of the last two terms, we obtain Eq.~\eqref{wxot}. This coincides with the leading behaviour that follows from a direct calculation of the Kontsevich integral as $t\to 0$.
\item As a small $z$ expansion we have:
\begin{align}
 \Tr \frac{1}{\hat{\Lambda}(z - \hat{\Lambda})} 
  = - \Tr \frac{1}{\hat{\Lambda}^2(1 - z \hat{\Lambda}^{-1})}  = - \sum_{k = 0}^\infty \Tr \hat{\Lambda}^{-k-2} z^{k} \ ,
\end{align}
leading to the announced result.
In the case $\Tr \Lambda^{-1} = 0$, we have $\hat{t}_k = t_k$, so that
\begin{align}
 x(z) = z + \frac{t}{2} \sum_{k=0}^\infty t_{k+2} z^k\ .
\end{align}

\end{enumerate}

\paragraph{Solution \ref{exo:sheets}.}

\begin{enumerate}
\item
The saddle-point equation is
\begin{align}
 \forall i \ , \quad \frac{\partial}{\partial \lambda_i} \log \left(\prod_{i<j} R(\lambda_i,\lambda_j) \prod_{i=1}^N e^{-N V(\lambda_i)}\right) = 0 \ .
\end{align}
Explicitly, this reads
\begin{align}
\forall i \ , \quad   2 \sum_{j\neq i} \frac{1}{\lambda_i - \lambda_j} + \sum_{s\in G\setminus \{\text{Id}\}} \sum_{j\neq i} \frac{a(s) s'(\lambda_i)}{s(\lambda_i) - \lambda_j} - N V'(\lambda_i) = 0 \ .
\end{align}
In the large $N$ limit, with $\overline{W}(x) = \frac{1}{N}\sum_{j} \frac{1}{x-\lambda_j}$, this reduces to Eq.~\eqref{eq:RHsG}. By our definition of a 1-cut solution, our ansatz $\overline W_0(x)$ is continuous across the cut,
so the functional saddle-point equation reduces to
\begin{align}
\sum_{s\in G} a(s) \overline W_0(s(x))s'(x) = V'(x) \ .
\end{align}
With the particular form of our ansatz, this equation becomes
\begin{align}
 \sum_{s,\tilde{s}\in G} a(\tilde{s}) b(s\circ \tilde{s}^{-1}) (V\circ s)' = V'\ .
\end{align}
This is satisfied if $b(s)$ obeys Eq.~\eqref{fsig}.

\item We notice that
\begin{align}
 \ell_g(\hat\alpha.e_g) = \sum_{\tilde g\in G} a(\tilde g) \ell_g(e_{\tilde g\circ g}) = a(\text{Id})=2 \ .
 \label{lgag}
\end{align}
It follows that $\hat T_g$ is an involution:
\begin{align}
 \hat T_g(\hat T_g(v)) = v -2\ell_g(v)\hat\alpha.e_g + \ell_g(v)\ell_g(\hat\alpha.e_g) = v\ .
\end{align}
By a direct calculation,
\begin{align}
 v\in \operatorname{Ker}(\text{Id}+\hat T_g) \iff v = \frac12 \ell_g(v) \hat\alpha.e_g\ .
\end{align}
And $v=\hat\alpha.e_g$ obeys this equation, due to Eq.~\eqref{lgag}.

\item
Near a point $x\in g^{-1}(\Gamma)$, and for any $\tilde{g}\in G$, the form $w\circ \tilde{g}$ behaves as $w(\tilde g(x-i0))=w(\tilde g(x+i0))$ if $\tilde g\neq g$, while the homogeneous equation implies
\begin{align}
 w(g(x-i0)) = -w(g(x+i0)) - \sum_{\tilde g\in G\setminus\{g \} } a(\tilde g\circ g^{-1}) w(\tilde g(x)))\ .
\end{align}
This allows us to work out the behaviour of $\phi_v$ across the cut $g^{-1}(\Gamma)$. We indeed have $\phi_v(x-i0)=\Phi(x+i0)$ where
\begin{align}
 \Phi = \sum_{\tilde g\in G \setminus \{ g\} } \ell_{\tilde g}(v) \  (w\circ \tilde g) - \ell_{g}(v) (w\circ g)
  - \ell_{g}(v)\sum_{\tilde g\in G \setminus \{ g\} } a(\tilde g\circ g^{-1}) (w\circ \tilde g)\ .
\end{align}
Making the sums run over the full group $G$, this becomes
\begin{align}
 \Phi = \sum_{\tilde g\in G} \ell_{\tilde g}(v) \  (w\circ \tilde g)
  - \ell_{g}(v)\sum_{\tilde g\in G} a(\tilde g\circ g^{-1}) (w\circ \tilde g)\ .
\end{align}
We then rewrite $a(\tilde g\circ g^{-1}) = \ell_{\tilde g}(\hat \alpha.e_g)$. Gathering the two terms, we obtain the combination
$\ell_{\tilde g}(v-\ell_g(v)\hat\alpha.e_g)=\ell_{\tilde g}(\hat T_g(v))$. This leads to $\Phi = \phi_{\hat T_g(v)}$, and therefore to Eq.~\eqref{fvftv}.

This means that when $x$ crosses the cut $g^{-1}(\Gamma)$, the function $\phi_v$ is analytically continued to $\phi_{\hat T_g(v)}$, whose parameter belongs to the orbit of $v$ by the group $\hat G$. Elements of this orbit therefore correspond to branches of $\phi_v$ continued to other sheets.
An algebraic function has finitely many sheets, so the orbit of $v$ must be finite.

Conversely, let us assume that the orbit of $v$ is finite. On the initial sheet, the only singularities of $\phi_v$ are the singularities of $dV(g(x))$ for $g\in G$. Since $dV$ is meromorphic, so is $\phi_v$ on that sheet. Then $\phi_v$ lives on a Riemann surface that is obtained by gluing a finite number of sheets, where it is meromorphic. So $\phi_v$ is algebraic.

\item 
Since $R(x,x')=R(x',x)$ is symmetric, we have $a(g^{-1})=a(g)$. This implies that our matrix is symmetric, and so is the corresponding quadratic form. We compute
\begin{align}
\hat T_g(\hat\alpha.v)
= \hat\alpha.v - 2 \left<v,e_g\right> \hat\alpha.e_g= \hat\alpha. T_g(v)\ .
\end{align}
This implies that the orbit of $\hat\alpha.v$ by the group $\hat G$ are the same as $\hat\alpha$ times the orbits of $v$ by the group $\hat G_{\text{reduced}} $.
If we want finite orbits, it is necessary and sufficient to have finite orbits of $\hat G_{\text{reduced}} $.
This is interesting because $\hat G_{\text{reduced}} $ is smaller than $\hat G$ in many examples. In particular, $\hat G_{\text{reduced}} $ can be finite in cases where $\hat G$ is infinite.

\item For the $O(n)$ model we have $R(x,x') = (x-x')^2 (x+x')^{-n}$, therefore  $G=\{\text{Id},-\text{Id}\}$ with $a(\text{Id})=2$ and $a(-\text{Id})=-n$. This leads to $
\hat\alpha = 2 e_+ - n e_-$, and to a quadratic form with the matrix $
\frac12 \left[\begin{smallmatrix}
2 & -n \\ -n & 2
\end{smallmatrix}\right]
$.
This matrix is non-degenerate iff $n\neq \pm 2$, and is positive definite iff $|n|<2$.
Then we compute the matrices
\begin{align}
T_+ = 
\begin{pmatrix}
-1 & n \\ 0 & 1
\end{pmatrix}
\ , \ \
T_- = 
\begin{pmatrix}
1 & 0 \\ n & -1
\end{pmatrix}
\ , \ \
R \equiv T_+ T_- = \begin{pmatrix}
n^2-1 & -n \\  n  & -1
\end{pmatrix} \ .
\end{align}
Its characteristic polynomial is $\det(\lambda-R) = \lambda^2-(n^2-2)\lambda+1$. Setting $n=2\cos\theta$, the eigenvalues are $\lambda = e^{\pm 2i\theta}$.

Since $T_+^2=T_-^2=\text{Id}$, the group $\hat G_{\text{reduced}} $ generated by $T_+,T_-$ is finite iff $R^q=\text{Id}$ for some $q\in\mathbb{N}$. This condition is equivalent to  $e^{2iq\theta}=1$ so that $n\in 2\cos \pi \mathbb{Q}$. This condition is necessary for $\hat G_{\text{reduced}}$ to have finite orbits, as can be seen considering the orbits of the eigenvectors of $R$,
\begin{align}
v_+ = e_+ + e^{i\theta} e_- \quad , \quad v_- = e^{i\theta} e_+ +  e_-\ .
\end{align}
Assuming $n\in 2\cos \pi \mathbb{Q}$, the vectors
$\hat\alpha.v_+$ and $\hat\alpha.v_-$ have finite orbits by $\hat T_G$, and the corresponding functions
\begin{align}
 \phi_{\hat\alpha.v_\pm}(x) = 2i\sin\theta \Big(w(\mp x)- e^{i\theta} w(\pm x)\Big)
\end{align}
are algebraic. This implies that $w$ is algebraic, as it is a linear combination of two algebraic functions.

\item In the example of the torus knot $\mathfrak K_{P,Q}$, we have 
\begin{align}
R(x,x') 
= (x'^P-x^P)(x'^Q-x^Q)
= (x'-x)^2 \prod_{k=1}^{P-1} (x'-\rho^{kQ} x) \prod_{k=1}^{Q-1} (x'-\rho^{kP} x)\ ,
\end{align}
where $\rho = e^{\frac{2i\pi}{PQ}}$.
The maps $x\mapsto \rho_{PQ}^k x$ form a group isomorphic to $G=\frac{\mathbb Z}{PQ \mathbb Z} $. 
The exponents are $a(k)=\delta_{P|k}+\delta_{Q|k}\in \{0,1,2\}$, leading to the stated expression for $\hat\alpha$.

Let us introduce $P+Q$ vectors $v_0,v_1,\dots, v_{P+Q-1}$, which will form an orbit under $\hat G$:
\begin{align}
 v_i\ \underset{0\leq i\leq Q-1}{=}\ \sum_{k=0}^{P-1}  e_{Pi+Qk} \quad , \quad v_{Q+j}\ \underset{0\leq j\leq Q-1}{=}\ -\sum_{k=0}^{Q-1}  e_{Qj+Pk} \ .
\end{align}
By a direct calculation, the action of $\hat T_k$ on such vectors is found to be:
\begin{align}
 \begin{cases}
T_k(v_i) = v_i \text{ if } k\notin Pi+Q\mathbb Z \ , \\
T_k(v_i) = v_{Q+j} \text{ if } k= Pi+Qj \ ,
\end{cases}
\begin{cases}
T_k(v_{Q+j}) = v_{Q+j} \text{ if } k\notin Qj+P\mathbb Z \ , \\
T_k(v_{Q+j}) = v_i \text{ if } j= Pi+Qj \ .
\end{cases}
\end{align}
For example, we have
\begin{align}
\hat T_{Pi+Qj}(v_i)
= \sum_{l=0}^{P-1}  e_{Pi+Ql} - \sum_{l=0}^{P-1}  e_{Pi+Qj+Ql} - \sum_{l=0}^{Q-1}  e_{Pi+Qj+Pl} =
 v_{Q+j}\ .
\end{align}
This implies that $f(x)$ can be continued to a Riemann surface with $P+Q$ sheets.
Moreover, $\overline{W}(x)$ behaves like $\log(x)/x$ at $\infty$, thus $f(x)$ is meromorphic at $\infty$.
This implies that $f(x)$ satisfies an algebraic equation of degree $P+Q$, given in \eqref{eq:knotsfPQ}.

\item 
The Torus  knot $\mathfrak K_{P,Q}$ corresponds to $r=2$ and $p_1=Q$, $p_2=P$, and $p=PQ$.

Let us write $\rho = e^{\frac{2i\pi}{p}}$, then we have
\begin{align}
 R(x',x) = \prod_{j=1}^r \prod_{k=0}^{p'_j-1}  (x'-\rho^{k p_j} x)  \prod_{k=0}^{p-1} (x'-\rho^{k}x)^{2-r}\ .
\end{align}
Therefore, $G$ is the
$\mathbb{Z}_p$ group of functions $x\mapsto \rho^n x$, with multiplicities
\begin{align}
 a(k) = 2-r+ \sum_{j=1}^r \delta_{p_j|k} \ ,
\end{align}
in particular $a(0)=2$, and the stated expression for $\hat \alpha $ follows.

If $\chi<0$, we compute
$n(\hat\alpha.e_k) = n(\hat\alpha) = (2-r)p + \sum_{j=1}^r p'_j = p \chi$, so that
\begin{align}
n(\hat T_k(v)) = n(v) - p \chi \ell_k(v)\ .
\end{align}
For any vector $v\neq 0$ such that $n(v)\geq 0$, there is $k$ such that $\ell_k(v)>0$. This implies $n(\hat T_k(v))>n(v)$. Repeating the operation, we find that the orbit of $v$ contains an infinite sequence on which the function $n$ is strictly increasing. Therefore, the orbit is infinite.

Let us now assume $\chi\geq 0$, then
\begin{align}
 2 \geq \sum_{j=1}^r \left(1-\frac{1}{p_j}\right) \geq \frac{r}{2}\ ,
\end{align}
therefore $r\leq 4$. If $r=4$ we must have $p_1=p_2=p_3=p_4=2$. If $r=2$, the positivity of $\chi$ does not constrain $p_1,p_2$. Let us focus on $r=3$, and assume $p_1\leq p_2\leq p_3$. We must have $p_1\leq 3$, and $p_1=3$ is only possible if $p_1=p_2=p_3=3$. Except in this case, we therefore have $p_1=2$, and $p_2\leq 4$. We display the possibilities in the following table:
\begin{align}
\renewcommand{\arraystretch}{1.4}
\begin{array}{|l|c|c|c|}
\hline
 (p_1,\dots,p_r) & \chi & \text{root system}  & \text{Seifert manifold} \\
\hline\hline
 (P,Q) &  \frac{1}{P}+\frac{1}{Q} & A_{\text{lcm}(P,Q)-1} & \text{lens space} L(P,Q)  \\
\hline
 (2,2,p_3) ,  p_3 \text{ odd} &  \frac{1}{p_3} & D_{p_3+1} & S^3/D_{p_3+2} \\
   (2,2,p_3) ,  p_3 \text{ even} &  \frac{1}{p_3} & A_{p_3} & S^3/D_{p_3+2} \\
  (2,3,3) &  \frac{1}{6} & D_4 & S^3/E_6 \\
  (2,3,4) &  \frac{1}{12} & E_6 & S^3/E_7 \\
  (2,3,5) &  \frac{1}{30} & E_8 & S^3/E_8 \\
  (2,3,6) &  0 &  A_5 & \\
 (2,4,4) &  0 & A_3 & \\
 (3,3,3) &  0 & A_2 & \\
\hline
 (2,2,2,2) &  0 & A_1 & \\
\hline
\end{array}
\end{align}
We let the motivated reader consult \cite{Borot:2014kda} for the corresponding finite orbits.
\end{enumerate}

\paragraph{Solution \ref{exo:burgers}.}

\begin{enumerate}
    \item
    Consider a chain of matrices with potentials $\tilde V_j(\tilde M_j)$, whose partition function is of the type $\mathcal{Z}=\int\left(\prod_{j=1}^k d\tilde M_j\right) e^{-NS}$, with the action
    \begin{align}
     S= \sum_{j=1}^k \Tr \left[   \tilde V_j(\tilde M_j) - \tilde M_j \tilde M_{j+1} \right] \ .
    \end{align}
    For $\epsilon>0$, we define
    \begin{subequations}
    \begin{align}
    \renewcommand{\arraystretch}{1.4}
     & t =  j\epsilon \ , \
    T =  k\epsilon \ , \
    M(t) = \sqrt{\epsilon}    \tilde M_j  \ ,
    \\
    & V(M,t) =  \begin{cases} \frac{1}{\epsilon}\Big(\tilde V_j(\tilde{M}_j) - \tilde{M}_j^2\Big)\ ,  \quad (j\neq 1,k)\ , \\
    \frac{1}{\epsilon}\Big(\tilde V_j(\tilde{M}_j) - \frac12\tilde{M}_j^2\Big)\ , \quad (j= 1,k) \ .
    \label{eq:exo3.7_pot_shift}
    \end{cases}
    \end{align}
    \end{subequations}
    Remarking that $\tilde M_j \tilde M_{j+1} = \frac12 (\tilde M_j^2 + \tilde M_{j+1}^2 -  (\tilde M_j-\tilde M_{j+1})^2)$, the action behaves as
    \begin{align}
    S =  \int_0^T  \Tr  \left[   V( M(t),t) + \frac1{2} M'(t)^2  \right] dt + O(\epsilon)\ .
    \end{align}    
 In the limit $\epsilon\to 0$, this yields the partition function \eqref{eq:matrix_chain_partition_fn}.

    \item
    In terms of our shifted potentials \eqref{eq:exo3.7_pot_shift}, the saddle-point equation becomes
    \begin{align}
    \frac{1}{\epsilon^2} \Big(Y(t-\epsilon) + Y(t+\epsilon)\Big) = V'(Y(t),t) + \frac{2}{\epsilon^2} Y(t) \ .
    \end{align}
    In the limit $\epsilon\to 0$, this becomes Eq.~\eqref{ptyzt}.

    \item
    According to Eq.~\eqref{ywyx}, the resolvent of the first matrix is $\tilde Y_0 = \frac{1}{N}\Tr (\tilde x-\tilde M_1)^{-1} $, i.,e.
\begin{align}
\frac{1}{\epsilon} Y(0) = \frac{1}{N}\Tr (x-M(\epsilon))^{-1} \qquad \text{ with }\  x=Y(\epsilon).
\end{align}
This implies
\begin{align}
 \frac{1}{N}\Tr (x-M(\epsilon))^{-1} 
 =&  \frac{1}{\epsilon} (Y(0) -Y(\epsilon))  + \frac{x}{\epsilon} =  - Y'(0) + O(\epsilon) + \frac{x}{\epsilon}.
\end{align}
Adding an analytic function to the resolvent does not change its discontinuity (the spectral density), therefore upon subtracting $x/\epsilon$ we have
\begin{align}
\frac{1}{N}\Tr (x-M(\epsilon))^{-1} = -Y'(0) + \text{analytic} + O(\epsilon).
\end{align}

 \item
Recall that $W(x,t) = \partial_t Y(z,t)$, where $z=z(x,t)$ is defined by $Y(z,t)=x$.
Taking a derivative with respect to $x$ at fixed $t$ gives
\begin{align}
1= \partial_x z \ \partial_z Y
\quad \implies \quad
\partial_x z = \frac{1}{\partial_z Y}.
\end{align}
Taking a derivative with respect to $t$ at fixed $x$ gives
\begin{align}
0 = \partial_t z \ \partial_z Y + \partial_t Y
\quad \implies \quad
\partial_t z = - \frac{\partial_t Y}{\partial_z Y}.
\end{align}
Therefore we have
\begin{align}
\partial_x W(x,t)
=  \partial_x z(x,t) \ \partial_t\partial_z Y(z(x,t),t) =  \frac{\partial_t\partial_z Y}{\partial_z Y}\ .
\end{align}
We deduce the non-homogeneous inviscid Burgers' equation by the following manipulations:
\begin{multline}
\partial_t W(x,t)
= \frac{d}{dt} \partial_t Y(z(x,t),t)
= \partial^2_t Y + \partial_t z \ \partial_t\partial_z Y
=  \partial^2_t Y - \frac{\partial_t Y}{\partial_z Y} \ \partial_t\partial_z Y   \ ,
\\
= \partial^2_t Y - \partial_t Y \ \partial_x W
=  \partial^2_t Y - W \ \partial_x W
=  V'(x,t) - W \ \partial_x W \ .
\end{multline}

\end{enumerate}

%============ section 4 Loop equations  =========

\section{Chapter 4}

\paragraph{Solution \ref{exo:loop_multi_matrix}.}

\begin{enumerate}
    \item
    The behaviour of the integration measure under our change of variables can be deduced from the split rule:
    \begin{align}
        dM_1 \to dM_1 + \epsilon \Tr \frac{1}{x-M_1} \Tr \left( \frac{1}{x-M_1} \frac{1}{y - M_2} \right) + O(\epsilon^2)\ .
    \end{align}
   Together with the behaviour of the integrand, this must make the partition function invariant at the order $O(\epsilon)$, leading to
    \begin{multline}
        \frac{1}{N} \left< \Tr \frac{1}{x - M_1} \Tr \frac{1}{x - M_1} \frac{1}{y - M_2} \right> \\
        - \left< \Tr \frac{V_1'(M_1)}{x - M_1} \frac{1}{y - M_2} \right> + \left< \Tr \frac{1}{x - M_1} \frac{M_2}{y - M_2} \right> = 0\ .
        \end{multline}
In this equation, we rewrite the second term using $\overline{P}(x,y)$ \eqref{eq:sdptdefP1Mext}.
In the perturbative $1/N$ topological asymptotic expansion the first term factorizes to leading order:
\begin{multline}
 \left< \Tr \frac{1}{x - M_1} \Tr \frac{1}{x - M_1} \frac{1}{y - M_2} \right>
 \\
 = \left< \Tr \frac{1}{x - M_1} \right>\left<\Tr \frac{1}{x - M_1} \frac{1}{y - M_2} \right> + O(N^{-2})\ .
\end{multline}
Using also the expression of the resolvent $\overline{W}(x)= \left<\Tr \frac{1}{x - M_1}\right>$, we obtain (to leading order in $1/N$)
    \begin{align}
        \left( \overline{W}(x) - V_1'(x) + y \right) \frac{1}{N} \left< \Tr \frac{1}{x - M_1} \frac{1}{y - M_2} \right> + \overline{P}(x,y) - \overline{W}(x) = 0 \ .
    \end{align}
   Choosing $y = Y_2(x) = V_1'(x) - \overline{W}(x)$, we deduce the loop equation
$
\overline{W}(x) =  \overline{P}(x,y)$.

    \item
 The first change of variables leads to the first loop equation,
    \begin{multline}
  \left< \Tr \frac{1}{x - M_1} \frac{V_2'(y) - V_2'(M_2)}{y - M_2} M_2 \right>- \left< \Tr \frac{V_1'(M_1)}{x - M_1} \frac{V_2'(y) - V_2'(M_2)}{y - M_2} \right>
        \\ + \frac{1}{N} \left< \Tr \frac{1}{x - M_1} \Tr \frac{1}{x - M_1} \frac{V_2'(y) - V_2'(M_2)}{y - M_2} \right> = 0
        \ .
    \end{multline}
  In the large $N$ perturbative topological expansion, to leading order the third term factorizes, leading to
    \begin{multline}
       \left( \overline{W}(x) - V_1'(x) + y \right) \left< \Tr \frac{1}{x - M_1} \frac{V_2'(y) - V_2'(M_2)}{y - M_2} \right>+ \left< \Tr \frac{1}{x - M_1} V_2'(M_2) \right>
       \\
        + \left< \Tr \frac{V_1'(x) - V_1'(M_1)}{x - M_1} \frac{V_2'(y) - V_2'(M_2)}{y - M_2} \right>
        - N V_2'(y) \overline{W}(x)=0\  .
    \end{multline}
    The second change of variables leads to the second loop equation,
    \begin{align}
        - \left< \Tr \frac{1}{x - M_1} V_2'(M_2) \right> + x\left< \Tr \frac{1}{x - M_1} \right> - N= 0\ .
    \end{align}
    Combining the two loop equations with setting $y = Y_2(x) = V_1'(x) - \overline{W}(x)$, we obtain the saddle-point equation \eqref{eq:sdpt2matrixmm}.

    \item 
Let us derive the large $N$ leading loop equations. For any random variable $X$ we introduce the notation
$\overline{X} = \lim_{N\to \infty}\frac{1}{N}\left< X \right>$.
For any $i=1,\dots,k+1$ we introduce the resolvent $w_i(y) = \frac{1}{y-M_i}$, a variable $y_i$, and $w_i=w_i(y_i)$. We define the functions $f_{i}=f_{i,k}$ as special cases of the functions $f_{i,j}(y_i,\dots,y_j)$ \eqref{fdet}, and also $f_{k+1}=1$ and $f_i=0$ if $i>k+1$.
They satisfy the recursion relation
\begin{align}
f_i=V'_i(y_i) f_{i+1} - y_i y_{i+1} f_{i+2}\ .
\end{align}
We write $\operatorname{Pol}_i=\operatorname{Pol}_{y_i,\dots, y_k}$ the operation of taking the polynomial part (near infinity) with respect to the last $k-i+1$ variables.
For $i\in \mathbb{Z}$ we define a function $U_i(y_{i},\dots,y_{k+1},y_1)$ by
\begin{subequations}
\begin{align}
 U_1 &= \operatorname{Pol}_1 f_1 \overline{\operatorname{Tr} w_1\dots w_{k+1}}\ ,
 \\
 U_i &= \operatorname{Pol}_{i} f_i \overline{\operatorname{Tr} w_i\dots w_{k+1}w_1} \qquad \text{if } 2\leq i\leq k-1 \ ,
 \\
 U_k & = \operatorname{Pol}_{k} V'_k(y_k)\overline{\operatorname{Tr} w_k w_{k+1}w_1} = \overline{\operatorname{Tr}\frac{V'_k(y_k)-V'_k(M_k)}{y_k-M_k}  w_{k+1} w_1} \ ,
 \\
 U_{k+1} &= \overline{\operatorname{Tr}w_{k+1}w_1}\ ,
 \\
 U_i &= 0 \qquad \text{if } i\leq 0 \text{ or } i\geq k+2\ .
\end{align}
\end{subequations}
Let us derive a few properties of these objects.
For $i=2,\dots,k$, we use the recursion relation for $f_i$ to make $U_i$ manifestly polynomial in $y_i$:
\begin{align}
 U_i = \operatorname{Pol}_{{i+1}} \overline{\operatorname{Tr}\frac{V'_i(y_i)-V'_i(M_i)}{y_i-M_i} w_{i+1} \dots w_{k+1} w_1 }-U_{i+2}\ .
\end{align}
We also consider the expression that is obtained from $U_i$ by inserting a factor of $y_i$ before taking the polynomial part, and apply the same treatment:
\begin{multline}
 \operatorname{Pol}_{i} y_if_i \overline{\operatorname{Tr} w_i\dots w_{k+1}w_1}
 = \operatorname{Pol}_{{i+1}} f_{i+1}\overline{\operatorname{Tr} V'_i(M_i)  w_{i+1} \dots w_{k+1} w_1}
 \\
 -  \operatorname{Pol}_{{i+2}} f_{i+2}\overline{ \operatorname{Tr}  M_{i}  w_{i+2} \dots w_{k+1} w_1 } + y_i U_i\ .
 \label{pypyu}
\end{multline}
We also consider the expression that is obtained from $U_i$ by omitting the factor $w_i$ in the trace, and apply the same treatment:
\begin{align}
 \operatorname{Pol}_{i} f_i \overline{\operatorname{Tr} w_{i+1}\dots w_{k+1}w_1} = V'_i(y_i) U_{i+1} -y_i U_{i+2}\ .
 \label{pyuyu}
\end{align}
We will also need to evaluate $U_2$ with an insertion of $V_1'(M_1)$.
Using $\operatorname{Pol}_{1}w_1=0$, we can show
\begin{align}
\operatorname{Pol}_{{2}} f_{2}\overline{ \operatorname{Tr} V'_1(M_1)  w_1 w_{2} \dots w_{k+1}  }
= V'_1(y_1) U_2  -  \operatorname{Pol}_{{1}} V'_1(y_1) f_{2}\overline{\operatorname{Tr} w_{2} \dots w_{k+1} w_1   }\ .
\end{align}
Using the recursion relation for $f_i$ as well as $\operatorname{Pol}_{2}y_2w_2=1$, this can be further transformed into
\begin{multline}
\operatorname{Pol}_{{2}} f_{2}\overline{ \operatorname{Tr} V'_1(M_1)  w_1 w_{2} \dots w_{k+1}  }
\\ =V'_1(y_1) U_2  - U_1 -y_1 U_3
+  \operatorname{Pol}_{{3}}  f_{3}\overline{  \operatorname{Tr}  M_1 w_{3} \dots w_{k+1}  w_1  }\ .
\label{u2wv1}
\end{multline}
We now come to loop equations, which follow from the changes of variables:
\begin{align}
M_i  \to   M_i  +\epsilon \operatorname{Pol}_{{i+1}}  f_{i+1}  w_{i+1} \dots w_{k+1} w_1  + O(\epsilon^2)\ .
\end{align}
For $i=2,\dots,k$,
this is an $M_i$-independent translation, which therefore does not modify the integration measure. The corresponding loop equation is
\begin{align}
 \operatorname{Pol}_{{i+1}}  f_{i+1} \overline{\operatorname{Tr}\left(V'_i(M_i)-M_{i-1}-M_{i+1}\right)w_{i+1} \dots w_{k+1} w_1} = 0\ .
\end{align}
Let us apply the identity $M_{i+1}w_{i+1} = y_{i+1}w_{i+1}-1$ to the last term. For $i=2\dots k-1$, we use Eqs.~\eqref{pyuyu} and \eqref{pypyu} to make the result manifestly polynomial in $y_{i+1}$:
\begin{multline}
 \operatorname{Pol}_{{i+1}} f_{i+1} \overline{ \operatorname{Tr} M_{i+1}  w_{i+1} \dots w_{k+1} w_1 } = y_{i+1} (U_{i+1}+U_{i+3}) - V'_{i+1}(y_{i+1}) U_{i+2}
\\ \hspace{-5mm} +
  \operatorname{Pol}_{{i+2}}f_{i+2} \overline{\operatorname{Tr}  V'_{i+1}(M_{i+1})  w_{i+2} \dots w_{k+1} w_1  }
 -  \operatorname{Pol}_{{i+3}} f_{i+3}\overline{  \operatorname{Tr}  M_{i+1}  w_{i+3} \dots w_{k+1} w_1  }
   \ .
\end{multline}
For $i=k$, the calculation is straightforward:
\begin{align}
f_{k+1}  \overline{\operatorname{Tr} M_{k+1}  w_{k+1}  w_1 } =  y_{k+1} U_{k+1}  - \overline{ \operatorname{Tr} w_1}\ .
\end{align}
The loop equations may thus be written as
\begin{multline}
 y_{i+1}U_{i+1} = V'_{i+1}(y_{i+1}) U_{i+2} - y_{i+1} U_{i+3}
 \\
 \hspace{-2mm} +
 \operatorname{Pol}_{{i+1}}   f_{i+1} \overline{ \operatorname{Tr} V'_i(M_i)  w_{i+1} \dots w_{k+1} w_1  }
  -  \operatorname{Pol}_{{i+2}}     f_{i+2} \overline{\operatorname{Tr}  V'_{i+1}(M_{i+1})  w_{i+2} \dots w_{k+1} w_1  }
 \\ -  \operatorname{Pol}_{{i+1}}   f_{i+1} \overline{\operatorname{Tr} M_{i-1}  w_{i+1} \dots w_{k+1} w_1  }
  +  \operatorname{Pol}_{{i+3}}    f_{i+3} \overline{\operatorname{Tr}  M_{i+1}  w_{i+3} \dots w_{k+1} w_1} \ .
  \label{yuvyu}
 \end{multline}
For $i=1$, our change of variable affects the integration measure $dM_1$, leading to a Jacobian term in the loop equation. Evaluating this term using the split rule yields $\overline{\operatorname{Tr}w_1}U_2$, which has to be added to the left-hand side of the $i=1$ case of Eq.~\eqref{yuvyu}.
 For $i=k$, the loop equation is
\begin{align}
\overline{ \operatorname{Tr} V'_k(M_k) w_{k+1} w_1}   -   \overline{  \operatorname{Tr}  M_{k-1}  w_{k+1} w_1}   - y_{k+1} U_{k+1}  + \overline{\operatorname{Tr}w_1}= 0\ .
 \end{align}
The idea is now to sum all loop equations for  $i=1,\dots,k$, which leads to the cancellation of all explicit traces but two:
\begin{multline}
\overline{\operatorname{Tr}w_1} U_2
=   \operatorname{Pol}_{{2}}   f_{2} \overline{ \operatorname{Tr} V'_1(M_1)  w_{2} \dots w_{k+1} w_1 }
  -  \operatorname{Pol}_{{3}}    f_{3} \overline{ \operatorname{Tr}  M_1  w_{3} \dots w_{k+1} w_1 }
  \\ \overline{\operatorname{Tr}w_1}
 - y_2 U_2 +  (V'_2(y_2)-y_3) U_3 + \sum_{i=3}^k(V'_i(y_i)-y_{i-1}-y_{i+1})U_{i+1} \ .
 \end{multline}
The two remaining explicit traces can be eliminated using Eq.~\eqref{u2wv1}, leading to
\begin{align}
 \overline{\operatorname{Tr}w_1} U_2
=  \overline{\operatorname{Tr}w_1} -U_1
+  (V'_1(y_1)-y_2) U_2 + \sum_{i=2}^k(V'_i(y_i)-y_{i-1}-y_{i+1})U_{i+1} \ .
\label{sumle}
\end{align}
Let us deduce an algebraic equation for $\overline{W}_1(y_1)=\overline{\operatorname{Tr}w_1}$, using a judicious choice of the variables $y_2,\dots, y_{k+1}$. We choose these variables recursively:
\begin{align}
y_2  = V'_1(y_1) -\overline{\operatorname{Tr}w_1} \quad , \quad y_{i+1} =  V'_i(y_i) - y_{i-1} \quad \text{for } i=2,\dots, k\ .
\end{align}
Then Eq.~\eqref{sumle} reduces to $\overline{W}_1(y_1)=U_1$, which is another notation for Eq.~\eqref{eq:sdptchainmatrices}. This is an algebraic equation for the resolvent, because $U_1$ is polynomial in $y_1,\dots, y_k$ by construction, while $U_1\det(y_{k+1}-M_{k+1})$ is polynomial in $y_{k+1}$.
\end{enumerate}

\paragraph{Solution \ref{exo:loop_On_matrix}.}

\begin{enumerate}
    \item Using the merge rule, we obtain
    \begin{multline}
  \frac{1}{N} \left< \Tr \frac{1}{x - M} \Tr \frac{1}{\tilde x- M} \right> =  \left< \Tr A_i \frac{1}{x - M} A_i \frac{1}{\tilde x - M} \right>
  \\ - \left< \Tr M A_i \frac{1}{x - M} A_i \frac{1}{\tilde x - M} \right>
        - \left< \Tr M \frac{1}{x - M} A_i \frac{1}{\tilde x - M} A_i \right>\ .
    \end{multline}
    This may be rewritten as
    \begin{multline}
     \frac{1}{N}\Big(W_1(x)W_1(\tilde x)+W_2(x,\tilde x)\Big)
     \\ = (1-x-\tilde x)\left< \Tr A_i \frac{1}{x - M} A_i \frac{1}{\tilde x - M} \right>
         + G(x)+G(\tilde x)\ .
    \end{multline}
Choosing  $\tilde x = 1-x$ leads to the announced result.

    \item
    The change of variable $M \to M + \epsilon \frac{1}{x - M}$ leads to a loop equation that almost coincides with Eq.~\eqref{eq:firstloop}. The only difference is that each term $\Tr A_i^2 M$ in the action gives rise to an extra term $-NG(x)$ in the right-hand side of the loop equation.

    \item
    Inserting the topological expansions of $W_1,W_2$ \eqref{eq:wgnexp} and $P_0$ \eqref{defPgn} in the loop equation, as well as $G(x)= N G_{0,1}(x)+ O(N^{-1})$, we obtain at the leading order
    \begin{subequations}
    \begin{align}
&    W_{0,1}(x) W_{0,1}(1-x) = G_{0,1}(x)+G_{0,1}(1-x) \ ,
\\
  &  W_{0,1}(x)^2 + n G_{0,1}(x) = V'(x) W_{0,1}(x) - P_{0,1}(x)\ .
    \end{align}
    \end{subequations}    
    We then use the second equation for eliminating  $G_{0,1}$ in the first equation.

\end{enumerate}

\paragraph{Solution \ref{exooopeqtrv}.}

\begin{enumerate}
    \item We obtain these relation from the loop equation \eqref{eq:1loop} with $k = 0$ and $k = 1$, respectively. The corresponding changes of variables are $M \to M + \epsilon$ and $M \to  M + \epsilon M$, which are infinitesimal translations and dilations respectively.

    \item We consider the change of variable $M \to M + \epsilon \frac{M - a}{x - M}$, and obtain the loop equation,
    \begin{align}
(x - a) \left< \left( \Tr \frac{1}{x - M} \right)^2 \right> - N \left< \Tr V'(M)  \frac{M-a}{x - M}  \right> = 0 \ .
\end{align}
This may be rewritten as
\begin{align}
\widehat{W}_2(x,x) - N V'(x) W_1(x) + N P_0(x) + \frac{N}{x - a} \left< \Tr V'(M) \right> = 0\ .
    \end{align}
The presence of a hard edge leads to a possible pole at $x=a$, whose residue vanishes iff translation invariance is unbroken.
But the change of variable $M\to M+\epsilon$ that leads to translation invariance does not vanish at $M=a$, and is therefore not allowed. The dilation $M-a\to e^\epsilon (M-a)$ is allowed, and leads to the stated loop equation.

In the formal large $N$ limit, we have $\widehat{W}_2(x,x) = NW_{0,1}(x)^2 +O(1)$, and we obtain a quadratic equation for the leading order resolvent $W_{0,1}(x)$. If $v\neq 0$, the solution behaves as $W_{0,1}(x)\underset{x\to a}{\sim} \frac{\sqrt{v}}{\sqrt{x-a}}$. The spectral density is the discontinuity of the resolvent, and also has a $(x-a)^{-\frac12}$ singularity, whose integral behaves like $(x-a)^{1/2}$ so converges as $x\to a$.

    \item We consider the change of variable, $M \to M + \epsilon \frac{\sigma(M)}{x - M}$, where we define
    \begin{align}
        \sigma(x) = (x - a)(x - b)  \implies \frac{\sigma(M)}{x - M} = \frac{\sigma(x)}{x - M} + a + b - x - M\ .
    \end{align}
    This leads to the loop equation
    \begin{align}
        \sigma(x) \left< \left( \Tr \frac{1}{x - M} \right)^2 \right> - N^2 - N \left< \Tr V'(M) \frac{\sigma(M)}{x - M}  \right>= 0\ ,
   \end{align}
   which may be rewritten as
   \begin{multline}
    \widehat{W}_2(x,x) - N V'(x) W_1(x) + N P_0(x)
        \\
        + \frac{N}{\sigma(x)} \Big[-N+ (x - a - b) \left< \Tr V'(M) \right> + \left< \Tr V'(M) M \right> \Big]= 0 \ .
    \end{multline}
    Both terms $\left< \Tr V'(M) \right>$ and $\left< \Tr V'(M) M \right>$ can produce poles at $x=a$ or $x=b$.
    Neither translations nor dilations are allowed as changes of variables.
In the formal large $N$ limit, the solution of the loop equation is
\begin{align}
W_{0,1}(x) = \frac{ V'(x)}{2}- \sqrt{\frac{V'(x)^2}{4} - P_{0,1}(x) + \frac{1 - (x-a-b) v - u}{\sigma(x)}  }  \ ,
\end{align}
where $u=\lim_{N\to\infty} \frac{1}{N} \left<\Tr M V'(M)\right>$.
    
    \item We consider the change of variable, $M \to \epsilon \frac{\sigma(M)}{x - M}$, where we now have
    \begin{align}
        \frac{\sigma(M)}{x - M} =  \frac{\sigma(x)}{x - M} - \sum_k \sum_{l=0}^{k-1} \sigma_k x^l M^{k-l-1}\ .
    \end{align}
   This leads to the loop equation,
    \begin{align}
            \widehat{W}_2(x,x) - N V'(x) W_1(x) + N P_0(x) - \frac{N}{\sigma(x)} Q(x)=0 \ ,
    \end{align}
    where we define
    \begin{multline}
     Q(x) =  \sum_{k=1} \sum_{0 \le i,j \le k-2} \sigma_k x^{k-i-j-2} \left< \Tr M^i \Tr M^{j} \right> \\ - N \left< \Tr V'(M) \frac{\sigma(x) - \sigma(M)}{x - M} \right> \ .
    \end{multline}
Then the formal large $N$ limit solution for $W_{0,1}(x)$ involves the leading term of $Q(x)=
 N^2 Q_{0,1}(x) + O(N^0)$.

\end{enumerate}

\paragraph{Solution \ref{exogauss}.}

\begin{enumerate}
    \item
    For the Gaussian matrix model we have
    $\Tr\frac{V'(x)-V'(M)}{x-M}=N$, therefore $P_0(x)=N$. For $n\geq 1$, the polynomial $P_n(x;x_1,\dots,x_n)$ is the cumulant of a function of $n+1\geq 2$ variables. But that function is in fact independent of one variable namely $x$, therefore $P_{n\geq 1}=0$.
    
    \item
    We obtain $W_{0,1}(x)$ by solving the quadratic equation Eq.~\eqref{eq:w01eq}.
    The Joukowsky map implies $\sqrt{x^2-4}=z-z^{-1}$ thus $W_{0,1}(x) = z^{-1}$.

    \item
	      As discussed in Section~\ref{sec:higherWgnJoukowsky}, $W_{0,2}$ coincides with $\overline{W}_2$ given by Eq.~\eqref{saddle2pt}. In the 1-cut case with the cut $[-2,2]$, we find
	      \begin{align}
	       W_{0,2}(x_1,x_2)  = \frac{-\frac{1}{2}}{(x_1 - x_2)^2} \left[ 1 - \frac{x_1 x_2 - 4}{\sqrt{x_1^2 - 4} \sqrt{x_2^2 - 4}} \right]\ .
	      \end{align}
	      In terms of the $z$ variables, we have~\eqref{eq:W2barBJ}:
	      \begin{align}
	      W_{0,2}(x_1,x_2) dx_1 dx_2  = - B(z_1,1/z_2) =  \frac{ dz_1 dz_2 }{(1- z_1 z_2)^2}\ .
	      \end{align}
	      or equivalently
	     \begin{align}
	      W_{0,2}(x_1,x_2)  =  \frac{ z_1^2 z_2^2 }{(z_1^2-1)(z_2^2-1)(1- z_1 z_2)^2}\ .
	      \end{align}
	 To determine $W_{1,1}$, we solve the loop equation \eqref{eq:w11} and find
	   \begin{align}
	    W_{1,1}(x) =\frac{W_{0,2}(x,x)}{\sqrt{x^2-4}} = \frac{1}{(x^2-4)^{\frac52}}\ .
	   \end{align}
     To determine $W_{0,3}$, we could consider the loop equation~\eqref{eq:loopconn} for $n = 2$ with $I = \{x_2,x_3\}$,
	   \begin{multline}
	    W_4(x_1,x_1,x_2,x_3) + W_1(x_1) W_3(x_1,x_2,x_3) + 2 W_2(x_1,x_2) W_2(x_1,x_3)
	   \\
	    + \frac{\partial}{\partial x_2} \frac{W_2(x_1,x_3) - W_2(x_2,x_3)}{x_1 - x_2}
	    + \frac{\partial}{\partial x_3} \frac{W_2(x_1,x_2) - W_2(x_2,x_3)}{x_1 - x_3}
	     \\
	    = N \Big(V'(x_1) W_3(x_1,x_2,x_3) - P_2(x_1;x_2,x_3)\Big)\ .
	   \end{multline}
	  To order $N^0$ in the topological expansion, the $W_4$ term does not contribute, and we can determine $W_{0,3}$ in terms of $W_{0,1}$ and $W_{0,2}$. Using the explicit form of $W_{0,1}$, we obtain
	   \begin{multline}
             W_{0,3}(x_1,x_2,x_3)  \sqrt{x_1^2-4}
            = 2 W_{0,2}(x_1,x_2) W_{0,2}(x_1,x_3)
            \\
             + \frac{\partial}{\partial x_2} \frac{W_{0,2}(x_2,x_3) - W_{0,2}(x_1,x_3)}{x_2 - x_1}
            + \frac{\partial}{\partial x_3} \frac{W_{0,2}(x_2,x_3) - W_{0,2}(x_1,x_2)}{x_3 - x_1}\ ,
	   \end{multline}
	   where it only remains to use the result for $W_{0,2}$. However, its is faster to use the explicit solution~\eqref{eq:omega03_TR} for $W_{0,3}$,
	   \begin{align}
	    \omega_{0,3}(z_1,z_2,z_3)
	    & = 2 \sum_{a = \pm 1} \underset{z = a}{\operatorname{Res}} \,
	    K_a(z_1,z) B(z,z_2) B(\sigma_a(z),z_3)\ ,
	    \nonumber \\
	    & =  \sum_{a = \pm 1} \operatorname*{Res}_{z = a}  \frac{dz_1}{(z_1-z)(zz_1-1)} \frac{z^3dz}{z^2-1}
	    \frac{dz_2}{(z - z_2)^2} \frac{dz_3}{(z-z_3)^2}\ .
	   \end{align}

\end{enumerate}

\paragraph{Solution \ref{exo:airy}.}

\begin{enumerate}
    \item
    The cut $(-\infty,0]$ is non-compact, and the eigenvalue density $\rho(x) = \frac{1}{\pi} \sqrt{-x} \mathbf{1}_{(-\infty,0]}$ is not normalizable.
    Thus, this spectral curve cannot be obtained from a matrix model.
    In fact, it appears in the edge microscopic limit of \autoref{sec:micro}.

    \item
    Since $dx = 2 z dz$, the branch point is $z = 0$, and the local Galois involution is $z \to -z$.

    \item
    The differentials produced by
    topological recursion can only have poles at the branch points. A meromorphic function on the Riemann sphere whose only pole is $z=0$ must be a polynomial of $z^{-1}$. Therefore, for $2g-2+n>0$, $\omega_{g,n}$ must be a polynomial of $z_i^{-1}$.
    Moreover, $\omega_{g,n}$ must be odd under $z_i\to -z_i$, therefore it must be a polynomial of $z_i^{-2}$, since the differential $dz_i$ is itself odd.
    
    If we rescale $z \to \lambda z$, we change $ydx \to \lambda^3 ydx$, and more generally $\omega_{g,n} \to \lambda^{3(2-2g-n)} \omega_{g,n}$. Therefore, $\omega_{g,n}$ must be a homogeneous polynomial of total degree
$\sum_{i=1}^n (2d_i+1) = 3(2g-2+n)$.

        \item
        The topological recursion kernel is
        \begin{align}
        K(z_1,z) 
 = \frac{z dz_1}{z_1^2-z^2} \ \frac{1}{4z^2 dz}  = \frac14 \frac{ dz_1}{z_1^2-z^2}  \frac{1}{z dz}\ .
        \end{align}
    This allows us to compute
    \begin{subequations}\label{eq:Airy_curve_omega}
    \begin{multline}
        \omega_{0,3}(z_1,z_2,z_3) 
        = \operatorname{Res}_{z\to 0} K(z_1,z) \ [ B(z,z_2)B(-z,z_3) + B(z,z_3)B(-z,z_2) ] \ ,
        \\
        = -\frac12 \operatorname{Res}_{z\to 0}  \frac{ dz_1}{z_1^2-z^2} \ \frac{dz}{z}  \frac{ dz_2}{(z_2-z)^2} \frac{ dz_3}{(z_3-z)^2}
         = -\frac{dz_1 dz_2 dz_3}{2 z_1^2 z_2^2 z_3^2}\ ,
    \end{multline}
    which corresponds to $\left<\tau_0 \tau_0 \tau_0\right>_0 = 1$.
    Similarly we get
    \begin{align}
        \omega_{0,4}(z_1,z_2,z_3,z_4) & = \frac{3}{4} \left( \frac{1}{z_1^2} + \frac{1}{z_2^2} + \frac{1}{z_3^2} + \frac{1}{z_4^2} \right) \frac{dz_1 dz_2 dz_3dz_4}{z_1^2 z_2^2 z_3^2 z_4^2}\ ,  \\
        \omega_{1,1}(z_1) & = -\frac{dz_1}{16 z_1^4}\ ,
    \end{align}
    \end{subequations}
    from which we deduce the results~\eqref{eq:Airy_curve_tau_av}.
    \item
    Let $Z_n=(z_1,\dots ,z_n)$.
  We start with the expression of $\omega_{g,n+1}(z_0,Z_n)$ according to the topological recursion equation \eqref{TRecursion}. Rewriting both sides of the equation in terms of $W$ instead of $\omega$, we obtain
   \begin{align}
W_{g,n+1}(z_0,Z_n)
    = - \frac1{2 z_0 } \mathop{\operatorname{Res}}_{z\to 0}  \frac{ z dz}{z_0^2-z^2} \   Q'_{g,n}(z;Z_n)\ ,
    \label{wgnres}
    \end{align}
  where we define
  \begin{multline}
Q'_{g,n}(z;Z_n)
=  W_{g-1,n+2}(z,z,Z_n)
+2 \sum_{j=1}^n \frac{B(z,z_j)}{dx(z)dx(z_j)} W_{g,n}(z,Z_n\setminus \{z_j\})
\\
+\sum_{g_1+g_2=g}\sum^{\text{stable}}_{I_1\sqcup I_2 = Z_n } W_{g_1,1+|I_1|}(z,I_1) W_{g_2,1+|I_2|}(z,I_2)
      \ .
\end{multline}
In the sum over bipartitions of $Z_n$, we have isolated the unstable terms, which involve $\omega_{0,2}=B$.
And we have used the fact that $\omega_{g,n}(Z_n)$ is odd under the Galois involution of one of its variables $z_j\mapsto -z_j$, according to Eq.~\eqref{ognz}.
For $2g-2+n\geq 0$, let us rewrite $Q'_{g,n}$ in terms of $P_{g,n}$ \eqref{eq:AiryLoopeqPgn}:
\begin{multline}
 Q'_{g,n}(z;Z_n)
 =  P_{g,n}(z;Z_n) - 2 W_{0,1}(z)  W_{g,n+1}(z,Z_n)+\sum_{j=1}^n  \frac{d}{dx_j}   \frac{W_{g,n}(Z_n)}{x-x_j}
 \\
 + 2\sum_{j=1}^n \left(\frac{B(z,z_j)}{dx(z)dx(z_j)}-W_{0,2}(z,z_j)\right) W_{g,n}(z,Z_n\setminus\{z_j\})    \ .
 \label{qpgn}
 \end{multline}
To take the residue at $z=0$ as in Eq.~\eqref{wgnres}, we rewrite it as minus the sum of the residues at the other poles. These other poles are at $z=\pm z_0$, and also at $z=z_j$ in the case of the term in $B(z,z_j)$. Therefore, we obtain
    \begin{multline}
 2z_0W_{g,n+1}(z_0,Z_n) =\sum_\pm
     \underset{z\to \pm z_0}{\operatorname{Res}} \frac{ z dz}{z_0^2-z^2}  Q'_{g,n}(z;Z_n)
     \\
     + \sum_{j=1}^n  \underset{z\to z_j}{\operatorname{Res}}  \frac{ 2z dz}{z_0^2-z^2}  \frac{B(z,z_j)}{dx(z)dx(z_j)}W_{g,n}(z,Z_n\setminus\{z_j\}) \ .
 \end{multline}
Evaluating the residues, this yields
\begin{align}
  2z_0W_{g,n+1}(z_0,Z_n)  =  -\frac12\sum_\pm  Q'_{g,n}(\pm z_0;Z_n) + \sum_{j=1}^n   \frac{d}{dx_j}   \frac{W_{g,n}(Z_n) }{x_0-x_j}   \ .
  \label{2zww}
    \end{align}
Let us use Eq.~\eqref{qpgn} for $Q'_{g,n}(\pm z_0;Z_n)$. Remember that $W_{g,n}(Z_n)$ is odd under  $z_j\mapsto -z_j$. So the first two terms of $Q'_{g,n}(\pm z_0;Z_n)$ are even in $z_0$, and survive the sum over signs. The third term cancels the derivative term in Eq.~\eqref{2zww}. The last term is odd in $z_0$ because $\frac{B(z_0,z_j)}{dx(z_0)dx(z_j)}-W_{0,2}(z_0,z_j)=\frac{1}{(x_0-x_j)^2}$ is even. We are left with
    \begin{align}
2z_0 W_{g,n+1}(z_0,Z_n) = - P_{g,n}(z_0;Z_n)  + 2z_0 W_{g,n+1}(z_0,Z_n) \ ,
    \end{align}
which implies $P_{g,n}(z_0;Z_n)
    = 0$ for $2g-2+n\geq 0$. The remaining cases are computed directly:
    \begin{subequations}
    \begin{align}
    P_{0,0}(z_0) &= W_{0,1}(z_0)^2 = x_0\ ,
    \\
    P_{0,1}(z_0,z_1) 
    & =  2W_{0,1}(z_0) W_{0,2}(z_0,z_1) - \frac{d}{dx_1} \frac{W_{0,1}(z_1)}{x_0-x_1}  = 0\ .
    \end{align}
    \end{subequations}

    \item
    The wave function is written in terms of $(F_{g,n})_{g \ge 0, n \ge 1}$ as follows,
    \begin{align}
        \log\psi(z) =  \sum_{n \ge 1} \sum_{g \ge 0} \frac{N^{2 - 2g - n}}{n!} F_{g,n}(z,\ldots,z) \ .
    \end{align}
    From the expressions~\eqref{eq:Airy_curve_omega}, we deduce
    \begin{align}
        F_{1,1}(z) = \frac{1}{48 z^3}
        \qquad , \qquad
        F_{0,3}(z,z,z) = \frac{1}{2 z^3}\ .
    \end{align}
    Together with the expressions of $F_{0,1}$ and $F_{0,2}$,
    this allows to compute the first three terms of $\log \psi(z)$, i.e. the terms in $N^1,N^0,N^{-1}$. This yields Eq.~\eqref{eq:Airy_wf_WKB}, which implies
    \begin{align}
        u(z) 
        =  z - N^{-1} \frac{1}{4z^2} - N^{-2} \frac{5}{32 z^5} + O(N^{-3})\ .
    \end{align}
    By an explicit calculation, we deduce
    \begin{subequations}        
    \begin{align}
    \frac{d u(z)}{dx(z)} 
    &= \frac{1}{2z} + N^{-1} \frac{1}{4 z^4}+ N^{-2} \frac{25}{64 z^7} + O(N^{-3}) \ ,
    \\
    u(z)^2 &= z^2 - N^{-1} \frac{1}{2z}  - N^{-2} \frac{1}{4 z^4} + O(N^{-3})\ .
    \end{align}
    \end{subequations}
    which gives the relation~\eqref{eq:uqdu}.

    \item For $\Phi_{0,1}$ we find
\begin{align}    
\Phi_{0,1}(x) = \frac23 z^3 \ , \
\Phi^{(1)}_{0,1}(z_1;x) =  z_1 \ , \
\Phi^{(2)}_{0,1}(z_1;x) =  \frac{1}{2 z_1} \ , \
\Phi^{(1,1)}_{0,1}(z_1,z_2;x) =  0 \ .
\end{align}
For $\Phi_{0, 2}$ we find
\begin{subequations}
\begin{align}    
& \Phi_{0,2}(x) =   -\frac12 \log{(2z)} \ , \
\Phi^{(1)}_{0,2}(z_1;x) =  \frac{-1}{2z_1(z_1+z)} \ ,
\\
& \Phi^{(2)}_{0,1}(z_1;x) =  \frac{2z_1+z}{4 z_1^3(z_1+z)^2} \ , \
\Phi^{(1,1)}_{0,1}(z_1,z_2;x) =   \frac{1}{4z_1 z_2(z_1+z_2)^2} \ .
\end{align}
\end{subequations}
For the stable cases $2g-2+n>0$ we find
\begin{subequations}
\begin{align}
F_{g,n}(z_1,\dots,z_n) = &  2^{2-2g-n} \sum_{d_1,\dots,d_n} \langle \tau_{d_1} \dots \tau_{d_n}\rangle_g \prod_{i=1}^n \frac{(2d_i-1)!!}{z_i^{2d_i+1}}\ , \\
\Phi_{g,n}(z) = &  \frac{2^{2-2g-n}}{n!z^{3(2g-2+n)}}  \sum_{d_1,\dots,d_n} \langle \tau_{d_1} \dots \tau_{d_n}\rangle_g \prod_{i=1}^n (2d_i-1)!!\ .
\end{align}
\end{subequations}

\item    
Integrating the loop equation \eqref{eq:AiryLoopeqPgn} with respect to $z_1,\dots,z_n$ from $\infty$ to $z$, we obtain equations for derivatives of $\Phi_{g,n}$:
\begin{multline}
 x_0 \delta_{g,0}\delta_{n,0}
= \Phi^{(1,1)}_{ g-1,n+2}(z_0,z_0,x)+  \frac{1}{x_0-x}  \left( \Phi^{(1)}_{g,n}(z_0,x) - \Phi^{(1)}_{g,n}(z_0,x)\right) \\
 + \sum_{g_1+g_2=g} \sum_{m=0,\cdots,n}  \Phi^{(1)}_{g_1,1+m}(z_0,x)  \Phi^{(1)}_{g_2,1+n-m}(z_0,x)
 \ .
\end{multline}
We then perform the limit $z_0\to z$, and rewrite the result in terms of $x$-derivatives of $\Phi_{g,n}$, using
\begin{align}
 \Phi_{g,n}'(x)=\Phi_{g,n}^{(1)}(z,x)\quad , \quad \Phi_{g,n}''(x)= \Phi_{g,n}^{(2)}(z,x)+\Phi^{(1,1)}_{g,n}(z,z,x)\ .
\end{align}
This leads to Eq.~\eqref{eq:Phi_recursion}.

\item
From Eq.~\eqref{eq:Phi_recursion} we deduce an equation for $\log \psi = \sum_{g=0}^\infty \sum_{n=0}^\infty N^{2-2g-n} \Phi_{g,n}$,
\begin{align}
 \frac{d^2}{dx^2}  \log\psi(x)  +   \frac{d}{dx}\log\psi(x)\frac{d}{dx}\log\psi(x) = N^2x\ .
\end{align}
This is the Airy equation.

\end{enumerate}

\paragraph{Solution \ref{exobackground}.}

\begin{enumerate}
    \item
    Using the definitions \eqref{eq:tuz} of the parameters $\tau,u_0$ and the definition \eqref{eq:ts} the the Theta function $\theta(u;\tau)$, we write derivatives of the  twisted Theta function as infinite sums:
    \begin{align}
    \Theta^{(k)}
    = \sum_{n\in \mathbb Z} e^{-N^2 F_0(\epsilon^*)-F_1(\epsilon^*)} e^{2i\pi \mu n} (n-N\epsilon^*)^k e^{-N(n-N\epsilon^*)F'_0(\epsilon^*)} e^{-\frac12 (n-N\epsilon^*)^2 F_0''(\epsilon^*)}\ .
    \end{align}
    Taking the derivative w.r.t. $\epsilon^*$, we obtain terms that correspond to the factors of the summand:
    \begin{multline}
   \partial_{\epsilon^*} \Theta^{(k)}
     =  -(N^2 F'_0(\epsilon^*)+F'_1(\epsilon^*)) \Theta^{(k)} - N k  \Theta^{(k-1)}  + N^2 F'_0(\epsilon^*)  \Theta^{(k)}
     \\
     -N F''_0(\epsilon^*)  \Theta^{(k+1)}
     + N F''_0(\epsilon^*)  \Theta^{(k+1)}  -\frac12  F'''_0(\epsilon^*)  \Theta^{(k+2)} \ .
    \end{multline}
    Due to cancellations between a few terms, this boils down to the announced result.

\item    
   We first express the terms $\mathcal{Z}_0,\mathcal{Z}_1,\mathcal{Z}_2$ in terms of the twisted Theta function $\Theta$, the free energies $F_g$, and their derivatives:
   \begin{subequations}
    \begin{align}
    &  \mathcal{Z}_0 = \Theta \qquad , \qquad
        \mathcal{Z}_1 = - F_1' \Theta^{(1)} - \frac{1}{6} F_0''' \Theta^{(3)}\ , \\
    &
    \mathcal{Z}_2 =
    -F_2 \Theta
    + \frac{\left(F_1'\right)^2-F_1''}{2}\Theta^{( 2)}
    + \frac{4F_0''' F_1'-F_0''''}{24}
    \Theta^{( 4)}
    + \frac{\left(F_0'''\right)^2}{72}\,\Theta^{( 6)}
    \ .
    \end{align}
   \end{subequations}
    We then compute the $\epsilon^*$-derivative of each term using the derivative \eqref{petk} of the twisted Theta function, which involves terms in $O(N^0)$ and in $O(N^1)$:
    \begin{subequations}
        \begin{align}
            \partial_{\epsilon^*} \mathcal{Z}_0 & = - F_1' \Theta - \frac{F_0'''}{2} \Theta^{(2)} , \\
            \partial_{\epsilon^*} \mathcal{Z}_1
            & =  N \left( F_1' \Theta + \frac{F_0'''}{2} \Theta^{(2)} \right)
            \nonumber \\ &
            + \left((F_1')^2- F_1''\right) \Theta^{(1)} + \frac{4F_0''' F_1'-F_0''''}{6} \Theta^{(3)} + \frac{(F_0''')^2}{12} \Theta^{(5)} , \\
            \partial_{\epsilon^*} \mathcal{Z}_2
            & = -N \left(\left((F_1')^2- F_1''\right) \Theta^{(1)} + \frac{4F_0''' F_1'-F_0''''}{6} \Theta^{(3)} + \frac{(F_0''')^2}{12} \Theta^{(5)}  \right)
            + O(1)\ .
        \end{align}
    \end{subequations}
    When computing $\partial_{\epsilon^*}\mathcal{Z}$, the term of order $N^0$ in $\partial_{\epsilon^*}\mathcal{Z}_N$ therefore cancels with the term of order $N$ in $\partial_{\epsilon^*}\mathcal{Z}_{N+1}$. We have therefore checked that $\partial_{\epsilon^*} \mathcal{Z} = O(N^{-2})$.
    In graphical representation, these calculations read
    \begin{subequations}
    \begin{align}
        \partial_{\epsilon^*} \mathcal{Z}_0 & = \
    \begin{tikzpicture}[baseline=(current  bounding  box.center),thick,scale=.7]
  \draw (0,0) circle [x radius = 1, y radius = .7];
  \begin{scope}
   \draw [clip] (-.5,.1) to [bend right] (.5,.1);
   \draw (-.5,-.1) to [bend left] (.5,-.1);
  \end{scope}
  \draw [] (0.8,0) -- ++(1,0) node (dot1) {};
  \draw (dot1) circle [radius=.1];
  \filldraw (1.5,1) circle [radius=.1];
 \end{tikzpicture}
 \ + \ \frac{1}{2} \
  \begin{tikzpicture}[baseline=(current  bounding  box.center),thick,scale=.7]
  \draw (0,0) circle [x radius = 1, y radius = .7];
   \begin{scope}
    \draw [clip] (-.5,.1) to [bend right] (.5,.1);
    \draw (-.5,-.1) to [bend left] (.5,-.1);
   \end{scope}
   \node (dot2) at (1.7,0) {};
   \node (dot3) at (1.2,1) {};
   \draw [] (.7,.1) to [out=30,in=180-45] (1.7,0);
   \draw [] (.7,-.3) to [out=-30,in=180+45] (1.7,0);
   \draw (.4,.4) -- (1.2,1);
   \filldraw (dot2) circle [radius=.1];
   \draw (dot3) circle [radius=.1];
  \end{tikzpicture}
  \\
  \partial_{\epsilon^*} \mathcal{Z}_1 & = \
  \begin{tikzpicture}[baseline=(current  bounding  box.center),thick,scale=.6]
  \draw (0,0) circle [x radius = 1, y radius = .7];
  \begin{scope}
   \draw [clip] (-.5,.1) to [bend right] (.5,.1);
   \draw (-.5,-.1) to [bend left] (.5,-.1);
  \end{scope}
  \draw [] (0.8,0) -- ++(1,0) node (dot1) {};
  \draw [] (0.5,.4) -- (1.5,1) node (dot2) {};
  \draw (dot1) circle [radius=.1];
  \filldraw (dot2) circle [radius=.1];
%  \filldraw (1.5,1) circle [radius=.1];
 \end{tikzpicture}
 \ + \
 \begin{tikzpicture}[baseline=(current bounding  box.center),thick,scale=.6]
  \draw (0,0) circle [x radius = 1, y radius = .7];
  \begin{scope}
   \draw [clip] (-.5,.1) to [bend right] (.5,.1);
   \draw (-.5,-.1) to [bend left] (.5,-.1);
  \end{scope}
  \draw [] (0.8,0) -- ++(1,0) node (dot1) {};
  \filldraw (dot1) circle [radius=.1];
  \begin{scope}[shift = {(.7,-1.7)}]
    \draw (0,0) circle [x radius = 1, y radius = .7];
  \begin{scope}
   \draw [clip] (-.5,.1) to [bend right] (.5,.1);
   \draw (-.5,-.1) to [bend left] (.5,-.1);
  \end{scope}
  \draw [] (0.8,0) -- ++(1,0) node (dot1) {};
  \draw (dot1) circle [radius=.1];
  \end{scope}
 \end{tikzpicture}
 \ + \ \frac{1}{2} \
 \begin{tikzpicture}[baseline=(current  bounding  box.center),thick,scale=.6]
   \draw (0,0) circle [x radius = 1, y radius = .7];
   \begin{scope}
    \draw [clip] (-.5,.1) to [bend right] (.5,.1);
    \draw (-.5,-.1) to [bend left] (.5,-.1);
   \end{scope}
   \begin{scope}[shift={(.7,-1.7)}]
    \draw (0,0) circle [x radius = 1, y radius = .7];
   \end{scope}
   \draw [] (0.8,0) -- ++(1,0) node (dot1) {};
%   \draw [] (1.2,-1.4) -- (1.8,0);
   \draw [] (1.1,-1.3) to [out=up,in=180+30] (1.8,0);
   \draw [] (1.3,-1.5) to [out=60,in=down] (1.8,0);
   \draw [] (1.4,-1.7) -- ++ (.8,0) node (dot2) {};
   \filldraw (1.8,0) circle [radius=.1];
   \draw (dot2) circle [radius=.1];
 \end{tikzpicture}
 \ - N \
\begin{tikzpicture}[baseline=(current  bounding  box.center),thick,scale=.6]
  \draw (0,0) circle [x radius = 1, y radius = .7];
  \begin{scope}
   \draw [clip] (-.5,.1) to [bend right] (.5,.1);
   \draw (-.5,-.1) to [bend left] (.5,-.1);
  \end{scope}
  \draw [] (0.8,0) -- ++(1,0) node (dot1) {};
  \draw (dot1) circle [radius=.1];
  \filldraw (1.5,1) circle [radius=.1];
 \end{tikzpicture}
 \nonumber \\
 & \  + \frac{1}{6} \
  \begin{tikzpicture}[baseline=(current  bounding  box.center),thick,scale=.6]
  \draw (0,0) circle [x radius = 1, y radius = .7];
  \draw [] (0.7,.2) to [bend left] (1.8,0) node (dot1) {};
  \draw [] (0.7,-.2) to [bend right] (1.8,0);
  \draw [] (0.7,0) -- (1.8,0);
  \draw [] (0.5,.4) -- (1.5,1) node (dot2) {};
  \filldraw (dot1) circle [radius=.1];
  \draw (dot2) circle [radius=.1];
 \end{tikzpicture}
 \ + \frac{1}{6} \
 \begin{tikzpicture}[baseline=(current  bounding  box.center),thick,scale=.6]
  \draw (0,0) circle [x radius = 1, y radius = .7];
  \begin{scope}
   \draw [clip] (-.5,.1) to [bend right] (.5,.1);
   \draw (-.5,-.1) to [bend left] (.5,-.1);
  \end{scope}
  \draw [] (0.8,0) -- ++(1,0) node (dot1) {};
  \draw (dot1) circle [radius=.1];
    \begin{scope}[shift={(.7,-1.7)}]
  \draw (0,0) circle [x radius = 1, y radius = .7];
  \draw [] (.7,0) -- ++(1,0) node (dot2) {};
  \draw [] (.7,.3) to [out=30,in=180-45] (1.7,0);
  \draw [] (.7,-.3) to [out=-30,in=180+45] (1.7,0);
  \filldraw (dot2) circle [radius=.1];
    \end{scope}
 \end{tikzpicture}
 \ + \frac{1}{12} \
 \begin{tikzpicture}[baseline=(current  bounding  box.center),thick,scale=.6]
   \draw (0,0) circle [x radius = 1, y radius = .7];
   \begin{scope}[shift={(.7,-1.7)}]
    \draw (0,0) circle [x radius = 1, y radius = .7];
   \end{scope}
   \draw [] (.7,0) -- ++(1.1,0) node (dot1) {};
   \draw [] (.7,.3) to [out=30,in=180-45] (1.8,0);
   \draw [] (.7,-.3) to [out=-30,in=180+45] (1.8,0);
%   \draw [] (1.2,-1.4) -- (1.8,0);
   \draw [] (1.1,-1.3) to [bend left] (1.8,0);
   \draw [] (1.3,-1.5) to [bend right] (1.8,0);
   \draw (1.4,-1.7) -- ++(.8,0) node (dot2) {};
   \filldraw (dot1) circle [radius=.1];
   \draw (dot2) circle [radius=.1];
 \end{tikzpicture}
 \ - \frac{N}{2} \
  \begin{tikzpicture}[baseline=(current  bounding  box.center),thick,scale=.6]
  \draw (0,0) circle [x radius = 1, y radius = .7];
   \begin{scope}
    \draw [clip] (-.5,.1) to [bend right] (.5,.1);
    \draw (-.5,-.1) to [bend left] (.5,-.1);
   \end{scope}
   \node (dot2) at (1.7,0) {};
   \node (dot3) at (1.2,1) {};
   \draw [] (.7,.1) to [out=30,in=180-45] (1.7,0);
   \draw [] (.7,-.3) to [out=-30,in=180+45] (1.7,0);
   \draw (.4,.4) -- (1.2,1);
   \filldraw (dot2) circle [radius=.1];
   \draw (dot3) circle [radius=.1];
  \end{tikzpicture}
    \end{align}
    \end{subequations}
    Notice that there are two different diagrams with one sphere and one torus. Both stand for $F_0'''F_1'\Theta^{(3)}$.

    \item
    Let $G$ be a graph with $\chi(G) = m$. We also call $G$ the corresponding contribution to the partition function. Since the partition function is linear in $\Theta$, $G$ has one vertex. The ${\epsilon^*}$-derivative of the partition function involves graph with one full vertex and also one empty vertex with one leg. Suppose $\partial_{\epsilon^*} G$ involves a graph of the type
    \begin{align}
  \begin{tikzpicture}[baseline=(current  bounding  box.center),thick,scale=1.2]
  \filldraw[draw=black,fill=gray!30] (0,0) circle [x radius = 1, y radius = .7]; % node (G') {$\partial_{\epsilon^*} G$};
  \draw [] (0.6,.2) to [out=45,in=180-60] (1.8,-.1) node (dot1) {};
  \draw [] (0.6,.1) to [bend left] (1.8,-.1);
  \draw [] (0.6,-.4) to [bend right=45] (1.8,-.1);
  \node at (1.2,-.05) {$\vdots$};
  \draw [] (0.5,.5) -- ++(1.3,0) node (dot2) {};
  \filldraw (dot1) circle [radius=.1];
  \draw (dot2) circle [radius=.1];
 \end{tikzpicture}
    \label{eq:G_e-der}
    \end{align}
    Then, there is always another graph $G'$ with $\chi(G') = m + 1$ of the form,
    \begin{align}
  \begin{tikzpicture}[baseline=(current  bounding  box.center),thick,scale=1.2]
  \filldraw[draw=black,fill=gray!30] (0,0) circle [x radius = 1, y radius = .7]; % node (G') {$\partial_{\epsilon^*} G$};
  \draw [] (0.6,.2) to [out=45,in=180-60] (1.8,-.1) node (dot1) {};
  \draw [] (0.6,.1) to [bend left] (1.8,-.1);
  \draw [] (0.6,-.4) to [bend right=45] (1.8,-.1);
  \node at (1.2,-.05) {$\vdots$};
  \draw [] (0.5,.5) to [out=30,in=90] (1.8,-.1);
  \filldraw (dot1) circle [radius=.1];
 \end{tikzpicture}
    \end{align}
    And the $\epsilon^*$-derivative of $G'$ cancels the previous contribution~\eqref{eq:G_e-der}.

\end{enumerate}

%============ section 5 Orthogonal polynomials  =========

\section{Chapter 5}

\paragraph{Solution \ref{exo:HOMFLY_Jones_unknot}.}

\begin{enumerate}
    \item
    Rewriting the hyperbolic sines in terms of exponential, we recognize a Vandermonde determinant:
    \begin{align}
       \prod_{1 \le i<j \le n} 2\sinh\frac{u_i-u_j}{2}= \prod_{i<j}^n \left( e^{\frac{u_i - u_j}{2}} - e^{-\frac{u_i - u_j}{2}} \right)
         =
        e^{-\frac{n-1}{2} \sum_{i=1}^n u_i} \prod_{i<j}^n \left( e^{u_i} - e^{u_j} \right)\ .
    \end{align}

    \item
    Rewriting the Schur polynomial  using the Weyl vector,
    \begin{align}
        \frac{\det_{1 \le i, j \le n} e^{(\lambda_i + n - i) u_j} }{\det_{1 \le i, j \le n} e^{(n - i) u_j} }
         = \frac{\det_{1 \le i, j \le n} e^{(\lambda + \rho)_i u_j} }{\det_{1 \le i, j \le n} e^{\rho_i u_j}} \ ,
    \end{align}
    the integrand becomes a product of two determinants:
    \begin{align}
        \mathcal{Z}_{\mathfrak{K}_{1,1},\lambda}(q)
        & = \int_{\mathbb{R}^n} du_1 \cdots du_n e^{- \sum_{i=1}^n \frac{u_i^2}{2g_s}} \det_{1 \le i, j \le n} e^{(\lambda + \rho)_i u_j} \det_{1 \le i, j \le n} e^{\rho_i u_j}\ .
     \end{align}
        The determinantal formula expresses this integral in terms of one-dimensional integrals of the type:
        \begin{align}
        \int_{\mathbb{R}} du \, e^{-\frac{u^2}{2g_s}} e^{(\lambda + \rho)_i u} e^{\rho_j u} = e^{\frac{g_s}{2}((\lambda + \rho)_i + \rho_j)^2}\ .
        \end{align}
        We therefore obtain
        \begin{align}
       \mathcal{Z}_{\mathfrak{K}_{1,1},\lambda}(q) =
        n! \det_{1 \le i, j \le n} \left( e^{\frac{g_s}{2}((\lambda + \rho)_i + \rho_j)^2} \right)
        \ .
    \end{align}
    Identifying $q = e^{-g_s}$, we obtain the coloured HOMFLY polynomial of the unknot,
    \begin{align}
        \text{HOMFLY}_{\mathfrak{K}_{1,1},\lambda}(q)
         = \frac{\mathcal{Z}_{\mathfrak{K}_{1,1},\lambda}(q) }{\mathcal{Z}_{\mathfrak{K}_{1,1},\emptyset}(q) }
        = q^{-\tfrac{1}{2} \sum_{i=1}^n \lambda_i(\lambda + 2 \rho)_i} \frac{\det_{1 \le i, j \le n} q^{-(\lambda + \rho)_i \rho_j} }{\det_{1 \le i, j \le n} q^{-\rho_i \rho_j} }\ .
    \end{align}

    \item Taking $n = 2$ and $\lambda = (N-1,0)$, the Weyl vector is given by $\rho = (\frac{1}{2},-\frac{1}{2})$.
    In this case, we obtain
    \begin{align}
        J_{\mathfrak{K}_{1,1},N}(q) & = \frac{\mathcal{Z}_{\mathfrak{K}_{1,1},(N-1,0)}(q)}{\mathcal{Z}_{\mathfrak{K}_{1,1},\emptyset}(q)} = q^{-\frac{N(N-1)}{2}} \frac{q^{\frac{N}{2}} - q^{-\frac{N}{2}}}{q^{\frac{1}{2}} - q^{-\frac{1}{2}}}\ ,
    \end{align}
	  which agrees with the result of \autoref{exo:jones}.
\end{enumerate}

\paragraph{Solution \ref{exo:volun}.}

\begin{enumerate}
    \item
    Writing the integrand in terms of determinants, we can use the determinantal formula to evaluate the integral as a determinant, with the result $(2 \pi)^N=\operatorname{Vol}((U_1)^N)$:
\begin{align}
\frac{\operatorname{Vol}(U_N)}{\operatorname{Vol}(U_N/U_1^N)}  & = \frac{1}{N!} \int_{(S^1)^N} \left( \det_{1 \le i, j \le N} z_i^{j-1} \right) \left( \det_{1 \le i, j \le N} z_i^{-j+1} \right) \prod_{i=1}^N \frac{dz_i}{iz_i}\ ,
  \nonumber \\
     & = \det_{0 \le i, j \le N-1} \left( \int_{S^1} z^{j-i} \frac{dz}{i z} \right)
     = \det_{0 \le i, j \le N-1} \left( 2 \pi \delta_{i,j} \right)
     = (2 \pi)^N\ .
\end{align}

    \item
    The integrand now has an extra factor $\det f(U) =  \prod_{i=1}^N f(z_i)$. The result is  $\det_{0 \le i, j \le N-1} \left( \int_{S^1} z^{j-i} f(z) \frac{dz}{2 \pi i z} \right)$, which coincides with the announced Toeplitz determinant.

    \item We again obtain a Toeplitz determinant, which can then be evaluated more explicitly as it is also a Vandermonde determinant:
    \begin{align}
        \left< \det \frac{\Theta(U;q)}{C_N(q)} \right>
         & = \frac{1}{C_N(q)^N} \det_{0 \le i, j \le N-1} q^{(i-j)^2}
        = \frac{q^{2 \sum_{i=1}^{N} (i-1)^2 }}{C_N(q)^N}  \det_{1 \le i, j \le N} q^{-2(i-1)(j-1)}
        \nonumber \\
        & = \frac{q^{2 \sum_{i=1}^{N} (i-1)^2 }}{C_N(q)^N}  \prod_{i<j}^N \left( q^{-2(j-1)} - q^{-2(i-1)} \right)
         = \prod_{j=1}^{N-1} \left( 1 - q^{2j} \right)^{-j}\ .
    \end{align}
    In the large $N$ limit, this tends to the MacMahon function,
    \begin{align}
        \lim_{N \to \infty} \left< \det \frac{\Theta(U;q)}{C_N(q)} \right> = \prod_{j=1}^{\infty} \left( 1 - q^{2j} \right)^{-j} ,
    \end{align}
    which is a generating function of the number of plane partitions.
\end{enumerate}

\paragraph{Solution \ref{exoHeinephi}.}

Let us compute the average of the inverse of the characteristic polynomial, for matrices of size $k$:
\begin{align}
    \left< \frac{1}{\det (x - M)} \right>
     = \frac{1}{k!\mathcal{Z}_k}  \int \frac{\Delta_{k}(\lambda_1,\ldots,\lambda_k)^2}{\prod_{i=1}^k (x - \lambda_i)} \prod_{i=1}^k e^{-V(\lambda_i)} d \lambda_i\ .
    \end{align}
We then decompose the Vandermonde determinant,
\begin{align}
    \Delta_k(\lambda_1,\ldots,\lambda_k)
     = \Delta_{k-1}(\lambda_1,\dots,\lambda_{k-1}) \prod_{j=1}^{k-1} (\lambda_k - \lambda_j) \ .
\end{align}
We also use the identity \eqref{oopxl}, where by permutation symmetry we may keep only the pole at $x=\lambda_k$. This yields:
\begin{align}
 \left< \frac{1}{\det (x - M)} \right>
    = \frac{1}{(k-1)!\mathcal{Z}_k}  \int \Delta_{k-1}(\lambda_1,\ldots,\lambda_{k-1})^2  \frac{\prod_{i=1}^{k-1}(\lambda_k - \lambda_i)}{x - \lambda_k} \prod_{i=1}^k e^{-V(\lambda_i)} d \lambda_i\ .
\end{align}
We perform the integrals over $\lambda_1,\dots,\lambda_{k-1}$, which by Heine's formula yields an orthogonal polynomial $p_{k-1}$, or equivalently a function $\psi_{k-1}$ \eqref{eq:psik}:
\begin{align}
   \left< \frac{1}{\det (x - M)} \right> = \frac{1}{h_{k-1}} \int \frac{p_{k-1}(\lambda_k)}{x - \lambda_k} e^{-V(\lambda_k)} d\lambda_k
     = \frac{1}{\sqrt{h_{k-1}}} \int \frac{\psi_{k-1}(\lambda_k)}{x - \lambda_k} e^{-\frac{1}{2}V(\lambda_k)} d\lambda_k
  \ .
\end{align}
This is proportional to $\varphi_{k-1}(x)$, leading to the Heine-type formula for that function.

\paragraph{Solution \ref{exodetD}.}

The recursion relation involes the matrix $\mathcal{R}_N$ \eqref{recPsiRN}, which obeys
\begin{align}
    \mathcal{R}_N^{-1} =
    \begin{pmatrix}
        \frac{x - S_N}{\gamma_N} & - \frac{\gamma_{N+1}}{\gamma_N} \\
        1 & 0
    \end{pmatrix}
    , \qquad
    \mathcal{R}_N^{-1} \partial_x \mathcal{R}_N =
    \begin{pmatrix}
        0 & - \frac{1}{\gamma_N} \\ 0 & 0
    \end{pmatrix} .
\end{align}
This allows us to write a recursion relation for $\operatorname{Tr} D_N^2$:
Hence, together with \eqref{eq:prp}, we obtain the recursion relation,
\begin{align}
    \operatorname{Tr} D_{N+1}^2 = \operatorname{Tr} D_N^2 - \frac{2}{\gamma_N} (D_N)_{N,N-1}=\operatorname{Tr} D_N^2 - 2 \left( \frac{V'(x) - V'(Q)}{x - Q} \right)_{N,N}\ .
\end{align}
Iterating on $N$ from $\operatorname{Tr} D_0^2 = \frac{V'(x)^2}{2}$, we obtain
\begin{align}
    \operatorname{Tr} D_{N}^2  = \frac{V'(x)^2}{2} - 2 \operatorname{Tr}\left( \Pi_N \frac{V'(x) - V'(Q)}{x - Q} \right)\ .
\end{align}
It only remains to recognize that the second term is the polynomial $P_N(x)$ \eqref{eq:P_N_traceform}.

\paragraph{Solution \ref{exoproofdetformula}.}

\begin{enumerate}
    \item
    By definition of the Tau function $\tau_N(\vec t)$, we have
%    \begin{multline}
%H(x_1,x_2;y_1,y_2) 
%=                  \frac{(x_1-x_2)(y_1-y_2)}{\prod_{i,j=1}^2 (x_i-y_j)}\left< \frac{\det(x_1 - M) \det(x_2 - M)}{\det(y_1 - M) \det(y_2 - M)} \right>
%        \\
%          - \frac{1}{(x_2 - y_1)(x_1 - y_2)} \left< \frac{\det(x_2 - M)}{\det(y_1 - M)} \right> \left< \frac{\det(x_1 - M)}{\det(y_2 - M)} \right>
%          \\
%        + \frac{1}{(x_1 - y_1)(x_2 - y_2)} \left< \frac{\det(x_1 - M)}{\det(y_1 - M)} \right> \left< \frac{\det(x_2 - M)}{\det(y_2 - M)} \right>\ .
%        \label{fhtn}
%    \end{multline}
%   Multiplying with $\prod_{i,j=1}^2 (x_i-y_j)$, we obtain a polynomial of $x_1,x_2$:
    \begin{multline}
\prod_{i,j=1}^2 (x_i-y_j) H(x_1,x_2;y_1,y_2) 
= (x_1-x_2)(y_1-y_2)\left< \frac{\det(x_1 - M) \det(x_2 - M)}{\det(y_1 - M) \det(y_2 - M)} \right>
        \\
          - (x_1-y_1)(x_2-y_2) \left< \frac{\det(x_2 - M)}{\det(y_1 - M)} \right> \left< \frac{\det(x_1 - M)}{\det(y_2 - M)} \right>
          \\
        + (x_1-y_2)(x_2-y_1)\left< \frac{\det(x_1 - M)}{\det(y_1 - M)} \right> \left< \frac{\det(x_2 - M)}{\det(y_2 - M)} \right> \ .
        \label{fhtn}
    \end{multline}
This is manifestly a polynomial of $x_1, x_2$. Its degree follows from $\det_N(x_i - M) = x_i^N + O(x_i^{N-1})$.

    \item
    In the limit $x_1\to y_1$, the second term of $H(x_1,x_2;y_1,y_2)$ \eqref{fhtn} vanishes, while the sum of the first and third term also vanishes. The same happens for all limits $x_i\to y_j$,  therefore the polynomial $ \prod_{i,j=1}^2 (x_i-y_j) H(x_1,x_2;y_1,y_2)$ is divisible by $(x_1-y_1)(x_1-y_2)$ as a polynomial of $x_1$, and is divisible by $(x_2-y_1)(x_2-y_2)$ as a polynomial of $x_2$.
    This implies that $H(x_1,x_2;y_1,y_2)$  itself is a polynomial of $x_1,x_2$, of degree at most $N-1$ in each variable, and it may be written in terms of orthogonal polynomials as
    \begin{align}
        H(x_1,x_2;y_1,y_2)
        & = \sum_{0 \le i, j \le N - 1} c_{i,j}(y_1,y_2) p_i(x_1) p_j(x_2)\ .
        \label{eq:H_orth_poly_exp}
    \end{align}

    \item
    The coefficients of the expansion \eqref{eq:H_orth_poly_exp} are given by
    \begin{align}
       h_i h_j c_{i,j}(y_1,y_2) =  \int dx \int dx' \, e^{- V(x) - V(x')} p_i(x) p_j(x')  H(x,x';y_1,y_2)\ .
    \end{align}
    
    Let us write the integral of the 1st term of $H$, renaming $x=x_{N+1}$ and $x'=x_{N+2}$ as
    \begin{multline}
     \int dx_1\dots dx_{N+2} \  \Delta(x_1,\dots,x_N)^2  \  p_i(x_{N+1}) p_j(x_{N+2}) \\
   \times    \frac{(x_{N+1}-x_{N+2})(y_1-y_2)}{(x_{N+1}-y_1)(x_{N+1}-y_2)(x_{N+2}-y_1)(x_{N+2}-y_2)}  \prod_{l=1}^N \frac{(x_{N+1}-x_l)(x_{N+2}-x_l)}{(y_1-x_l)(y_2-x_l)}  \\
=  (y_1-y_2)   \int dx_1\dots dx_{N+2} \  \Delta(x_1,\dots,x_{N+2}) \prod_{l=1}^{N+2} \frac{e^{-V(x_l)}}{(y_1-x_l)(y_2-x_l)}  \\  \times
\Delta(x_1,\dots,x_N) \  p_i(x_{N+1}) p_j(x_{N+2})    \ .
    \end{multline}
Since the 1st line of the integrand is totally antisymmetric, we can antisymmetrize the second factor $\Delta(x_1,\dots,x_N) \  p_i(x_{N+1}) p_j(x_{N+2})$. However, since $\deg p_i\leq N-1$ and $\deg p_j\leq N-1$, the last 2 lines of the  determinant will be linear combinations of the $N$ first, and thus the antisymmetrization gives 0.

Similarly, consider the integral of the 2nd term of $H$ with respect to $x$, renamed $x_{N+1}$:
    \begin{multline}
    \int dx_{N+1}  \frac{p_i(x_{N+1})e^{-V(x_{N+1})}}{(x_{N+1}-y_2)}   \left< \frac{\det(x_{N+1} - M)}{\det(y_2 - M)} \right> \\
=     \int dx_1\dots dx_{N+1} \  \Delta(x_1,\dots,x_N)^2
       \frac{p_i(x_{N+1})e^{-V(x_{N+1})}}{(x_{N+1}-y_2)}  \prod_{l=1}^N \frac{(x_{N+1}-x_l)e^{-V(x_l)}}{(y_2-x_l)}  \\
=  -   \int dx_1\dots dx_{N+1} \  \Delta(x_1,\dots,x_{N+1}) \prod_{l=1}^{N+1} \frac{e^{-V(x_l)}}{(y_2-x_l)}  \\ 
\times \Delta(x_1,\dots,x_N) \  p_i(x_{N+1})  \ .
    \end{multline}
Again, since the 1st line of the integrand is totally antisymmetric, we can antisymmetrize the second factor $\Delta(x_1,\dots,x_N) \  p_i(x_{N+1}) $. However, since $\deg p_i\leq N-1$, the last line of the  determinant will be linear combinations of the $N$ first, and thus the antisymmetrization gives 0.
The same happens for the 3rd term of $H$.
Therefore $c_{i,j}(y_1,y_2)=0$, thus $H=0$.
\end{enumerate}

\paragraph{Solution \ref{exproofKdetbio}.}

\begin{enumerate}
    \item
    $\left<\det(x-M)\det(y-\tilde M)\right>$ is manifestly a polynomial of $x$ of degree $N$.
    Therefore it is a linear combination of the first $N+1$ biorthogonal polynomial. By Eq.~\eqref{tpkpk}, the coefficients are obtained by taking scalar products with the polynomials $\tilde{p}_k$, leading to Eq. \eqref{hkcky}.

\item
This can be rewritten as a matrix integral. After renaming $x'=\lambda_N$ and $y'=\tilde \lambda_N$, we have
    \begin{align}
N!^2 Z_N h_k c_k(y)
&=\int d\lambda_0\dots d\lambda_{N} d\tilde \lambda_0\dots d\tilde \lambda_{N} \Delta_{N}(\lambda_0,\dots,\lambda_{N-1}) \Delta_{N}(\tilde \lambda_0,\dots,\tilde \lambda_{N-1}) \nonumber\\ 
& \qquad \prod_{i=0}^{N-1} (\lambda_{N}- \lambda_i)  \prod_{i=0}^{N-1} (y-\tilde \lambda_i) 
\left[ \det_{0 \le i, j \le N-1} \omega(\tilde\lambda_i,{\lambda}_j) \right]  \omega(\tilde \lambda_{N},\lambda_{N}) \ \tilde p_k(\tilde \lambda_N) \nonumber\\
&=\int d\lambda_0\dots d\lambda_{N} d\tilde \lambda_0\dots d\tilde \lambda_{N} \Delta_{N+1}(\lambda_0,\dots,\lambda_{N}) \Delta_{N+1}(\tilde \lambda_0,\dots,\tilde \lambda_{N-1},y) \nonumber\\ 
& \qquad \left[ \det_{0 \le i, j \le N-1} \omega(\tilde\lambda_i,{\lambda}_j) \right]  \omega(\tilde \lambda_{N},\lambda_{N}) \ \tilde p_k(\tilde \lambda_N) \ .
    \end{align}
We then decompose the Vandermonde determinants in terms of the biorthogonal polynomials:    
    \begin{align}
& N!^2 Z_N h_k c_k(y) \nonumber \\
& = \sum_{\sigma\in \mathfrak S_{N+1}} \sum_{\tilde\sigma\in \mathfrak S_{N+1}} \sum_{\tau\in \mathfrak S_{N}} (-1)^\sigma (-1)^{\tilde \sigma} (-1)^\tau 
\int d\lambda_0\dots d\lambda_{N} d\tilde \lambda_0\dots d\tilde \lambda_{N} \nonumber\\
& \qquad \left(\prod_{i=0}^N p_{\sigma(i)}(\lambda_i)\right) \left(\prod_{i=0}^{N-1} \tilde p_{\tilde \sigma(i)}(\tilde \lambda_i) \right) \tilde p_{\tilde \sigma(N)}(y) 
\left(\prod_{i=0}^{N-1} \omega(\tilde \lambda_i,\lambda_{\tau(i)}) \right)  \omega(\tilde \lambda_{N},\lambda_{N}) \ \tilde p_k(\tilde \lambda_N) \nonumber\\
& = \sum_{\sigma\in \mathfrak S_{N+1}} \sum_{\tilde\sigma\in \mathfrak S_{N+1}} \sum_{\tau\in \mathfrak S_{N}} (-1)^\sigma (-1)^{\tilde \sigma} (-1)^\tau 
\prod_{i=0}^{N-1} \left<\tilde p_{\tilde \sigma(i)},p_{\sigma\circ\tau(i)}\right>  \nonumber\\
& \qquad\qquad \int d\lambda_{N} d\tilde \lambda_{N} p_{\sigma(N)}(\lambda_N) \ \tilde p_{\tilde \sigma(N)}(y)  \omega(\tilde \lambda_{N},\lambda_{N}) \ \tilde p_k(\tilde \lambda_N) \nonumber\\
& = \sum_{\sigma\in \mathfrak S_{N+1}} \sum_{\tilde\sigma\in \mathfrak S_{N+1}} \sum_{\tau\in \mathfrak S_{N}} (-1)^\sigma (-1)^{\tilde \sigma} (-1)^\tau \prod_{i=0}^{N-1} \left<\tilde p_{\tilde \sigma(i)},p_{\sigma\circ\tau(i)}\right> 
\left<\tilde p_k,p_{\sigma(N)}\right>  p_{\tilde \sigma(N)}(y) \nonumber\\
& = \sum_{\sigma\in \mathfrak S_{N+1}} \sum_{\tilde\sigma\in \mathfrak S_{N+1}} \sum_{\tau\in \mathfrak S_{N}} (-1)^\sigma (-1)^{\tilde \sigma} (-1)^\tau  
\left( \prod_{i=0}^{N-1} \delta_{\tilde \sigma(i),\sigma\circ\tau(i)} h_{\tilde\sigma(i)} \right) \delta_{k,\sigma(N)} h_k \ p_{\tilde \sigma(N)}(y) \ .
    \end{align}
The only possibility to have non-zero contribution is $\sigma(N)=k$, and for $i=0,\dots,N-1$, we can replace $\sigma$ by $\sigma\circ\tau$, 
    \begin{multline}
N!^2 Z_N h_k c_k(y)
= N! \sum_{\sigma\in \mathfrak S_{N+1}, \ \sigma(N)=k} \sum_{\tilde\sigma\in \mathfrak S_{N+1}}  (-1)^\sigma (-1)^{\tilde \sigma} 
h_k \ \tilde p_{\tilde \sigma(N)}(y) \prod_{i=0}^{N-1} \delta_{\tilde \sigma(i),\sigma(i)} h_{\tilde\sigma(i)} \ .
\end{multline}
Since  $\tilde \sigma(i)=\sigma(i)$ for all $i=0,\dots,N-1$, this implies that $\tilde\sigma=\sigma$, and thus $\tilde\sigma(N)=k$, therefore
    \begin{multline}
N!^2 Z_N h_k c_k(y)
= N! \sum_{\sigma\in \mathfrak S_N, \ \sigma(N)=k} 
h_k \ \tilde p_{k}(y) \prod_{i=0}^{N-1}  h_{\sigma(i)} 
= N!^2 Z_N \tilde p_{k}(y). 
\end{multline}
This fixes the coefficient $c_k(y) = \tilde p_k(y) / h_k$, and obtain the result.
\end{enumerate}

\paragraph{Solution \ref{exo:cft1}.}

\begin{enumerate}
    \item
    Applying the Lax matrix formula~\eqref{formulaDN} to our particular potential, we find
    \begin{align}
        D_N(x) = \sum_{j=1}^n \frac{A_j}{x - z_j}\quad \text{ with }\quad  A_j  =  -2i\alpha_j \left( \frac12 \sigma_3 +    \left( \frac{1}{Q-z_j} A^N \right)_{[N-1,N]^2}  \right)\ .
    \end{align}
    From the trace and determinant of $D_N(x)$ \eqref{dn_inv}, let us deduce those of $A_j$ by considering the limit $x\to z_j$. We immediately find $\Tr A_j=0$. Then  in $\det D_N(x)=P_N(x)-\frac14 V'(x)^2$, the term $P_N(x)$ only has a simple pole at $x=z_j$, so that $\det D_N(x)\underset{x\to z_j}{\sim} \frac{\alpha_j^2}{(x-z_j)^2}$, leading to $\det A_j = \alpha_j^2$. Since $A_j$ is a matrix of size two, its eigenvalues are determined by its trace and determinant.

    Using again the formula~\eqref{formulaDN}, we obtain
    \begin{align}
     \sum_{j=1}^n A_j = \lim_{x\to\infty} xD_N(x) = -N\sigma_3 - \left( V'(Q) A^N \right)_{[N-1,N]^2}\ .
    \end{align}
   We have $A^N = \gamma_N\left[\begin{smallmatrix} 0 & 1 \\ -1 & 0 \end{smallmatrix}\right]$ from its definition \eqref{anpnq}, and $V'(Q)_{[N-1,N]^2} = \frac{N}{\gamma_N}\left[\begin{smallmatrix} 0 & 1 \\ 1 & 0 \end{smallmatrix}\right]$ by Eq.~\eqref{eq:recurPQ}, leading to $\left( V'(Q) A^N \right)_{[N-1,N]^2}=-N\sigma_3$.

    \item Let us study the asymptotic behaviour of $\Psi_N(x)$ near $x=z_j$ from its definition \eqref{eq:psindef} in terms of the functions $\psi_N,\varphi_N$. The asymptotic behaviour of $\psi_N$ \eqref{eq:psik} is dominated by a factor $e^{-\frac12 V}$, and the behaviour of $\varphi_N$ \eqref{defvarphi} by a factor $e^{\frac12 V}$, therefore
    \begin{align}
\Psi_N(x) \underset{x\to z_j}{\sim}   \hat\Psi_N(x) e^{-\frac12 V(x) \sigma_3} \sim     \hat\Psi_N(x) (x-z_j)^{-i\alpha_j  \sigma_3}\ ,
    \end{align}
    where $\hat\Psi_N(x)$ is analytic at $x=z_j$. This leads to the monodromy matrix around $z_j$,
   \begin{align}
S_j =  e^{2\pi\alpha_j \sigma_3} = \operatorname{diag}(e^{2 \pi \alpha_j},e^{-2 \pi \alpha_j})\ .
   \end{align}
 Alternatively, we could use the matrix differential equation $\Psi_N'=D_N\Psi_N$ \eqref{Lax_eq}, and show that solutions behave as $\Psi_N(x) \underset{x\to z_j}{\sim} \hat \Psi_N(x) (x-z_j)^{A_j}C$, where $\hat \Psi_N(x)$ is analytic at $x=z_j$ with $\Psi_N(z_j)= \left[\begin{smallmatrix}
1 & 0 \\ 0 & 1                                                                                                                                                                                                                                                                 \end{smallmatrix}
\right]$. Then the monodromy matrix is $S_j= C^{-1}e^{2\pi i A_j}C$.

   Let $R_j =  \partial_{z_j} \Psi_N  \Psi^{-1}_N$. From the asymptotic behaviour of $\Psi_N$ near $x=z_j$, it follows that $R_j(x)$ has a simple pole at $x=z_j$. From the asymptotic behaviour of $\Psi_N$ near $x=z_{i\neq j}$, it follows that $R_j(x)$ is analytic near $x=z_{i\neq j}$. And from the asymptotic behaviour of $\Psi_N$ near $x=\infty$, we deduce $\lim_{x\to\infty} R_j(x)=0$. Therefore, $R_j(x)$ is a rational function with one simple pole at $x=z_j$. Calling $-\tilde{A}_j$ the residue, we have
   \begin{align}
   R_j = \frac{\tilde A_j}{z_j-x}\ .
   \end{align}
   The compatibility of $\Psi_N'=D_N\Psi_N$ \eqref{Lax_eq} with $\partial_{z_j} \Psi_N = R_j\Psi_N$ implies
   \begin{align}
    \partial_x R_j(x) - \partial_{z_j} D_N(x)  - [D_N(x),R_j(x)] = 0\ .
   \end{align}
   Near $x=z_j$, the terms of this equation have double poles. The sum of their residues must vanish, leading to $\tilde{A}_j - A_j + [A_j,\tilde{A}_j]=0$. This implies $\tilde A_j=A_j$.
   
    \item
    Near $x=z_j$, the terms of the compatibility equation have simple poles. The sum of their residues must vanish, leading to the Schlesinger equation.
    
\end{enumerate}

%============ section 6 Angular integrals  =========

\section{Chapter 6}

\paragraph{Solution \ref{exo:izuto}.}

\begin{enumerate}
    \item For diagonal matrices of size two $X$ and $Y$, we have
    \begin{align}
        \Tr U X U^{-1} Y
        =
        \frac{1}{2} (X_1 + X_2)(Y_1 + Y_2)
        + \frac{\cos (2\theta)}{2} (X_1 - X_2)(Y_1 - Y_2)\ .
    \end{align}
    The Itzykson--Zuber integral therefore takes the form
    \begin{align}
        \mathcal{Z}(X,Y)
        & = e^{- \frac{1}{2} (X_1 + X_2)(Y_1 + Y_2)} \int_0^{2\pi} d \theta \, e^{- z \cos (2 \theta)}\ .
    \end{align}
    Applying the change of variable $s = - \cos (2 \theta)$, this can be computed in terms of the modified Bessel function.

    \item
    The Itzykson--Zuber integral for $U_{N=1}^{\beta=1}$ is given by
    \begin{align}
        \mathcal{Z}_{N=1}(-i \Lambda, Y_1) = 2 e^{i \Lambda Y_1}\ ,
    \end{align}
    where we have the volume factor, $\operatorname{Vol}(U_{N=1}^{\beta=1}) = \operatorname{Vol}(\{ \pm 1 \}) = 2$.
    Applying the recursion relation \eqref{eq:recurIZ}, we obtain
    \begin{align}
        \mathcal{Z}_{N=2}(X,Y) \propto e^{-Y_2 \Tr X} \int_{\mathbb{R}} d\Lambda \frac{e^{i (Y_1 - Y_2) \Lambda}}{\sqrt{(i\Lambda + X_1)(i\Lambda + X_2)}} .
    \end{align}
    Let us change the integration variable to
    \begin{align}
        s = \frac{2}{X_1 - X_2} \left( i \Lambda + \frac{X_1 + X_2}{2} \right)\ ,
    \end{align}
    so that the denominator of the integrand becomes
    \begin{align}
        (i\Lambda + X_1)(i\Lambda + X_2)
         = - \left( \frac{X_1 - X_2}{2} \right)^2 (1 - s^2)
    \end{align}
This leads to
    \begin{align}
        \mathcal{Z}_{N=2}(X,Y) \propto e^{- \frac{1}{2} (X_1 + X_2)(Y_1 + Y_2)} \int_\gamma ds \, \frac{e^{sz}}{\sqrt{1 - s^2}} \ ,
    \end{align}
    where the integration contour is
    \begin{align}
        s\in \gamma = \frac{X_1+X_2}{X_1-X_2} + i\mathbb R\ .
    \end{align}
    Since the integrand decreases exponentially as $s\to-\infty$, we can deform $\gamma$ until it becomes a  closed contour around $[-1,1]$.
    Then since the value of the square-root above and below $[-1,1]$ are the same up to a sign, we recover $\mathcal{Z}(X,Y)$ up to an $X,Y$-independent factor.

    \item
    The differential equation for $\hat{\mathcal{Z}}(z)$ follows from the identity $\int_{-1}^1 ds \left( e^{sz} \sqrt{1 - s^2} \right)'=0$. In the case $N=2$, the Calogero--Moser equation \eqref{eq:CalogeroMoser} is
    \begin{align}
     \left( \frac{\partial^2}{\partial X_1^2} + \frac{\partial^2}{\partial X_2^2} + \frac{1}{X_1 - X_2} \left( \frac{\partial}{\partial X_1} - \frac{\partial}{\partial X_2} \right) \right) \mathcal{Z}(X,Y) = \left( Y_1^2 + Y_2^2 \right) \mathcal{Z}(X,Y)\ .
    \end{align}
  When applied to $\mathcal{Z}(X,Y)$ \eqref{eq:zis}, the second-order derivatives yield
    \begin{align}
        \left( \frac{\partial^2}{\partial X_1^2} + \frac{\partial^2}{\partial X_2^2} \right) \mathcal{Z}(X,Y)
        & = \left( \frac{(Y_1 - Y_2)^2}{2} \hat{\mathcal{Z}}''(z) + \frac{(\Tr Y)^2}{2} \hat{\mathcal{Z}}(z) \right) e^{-\frac{1}{2} \Tr X \Tr Y}\ ,
    \end{align}
    while the first-order term yields
    \begin{align}
        \frac{1}{X_1 - X_2} \left( \frac{\partial}{\partial X_1} - \frac{\partial}{\partial X_2} \right) \mathcal{Z}(X,Y)
        = \frac{(Y_1 - Y_2)^2}{2} \frac{\hat{\mathcal{Z}}'(z)}{z} e^{-\frac{1}{2} \Tr X \Tr Y}\ .
    \end{align}
    Summing the second-order and first-order terms, and using the differential equation that is obeyed by $\hat{\mathcal{Z}}(z)$, we obtain the expected Calogero--Moser equation.
\end{enumerate}

\paragraph{Solution \ref{exodunkl}.}

\begin{enumerate}
    \item
    The method of Lagrange multipliers leads to the expression \eqref{zxylag} for the partition function $\mathcal{Z}(X,Y)$. For $M_{i,j}$, the analogous expression is
    \begin{align}
    M_{i,j}
     = \frac{1}{(2\pi)^{\dim E_N^\beta}} \int_{E_N^\beta}dS\ e^{i\operatorname{Tr} S} \prod_{k=1}^N \int_{\mathbb{R}^{\beta N}} dv_k\ e^{- v_k^\dagger (iS + X_k Y) v_k } |v_{j,i}|^{2 \delta_{k,j}} ,
    \end{align}
    where $v_{j,i}$ is the $i$-th component of the vector $v_j$.
   Using the Gaussian integral
    \begin{align}
        \int_{\mathbb{R}^{\beta N}} dv \, e^{-v^\dag A v} v_i \bar{v}_j \propto (A^{-1})_{i,j}\ ,
    \end{align}
    we obtain the formula~\eqref{eq:Mijintexo}. Then let us compute the $X_j$-derivative of $\mathcal{Z}(X,Y)$, as expressed in Eq.~\eqref{zxyie}. Using
    \begin{align}
        \frac{1}{\det A} \frac{\partial}{\partial x} \det A = \sum_{i,j} (A^{-1})_{i,j} \frac{\partial A_{j,i}}{\partial x}\ ,
        \label{eq:ex6.2.det_der}
    \end{align}
    and recalling that the matrix $Y$ is diagonal,
    we compute derivatives of the determinant factor
    \begin{align}
        \frac{\partial}{\partial X_j} \det (iS + X_j Y)
         = \det (iS + X_j Y) \sum_{i=1}^N ((iS + X_jY)^{-1})_{i,i} Y_i\ .
    \end{align}
    Hence, we obtain
    \begin{align}
        \frac{\partial}{\partial X_j} \mathcal{Z}(X, Y)
        & \propto - \frac{\beta}{2} \int_{E_N^\beta} dS\
    \frac{e^{i\operatorname{Tr} S}}{\prod_{k=1}^N\det (iS+X_k Y)^{\frac{\beta}{2}}} \sum_{i=1}^N ((iS + X_j Y)^{-1})_{i,i} Y_i\ ,
    \end{align}
    which is the last equation of \eqref{eq:IZMijrels}.

    \item
    Using again the formula~\eqref{eq:ex6.2.det_der}, we compute the integrand of the loop equation:
    \begin{multline}
        \frac{\partial}{\partial S_{j,j}} \left( \frac{e^{i\operatorname{Tr} S}}{\prod_{k=1}^N\det (iS+X_k Y)^{\frac{\beta}{2}}} \right)
     \\
        = i \left( 1 - \frac{\beta}{2} \sum_{k=1}^N ((iS + X_k Y)^{-1})_{j,j} \right) \frac{e^{i\operatorname{Tr} S}}{\prod_{k=1}^N\det (iS+X_k Y)^{\frac{\beta}{2}}}\ .
    \end{multline}
    This leads to the first relation in \eqref{eq:IZMijrels}.

    \item
    Let us introduce the notation
     \begin{align}
            \left<\!\left< \mathcal{O}(M) \right>\!\right> := \det Y^{1 - \frac{\beta}{2}} \int dM \, \frac{e^{-\Tr M Y}}{ \prod_{k=1}^N \det(X_k - M)^{\frac{\beta}{2}} } \mathcal{O}(M)\ .
        \end{align}
    After the change of variable $M = - i S Y^{-1}$, the Itzykson--Zuber integral and the quadratic moments may be written as
    \begin{align}
        \mathcal{Z}(X,Y) = \left<\!\left< 1 \right>\!\right> \quad , \quad M_{i,j}= \frac{\beta}{2Y_i}\left<\!\!\!\left< \left( \frac{1}{X_j - M} \right)_{i,i} \right>\!\!\!\right>\  ,
        \label{eq:ex6.3.Mij}
    \end{align}
    where we neglect $X,Y$-independent factors.
        The $X_j$-derivative of the quadratic moment is
        \begin{align}
        \frac{\partial M_{i,j}}{\partial X_j}
        =
        - \frac{\beta}{2} \frac{1}{Y_i} \left<\!\!\!\left< \left(\frac{\beta}{2} \frac{\Tr \left(X_j - M \right)^{-1} }{X_j - M} + \frac{1}{(X_j - M)^2} \right)_{i,i} \right>\!\!\!\right> .
        \label{eq:ex6.3.M}
        \end{align}
        On the other hand, we compute
        \begin{align}
            \sum_{k (\neq j)} \frac{M_{i,j} - M_{i,k}}{X_j - X_k}
            & =
            - \frac{\beta}{2}  \frac{1}{Y_i} \sum_{k (\neq j)}
            \left<\!\!\!\left< \left( \frac{1}{X_j - M} \frac{1}{X_k - M} \right)_{i,i} \right>\!\!\!\right>.
            \label{eq:ex6.3.MM/XX}
        \end{align}
        Applying the split rule to the expression $\mathcal{Z}(X,Y)$~\eqref{eq:ex6.3.M}, we obtain
        \begin{multline}
         \left<\!\!\!\left< \left(\frac{\beta}{2} \frac{\Tr \left(X_j - M \right)^{-1} }{X_j - M} + \frac{1}{(X_j - M)^2} \right)_{i,i} \right>\!\!\!\right>- Y_i
         \left<\!\!\!\left< \left( \frac{1}{X_j - M} \right)_{i,i}  \right>\!\!\!\right>
         \\
         + \frac{\beta}{2} \sum_{k (\neq j)} \left<\!\!\!\left< \left( \frac{1}{X_j - M} \frac{1}{X_k - M} \right)_{i,i}  \right>\!\!\!\right> = 0\ .
        \end{multline}
        This amounts to the first-order differential equation~\eqref{eq:dunkl} for $M_{i,j}$.

\end{enumerate}

%%%%%%%%%%%%%%%% List of exercises %%%%%%%%%%%%%

\hypersetup{linkcolor=black}

\renewcommand{\listtheoremname}{List of exercises}
\listoftheorems[ignoreall, show={exo}]

\hypersetup{linkcolor=blue}

\if0
%%%%%%%%%%%%%%%% List of exercises %%%%%%%%%%%%%

\hypersetup{linkcolor=black}

\renewcommand{\listtheoremname}{List of exercises}
\listoftheorems[ignoreall, show={exo}]

\hypersetup{linkcolor=blue}
\fi

%%%%%%%%%%%%%%%%%  References  %%%%%%%%%%%%%%%%%

\bibliographystyle{ytphys}
\bibliography{RM}

\providecommand{\href}[2]{#2}\begingroup\raggedright\begin{thebibliography}{10}

\bibitem{Mehta:2004RMT}
M.~L. Mehta, \href{http://dx.doi.org/10.1016/S0079-8169(04)80088-6}{{\em
  {Random Matrices}}}, vol.~142 of {\em Pure and Applied Mathematics}.
\newblock Academic Press, 3rd~ed., 2004.

\bibitem{Bleher:2001random}
P.~Bleher and A.~Its, eds., \href{http://dx.doi.org/10.1017/9781009701440}{{\em
  {Random Matrix Models and their Applications}}}.
\newblock Cambridge Univ. Press, 2001.

\bibitem{Anderson:2009RMT}
G.~W. Anderson, A.~Guionnet, and O.~Zeitouni,
  \href{http://dx.doi.org/10.1017/CBO9780511801334}{{\em {An Introduction to
  Random Matrices}}}.
\newblock Cambridge Studies in Advanced Mathematics. Cambridge Univ. Press,
  2009.

\bibitem{Forrester:2010log}
P.~J. Forrester, {\em {Log-Gases and Random Matrices}}, vol.~34 of {\em London
  Mathematical Society Monographs}.
\newblock Princeton Univ. Press, 2010.

\bibitem{Akemann:2011RMT}
G.~Akemann, J.~Baik, and P.~Di~Francesco, eds.,
  \href{http://dx.doi.org/10.1093/oxfordhb/9780198744191.001.0001}{{\em {The
  Oxford Handbook of Random Matrix Theory}}}.
\newblock Oxford Handbooks in Mathematics. Oxford Univ. Press, 2011.

\bibitem{Harnad:2011RM}
J.~Harnad, ed., \href{http://dx.doi.org/10.1007/978-1-4419-9514-8}{{\em {Random
  Matrices, Random Processes and Integrable Systems}}}.
\newblock CRM Series in Mathematical Physics. Springer New York, 2011.

\bibitem{Kazakov:1996}
V.~A. Kazakov, M.~Staudacher, and T.~Wynter, ``{Character expansion methods for
  matrix models of dually weighted graphs},''
  \href{http://dx.doi.org/10.1007/BF02101902}{{\em Commun. Math. Phys.}
  {\bfseries 177} (1996) 451--468},
\href{http://arxiv.org/abs/hep-th/9502132}{{\ttfamily arXiv:hep-th/9502132
  [hep-th]}}.
%%CITATION = HEP-TH/9502132;%%.

\bibitem{Efetov:1996SUSY}
K.~Efetov, \href{http://dx.doi.org/10.1017/CBO9780511573057}{{\em
  {Supersymmetry in Disorder and Chaos}}}.
\newblock Cambridge Univ. Press, 1996.

\bibitem{Tao:2010CMP}
T.~Tao and V.~Vu, ``{Random Matrices: Universality of Local Eigenvalue
  Statistics up to the Edge},''
  \href{http://dx.doi.org/10.1007/s00220-010-1044-5}{{\em Commun. Math. Phys.}
  {\bfseries 298} (2010) 549--572},
  \href{http://arxiv.org/abs/0908.1982}{{\ttfamily arXiv:0908.1982 [math.PR]}}.

\bibitem{Brezin:2016eax}
E.~Br\'ezin and S.~Hikami,
  \href{http://dx.doi.org/10.1007/978-981-10-3316-2}{{\em {Random Matrix Theory
  with an External Source}}}, vol.~19 of {\em SpringerBriefs in Mathematical
  Physics}.
\newblock Springer Singapore,
2016.
\newblock
%%CITATION = INSPIRE-1515259;%%.

\bibitem{Wishart:1928Bio}
J.~Wishart, ``{The generalised product moment distribution in samples from a
  normal multivariate population},''
  \href{http://dx.doi.org/10.1093/biomet/20A.1-2.32}{{\em Biometrika}
  {\bfseries 20A} (1928) 32--52}.

\bibitem{Ginibre:1965zz}
J.~Ginibre, ``{Statistical Ensembles of Complex, Quaternion and Real
  Matrices},''
\href{http://dx.doi.org/10.1063/1.1704292}{{\em J. Math. Phys.} {\bfseries 6}
  (1965) 440--449}.
%%CITATION = JMAPA,6,440;%%.

\bibitem{Marchenko:1967MS}
V.~A. Marchenko and L.~A. Pastur, ``{Distribution of eigenvalues for some sets
  of random matrices},''
  \href{http://dx.doi.org/10.1070/SM1967v001n04ABEH001994}{{\em Mat. Sb.}
  {\bfseries 72} (1967) 507--536}.

\bibitem{Berry:1977PRS}
M.~V. Berry and M.~Tabor, ``{Level Clustering in the Regular Spectrum},''
  \href{http://dx.doi.org/10.1098/rspa.1977.0140}{{\em Proc. Roy. Soc.}
  {\bfseries A356} (1977) 375--394}.

\bibitem{Bohigas:1984PRL}
O.~Bohigas, M.~J. Giannoni, and C.~Schmit, ``{Characterization of Chaotic
  Quantum Spectra and Universality of Level Fluctuation Laws},''
  \href{http://dx.doi.org/10.1103/PhysRevLett.52.1}{{\em Phys. Rev. Lett.}
  {\bfseries 52} (1984) 1--4}.

\bibitem{'tHooft:1973jz}
G.~'t~Hooft, ``{A Planar Diagram Theory for Strong Interactions},''
\href{http://dx.doi.org/10.1016/0550-3213(74)90154-0}{{\em Nucl. Phys.}
  {\bfseries B72} (1974) 461--473}.
%%CITATION = NUPHA,B72,461;%%.

\bibitem{Brezin:1977sv}
E.~Br\'ezin, C.~Itzykson, G.~Parisi, and J.-B. Zuber, ``{Planar Diagrams},''
\href{http://dx.doi.org/10.1007/BF01614153}{{\em Commun. Math. Phys.}
  {\bfseries 59} (1978) 35--51}.
%%CITATION = CMPHA,59,35;%%.

\bibitem{Verbaarschot:2000dy}
J.~J.~M. Verbaarschot and T.~Wettig, ``{Random matrix theory and chiral
  symmetry in QCD},''
  \href{http://dx.doi.org/10.1146/annurev.nucl.50.1.343}{{\em Ann. Rev. Nucl.
  Part. Sci.} {\bfseries 50} (2000) 343--410},
\href{http://arxiv.org/abs/hep-ph/0003017}{{\ttfamily arXiv:hep-ph/0003017
  [hep-ph]}}.
%%CITATION = HEP-PH/0003017;%%.

\bibitem{Dijkgraaf:2002fc}
R.~Dijkgraaf and C.~Vafa, ``{Matrix models, topological strings, and
  supersymmetric gauge theories},''
  \href{http://dx.doi.org/10.1016/S0550-3213(02)00766-6}{{\em Nucl. Phys.}
  {\bfseries B644} (2002) 3--20},
\href{http://arxiv.org/abs/hep-th/0206255}{{\ttfamily arXiv:hep-th/0206255}}.
%%CITATION = HEP-TH/0206255;%%.

\bibitem{Kontsevich:1992ti}
M.~Kontsevich, ``{Intersection theory on the moduli space of curves and the
  matrix Airy function},''
\href{http://dx.doi.org/10.1007/BF02099526}{{\em Commun. Math. Phys.}
  {\bfseries 147} (1992) 1--23}.
%%CITATION = CMPHA,147,1;%%.

\bibitem{Bouchard:2007ys}
V.~Bouchard, A.~Klemm, M.~Mari\~no, and S.~Pasquetti, ``{Remodeling the
  B-model},'' \href{http://dx.doi.org/10.1007/s00220-008-0620-4}{{\em Commun.
  Math. Phys.} {\bfseries 287} (2009) 117--178},
\href{http://arxiv.org/abs/0709.1453}{{\ttfamily arXiv:0709.1453 [hep-th]}}.
%%CITATION = ARXIV:0709.1453;%%.

\bibitem{Eynard:2012nj}
B.~Eynard and N.~Orantin, ``{Computation of open Gromov--Witten invariants for
  toric Calabi--Yau 3-folds by topological recursion, a proof of the BKMP
  conjecture},'' \href{http://dx.doi.org/10.1007/s00220-015-2361-5}{{\em
  Commun. Math. Phys.} {\bfseries 337} (2015) 483--567},
\href{http://arxiv.org/abs/1205.1103}{{\ttfamily arXiv:1205.1103 [math-ph]}}.
%%CITATION = ARXIV:1205.1103;%%.

\bibitem{Brini:2011}
A.~Brini, B.~Eynard, and M.~Mari\~{n}o, ``{Torus knots and mirror symmetry},''
  \href{http://dx.doi.org/10.1007/s00023-012-0171-2}{{\em Ann. Henri
  Poincar\'e} {\bfseries 13} (2012) 1873--1910},
\href{http://arxiv.org/abs/1105.2012}{{\ttfamily arXiv:1105.2012 [hep-th]}}.
%%CITATION = ARXIV:1105.2012;%%.

\bibitem{Penner:1988JDG}
R.~C. Penner, ``{Perturbative series and the moduli space of Riemann
  surfaces},'' \href{http://projecteuclid.org/euclid.jdg/1214441648}{{\em J.
  Differential Geom.} {\bfseries 27} (1988) 35--53}.

\bibitem{Dotsenko:1984nm}
V.~S. Dotsenko and V.~A. Fateev, ``{Conformal algebra and multipoint
  correlation functions in 2D statistical models},''
\href{http://dx.doi.org/10.1016/0550-3213(84)90269-4}{{\em Nucl. Phys.}
  {\bfseries B240} (1984) 312--348}.
%%CITATION = NUPHA,B240,312;%%.

\bibitem{Dotsenko:1984ad}
V.~S. Dotsenko and V.~A. Fateev, ``{Four-point correlation functions and the
  operator algebra in 2D conformal invariant theories with central charge $c
  \le 1$},''
\href{http://dx.doi.org/10.1016/S0550-3213(85)80004-3}{{\em Nucl. Phys.}
  {\bfseries B251} (1985) 691--734}.
%%CITATION = NUPHA,B251,691;%%.

\bibitem{Eynard:2009nd}
B.~Eynard, ``{A Matrix model for plane partitions and (T)ASEP},''
  \href{http://dx.doi.org/10.1088/1742-5468/2009/10/P10011}{{\em J. Stat.
  Mech.} {\bfseries 0910} (2009) P10011},
\href{http://arxiv.org/abs/0905.0535}{{\ttfamily arXiv:0905.0535 [math-ph]}}.
%%CITATION = ARXIV:0905.0535;%%.

\bibitem{Logan:1977AM}
B.~F. Logan and L.~A. Shepp, ``{A variational problem for random Young
  tableaux},'' \href{http://dx.doi.org/10.1016/0001-8708(77)90030-5}{{\em Adv.
  Math.} {\bfseries 26} (1977) 206--222}.

\bibitem{Vershik:1977SMD}
A.~Vershik and S.~Kerov, ``{Asymptotics of the Placherel measure of the
  symmetric group and the limit form of Young tableaux},'' {\em Soviet Math.
  Dokl.} {\bfseries 18} (1977) 527--531.

\bibitem{MineevWeinstein:2000ie}
M.~Mineev-Weinstein, P.~B. Wiegmann, and A.~Zabrodin, ``{Integrable Structure
  of Interface Dynamics},''
  \href{http://dx.doi.org/10.1103/PhysRevLett.84.5106}{{\em Phys. Rev. Lett.}
  {\bfseries 84} (2000) 5106--5109},
\href{http://arxiv.org/abs/nlin/0001007}{{\ttfamily arXiv:nlin/0001007
  [nlin-si]}}.
%%CITATION = NLIN/0001007;%%.

\bibitem{Kostov:2000ed}
I.~Kostov, I.~Krichever, M.~Mineev-Weinstein, P.~Wiegmann, and A.~Zabrodin,
  ``{$\tau$-function for analytic curves},'' in {\em {Random Matrix Models and
  their Applications}}, vol.~40 of {\em Math. Sci. Res. Inst. Publ.},
  pp.~285--299.
\newblock Cambridge Univ. Press, 2001.
\newblock
\href{http://arxiv.org/abs/hep-th/0005259}{{\ttfamily arXiv:hep-th/0005259
  [hep-th]}}.
\newblock
%%CITATION = HEP-TH/0005259;%%.

\bibitem{Montgomery:1973ANT}
H.~L. Montgomery, ``{The pair correlation of zeros of the zeta function},''
  {\em Analytic Number Theory, Proc. Sympos. Pure Math.} {\bfseries XXIV}
  (1973) 181--193.

\bibitem{Odlyzko:1987MC}
A.~M. Odlyzko, ``{On the distribution of spacings between zeros of the zeta
  function},'' \href{http://dx.doi.org/10.1090/S0025-5718-1987-0866115-0}{{\em
  Math. Comp.} {\bfseries 48} (1987) 273--308}.

\bibitem{Orland:2001ba}
H.~Orland and A.~Zee, ``{RNA Folding and Large $N$ Matrix Theory},''
  \href{http://dx.doi.org/10.1016/S0550-3213(01)00522-3}{{\em Nucl. Phys.}
  {\bfseries B620} (2002) 456--476},
\href{http://arxiv.org/abs/cond-mat/0106359}{{\ttfamily arXiv:cond-mat/0106359
  [cond-mat]}}.
%%CITATION = COND-MAT/0106359;%%.

\bibitem{DUMITRIU2002}
I.~{Dumitriu} and A.~{Edelman}, ``{Matrix models for beta ensembles},''
  \href{http://dx.doi.org/10.1063/1.1507823}{{\em Journal of Mathematical
  Physics} {\bfseries 43} (Nov., 2002) 5830--5847},
  \href{http://arxiv.org/abs/math-ph/0206043}{{\ttfamily math-ph/0206043}}.

\bibitem{DUMITRIU2007587}
I.~Dumitriu, A.~Edelman, and G.~Shuman, ``Mops: Multivariate orthogonal
  polynomials (symbolically),''
  \href{http://www.sciencedirect.com/science/article/pii/S0747717107000090}{{\em
  Journal of Symbolic Computation} {\bfseries 42} no.~6, (2007) 587 -- 620}.

\bibitem{Nekrasov:2002qd}
N.~Nekrasov, ``{Seiberg--Witten Prepotential from Instanton Counting},''
  \href{http://dx.doi.org/10.4310/ATMP.2003.v7.n5.a4}{{\em Adv. Theor. Math.
  Phys.} {\bfseries 7} (2004) 831--864},
\href{http://arxiv.org/abs/hep-th/0206161}{{\ttfamily arXiv:hep-th/0206161}}.
%%CITATION = HEP-TH/0206161;%%.

\bibitem{Alday:2009aq}
L.~F. Alday, D.~Gaiotto, and Y.~Tachikawa, ``{Liouville Correlation Functions
  from Four-dimensional Gauge Theories},''
  \href{http://dx.doi.org/10.1007/s11005-010-0369-5}{{\em Lett. Math. Phys.}
  {\bfseries 91} (2010) 167--197},
\href{http://arxiv.org/abs/0906.3219}{{\ttfamily arXiv:0906.3219 [hep-th]}}.
%%CITATION = 0906.3219;%%.

\bibitem{Alvarez-Gaume:1991ozb}
L.~Alvarez-Gaume and J.~L. Manes, ``{Supermatrix models},''
  \href{http://dx.doi.org/10.1142/S0217732391002219}{{\em Mod. Phys. Lett. A}
  {\bfseries 6} (1991) 2039--2050}.

\bibitem{Yost:1991ht}
S.~A. Yost, ``{Supermatrix models},''
  \href{http://dx.doi.org/10.1142/S0217751X92002775}{{\em Int. J. Mod. Phys. A}
  {\bfseries 7} (1992) 6105--6120},
  \href{http://arxiv.org/abs/hep-th/9111033}{{\ttfamily arXiv:hep-th/9111033}}.

\bibitem{Kimura:2023iup}
T.~Kimura, ``{Aspects of supergroup gauge theory},''
  \href{http://dx.doi.org/10.1142/S0217751X23300016}{{\em Int. J. Mod. Phys. A}
  {\bfseries 38} no.~03, (2023) 2330001},
  \href{http://arxiv.org/abs/2301.05927}{{\ttfamily arXiv:2301.05927
  [hep-th]}}.

\bibitem{Caselle:2003qa}
M.~Caselle and U.~Magnea, ``{Random matrix theory and symmetric spaces},''
  \href{http://dx.doi.org/10.1016/j.physrep.2003.12.004}{{\em Phys. Rept.}
  {\bfseries 394} (2004) 41--156},
\href{http://arxiv.org/abs/cond-mat/0304363}{{\ttfamily arXiv:cond-mat/0304363
  [cond-mat]}}.
%%CITATION = COND-MAT/0304363;%%.

\bibitem{Zirnbauer:2011RMT}
M.~R. Zirnbauer, {\em {The Oxford Handbook of Random Matrix Theory}},
  ch.~{Symmetry Classes}, pp.~43--65.
\newblock Oxford Univ. Press, 2011.
\newblock \href{http://arxiv.org/abs/1001.0722}{{\ttfamily arXiv:1001.0722
  [math-ph]}}.

\bibitem{Bernard:2002SFT}
D.~Bernard and A.~LeClair,
  \href{http://dx.doi.org/10.1007/978-94-010-0514-2_19}{``{A Classification of
  Non-Hermitian Random Matrices},''} in {\em Statistical Field Theories},
  vol.~73 of {\em NATO Science Series}, pp.~207--214.
\newblock 2002.
\newblock \href{http://arxiv.org/abs/cond-mat/0110649}{{\ttfamily
  arXiv:cond-mat/0110649 [cond-mat.dis-nn]}}.

\bibitem{Eynard:2006:JSM}
B.~Eynard, ``{Universal distribution of random matrix eigenvalues near the
  `birth of a cut' transition},''
  \href{http://dx.doi.org/10.1088/1742-5468/2006/07/P07005}{{\em J. Stas.
  Mech.} {\bfseries 2006} (2006) P07005},
  \href{http://arxiv.org/abs/math-ph/0605064}{{\ttfamily arXiv:math-ph/0605064
  [math-ph]}}.

\bibitem{Bertola:2009CA}
M.~Bertola and S.~Y. Lee, ``{First Colonization of a Spectral Outpost in Random
  Matrix Theory},'' \href{http://dx.doi.org/10.1007/s00365-008-9026-y}{{\em
  Const. Approx.} {\bfseries 30} (2009) 225--263},
  \href{http://arxiv.org/abs/0711.3625}{{\ttfamily arXiv:0711.3625 [math-ph]}}.

\bibitem{EynBook}
B.~Eynard, \href{http://dx.doi.org/10.1007/978-3-7643-8797-6}{{\em {Counting
  Surfaces}}}, vol.~70 of {\em Progress in Mathematical Physics}.
\newblock Springer, 2016.

\bibitem{Eynard:2006bs}
B.~Eynard, {\em {Random Matrices, Random Processes and Integrable Systems}},
  \href{http://dx.doi.org/10.1007/978-1-4419-9514-8_6}{ch.~{Formal Matrix
  Integrals and Combinatorics of Maps}, pp.~415--442}.
\newblock CRM Series in Mathematical Physics.
\newblock Springer New York, 2011.
\newblock
\href{http://arxiv.org/abs/math-ph/0611087}{{\ttfamily arXiv:math-ph/0611087
  [math-ph]}}.
\newblock
%%CITATION = MATH-PH/0611087;%%.

\bibitem{Bershadsky:1986ws}
M.~A. Bershadsky and A.~A. Migdal, ``{Ising Model of a Randomly Triangulated
  Random Surface as a Definition of Fermionic String Theory},''
\href{http://dx.doi.org/10.1016/0370-2693(86)91023-3}{{\em Phys. Lett.}
  {\bfseries B174} (1986) 393--398}.
%%CITATION = PHLTA,B174,393;%%.

\bibitem{Kazakov:1986hu}
V.~A. Kazakov, ``{Ising model on a dynamical planar random lattice: Exact
  solution},''
\href{http://dx.doi.org/10.1016/0375-9601(86)90433-0}{{\em Phys. Lett.}
  {\bfseries A119} (1986) 140--144}.
%%CITATION = PHLTA,A119,140;%%.

\bibitem{Kostov:1988fy}
I.~K. Kostov, ``{O($n$) Vector Model on a Planar Random Lattice: Spectrum of
  Anomalous Dimensions},''
\href{http://dx.doi.org/10.1142/S0217732389000289}{{\em Mod. Phys. Lett.}
  {\bfseries A4} (1989) 217--226}.
%%CITATION = MPLAE,A4,217;%%.

\bibitem{Gaudin:1989vx}
M.~Gaudin and I.~Kostov, ``{O($N$) Model on a Fluctuating Planar Lattice: Some
  Exact Results},''
\href{http://dx.doi.org/10.1016/0370-2693(89)90037-3}{{\em Phys. Lett.}
  {\bfseries B220} (1989) 200--206}.
%%CITATION = PHLTA,B220,200;%%.

\bibitem{Eynard:1995nv}
B.~Eynard and C.~Kristjansen, ``{Exact solution of the $O(n)$ model on a random
  lattice},'' \href{http://dx.doi.org/10.1016/0550-3213(95)00469-9}{{\em Nucl.
  Phys.} {\bfseries B455} (1995) 577--618},
\href{http://arxiv.org/abs/hep-th/9506193}{{\ttfamily arXiv:hep-th/9506193
  [hep-th]}}.
%%CITATION = HEP-TH/9506193;%%.

\bibitem{Eynard:1998he}
B.~Eynard and C.~Kristjansen, ``{An iterative solution of the three-colour
  problem on a random lattice},''
  \href{http://dx.doi.org/10.1016/S0550-3213(98)00042-X}{{\em Nucl. Phys.}
  {\bfseries B516} (1998) 529--542},
\href{http://arxiv.org/abs/cond-mat/9710199}{{\ttfamily arXiv:cond-mat/9710199
  [cond-mat]}}.
%%CITATION = COND-MAT/9710199;%%.

\bibitem{Kostov:2000vn}
I.~K. Kostov, ``{Exact solution of the three color problem on a random
  lattice},'' \href{http://dx.doi.org/10.1016/S0370-2693(02)02887-3}{{\em Phys.
  Lett.} {\bfseries B549} (2002) 245--252},
\href{http://arxiv.org/abs/hep-th/0005190}{{\ttfamily arXiv:hep-th/0005190
  [hep-th]}}.
%%CITATION = HEP-TH/0005190;%%.

\bibitem{Ginsparg:1991bi}
P.~H. Ginsparg, \href{http://dx.doi.org/10.1142/9789814537445}{``{Matrix models
  of 2d gravity},''} in {\em {High Energy Physics and Cosmology}}, vol.~8 of
  {\em The ICTP Series in Theoretical Physics}, pp.~785--826.
\newblock 1991.
\newblock
\href{http://arxiv.org/abs/hep-th/9112013}{{\ttfamily arXiv:hep-th/9112013
  [hep-th]}}.
\newblock
%%CITATION = HEP-TH/9112013;%%.

\bibitem{Kostov:1999qx}
I.~K. Kostov, ``{Exact solution of the six vertex model on a random lattice},''
  \href{http://dx.doi.org/10.1016/S0550-3213(00)00060-2}{{\em Nucl. Phys.}
  {\bfseries B575} (2000) 513--534},
\href{http://arxiv.org/abs/hep-th/9911023}{{\ttfamily arXiv:hep-th/9911023
  [hep-th]}}.
%%CITATION = HEP-TH/9911023;%%.

\bibitem{Bertola:2002}
M.~Bertola, B.~Eynard, and J.~Harnad, ``{Differential systems for biorthogonal
  polynomials appearing in 2-matrix models and the associated Riemann-Hilbert
  problem},'' \href{http://dx.doi.org/10.1007/s00220-003-0934-1}{{\em Commun.
  Math. Phys.} {\bfseries 243} (2003) 193--240},
\href{http://arxiv.org/abs/nlin/0208002}{{\ttfamily arXiv:nlin/0208002
  [nlin.SI]}}.
%%CITATION = NLIN/0208002;%%.

\bibitem{Fay:1973}
J.~D. Fay, \href{http://dx.doi.org/10.1007/BFb0060090}{{\em {Theta Functions on
  Riemann Surfaces}}}, vol.~352 of {\em Lecture Notes in Mathematics}.
\newblock Springer Berlin Heidelberg, 1973.

\bibitem{Eynard:2018gor}
B.~Eynard, ``{Combinatorial expression of the fundamental second kind
  differential on an algebraic curve},''
  \href{http://dx.doi.org/10.4171/aihpd/116}{{\em Ann. Inst. H. Poincare D
  Comb. Phys. Interact.} {\bfseries 9} no.~2, (2022) 219--238},
  \href{http://arxiv.org/abs/1805.07247}{{\ttfamily arXiv:1805.07247
  [math-ph]}}.

\bibitem{Borot:2014kda}
G.~Borot, B.~Eynard, and A.~Weisse, ``{Root systems, spectral curves, and
  analysis of a Chern--Simons matrix model for Seifert fibered spaces},''
  \href{http://dx.doi.org/10.1007/s00029-016-0266-6}{{\em Sel. Math. New Ser.}
  (2016) 915--1025},
\href{http://arxiv.org/abs/1407.4500}{{\ttfamily arXiv:1407.4500 [math-ph]}}.
%%CITATION = ARXIV:1407.4500;%%.

\bibitem{Nekrasov:2012xe}
N.~Nekrasov and V.~Pestun, ``{Seiberg-Witten Geometry of Four-Dimensional
  $\mathcal N=2$ Quiver Gauge Theories},''
  \href{http://dx.doi.org/10.3842/SIGMA.2023.047}{{\em SIGMA} {\bfseries 19}
  (2023) 047}, \href{http://arxiv.org/abs/1211.2240}{{\ttfamily arXiv:1211.2240
  [hep-th]}}.

\bibitem{Eynard:2009zz}
B.~Eynard and A.~Prats~Ferrer, ``{Topological expansion of the chain of
  matrices},'' \href{http://dx.doi.org/10.1088/1126-6708/2009/07/096}{{\em
  JHEP} {\bfseries 0907} (2009) 096},
\href{http://arxiv.org/abs/0805.1368}{{\ttfamily arXiv:0805.1368 [math-ph]}}.
%%CITATION = JHEPA,0907,096;%%.

\bibitem{Migdal:1984gj}
A.~A. Migdal, ``{Loop Equations and $1/N$ Expansion},''
\href{http://dx.doi.org/10.1016/0370-1573(83)90076-5}{{\em Phys. Rept.}
  {\bfseries 102} (1983) 199--290}.
%%CITATION = PRPLC,102,199;%%.

\bibitem{Eynard:2007kz}
B.~Eynard and N.~Orantin, ``{Invariants of algebraic curves and topological
  expansion},'' \href{http://dx.doi.org/10.4310/CNTP.2007.v1.n2.a4}{{\em
  Commun. Num. Theor. Phys.} {\bfseries 1} (2007) 347--452},
\href{http://arxiv.org/abs/math-ph/0702045}{{\ttfamily arXiv:math-ph/0702045
  [math-ph]}}.
%%CITATION = MATH-PH/0702045;%%.

\bibitem{Chekhov:2010zg}
L.~O. Chekhov, B.~Eynard, and O.~Marchal, ``{Topological expansion of
  $\beta$-ensemble model and quantum algebraic geometry in the sectorwise
  approach},'' \href{http://dx.doi.org/10.1007/s11232-011-0012-3}{{\em Theor.
  Math. Phys.} {\bfseries 166} (2011) 141--185},
\href{http://arxiv.org/abs/1009.6007}{{\ttfamily arXiv:1009.6007 [math-ph]}}.
%%CITATION = ARXIV:1009.6007;%%.

\bibitem{Bouchard:2012yg}
V.~Bouchard and B.~Eynard, ``{Think globally, compute locally},''
  \href{http://dx.doi.org/10.1007/JHEP02(2013)143}{{\em JHEP} {\bfseries 02}
  (2013) 143},
\href{http://arxiv.org/abs/1211.2302}{{\ttfamily arXiv:1211.2302 [math-ph]}}.
%%CITATION = ARXIV:1211.2302;%%.

\bibitem{Borot:2013}
G.~Borot and A.~Guionnet, ``Asymptotic expansion of beta matrix models in the
  multi-cut regime,'' \href{http://dx.doi.org/10.1017/fms.2023.129}{{\em Forum
  Math. Sigma} {\bfseries 12} no.~e13, (2024) 1--93},
  \href{http://arxiv.org/abs/1303.1045}{{\ttfamily arXiv:1303.1045 [math-ph]}}.

\bibitem{Bergere:2006}
M.~Berg\`ere and B.~Eynard, ``{Mixed correlation function and spectral curve
  for the 2-matrix model},''
  \href{http://dx.doi.org/10.1088/0305-4470/39/49/004}{{\em J. Phys.}
  {\bfseries A39} (2006) 15091--15134},
  \href{http://arxiv.org/abs/math-ph/0605010}{{\ttfamily
  arXiv:math-ph/0605010}}.

\bibitem{Deift:1993steepest}
P.~Deift and X.~Zhou, ``{A Steepest Descent Method for Oscillatory
  Riemann--Hilbert Problems. Asymptotics for the MKdV Equation},''
  \href{http://dx.doi.org/10.2307/2946540}{{\em Annal. Math.} {\bfseries 137}
  (1993) 295--368}.

\bibitem{Deift:2000orthogonal}
P.~Deift, \href{http://dx.doi.org/10.1090/cln/003}{{\em {Orthogonal polynomials
  and random matrices: a Riemann--Hilbert approach}}}, vol.~3 of {\em Courant
  Lecture Notes}.
\newblock American Mathematical Soc., 2000.

\bibitem{Jimbo:1983if}
M.~Jimbo and T.~Miwa, ``{Solitons and Infinite Dimensional Lie Algebras},''
\href{http://dx.doi.org/10.2977/prims/1195182017}{{\em Publ. Res. Inst. Math.
  Sci. Kyoto} {\bfseries 19} (1983) 943--1001}.
%%CITATION = PBMAA,19,943;%%.

\bibitem{Ueno:1984zz}
K.~Ueno and K.~Takasaki, ``{Toda lattice hierarchy},''
\href{http://dx.doi.org/10.2969/aspm/00410001}{{\em Adv. Stud. Pure Math.}
  {\bfseries 4} (1984) 1--95}.
%%CITATION = ASPME,4,1;%%.

\bibitem{Babelon:2003CIS}
O.~Babelon, D.~Bernard, and M.~Talon,
  \href{http://dx.doi.org/10.1017/CBO9780511535024}{{\em {Introduction to
  Classical Integrable Systems}}}.
\newblock Cambridge Monographs on Mathematical Physics. Cambridge Univ. Press,
  2003.

\bibitem{Eynard:1998JPA}
B.~Eynard and M.~L. Mehta, ``{Matrices coupled in a chain: I. Eigenvalue
  correlations},'' \href{http://dx.doi.org/10.1088/0305-4470/31/19/010}{{\em J.
  Phys.} {\bfseries A31} (1998) 4449--4456},
  \href{http://arxiv.org/abs/cond-mat/9710230}{{\ttfamily
  arXiv:cond-mat/9710230 [cond-mat]}}.

\bibitem{Okuda:2004mb}
T.~Okuda, ``{Derivation of Calabi-Yau crystals from Chern-Simons gauge
  theory},'' \href{http://dx.doi.org/10.1088/1126-6708/2005/03/047}{{\em JHEP}
  {\bfseries 03} (2005) 047},
  \href{http://arxiv.org/abs/hep-th/0409270}{{\ttfamily arXiv:hep-th/0409270}}.

\bibitem{Itzykson:1979fi}
C.~Itzykson and J.-B. Zuber, ``{The Planar Approximation. II},''
\href{http://dx.doi.org/10.1063/1.524438}{{\em J. Math. Phys.} {\bfseries 21}
  (1980) 411--421}.
%%CITATION = JMAPA,21,411;%%.

\bibitem{HarishChandra:1957AJM}
Harish-Chandra, ``{Differential Operators on a Semisimple Lie Algebra},''
  \href{http://dx.doi.org/10.2307/2372387}{{\em Amer. J. Math.} {\bfseries 79}
  (1957) 87--120}.

\bibitem{Duistermaat:1982vw}
J.~J. Duistermaat and G.~J. Heckman, ``{On the Variation in the cohomology of
  the symplectic form of the reduced phase space},''
\href{http://dx.doi.org/10.1007/BF01399506}{{\em Invent. Math.} {\bfseries 69}
  (1982) 259--268}.
%%CITATION = INVMB,69,259;%%.

\bibitem{Bertola:2008}
M.~Bertola and A.~Prats~Ferrer, ``{Harish-Chandra integrals as nilpotent
  integrals},'' \href{http://dx.doi.org/10.1093/imrn/rnn062}{{\em Int. Math.
  Res. Not.} (2008) rnn062}, \href{http://arxiv.org/abs/0801.3452}{{\ttfamily
  arXiv:0801.3452 [math.GR]}}.

\bibitem{Morozov:1992zb}
A.~Morozov, ``{Pair correlator in the Itzykson--Zuber integral},''
  \href{http://dx.doi.org/10.1142/S0217732392002913}{{\em Mod. Phys. Lett.}
  {\bfseries A7} (1992) 3503--3508},
\href{http://arxiv.org/abs/hep-th/9209074}{{\ttfamily arXiv:hep-th/9209074
  [hep-th]}}.
%%CITATION = HEP-TH/9209074;%%.

\bibitem{Eynard:2005wg}
B.~Eynard and A.~Prats~Ferrer, ``{2-matrix versus complex matrix model,
  integrals over the unitary group as triangular integrals},''
  \href{http://dx.doi.org/10.1007/s00220-006-1541-8}{{\em Commun. Math. Phys.}
  {\bfseries 264} (2006) 115--144},
\href{http://arxiv.org/abs/hep-th/0502041}{{\ttfamily arXiv:hep-th/0502041
  [hep-th]}}.
%%CITATION = HEP-TH/0502041;%%.

\bibitem{Bergere:2008da}
M.~Berg\`ere and B.~Eynard, ``{Some properties of angular integrals},''
  \href{http://dx.doi.org/10.1088/1751-8113/42/26/265201}{{\em J. Phys.}
  {\bfseries A42} (2009) 265201},
\href{http://arxiv.org/abs/0805.4482}{{\ttfamily arXiv:0805.4482 [math-ph]}}.
%%CITATION = ARXIV:0805.4482;%%.

\bibitem{Macdonald:1997}
I.~G. Macdonald,
  \href{http://dx.doi.org/10.1093/oso/9780198534891.001.0001}{{\em {Symmetric
  Functions and Hall Polynomials}}}.
\newblock Oxford University Press, 2nd~ed., 1997.

\bibitem{Desrosiers:2008tp}
P.~Desrosiers, ``{Duality in random matrix ensembles for all $\beta$},''
  \href{http://dx.doi.org/10.1016/j.nuclphysb.2009.02.019}{{\em Nucl. Phys.}
  {\bfseries B817} (2009) 224--251},
\href{http://arxiv.org/abs/0801.3438}{{\ttfamily arXiv:0801.3438 [math-ph]}}.
%%CITATION = ARXIV:0801.3438;%%.

\end{thebibliography}\endgroup

%%%%%%%%%%%%%%%%%  Index  %%%%%%%%%%%%%%%%%

\printindex

\end{document}